\documentclass[amsmath,amssymb,onecolumn,superscriptaddress]{revtex4}

\usepackage{graphicx}
\usepackage{dcolumn}

\def\be{\begin{equation}}
\def\ee{\end{equation}}
\def\beq{\begin{eqnarray}}
\def\eeq{\end{eqnarray}}

\usepackage{bm}
\usepackage{graphicx}
\usepackage{dcolumn}
\usepackage{amsmath}
\usepackage[latin1]{inputenc}
\usepackage{graphicx, psfrag}
\usepackage{amssymb}
\usepackage[colorlinks=true, citecolor=blue, urlcolor = blue, linkcolor= red, bookmarks=true]{hyperref}
\usepackage{float}
\usepackage{amsmath}
\usepackage{tensor}
\usepackage{amsfonts}
\usepackage{dcolumn}
\usepackage{hyperref}
\usepackage{subfigure}
\usepackage{pgfplots}
\usepackage{epstopdf}
\usepackage{booktabs}
\usepackage{pdfpages}

\begin{document}
\title{Testing the Metric-Affine Gravity Using Particle Dynamics and Photon Motion}

\author{Allah Ditta}
\email{mradshahid01@gmail.com}
\affiliation{Department of Mathematics, Shanghai University,
Shanghai, 200444, Shanghai, People's Republic of China.}

\author{Xia Tiecheng}
\email{xiatc@shu.edu.cn (Corresponding Authors)}
\affiliation{Department of Mathematics, Shanghai University,
Shanghai, 200444, Shanghai, People's Republic of China.}

\author{Saadia Mumtaz}
\email{saadia.icet@pu.edu.pk}
\affiliation{Institute of Chemical
Engineering and Technology, University of the Punjab, Quaid-e-Azam
Campus, Lahore-54590, Pakistan}

\author{Farruh~Atamurotov}
\email{atamurotov@yahoo.com}
\affiliation{Inha University in Tashkent, Ziyolilar 9, Tashkent 100170, Uzbekistan}
\affiliation{Akfa University, Milliy Bog' Street 264, Tashkent 111221, Uzbekistan}
\affiliation{National University of Uzbekistan, Tashkent 100174, Uzbekistan}

\author{G. Mustafa}
\email{gmustafa3828@gmail.com(Corresponding Authors)}\affiliation{Department of Physics, Zhejiang Normal University, Jinhua 321004, People's Republic of China}

\author{Ahmadjon~Abdujabbarov}
\email{ahmadjon@astrin.uz}

\affiliation{Ulugh Beg Astronomical Institute, Astronomy St 33, Tashkent 100052, Uzbekistan}
\affiliation{National University of Uzbekistan, Tashkent 100174, Uzbekistan}
\affiliation{Institute of Nuclear Physics, Tashkent 100214, Uzbekistan}
\affiliation{Tashkent State Technical University, Tashkent 100095, Uzbekistan}

\begin{abstract}
This work mainly focuses to unveil the optical features of a black hole. For this objective, we utilize the metric-affine black hole geometry with the inclusion of dilation, spin, and shear charge. The Lagrangian coefficients $f_1$ and $d_1$ are the main parameters, where $f_1<0$, which differentiate the solutions by $d_1=8f_1,\;d_1=-8f_1,\;\&\;d_1=\pm8f_1$. Based on these parameters, we carry out this work in two cases, i.e., $d_1=8f_1,\;\&\;d_1=-8f_1$. We forecast the detailed impact of dilation, spin, and shear charges on the optical properties of black hole in both cases. To unreveal the optical features, we calculate horizon radius, inner stable circular orbit, photon sphere radius, BH shadows, quasi-periodic oscillations, the red-blue shift of photon particles, effective force, weak gravitational lensing, and image magnification by using metric-affine gravity black hole geometry.
\end{abstract}

\date{\today}

\maketitle

\section{introduction}

The observations of the shadow of the supermasive black hole (SMBH) M87 and Sgr A* by the Event Horizon Telescope (EHT) collaboration~\cite{1,2,3,4,5,6,7} have led scientists to probe the modified theories of gravity using optical features of spacetime. On the other hand the gravitational field of galaxy clusters and BHs has the capability to lens the nearby passing light emitted from the distant source. The process of gravitational lensing originates from the bending and magnification of photons in the domain of gravitational perturbations of dark matter and BHs. This particular phenomenon has been used to guarantee the presence of BHs and comprehend the geometrical composition of our Universe. Different techniques have been explored in the literature to study the lensing process in BH vicinity \cite{8,9,10,11}.

Gravitational lensing in plasma is a fascinating phenomenon in which light rays propagate through an inhomogeneous medium, causing the motion of photons to be observed along curved trajectories. The nature of this phenomenon may be explained by the dispersive property of the plasma medium. In this process, the deflection will be caused due to both gravity and non-homogeneity of the plasma medium. While the first effect is neutral, the second effect depends on the frequency of photons in a dispersive medium and approaches zero in a homogeneous medium. Based on the general theory of geometrical optics in curved spacetime \cite{8}, a gravitational lensing model in a plasma field was developed by various authors \cite{12,13}. At the same time, due to nonlinear processes the deflection angle also depends on the frequency of photons in a homogeneous plasma \cite{14}. This deflection angle differs significantly from the vacuum case. Several authors have studied the impact of different plasma topologies and works on light propagation, providing deep insight into this fascinating area of research \cite{15,16,17,18,19,20,21,22,23,24,25,26,27,28,29,30}. Understanding gravitational lensing in plasma has significant implications for our understanding of the universe and the behavior of light in different environments.

In the pioneering work of Synge \cite{31} it has been proposed possibility of observing the black spot as a result of photons captured by a black hole. When photons fall into the black hole located between the bright source and observer, a dark region will be generated in the source image which is referred as the BH shadow. Extensive studies in the literature \cite{32,33,34} suggest that BH shadow can be effectively produced by means of gravitational lensing \cite{35,36,37,38,39,40,41,42,43,44,45}. BH shadow has been widely discussed in the works of various authors \cite{46,47,48,49,50,51,52,53,54,55,56,57,58,59,60,61,62,63,64,65,66,67,68,69,70,71,72,73,74,75,76,77,78,79,80}. The boundary of the BH shadow is determined by the structure of the black hole. The presence of a plasma field in the black hole domain may also alter the equation of motion of photons.
Consequently, the form and size of shadow is sensitive to the plasma medium surrounding the black hole (see, e.g. \cite{81,82,83,84,85,86,87,88}).

The dynamical survey of test particles appears as a strong tool to examine the presence of BH solution in a strong gravitational field regime around the gravitating compact objects. Benchmark BH solutions like Schwarzschild and Kerr BH have been tested in a successful way in both the strong \cite{89,90,91,92} and weak-field \cite{83,94} regimes. However, the margin to test the other theories of gravity still exists with the dilaton scalar field including the cosmological constant. Specifically, alternative and modified theories of gravity have been tested by using the x-ray observation data from
astrophysical compact objects \cite{95,96,97}. The circular orbits of the test particle, particularly, the inner stable circular orbits $(ISCOs)$ have gained remarkable attention in the current arena of research. Estimation and constraints on parameters of BH may be obtained utilizing the observation of accretion disc \cite{98,99,100,101}. Specifically, the magnetic field surrounding the BH alters the network of charged particle motion \cite{102,103,104,105}. One can consult some of the studies for more details about spacetime structure and particle dynamics around BHs \cite{106,107,108,109,110,111,112}.

It is believed that the quasi-periodic oscillations $(QPOs)$ established around the self-gravitating compact objects are amongst the privileged tools to probe the phenomena emerging in the strong gravitational field of BH noticed as x-ray micro-quasars. Amidst this type of structural formalism to portray the observational phenomena are the epicyclic frequencies governed by the motion of a neutral test particle orbiting around the BH \cite{113,114,115,116}, and the charged particle motion \cite{117,118}. In the low-mass X-ray binary systems like a neutron star $(NS)$ or a BH, $(QPOs)$ are perceived in their power spectra. Generally, $QPOs$ are categorized by low-frequency $(LF)$ or high-frequency $(HF)$ $QPOs$. Frequencies that are commonly generated in the form of
pairs are named twin peak $HF\;\; QPOs$. The $HF\;\; QPOs$ possess information about matter falling and/or moving in the strong gravitational field surrounding the compact body. On the other hand, $LF \; QPOs$ are strong, steady, and tend to shift in the frequency domain, whereas
$HF\; QPOs$ yield a weak and transparent behavior but do not illustrate significant drift in their frequencies \cite{119,120}. Both $LF$ and $HF\; QPOs$ come up together in the case of some X-ray binaries. By grasping the accessible astrophysical knowledge, one can argue that $HF$ and $LF\; QPOs$ are induced in distinct
parts of the accretion disk. $QPOs$ have already been explored for various models in literature  \cite{121,122,123,124,125,126,127}.

The phenomenological movement of the photon away from the gravitational well is explained by gravitational redshift in General Relativity (GR). During this process, the photon loses energy which results in a decrease in frequency and an increase in wavelength. The respective phenomenon is known as frequency red-shift. Contrarily, if the photon moves towards the gravitational well, it results in a gain of energy. This gain of energy results in an increase in the frequency of the ongoing photon, consequently its wavelength decreases, and the phenomenon is famed as gravitational blue-shift. The phenomenon of gravitational red blue shifts of photons moving in $ISCO$ around BH is a genuine cause of attraction for physicists \cite{128,129,130,131,132}.

In this manuscript, our prime focus is to test the Metric-Affine Gravity (MAG) using particle dynamics and photon motion. In short, MAG is the intuitive extension of GR. Indeed, our current perception of the gravitational interaction depends upon the physical relation between the space-time curvature and the energy-momentum tensor of matter, as used in Einstein's field equations (EFEs) \cite{133}. In this context, the structure of MAG comprises the notions of torsion
and nonmetricity in a supplemented space-time geometry \cite{134}. Specifically, the spin angular momentum of matter acts as a source of torsion \cite{135,136,137,138}, whereas the dilation and shear charges of matter turn out to be sources of nonmetricity \cite{139,140,141}. The insufficient insights or evidence for these physical entities lead us to the concept of new MAG models incorporating exact solutions with torsion and nonmetricity beyond GR.

We proceed with our study according to the following format. In section \ref{A2}, we discuss the necessary preliminary setup for MAG. In section \ref{A3}, we study the $ISCO$ and photon radius related to the particle dynamics. Section \ref{A4} is reserved for the BH shadows. In section \ref{A5}, we discuss the $QPOs$ of massive particles while section \ref{A6} deals with the red-blue shift of the photon released by the particle. Section \ref{A7} contains the details about effective force. Section \ref{A8} corresponds to the discussion about plasma lensing, i.e., uniform plasma, non-uniform ($SIS$) plasma, and isothermal sphere $NSIS$ plasma. Similarly, section \ref{A9} comprises formalism about image magnification further categorized into two subsections (subsection \ref{A9.1} for uniform plasma and subsection \ref{A9.2} for $SIS$ plasma). In the last section, we conclude our research in a short summary.

\section{Metric-Affine Gravity: Reissner-Nordstr\"{o}m-like geometry with spin, dilation and shear charges}\label{A2}

The study of a non-holonomic connection $w_{\mu} \in g[(4,\mathrm{R})$ certifies the dilation, spin and shear charges of matter, and can be justified through the vector isomorphism, or a coframe field $e^a_{\mu}$. The fundamental relation of an affine-connected metric spacetime is given as follows \cite{134,139}:
\begin{equation}\label{1}
    w^a{}_{b\mu}=e^a{}_{\lambda} e_b{}^{\rho}\tilde{\Gamma}^{\lambda}_{\rho\mu}+e^{a}{}_{\lambda}\partial_{\mu}e_{b}{}^{\lambda},
\end{equation}
which is fully accessorized by the torsion and nonmetricity tensors \cite{143}.
\begin{eqnarray*}
    T^\lambda_{\mu\nu} = 2\tilde{\Gamma}^\lambda_{[\mu\nu]}, \quad Q_{\lambda\mu\nu} = \tilde{\nabla}\lambda g{\mu\nu}.
\end{eqnarray*}

Let us start our discussion with a quadratic parity preserving action that provides geometrical corrections to $\mathrm{GR}^3$ in the form of dynamical nonmetricity tensor \cite{134}
\begin{equation}\label{2}
S=\int d^4 x \sqrt{-g}\left\{\mathcal{L}_{\mathrm{m}}+\frac{1}{16 \pi}\left[-R+2 f_1 \tilde{R}_{(\lambda \rho) \mu \nu} \tilde{R}^{(\lambda \rho) \mu \nu}+2 f_2\left(\tilde{R}_{(\mu \nu)}-\hat{R}_{(\mu \nu)}\right)\left(\tilde{R}^{(\mu \nu)}-\hat{R}^{(\mu \nu)}\right)\right]\right\},
\end{equation}
where $\mathcal{L}_{\mathrm{m}}$ defines the matter Lagrangian. The nonmetricity field reported in the action propagates by the means of symmetric part of the curvature tensor and its contraction \cite{134}
\begin{eqnarray}
\begin{aligned}\label{3}
\tilde{R}^{(\lambda \rho)}{ }_{\mu \nu} & =\tilde{\nabla}_{[\nu} Q_{\mu]}{ }^{\lambda \rho}+\frac{1}{2} T_{\mu \nu}^\sigma Q_\sigma^{\lambda \rho}, \\
\tilde{R}_{(\mu \nu)}-\hat{R}_{(\mu \nu)} & =\tilde{\nabla}_{(\mu} Q_{\nu) \lambda}^\lambda-\tilde{\nabla}_\lambda Q_{(\mu \nu)}{ }^\lambda-Q^{\lambda \rho}{ }_\lambda Q_{(\mu \nu) \rho}+Q_{\lambda \rho(\mu} Q_{\nu)}{ }^{\lambda \rho}+T_{\lambda \rho(\mu} Q_{\nu \rho}^{\lambda \rho},
\end{aligned}
\end{eqnarray}
which consequently tend to abandon the third Bianchi of GR. Varying the action (\ref{2}) with respect to the coframe field as well as  the anholonomic connection, one can derive the following set of equations  (see appendix in ref. \cite{134} for detailed explanations).
\begin{eqnarray}
Y 1_\mu^\nu =8 \pi \theta_\mu{ }^\nu,\label{4}\\
Y 2^{\lambda \mu \nu} =4 \pi \Delta^{\lambda \mu \nu},\label{5}
\end{eqnarray}
where $Y 1_\mu{ }^\nu$ and $Y 2^{\lambda \mu \nu}$ correspond to the tensor quantities. Also, $\theta_\mu{ }^\nu$ is the canonical energy-momentum tensor and $\Delta^{\lambda \mu\nu}$ represents the hypermomentum density tensor of the matter as
\begin{eqnarray}
\begin{aligned}\label{6}
\theta_\mu{ }^\nu & =\frac{e^a{ }_\mu}{\sqrt{-g}} \frac{\delta\left(\mathcal{L}_m \sqrt{-g}\right)}{\delta e^a \nu}, \\
\Delta^{\lambda \mu \nu} & =\frac{e^{a \lambda} e_b \mu}{\sqrt{-g}} \frac{\delta\left(\mathcal{L}_m \sqrt{-g}\right)}{\delta w^a{ }_{b \nu}}.
\end{aligned}
\end{eqnarray}
Thus, both matter currents serve as the source of developing the extended gravitational field. For metric-affine geometries, the anholonomic connection (\ref{1}) in the Lie algebra of the linear group $GL(4,\mathcal{R})$ decomposes into hypermomentum showing its formal decay into spin, dilation, and shear currents \cite{139}. Specifically, the essential appearance of shears gives rise to dynamical traceless nonmetricity tensor\cite{142} as per the interior composition of the special linear group $SL(4,\mathcal{R})\subset GL(4,\mathcal{R})$.

\begin{figure}
    \centering
    \includegraphics[scale=0.66]{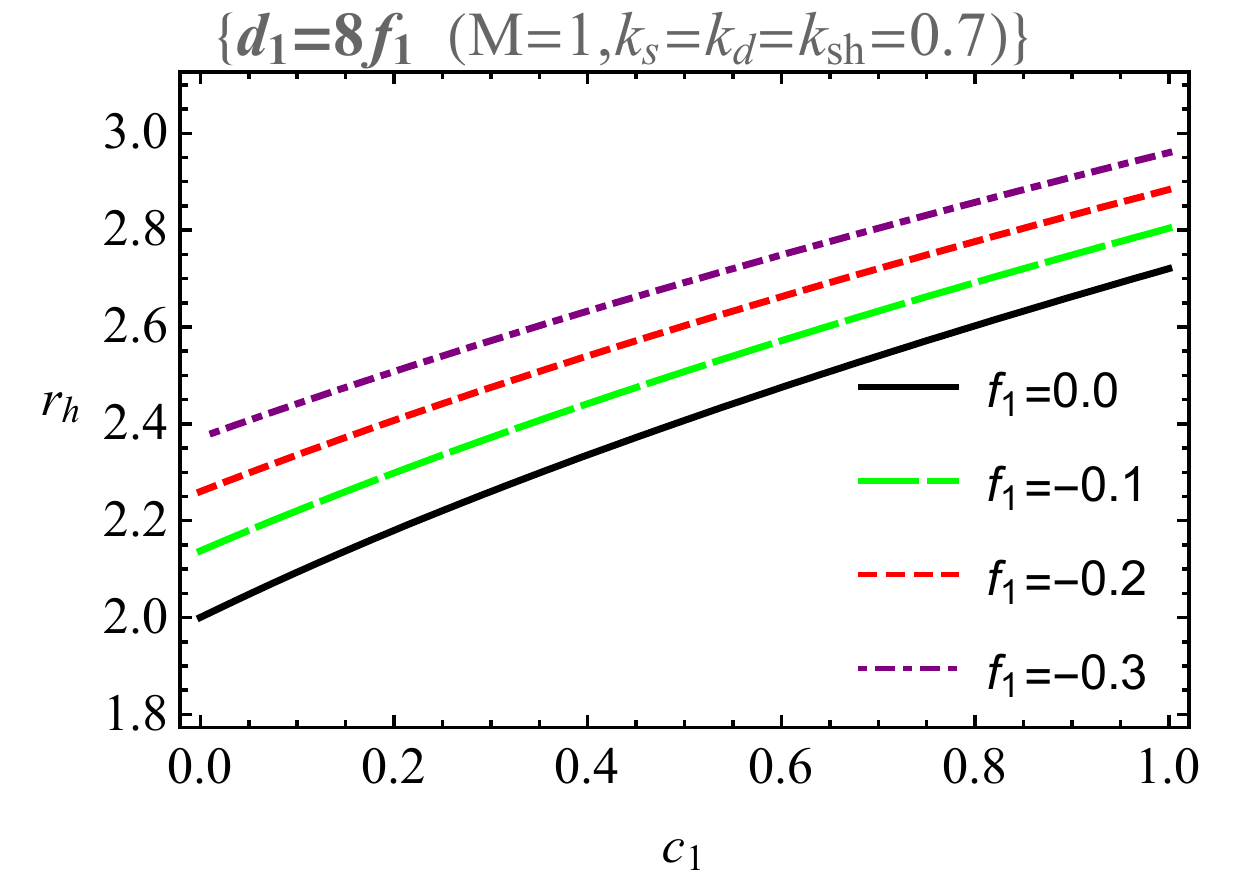}
    \includegraphics[scale=0.66]{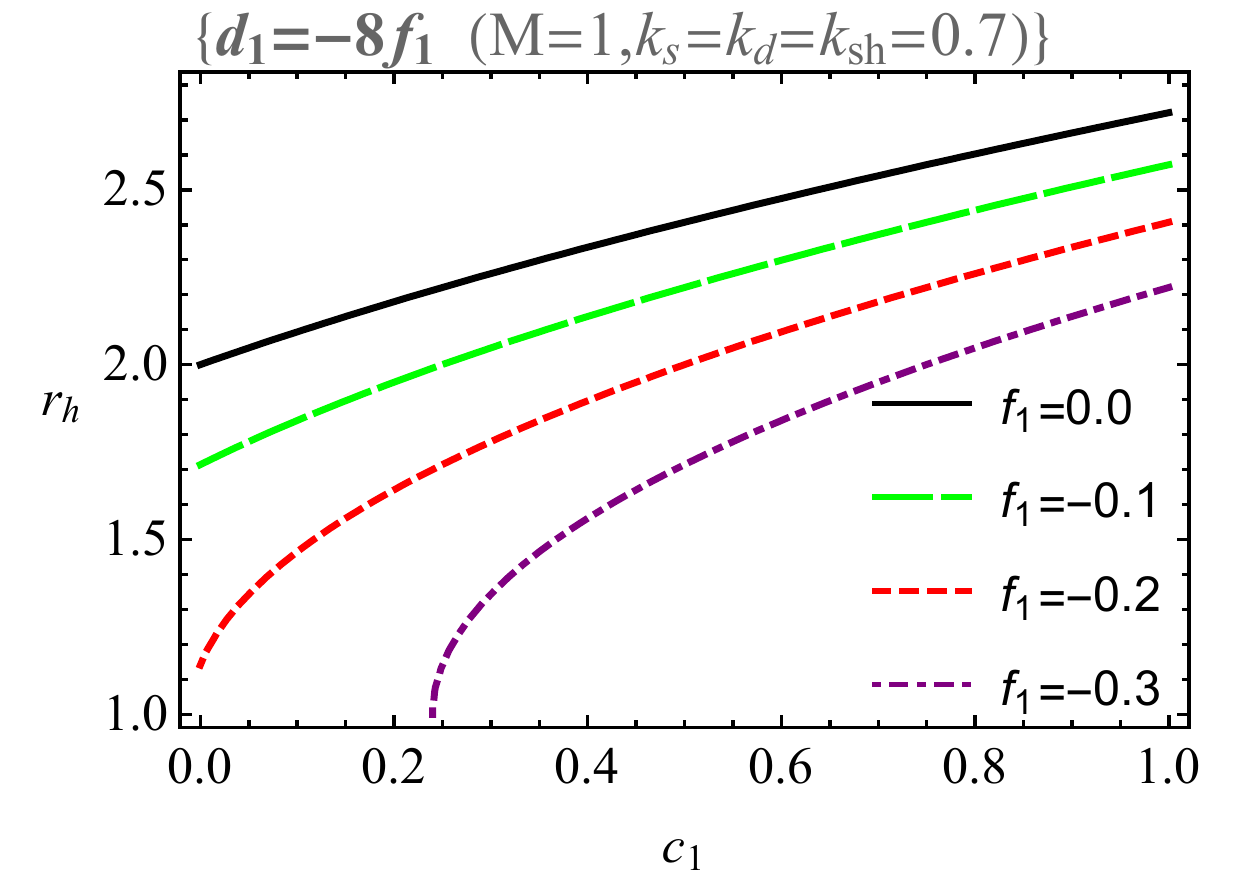}
    \includegraphics[scale=0.66]{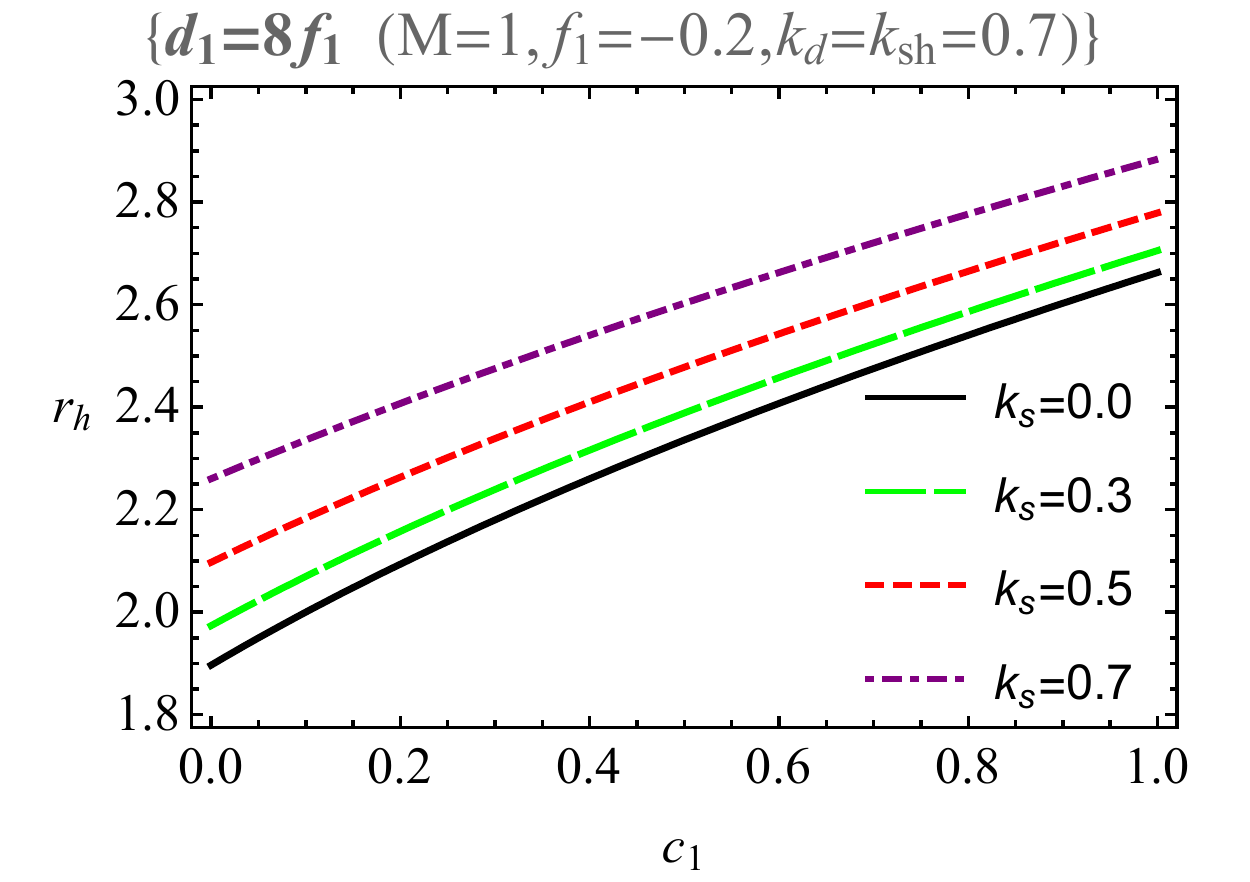}
    \includegraphics[scale=0.66]{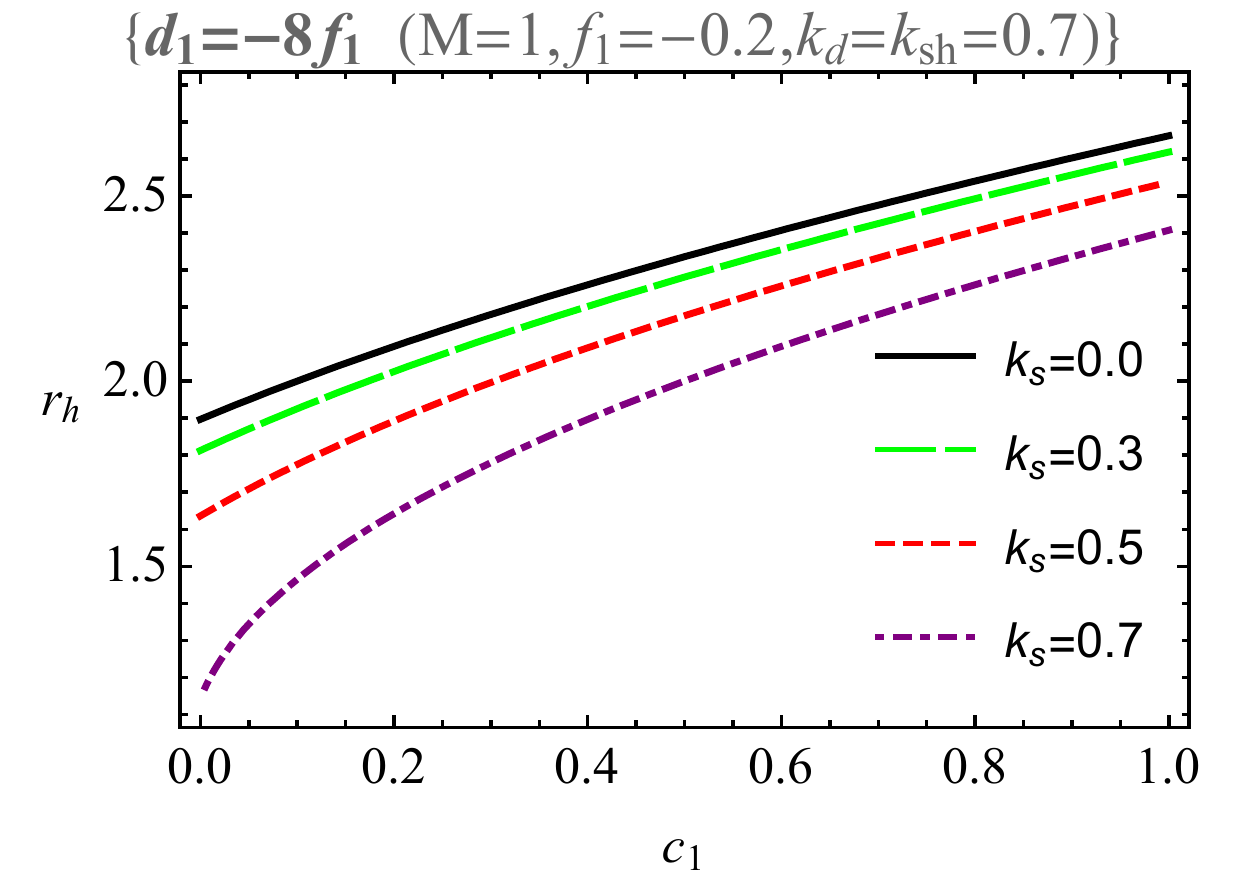}
    \includegraphics[scale=0.66]{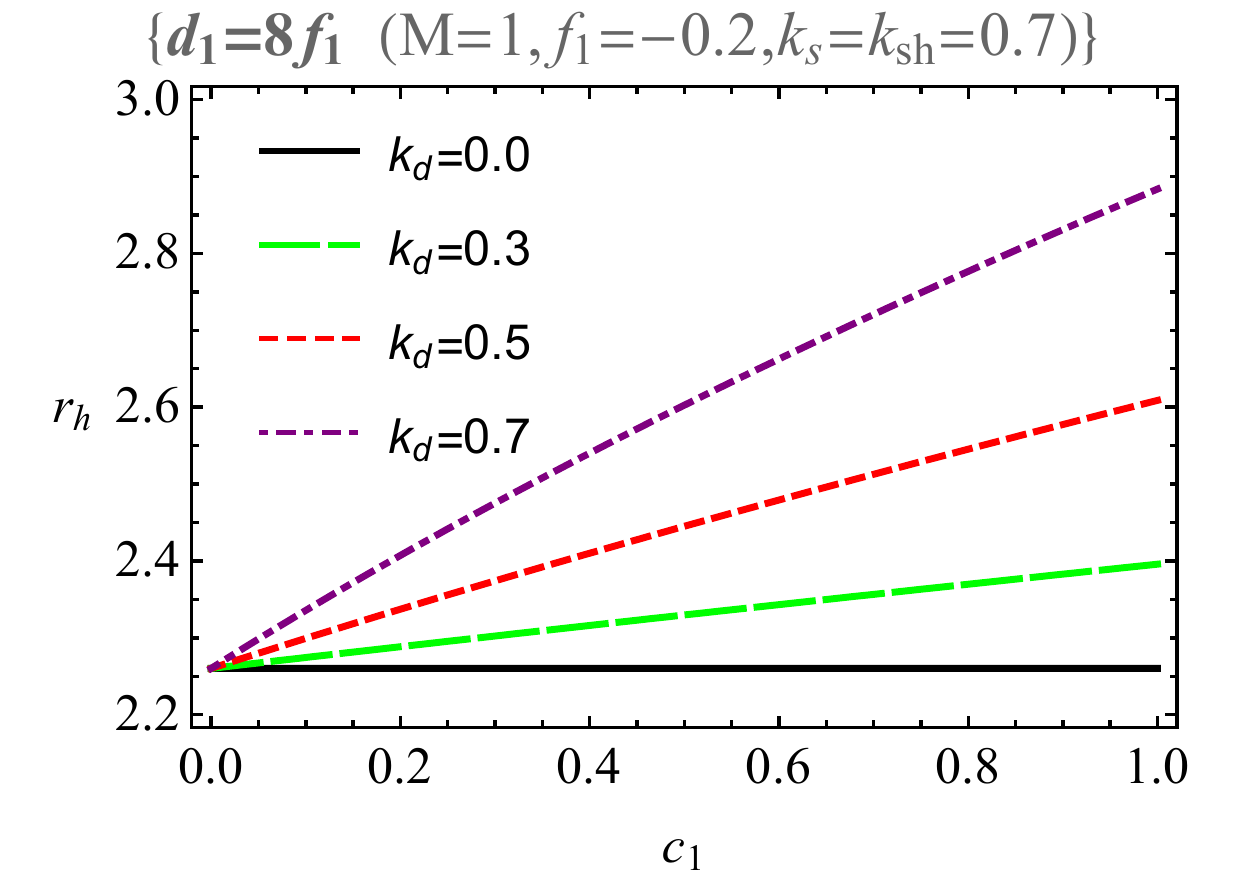}
    \includegraphics[scale=0.66]{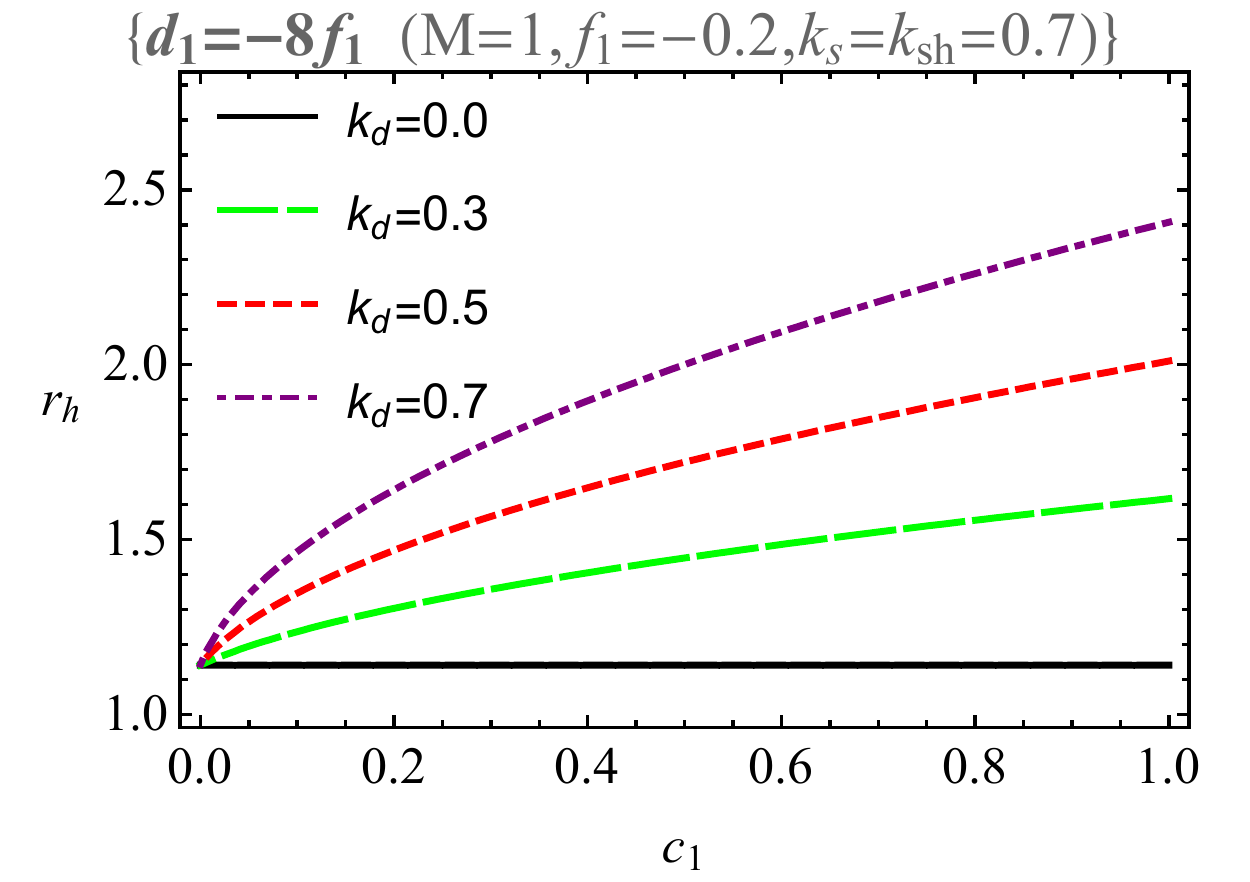}
    \includegraphics[scale=0.66]{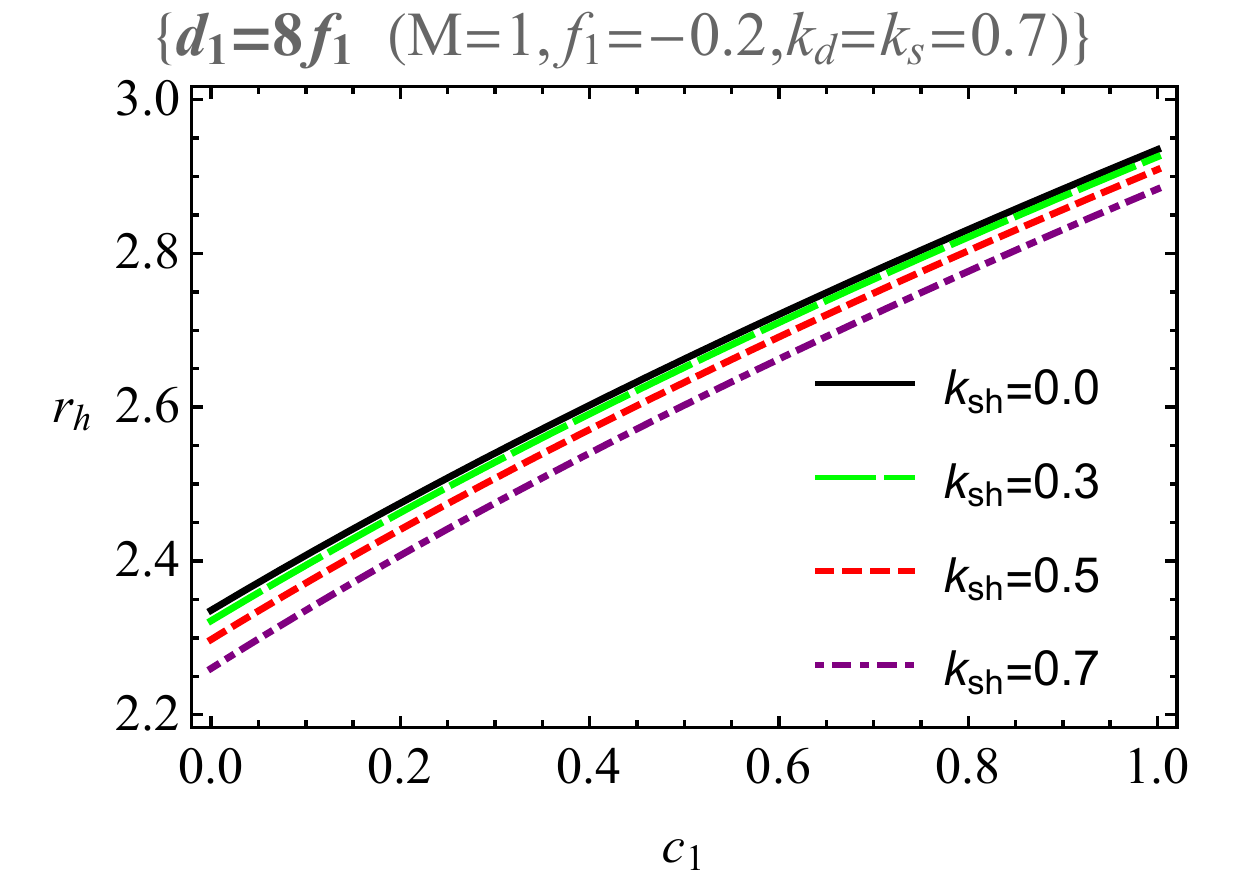}
    \includegraphics[scale=0.66]{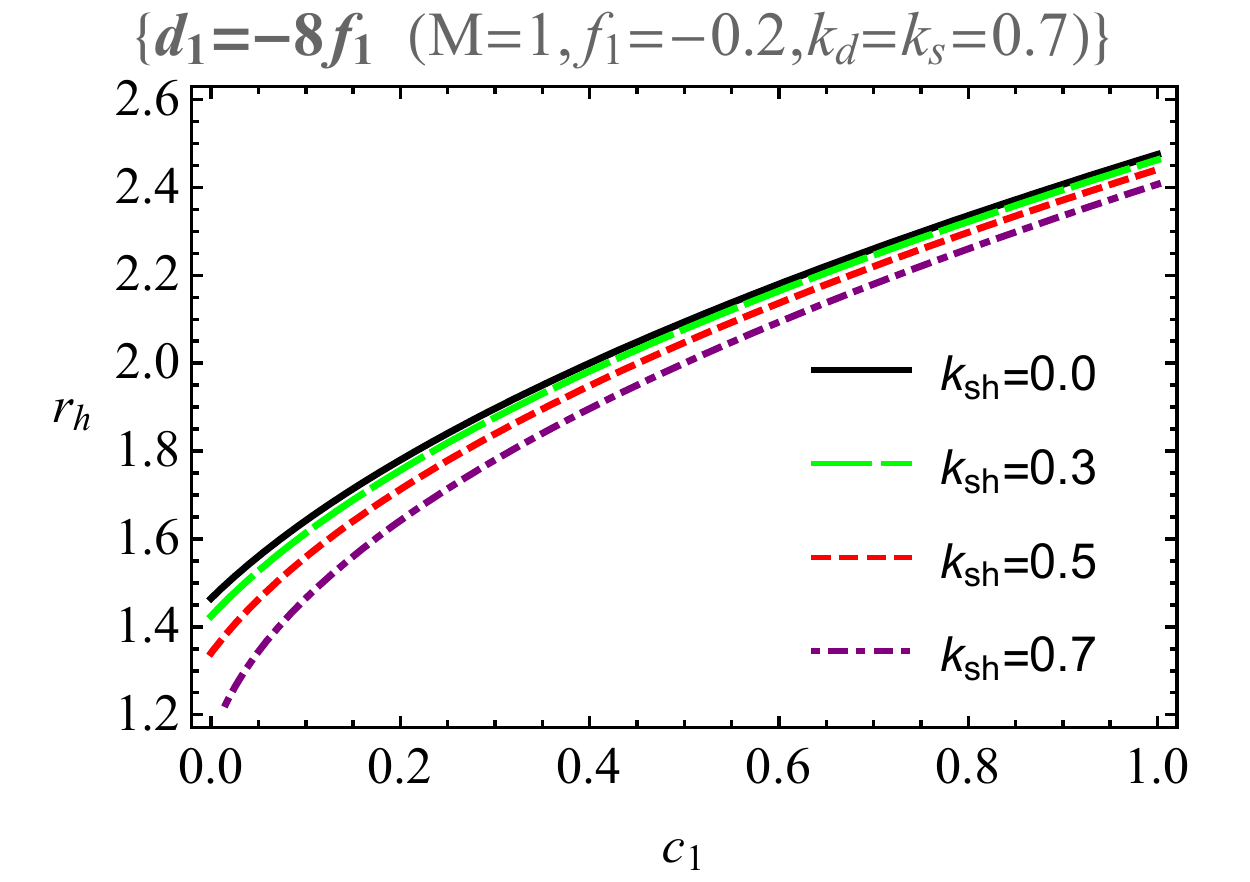}
    \caption{Horizon radius $r_h$ for the cases $d_1=8f_1$ (Left panel) and $d_1=-8f_1$ (Right panel) along with $c_1$ taking different values of $f_1,\; k_s,\; k_d, \;\&\; k_{sh}$.}
    \label{plot:1}
\end{figure}

Furthermore, the effective gravitational action of the model expressed in terms of these quantities acquires the following form \cite{134}:
\begin{eqnarray}
\begin{aligned} \label{7}
S= & \frac{1}{64 \pi} \int d^4 x \sqrt{-g}\left[-4 R-6 d_1 \bar{R}_{\lambda[\rho \mu \nu]} \bar{R}^{\lambda[\rho \mu \nu]}-9 d_1 \bar{R}_{\lambda[\rho \mu \nu]} \bar{R}^{\mu[\lambda \nu \rho]}+8 d_1 \bar{R}_{[\mu \nu]} \bar{R}^{[\mu \nu]}+4 e_1 \tilde{R}^\lambda{ }_{\lambda \mu \nu} \tilde{R}^\rho{ }_\rho \mu \nu\right. \\
& +8 f_1 \tilde{R}_{(\lambda \rho) \mu \nu} \tilde{R}^{(\lambda \rho) \mu \nu}-2 f_1\left(\tilde{R}_{(\mu \nu)}-\hat{R}_{(\mu \nu)}\right)\left(\tilde{R}^{(\mu \nu)}-\hat{R}^{(\mu \nu)}\right)+18 d_1 \bar{R}_{\mu[\lambda \rho \nu]} \tilde{R}^{(\mu \nu) \lambda \rho}-6 d_1 \bar{R}_{[\mu \nu]} \bar{R}^{\mu \nu} \\
& \left.-3 d_1 \tilde{R}_{(\lambda \rho) \mu \nu} \tilde{R}^{(\lambda \rho) \mu \nu}+6 d_1 \tilde{R}_{(\lambda \rho) \mu \nu} \tilde{R}^{(\lambda \mu) \rho \nu}+\frac{9}{2} d_1 \tilde{R}_{\mu \nu} \tilde{R}^{\mu \nu}+3\left(1-2 a_2\right) T_{[\lambda \mu \nu]} T^{[\lambda \mu \nu]}\right],
\end{aligned}
\end{eqnarray}
which lead to the following independent field equations  (see appendix in ref. \cite{134} for detailed explanations):
\begin{eqnarray}
      X1_{\mu}^{\nu}=0,\label{8}\\
    X2^{\lambda\mu\nu}=0,\label{9}
\end{eqnarray}
where $X1_{\mu}^{\nu}$ and $ X2^{\lambda\mu\nu}$ denote the tensor quantities \cite{134}. As a consequence, the tetrad field equations (\ref{8}-\ref{9}) for any value of the Lagrangian coefficients enable to compute the metric function $\Psi(r)$ in the form of Reissner-Nordstrom-like geometry with spin, dilation and shear charges \cite{134}
\begin{eqnarray}\label{10}
    \Psi(r)&=&1-\frac{2 m}{r}+\frac{d_1 \kappa_{\rm s}^2-4 c_1 \kappa_{\rm d}^2-2f_1 \kappa_{\rm sh}^2}{r^2}\,
\end{eqnarray}
where $d_{1}$ and $f_{1}$ represent the Lagrangian coefficients, $c_{1}$ is a constant whereas $k_{s}^2$, $k_{d}^2$ and $k_{sh}^2$ are the spin, dilation and shear charges, respectively. In this scenario, one can generate three kinds of solutions, i.e., ($i$) $d_{1}=8f_{1}$, ($ii$) $d_{1}=-8f_{1}$ and ($iii$) $d_{1}\neq 8f_{1}$ for which $f_{1}\leq 0$ sets the geometry as the one described by the standard Reissner-Nordstrom solution of GR, which is induced by a traceless nonmetricity field in the current case rather than assuming an electromagnetic part. In order to discuss the optical properties of BHs, we just consider two cases of solutions, i.e., $d_{1}=8f_{1}$ and $d_{1}=-8f_{1}$. The horizon radii for the respective metric function (\ref{10}) can be determined as
\begin{eqnarray}\label{11}
   r_{h}= M-\sqrt{4 c_{1} k_{d}^2-d_{1} k_{s}^2+2 f_{1} k_{sh}^2+M^2},\;\;\&\;\;r_{h}= M+\sqrt{4 c_{1} k_{d}^2-d_{1} k_{s}^2+2 f_{1} k_{sh}^2+M^2}.
\end{eqnarray}
We plot $ M+\sqrt{4 c_{1} k_{d}^2-d_{1} k_{s}^2+2 f_{1} k_{sh}^2+M^2}$ by varying the values of $f_1,\;k_s,\;k_d,\;k_{sh}$ alongwith $c_1$ for both cases $d_1=\pm 8f_1$ as shown in Fig. (\ref{plot:1}). It is significant to note that $f_1,\; k_s,\;\&\;k_{sh}$ reverse the impact in both cases. If the radius is increasing with an increase in parameters for $d_1=8f_1$ then it tends to decrease from the same point by increasing the parameters for the case $d_1=-8f_1$ whereas $k_d$ only changes the horizon radius with similar behavior.

\section{ISCO and Photon Sphere Radius}\label{A3}
\begin{figure}
    \centering
    \includegraphics[scale=0.66]{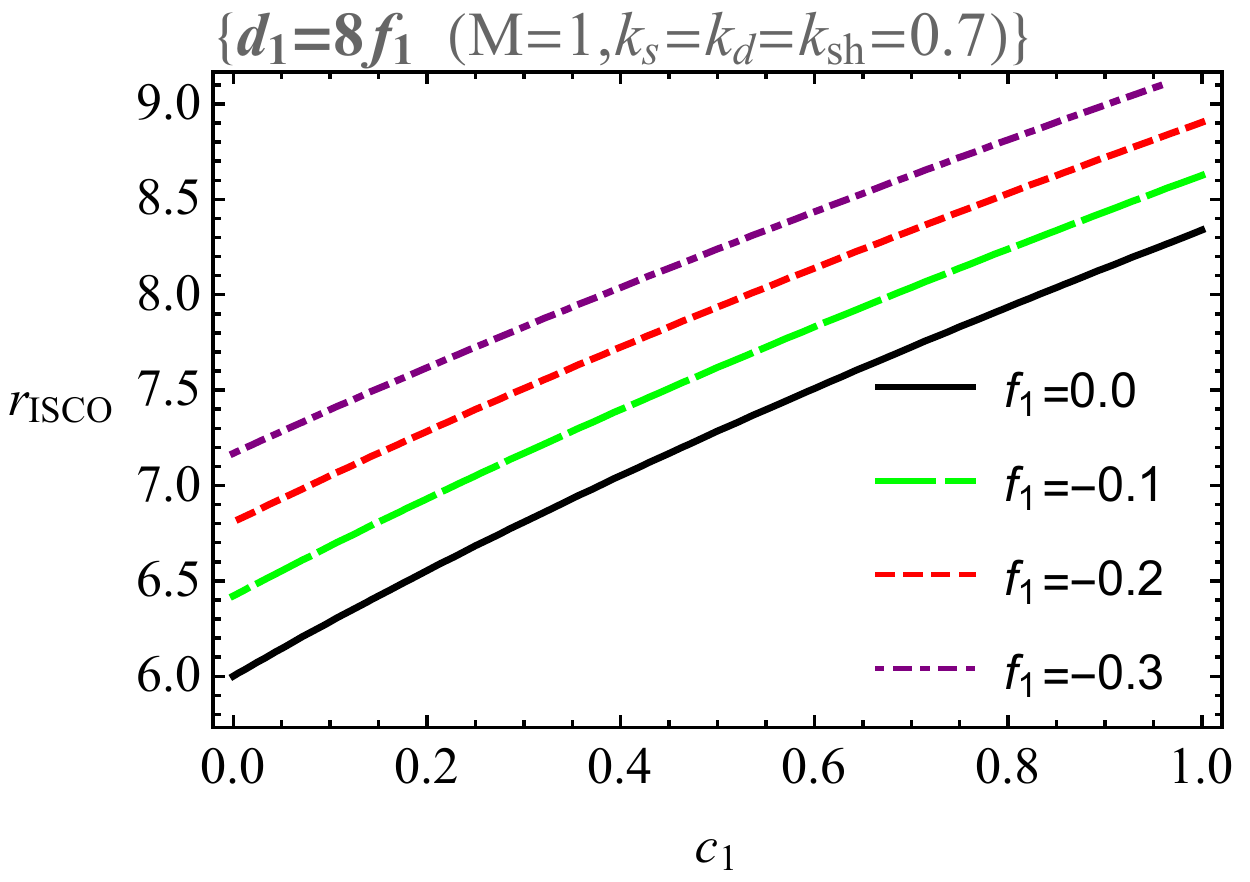}
    \includegraphics[scale=0.66]{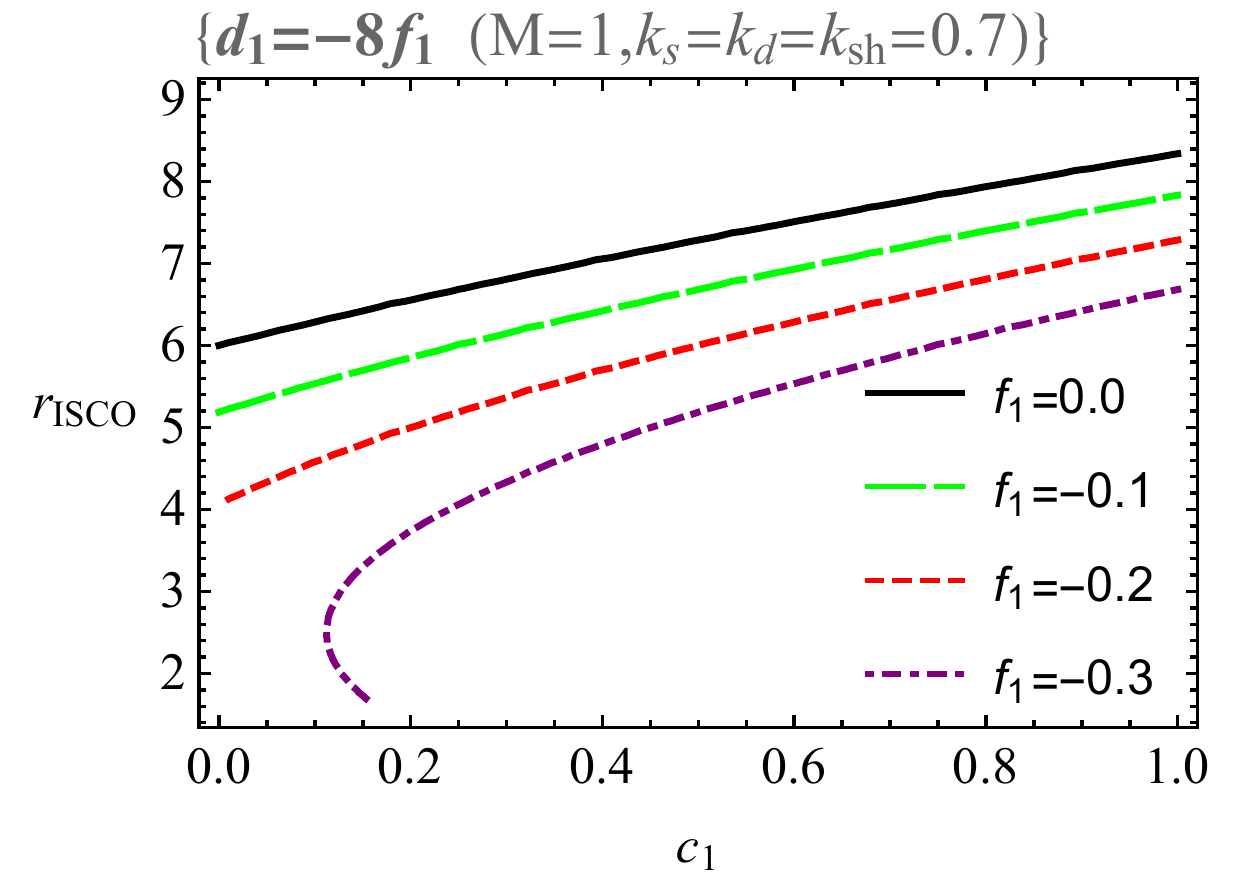}
    \includegraphics[scale=0.66]{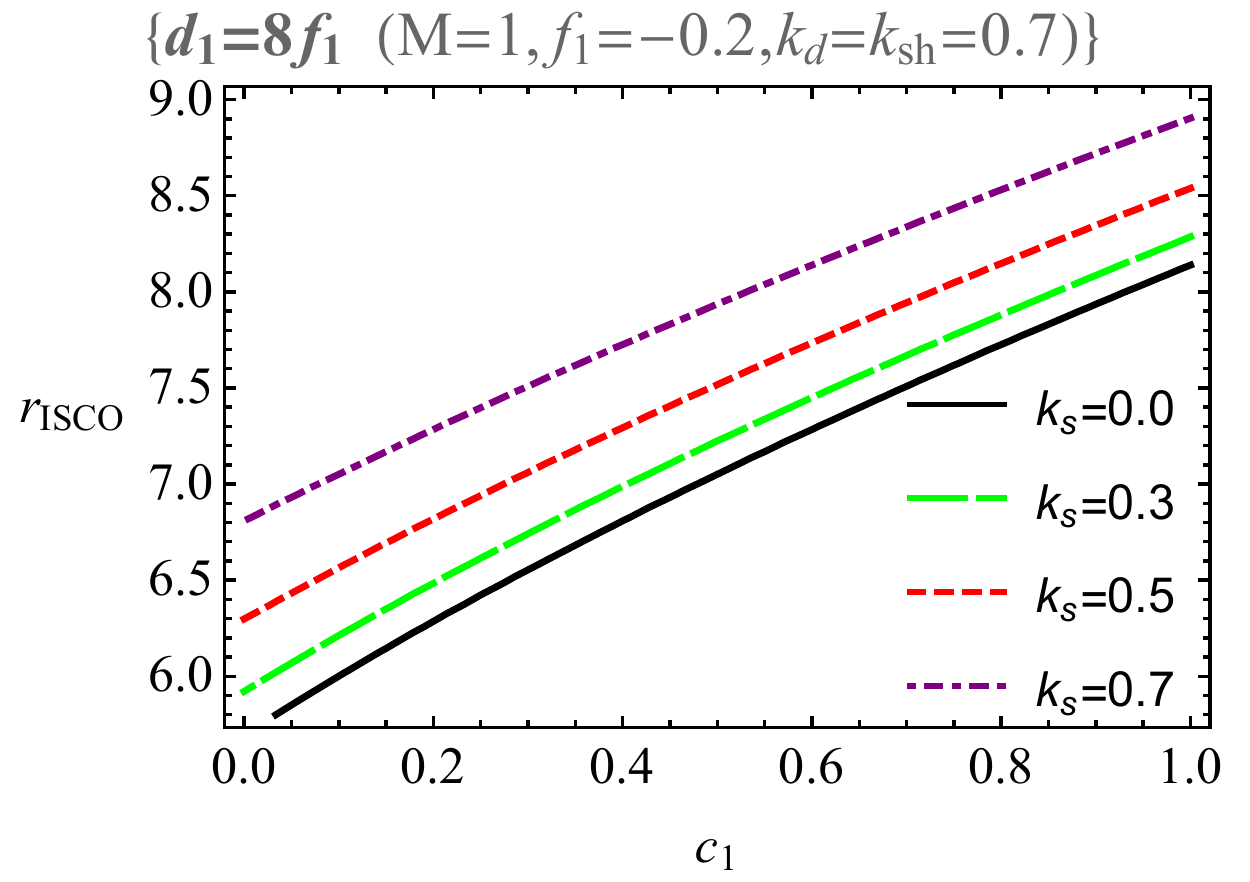}
    \includegraphics[scale=0.66]{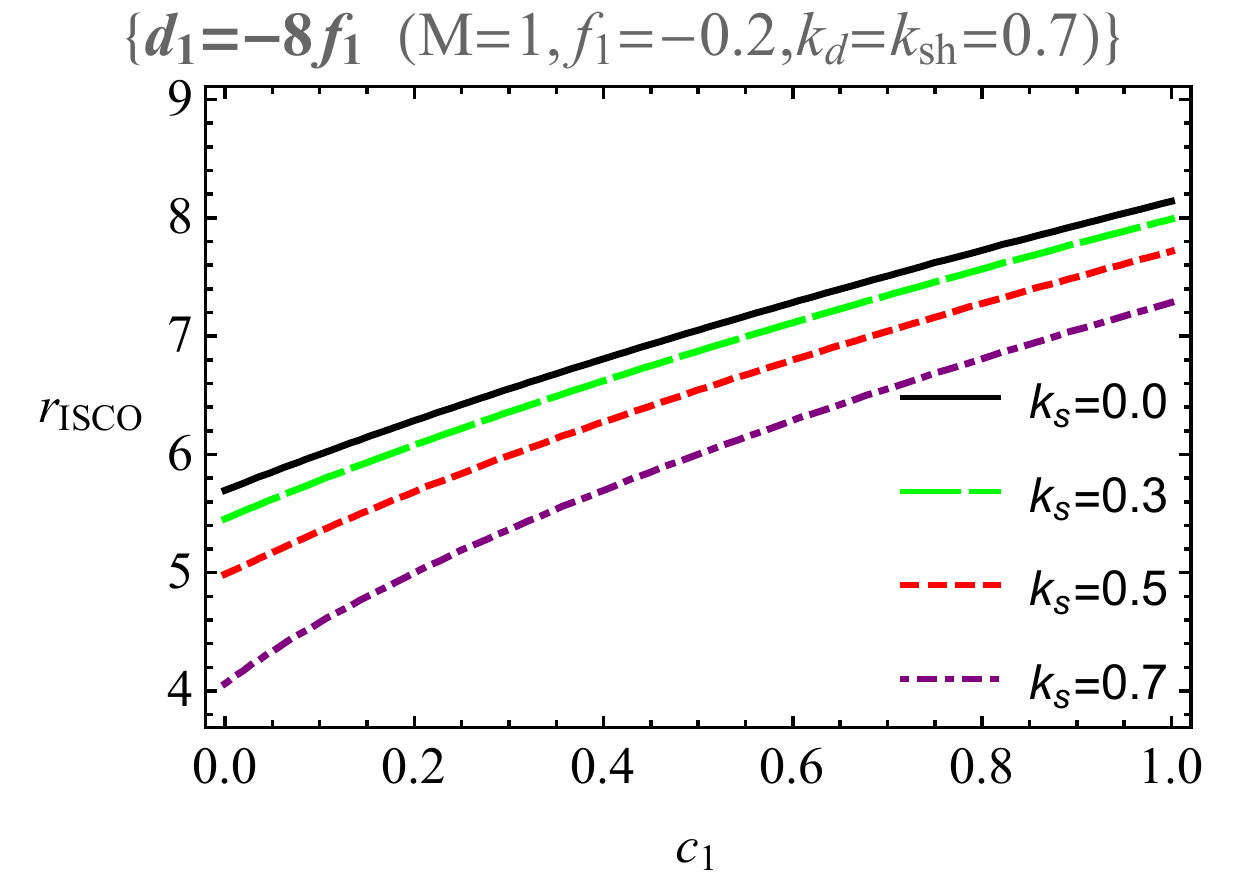}
    \includegraphics[scale=0.66]{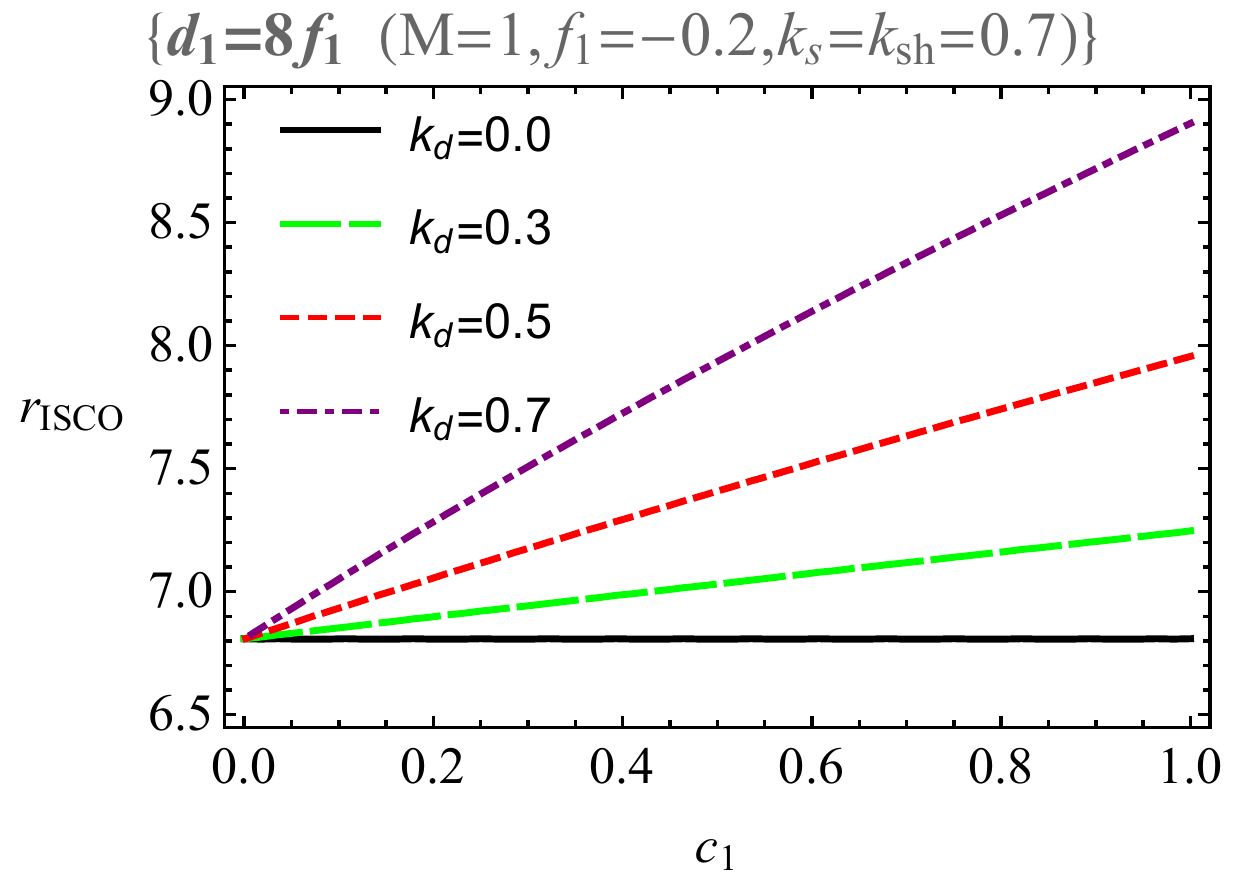}
    \includegraphics[scale=0.66]{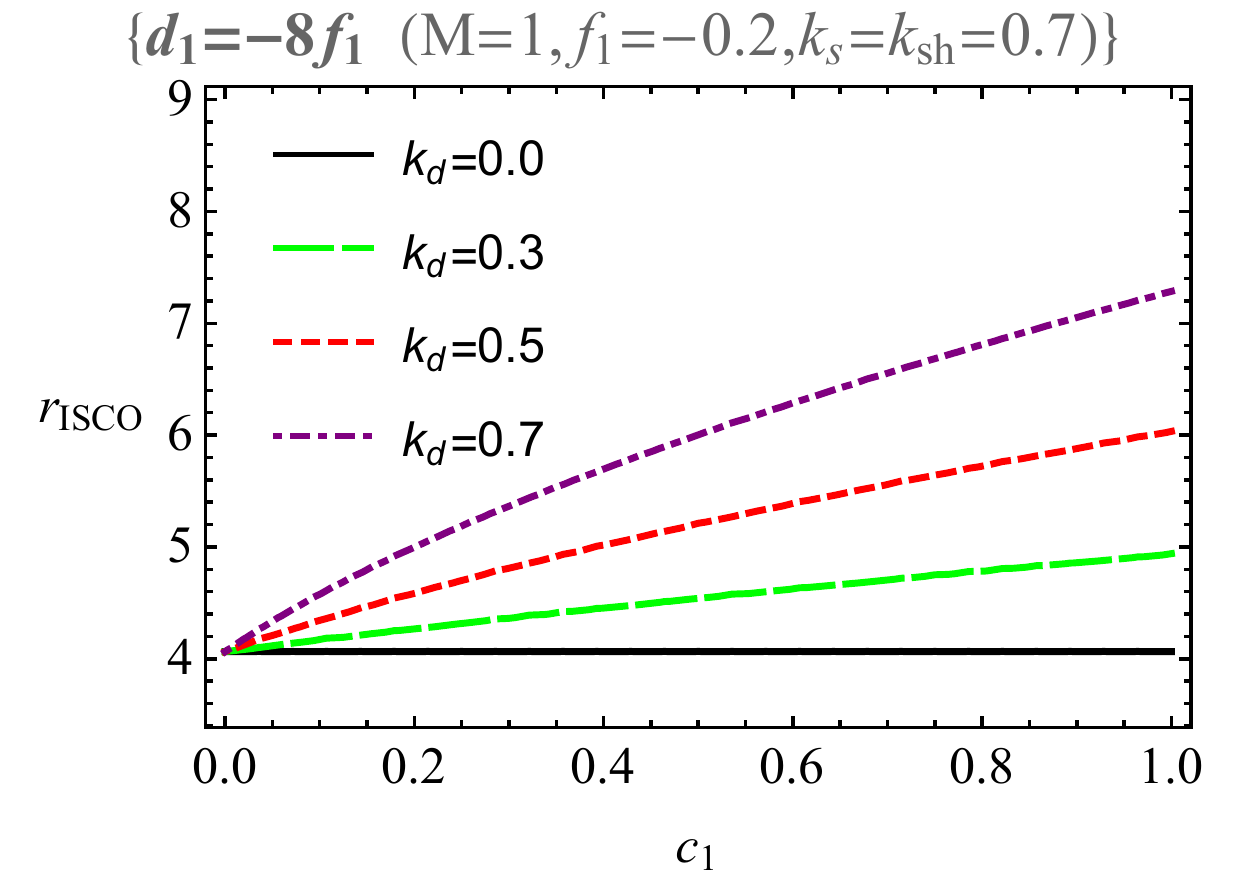}
    \includegraphics[scale=0.66]{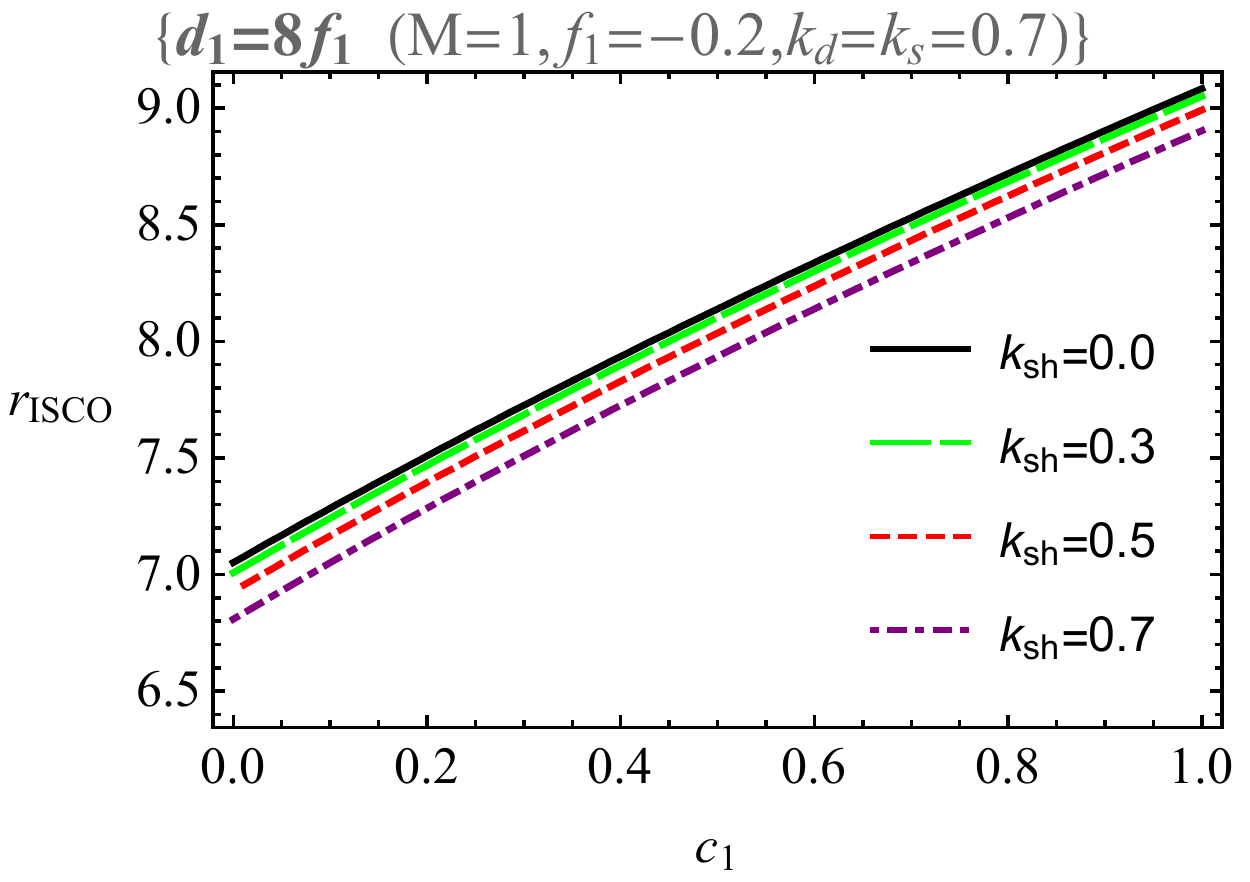}
    \includegraphics[scale=0.66]{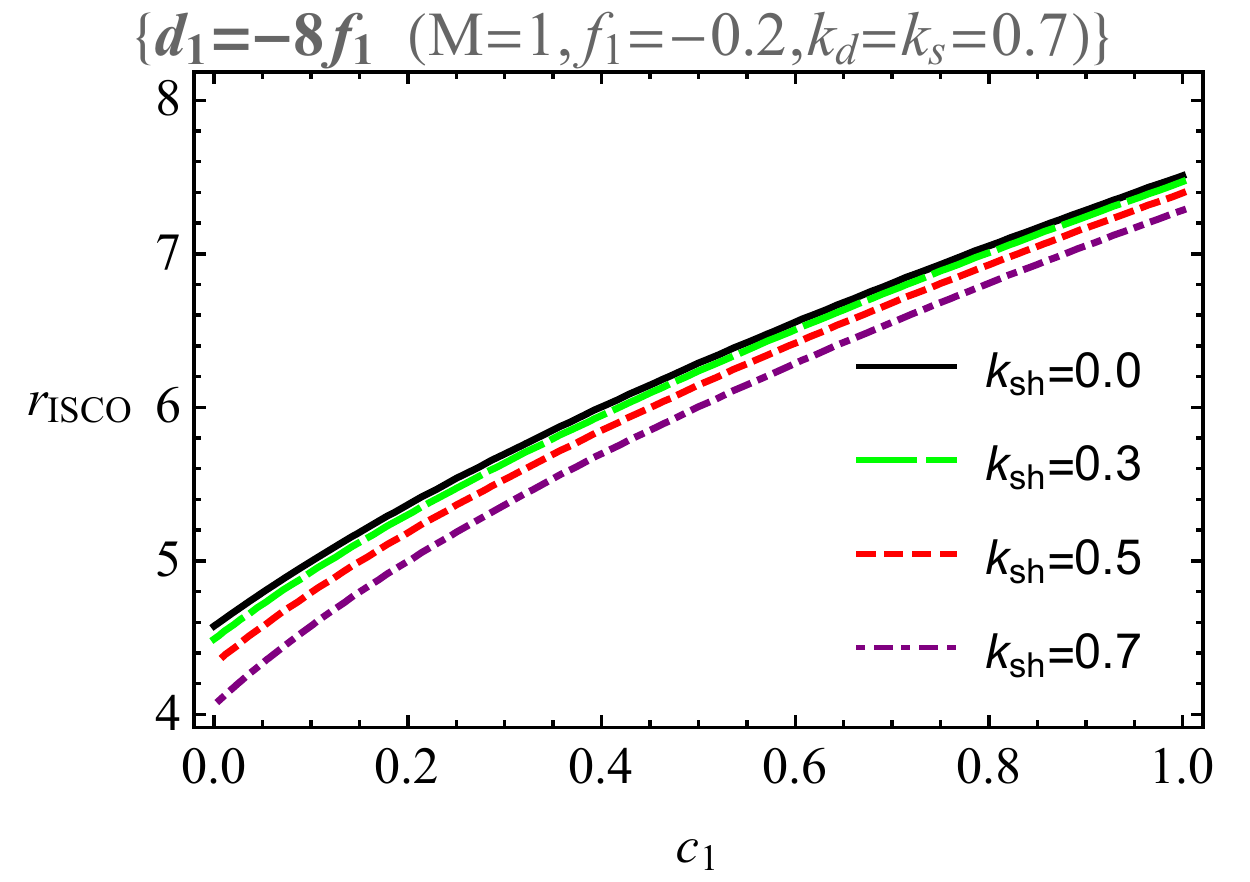}
    \caption{$ISCO$ radius $r_{ISCO}$ for the cases $d_1=8f_1$ (Left panel) and $d_1=-8f_1$ (Right panel) along with $c_1$ taking different values of $f_1,\;k_s,\;k_d,\;\&\; k_{sh}$.}
    \label{plot:2}
\end{figure}
\begin{figure}
    \centering
    \includegraphics[scale=0.66]{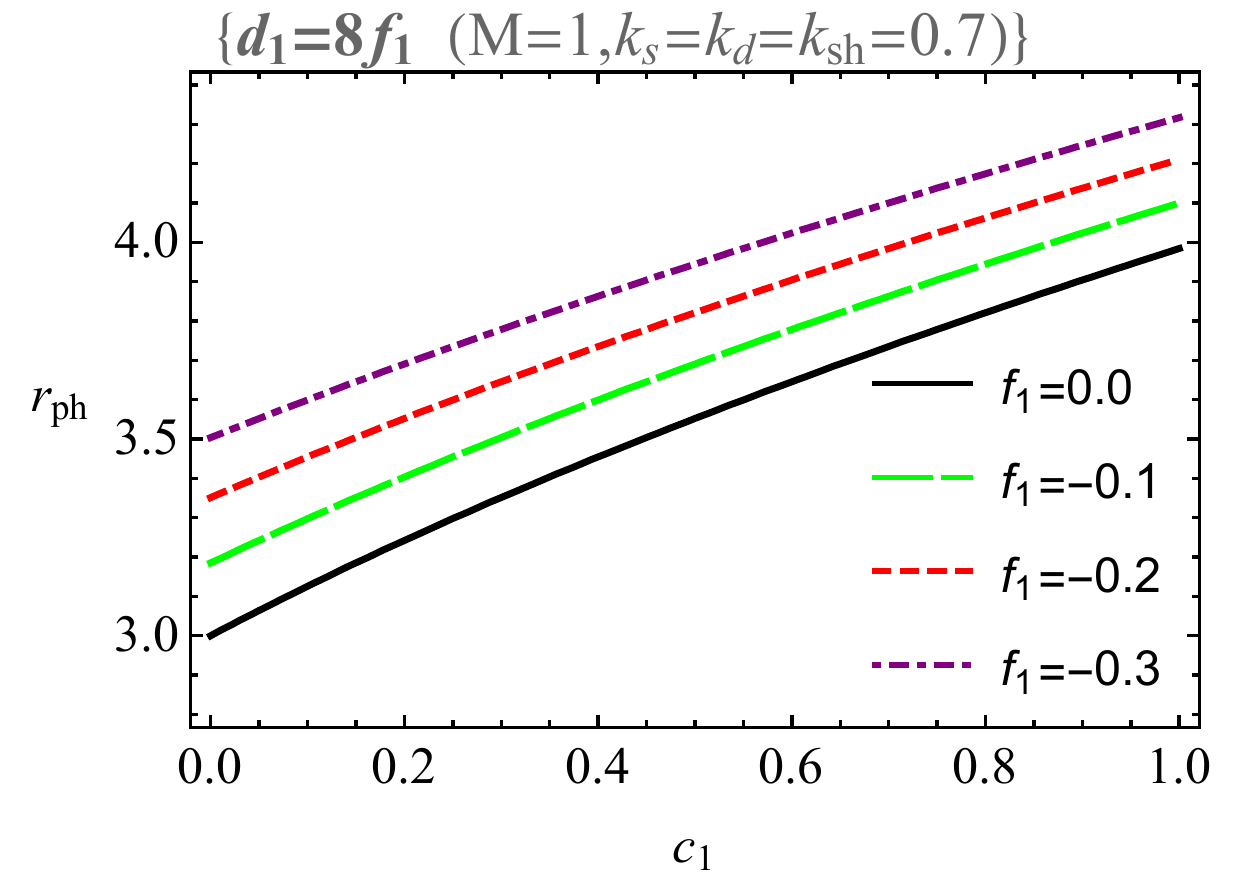}
    \includegraphics[scale=0.66]{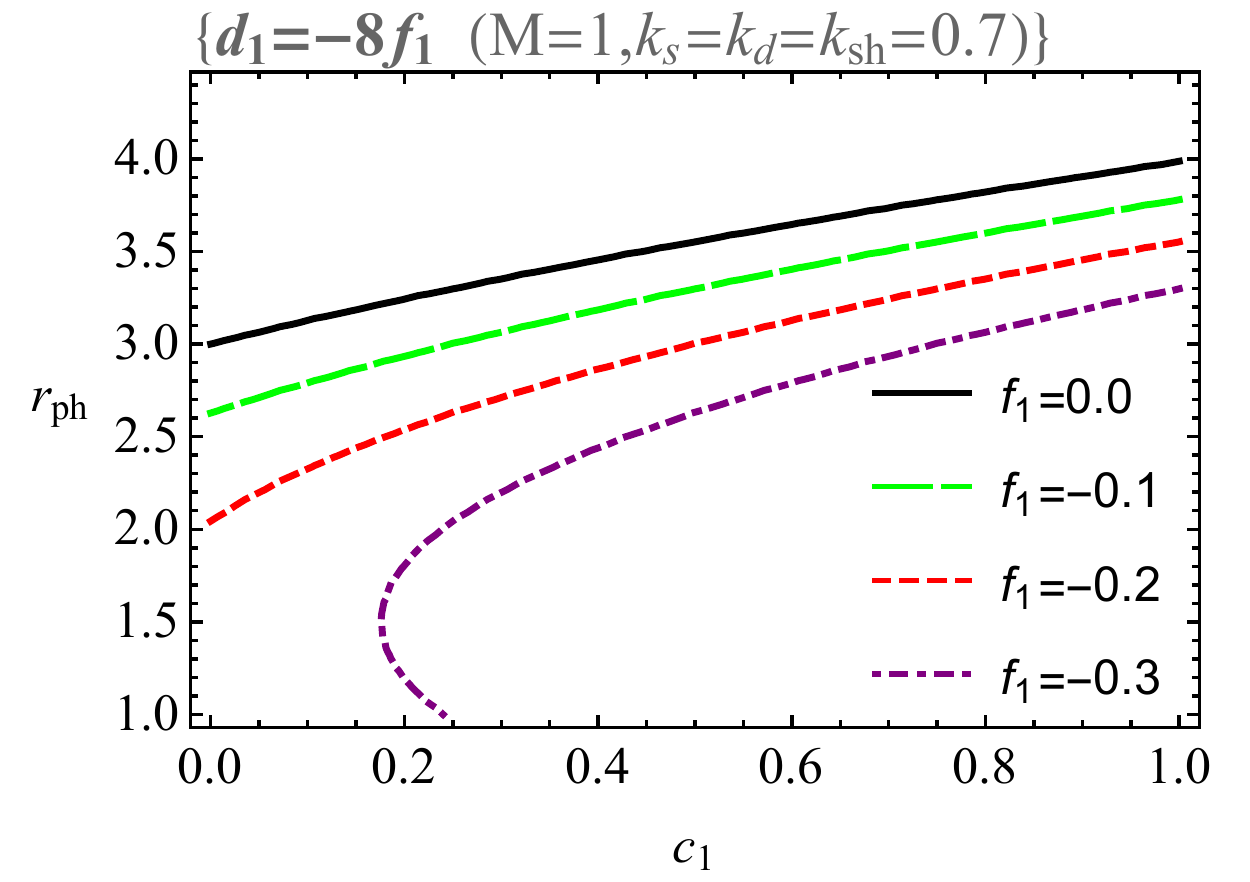}
    \includegraphics[scale=0.66]{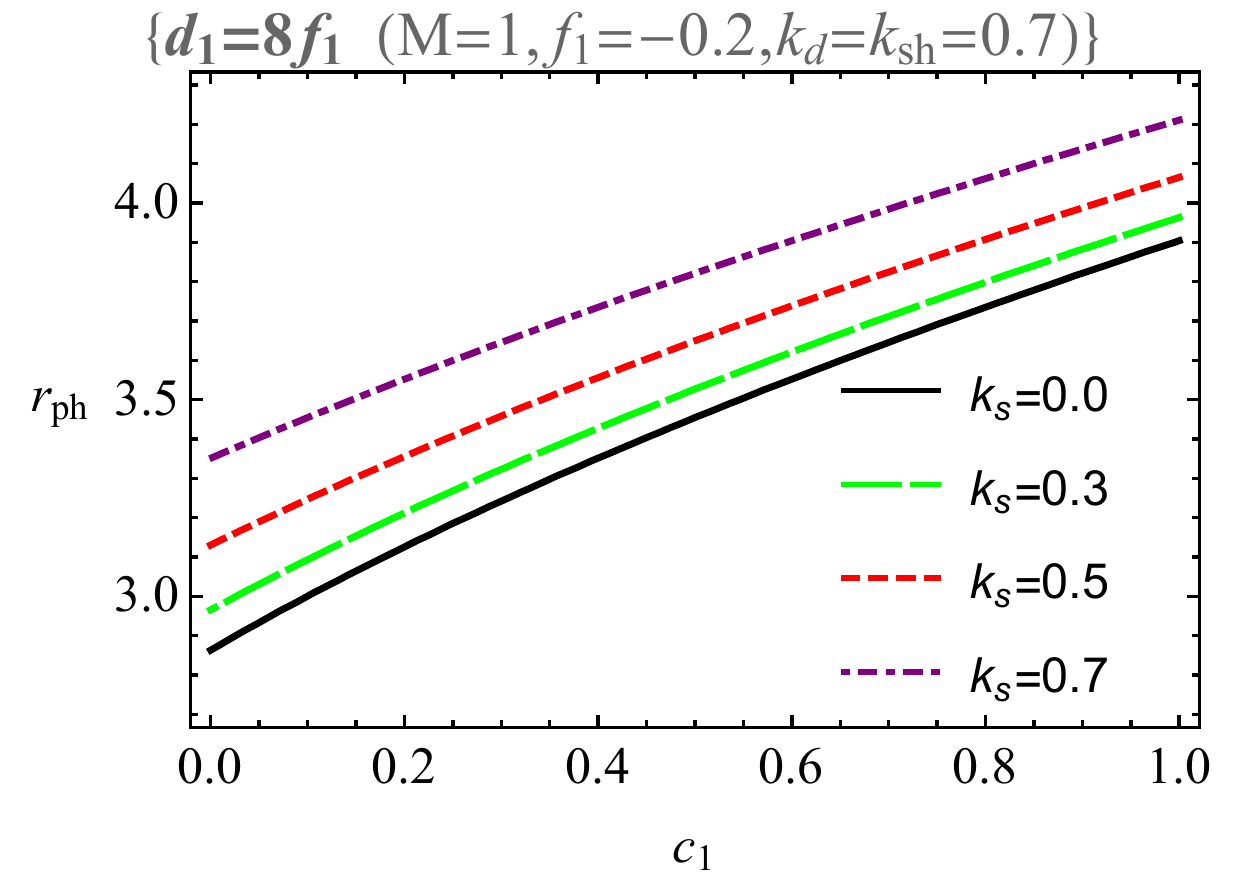}
    \includegraphics[scale=0.66]{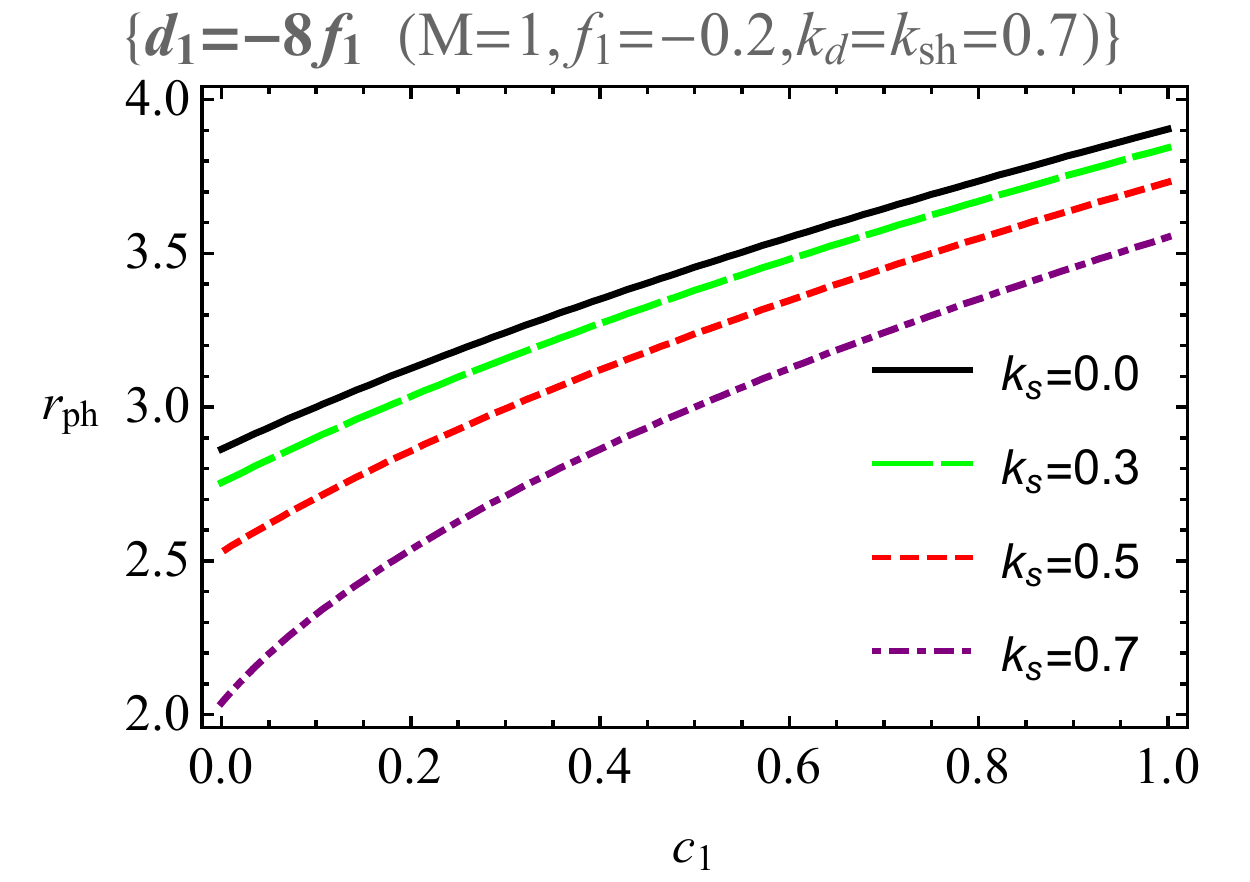}
    \includegraphics[scale=0.66]{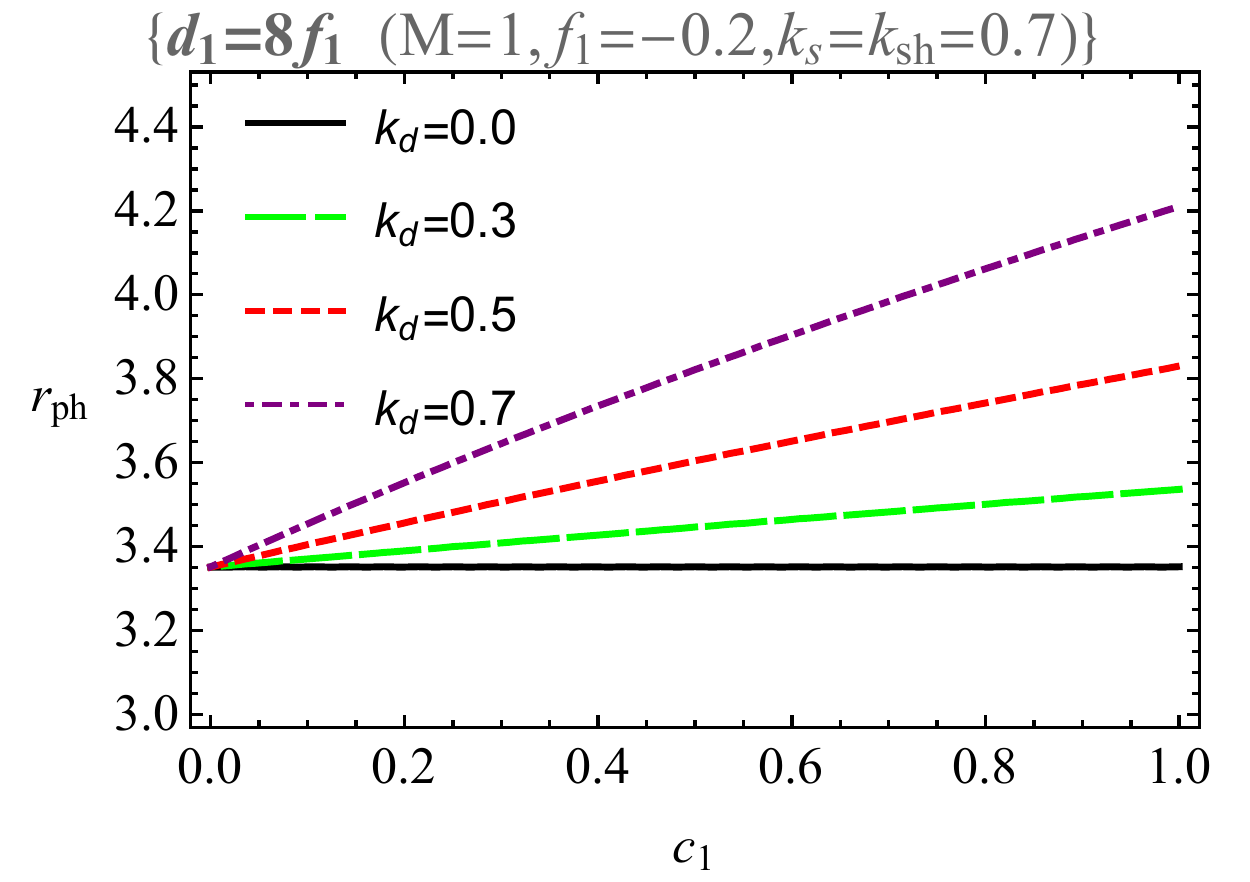}
    \includegraphics[scale=0.66]{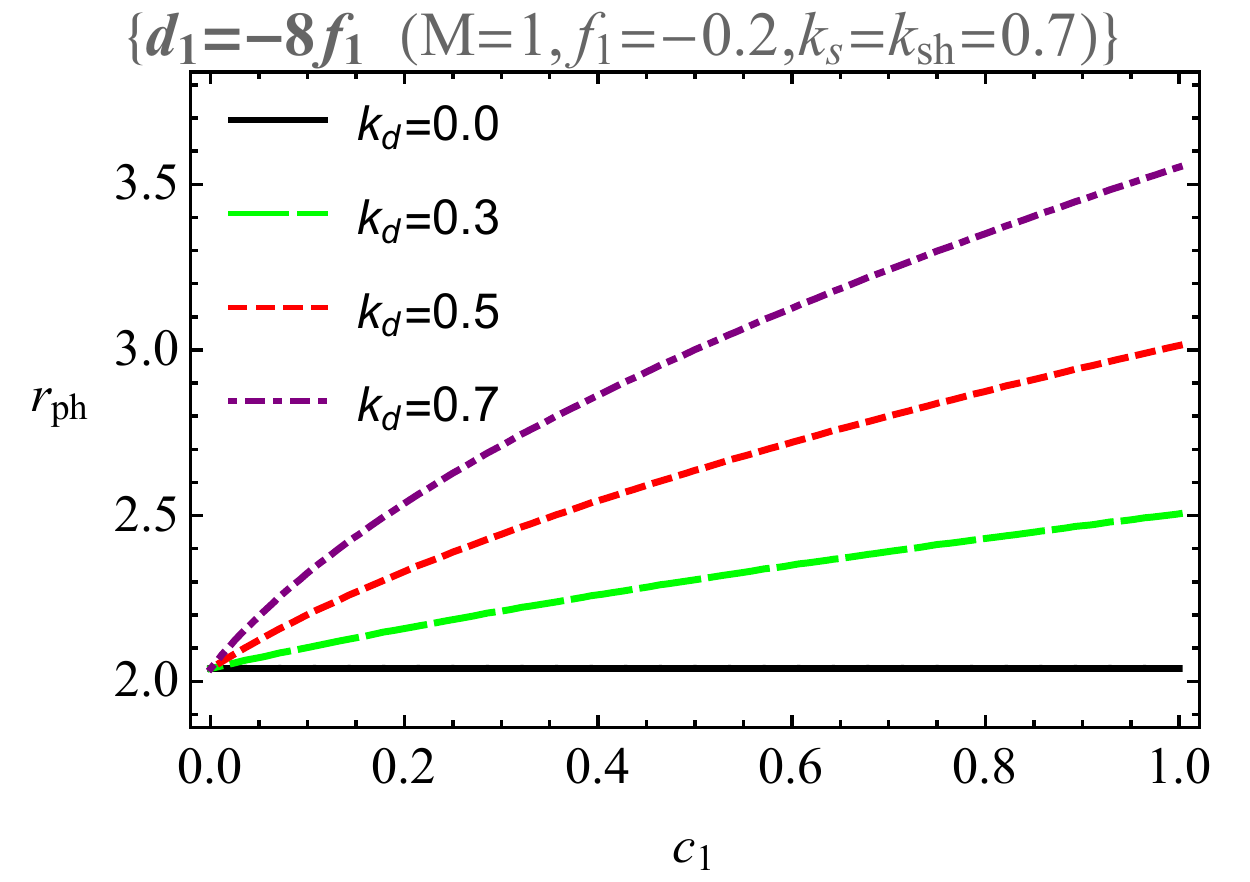}
    \includegraphics[scale=0.66]{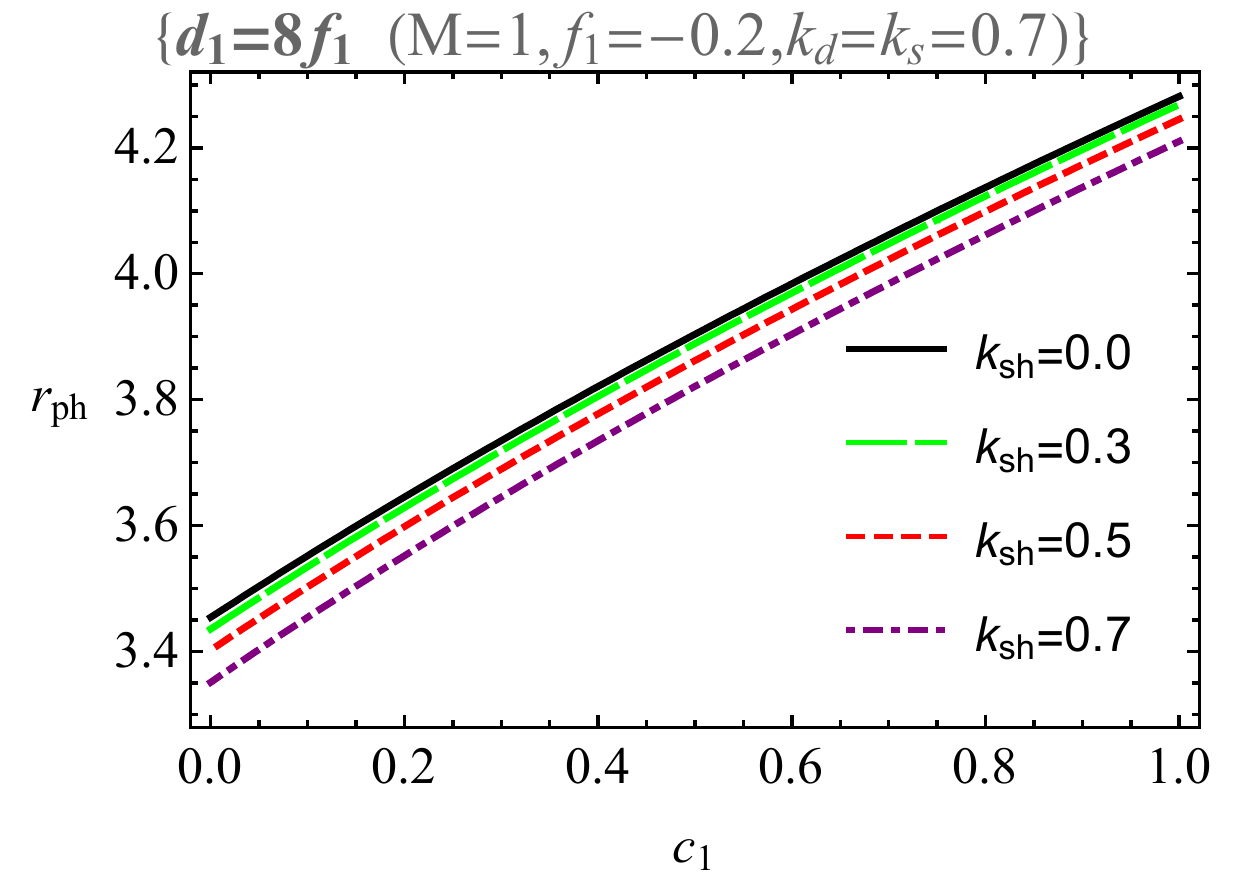}
    \includegraphics[scale=0.66]{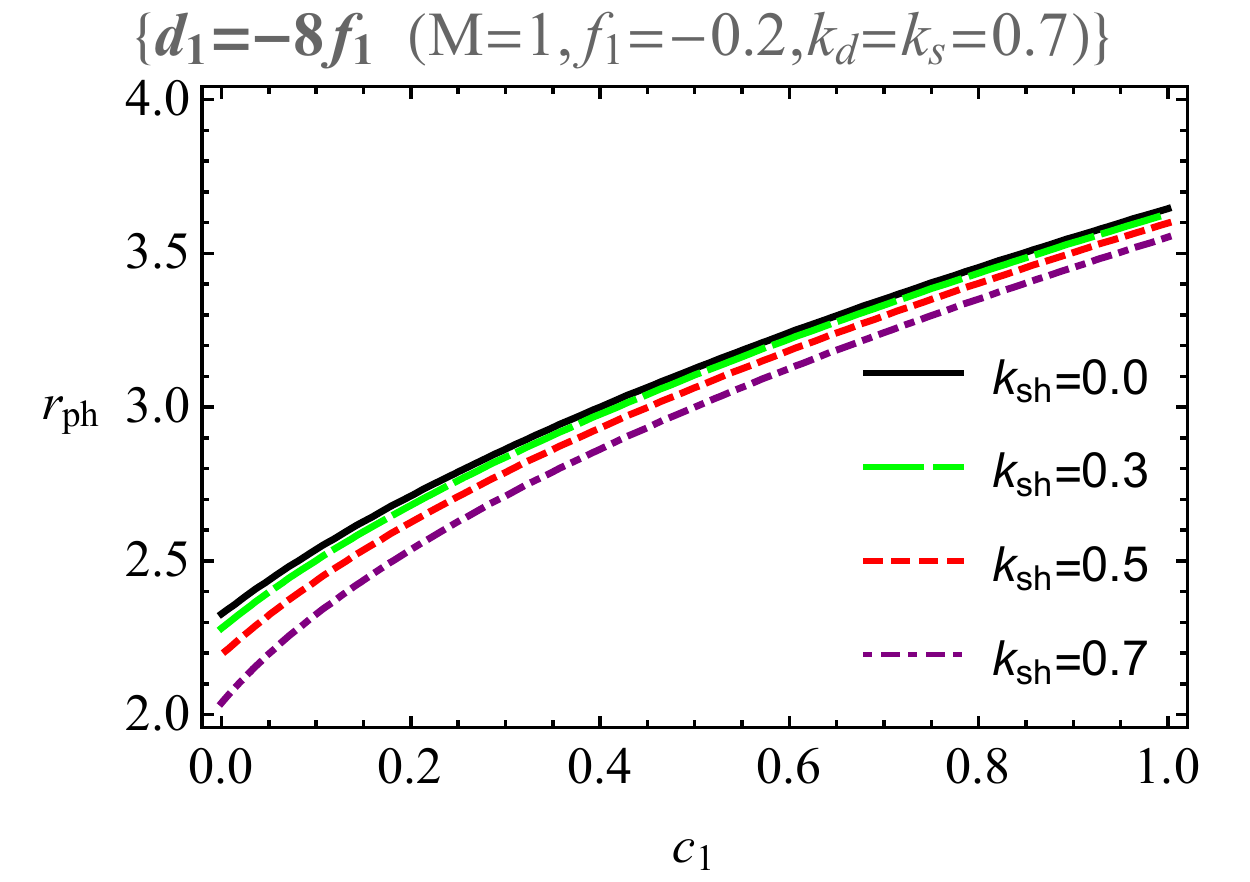}
    \caption{Photon radius $r_{ph}$ for the cases $d_1=8f_1$ (Left panel) and $d_1=-8f_1$ (Right panel) along with $c_1$ taking different values of $f_1,\;k_s,\;k_d,\;\&\;k_{sh}$.}
    \label{plot:3}
\end{figure}
Here, we start with the Lagrangian for the test particle of mass $m$ given by
\begin{equation}\label{12}
   \mathcal{L}^{'}=\frac{1}{2}g_{\mu\nu}u^{\mu}u^{\nu},\;\;\; u^{\mu}=\frac{dx^{\mu}}{d\tau},
\end{equation}
 where $\tau$, $x^{\mu}$ and $u^{\mu}$ correspond to the affine parameter, coordinate, and the four-velocity of the test particle, respectively. The angular momentum $\mathcal{L}$ and energy $\mathcal{E}$ appear as the key source causing motion of the test particle whose mathematical expressions yield
\begin{equation*}
\mathcal{E}=\frac{\partial \mathcal{L}^{'}}{\partial u^t}=- \Psi(r)\frac{dT}{d\tau},
\end{equation*}
\begin{equation}\label{13}
\mathcal{L}=\frac{\partial \mathcal{L}^{'}}{\partial u^{\phi}}=r^2 \sin^2\theta \frac{d\phi}{d\tau}.
\end{equation}
Using the normalization condition $g_{\mu\nu}u^{\mu}u^{\nu} = -\epsilon$ in Eq.(\ref{13}), we get the dynamical equation for test particle in the equatorial plane as
\begin{eqnarray}
   \frac{dr}{d\tau}&=&\sqrt{\mathcal{E}^2- \Psi(r)\left(\epsilon+\frac{\mathcal{L}^2}{r^2}\right)},\label{14}\\
     \frac{d\phi}{d\tau}&=&\frac{\mathcal{L}^2}{r^2 \sin^2 \theta},\label{16}\\
     \frac{dt}{d\tau}&=&\frac{\mathcal{E}}{ \Psi(r)},\label{17}
\end{eqnarray}
with parameter $\epsilon$ has the form
\begin{equation}\label{18}
\epsilon=
\begin{cases}
1,\hspace{0.5cm}  {\rm for\ \ timelike \ \ geodesics} \\
0,\hspace{0.5cm} {\rm for\ \ null\ \  geodesics} \\
-1,\hspace{0.5cm} {\rm for \ \  spacelike\ \  geodesics}.
\end{cases}
\end{equation}
The geodesics equation of massive particle depending on the radial coordinate, thus we have
\begin{equation}\label{19}
    \left(\frac{dr}{d\tau}\right)^2=\mathcal{E}^2-V_{eff}(r)=\mathcal{E}^2- \Psi(r)\left(1+\frac{\mathcal{L}^2}{r^2}\right),
\end{equation}
for which
\begin{equation}\label{20}
    V_{eff}(r)= \Psi(r)\left(1+\frac{\mathcal{L}^2}{r^2}\right).
\end{equation}
Here $V_{eff}(r)\;\&\;\mathcal{L}$ represent the effective potential for test particle in radial motion, geodesic motion and angular momentum, respectively.
Next, we intend to study the stability conditions for the inner circular orbit $r_{ISCO}$. In the $ISCO$ formalism, one can incorporate the following conditions
\begin{equation}\label{21}
\begin{cases}
V^{'}_{eff}=0,\\
V^{''}_{eff}=0.
\end{cases}
\end{equation}
This system can be applied to calculate the $ISCO$ radius for MAGBH geometry given by
\begin{equation}\label{22}
    r_{ISCO}=\frac{12 c_1 k_{d}^2 M^2-3 d_{1} k_{s}^2 M^2+6 f_1 k_{sh}^2 M^2+4 M^4+2 M^2 \sqrt[3]{r_1+r_2}+\left(r_1+r_2\right){}^{2/3}}{M \sqrt[3]{r_1+r_2}},
\end{equation}
where
\begin{eqnarray*}
    r_1&=&\Big[M^4 \left(4 c_1 k_{d}^2-d_{1} k_{s}^2+2 f_1 k_{sh}^2\right)^2 \Big[64 c_1^2 k_{d}^4+4 c_1 k_{d}^2 \left(-8 d_{1} k_{s}^2+16 f_1 k_{sh}^2+9 M^2\right)+4 d_{1}^2 k_{s}^4\\&-&2 f_1 k_{sh}^2 \left(8 d_{1} k_{s}^2-9 M^2\right)-9 d_{1} k_{s}^2 M^2+16 f_1^2 k_{sh}^4+5 M^4\Big]\Big]^{\frac{1}{2}},\\
    r_2&=&32 c_1^2 k_{d}^4 M^2+4 c_1 k_{d}^2 M^2 \left(-4 d_{1} k_{s}^2+8 f_1 k_{sh}^2+9 M^2\right)+2 d_{1}^2 k_{s}^4 M^2-8 d_{1} f_1 k_{sh}^2 k_{s}^2 M^2\\&-&9 d_{1} k_{s}^2 M^4+8 f_1^2 k_{sh}^4 M^2+18 f_1 k_{sh}^2 M^4+8 M^6.
\end{eqnarray*}
We plot $r_{ISCO}$ in Fig.~\ref{plot:2} taking $c_1$ for different values of $f_1, \;k_d,\;k_s,\;k_{sh}$. Though the trending behavior of $r_{ISCO}$ is similar to $r_{h}$, we can observe $r_{ISCO}>r_{h}$.

The equation of photon motion around BH through Eq.(\ref{18}) with $\epsilon=0$ in the equatorial plane can be expressed as
\begin{eqnarray}
    \Dot{r}^2&=&\mathcal{E}^2- \Psi(r)\frac{\mathcal{L}^2}{r^2},\label{17a}\\
    \Dot{\phi}&=&\frac{\mathcal{L}}{r^2},\label{17b}\\
    \Dot{t}&=&\frac{\mathcal{E}}{ \Psi(r)}.\label{17c}
\end{eqnarray}
Equation (\ref{17a}) leads to an effective potential $V_{eff}$ for the photon's motion around the BH radius which has the form
\begin{equation}\label{17d}
V_{eff}= \Psi(r)\frac{\mathcal{L}^2}{r^2}.
\end{equation}
Photon's circular orbit radius $r_{ph}$ around the BH can be calculated by applying the given below conditions:
\begin{equation}
    V^{'}_{eff}=0,
\end{equation}
The orbit radius of photon for MAGBH geometry is given by
\begin{eqnarray}
r_{ph}=\frac{1}{2} \left(3 M+\sqrt{32 c_{1} k_{d}^2-8 d_{1} k_{s}^2+16 f_{1} k_{sh}^2+9 M^2}\right)\;\&\;r_{ph}=\frac{1}{2} \left(3M-\sqrt{32 c_{1} k_{d}^2-8 d_{1} k_{s}^2+16 f_{1} k_{sh}^2+9 M^2}\right).
\end{eqnarray}
We provide a graphical illustration of the radius $r_{ph}$ in Fig.~\ref{plot:3}. We can observe an identical trend in the propagation of $r_{ph}$ to ISCO and horizon radius, but $r_{h}<r_{ph}<r_{ISCO}$ authentically seems accurate in this case study.

\section{Black Hole Shadows}\label{A4}
In this section, we study shadows of MAGBH. The angular radius of the BH shadow yields \cite{143,144}
\begin{eqnarray}\label{25}
\sin^2 \alpha_{sh}=\frac{Y(r_{ph})^2}{Y(r_{obs})^2},
\end{eqnarray}
where
\begin{eqnarray}\label{26}
Y(r)^2=\frac{g_{22}}{g_{00}}=\frac{r^2}{ \Psi(r)},
\end{eqnarray}
where $\alpha_{sh}$ is the notation for angular radius of BH shadow, $r_{obs}$ shows the observed distance observed while $r_{p}$ expresses the radius for photon sphere. The combination of Eqs.(\ref{25} and \ref{26}) gives the following expression
\begin{equation}\label{27}
\sin^2 \alpha_{sh}=\frac{r_{ph}^2}{ \Psi(r_{ph})}\frac{\Psi(r_{obs})}{r^2_{obs}}.
\end{equation}

Using Eq.(\ref{27}), it is convenient to explore radius of the BH shadow from an observer at a far distance \cite{144}
\begin{eqnarray}\label{28}
R_{sh}&\simeq&r_{obs} \sin \alpha_{sh} \simeq \frac{r_{ph}}{\sqrt{ \Psi(r_{ph})}}.
\end{eqnarray}

\begin{figure}
    \centering
    \includegraphics[scale=0.65]{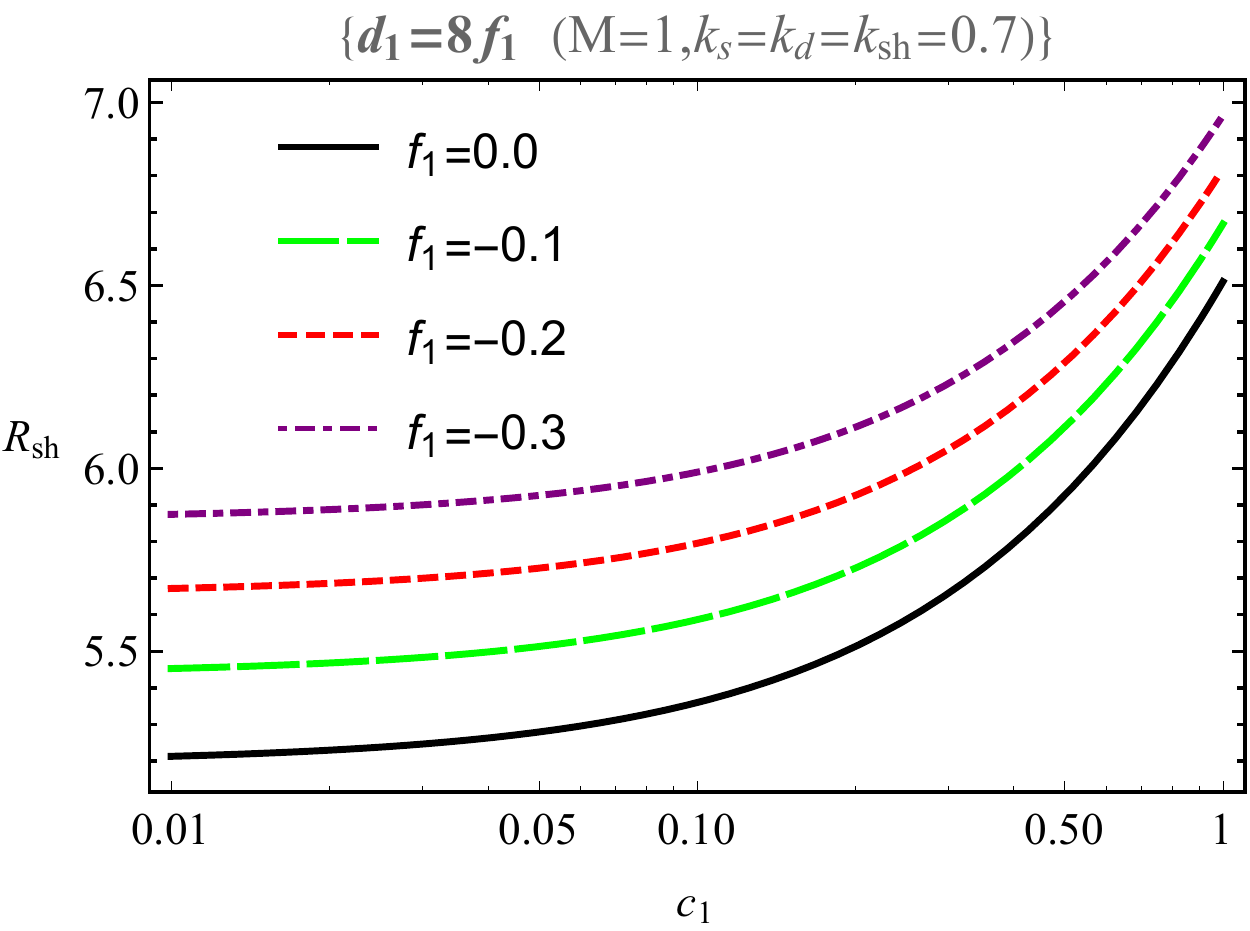}
    \includegraphics[scale=0.65]{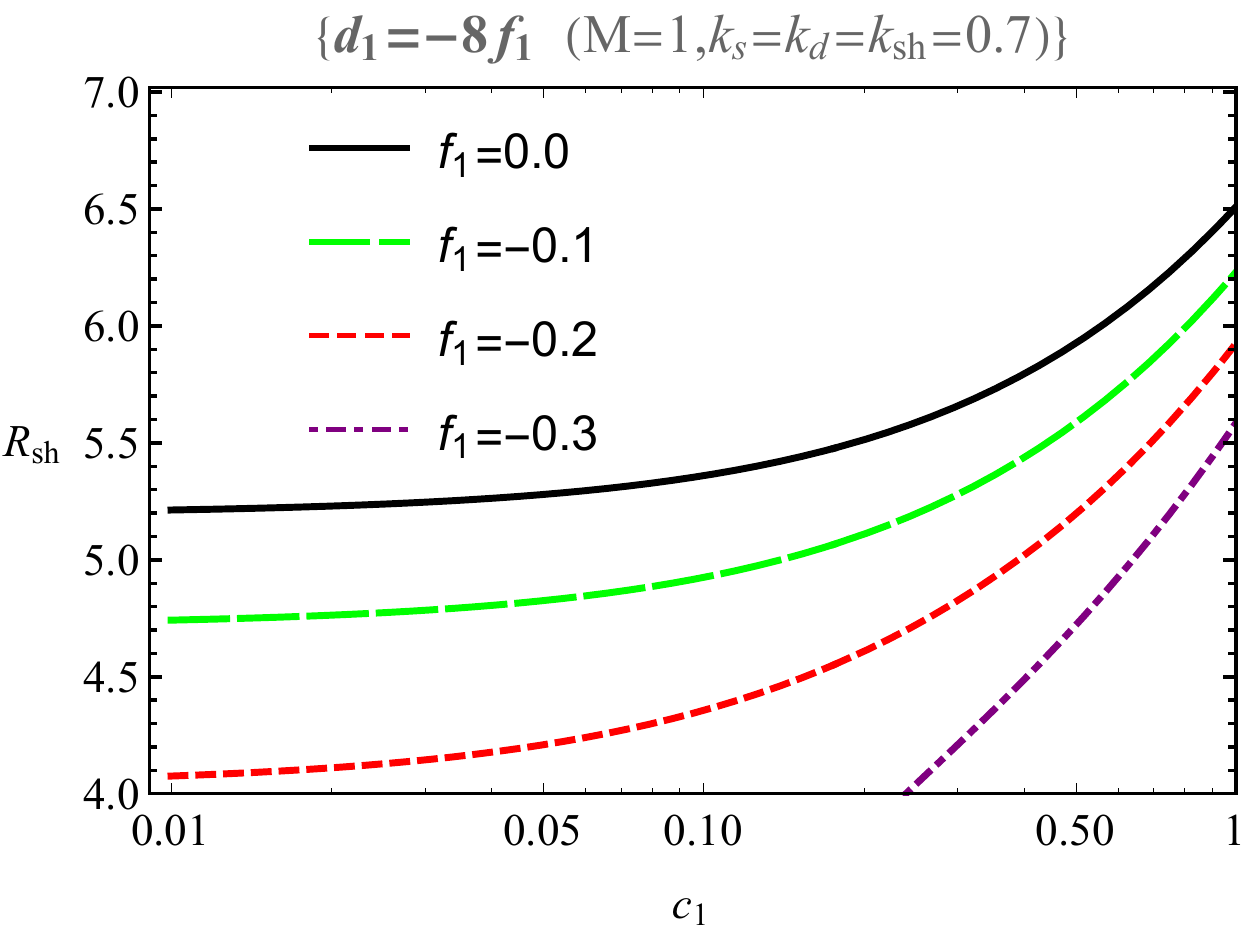}
    \includegraphics[scale=0.65]{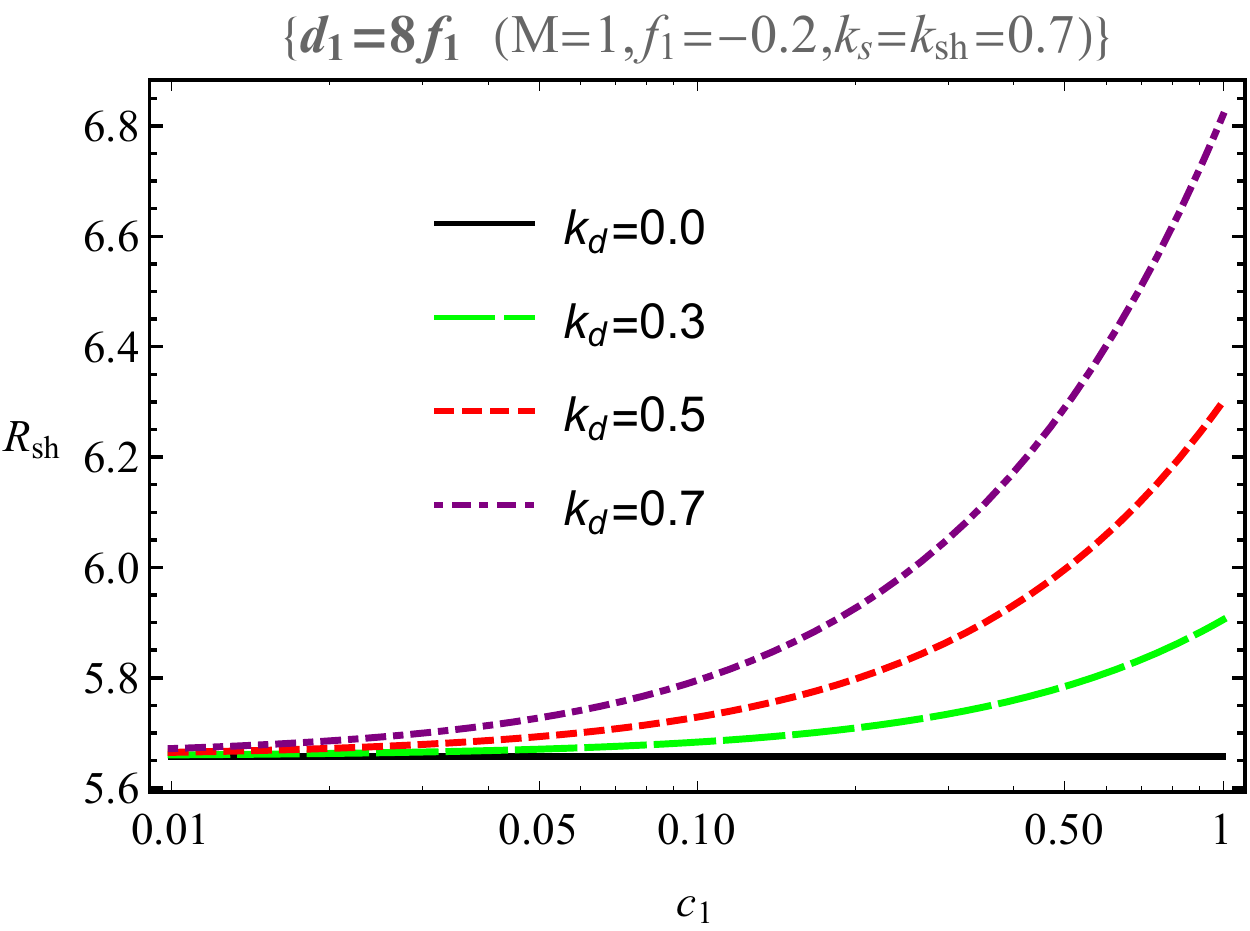}
    \includegraphics[scale=0.65]{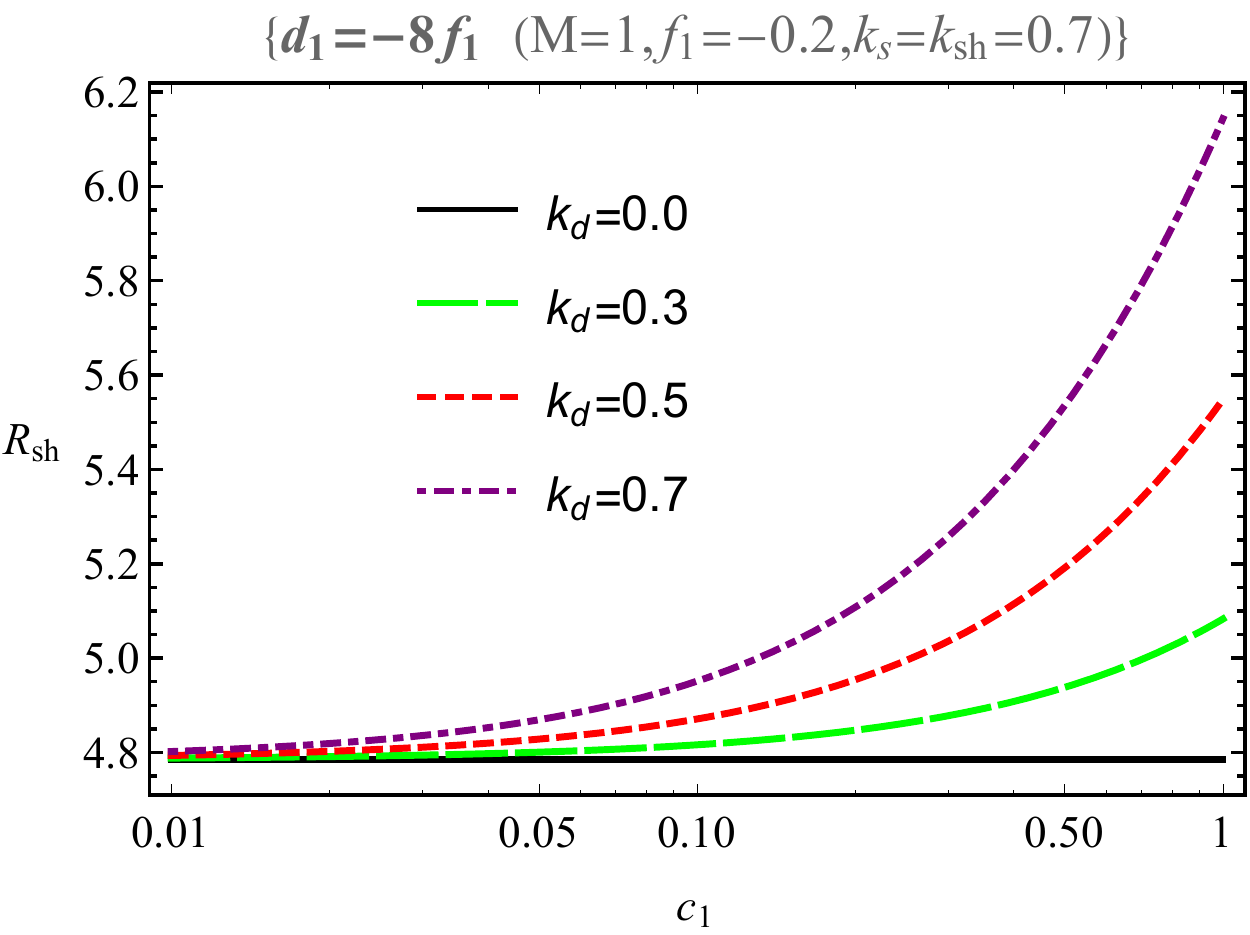}
    \includegraphics[scale=0.65]{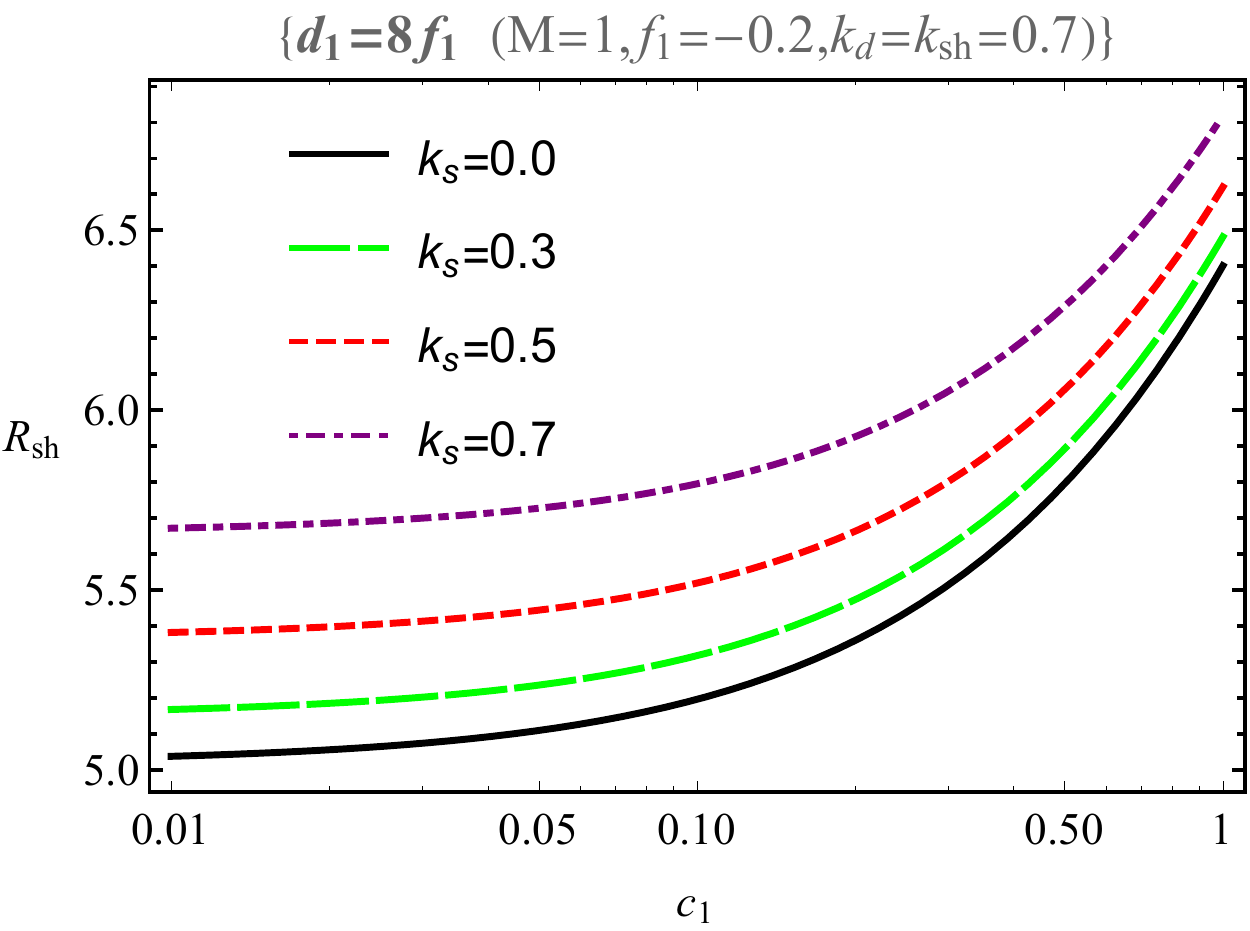}
    \includegraphics[scale=0.65]{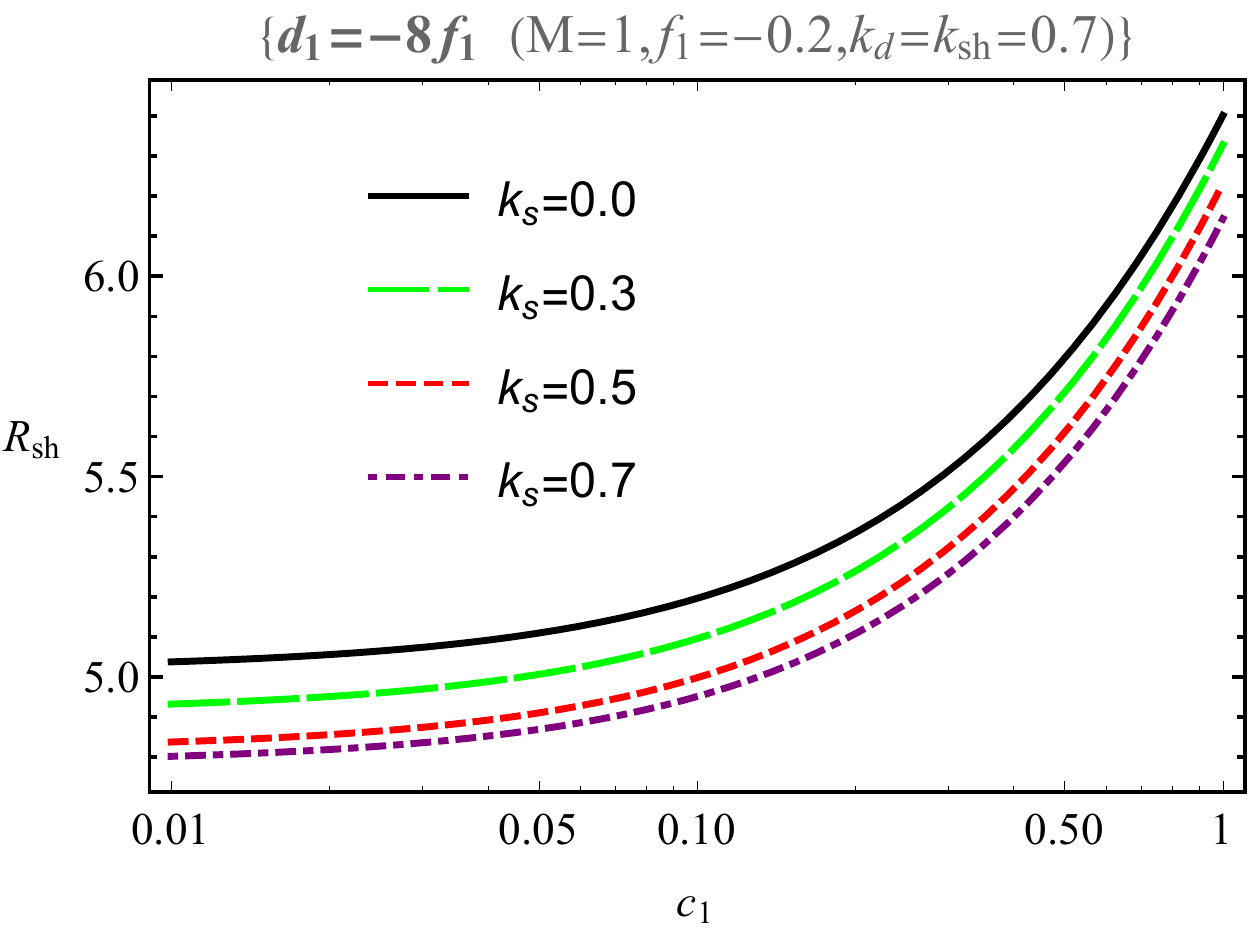}
    \includegraphics[scale=0.65]{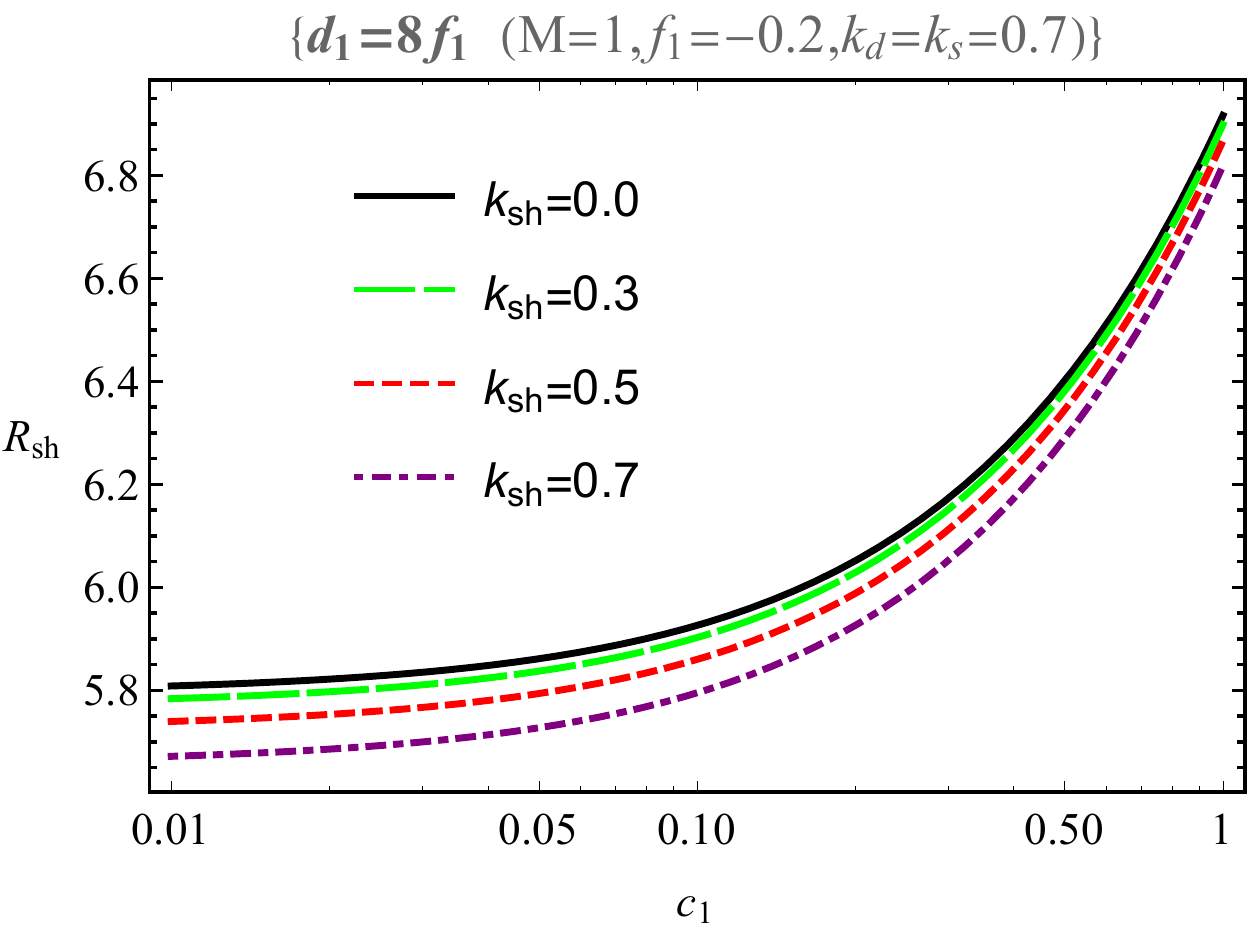}
    \includegraphics[scale=0.65]{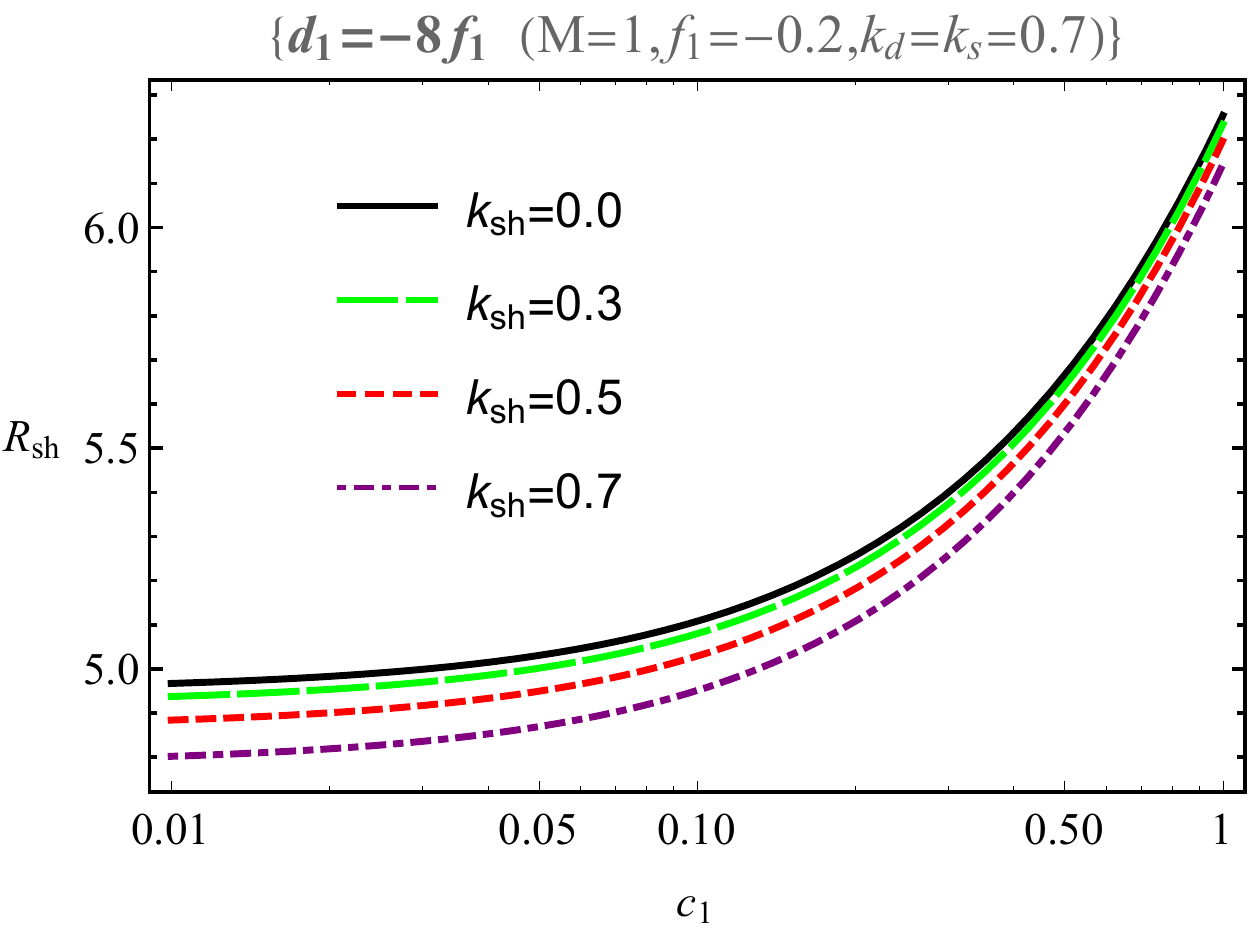}
    \caption{Black hole shadow's radius $r_h$ for cases $d_1=8f_1$ (Left panel) and for $d_1=-8f_1$ (Right panel) alongwith $c_1$ taking different values of $f_1,\; k_s,\; k_d, \;\&\; k_{sh} $}
    \label{plot:4}
\end{figure}

\begin{figure}
    \centering
    \includegraphics[scale=0.58]{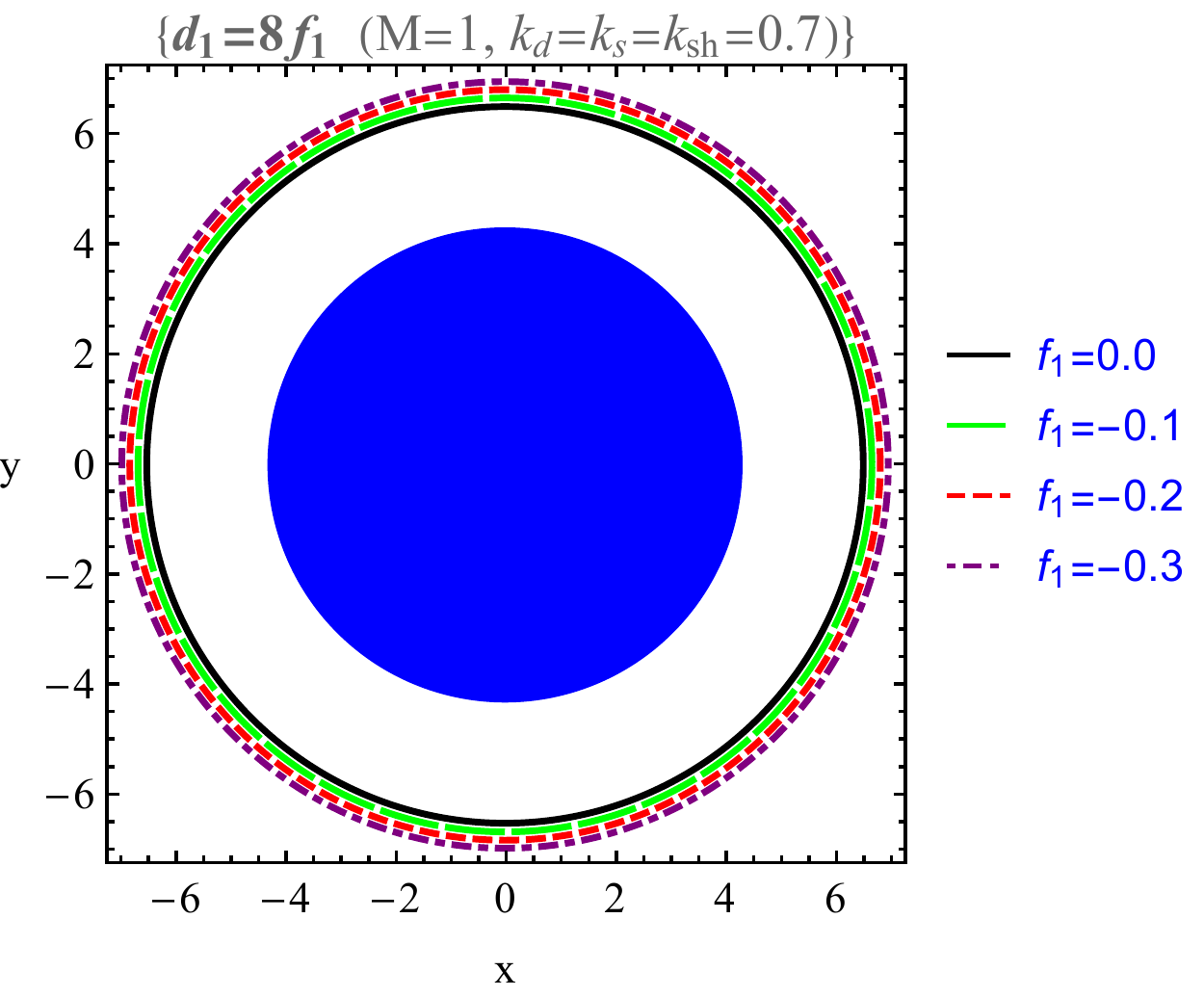}
    \includegraphics[scale=0.58]{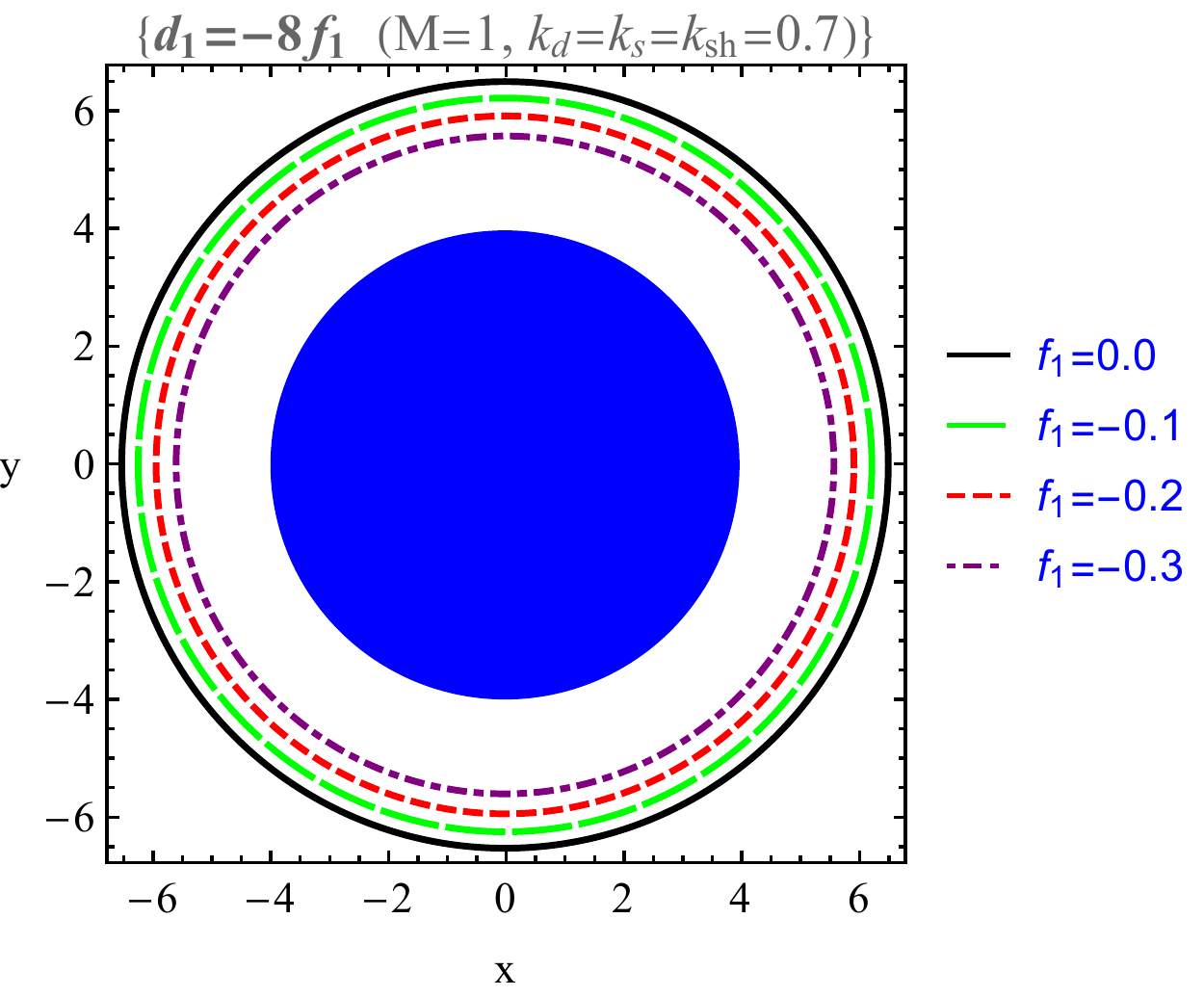}
    \includegraphics[scale=0.58]{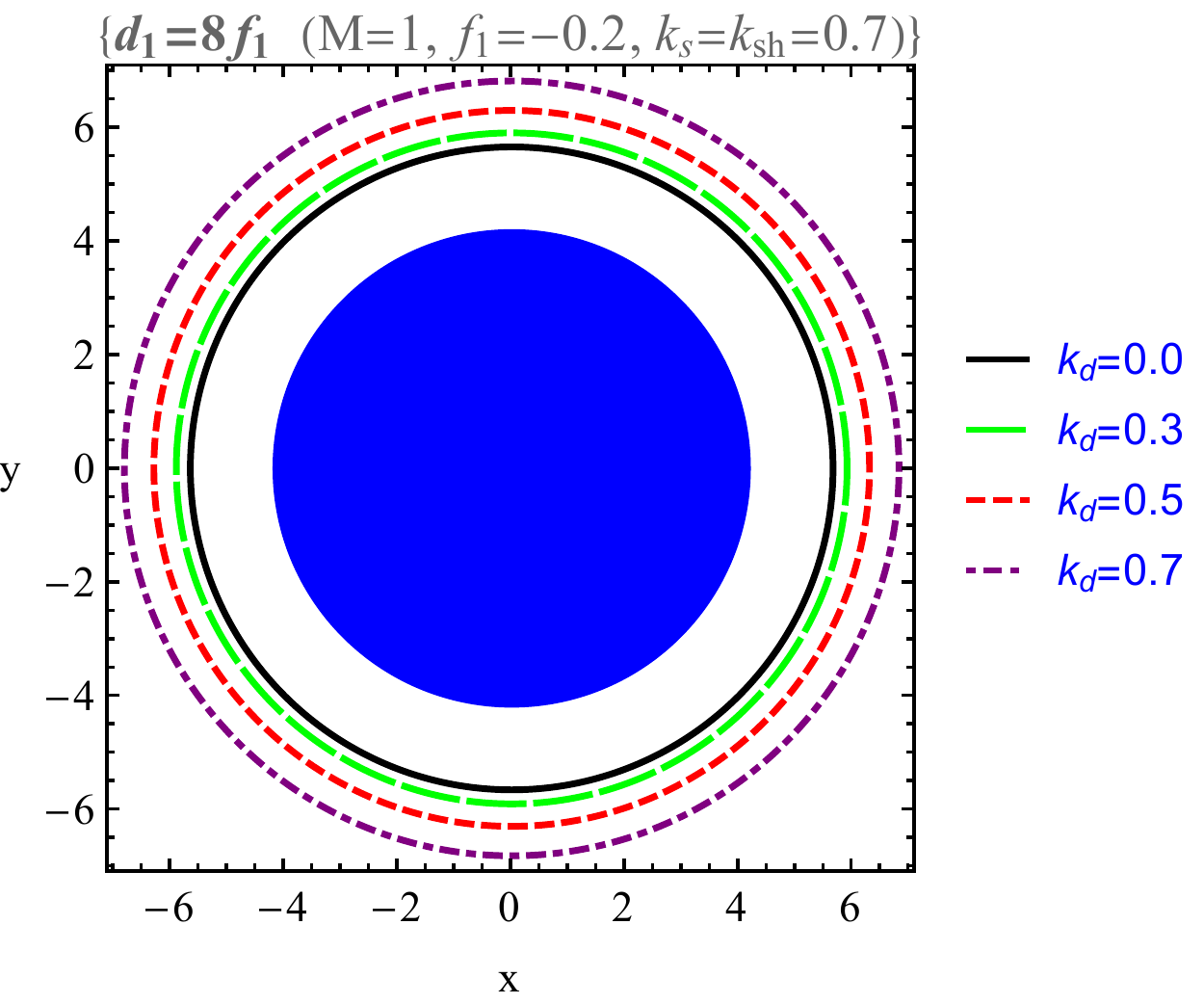}
    \includegraphics[scale=0.58]{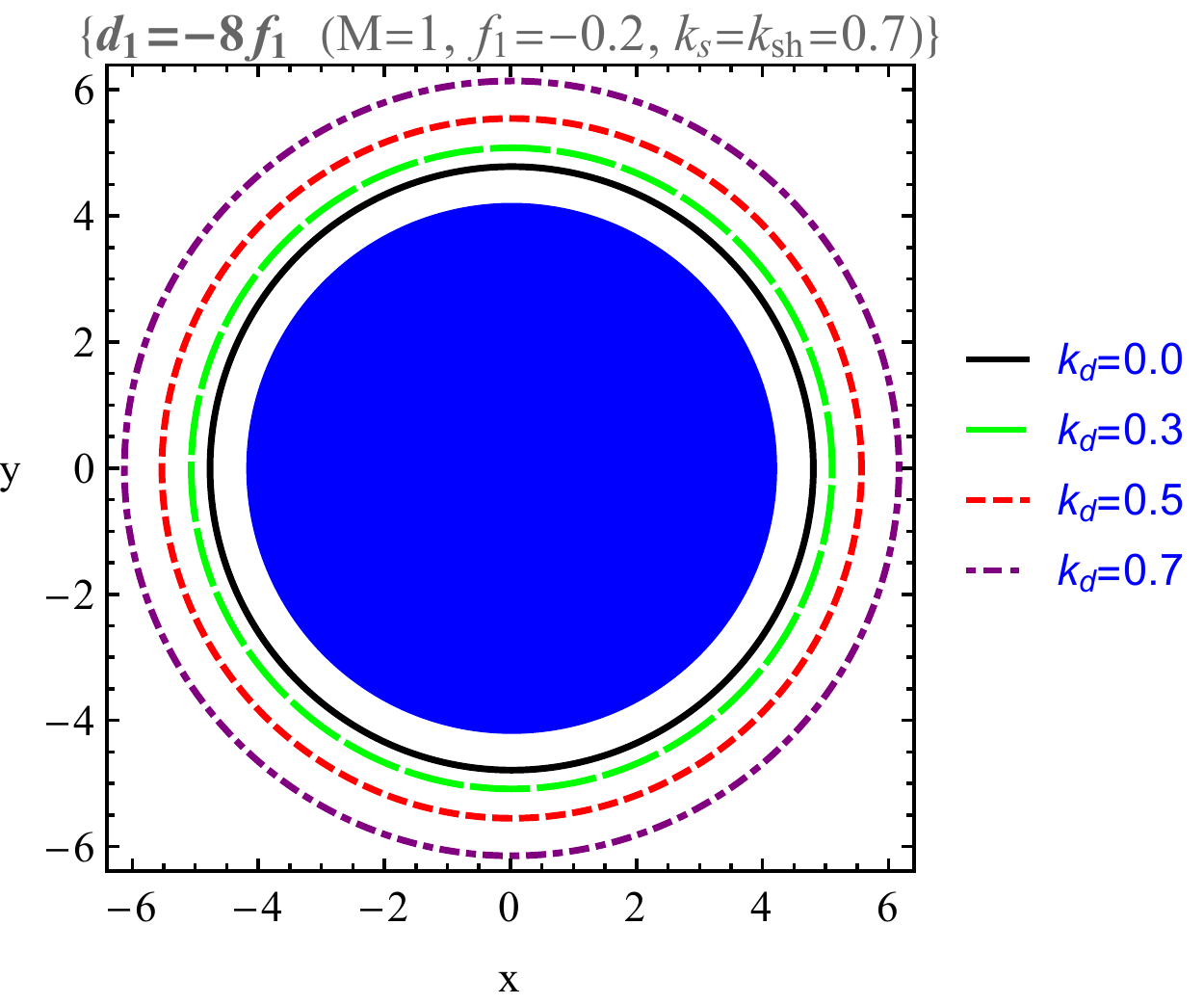}
    \includegraphics[scale=0.58]{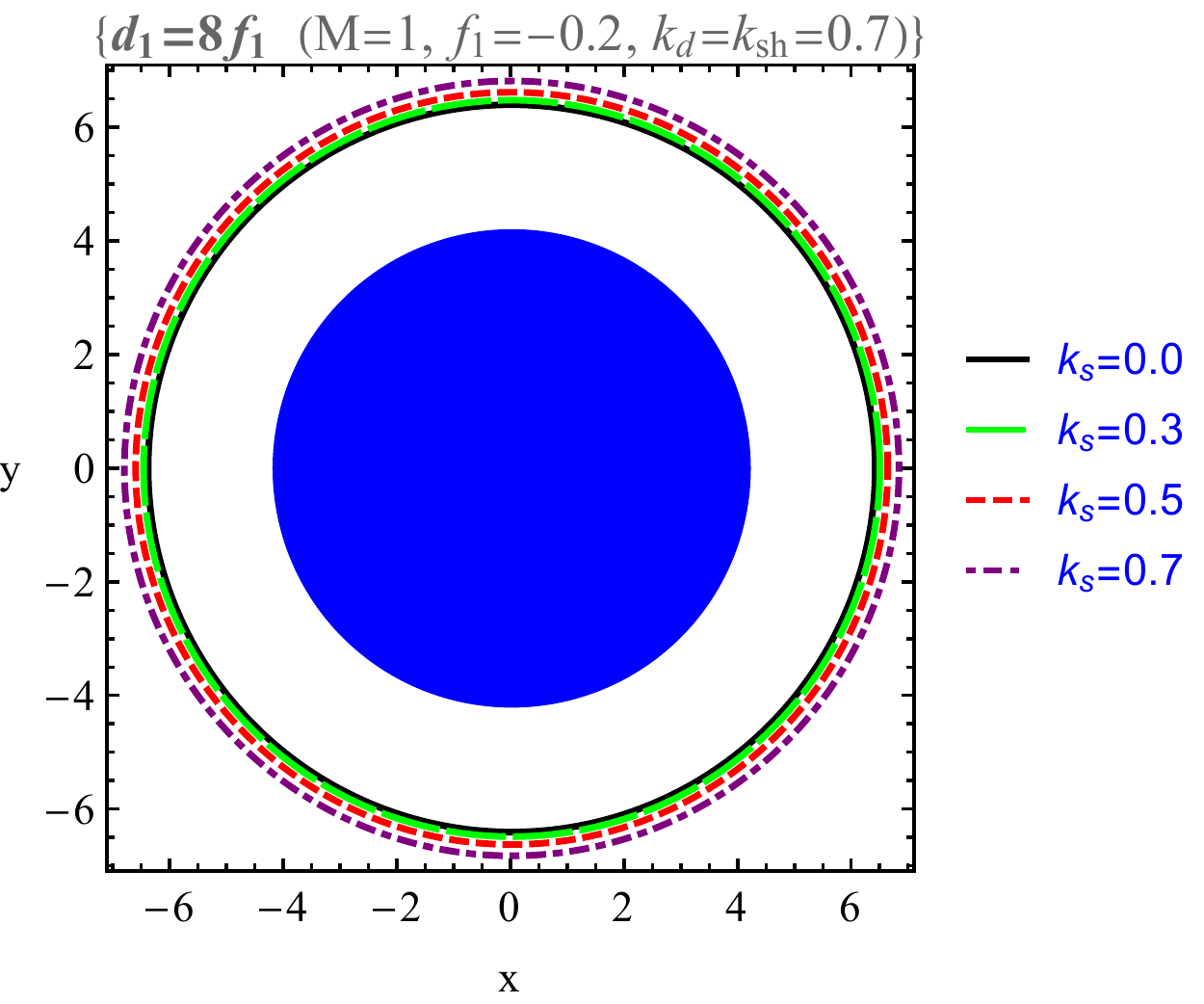}
    \includegraphics[scale=0.58]{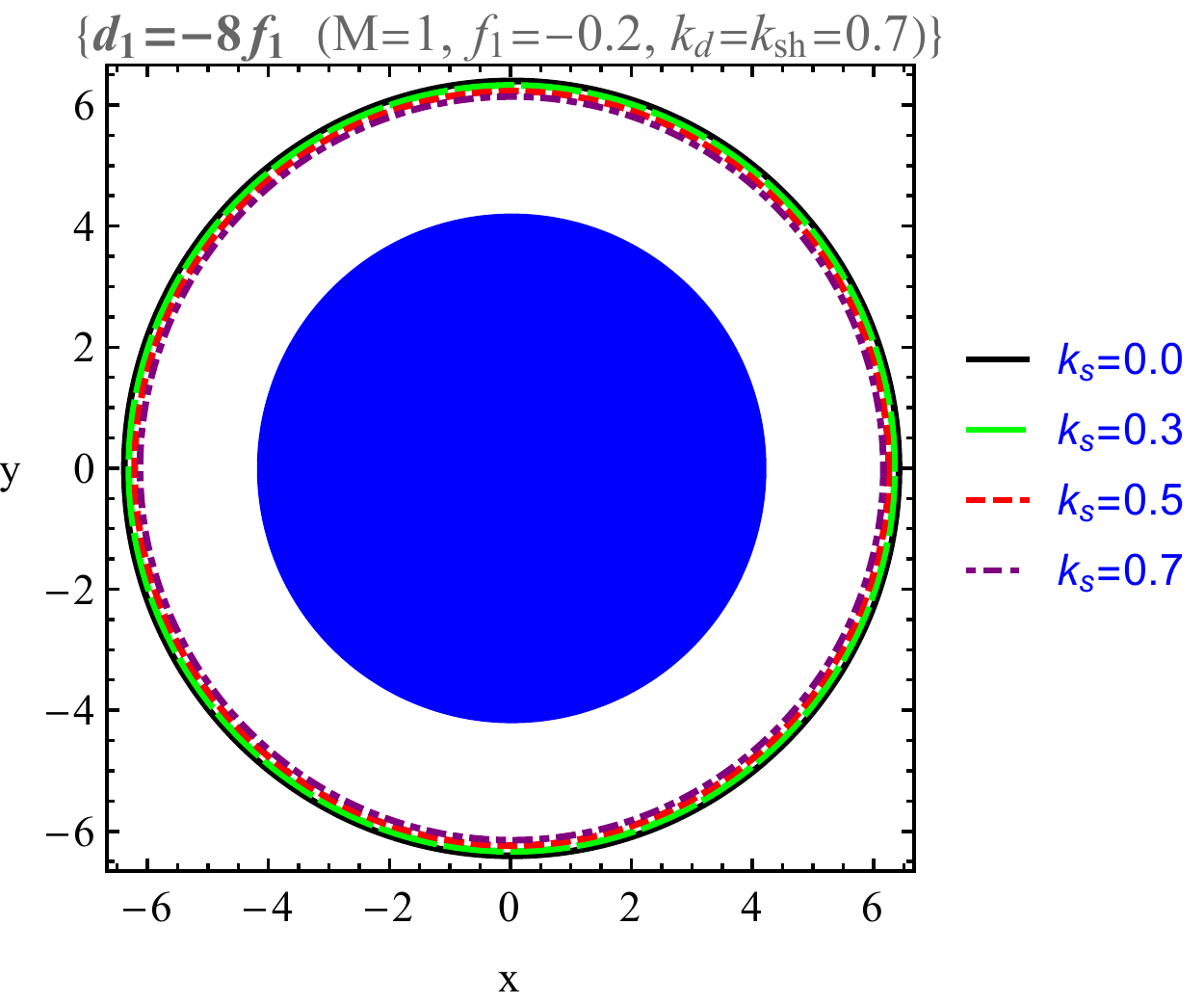}
    \includegraphics[scale=0.58]{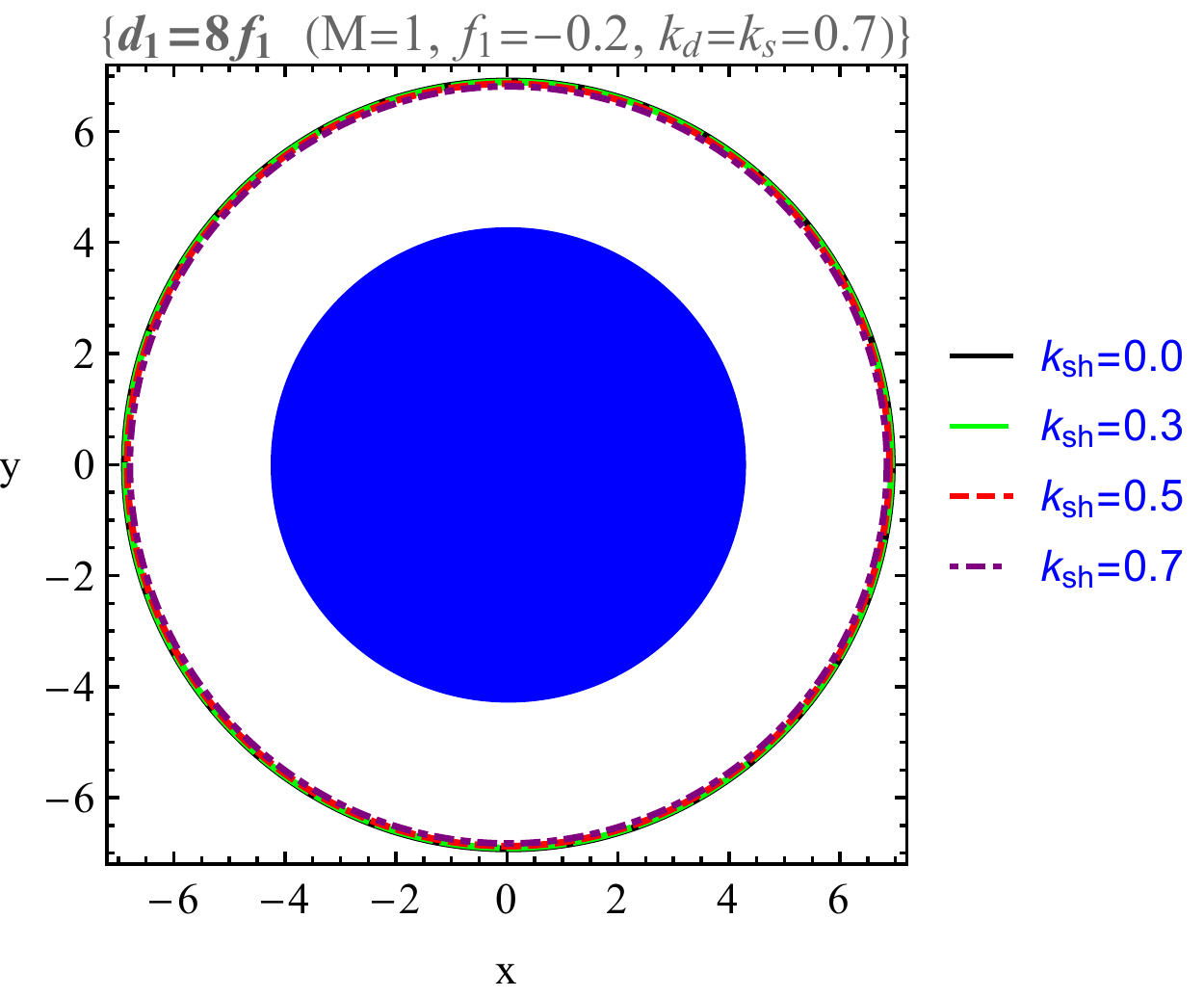}
    \includegraphics[scale=0.58]{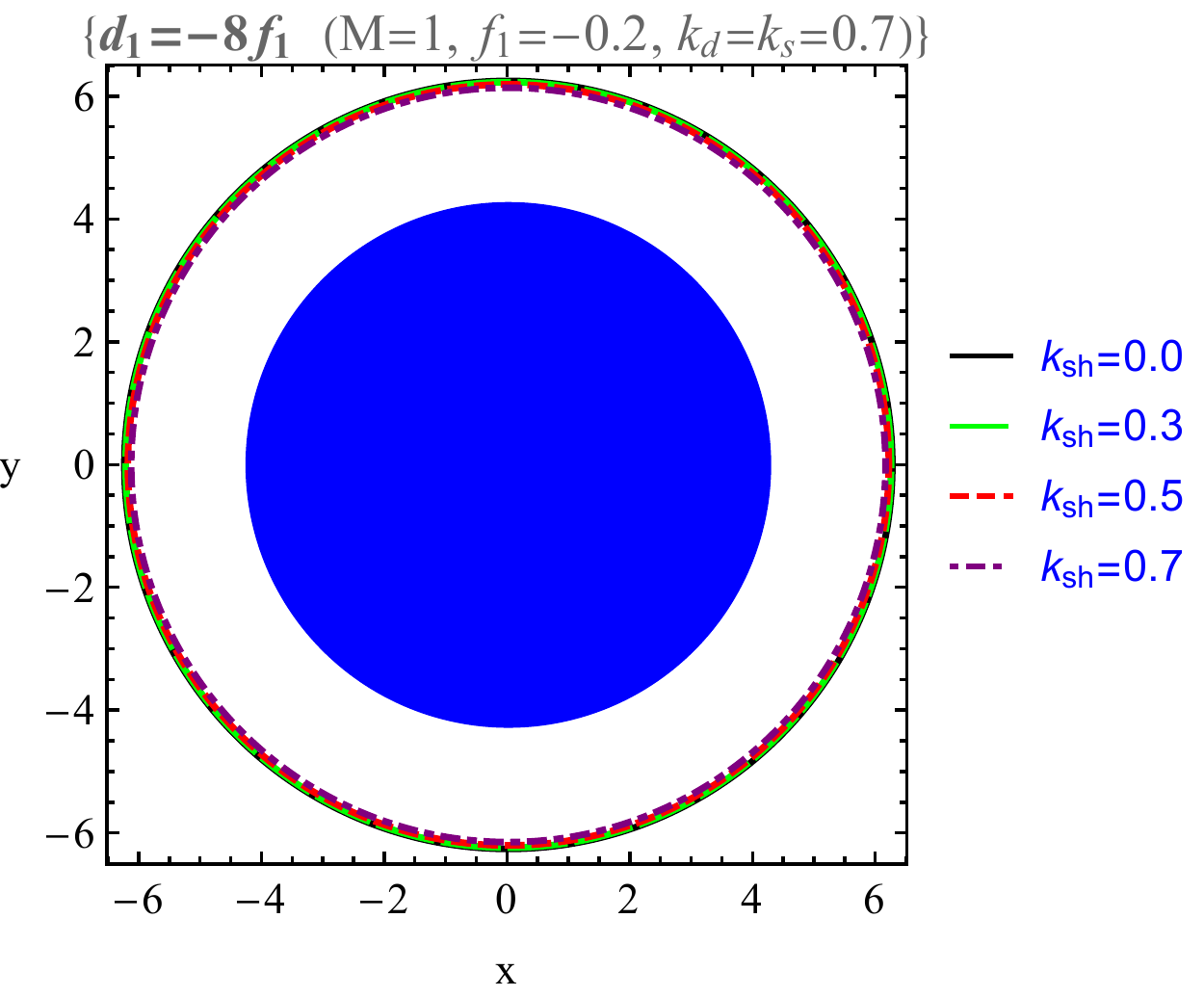}
    \caption{Black hole shadow's radius $r_h$ for cases $d_1=8f_1$ (Left panel) and $d_1=-8f_1$ (Right panel) along $c_1$ taking different values of $f_1,\; k_s,\; k_d, \;\&\; k_{sh} $}.
    \label{plot:5}
\end{figure}

The non-rotating case of BH shadow is expressed in Eq. (\ref{28}) whose graphical analysis shows the radius variations as given in Figs.~\ref{plot:4} and \ref{plot:5}. The propagation of the shadow's radius also follows a similar trend like the horizon radius for both cases. The shadow's radius for $d_1=8f_1$ is more than $d_1=-8f_1$.

\section{Oscillations of massive particles near the circular orbit}\label{A5}
\begin{figure}
    \centering
    \includegraphics[scale=0.52]{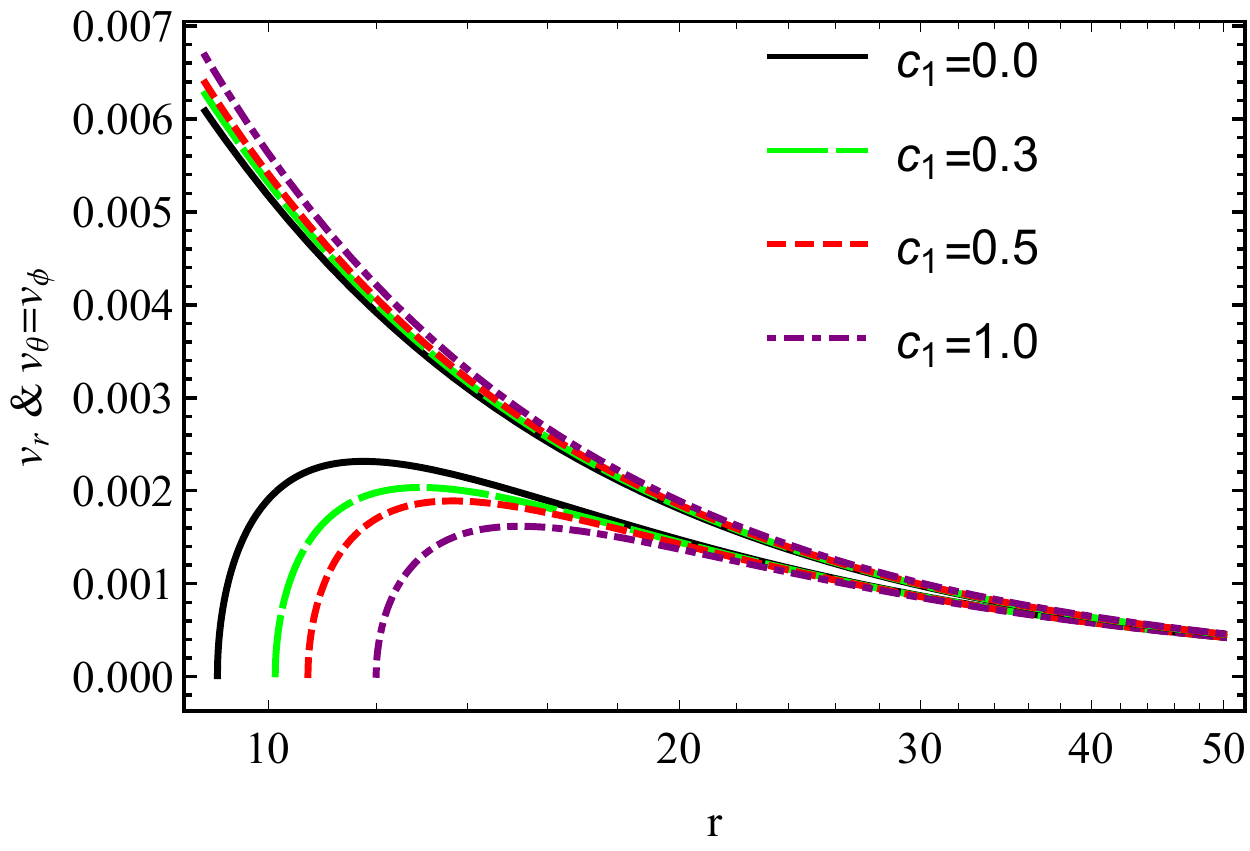}
    \includegraphics[scale=0.52]{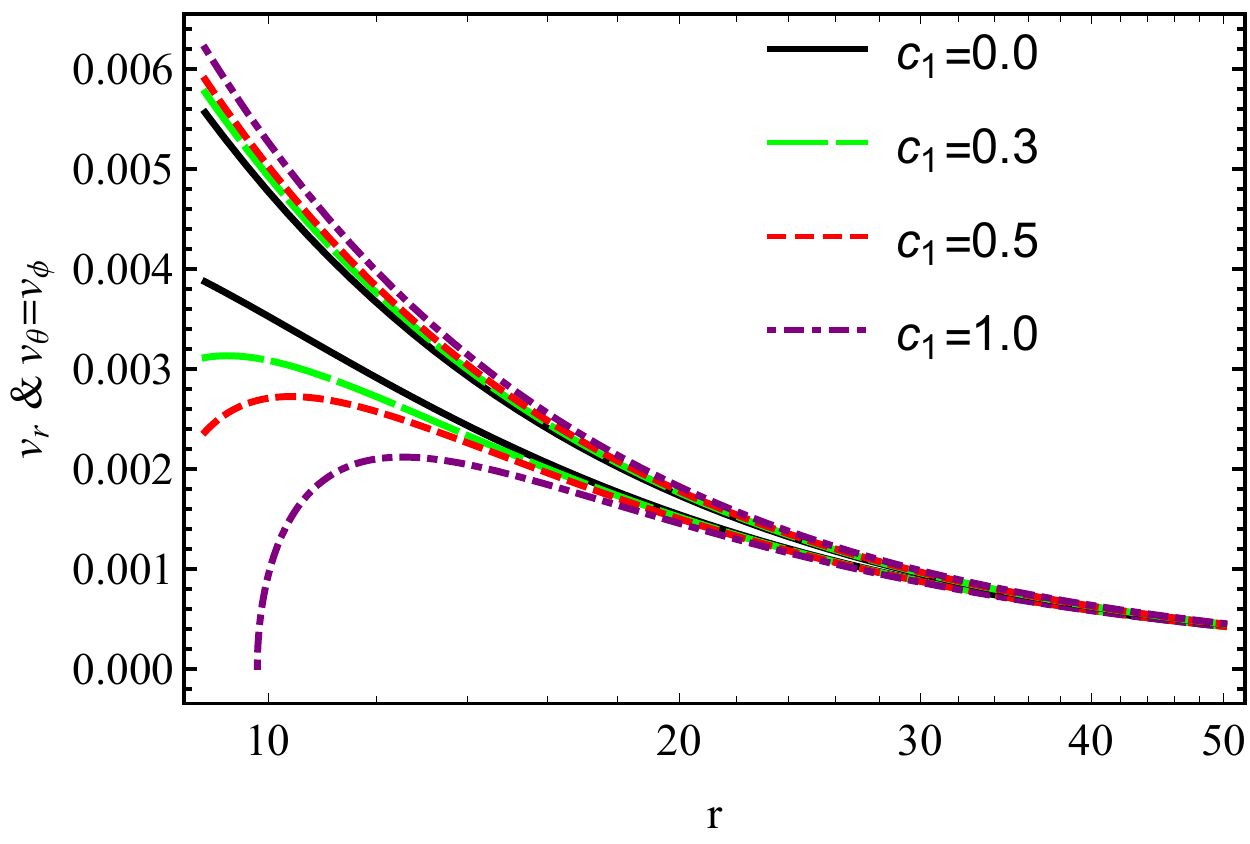}
    \includegraphics[scale=0.52]{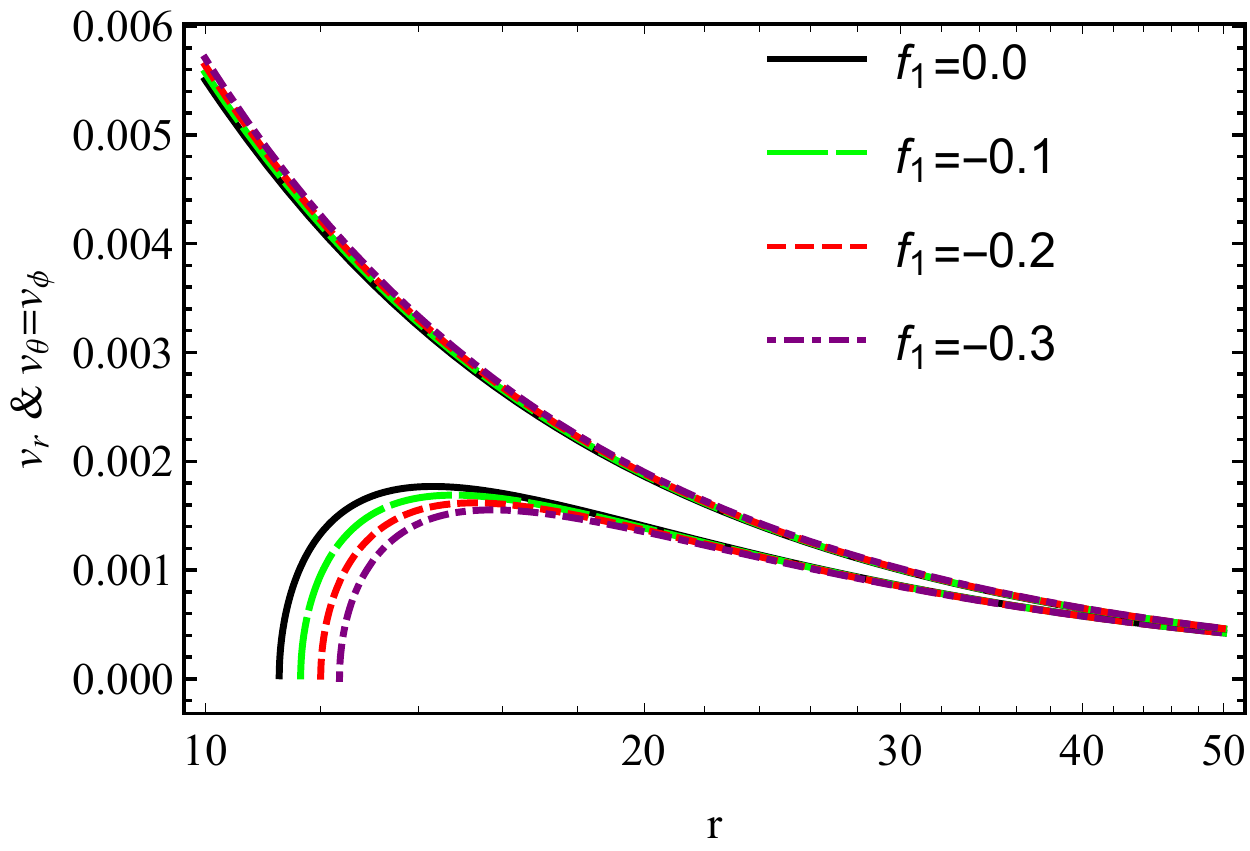}
    \includegraphics[scale=0.52]{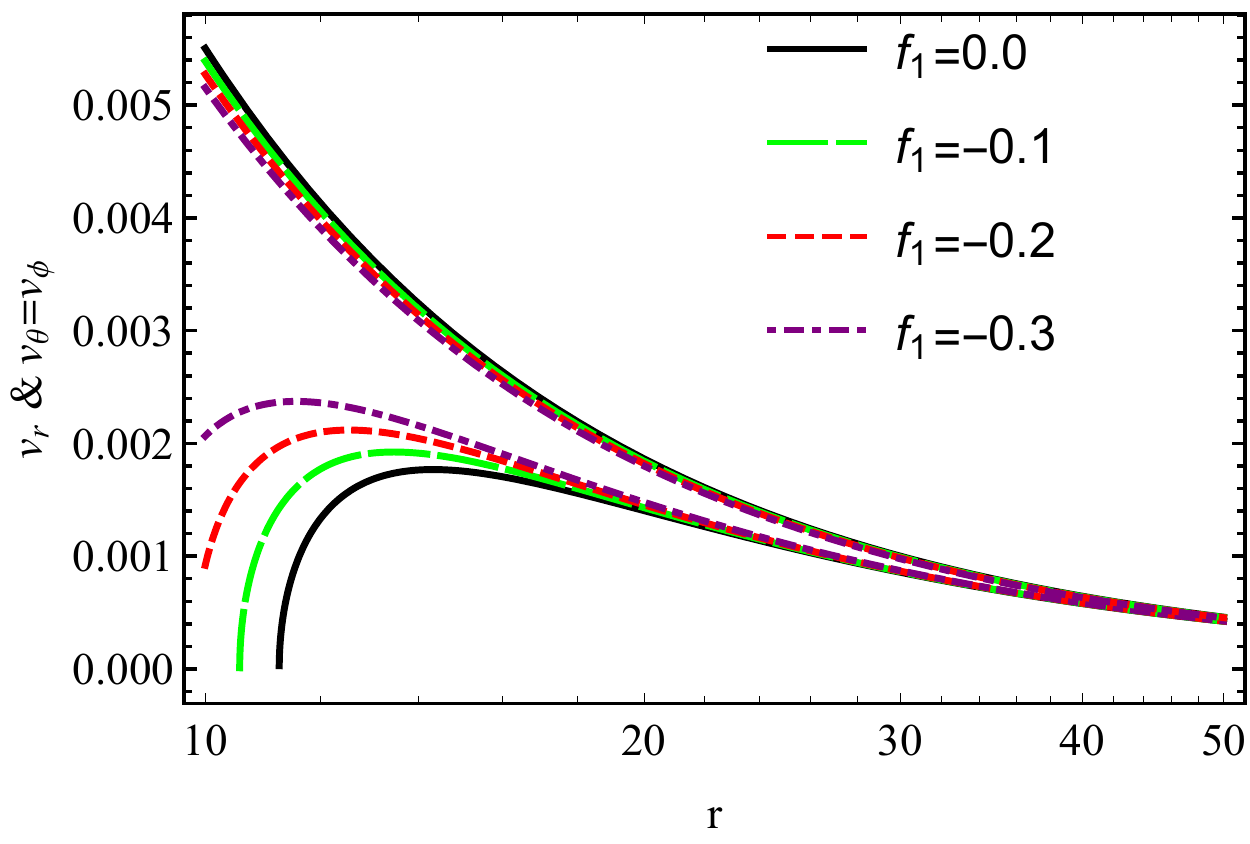}
    \includegraphics[scale=0.52]{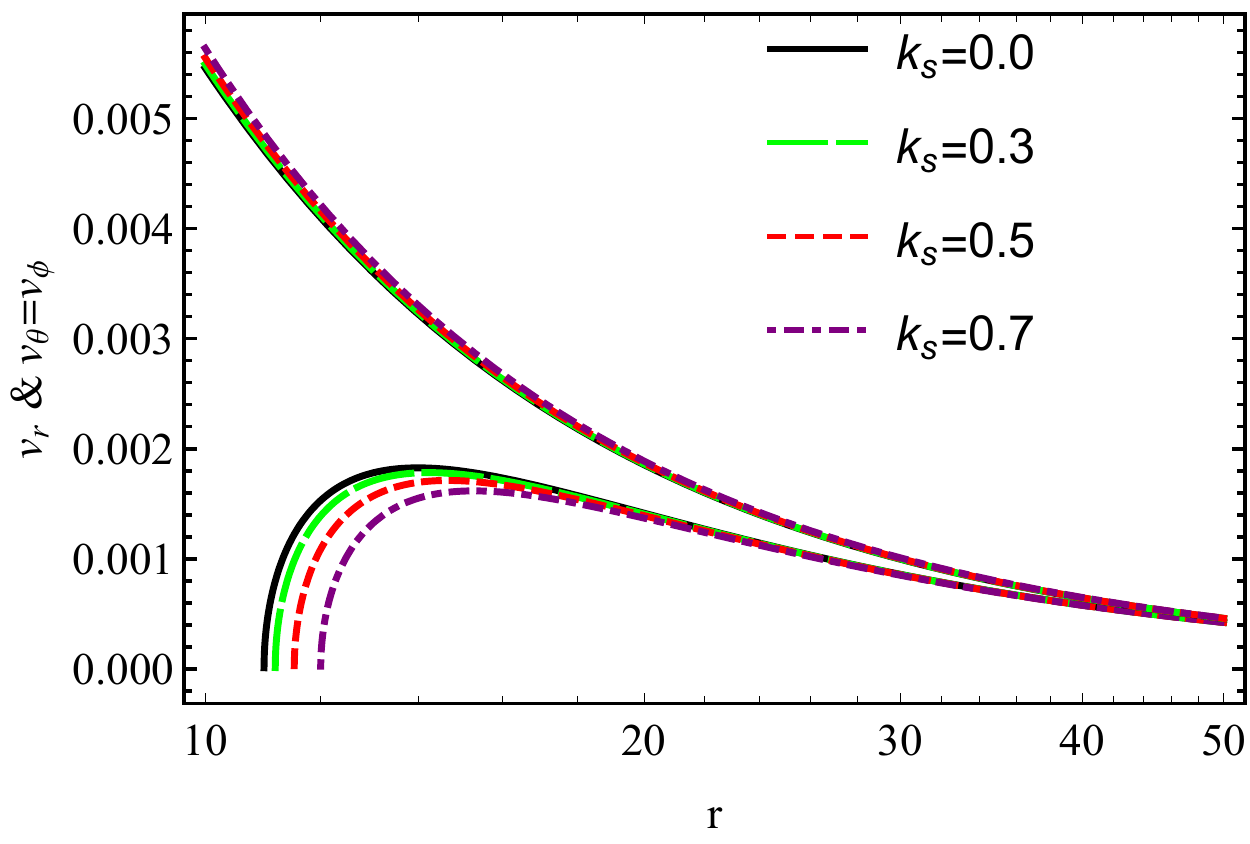}
    \includegraphics[scale=0.52]{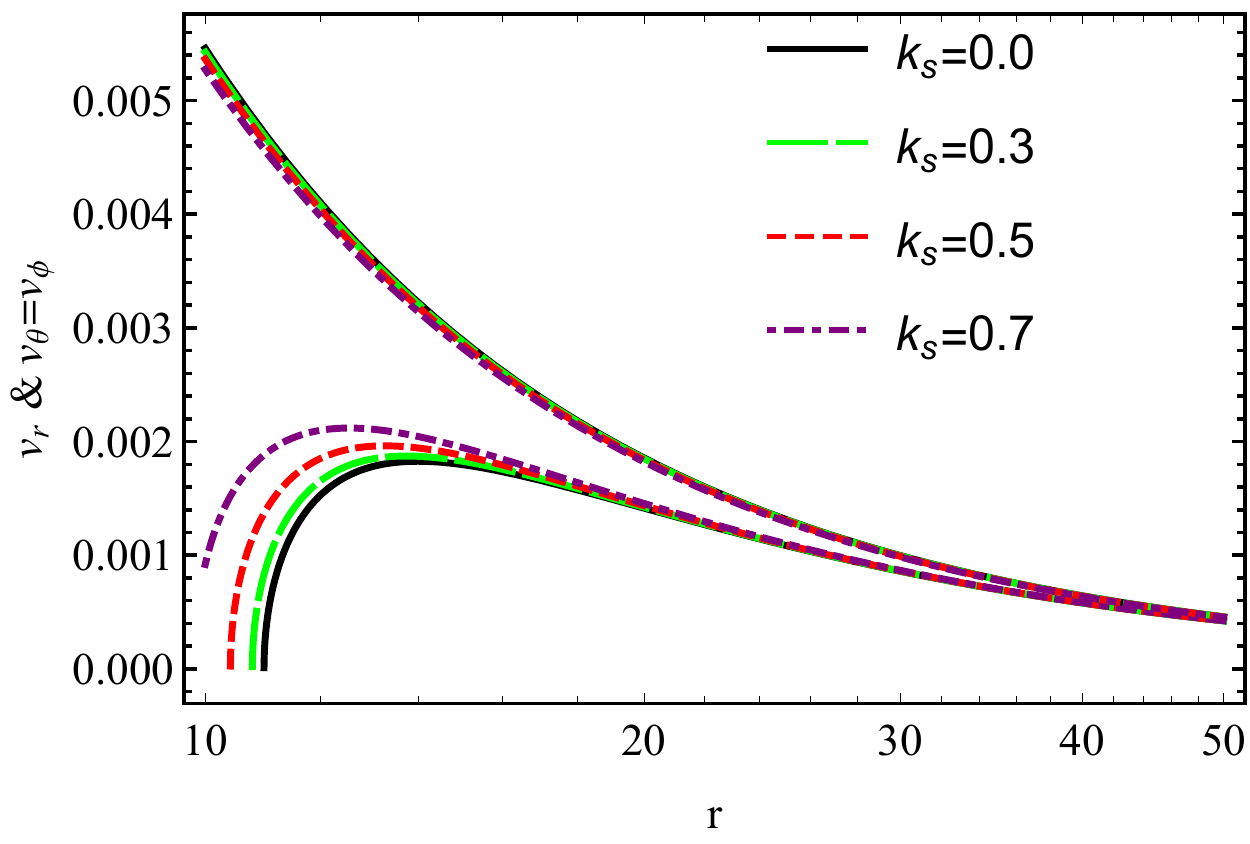}
    \includegraphics[scale=0.52]{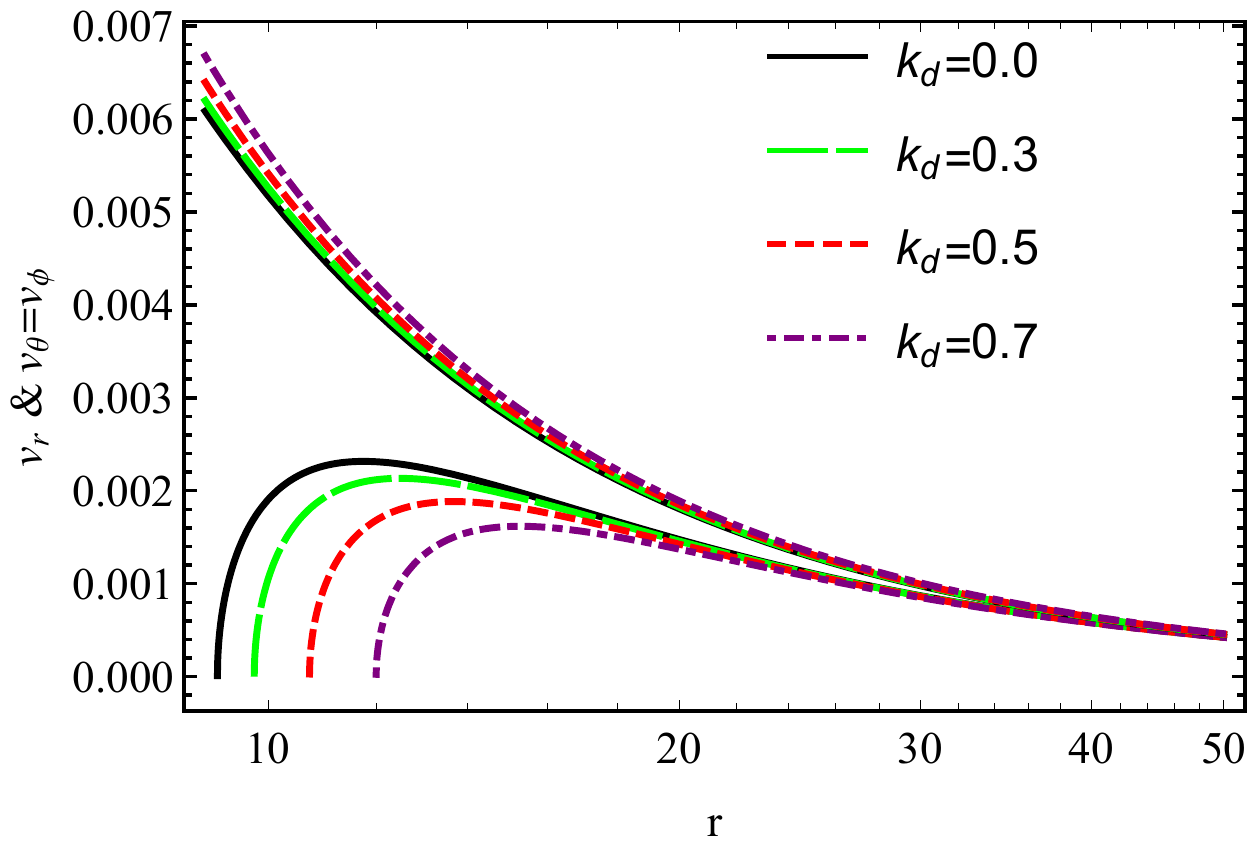}
    \includegraphics[scale=0.52]{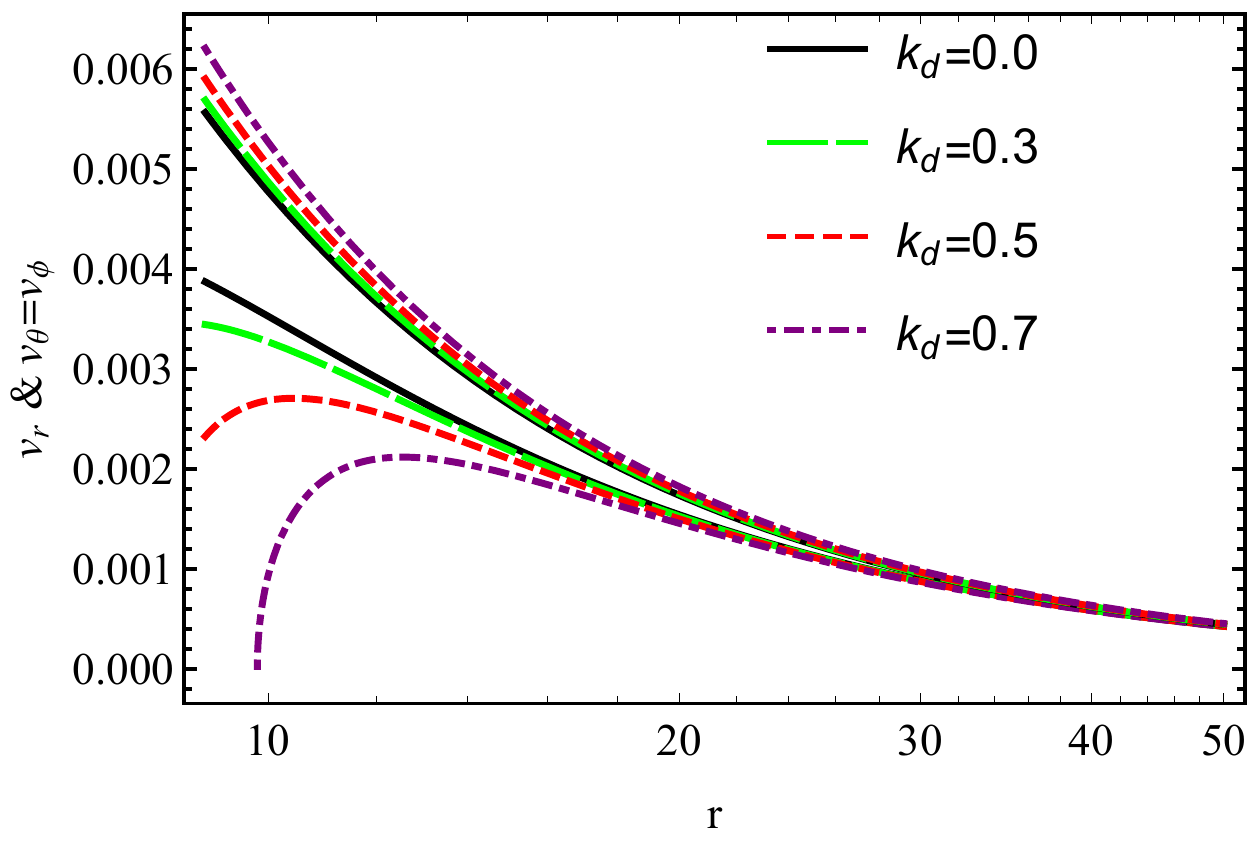}
    \includegraphics[scale=0.52]{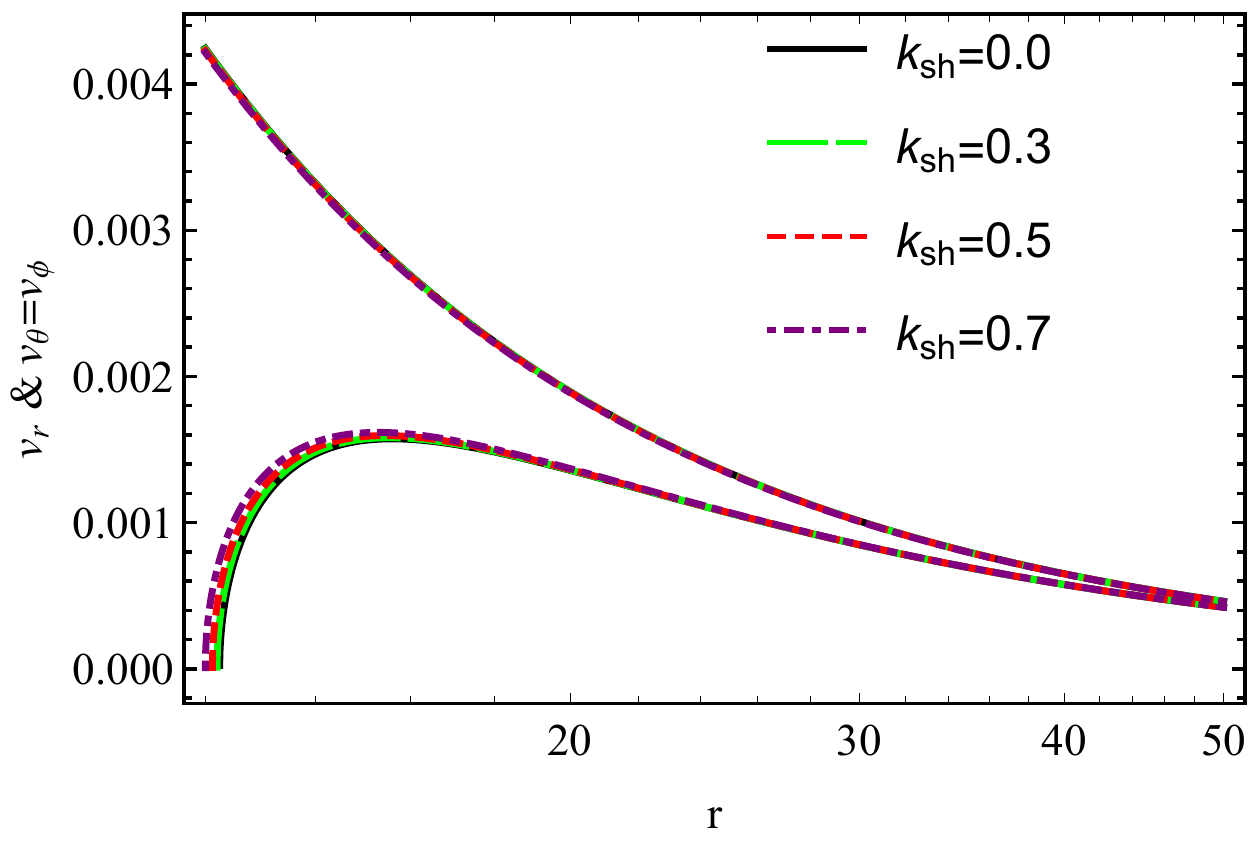}
    \includegraphics[scale=0.52]{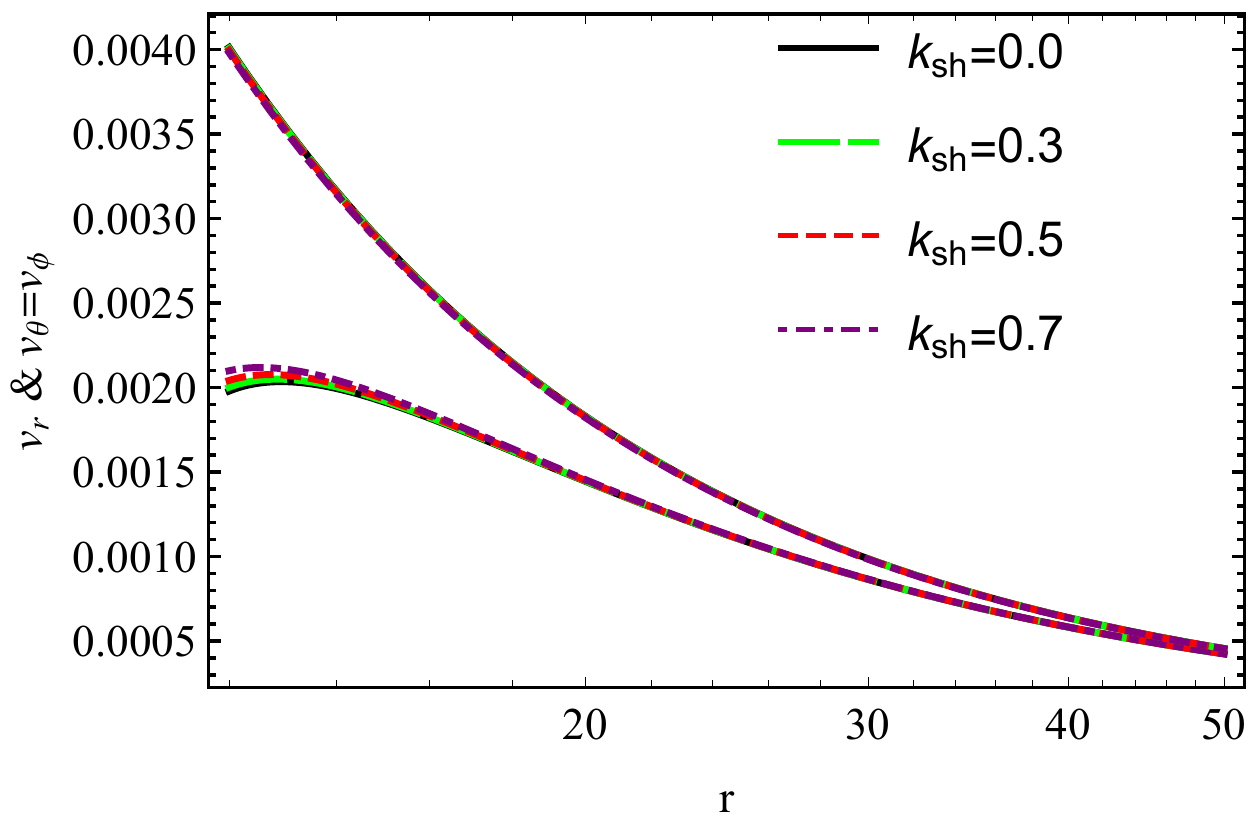}
    \caption{Radial and tangential frequencies $v_r\;\&\;v_\theta=v_\phi$ for case $d_1=8f_1$ (Left panel) and for case $d_1=-8f_1$ (Right panel) along $c_1$ for different values of $f_1,\; k_s,\; k_d, \;\&\; k_{sh} $. Here we consider the choice for fix values $M=1,c_1=1,\;f_1=-0.2,\;k_s=k_{sh}=k_d=0.7$.}
    \label{plot:6}
\end{figure}

In this section, we discuss the epicyclic frequencies related to the QPOs of test particle in the effective region of ISCO radius. The detailed formulation of these frequencies can be obtained from \cite{95}. Three forms of fundamental frequencies $(i)$ the Keplerian frequency or orbital frequency $v_\phi={w_\phi}/{2 \pi}$, $(ii)$ the radial epicyclic frequency $v_r={w_r}/{2 \pi}$ which is the the radial oscillations frequency in the surrounding of the mean orbit and $(iii)$ the vertical epicyclic frequency $v_\theta={w_\theta}/{2 \pi}$ which is the vertical oscillations frequency around the mean orbit, characterize the time like equatorial circular orbits. We have:
\begin{eqnarray}
w_\phi &=& \frac{d \phi}{d t}=\frac{-\frac{\partial g_{t \phi}}{\partial r} \pm \sqrt{\left(\frac{\partial g_{t \phi}}{\partial r}\right)^2-\frac{\partial g_H}{\partial r} \frac{\partial g_{\phi \phi}}{\partial r}}}{\frac{\partial g_{\phi \phi}}{\partial r}},\\
w_{r}^{2} &=& -\frac{1}{2 i^2 g_{r r}} \frac{\partial^2 V_{e f f}}{\partial r^2},\\
 w_{\theta}^{2}&=& -\frac{1}{2 t^2 g_{\theta \theta}} \frac{\partial^2 V_{e f f}}{\partial \theta^2} .
\end{eqnarray}
From the metric space given in Eq. (\ref{10}) give the following form of fundamental frequencies of the massive particles around the circular orbits
\begin{eqnarray}
    v_r=\frac{\sqrt{v_{1r}}}{2 \pi },\;\Rightarrow
    v_{1r}&=&\frac{\left(4 c_1 k_{d}^2-d_{1} k_{s}^2+2 f_{1} k_{sh}^2+2 M r-r^2\right)}{r^6 \left(8 c_1 k_{d}^2-2 d_{1} k_{s}^2+4 f_{1} k_{sh}^2+3 M r-r^2\right)^3} \Big[r^2 \left(4 c_1 k_{d}^2-d_{1} k_{s}^2+2 f_{1} k_{sh}^2+2 M r-r^2\right)\nonumber\\&& \Big[32 c_1^2 k_{d}^4+4 c_1 k_{d}^2 \left(-4 d_{1} k_{s}^2+8 f_{1} k_{sh}^2+3 \left(M^2-M r+r^2\right)\right)+2 d_{1}^2 k_{s}^4+2 f_{1} k_{sh}^2 \nonumber\\&&\left(3 \left(M^2-M r+r^2\right)-4 d_{1} k_{s}^2\right)-3 d_{1} k_{s}^2 M^2+3 d_{1} k_{s}^2 M r-3 d_{1} k_{s}^2 r^2+8 f_{1}^2 k_{sh}^4+M r^3\Big]\nonumber\\&+&2 r \left(4 c_1 k_{d}^2-d_{1} k_{s}^2+2 f_{1} k_{sh}^2+M r\right) \left(-8 c_1 k_{d}^2+2 d_{1} k_{s}^2-4 f_{1} k_{sh}^2-3 M r+r^2\right)\nonumber\\&& \left(-4 c_1 k_{d}^2 (M-2 r)+d_{1} k_{s}^2 M-2 d_{1} k_{s}^2 r-2 f_{1} k_{sh}^2 (M-2 r)+M r^2\right)-\Big[12 c_1 k_{d}^2-3 d_{1} k_{s}^2\nonumber\\&+&6 f_{1} k_{sh}^2+2 M r\Big] \left(4 c_1 k_{d}^2-d_{1} k_{s}^2+2 f_{1} k_{sh}^2+2 M r-r^2\right) \Big[-8 c_1 k_{d}^2+2 d_{1} k_{s}^2-4 f_{1} k_{sh}^2\nonumber\\&-&3 M r+r^2\Big]^2\Big],\\
v_\phi=v_\theta &=& \frac{\sqrt{4 c_1 k_d^2-d_1k_s^2+2 f_1k_{sh}^2+M r}}{2 \pi  r^2}.
\end{eqnarray}
These frequencies depend upon the spacetime parameters $f_1,\;d_1,\;c_1$, dilation, spin and sharecharges $k_d,\;k_s,\;\&\;k_{sh}$. We plot these frequencies in Fig.~\ref{plot:6}. Overall values of these frequencies are more for $d_1=8f_1$ than $d_1=-8f_1$. Specifically one can note that, values of $v_r,\;v_{\theta}$ increase and decrease with increase in $c_1,\;k_d,\;k_{sh}$, values of $v_r,\;v_{\theta}$ increase and decrease with increase in $f_1,\;k_s$ for $d_1=8f_1$ but decrease and increase in $d_1=-8f_1$.

\begin{figure}
    \centering
    \includegraphics[scale=0.52]{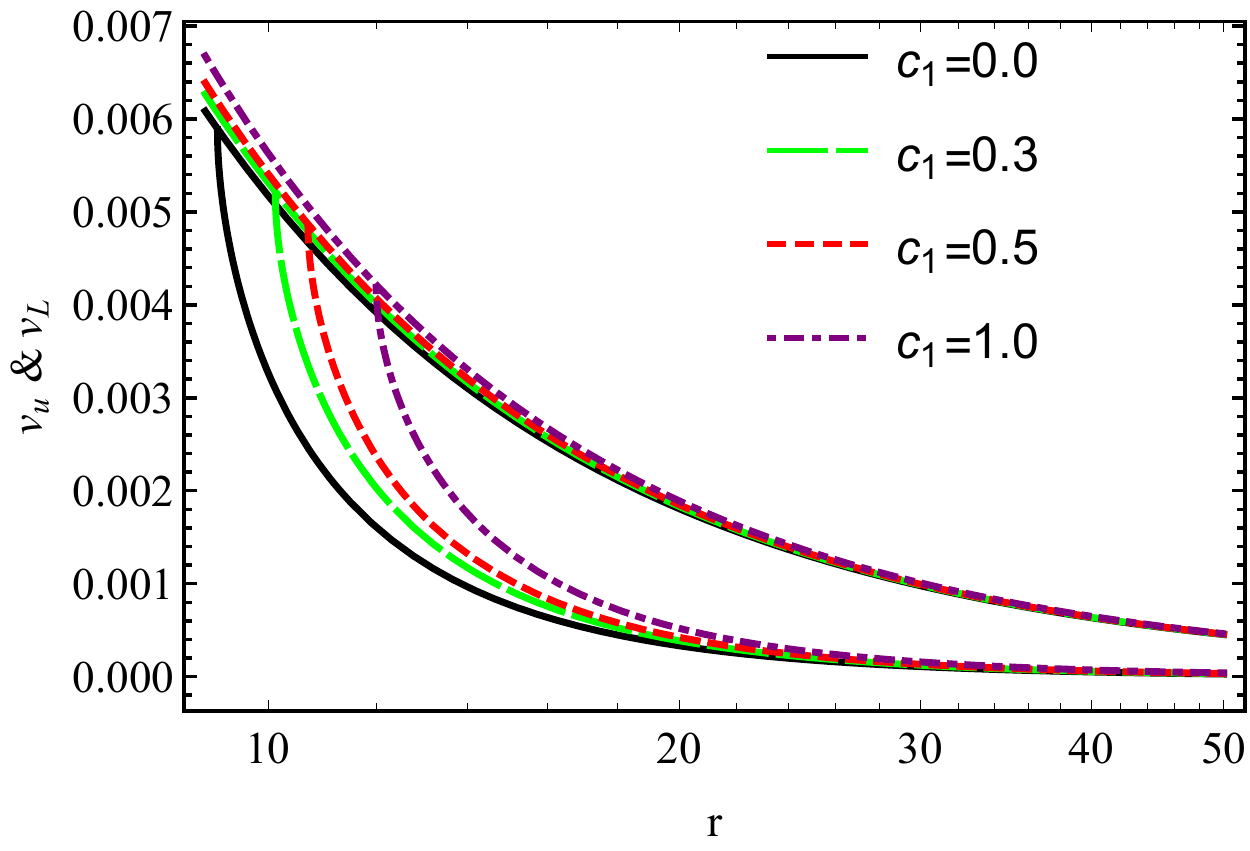}
    \includegraphics[scale=0.52]{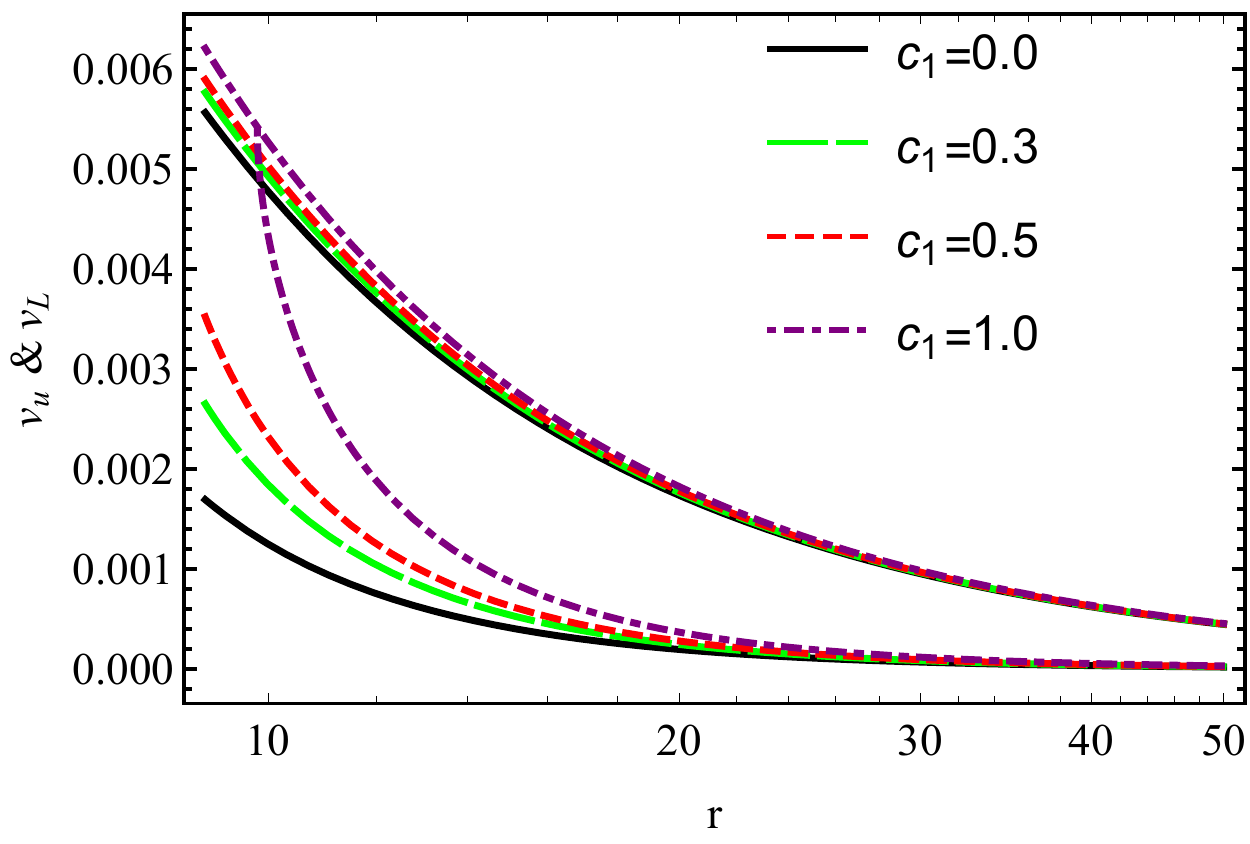}
    \includegraphics[scale=0.52]{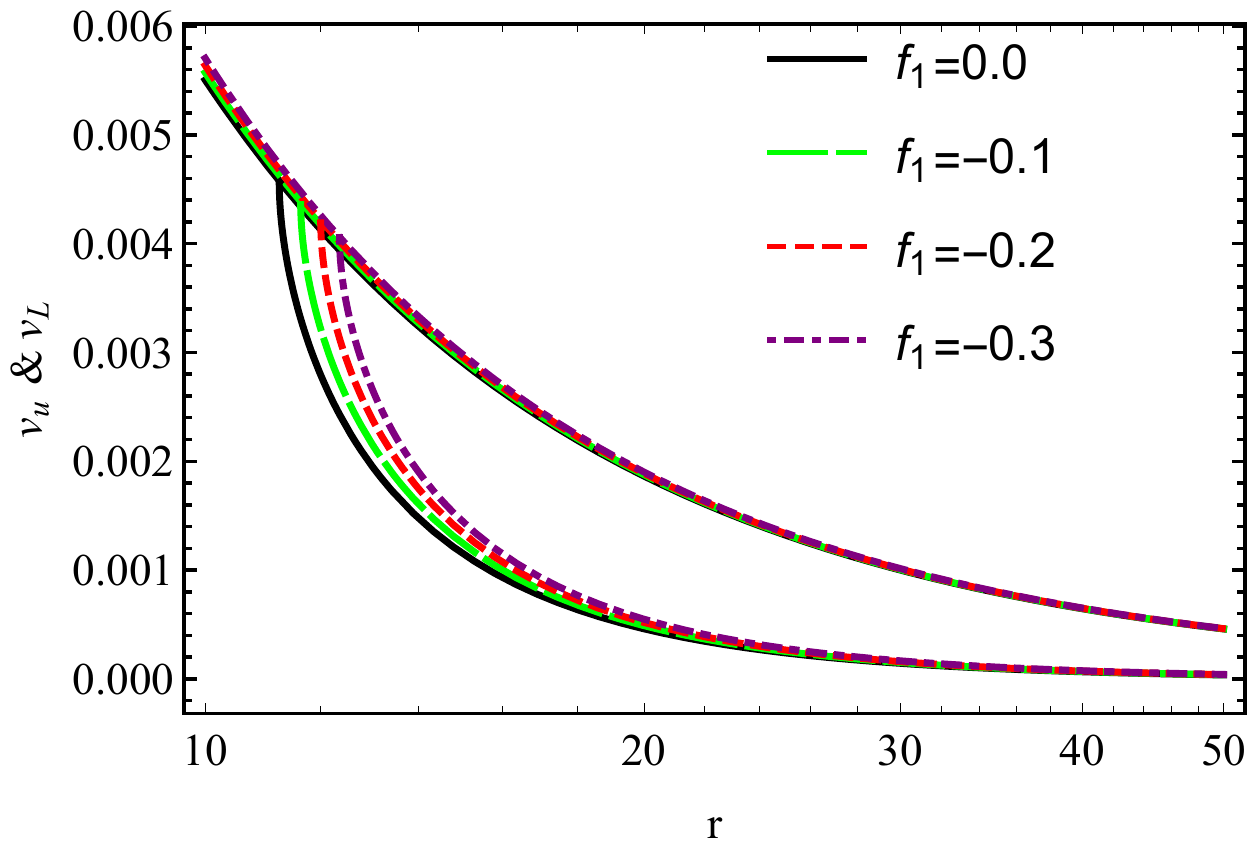}
    \includegraphics[scale=0.52]{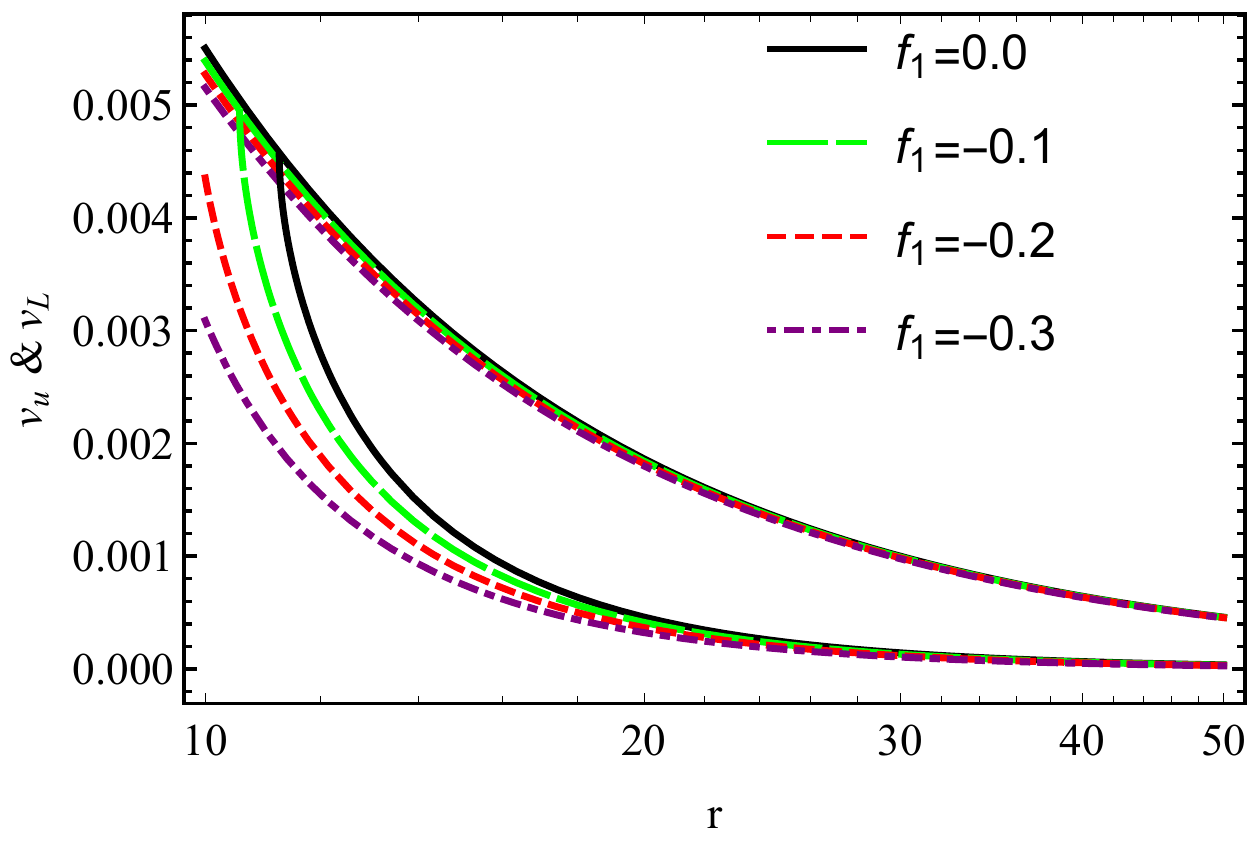}
    \includegraphics[scale=0.52]{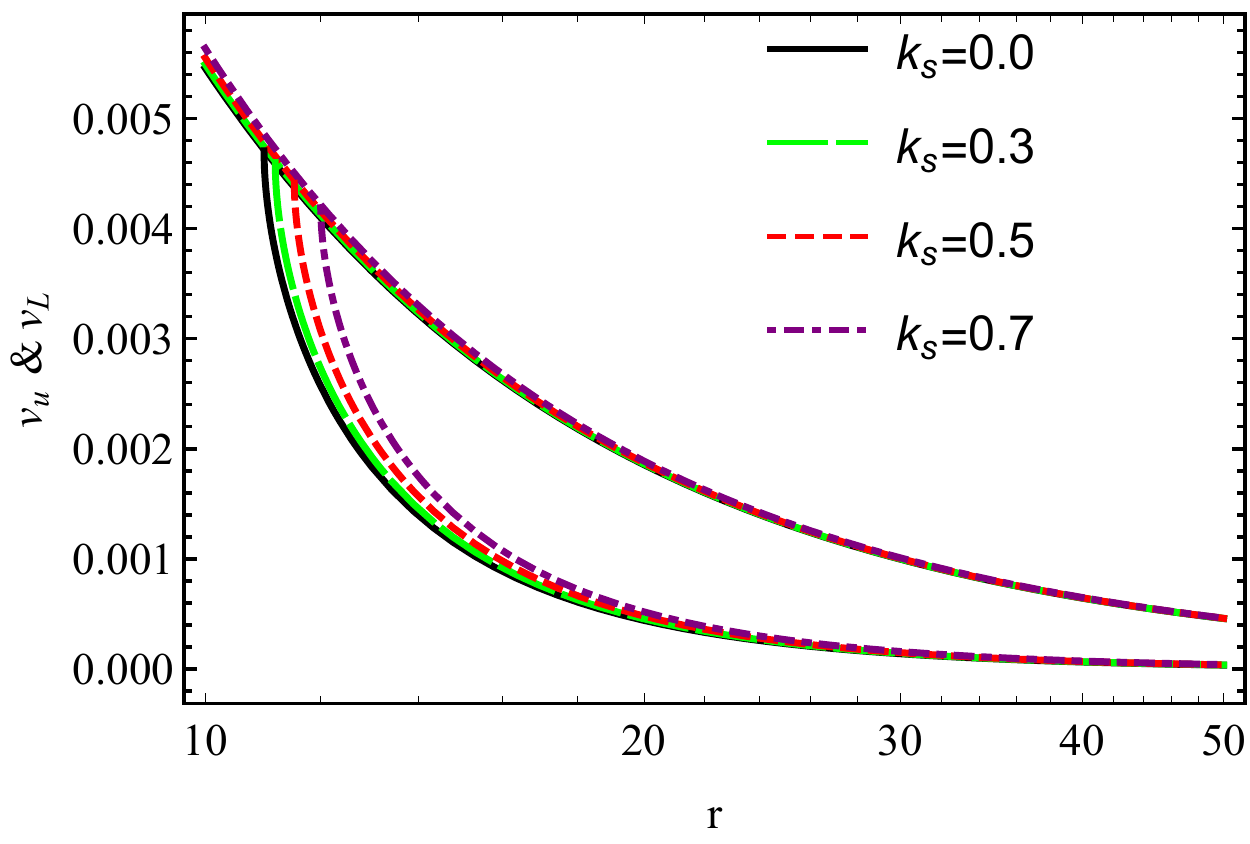}
    \includegraphics[scale=0.52]{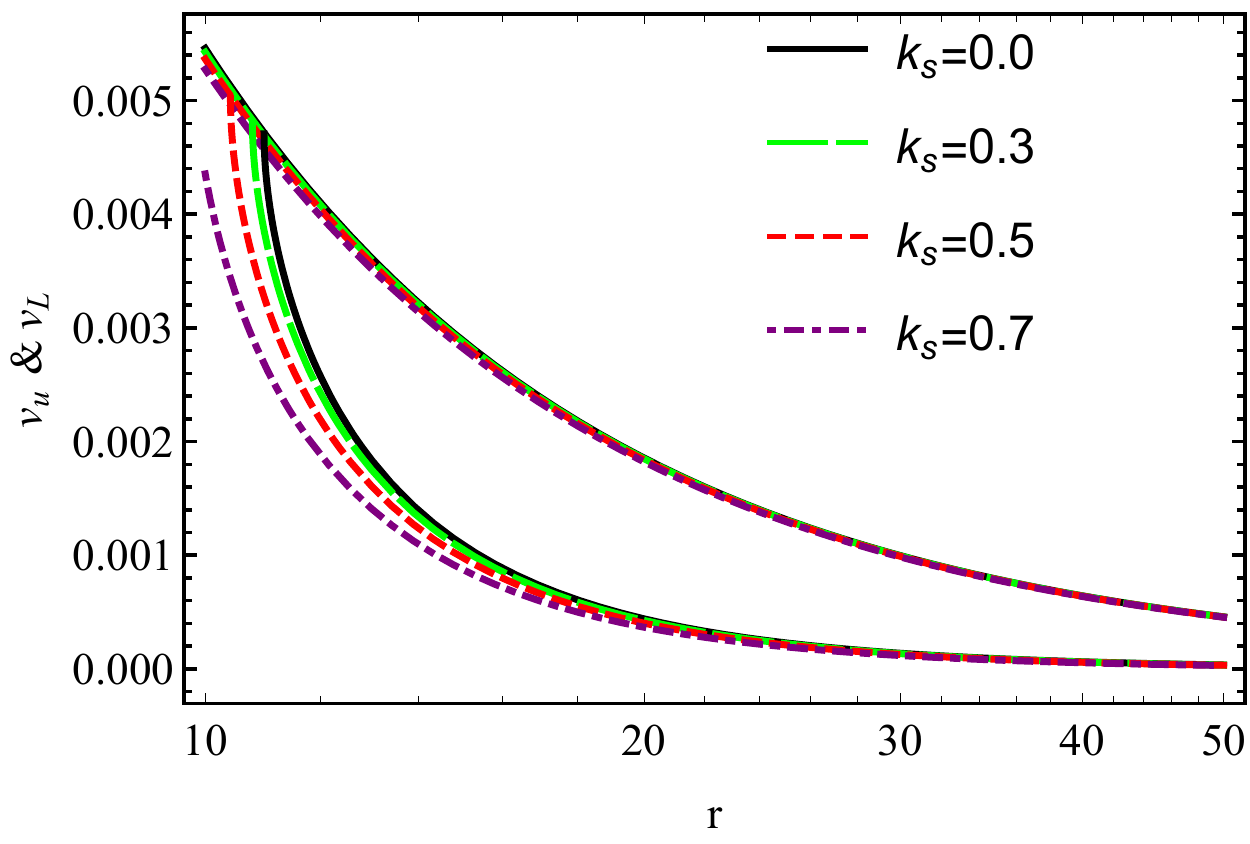}
    \includegraphics[scale=0.52]{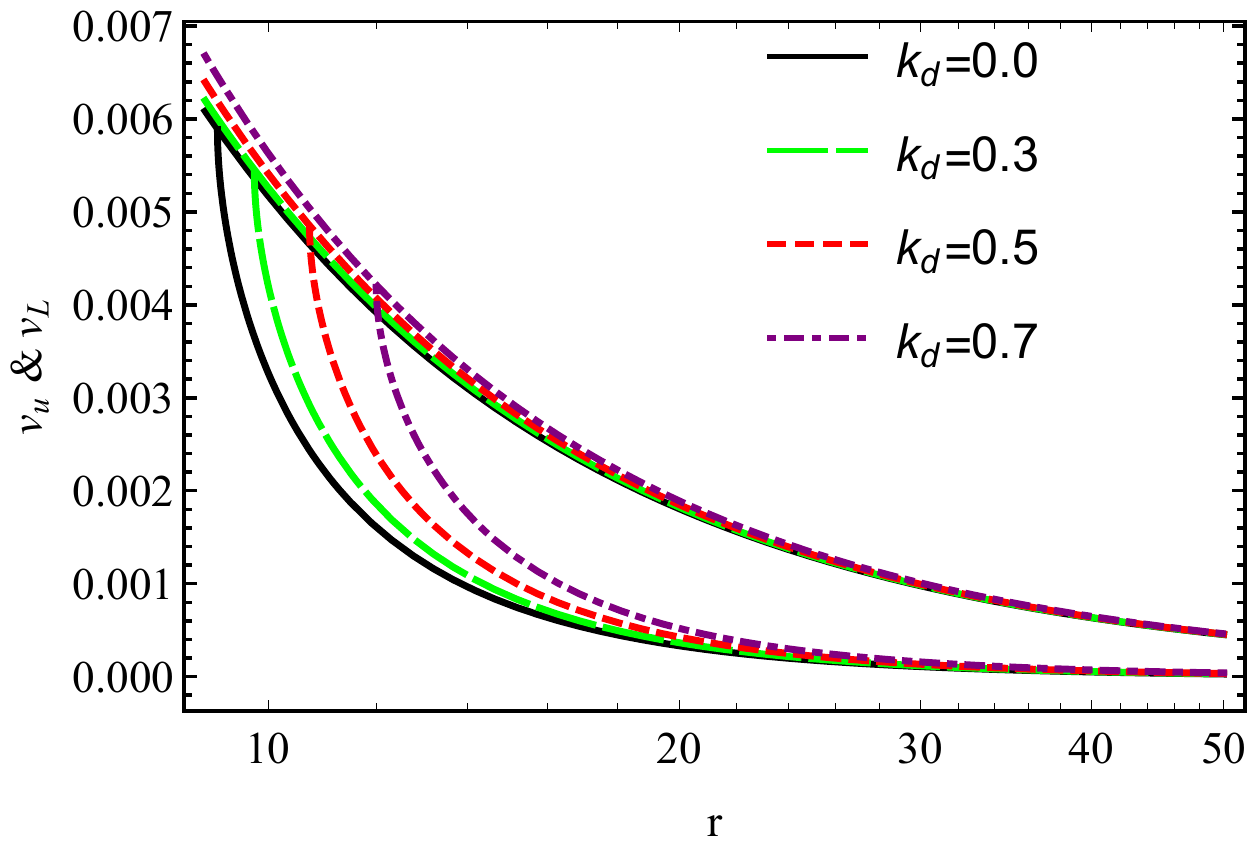}
    \includegraphics[scale=0.52]{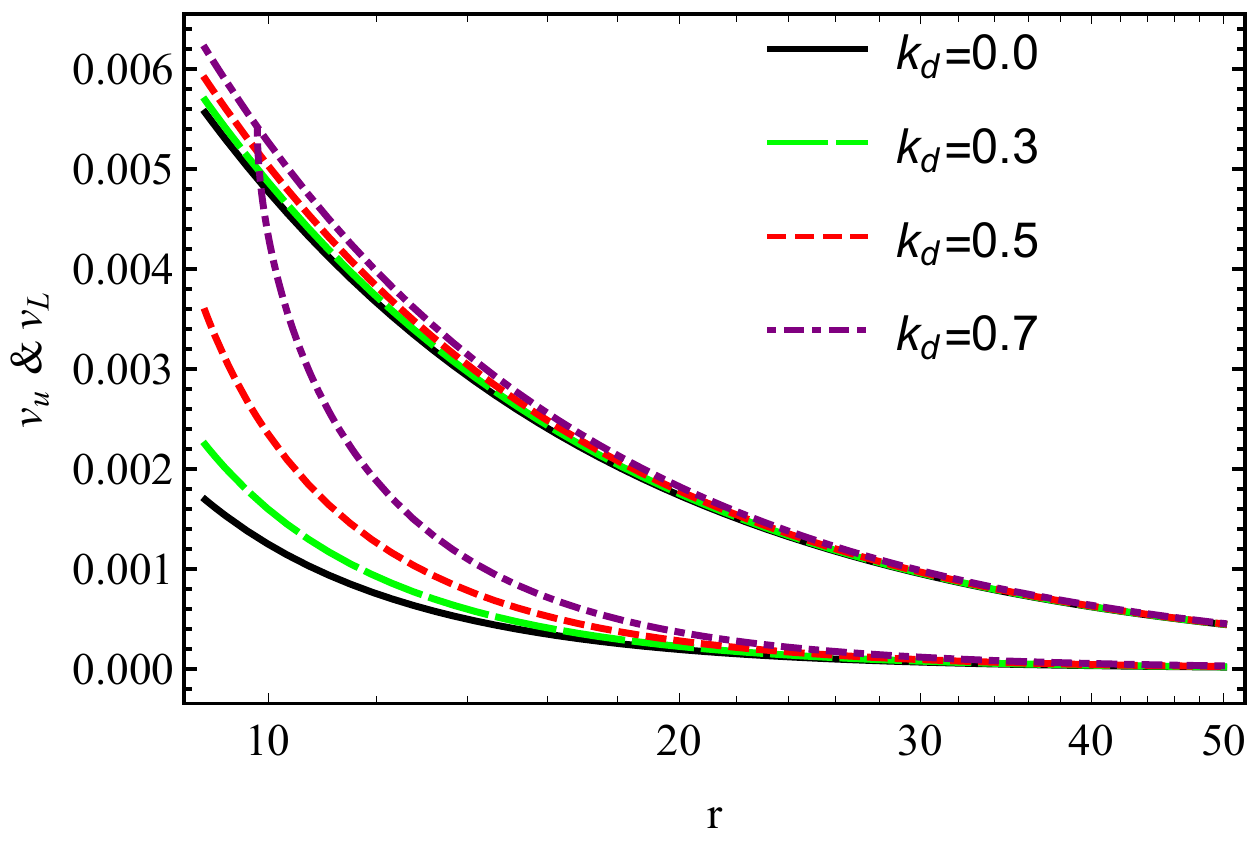}
    \includegraphics[scale=0.52]{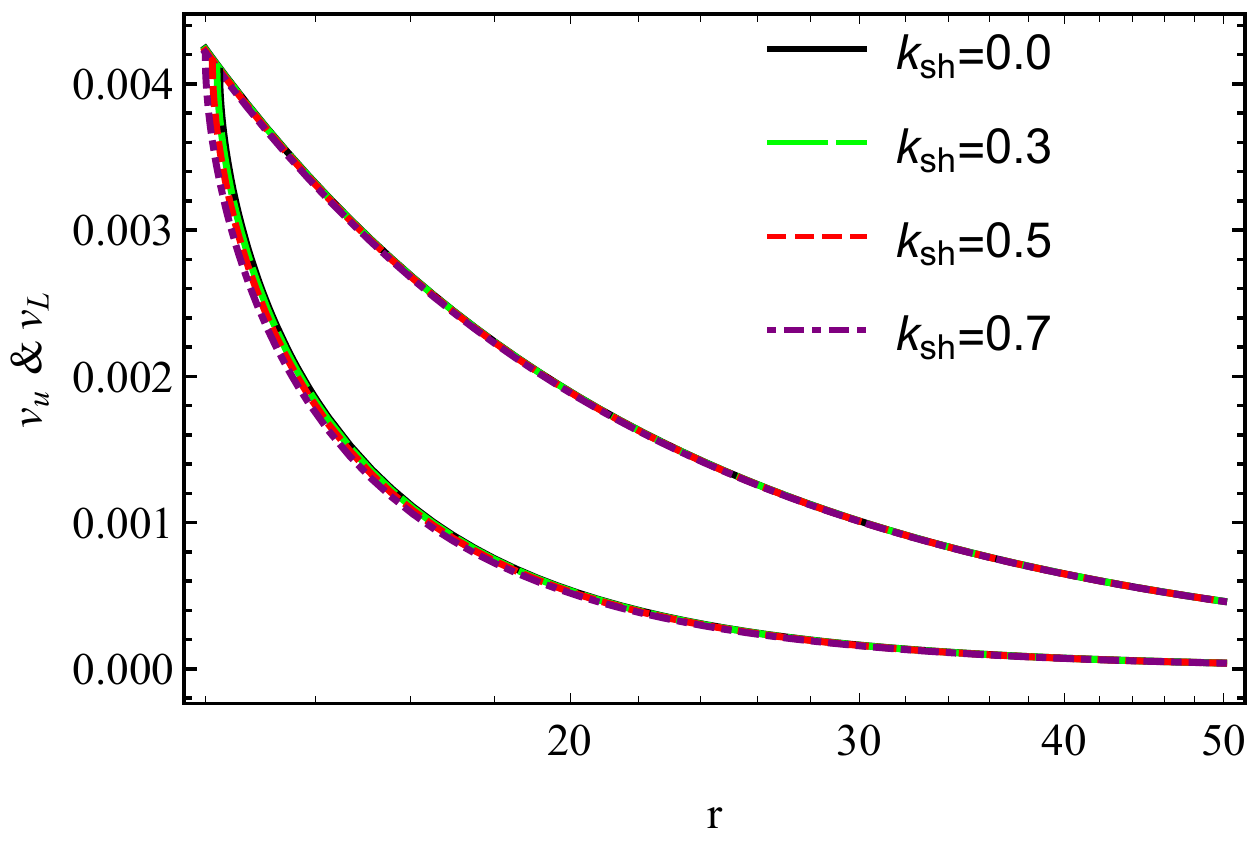}
    \includegraphics[scale=0.52]{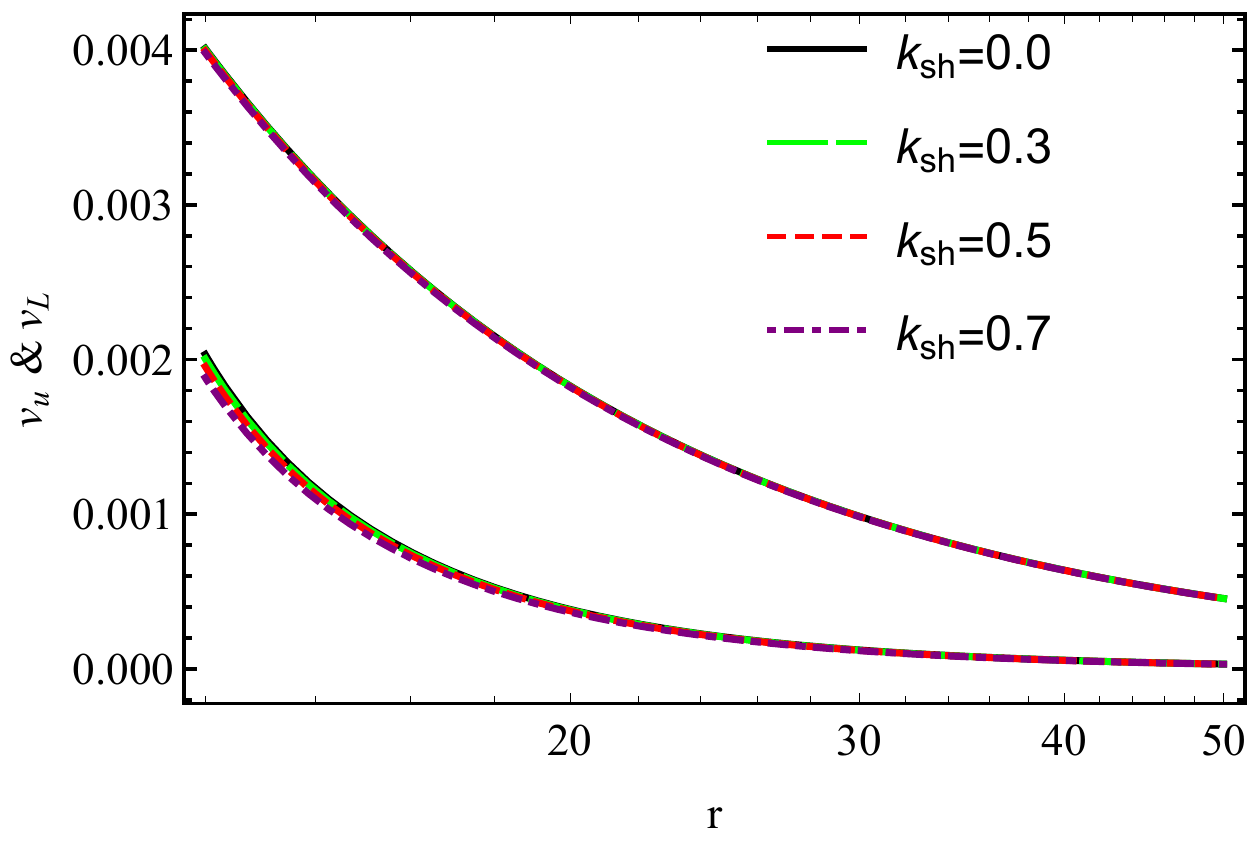}
    \caption{RP Model: Upper $v_u$ and lower $v_L$ frequencies for case $d_1=8f_1$ (Left panel) and for case $d_1=-8f_1$ (Right panel) along $c_1$ for different values of $f_1,\; k_s,\; k_d, \;\&\; k_{sh} $. Here we consider the choice for fix values $M=1,c_1=1,\;f_1=-0.2,\;k_s=k_{sh}=k_d=0.7$.}
    \label{plot:7}
\end{figure}

\begin{figure}
    \centering
    \includegraphics[scale=0.52]{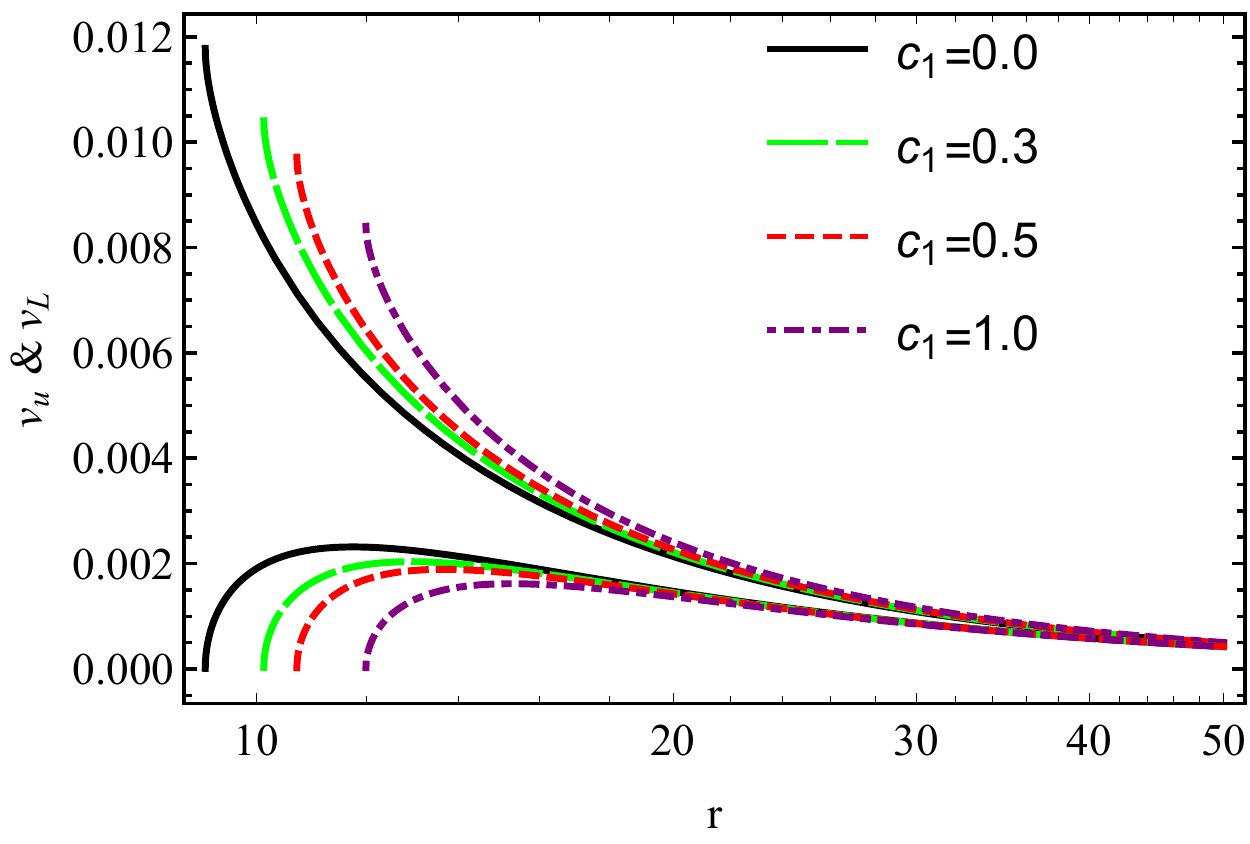}
    \includegraphics[scale=0.52]{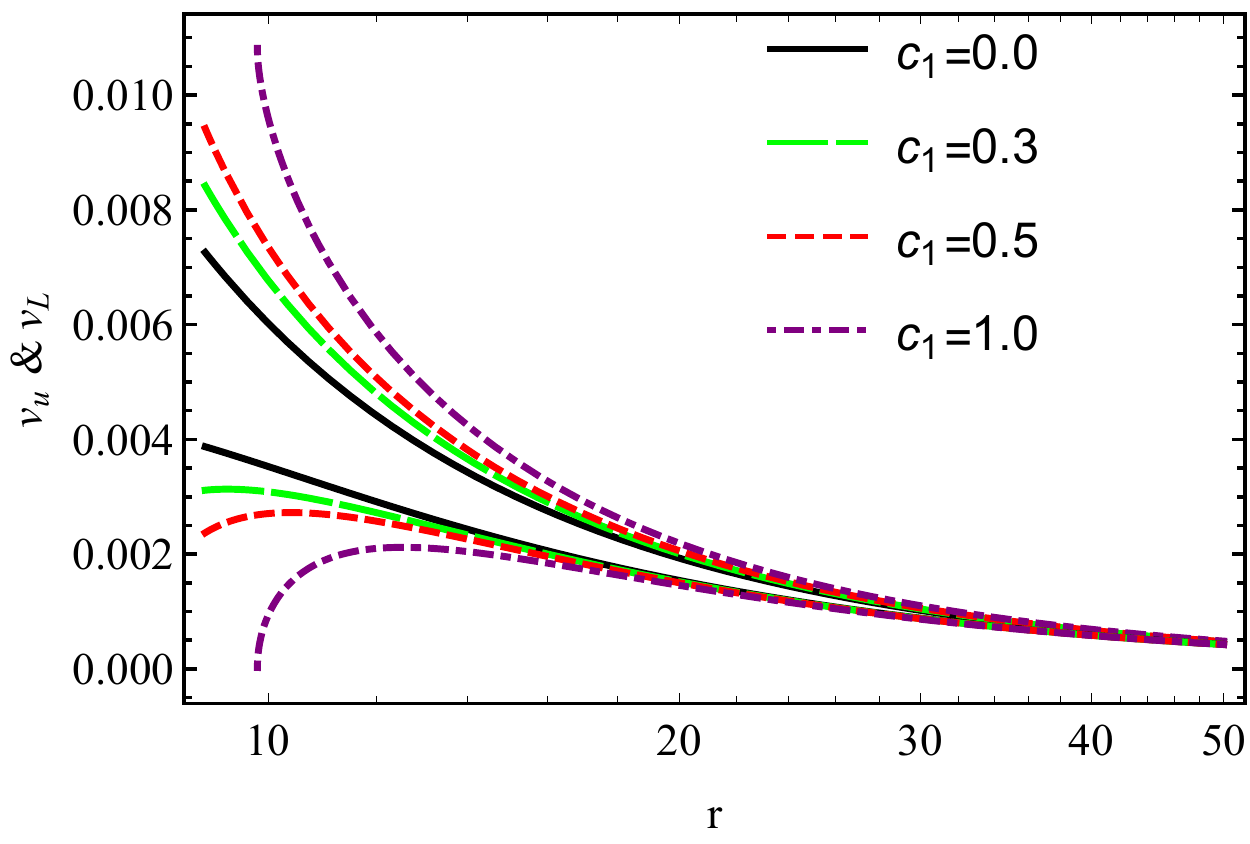}
    \includegraphics[scale=0.52]{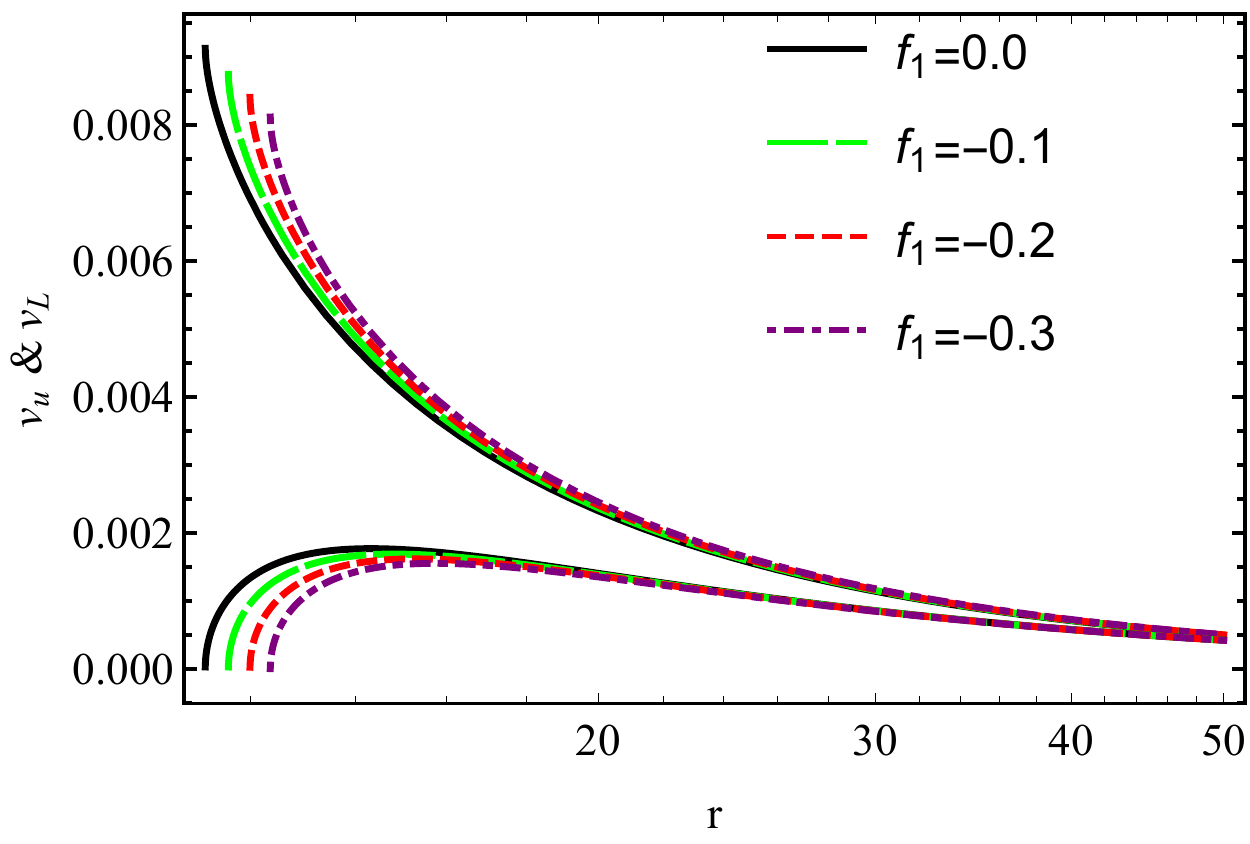}
    \includegraphics[scale=0.52]{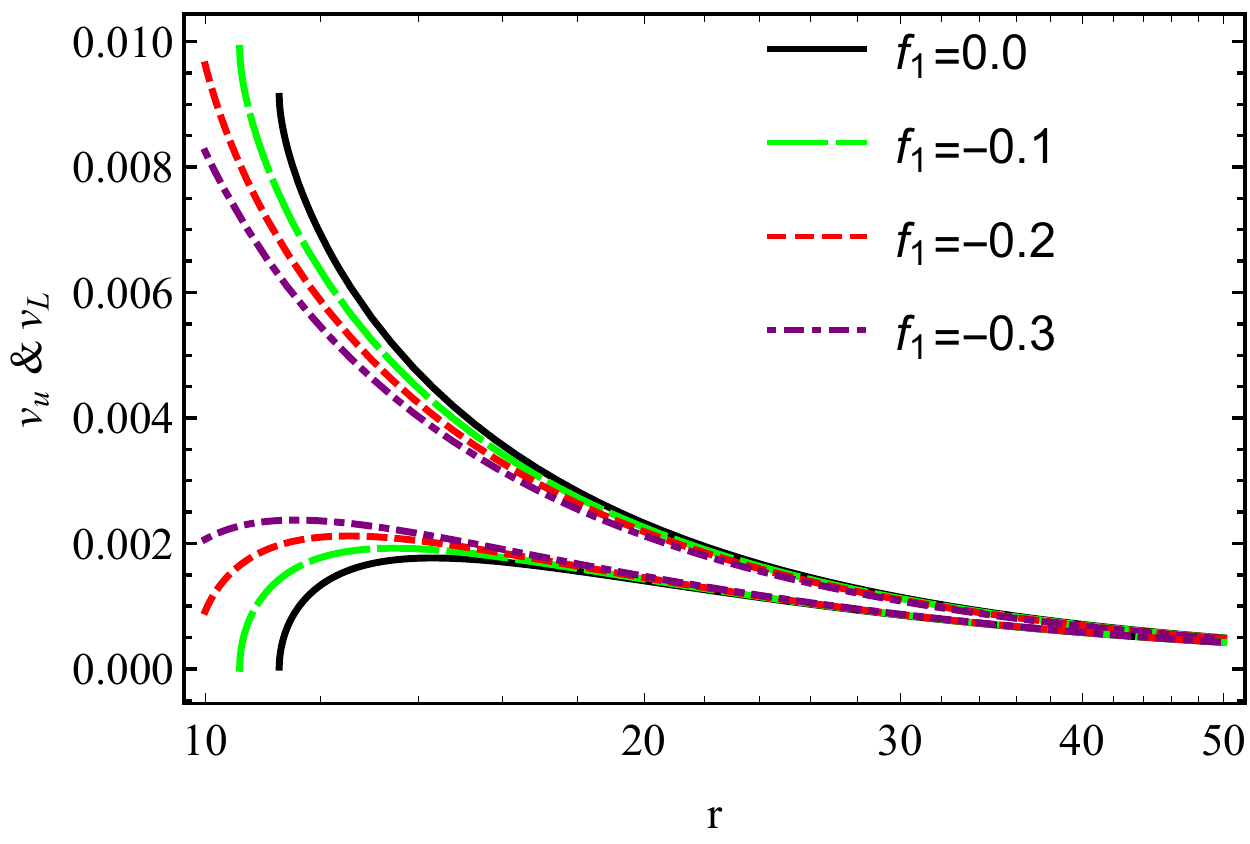}
    \includegraphics[scale=0.52]{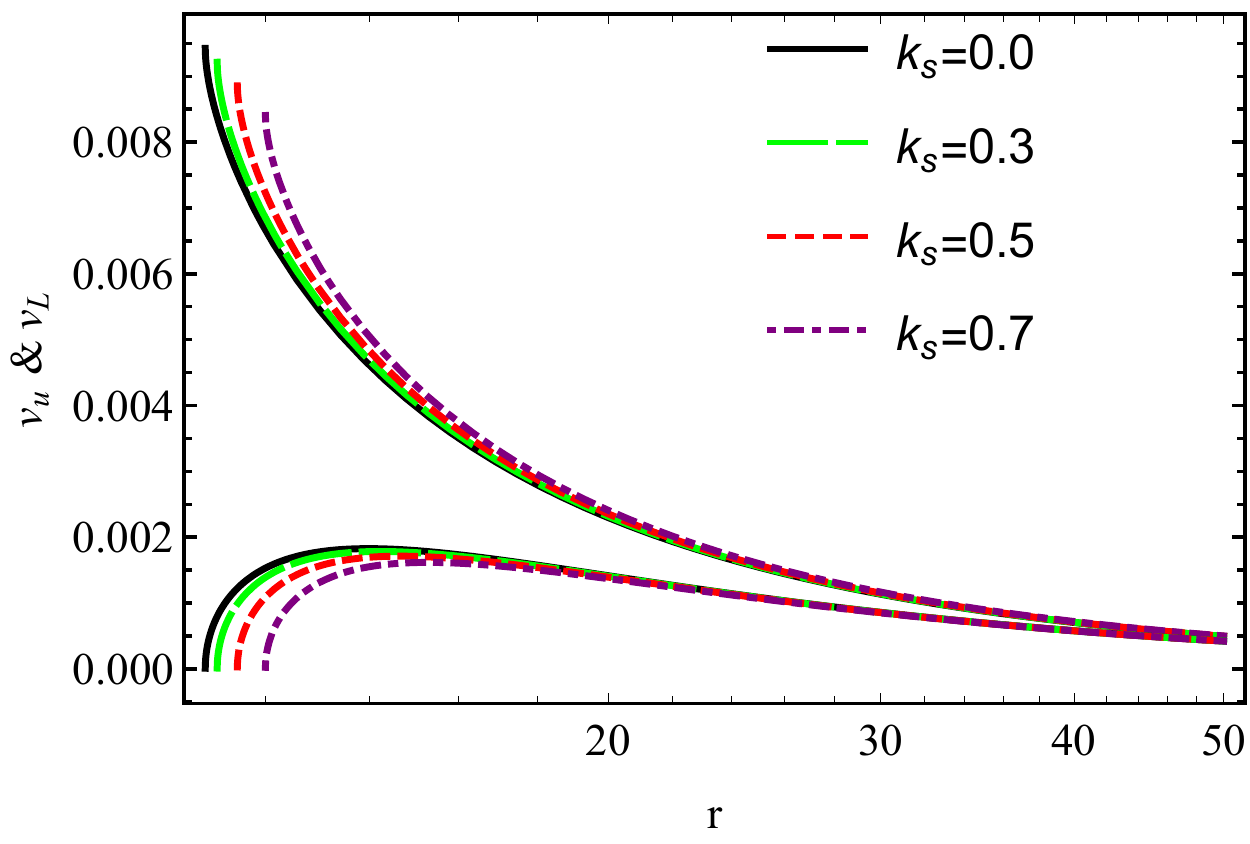}
    \includegraphics[scale=0.52]{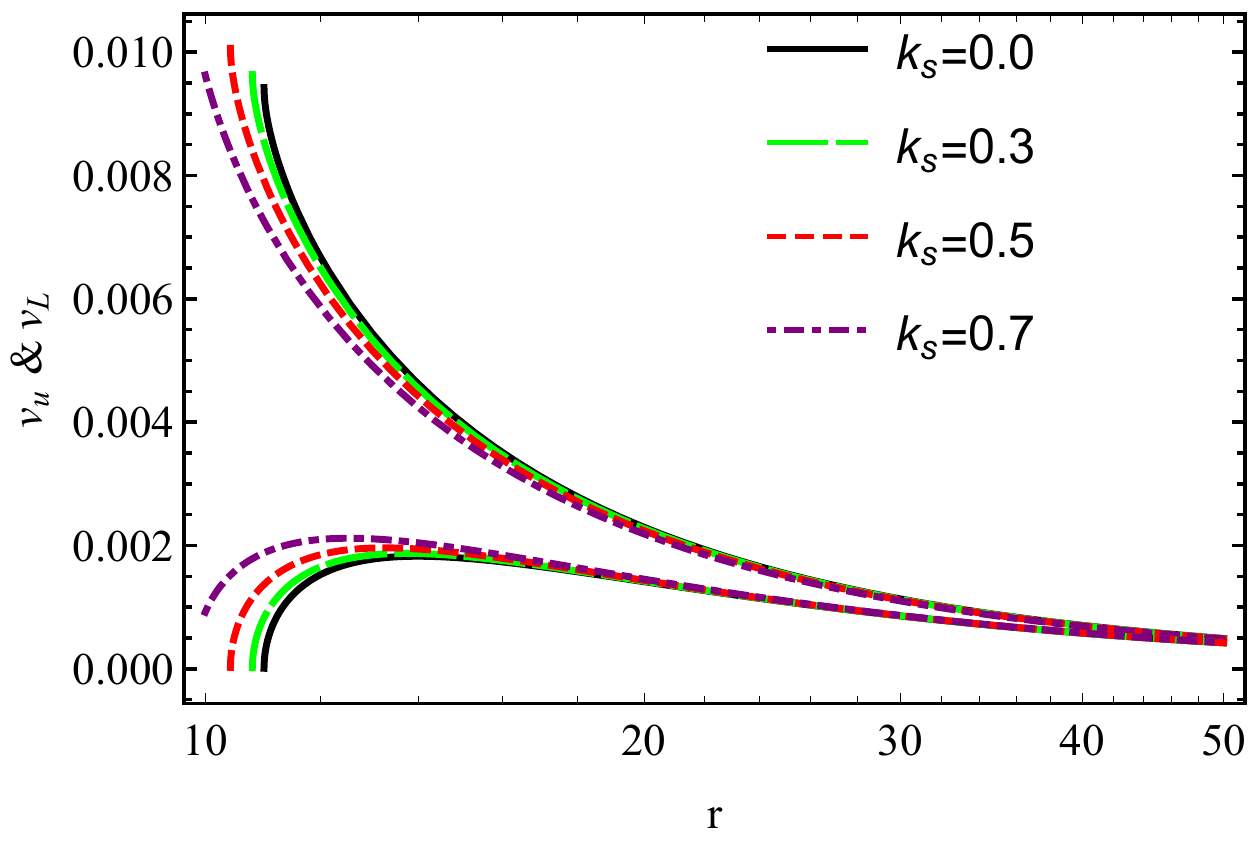}
    \includegraphics[scale=0.52]{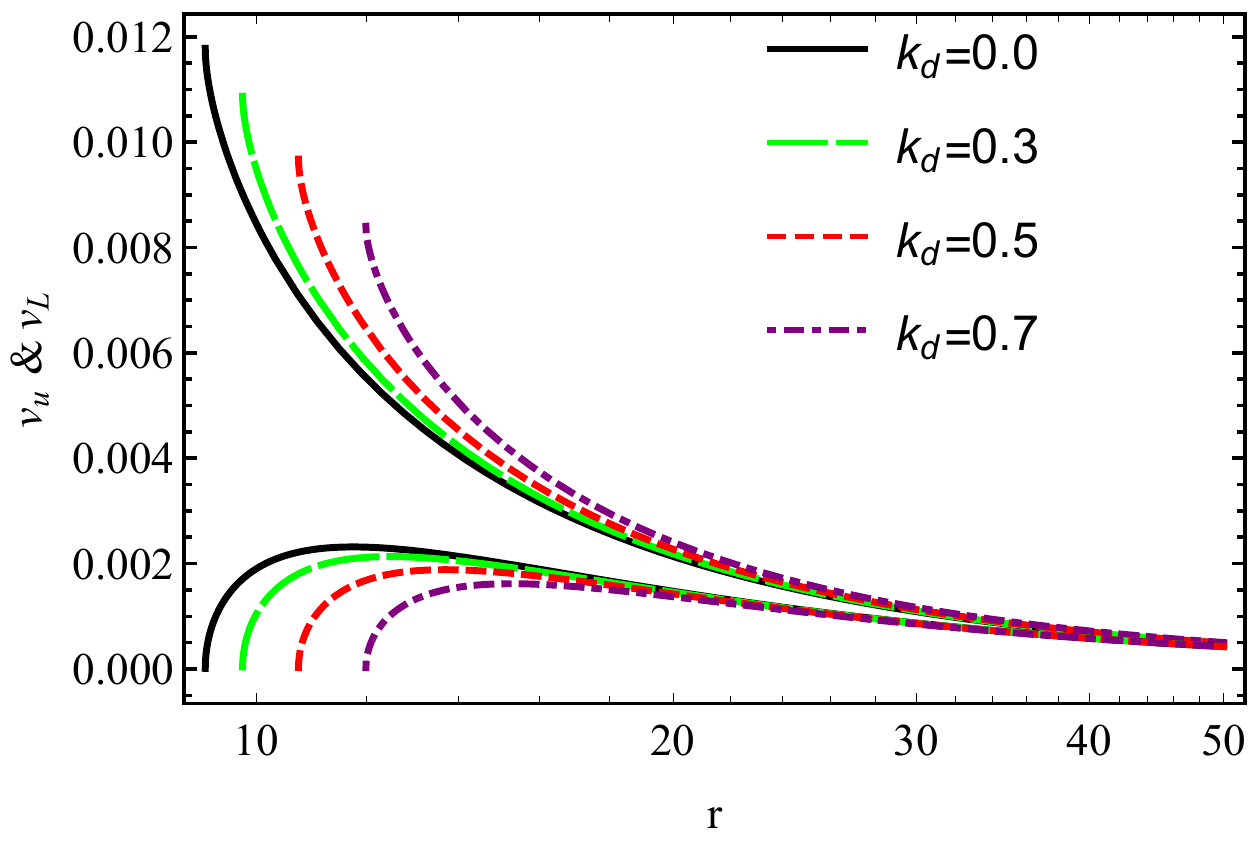}
    \includegraphics[scale=0.52]{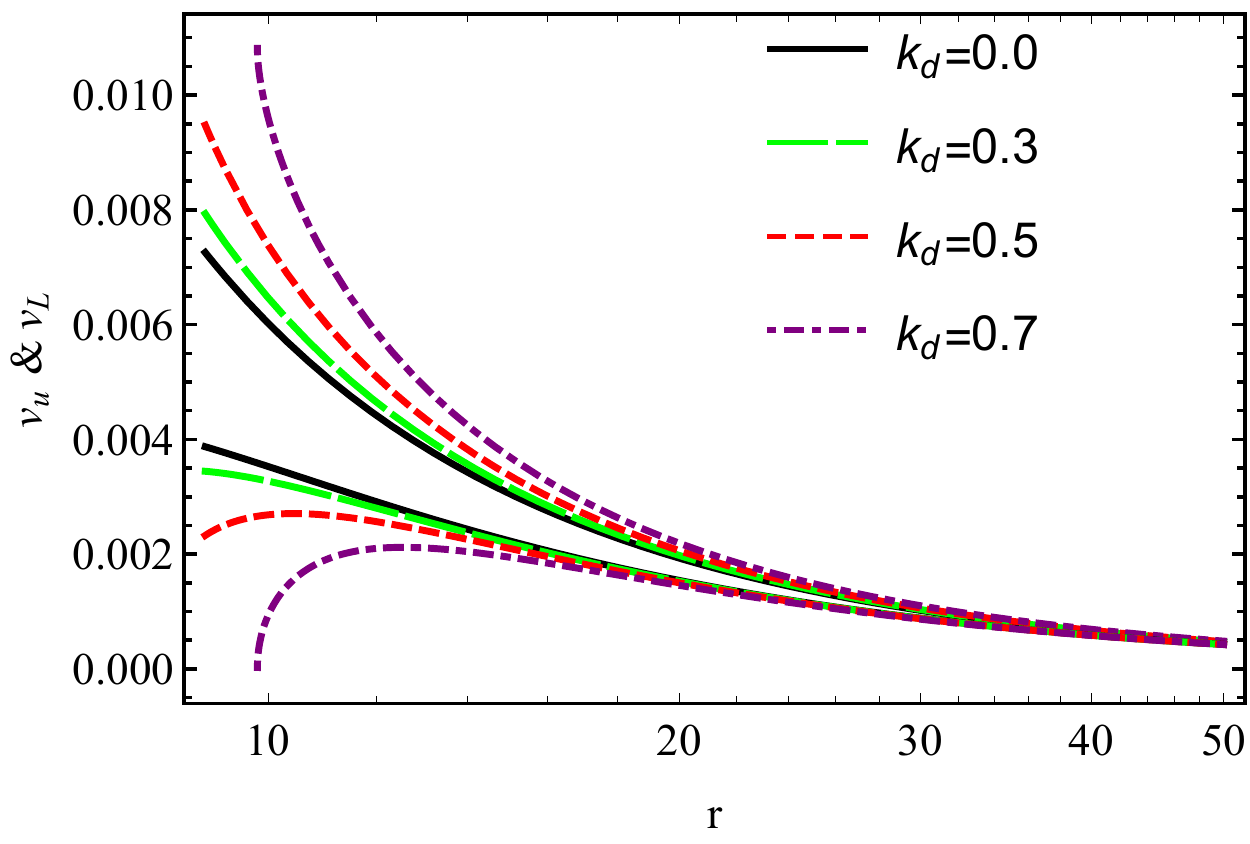}
    \includegraphics[scale=0.52]{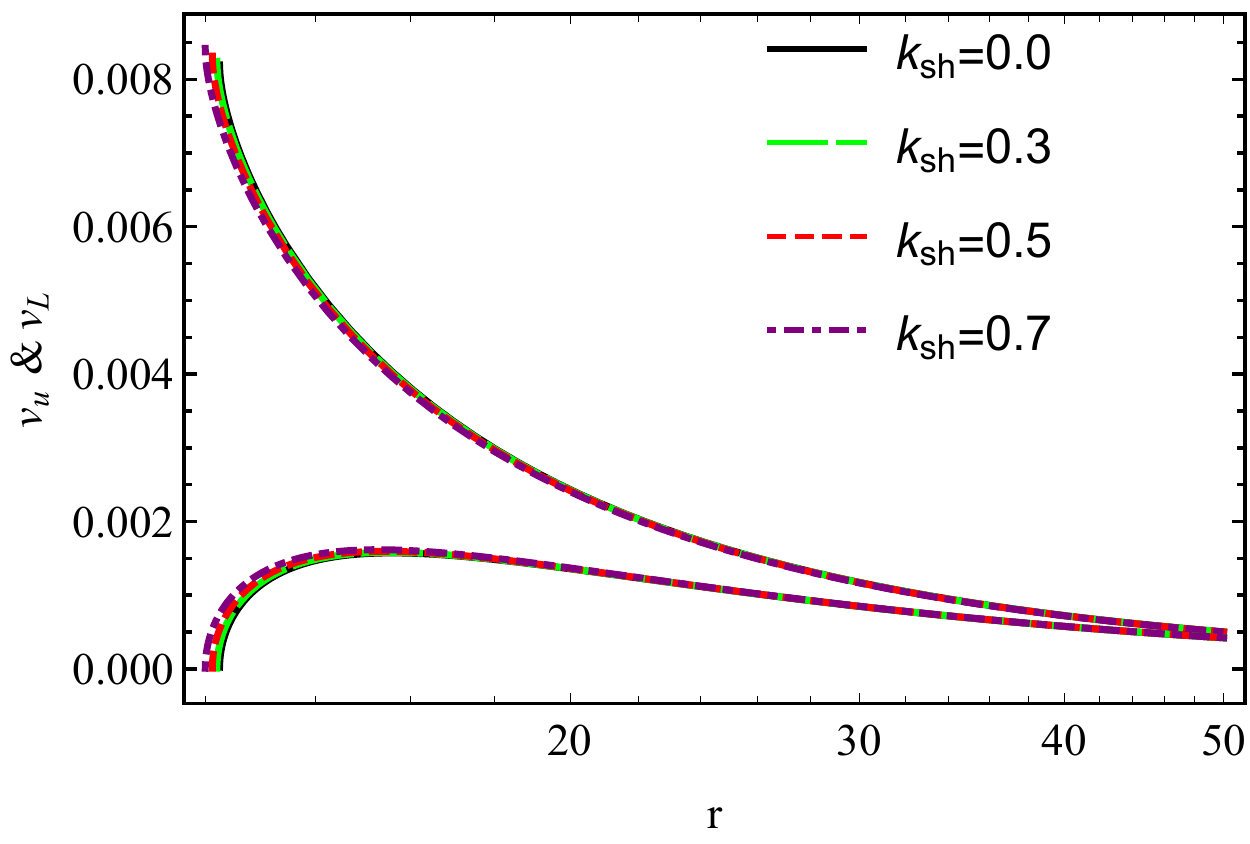}
    \includegraphics[scale=0.52]{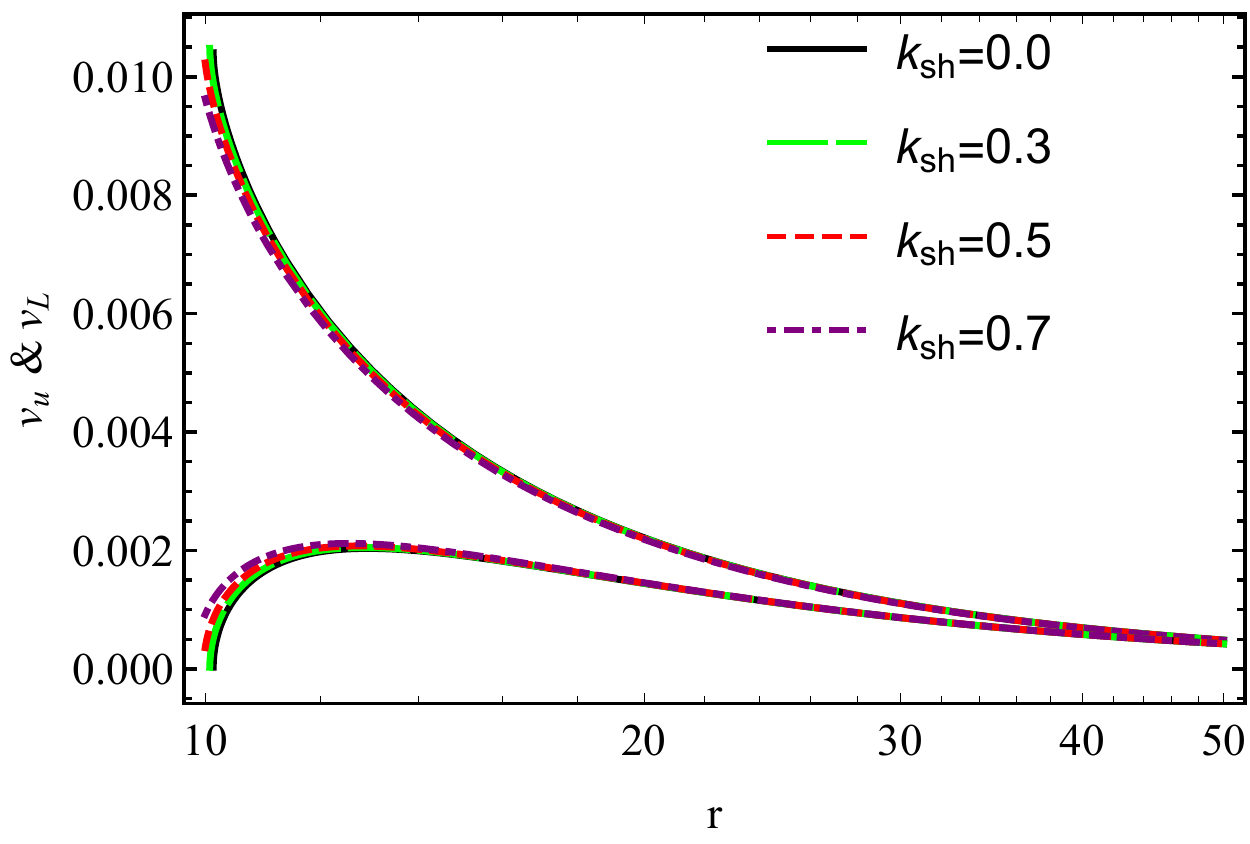}
    \caption{ER2 Model: Upper $v_u$ and lower $v_L$ frequencies for case $d_1=8f_1$ (Left panel) and for case $d_1=-8f_1$ (Right panel) along $c_1$ for different values of $f_1,\; k_s,\; k_d, \;\&\; k_{sh} $. Here we consider the choice for fix values $M=1,c_1=1,\;f_1=-0.2,\;k_s=k_{sh}=k_d=0.7$.}
    \label{plot:8}
\end{figure}

\begin{figure}
    \centering
    \includegraphics[scale=0.52]{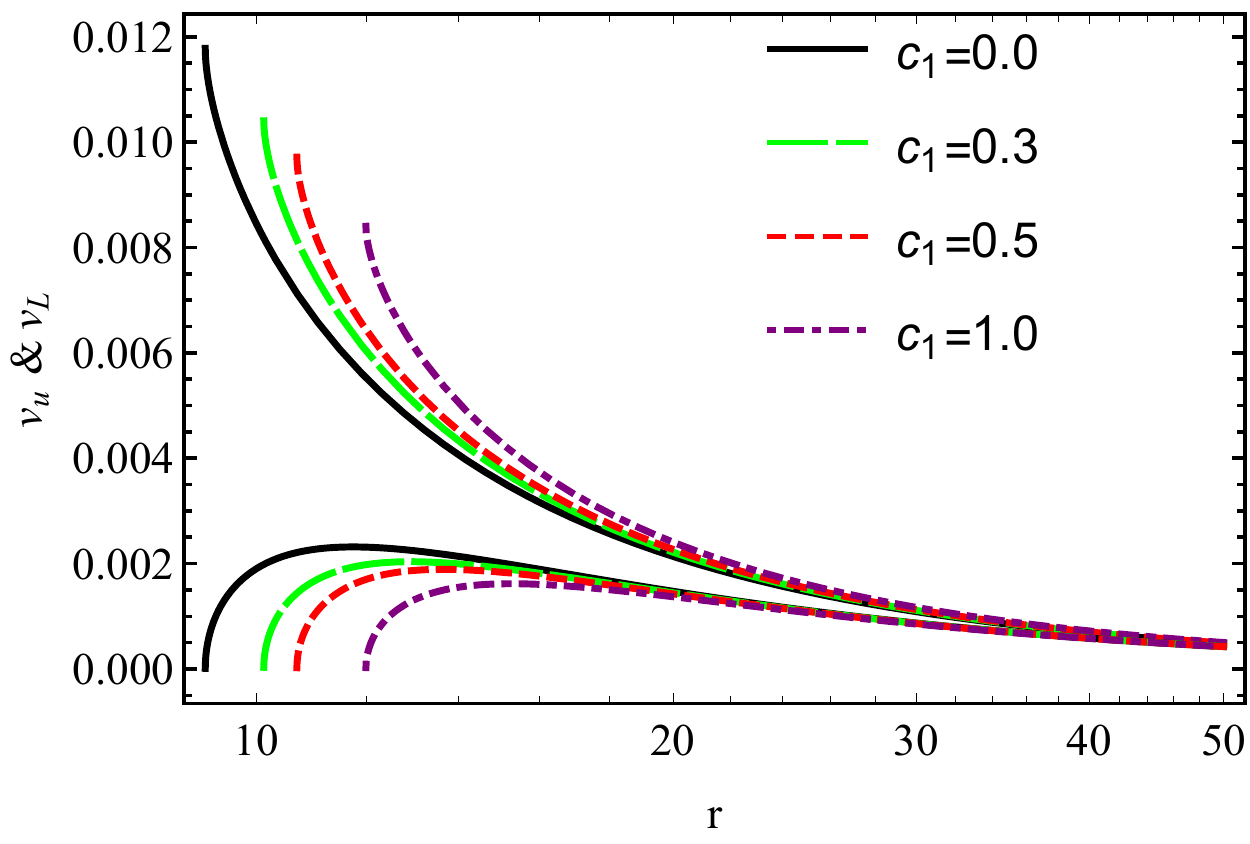}
    \includegraphics[scale=0.52]{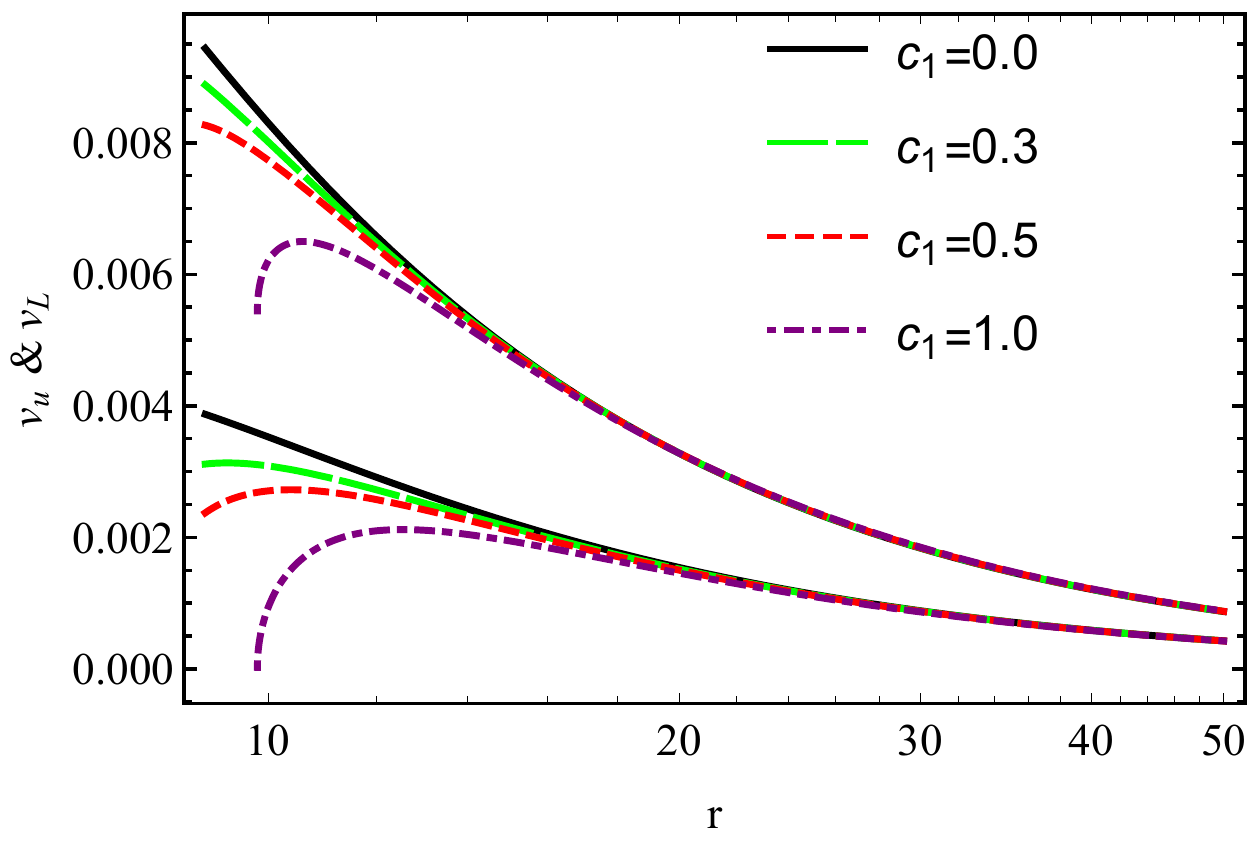}\\
    \includegraphics[scale=0.52]{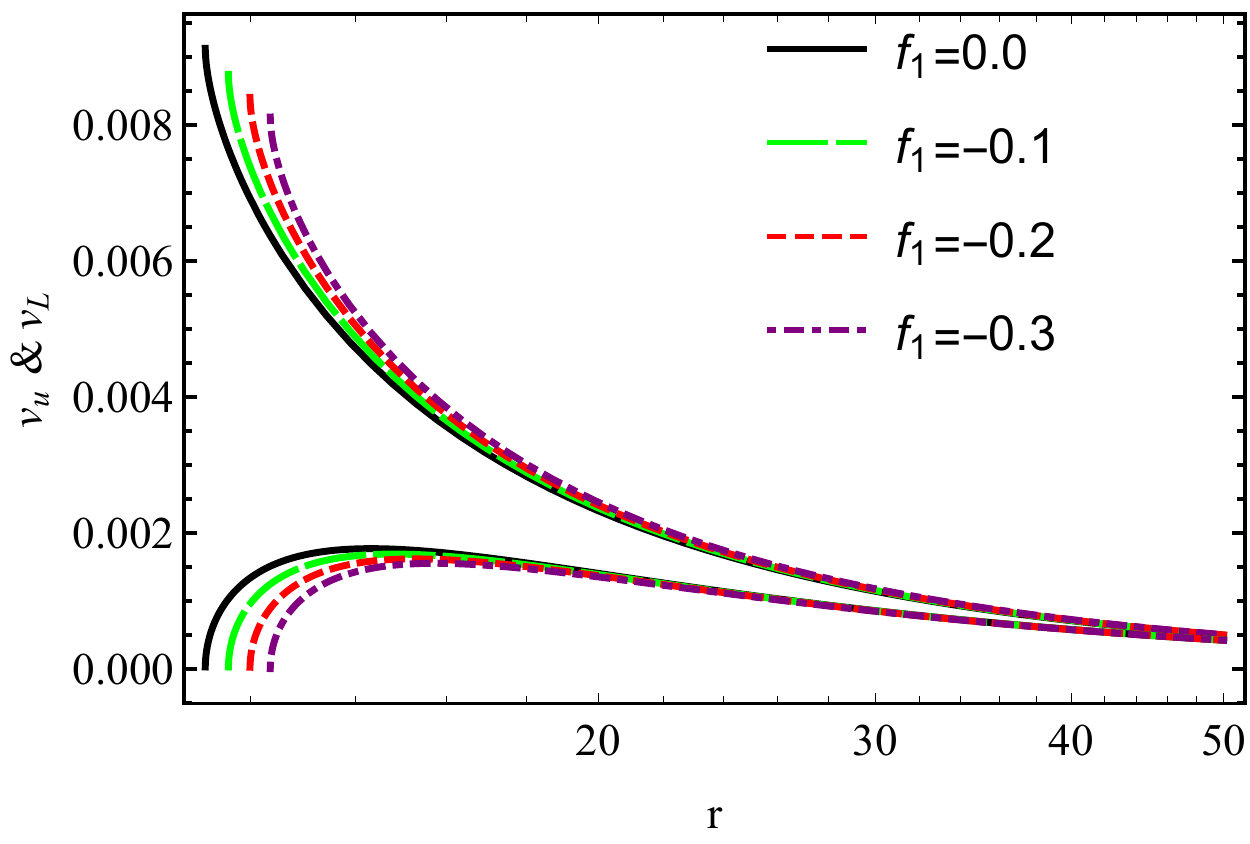}
    \includegraphics[scale=0.52]{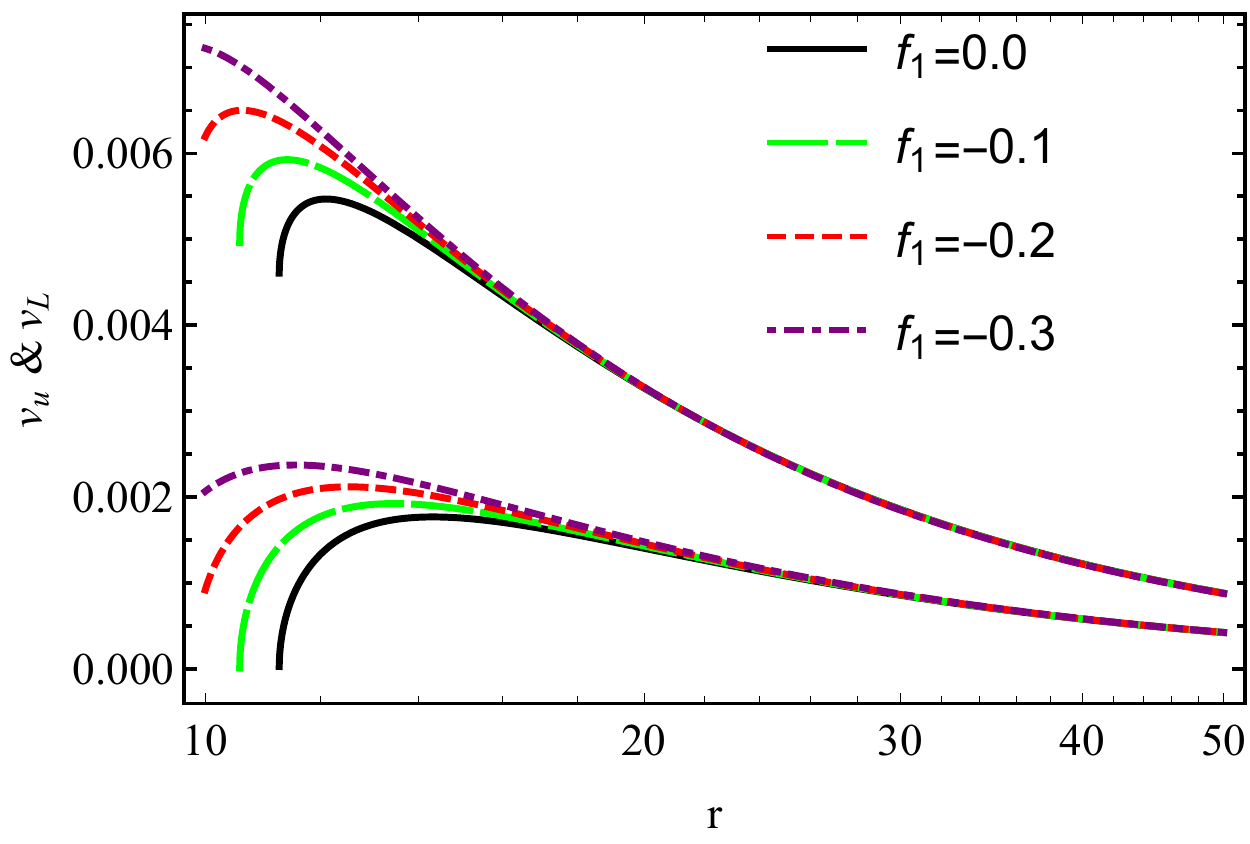}\\
    \includegraphics[scale=0.52]{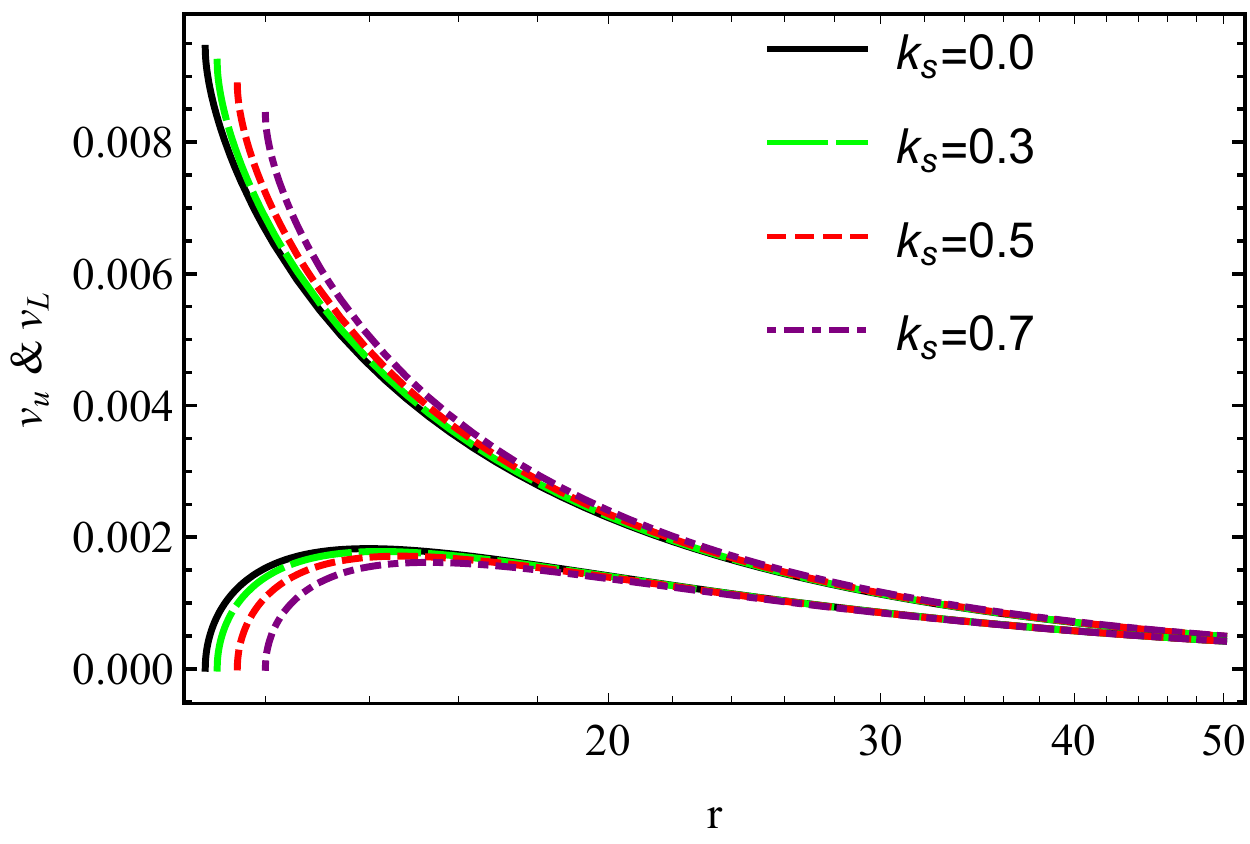}
    \includegraphics[scale=0.52]{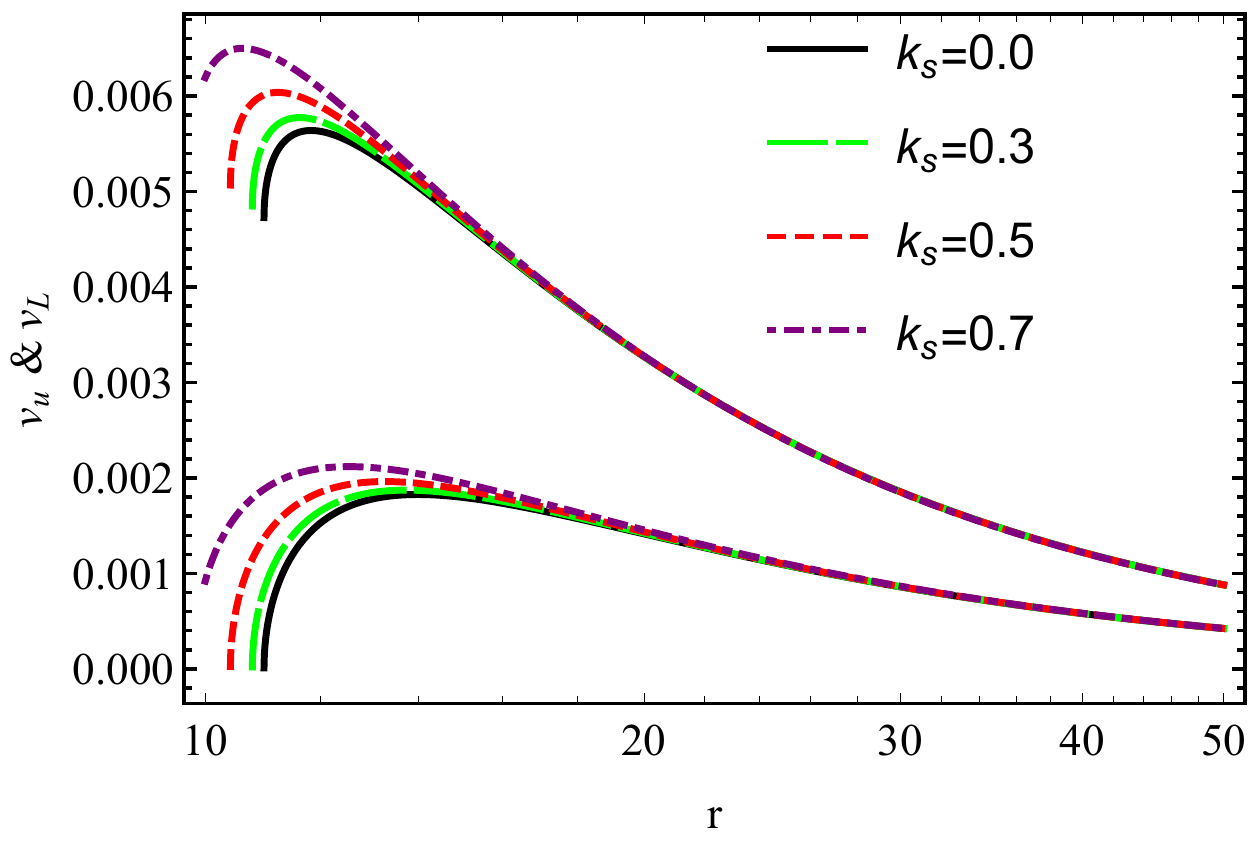}\\
    \includegraphics[scale=0.52]{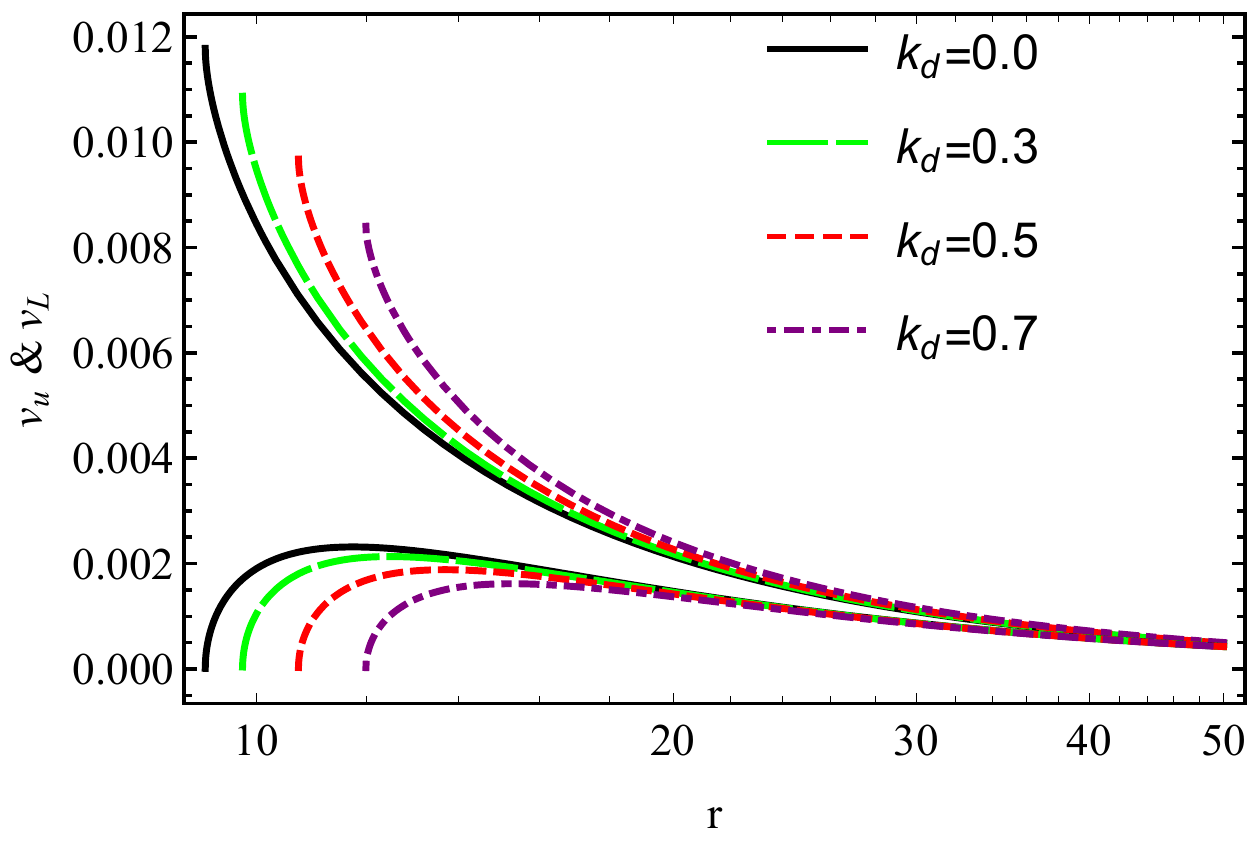}
    \includegraphics[scale=0.52]{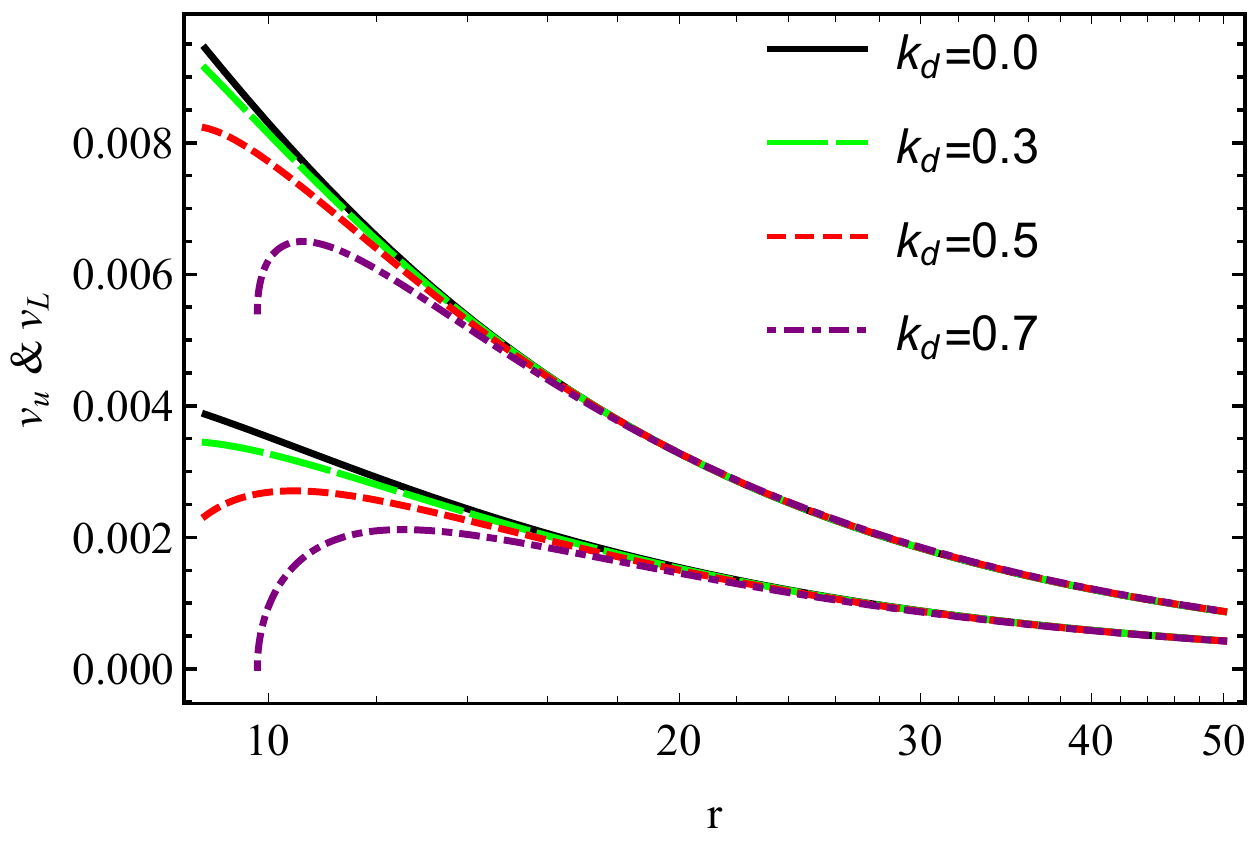}\\
    \includegraphics[scale=0.52]{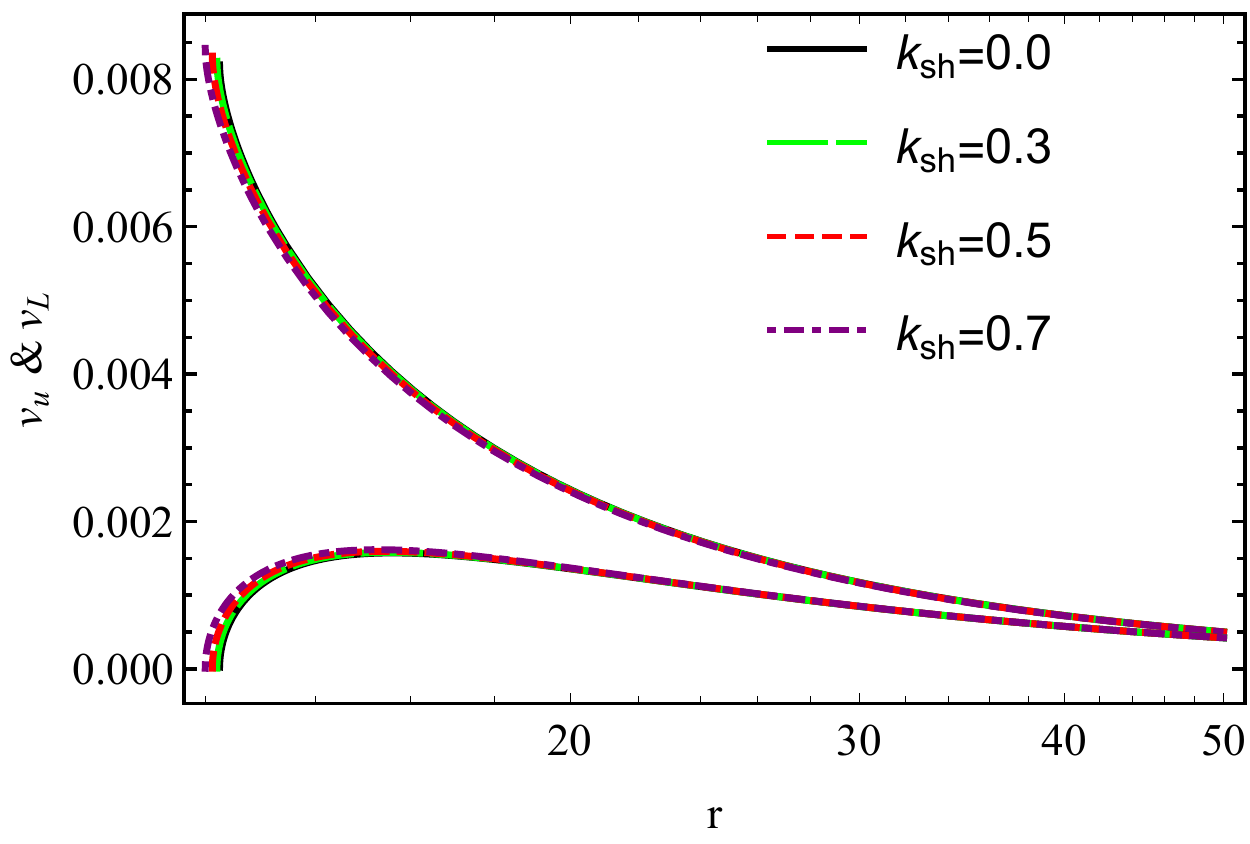}
    \includegraphics[scale=0.52]{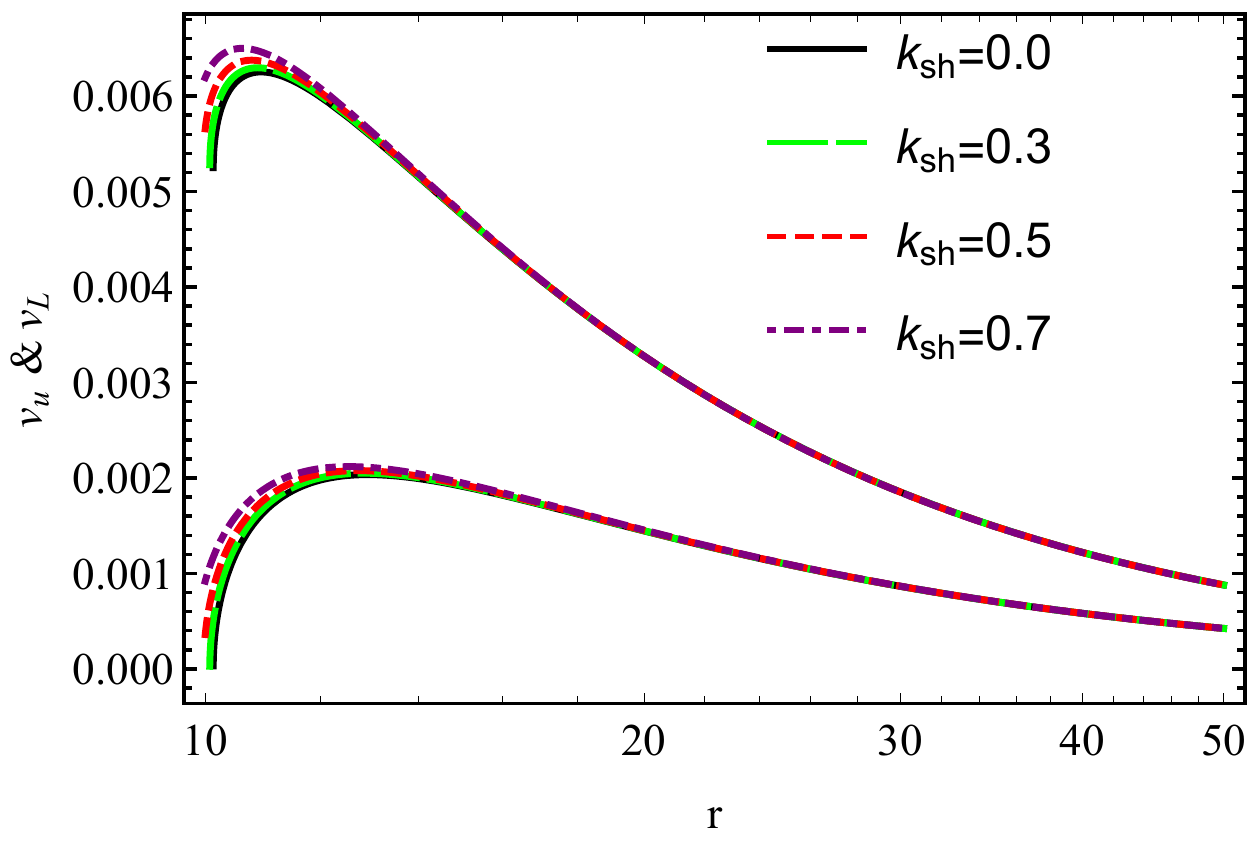}\\
    \caption{ER3 Model: Upper $v_u$ and lower $v_L$ frequencies for case $d_1=8f_1$ (Left panel) and for case $d_1=-8f_1$ (Right panel) along $c_1$ for different values of $f_1,\; k_s,\; k_d, \;\&\; k_{sh} $. Here we consider the choice for fix values $M=1,c_1=1,\;f_1=-0.2,\;k_s=k_{sh}=k_d=0.7$.}
    \label{plot:9}
\end{figure}

\begin{figure}
    \centering
    \includegraphics[scale=0.52]{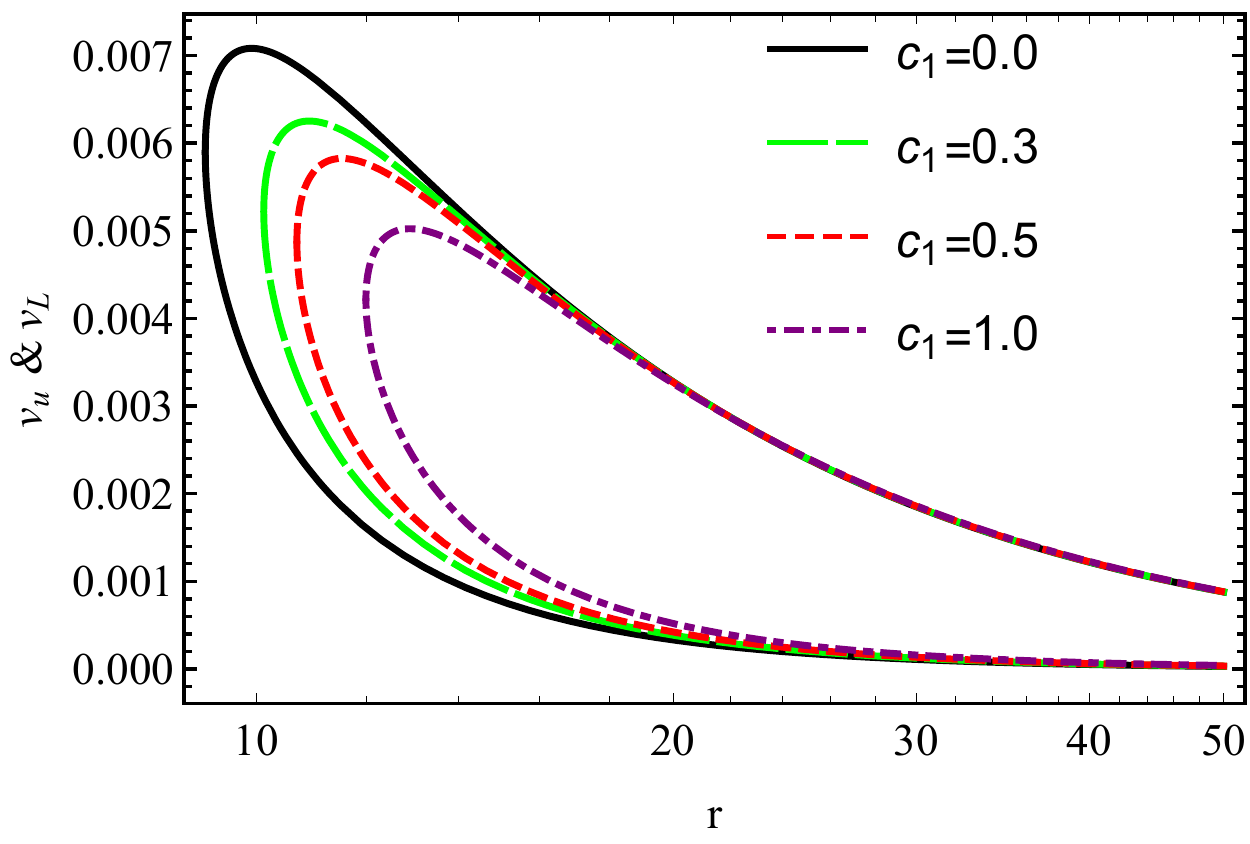}
    \includegraphics[scale=0.52]{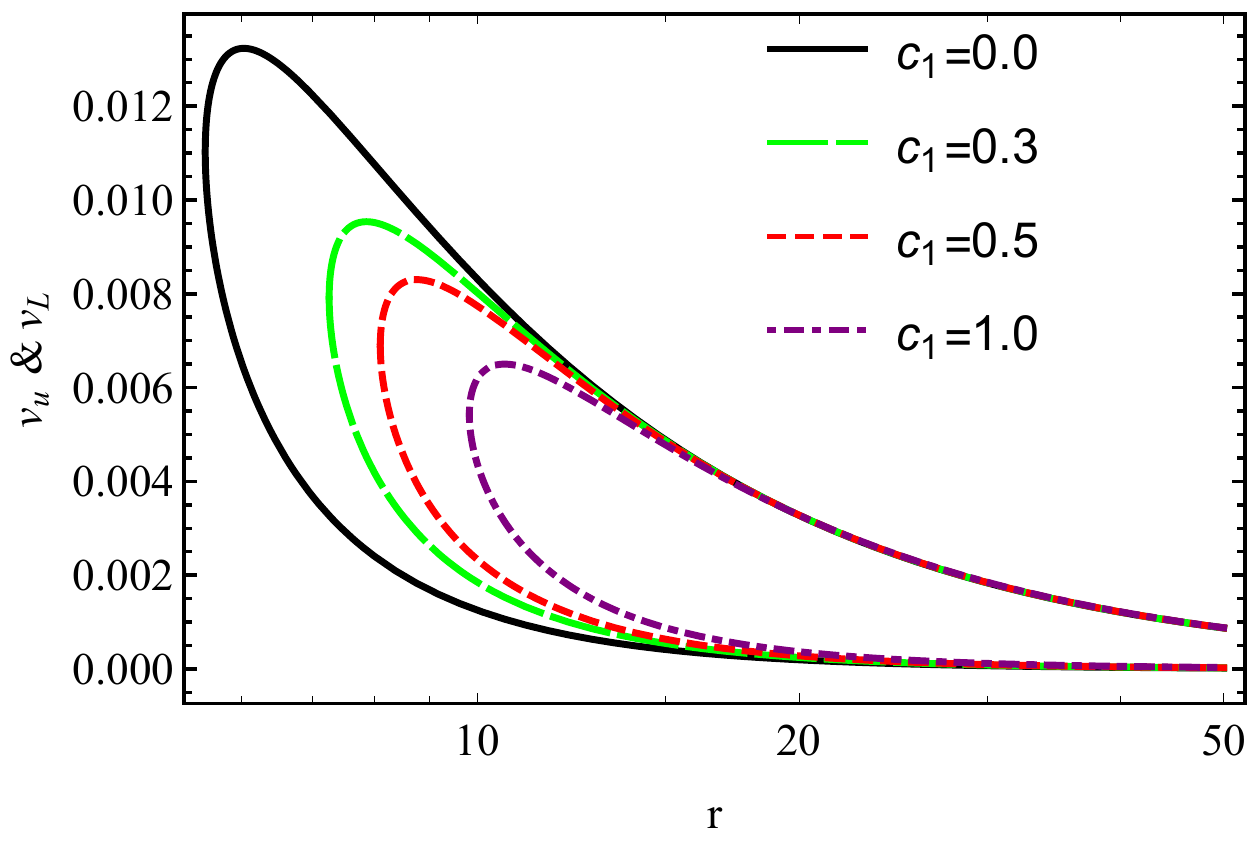}\\
    \includegraphics[scale=0.52]{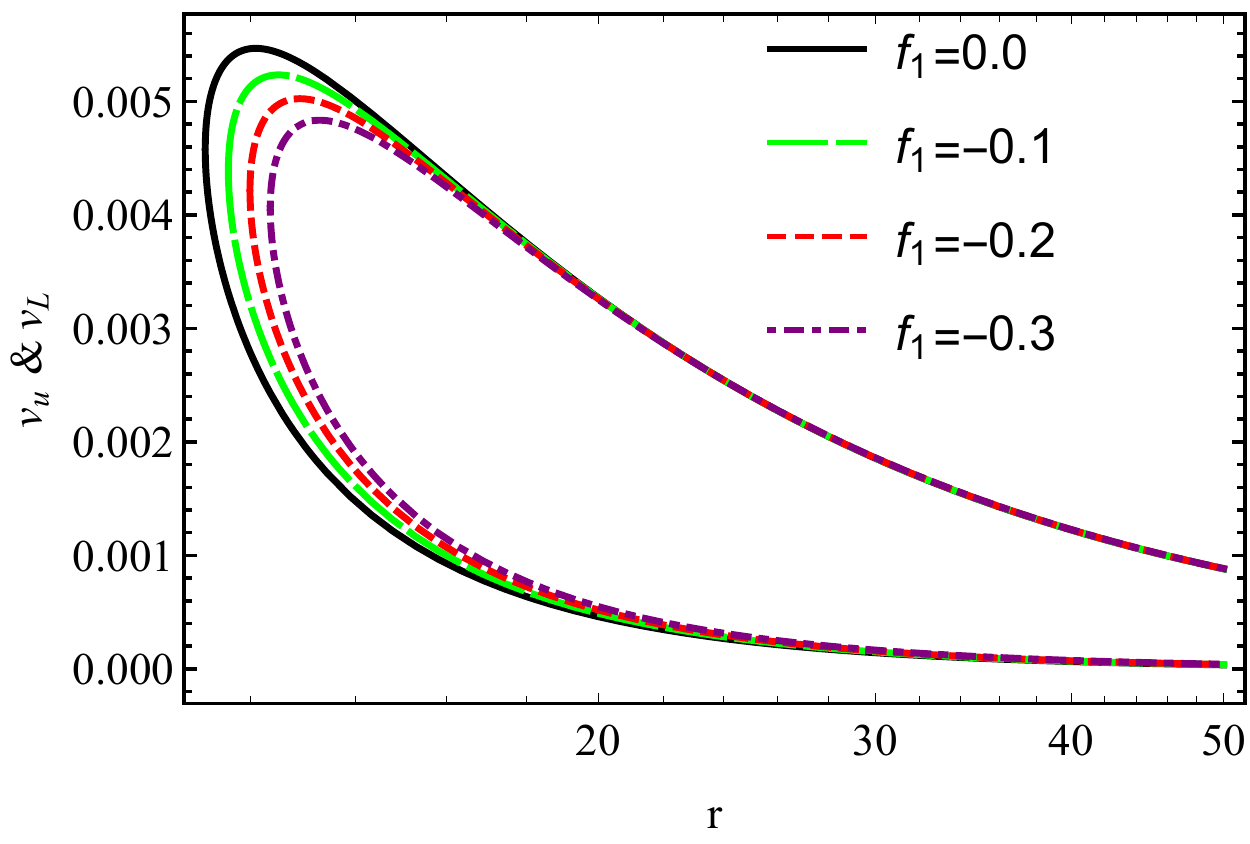}
    \includegraphics[scale=0.52]{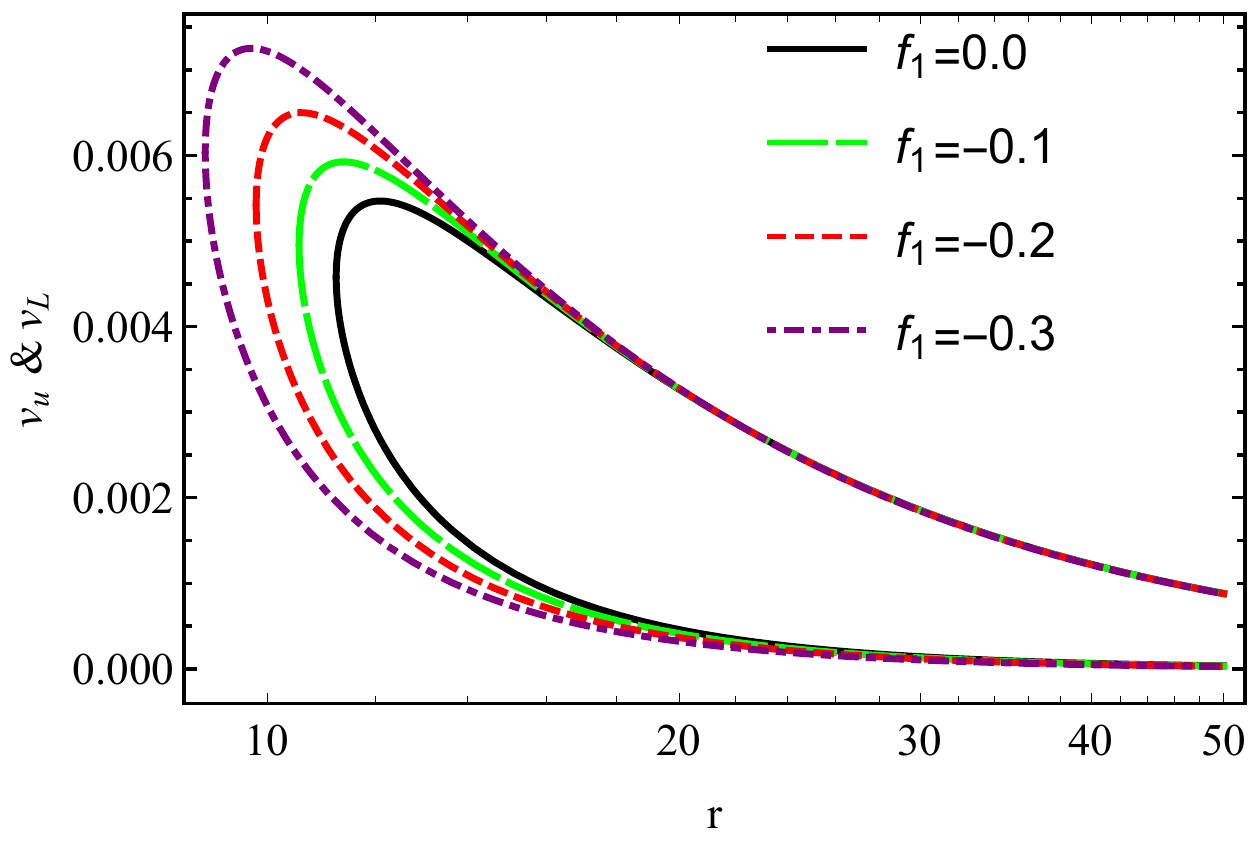}\\
    \includegraphics[scale=0.52]{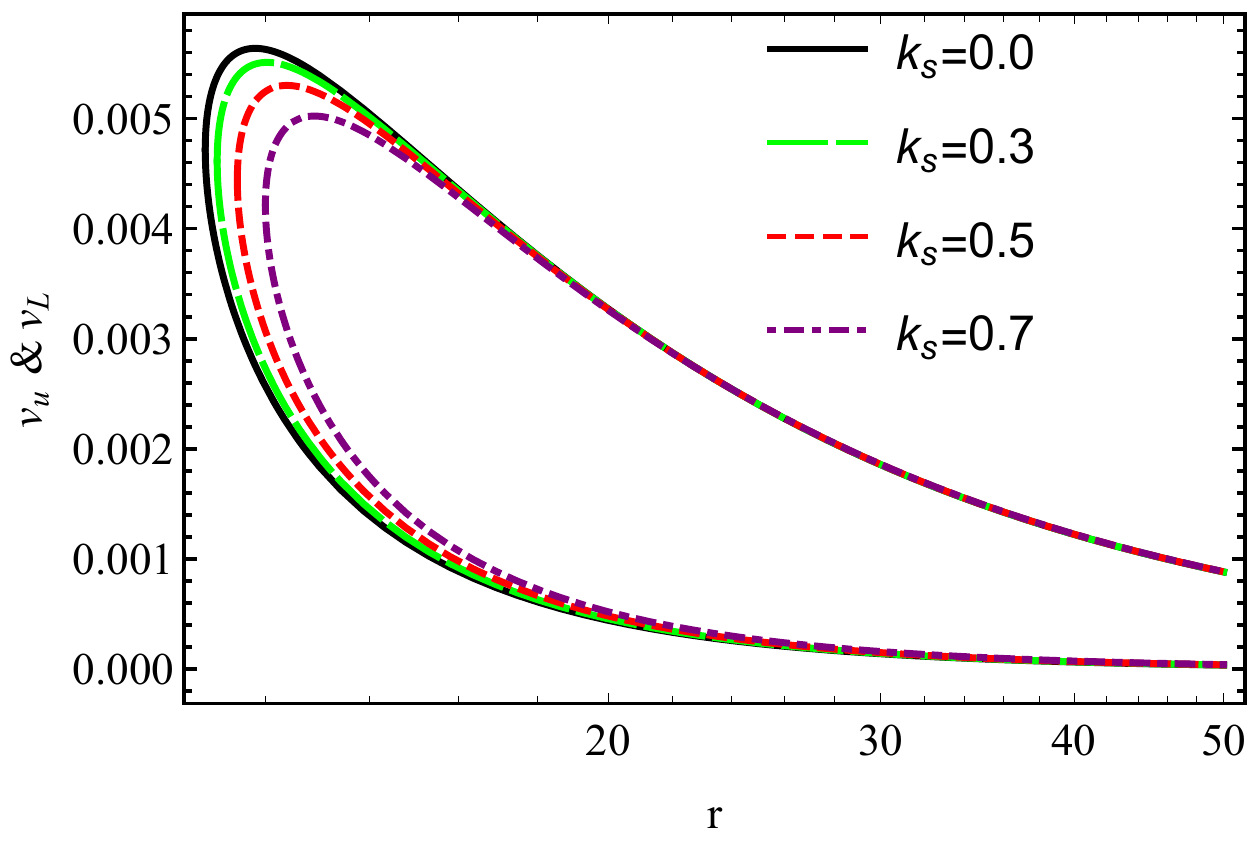}
    \includegraphics[scale=0.52]{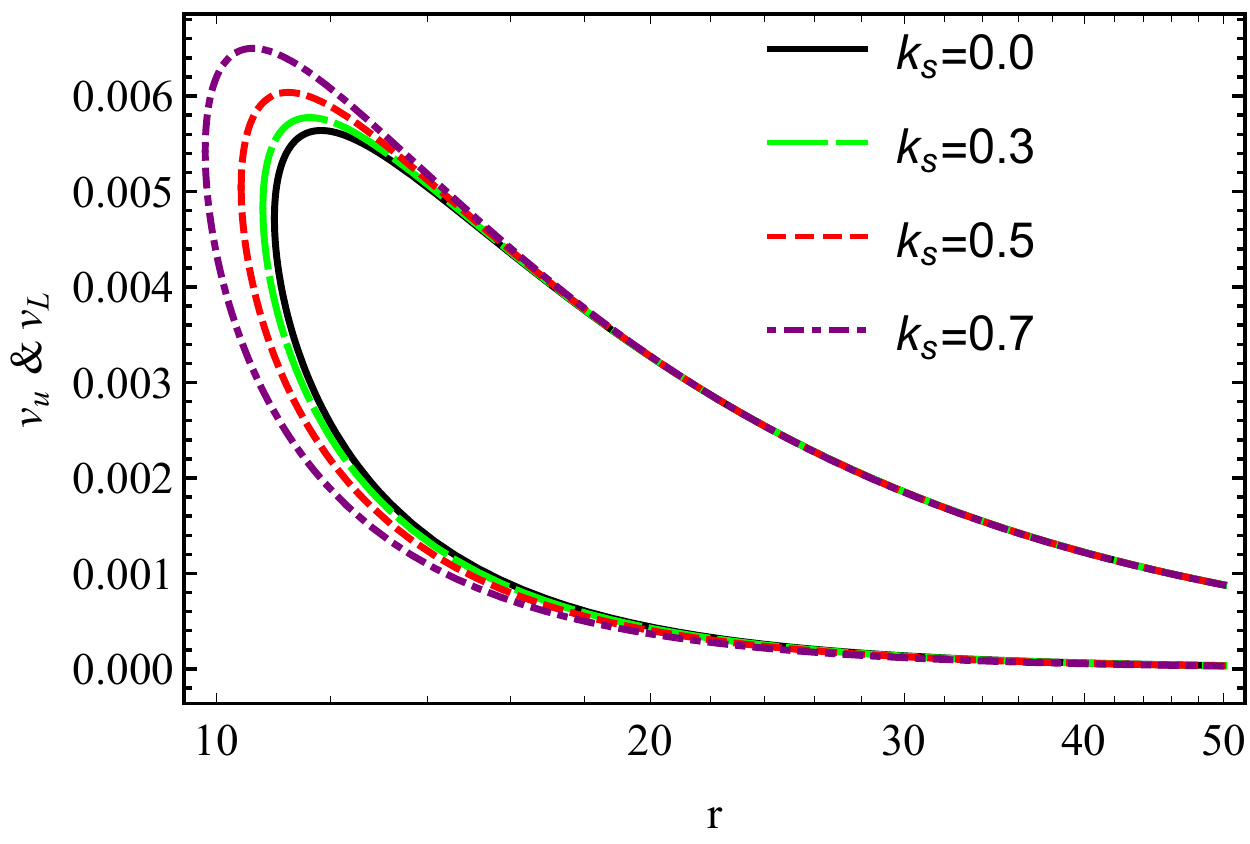}\\
    \includegraphics[scale=0.52]{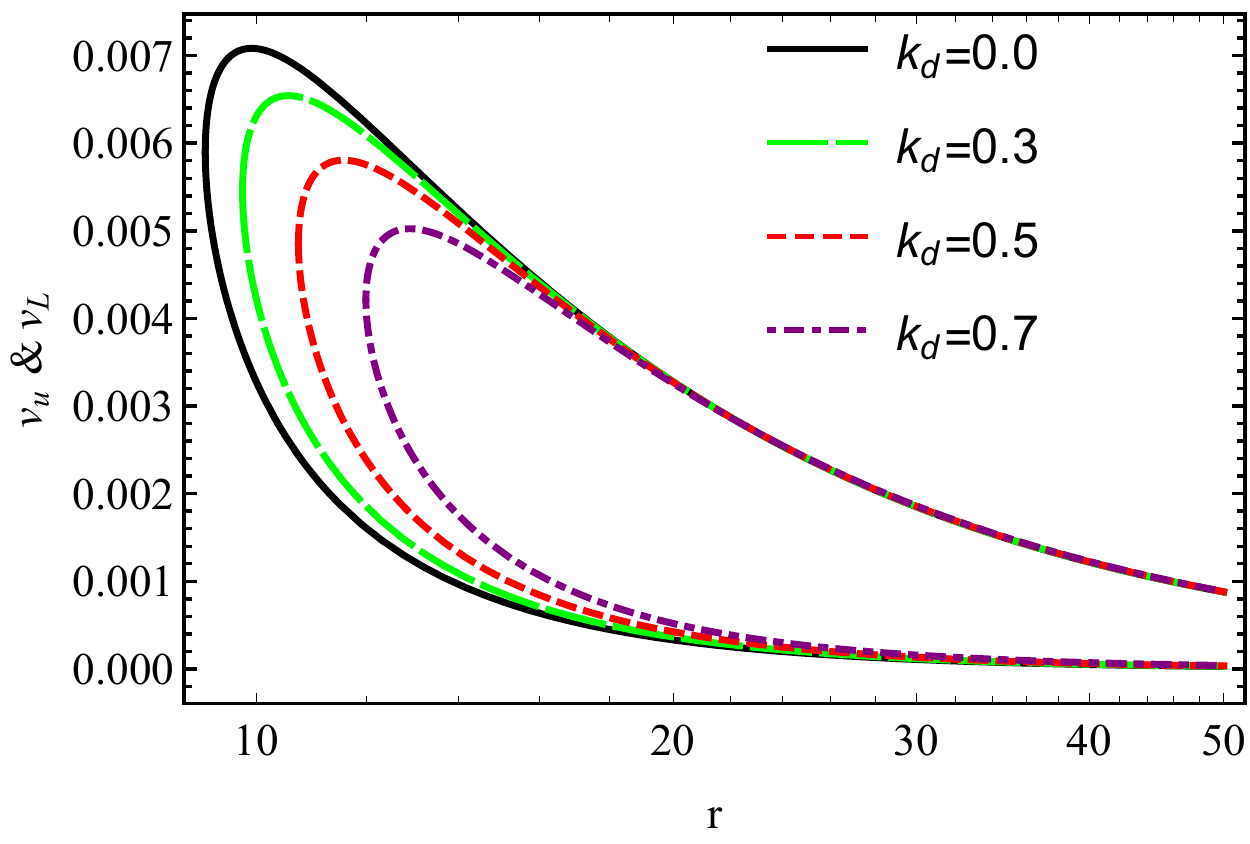}
    \includegraphics[scale=0.52]{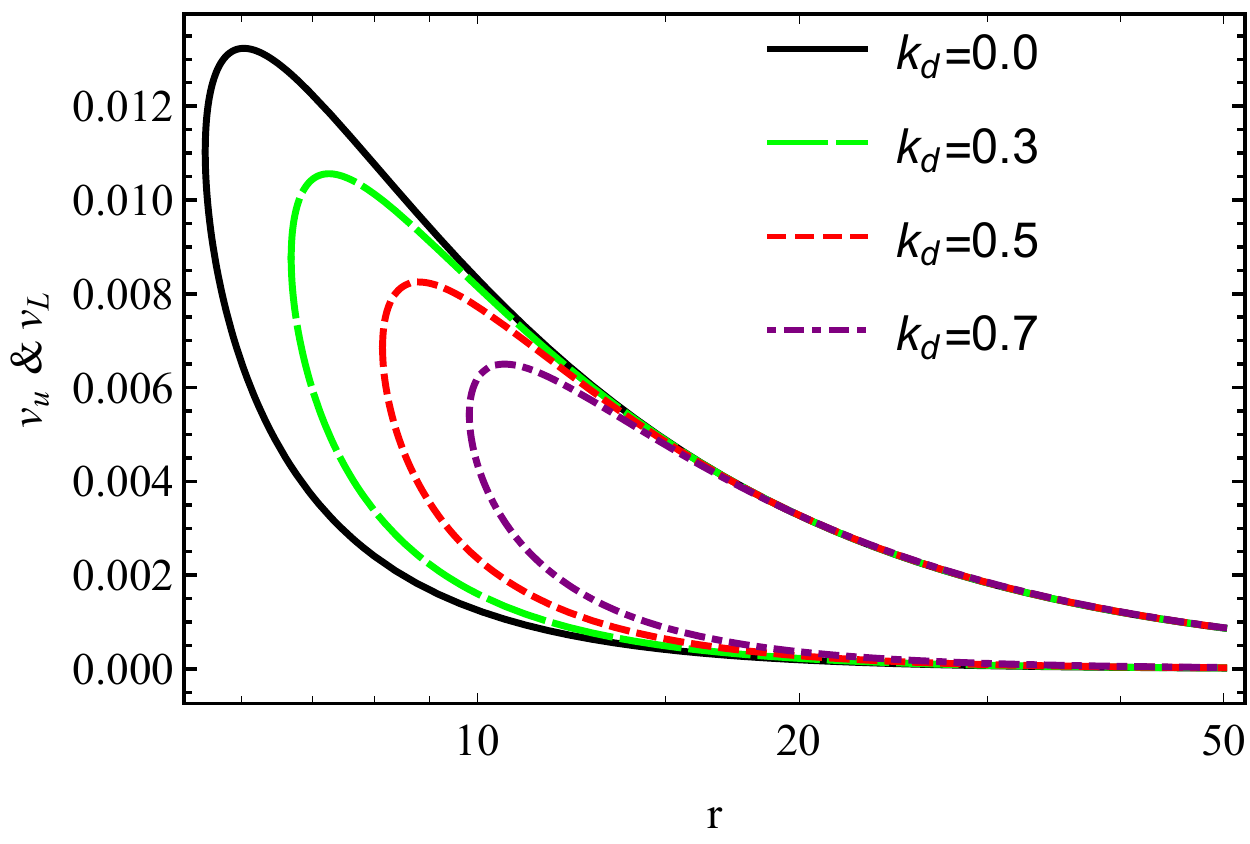}\\
    \includegraphics[scale=0.52]{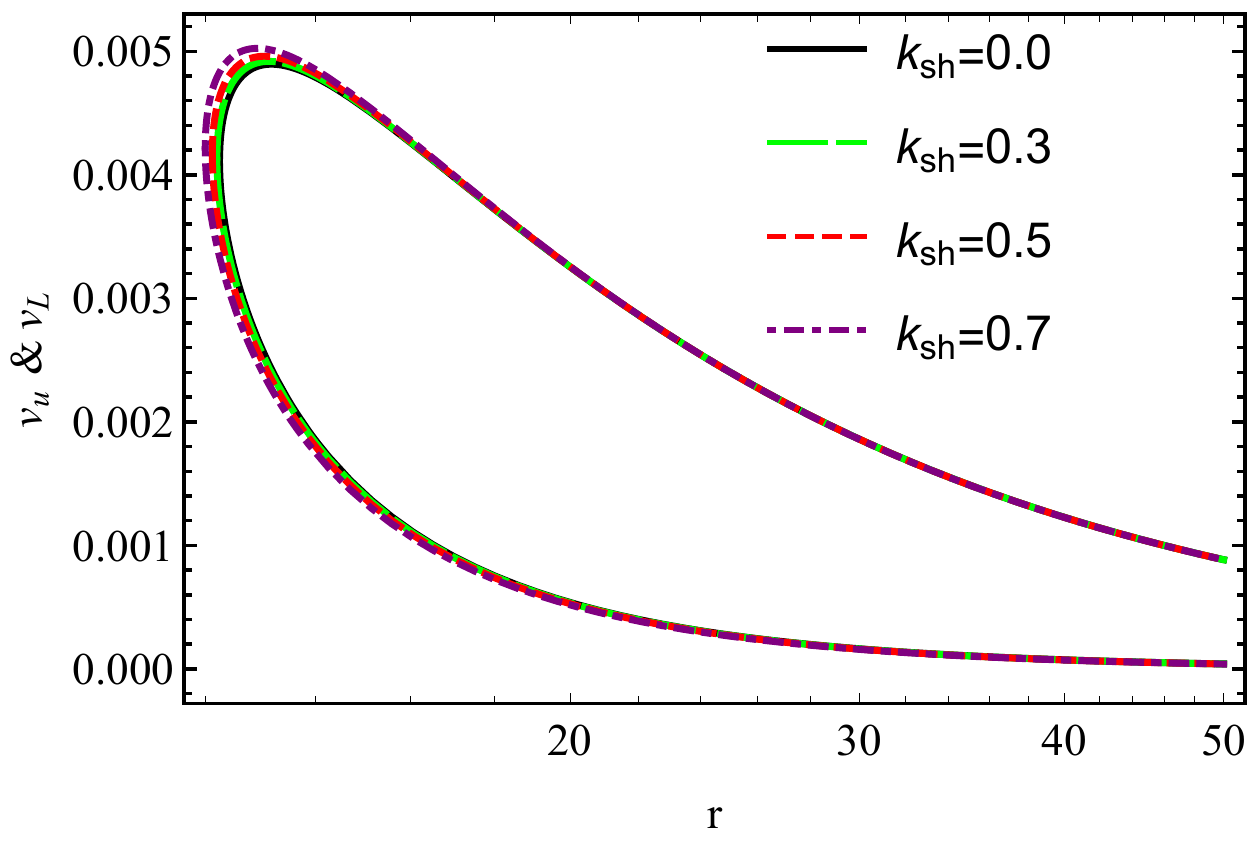}
    \includegraphics[scale=0.52]{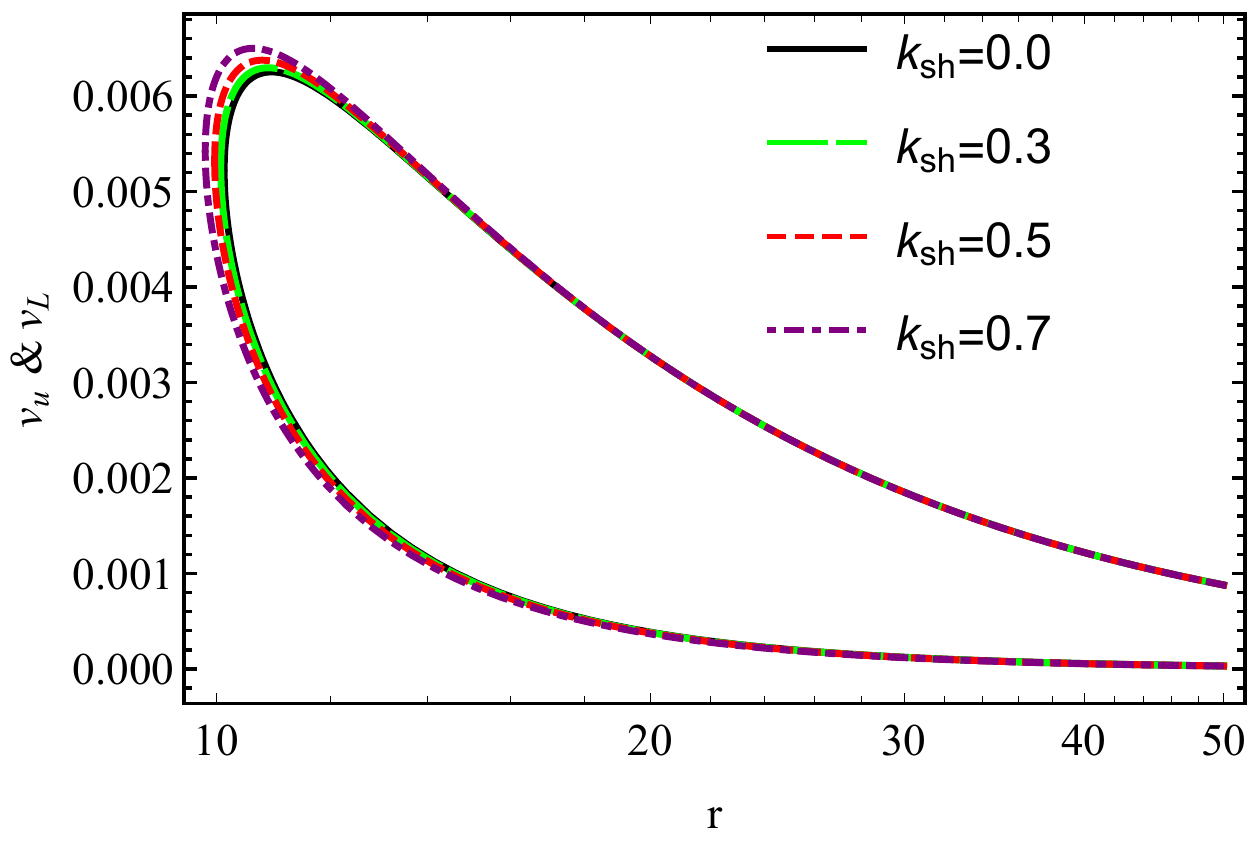}
    \caption{ER4 Model: Upper $v_u$ and lower $v_L$ frequencies for case $d_1=8f_1$ (Left panel) and for case $d_1=-8f_1$ (Right panel) along $c_1$ for different values of $f_1,\; k_s,\; k_d, \;\&\; k_{sh} $. Here we consider the choice for fix values $M=1,c_1=1,\;f_1=-0.2,\;k_s=k_{sh}=k_d=0.7$.}
    \label{plot:10}
\end{figure}

In this portion of the manuscript we account for the possible frequencies of twin peak QPOs  \cite{145} surrounding the $MAGBHs$ (metric affine gravity black holes).
\begin{itemize}

\item The standard relativistic procession (RP) model \cite{146} identifies the upper and lower frequencies based on the upper and lower frequencies respectively as $v_{U}=v_{\phi}$ and $v_{L}=v_{\phi}-v_{r}$. In modified RP1 model frequencies are represented as $v_{U}=v_{\theta},\;v_{L}=v_{\phi}-v_{r}$, and in modified RP2 model frequencies are represented as $v_{U}=v_{\phi},\;v_{L}=v_{\theta}-v_{r}$. The graphical formation of RP model with increasing and decreasing behaviours with change in MAGBHs parameters in both $d_{1}=8f_{1}\;\&\;d_{1}=-8f_{1}$  are plotted in Fig.~\ref{plot:7}.

\item The three epicyclic resonance models are represented by ER2, ER2, ER3 \cite{147}. ER models hypothesis the amassing disks to be thick, and the resonance of regularly radiation particles along geodesics orbits produce the QPOs. The upper and lower frequencies of
orbital and epicyclic oscillations  for ER models are defined as: for ER2 model $v_U=2v_{\theta}-v_r,\;\&\;v_L=v_r$, for ER3 model $v_U=v_{\theta}+v_r,\;\&\;v_L=v_{\theta}$, and for ER4 model $v_U=v_{\theta}+v_r,\;\&\;v_L=v_{\theta}-v_r$. We plot the ER models (ER2, ER3, ER4) with all the effects of MAGBH geometry parameters ($c_1,\;f_1,\;d_1,\;k_d,\;k_s,\;\&\;k_{sh}$) in both $d_1=8f_1\;\&\;d_1=-8f_1$ in Figs.~\ref{plot:8}, \ref{plot:9}, \ref{plot:10}.
\end{itemize}


\section{red-blue shifts of the photons emitted by the particles}\label{A6}
This section of the article is concerned to the red-blue shift of the light rays advancing from moving the test particles in the domain of MAGBH. By the virtue of being angular momentum and energy conserved quantities, one can obtain the photon's impact parameter $b$ as:
\begin{equation}
    b=\frac{L}{E}\pm\left(-\frac{g_{\phi\phi}}{g_{tt}}\right)^{\frac{1}{2}}
\end{equation}
The frequency shift $z$ in the context of photons emission and detection with circular geodesics $\left(\gimel^r=0\right)$ in equatorial motion $\left(\gimel^\theta=0\right)$ can be calculated by the formula:
\begin{equation}
    1+z=\frac{\gimel_e^t-b_e \gimel_e^\phi}{\gimel_d^t-b_d \gimel_d^\phi} .
\end{equation}
The kinematic frequency shift form of observational red shift is: $z_k=z-z_c$, where $z_c$ notions the shift at $b=0$ as given below
\begin{equation}
    1+z_c=\frac{\gimel_e^t}{\gimel_d^t} .
\end{equation}
The expression realizing the kinematic frequency shift is as follows
\begin{equation}
    z_{\text {kin }}=\frac{\gimel_e^t \gimel_d^\phi b_d-\gimel_d^t \gimel_e^\phi b_e}{\gimel_d^t\left(\gimel_d^t-b_d \gimel_d^\phi\right)} .
\end{equation}
Since the location of  detector far away from the BH, so
\begin{equation}
    z=\left.\gimel_e^\phi b_{+}\right|_{r_c}=\left.\sqrt{-\frac{g_{\phi \phi}}{g_{t t}}}\left(\frac{L}{g_{\phi \phi}}-\frac{k A_\phi}{g_{\phi \phi}}\right)\right|_{r_c} .
\end{equation}
The redshift $z$, calculated for the light emitted from the test particles in a stable circular orbit in the equatorial plane for MAGBH geometry is as given below:
\begin{equation}
    z=-\sqrt{\frac{\left(M^2 \left(12 c_1 k_d^2-3 d_1 k_s^2+6 f_1 k_{\text{sh}}^2+4 M^2\right)+2 M^2 \sqrt[3]{z_2}+z_2^{2/3}\right){}^2}{M^2 z_2^{2/3} \left(\frac{-4 c_1 k_d^2+d_1 k_s^2-2 f_1 k_{\text{sh}}^2}{\sqrt[3]{M \left(\frac{z_3}{z_2}+2\right)}}+\frac{\sqrt[3]{3} \sqrt[3]{z_2}}{M}+1\right){}^2-\frac{2 M}{z_4}}} \left(\frac{4 c_1 k_d^2-d_1 k_s^2+2 f_1 k_{\text{sh}}^2+M z_4}{8 c_1 k_d^2-2 d_1 k_s^2+4 f_1 k_{\text{sh}}^2+3 M z_4-z_4^2}\right)
\end{equation}
where
\begin{eqnarray*}
    z_1&=&\sqrt{M^4 z_5 \left(4 c_1 k_d^2 z_6+64 c_1^2 k_d^4-2 f_1 k_{\text{sh}}^2 \left(8 d_1 k_s^2-9 M^2\right)-9 d_1 M^2 k_s^2+4 d_1^2 k_s^4+16 f_1^2 k_{\text{sh}}^4+5 M^4\right)},\\
    z_2&=&4 c_1 M^2 k_d^2 \left(-4 d_1 k_s^2+8 f_1 k_{\text{sh}}^2+9 M^2\right)+32 c_1^2 M^2 k_d^4-8 d_1 f_1 M^2 k_s^2 k_{\text{sh}}^2-9 d_1 M^4 k_s^2+2 d_1^2 M^2 k_s^4\\&+&18 f_1 M^4 k_{\text{sh}}^2+8 f_1^2 M^2 k_{\text{sh}}^4+8 M^6+z_1,\;\;\;
    z_3=12 c_1 k_d^2-3 d_1 k_s^2+6 f_1 k_{\text{sh}}^2+4 M^2,\\
    z_4&=&\frac{\sqrt[3]{z_2}}{M}+\frac{M z_3}{\sqrt[3]{z_2}}+2 M,\;\;\;
    z_5=\left(4 c_1 k_d^2-d_1 k_s^2+2 f_1 k_{\text{sh}}^2\right){}^2,\;\;
    z_6=\left(-8 d_1 k_s^2+16 f_1 k_{\text{sh}}^2+9 M^2\right).
\end{eqnarray*}

\begin{figure}
    \centering
    \includegraphics[scale=0.485]{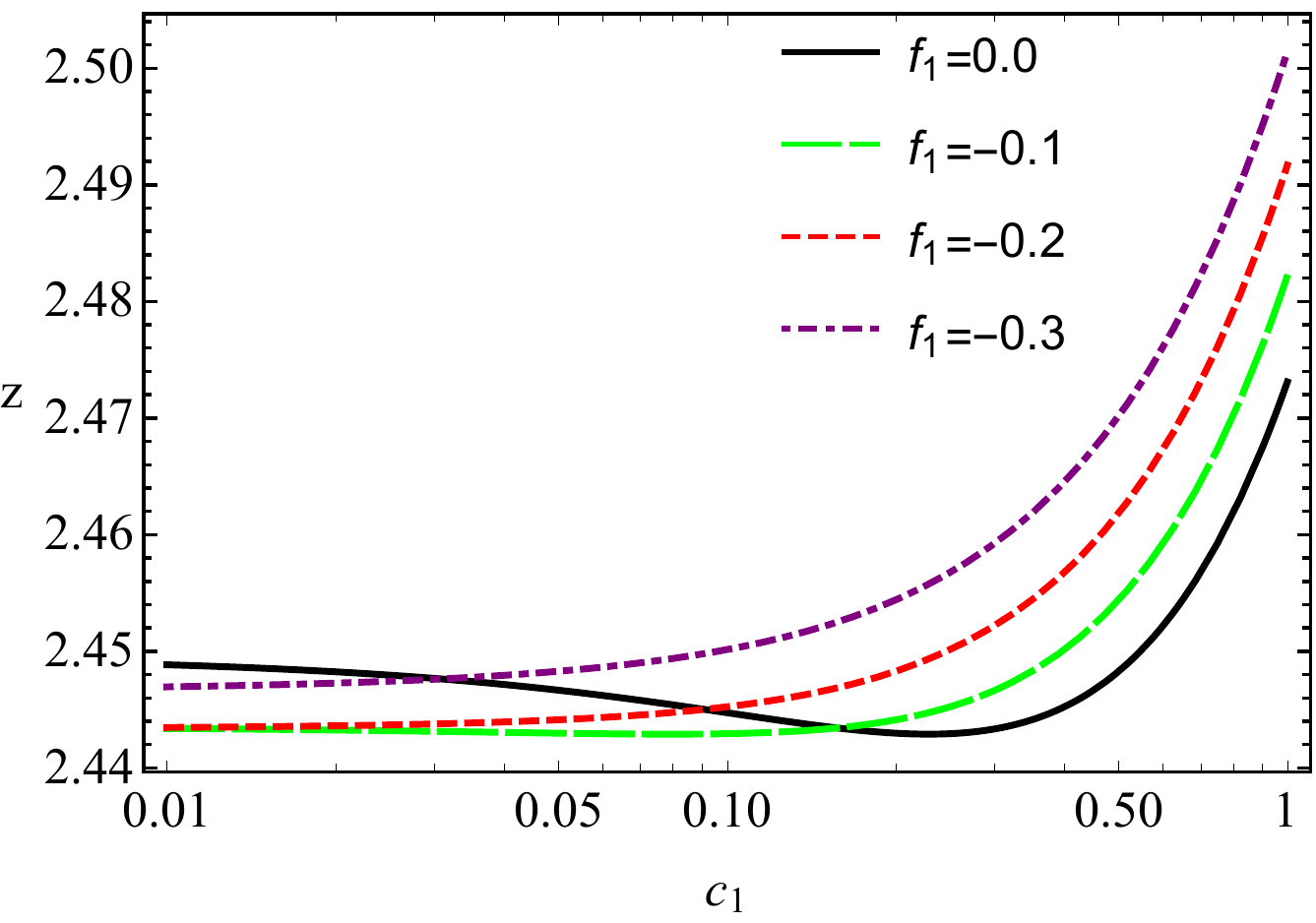}
    \includegraphics[scale=0.52]{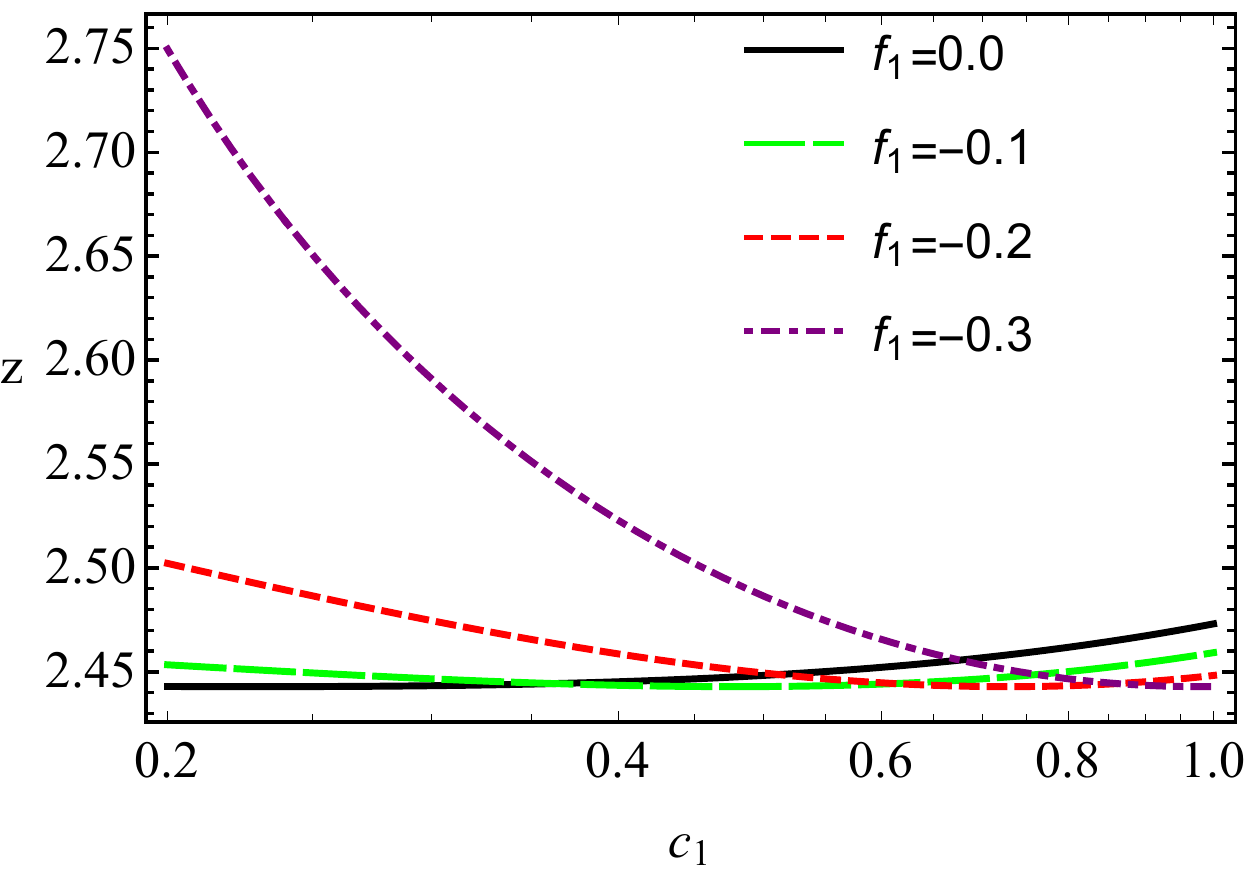}
    \includegraphics[scale=0.52]{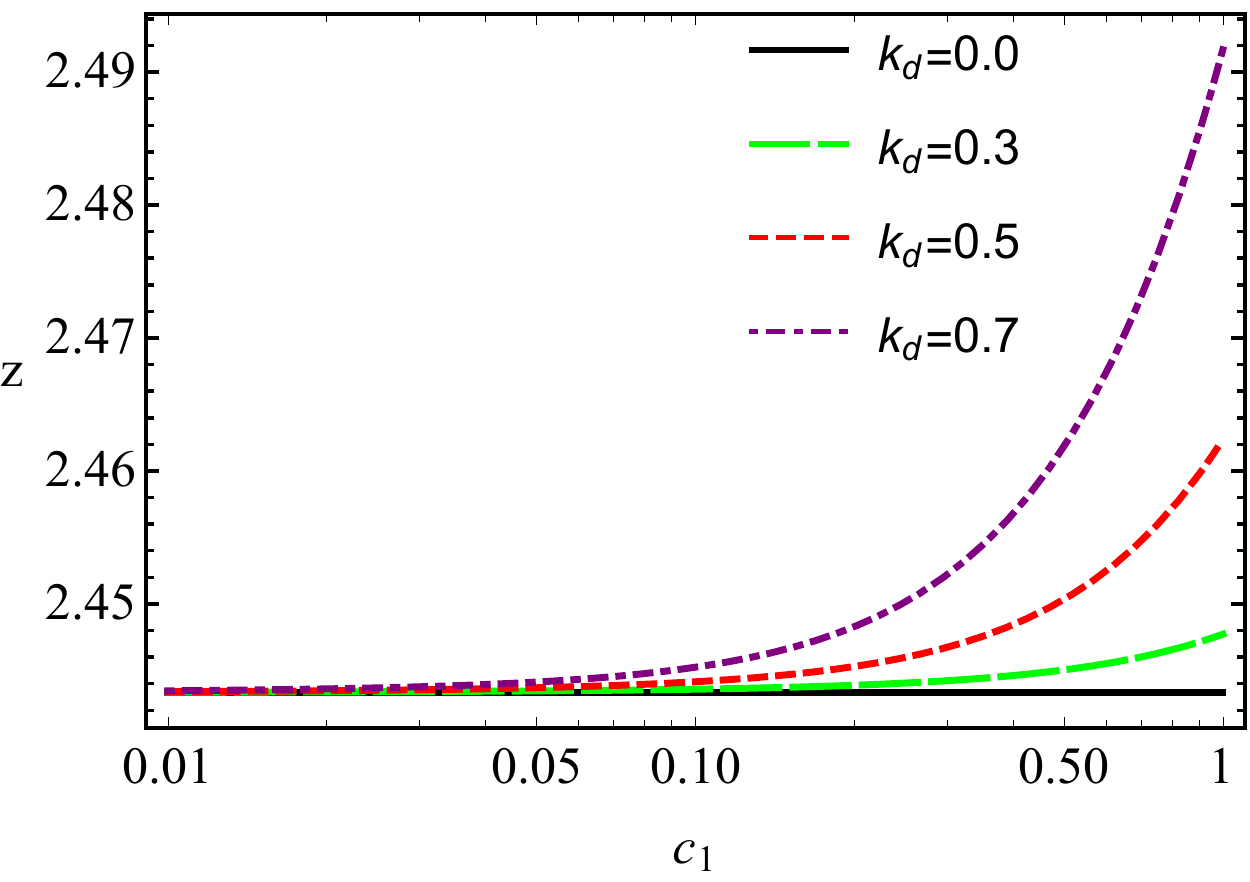}
    \includegraphics[scale=0.52]{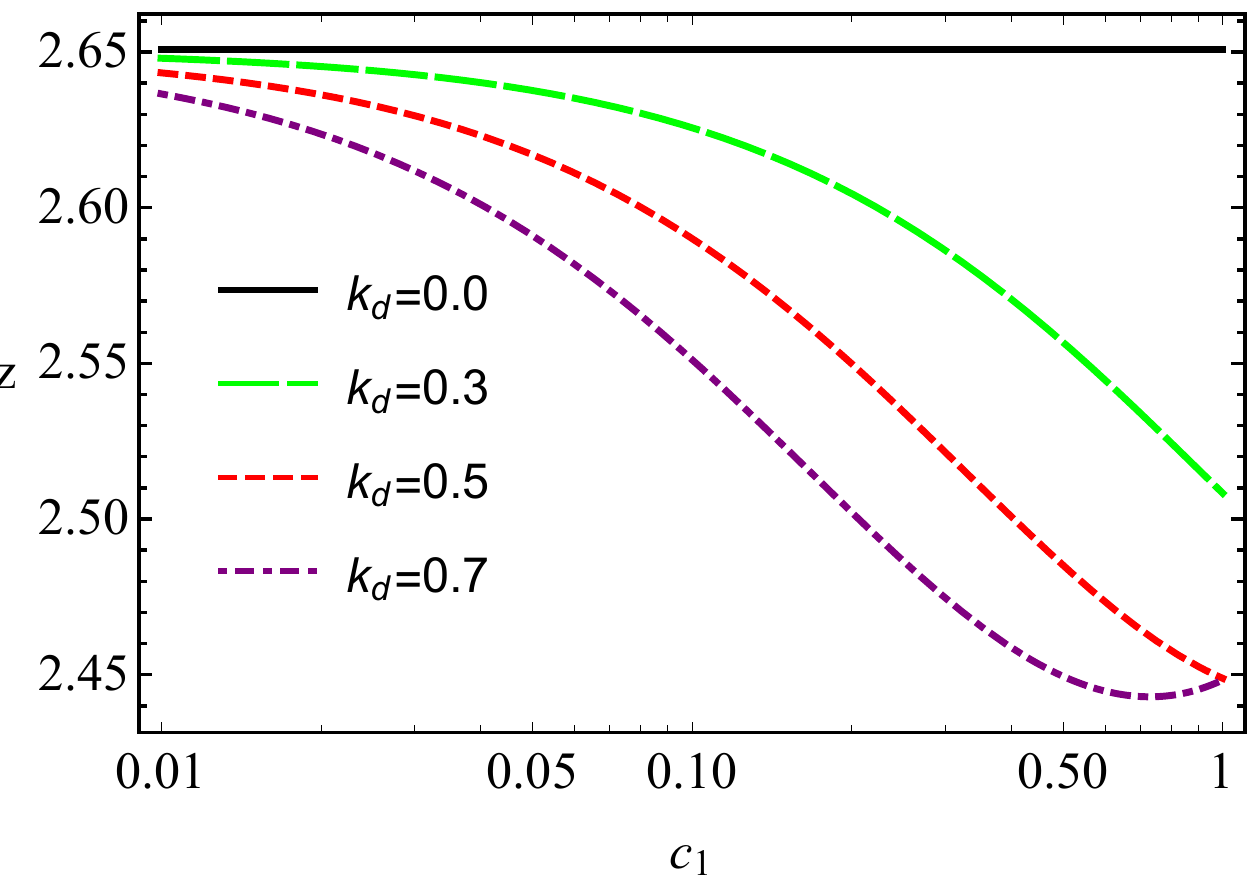}
    \includegraphics[scale=0.52]{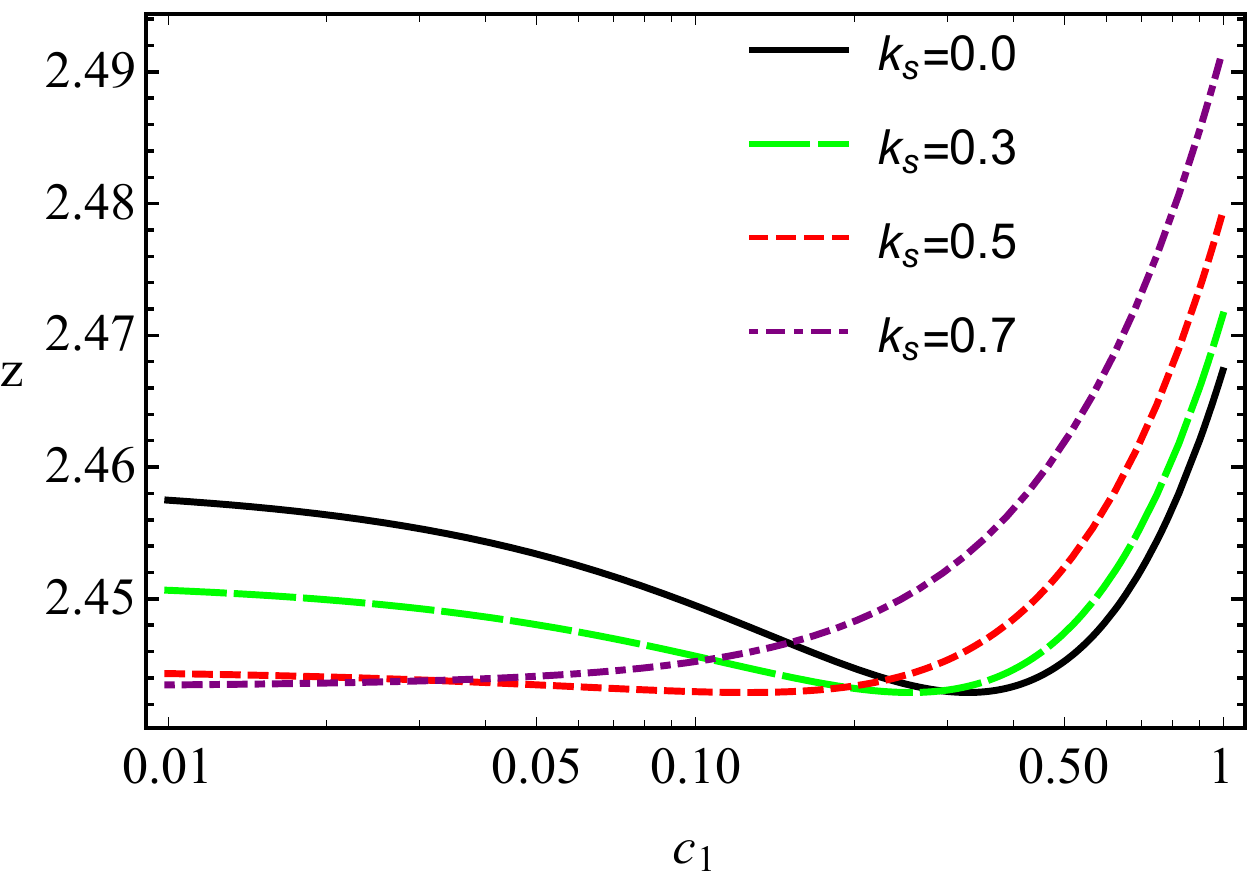}
    \includegraphics[scale=0.52]{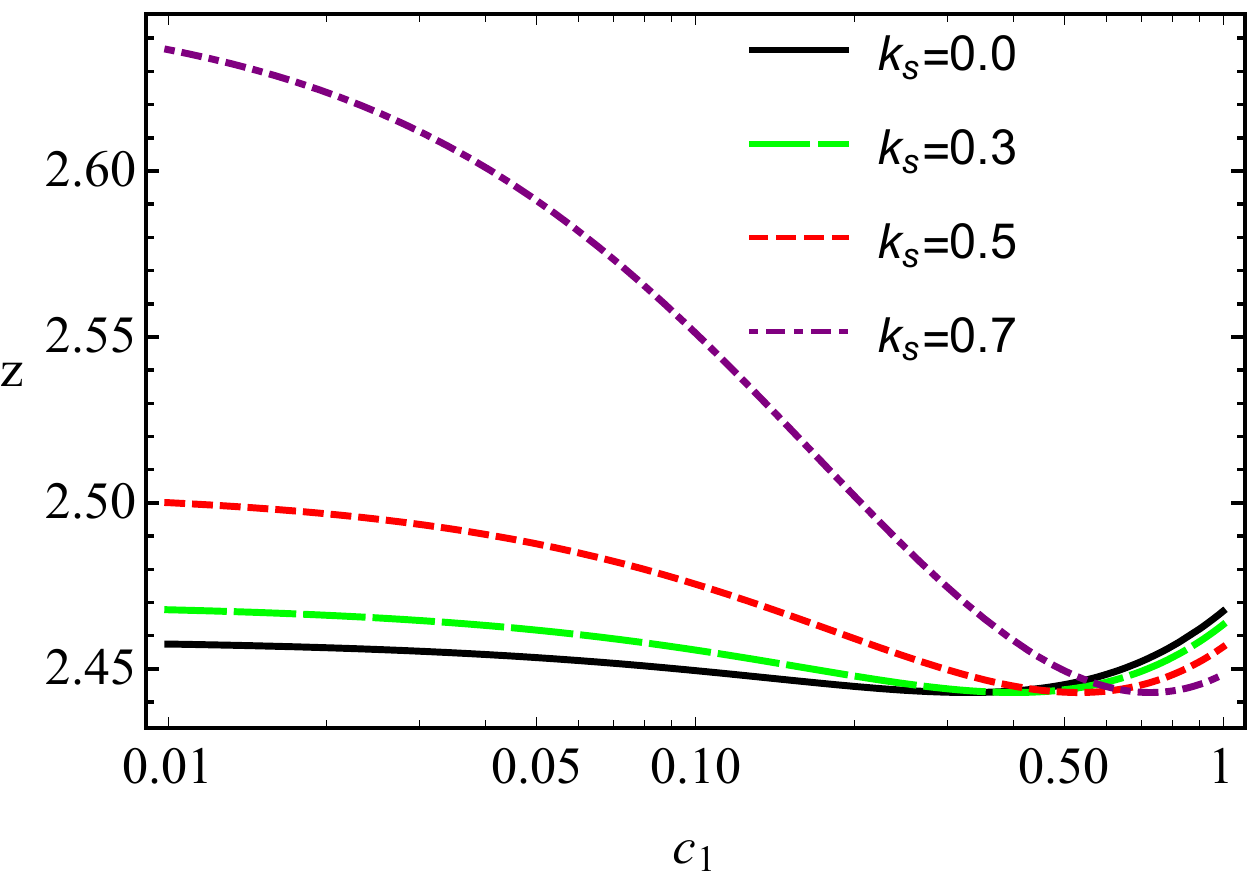}
    \includegraphics[scale=0.52]{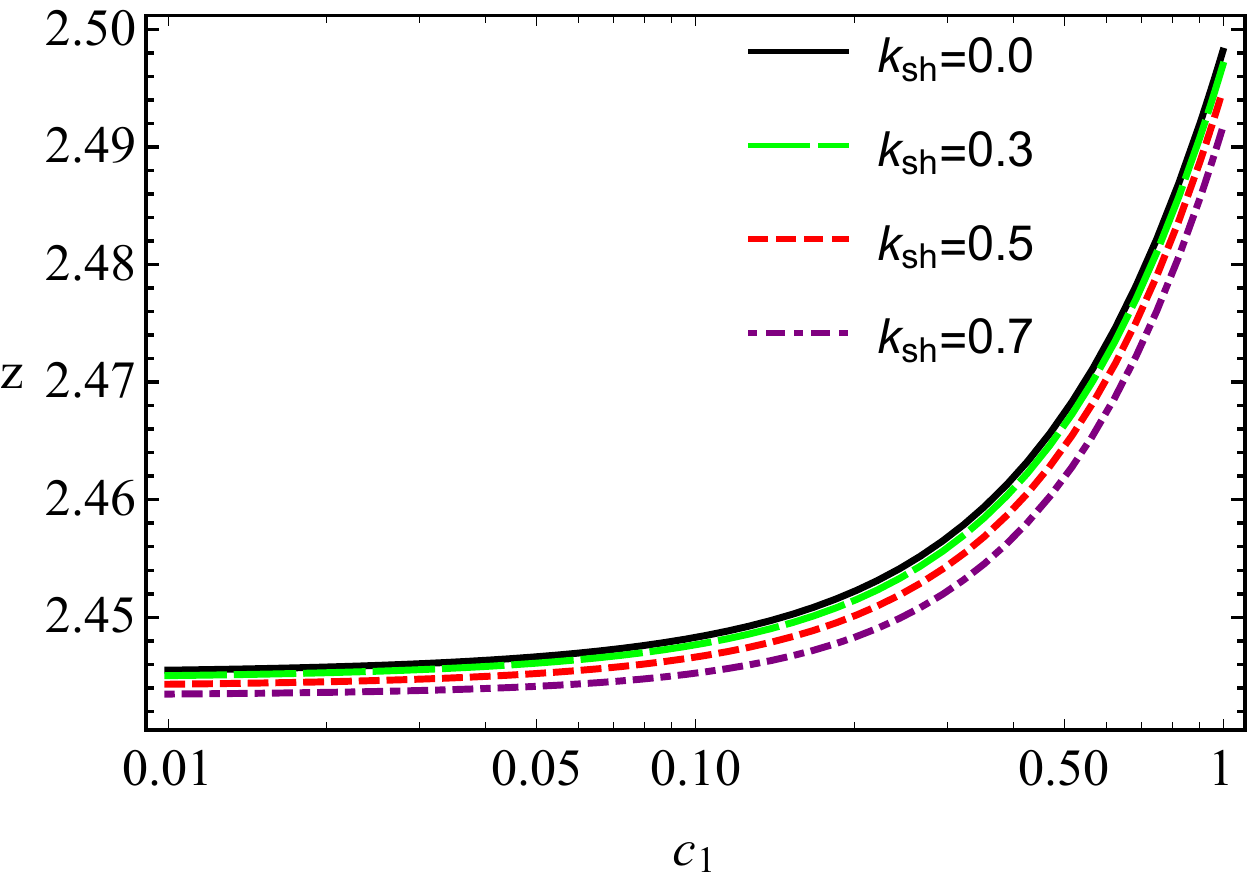}
    \includegraphics[scale=0.52]{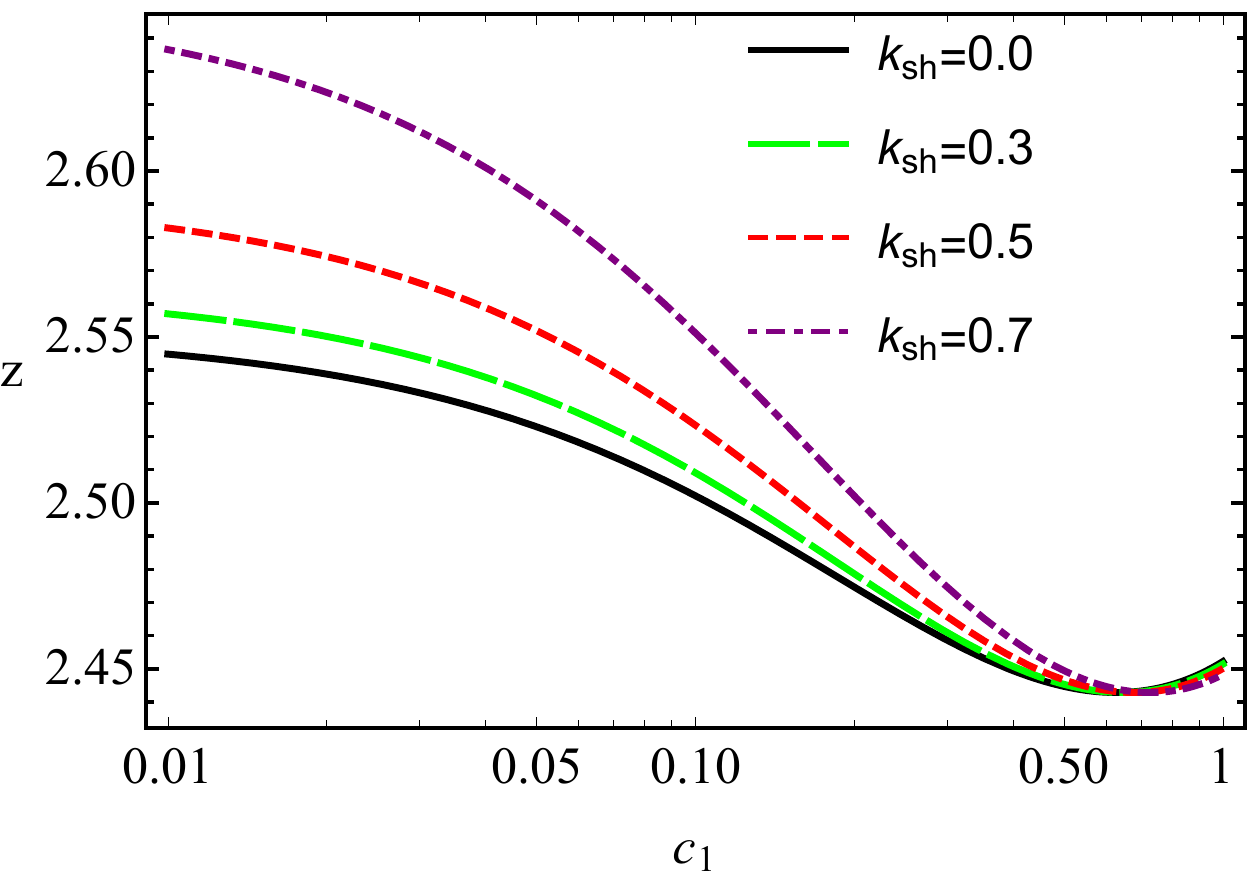}
    \caption{Blue redshift $z$ for case $d_1=8f_1$ (Left panel) and for case $d_1=-8f_1$ (Right panel) along $c_1$ for different values of $f_1,\; k_s,\; k_d, \;\&\; k_{sh} $. Here we consider the choice for fix values $M=1,c_1=1,\;f_1=-0.2,\;k_s=k_{sh}=k_d=0.7$.}
    \label{plot:11}
\end{figure}

Plots in Fig.~\ref{plot:11} show that values of $z$ along $c_1$ increase with increase in $f_1,\;k_d,\;k_s,\;k_{sh}$ in case of $d_1=8f_1$ while decrease in case of $d_1=-8f_1$. Thus the observer may experience with the change of the horizons depending upon the values of $f_1,\;k_d,\;k_s,\;\&\;k_{sh}$.

\section{Effective force}\label{A7}
One may guess whether the movement of test particle is away or towards the central source from the behaviour of effective force. Effective force a test particle may experience in the field of gravitating source \cite{148} can be determined as given below:
\begin{equation}
    F=-\frac{1}{2}\frac{\partial\; V_{eff}(r)}{\partial r}.
\end{equation}
The expression of effective force calculated in case of MAGBH geometry is as:
\begin{eqnarray}
    F&=&-\frac{\left(4 c_{1} k_{d}^2-d_{1} k_{s}^2+2 f_{1} k_{sh}^2+2 M r-r^2\right)}{2 r^3 \left(-8 c_{1} k_{d}^2+2 d_{1} k_{s}^2-4 f_{1} k_{sh}^2-3 M r+r^2\right)^2} \Big[64 c_{1}^2 k_{d}^4+4 c_{1} k_{d}^2 \left(-8 d_{1} k_{s}^2+16 f_{1} k_{sh}^2+9 M r\right)\nonumber\\&+&4 d_{1}^2 k_{s}^4-2 f_{1} k_{sh}^2 \left(8 d_{1} k_{s}^2-9 M r\right)-9 d_{1} k_{s}^2 M r+16 f_{1}^2 k_{sh}^4+6 M^2 r^2-M r^3\Big]\ .
\end{eqnarray}

\begin{figure}
    \centering
    \includegraphics[scale=0.52]{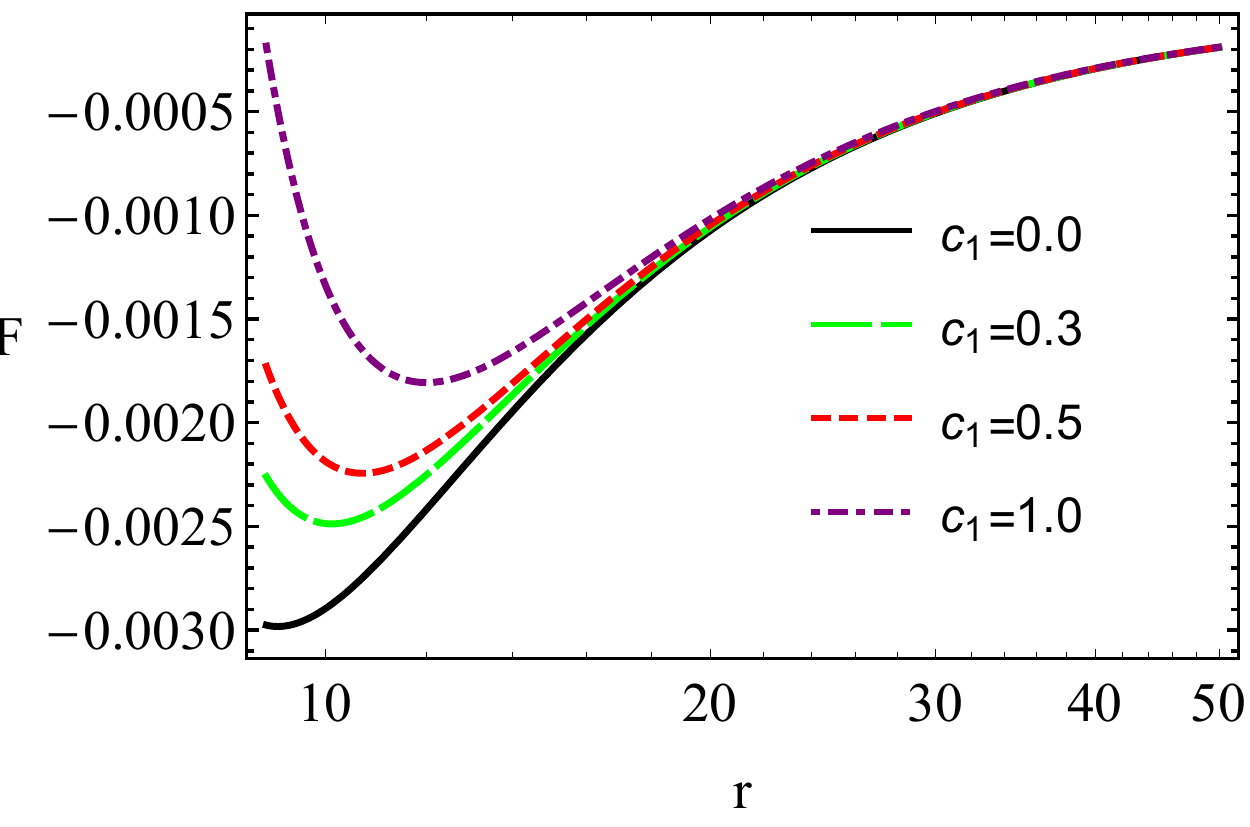}
    \includegraphics[scale=0.52]{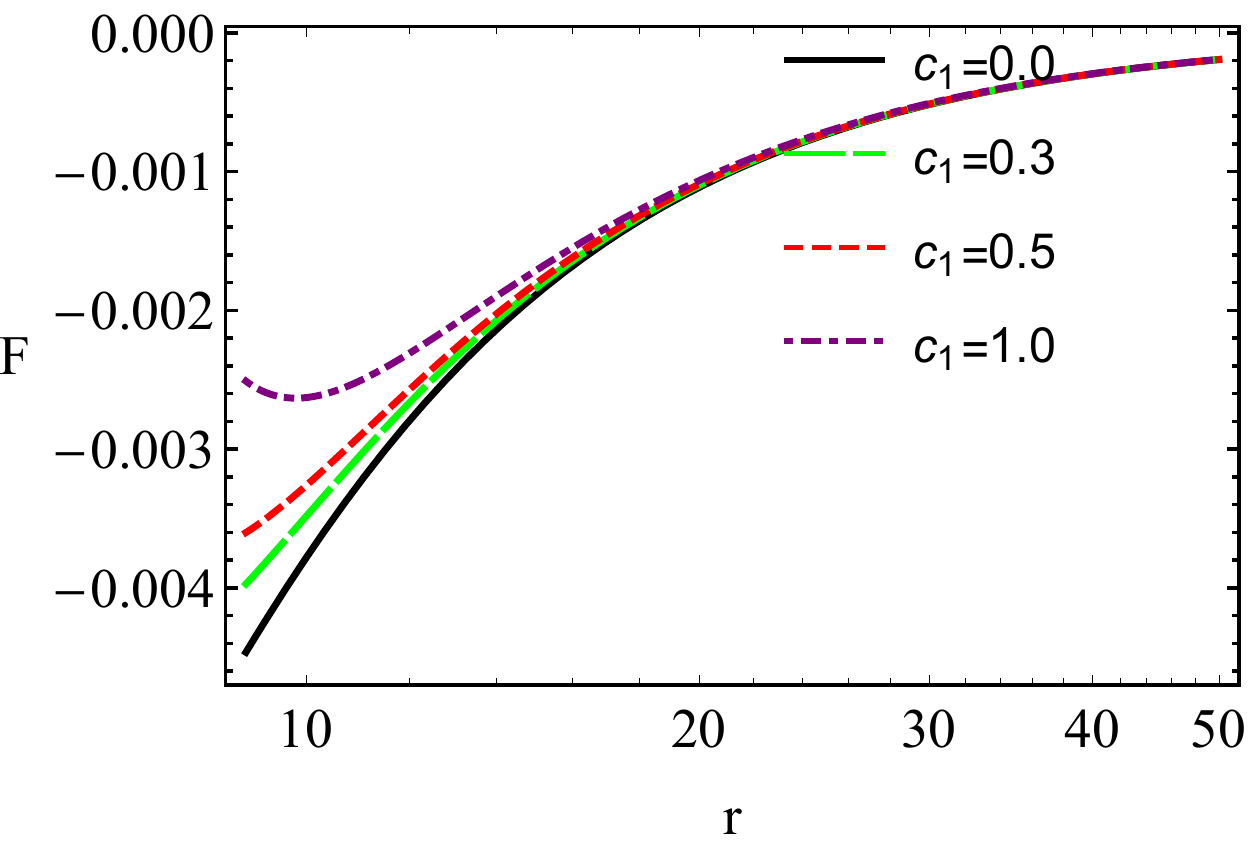}
    \includegraphics[scale=0.52]{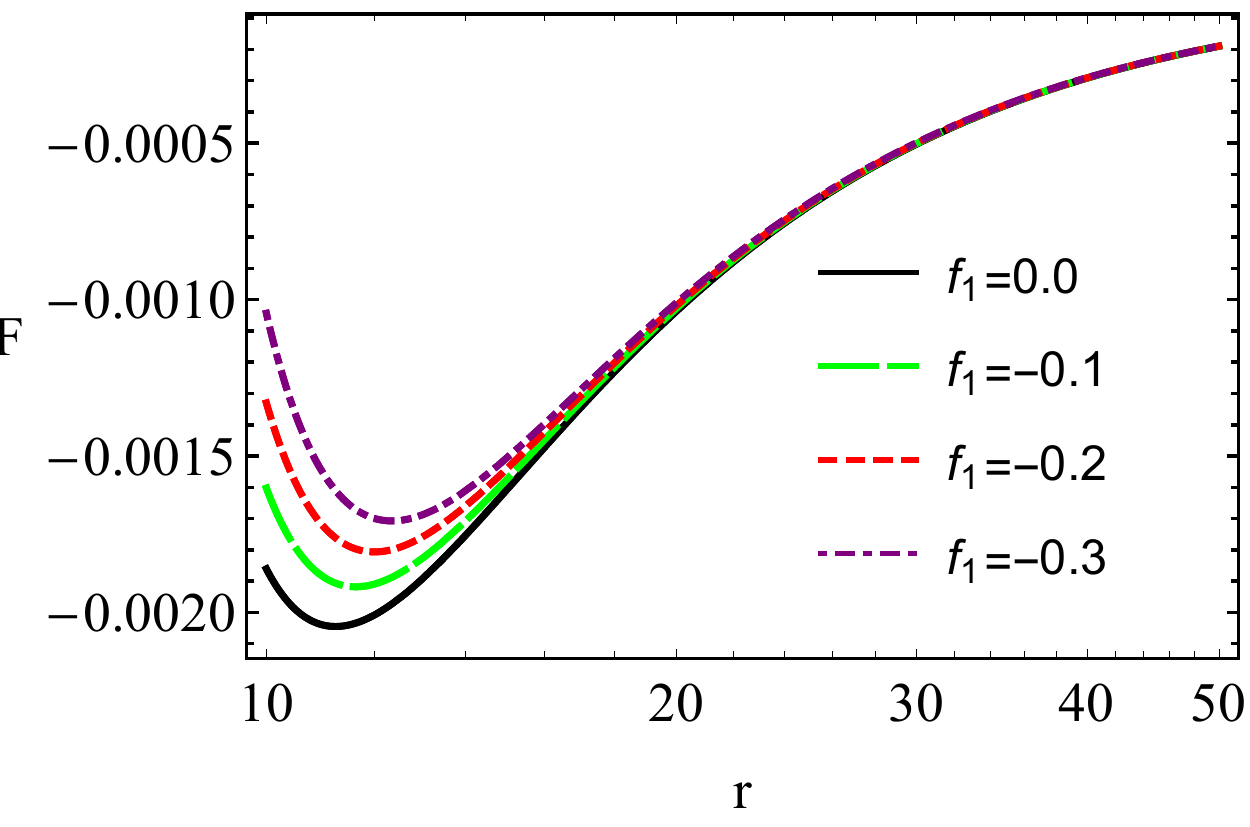}
    \includegraphics[scale=0.52]{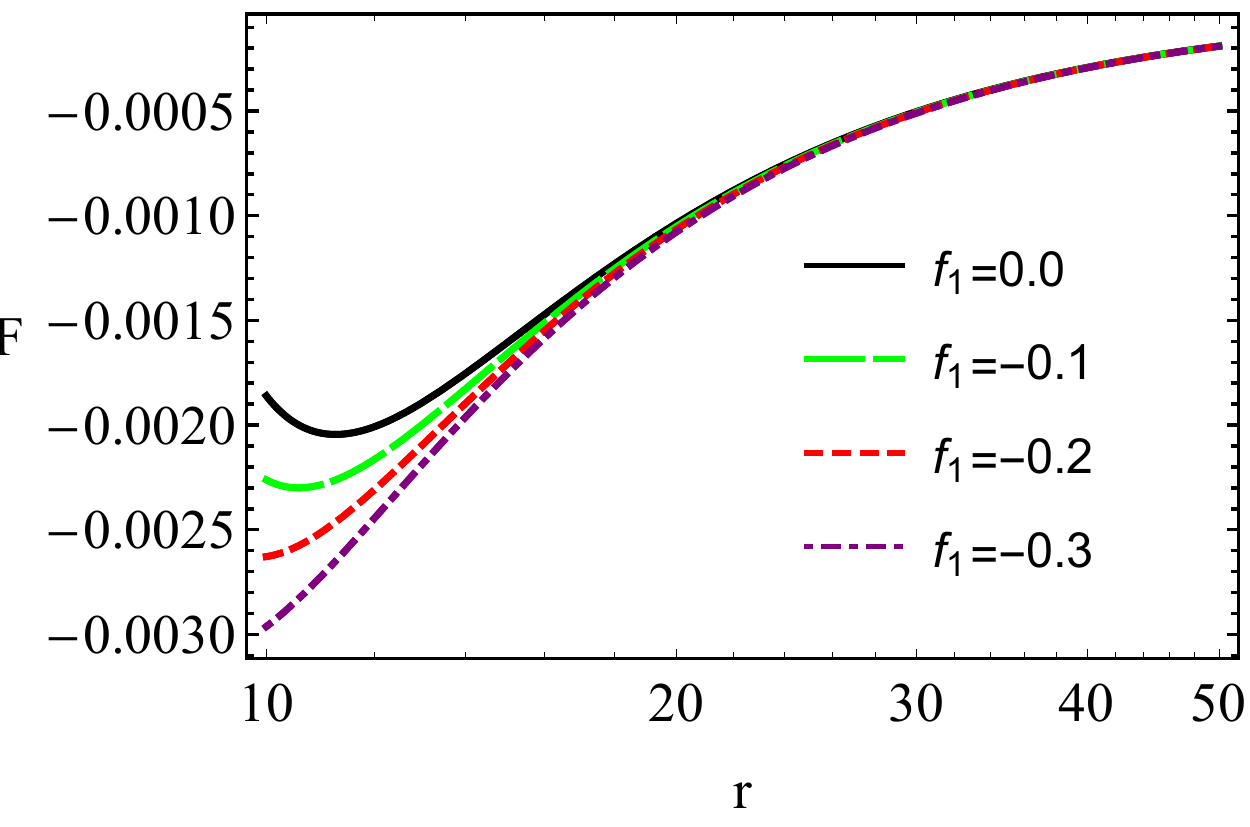}
    \includegraphics[scale=0.52]{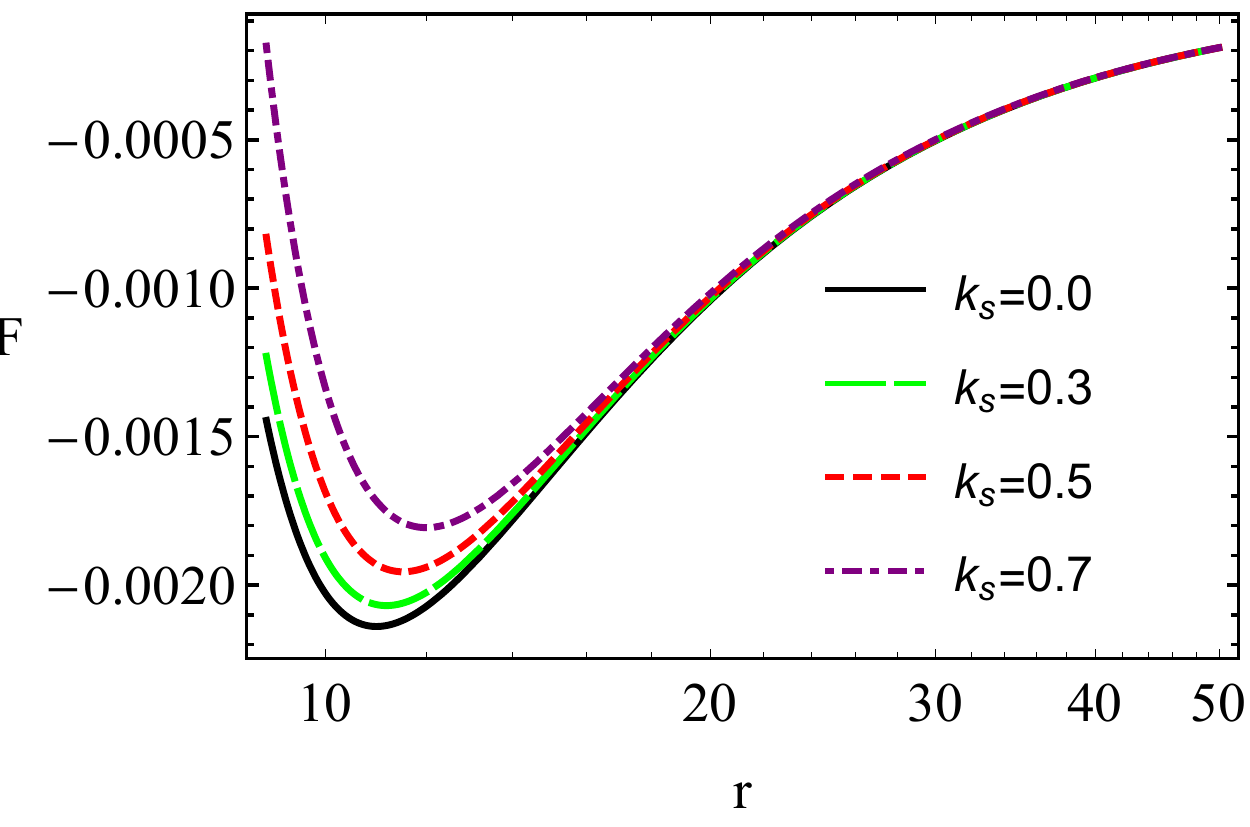}
    \includegraphics[scale=0.52]{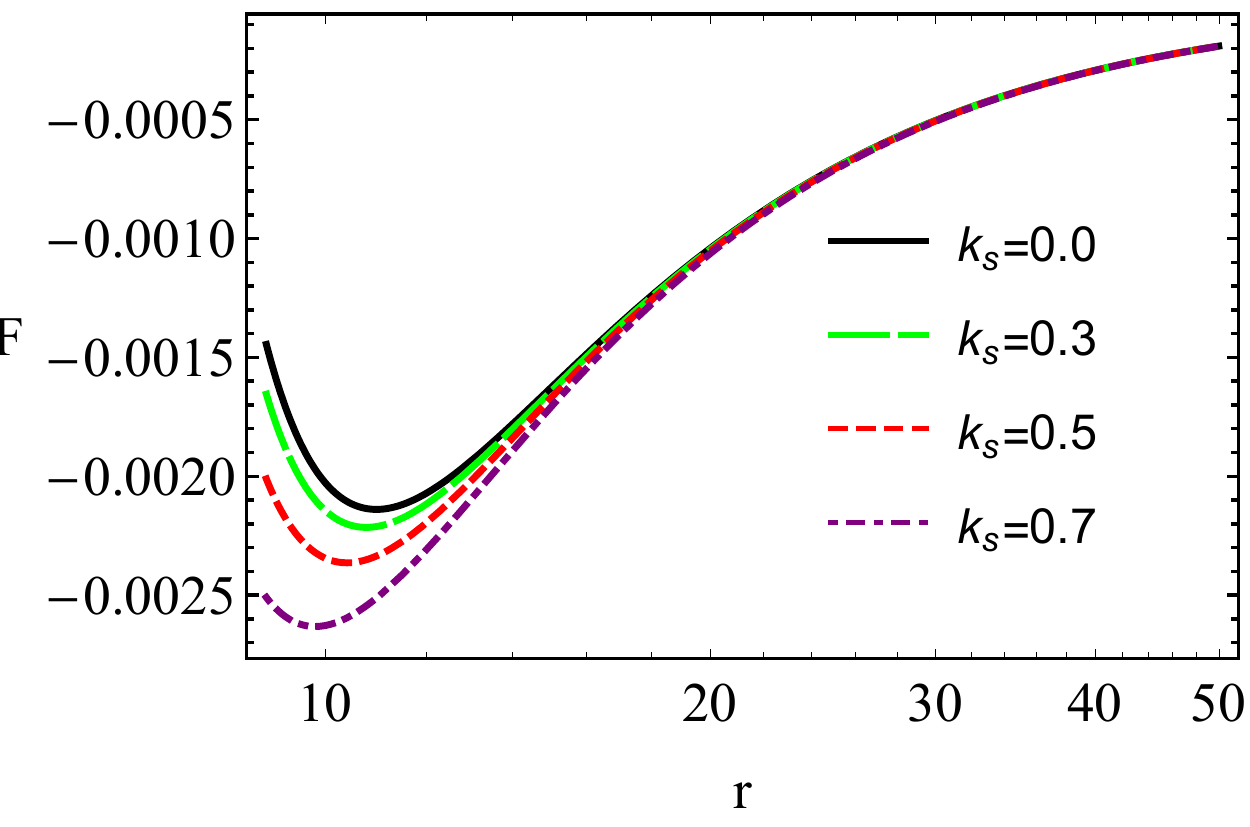}
    \includegraphics[scale=0.52]{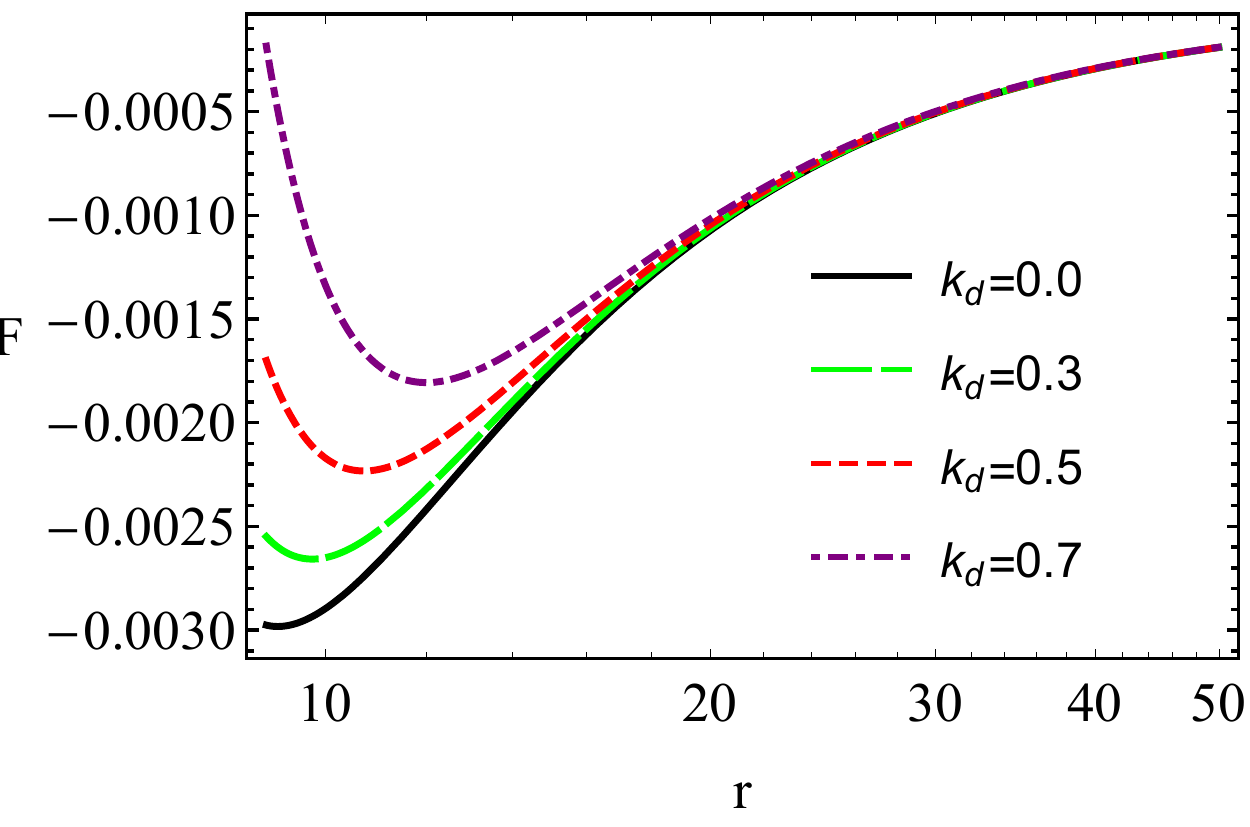}
    \includegraphics[scale=0.52]{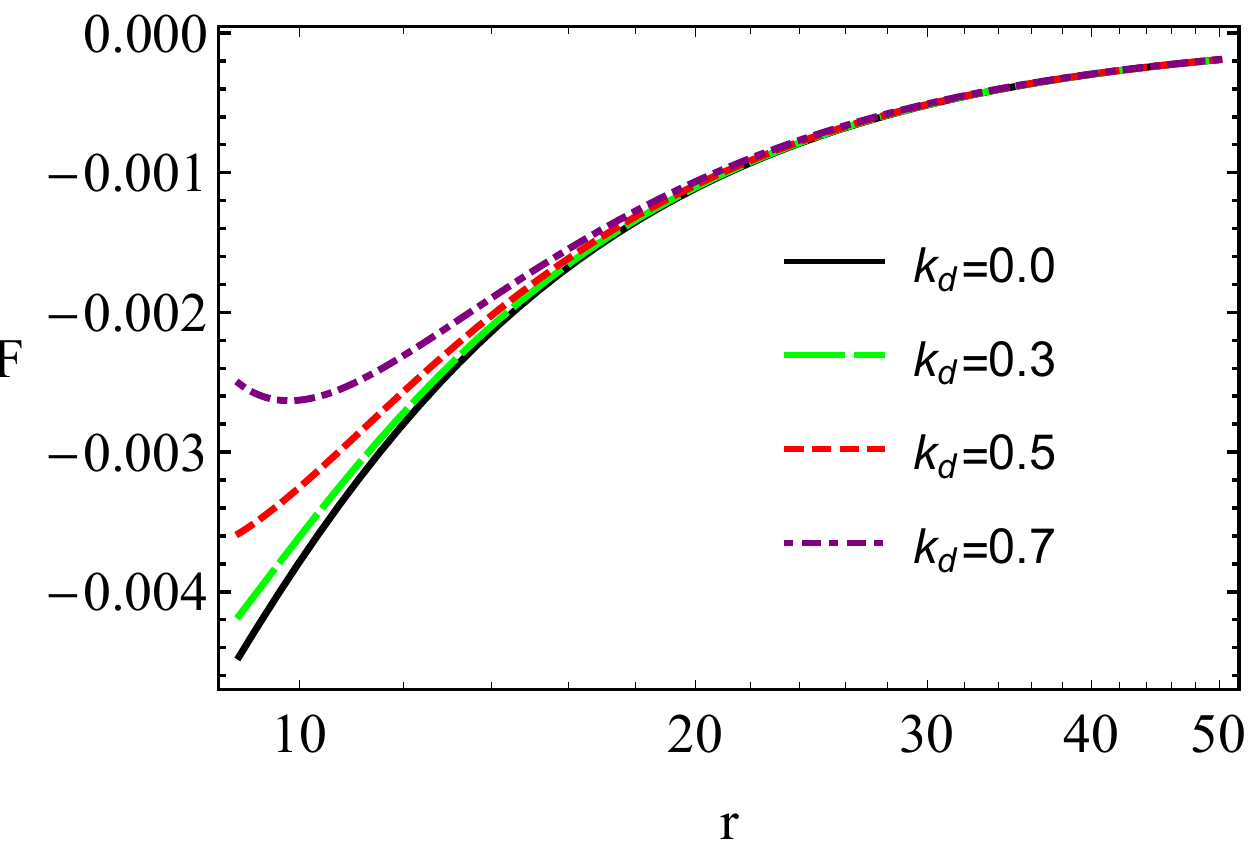}
    \includegraphics[scale=0.52]{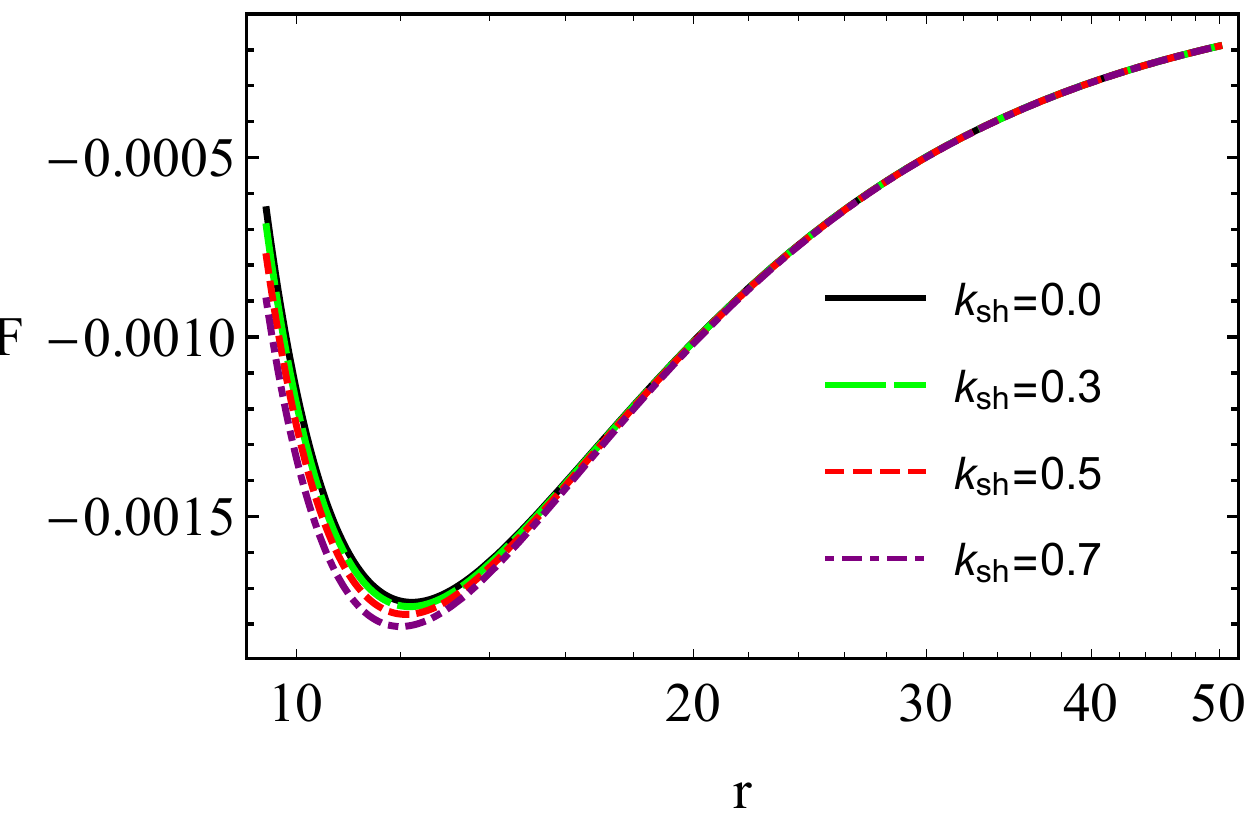}
    \includegraphics[scale=0.52]{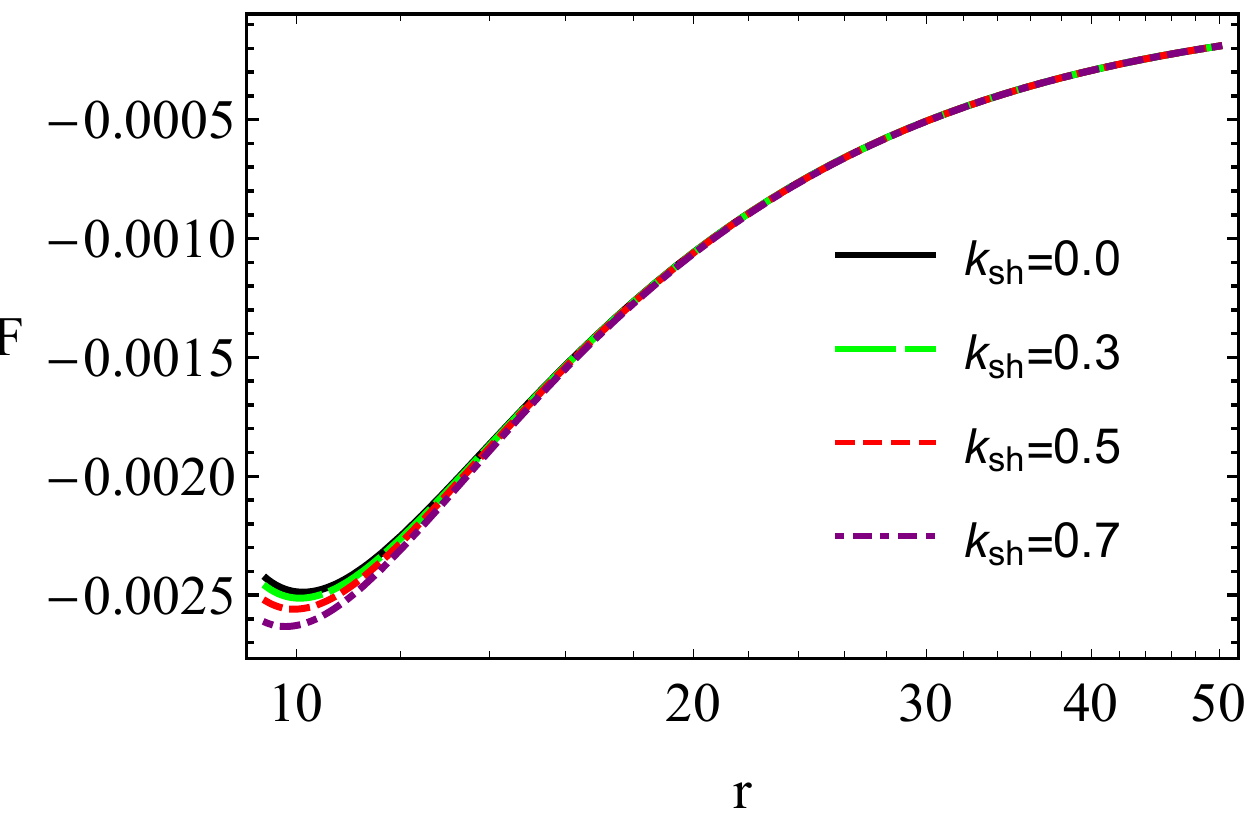}
    \caption{Effective force $F$ for the cases $d_1=8f_1$ (Left panel) and $d_1=-8f_1$ (Right panel) along $c_1$ taking different values of $f_1,\; k_s,\; k_d, \;\&\; k_{sh}$. Here we consider the choice for fix values $M=1,c_1=1,\;f_1=-0.2,\;k_s=k_{sh}=k_d=0.7$.}
    \label{plot:12}
\end{figure}

The impact of effective force can be better interpreted from the graphical analysis as provided in Fig.~\ref{plot:12}. One can see the fluctuating behavior of effective force from attractive to repulsive and ultimately attractive along the radial motion. In case of $d_1=8f_1$, the force becomes more attractive by increasing $c_1,\;f_1,\;k_d,\;k_s$ whereas its repulsion tends to increase with increase in $k_{sh}$. Secondly, in case of $d_1=-8f_1$, attraction increases positively with $c_1,\;\&\;k_d$ while repulsion increases for $f_1,\;k_s\;\&\;k_{sh}$. But the overall movement of the test particle falls inward along the radial direction.
\section{Weak gravitational lensing in metric affine gravity black hole}\label{A8}
Weak gravitational lensing has been one of the remarkable optical features in the study of BHs. Currently, we intend to discuss gravitational lensing in the context of the plasma field. The metric tensor governing the weak field approximation is given as follows \cite{149}
\begin{equation}
    g_{\alpha \beta}=\eta_{\alpha \beta}+h_{\alpha \beta}\,
\end{equation}
where $\eta_{\alpha \beta}$ and $h_{\alpha \beta}$ represent the Minkowski spacetime and perturbation gravity field, respectively. These terms must satisfy the following properties:
\begin{eqnarray}
 &&   \eta_{\alpha \beta}=diag(-1,1,1,1)\ , \nonumber\\
 &&   h_{\alpha \beta} \ll 1, \hspace{0.5cm} h_{\alpha \beta} \rightarrow 0 \hspace{0.5cm} under\hspace{0.2cm}  x^{\alpha}\rightarrow \infty \ ,\nonumber\\
 &&     g^{\alpha \beta}=\eta^{\alpha \beta}-h^{\alpha \beta}, \hspace{0,5cm} h^{\alpha \beta}=h_{\alpha \beta}.
\end{eqnarray}
The angle of deflection around MAGBH can be obtained by varying the above basic equations given by
\begin{equation}
    \hat{\alpha }_{\text{b}}=\frac{1}{2}\int_{-\infty}^{\infty}\frac{b}{r}\left(\frac{dh_{33}}{dr}+\frac{1}{1-w^2_e/ w}\frac{dh_{00}}{dr}-\frac{K_e}{w^2-w^2_e}\frac{dN}{dr} \right)dz\ ,
\end{equation}
where $w$ and $w_{e}$ subsequently stand for the photon and plasma frequencies. We rewrite the line element in terms of MAGBH geometry as
\begin{eqnarray}
  ds^2&=&ds^2_0+\Big[ \frac{16 c_{1}^2 k_{d}^4}{r^4}-\frac{8 c_{1} d_{1} k_{d}^2 k_{s}^2}{r^4}+\frac{16 c_{1} f_{1} k_{d}^2 k_{sh}^2}{r^4}+\frac{8 c_{1} k_{d}^2 R_s}{r^3}+\frac{4 c_{1} k_{d}^2}{r^2}+\frac{d_{1}^2 k_{s}^4}{r^4}-\frac{4 d_{1} f_{1} k_{sh}^2 k_{s}^2}{r^4}-\frac{2 d_{1} k_{s}^2 R_s}{r^3}\nonumber\\&-&\frac{d_{1} k_{s}^2}{r^2}+\frac{4 f_{1}^2 k_{sh}^4}{r^4}+\frac{4 f_{1} k_{sh}^2 R_s}{r^3}+\frac{2 f_{1} k_{sh}^2}{r^2}+\frac{R_s}{r} \Big]dt^2 +\Big[\frac{16 c_{1}^2 k_{d}^4}{r^4}-\frac{8 c_{1} d_{1} k_{d}^2 k_{s}^2}{r^4}+\frac{16 c_{1} f_{1} k_{d}^2 k_{sh}^2}{r^4}+\frac{8 c_{1} k_{d}^2 R_s}{r^3}\nonumber\\&+&\frac{4 c_{1} k_{d}^2}{r^2}+\frac{d_{1}^2 k_{s}^4}{r^4}-\frac{4 d_{1} f_{1} k_{sh}^2 k_{s}^2}{r^4}-\frac{2 d_{1} k_{s}^2 R_s}{r^3}-\frac{d_{1} k_{s}^2}{r^2}+\frac{4 f_{1}^2 k_{sh}^4}{r^4}+\frac{4 f_{1} k_{sh}^2 R_s}{r^3}+\frac{2 f_{1} k_{sh}^2}{r^2}+\frac{R_s}{r})^{-1}\Big]dr^2\ ,
\end{eqnarray}
where $ds^2_0=-dt^2+dr^2+r^2(d\theta^2+\sin^2\theta d\phi^2)$. The components $h_{\alpha \beta}$ in the form of Cartesian coordinates yield
\begin{eqnarray}
     h_{00}&=&\frac{16 c_{1}^2 k_{d}^4}{r^4}-\frac{8 c_{1} d_{1} k_{d}^2 k_{s}^2}{r^4}+\frac{16 c_{1} f_{1} k_{d}^2 k_{sh}^2}{r^4}+\frac{8 c_{1} k_{d}^2 R_s}{r^3}+\frac{4 c_{1} k_{d}^2}{r^2}+\frac{d_{1}^2 k_{s}^4}{r^4}-\frac{4 d_{1} f_{1} k_{sh}^2 k_{s}^2}{r^4}-\frac{2 d_{1} k_{s}^2 R_s}{r^3}\nonumber\\&-&\frac{d_{1} k_{s}^2}{r^2}+\frac{4 f_{1}^2 k_{sh}^4}{r^4}+\frac{4 f_{1} k_{sh}^2 R_s}{r^3}+\frac{2 f_{1} k_{sh}^2}{r^2}+\frac{R_s}{r} ,\label{h3}\\
    h_{ik}&=&\Big[\frac{16 c_{1}^2 k_{d}^4}{r^4}-\frac{8 c_{1} d_{1} k_{d}^2 k_{s}^2}{r^4}+\frac{16 c_{1} f_{1} k_{d}^2 k_{sh}^2}{r^4}+\frac{8 c_{1} k_{d}^2 R_s}{r^3}+\frac{4 c_{1} k_{d}^2}{r^2}+\frac{d_{1}^2 k_{s}^4}{r^4}-\frac{4 d_{1} f_{1} k_{sh}^2 k_{s}^2}{r^4}-\frac{2 d_{1} k_{s}^2 R_s}{r^3}\nonumber\\&-&\frac{d_{1} k_{s}^2}{r^2}+\frac{4 f_{1}^2 k_{sh}^4}{r^4}+\frac{4 f_{1} k_{sh}^2 R_s}{r^3}+\frac{2 f_{1} k_{sh}^2}{r^2}+\frac{R_s}{r}\Big]n_i n_k,\label{h2} \\
    h_{33}&=&\Big[\frac{16 c_{1}^2 k_{d}^4}{r^4}-\frac{8 c_{1} d_{1} k_{d}^2 k_{s}^2}{r^4}+\frac{16 c_{1} f_{1} k_{d}^2 k_{sh}^2}{r^4}+\frac{8 c_{1} k_{d}^2 R_s}{r^3}+\frac{4 c_{1} k_{d}^2}{r^2}+\frac{d_{1}^2 k_{s}^4}{r^4}-\frac{4 d_{1} f_{1} k_{sh}^2 k_{s}^2}{r^4}-\frac{2 d_{1} k_{s}^2 R_s}{r^3}\nonumber\\&-&\frac{d_{1} k_{s}^2}{r^2}+\frac{4 f_{1}^2 k_{sh}^4}{r^4}+\frac{4 f_{1} k_{sh}^2 R_s}{r^3}+\frac{2 f_{1} k_{sh}^2}{r^2}+\frac{R_s}{r}\Big]\cos^2\chi \label{h},\label{h3}
\end{eqnarray}
The deflection angle can be expressed by the following relation \cite{150}
\begin{equation}
    \hat{\alpha}_{b}=\hat{\alpha}_{1}+\hat{\alpha}_{2}+\hat{\alpha}_{3}\ , \label{h6}
\end{equation}
where
 \begin{eqnarray}
\hat{\alpha}_{1}&=&\frac{1}{2}\int_{-\infty}^{\infty} \frac{b}{r}\frac{dh_{33}}{dr}dz\ ,\nonumber\\
\hat{\alpha}_{2}&=&\frac{1}{2}\int_{-\infty}^{\infty} \frac{b}{r}\frac{1}{1-w^2_e/ w}\frac{dh_{00}}{dr}dz\ ,\nonumber\\
\hat{\alpha}_{3}&=&\frac{1}{2}\int_{-\infty}^{\infty} \frac{b}{r}\left(-\frac{K_e}{w^2-w^2_e}\frac{dN}{dr} \right)dz\ .  \label{h7}
\end{eqnarray}

Furthermore, we compute the deflection angle for both uniform and non-uniform plasma density distributions.

It may be of worth noting that in our further discussions, we used $w$ instead of $w(\infty)$ and $w_0$ instead of $w_e(\infty)$ \cite{13,151}.

\subsection{Uniform plasma}\label{A8.1}
\begin{figure}
    \centering
    \includegraphics[scale=0.26]{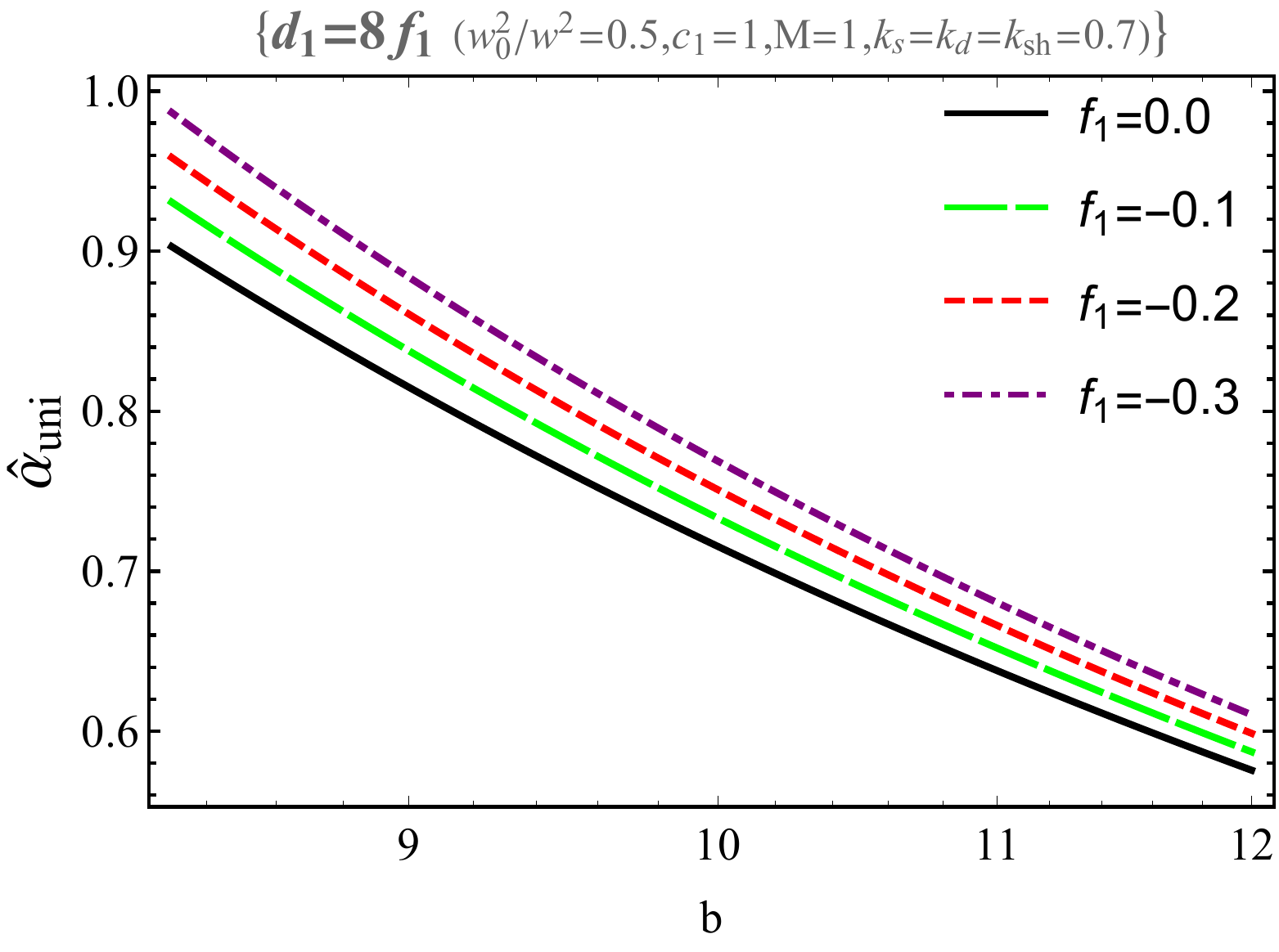}
    \includegraphics[scale=0.26]{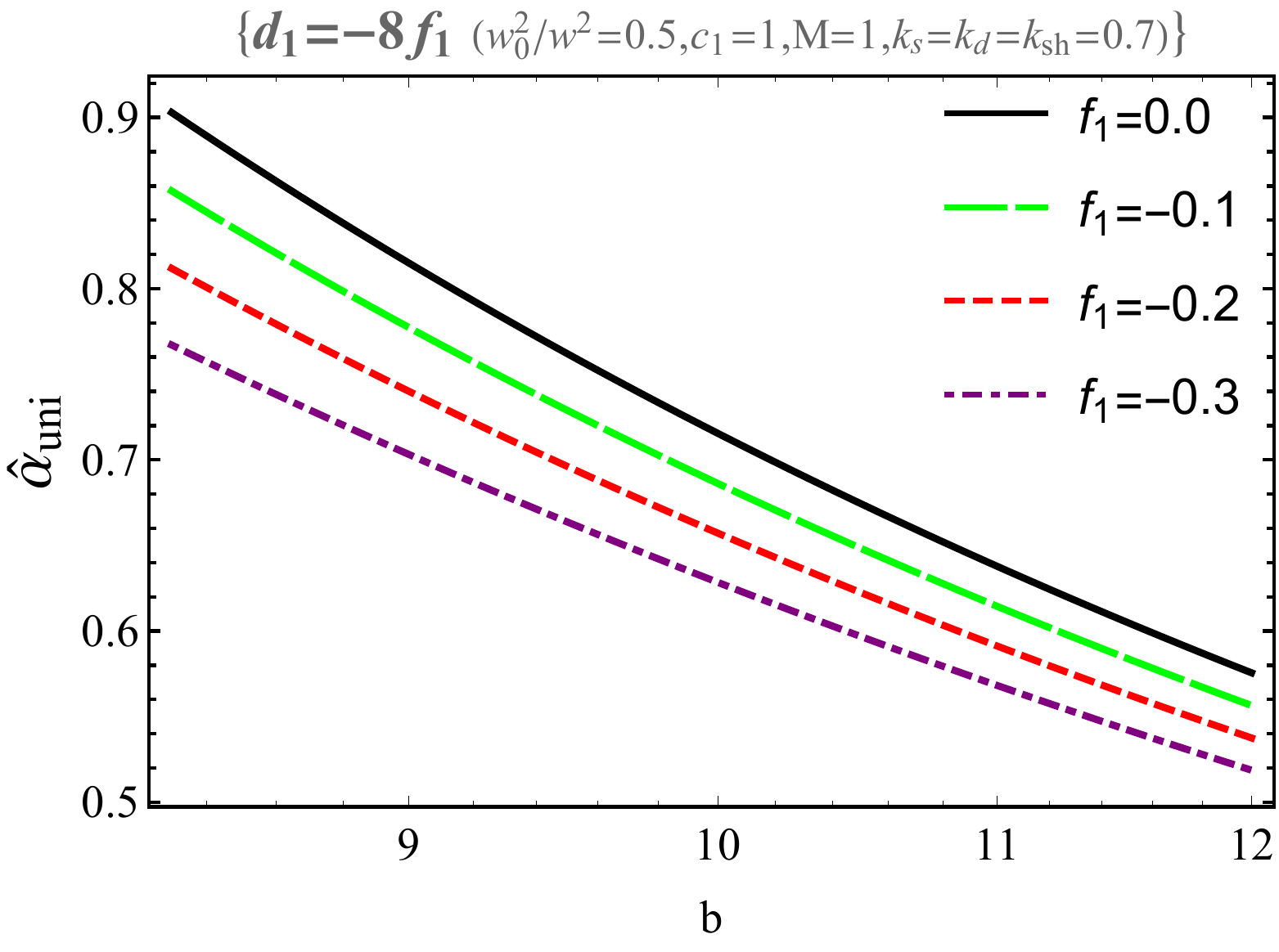}
    \includegraphics[scale=0.26]{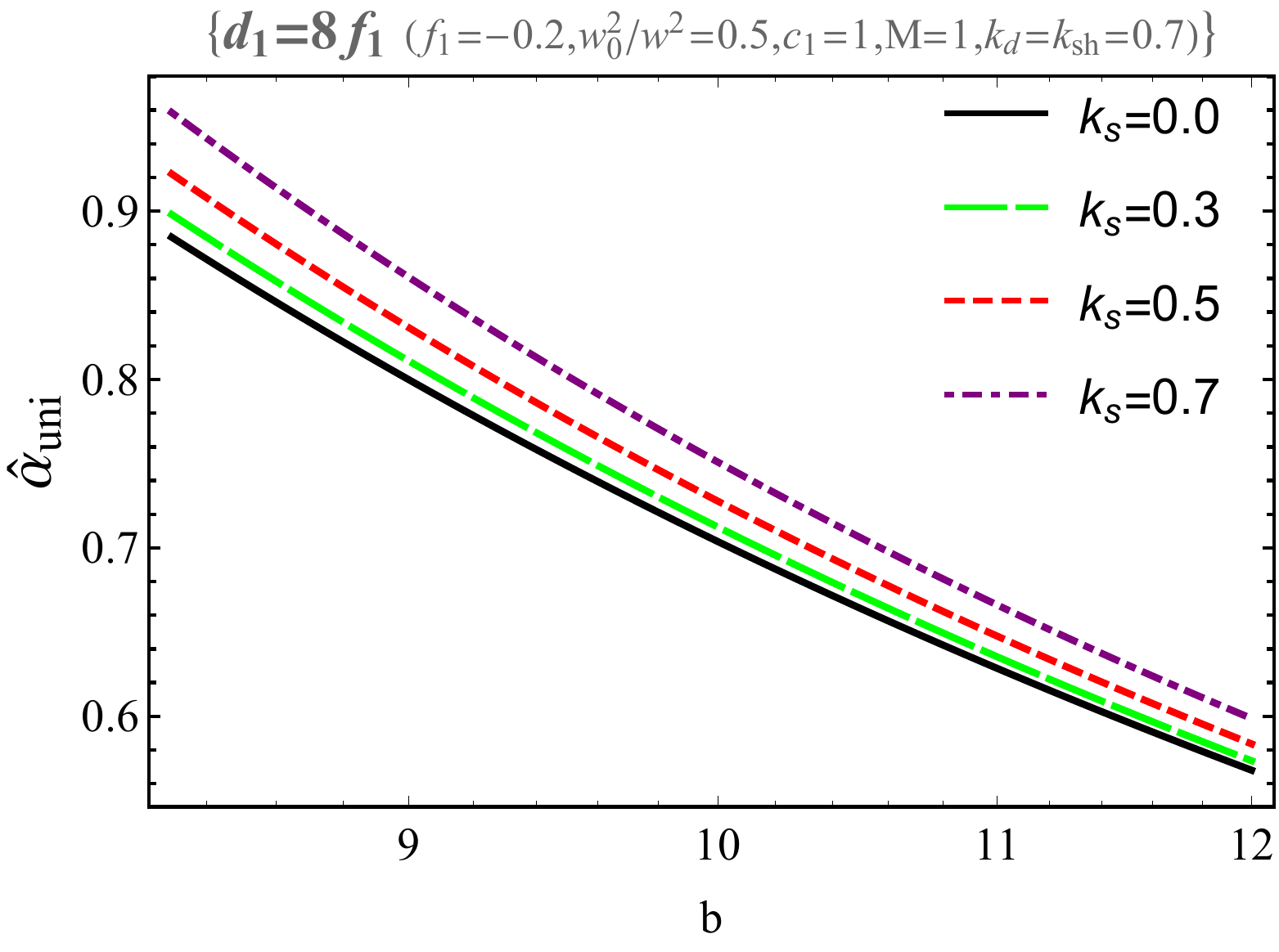}
    \includegraphics[scale=0.26]{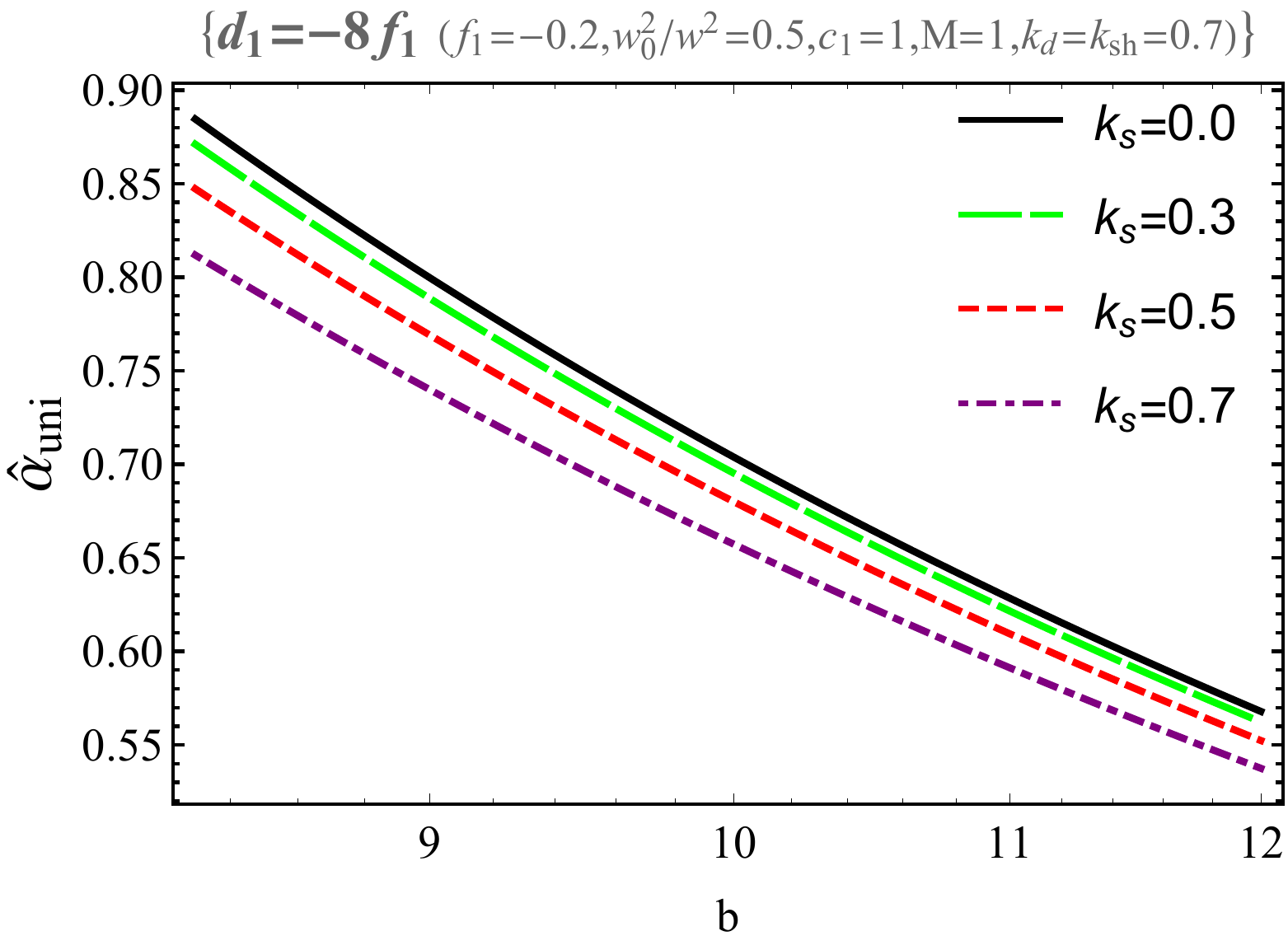}
    \includegraphics[scale=0.26]{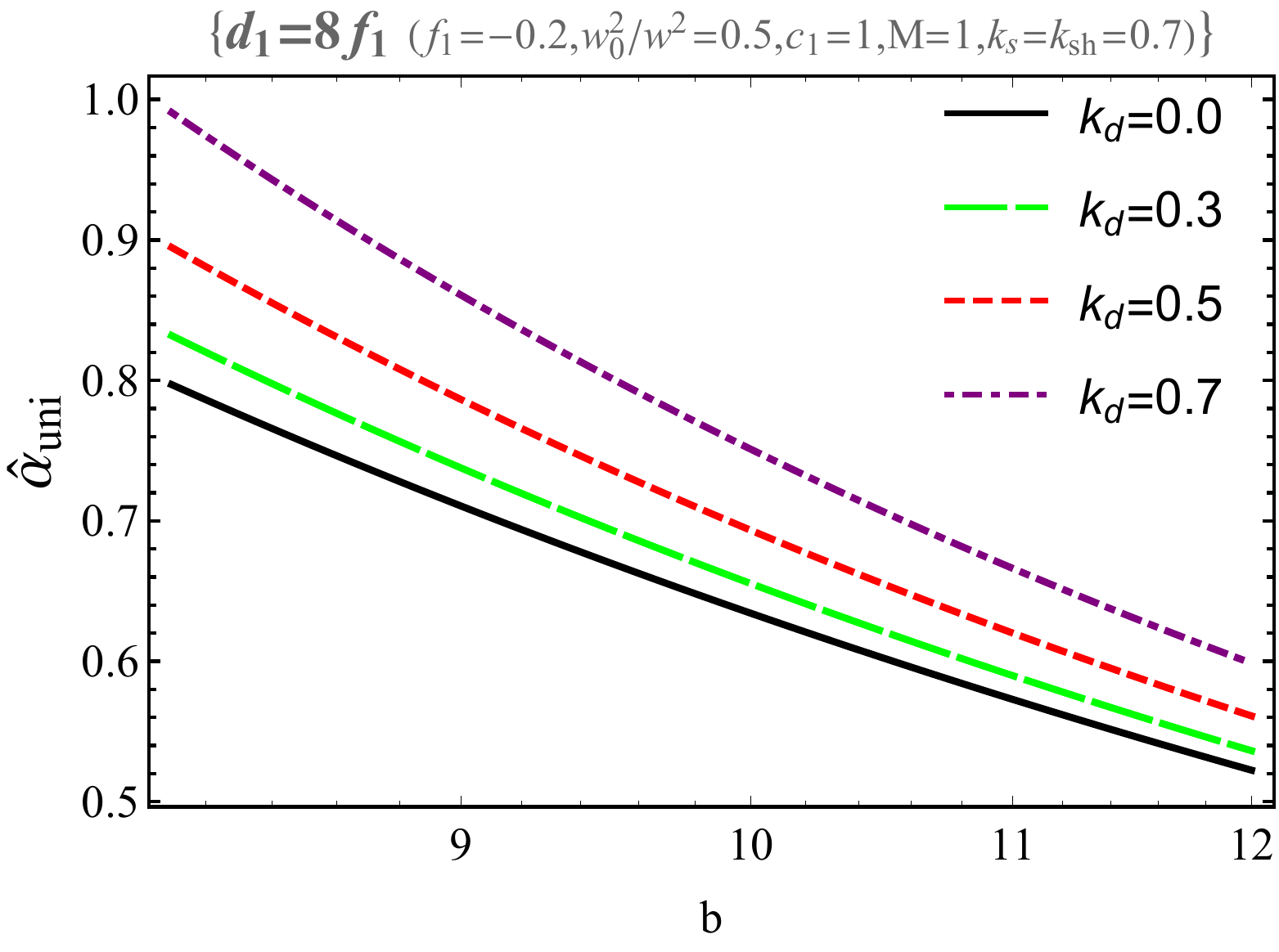}
    \includegraphics[scale=0.26]{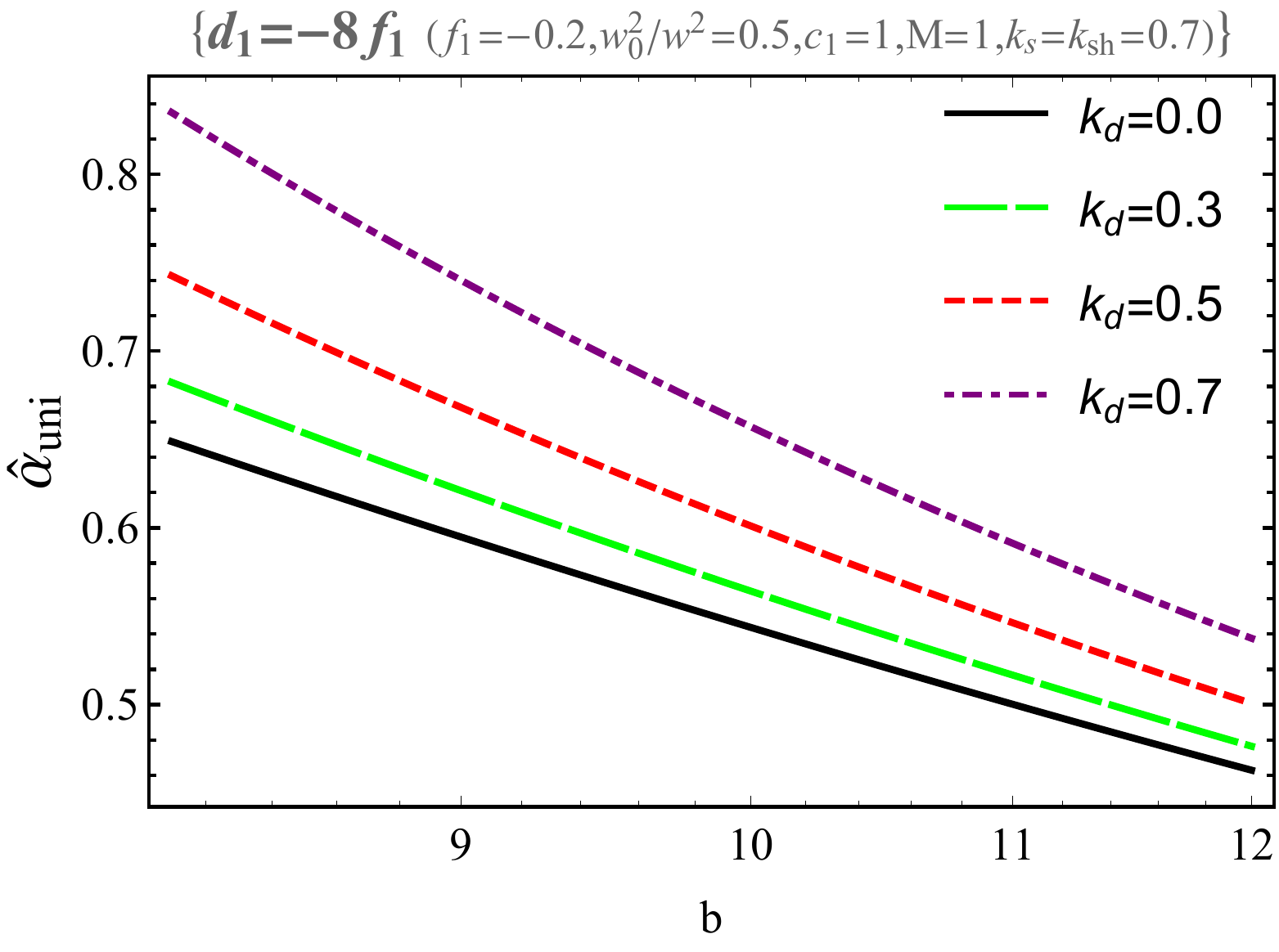}
    \includegraphics[scale=0.26]{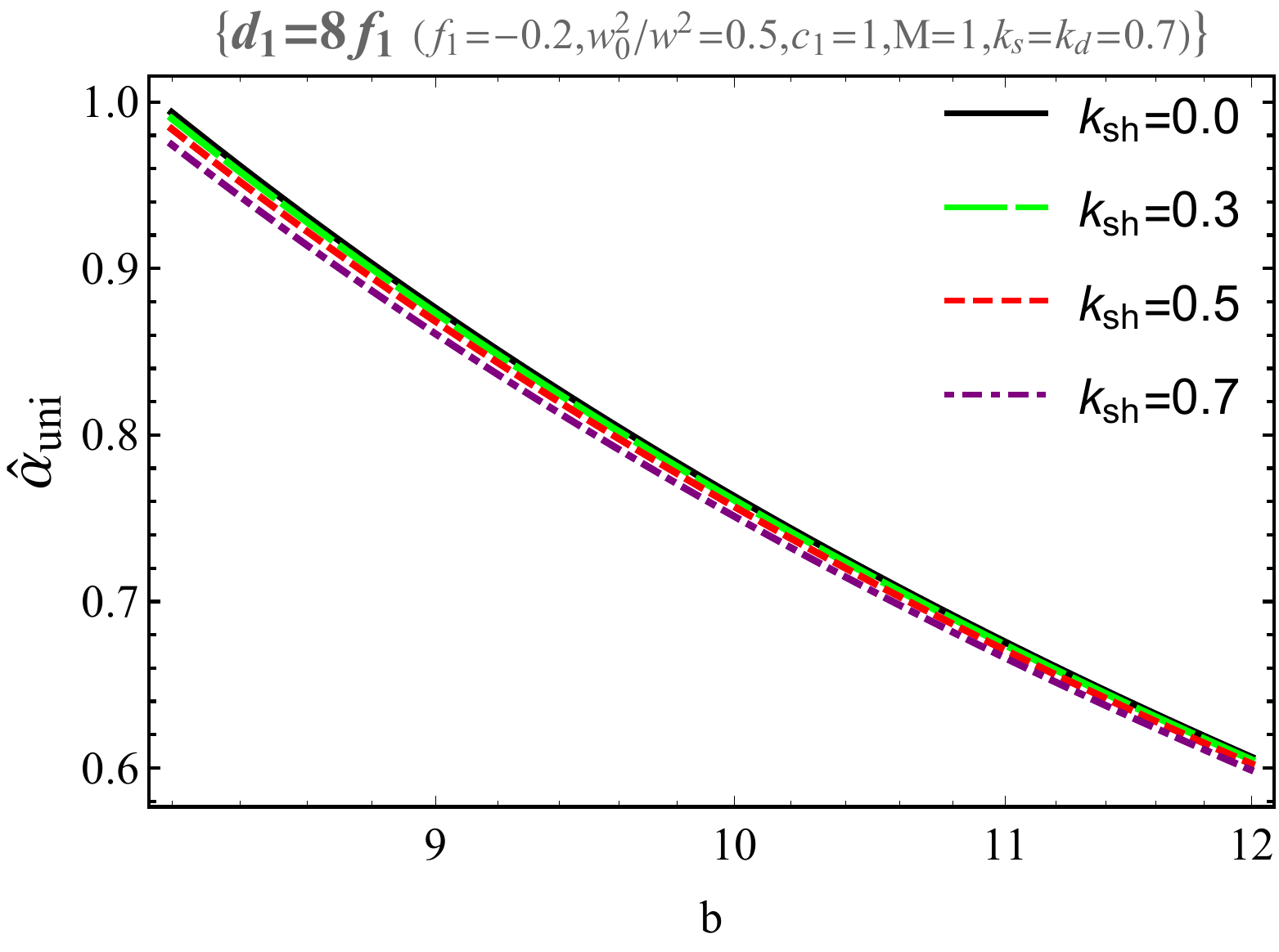}
    \includegraphics[scale=0.26]{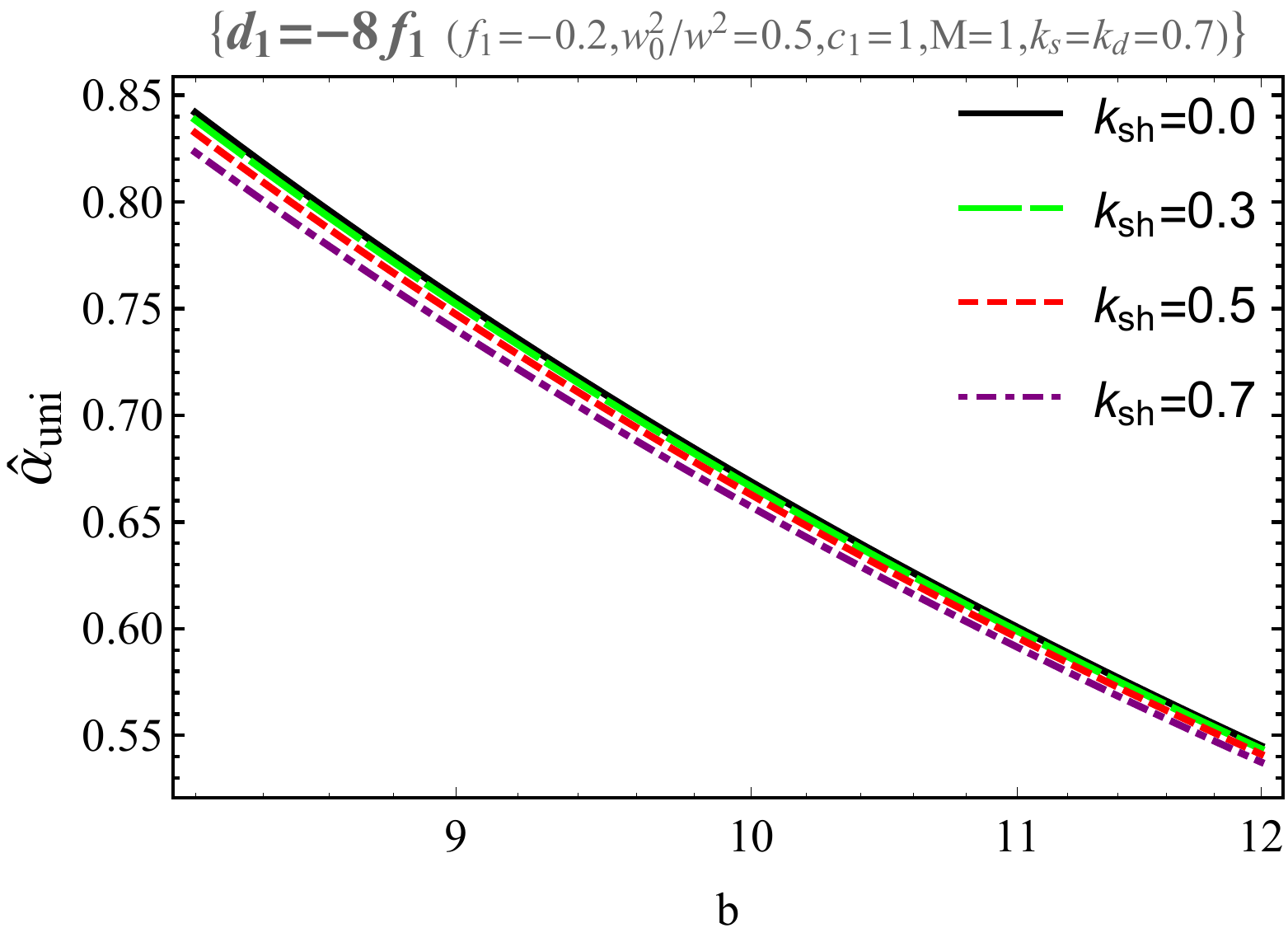}
    \includegraphics[scale=0.26]{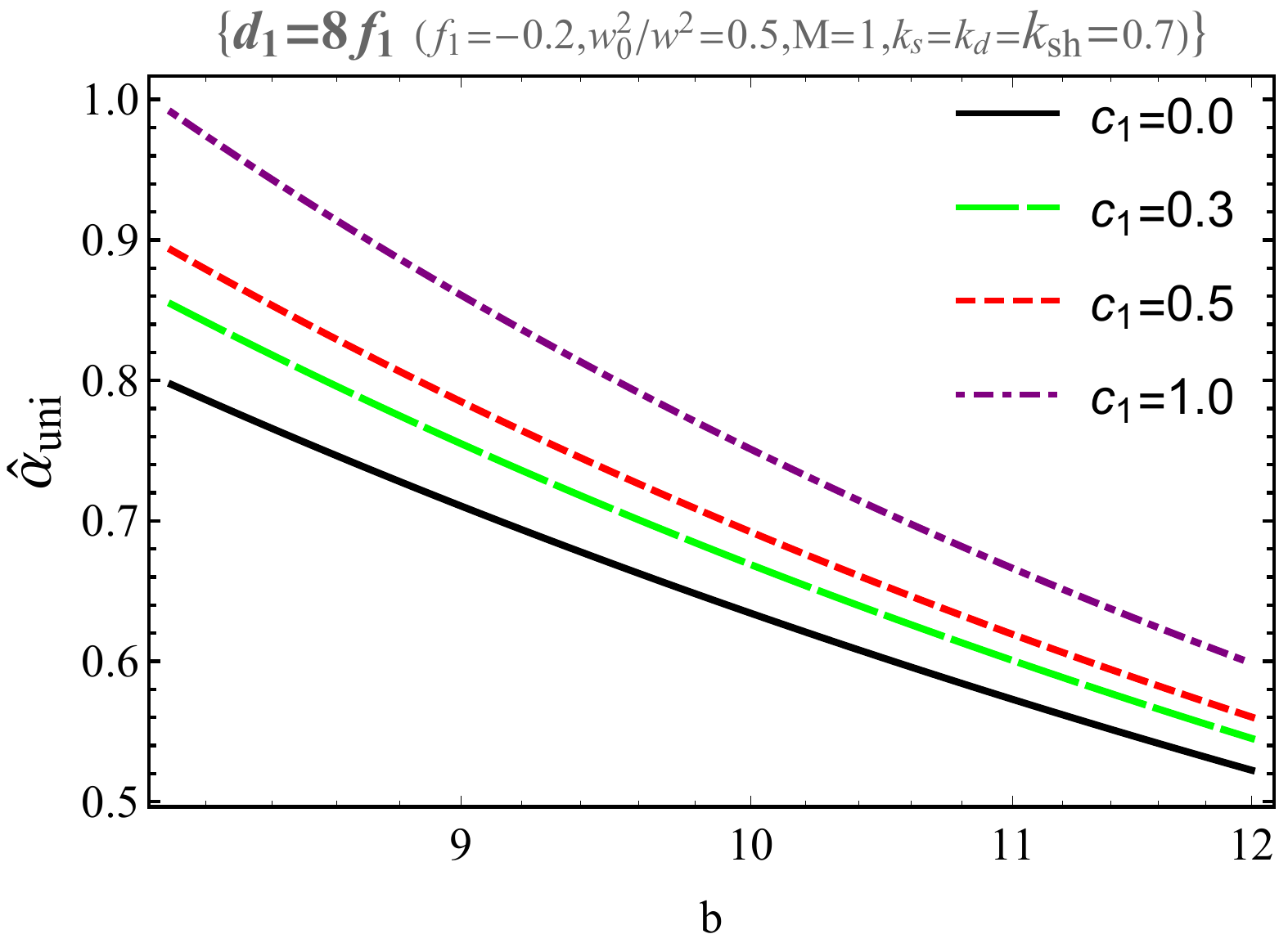}
    \includegraphics[scale=0.26]{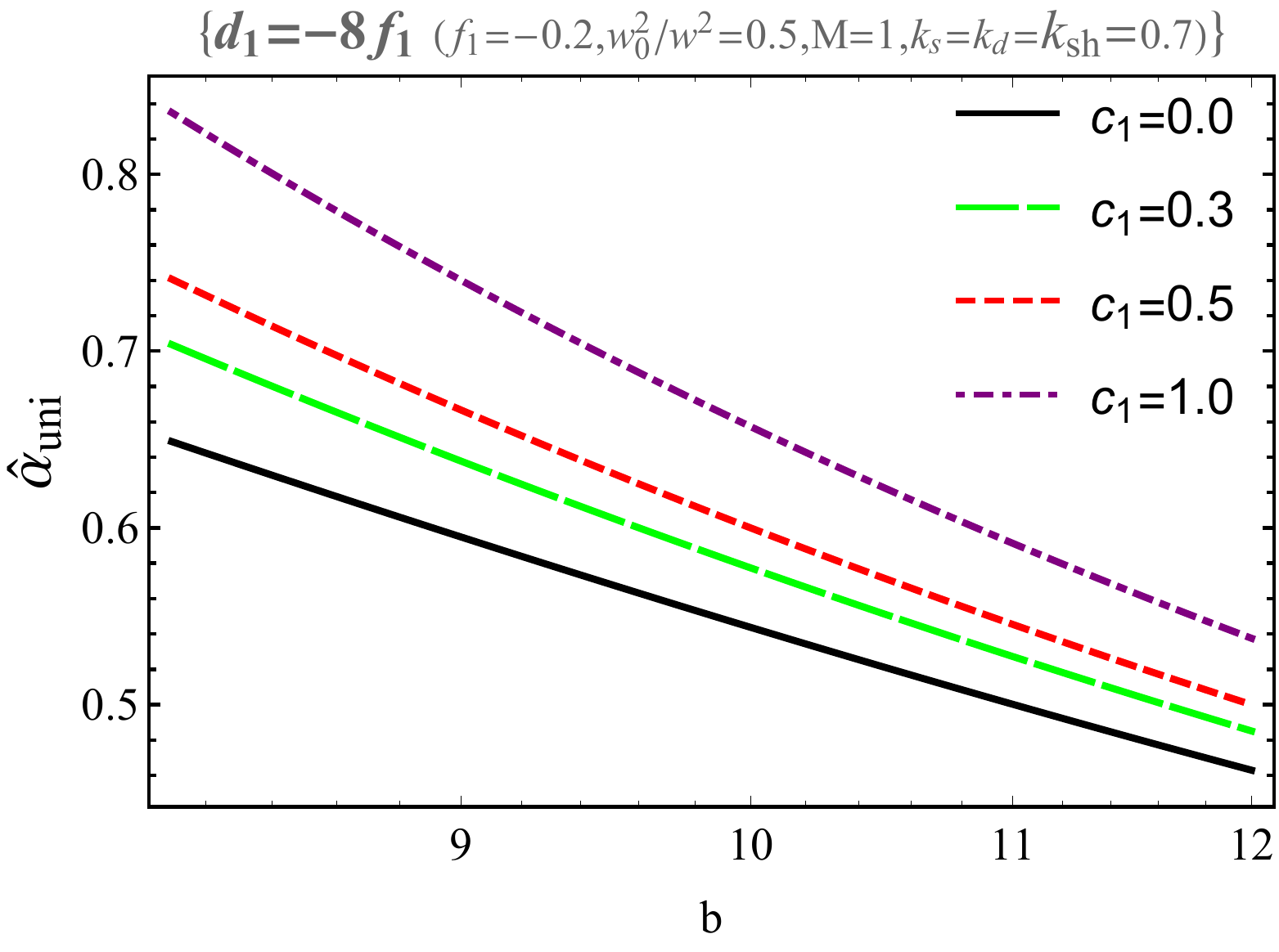}
    \includegraphics[scale=0.26]{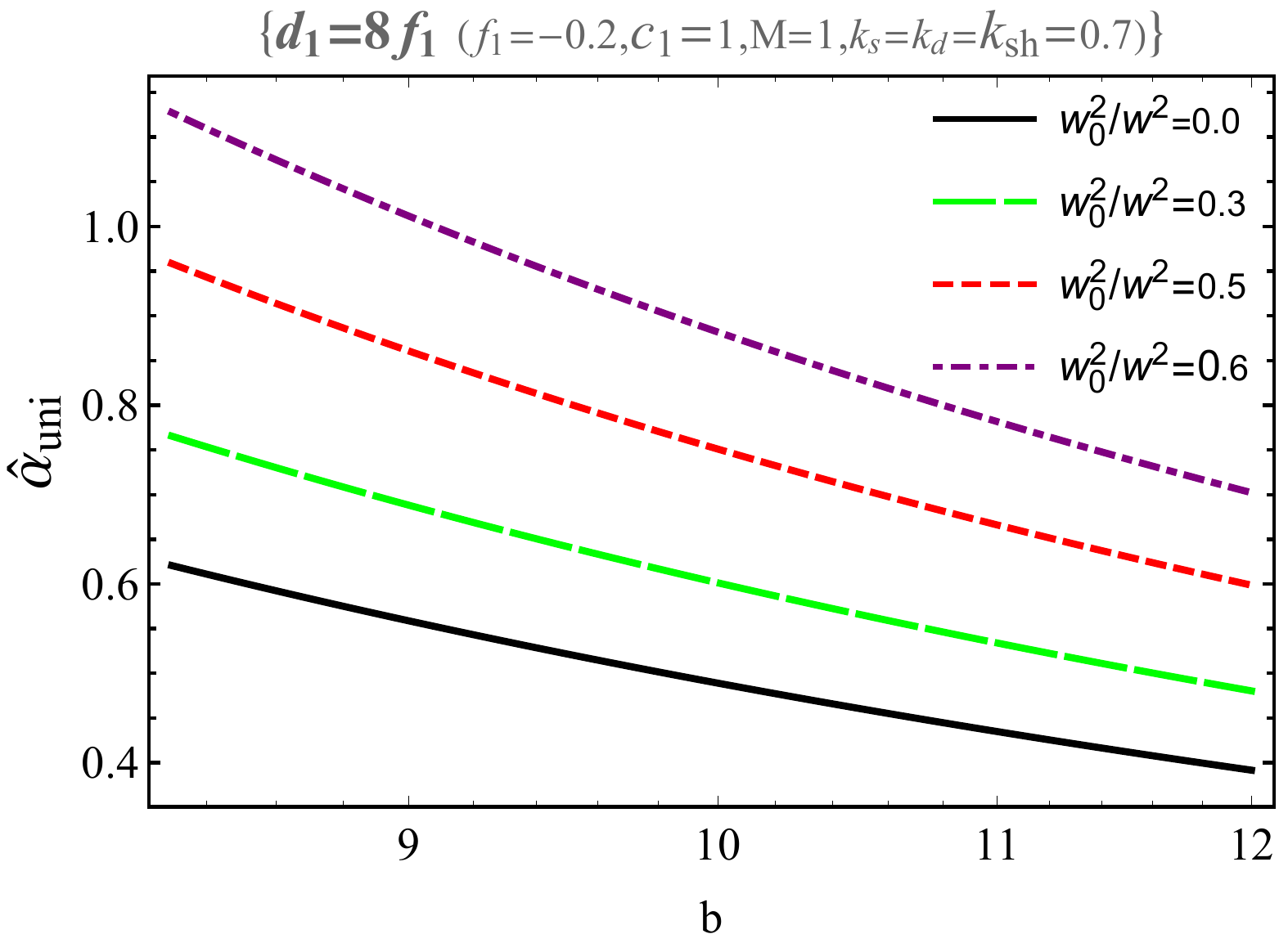}
    \includegraphics[scale=0.26]{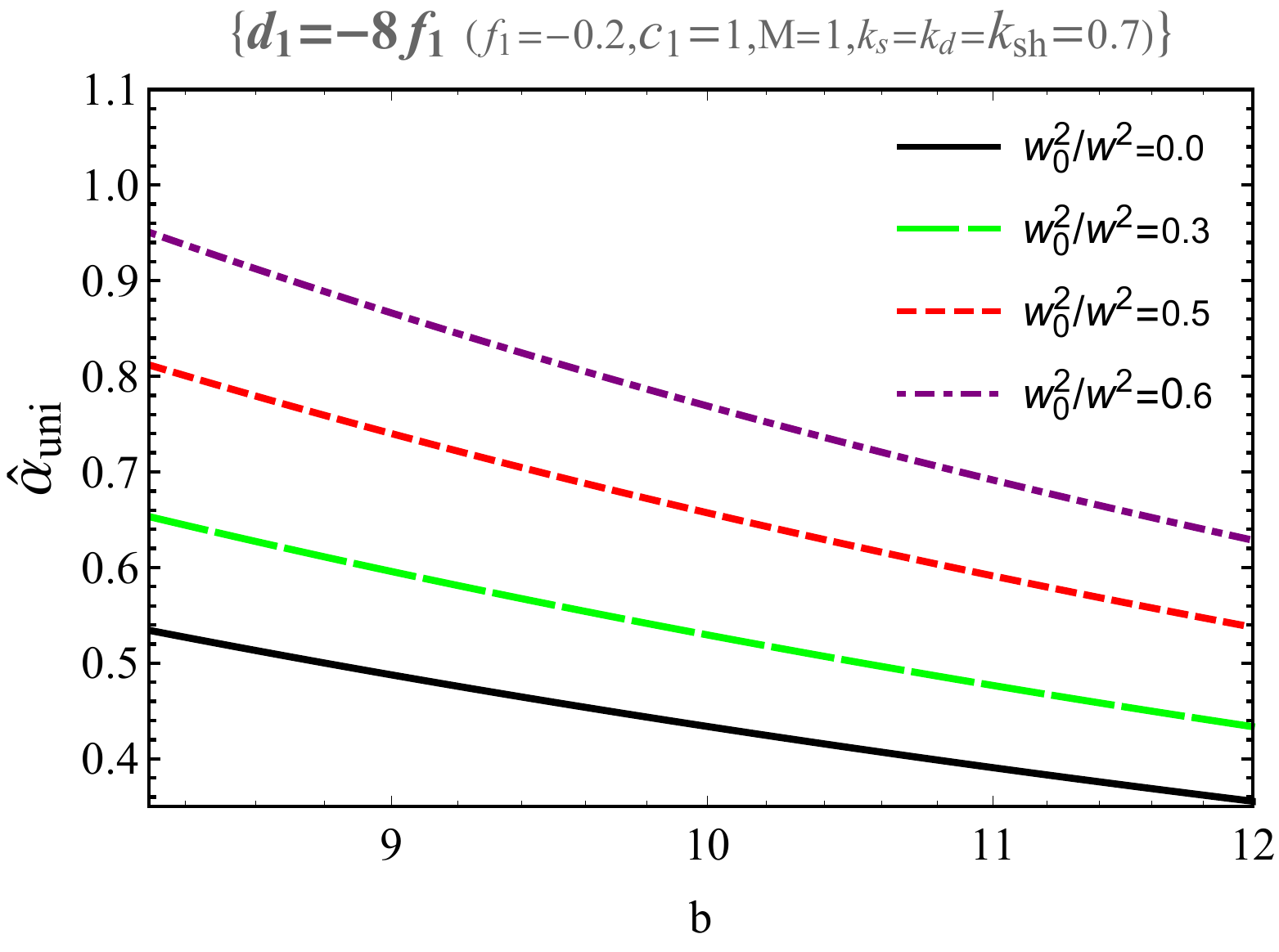}
    \includegraphics[scale=0.26]{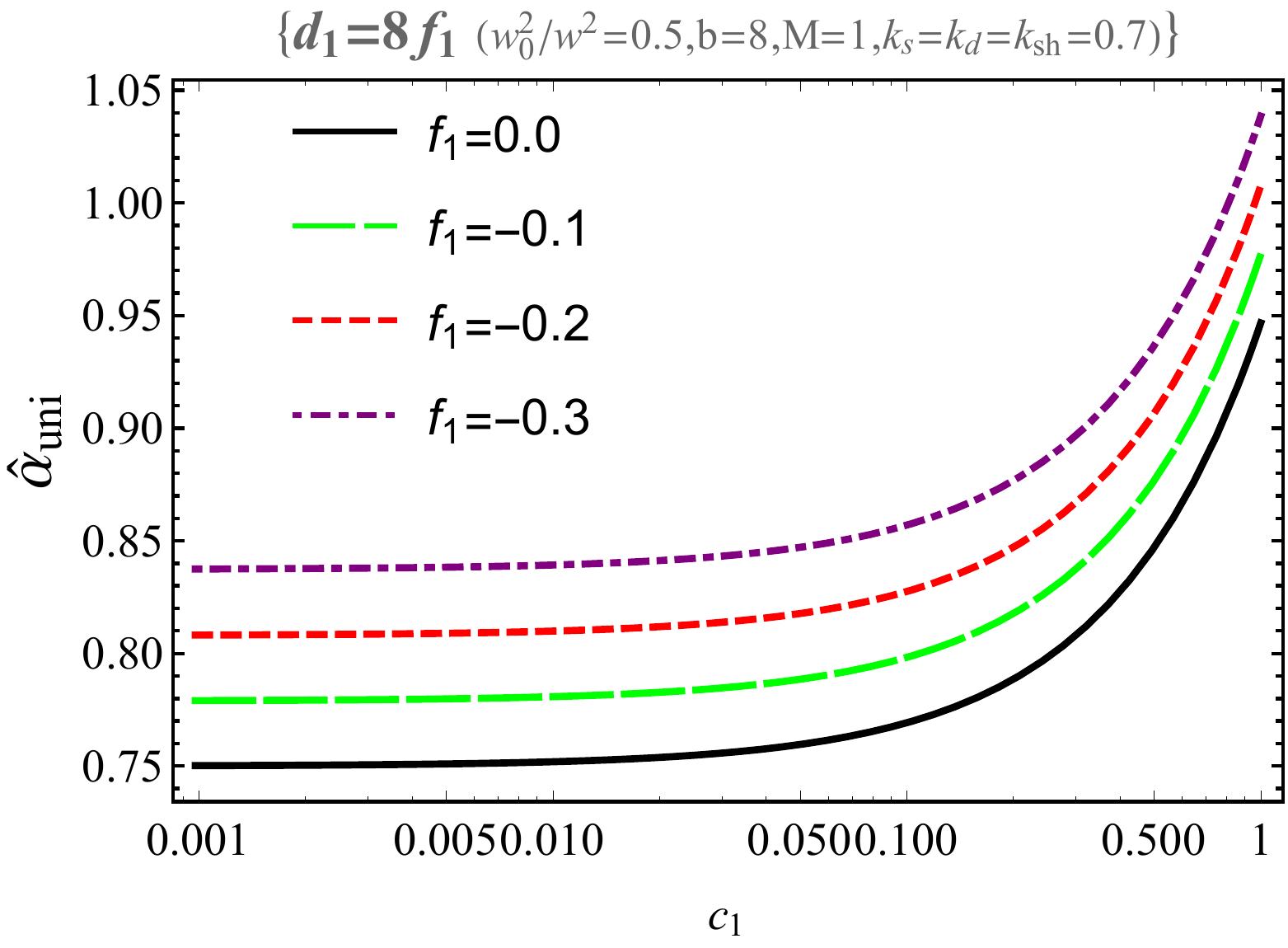}
    \includegraphics[scale=0.26]{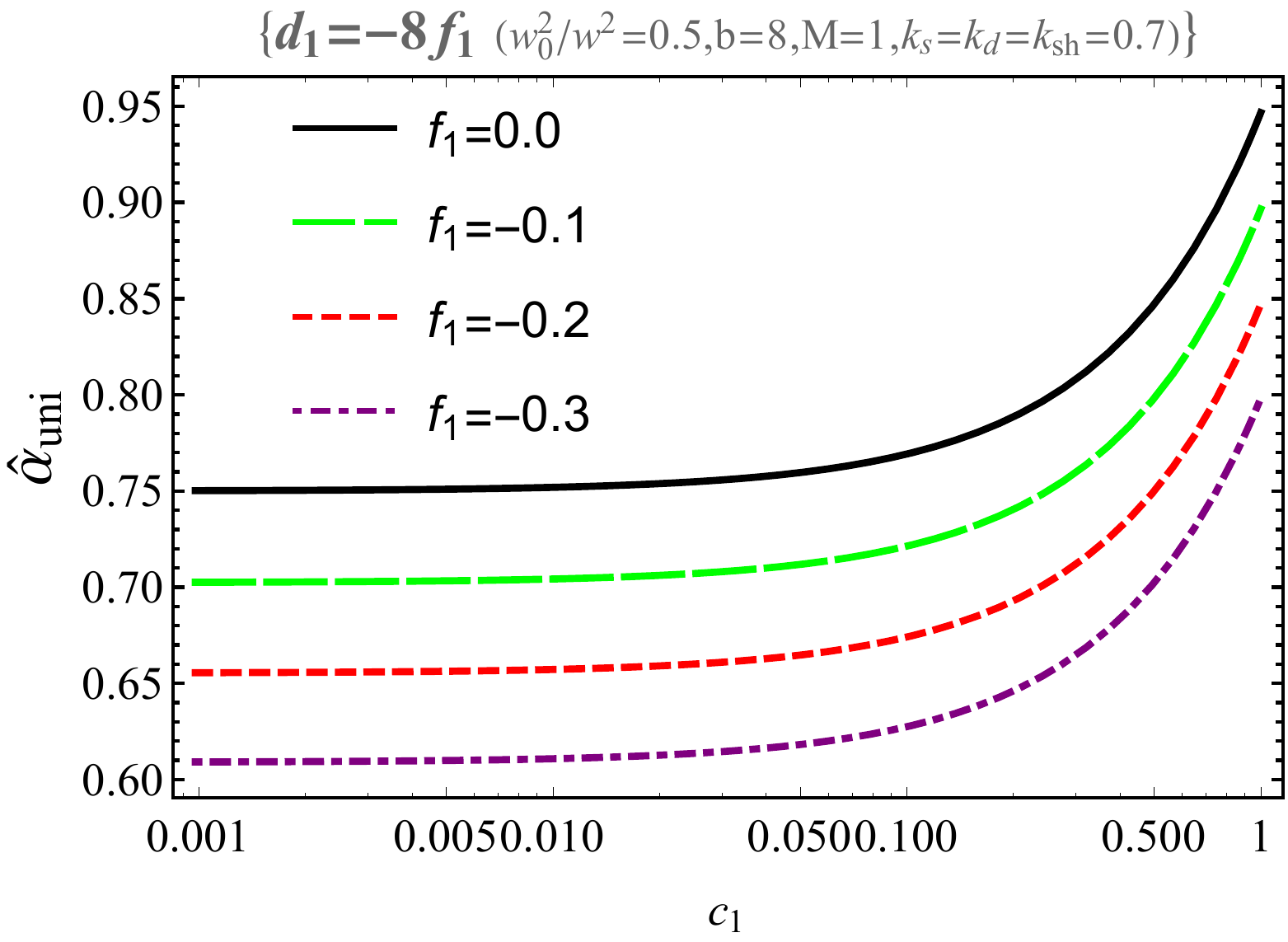}
    \includegraphics[scale=0.26]{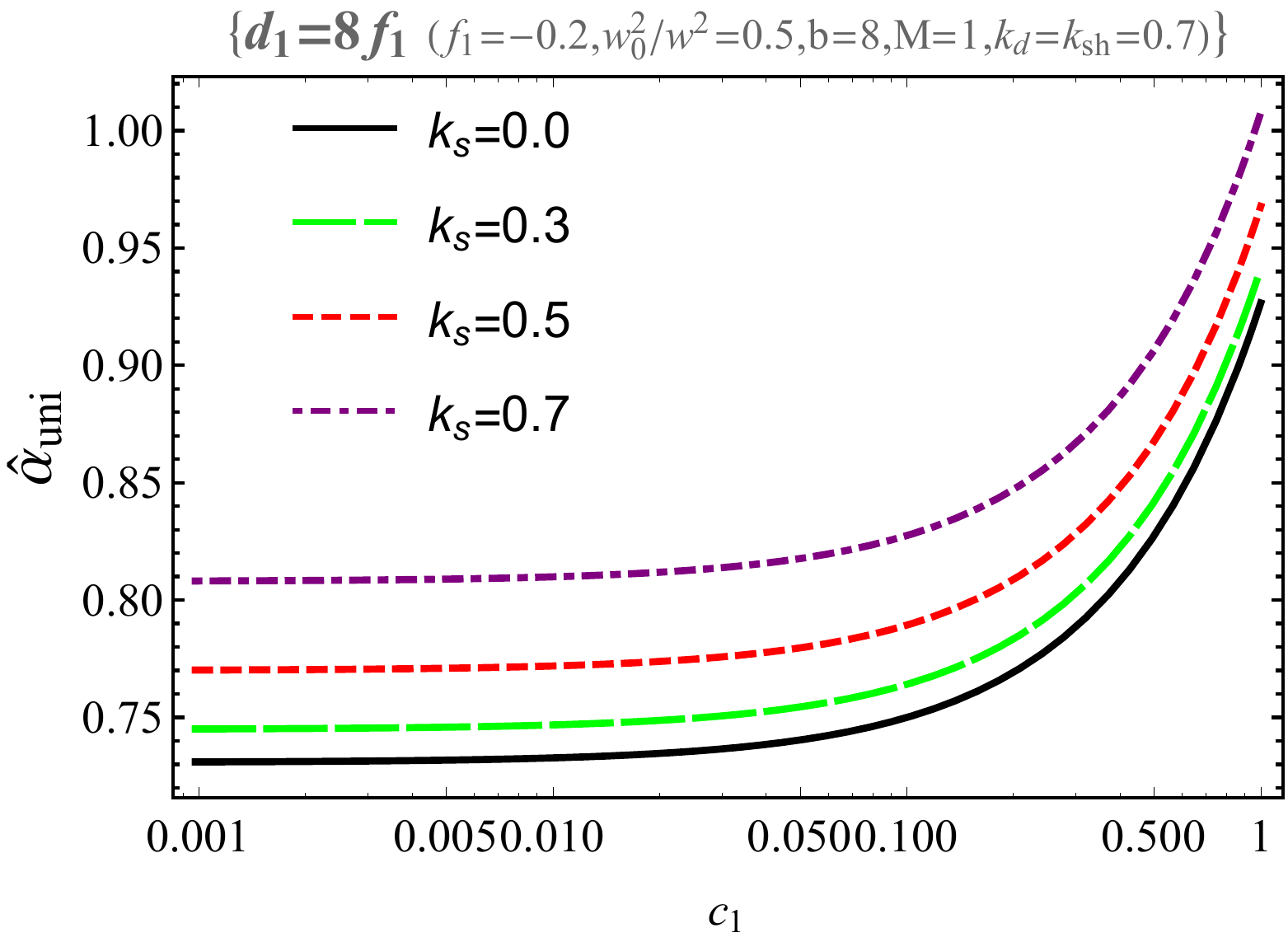}
    \includegraphics[scale=0.26]{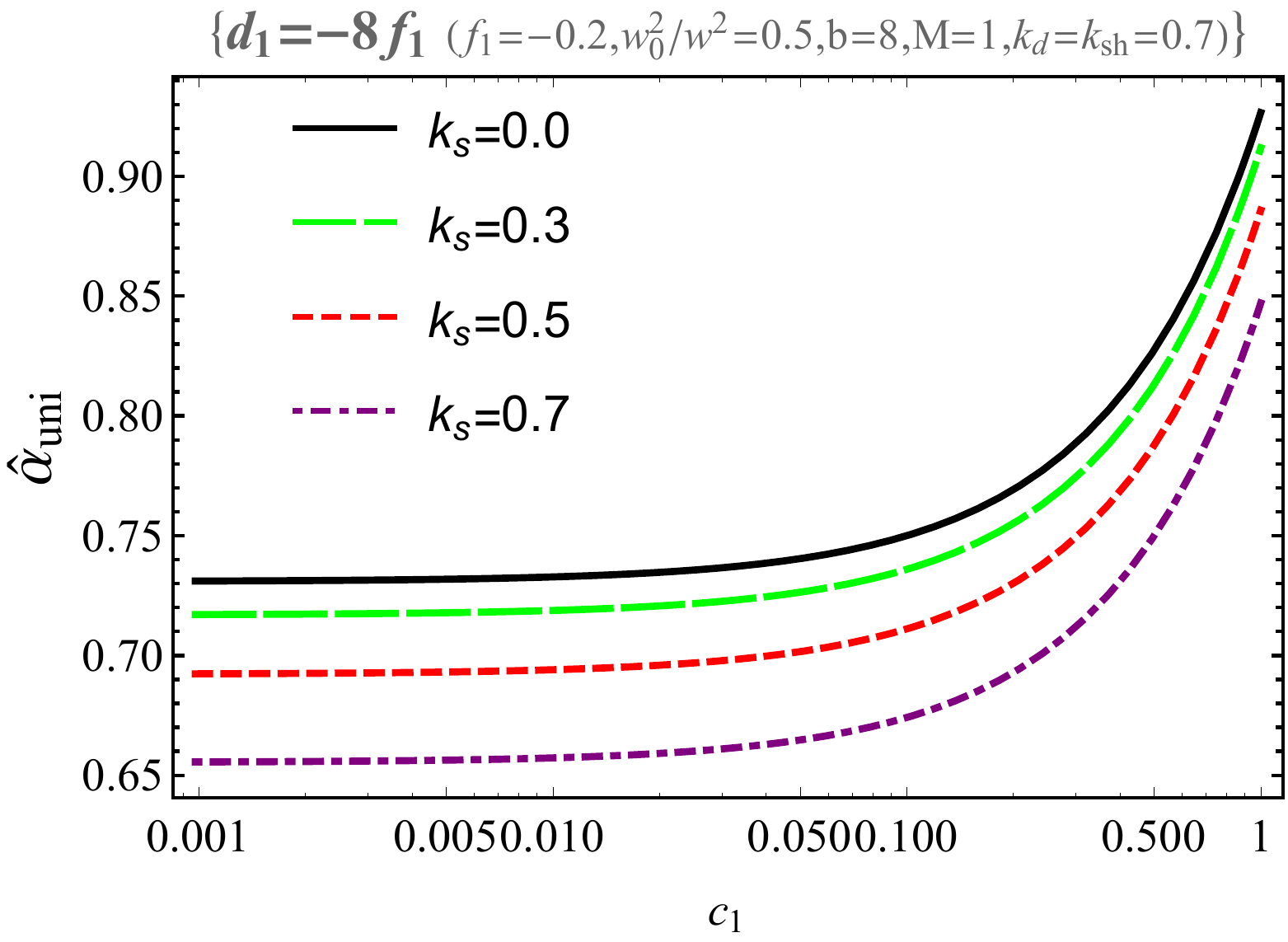}
    \includegraphics[scale=0.26]{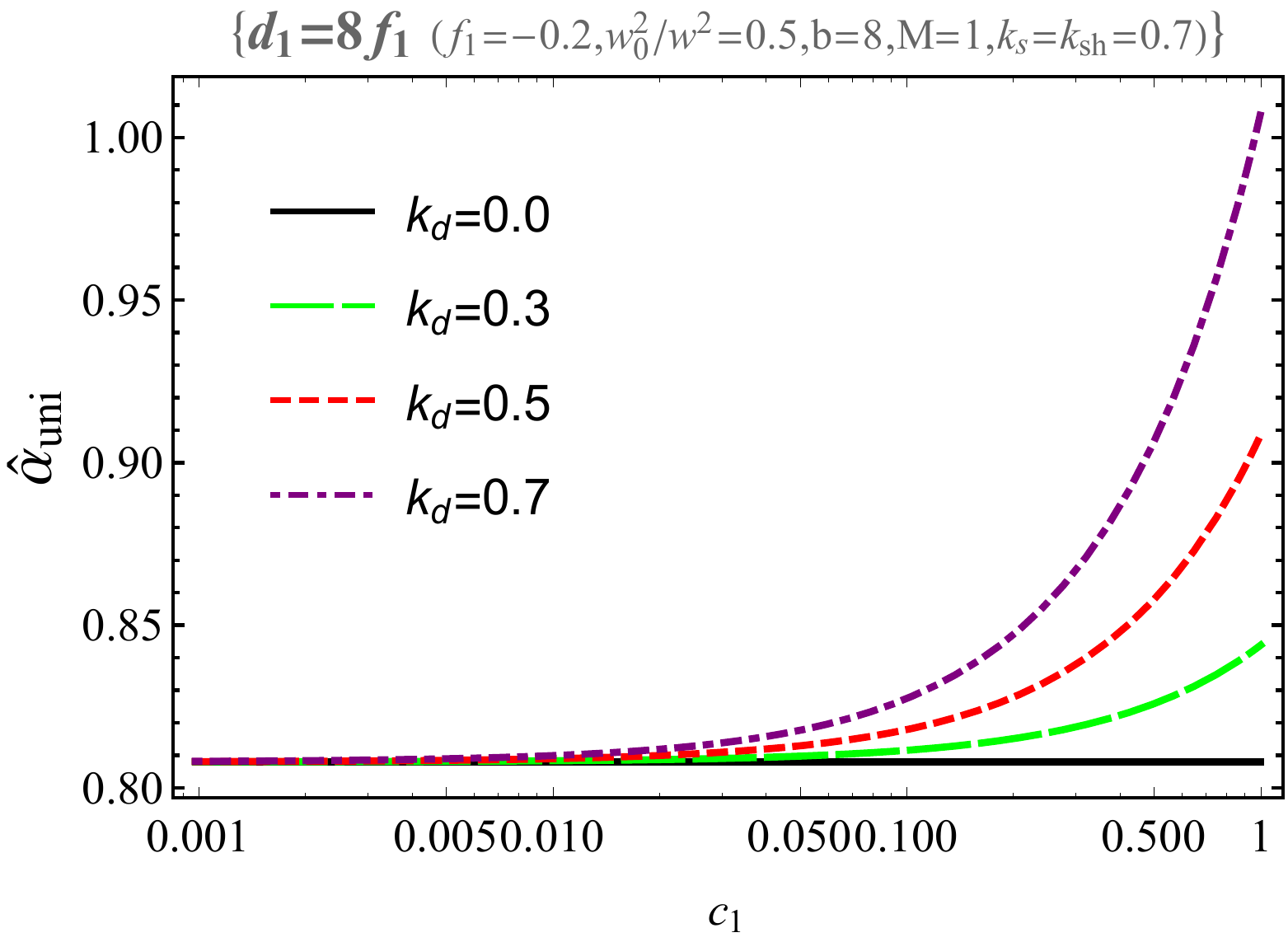}
    \includegraphics[scale=0.26]{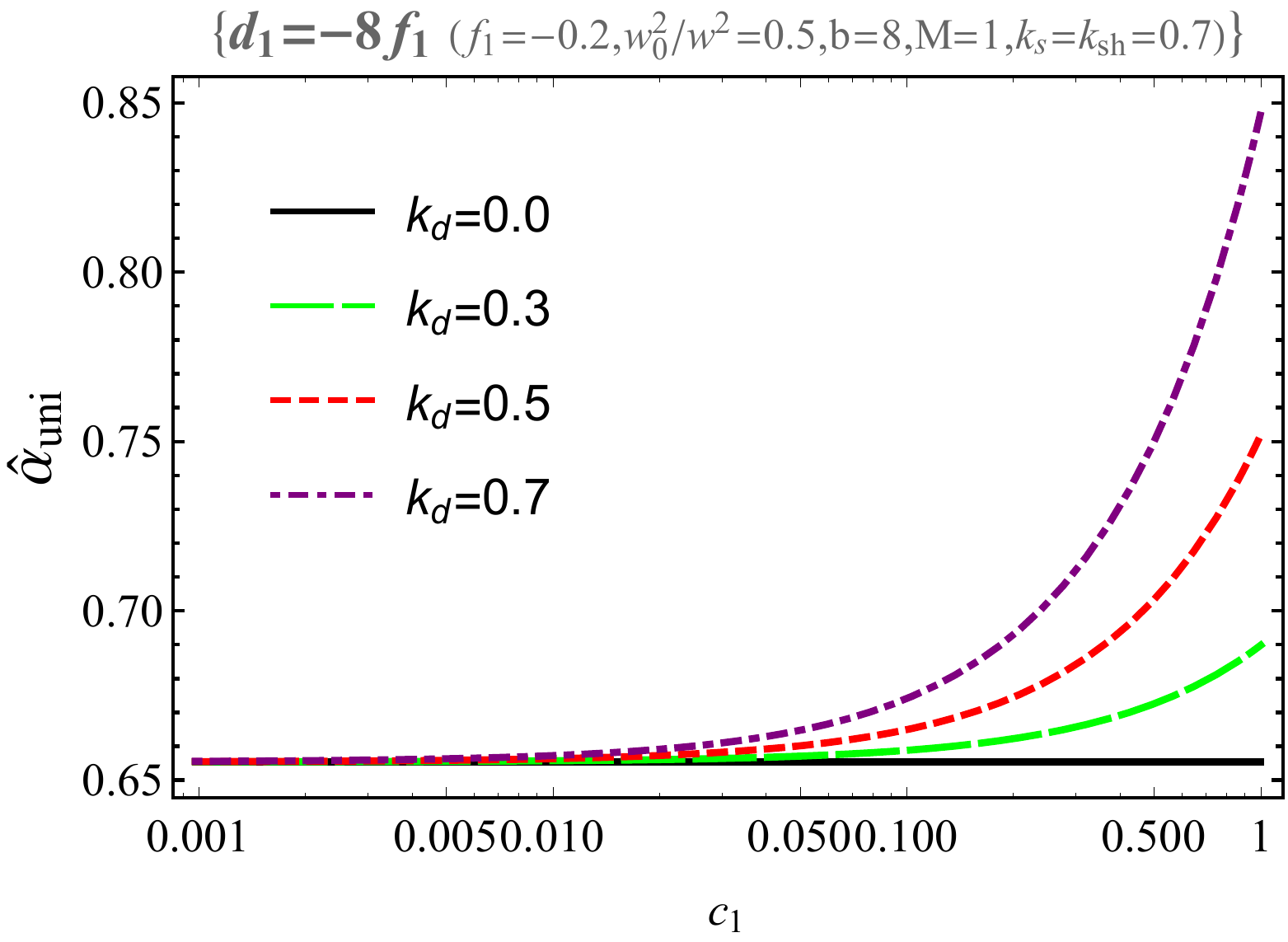}
    \includegraphics[scale=0.26]{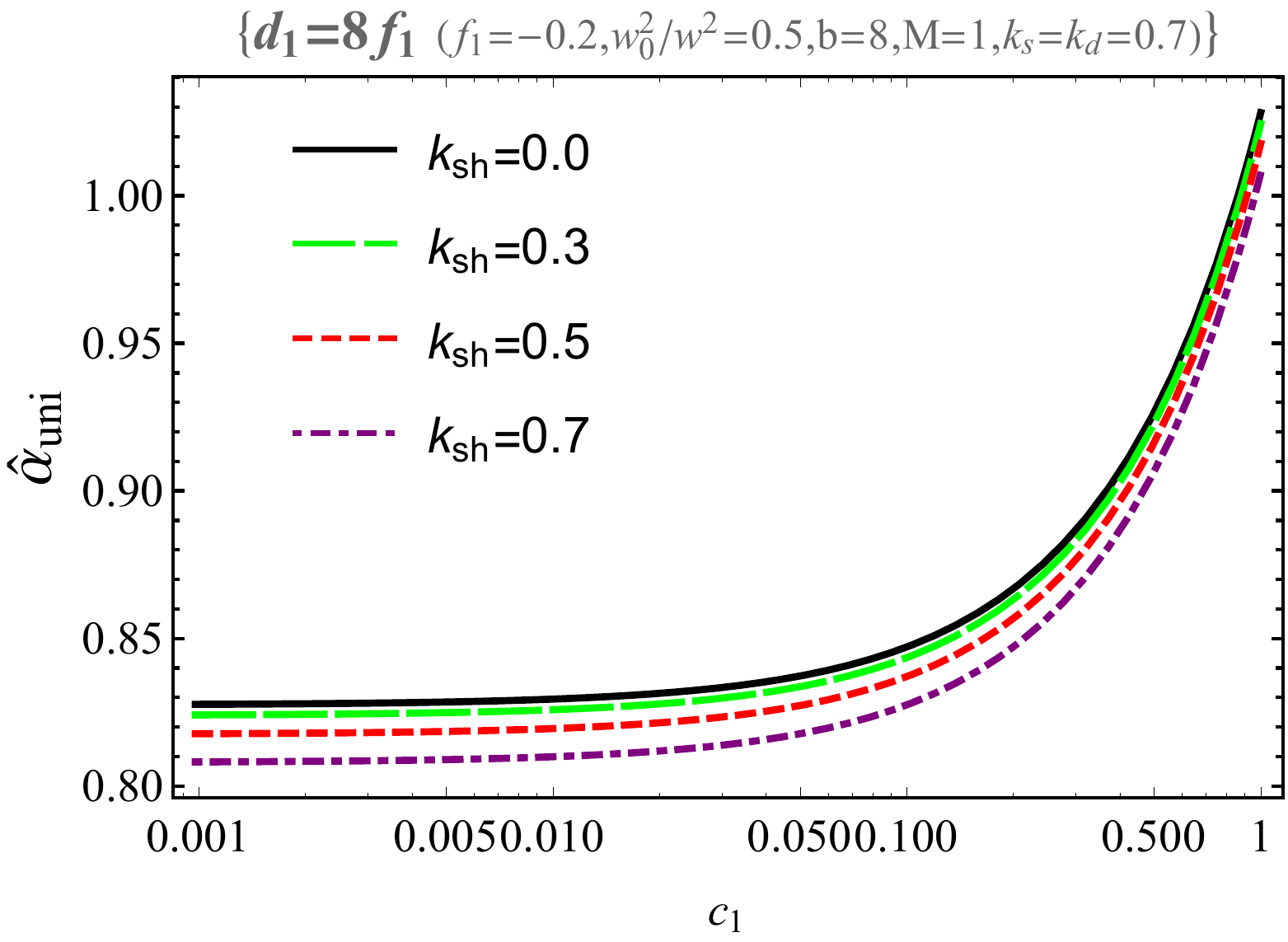}
    \includegraphics[scale=0.26]{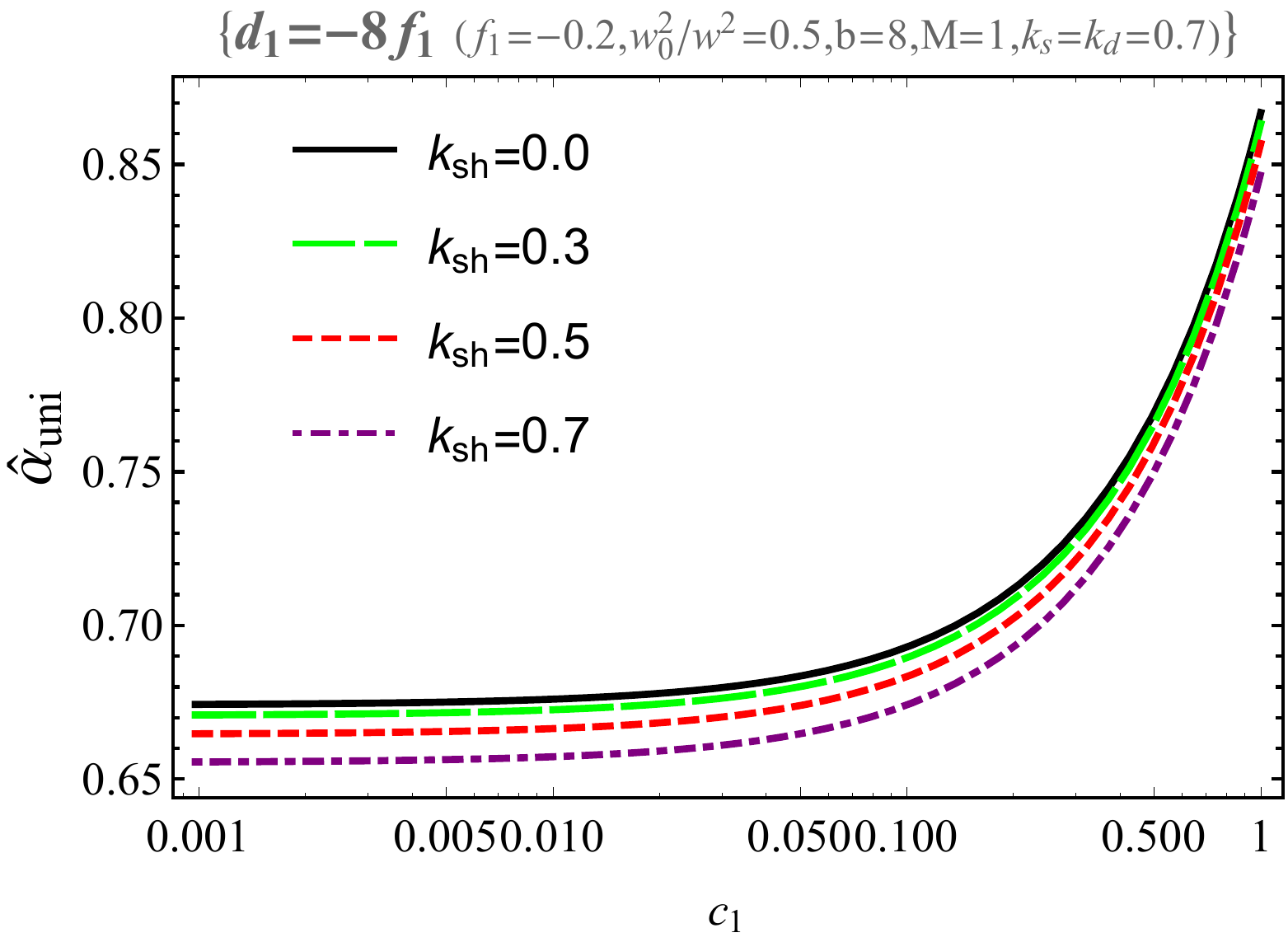}
    \includegraphics[scale=0.26]{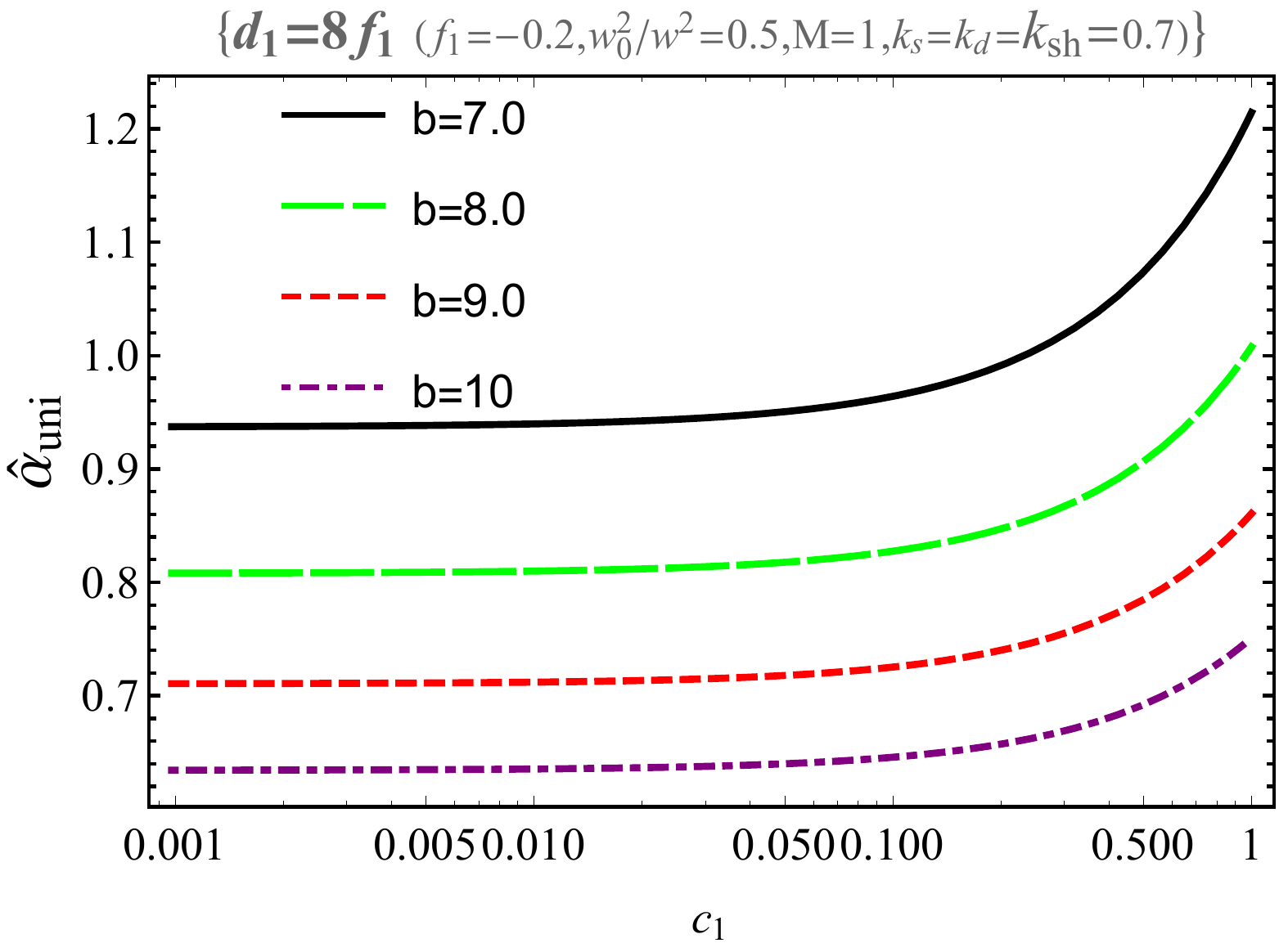}
    \includegraphics[scale=0.26]{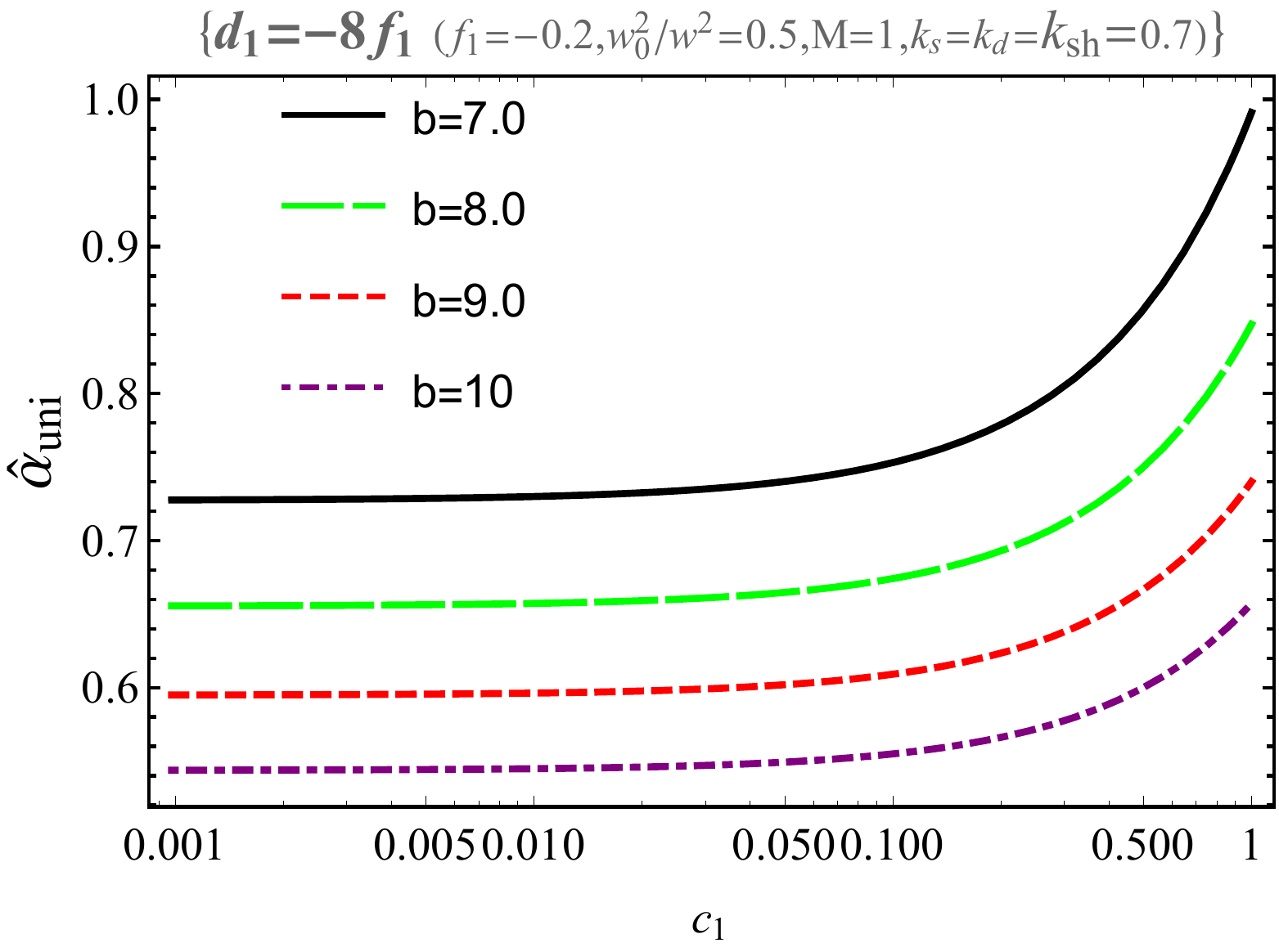}
    \includegraphics[scale=0.26]{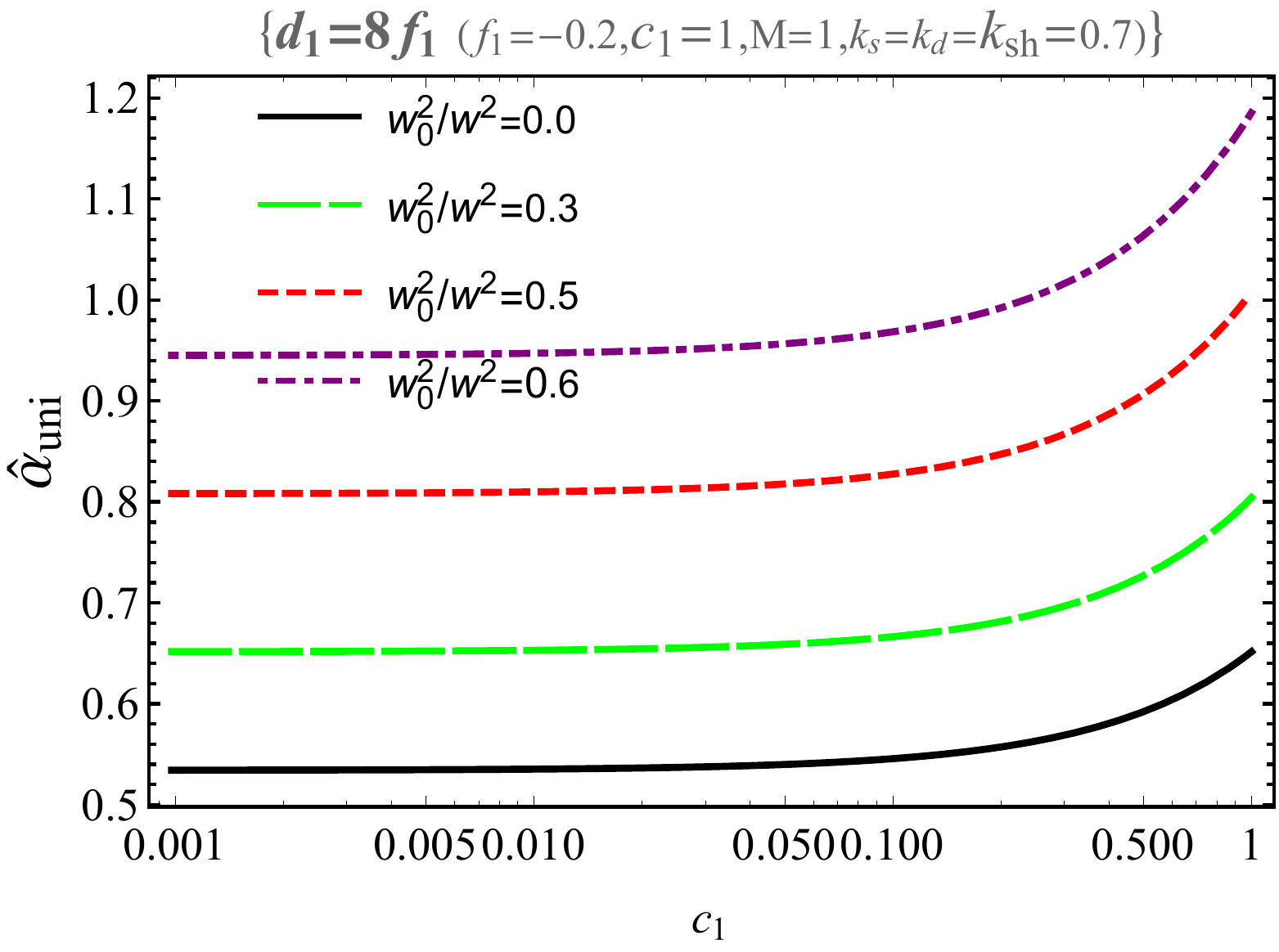}
    \includegraphics[scale=0.26]{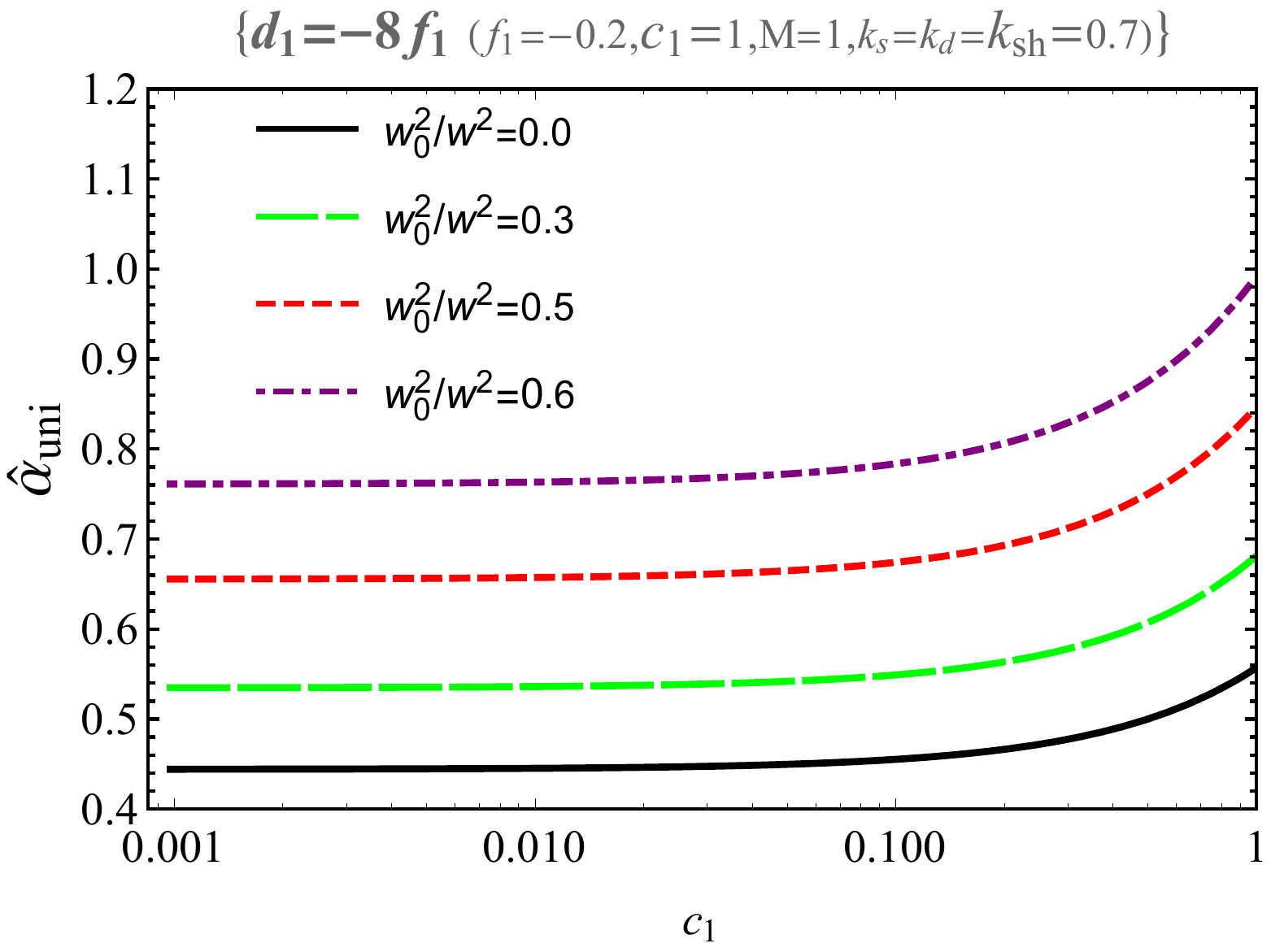}
    \caption{The deflection angle $\hat{\alpha}_{uni}$ in uniform plasma for $d_1=8f_1$ (Left panel) and $d_1=-8f_1$ (Right panel) along $c_1$ taking different values of $f_1,\; k_s,\; k_d, \;\&\; k_{sh}. $}
    \label{plot:13}
    \end{figure}

\begin{figure}
    \centering
    \includegraphics[scale=0.26]{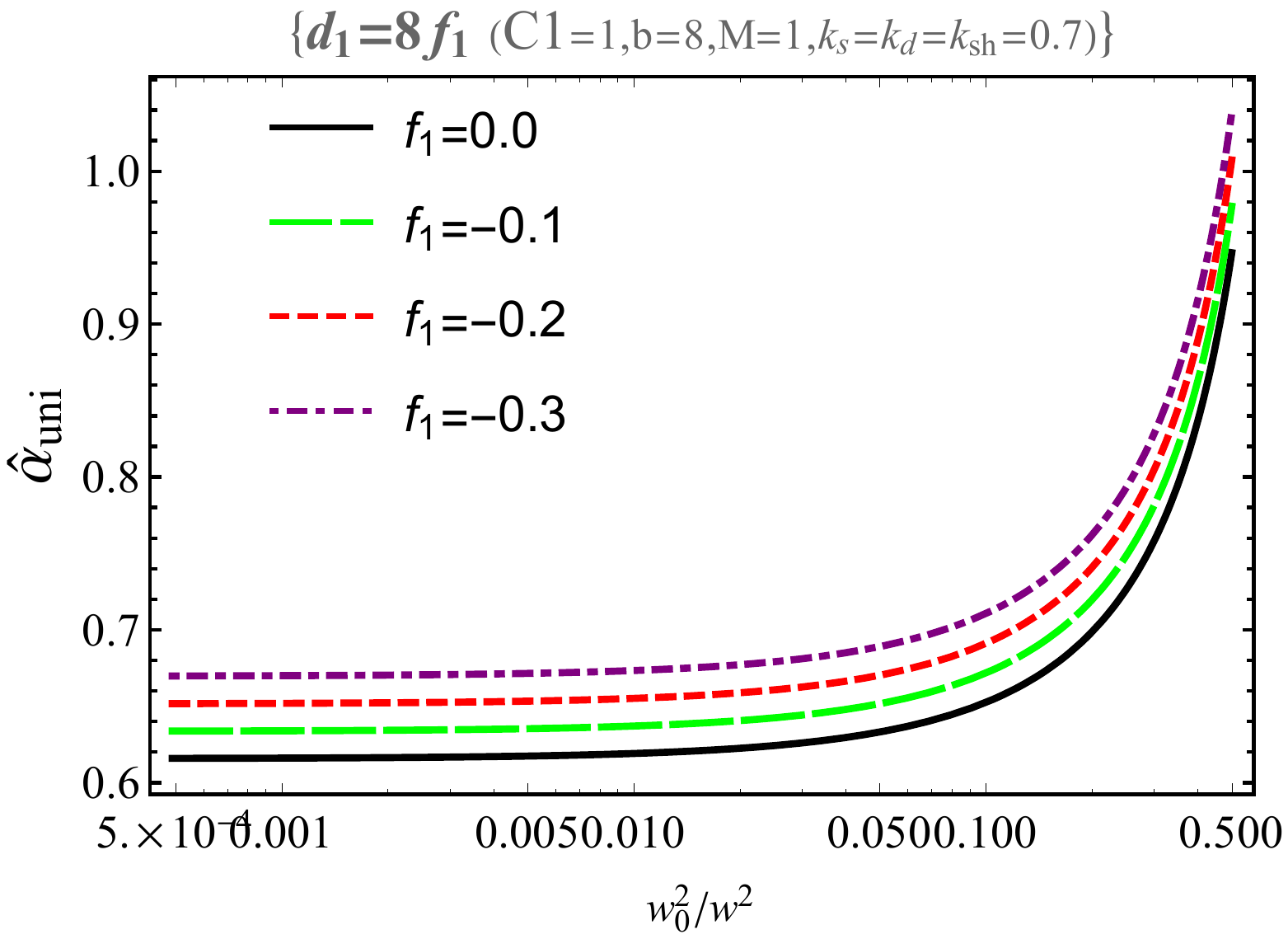}
    \includegraphics[scale=0.26]{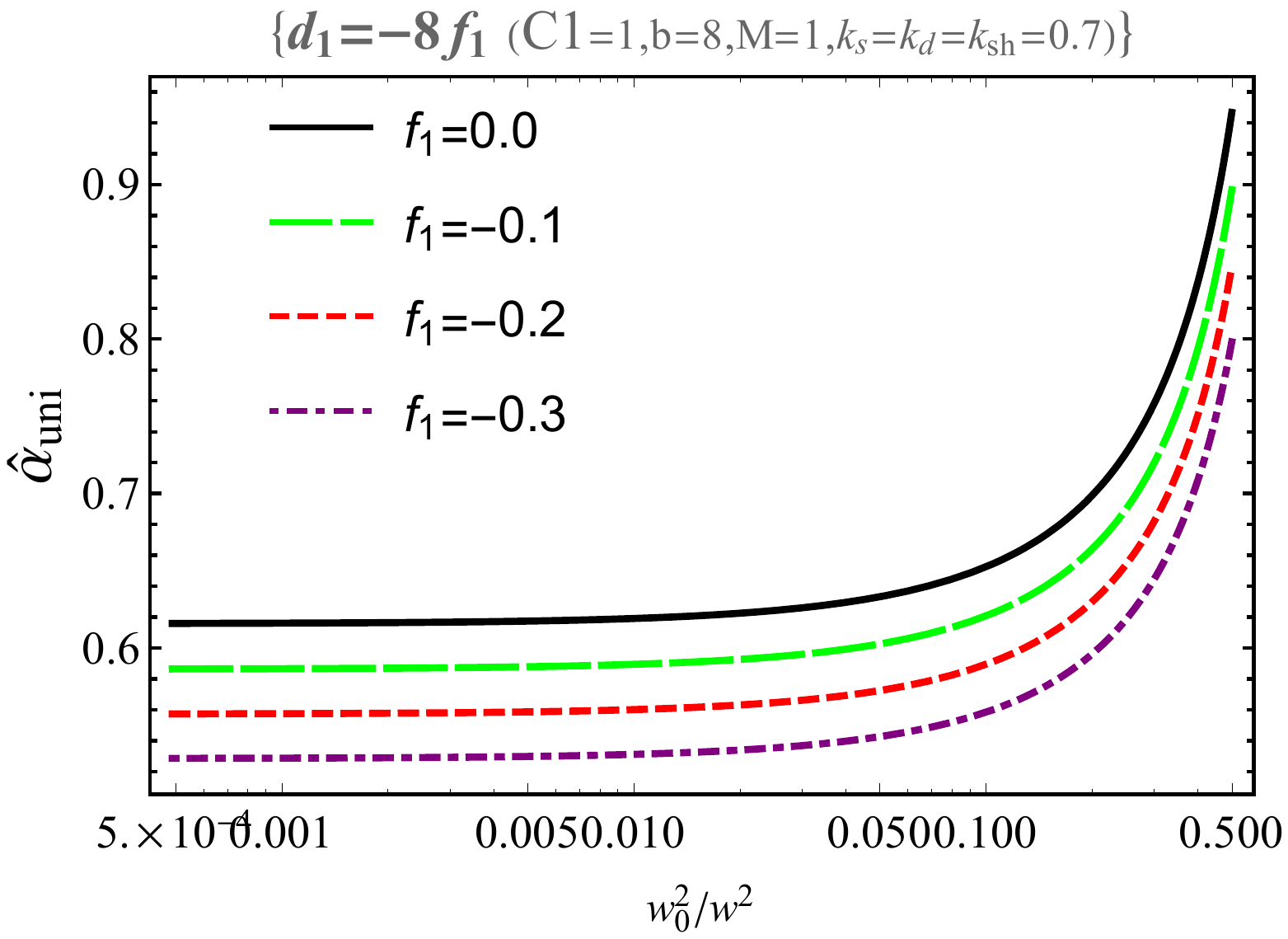}
    \includegraphics[scale=0.26]{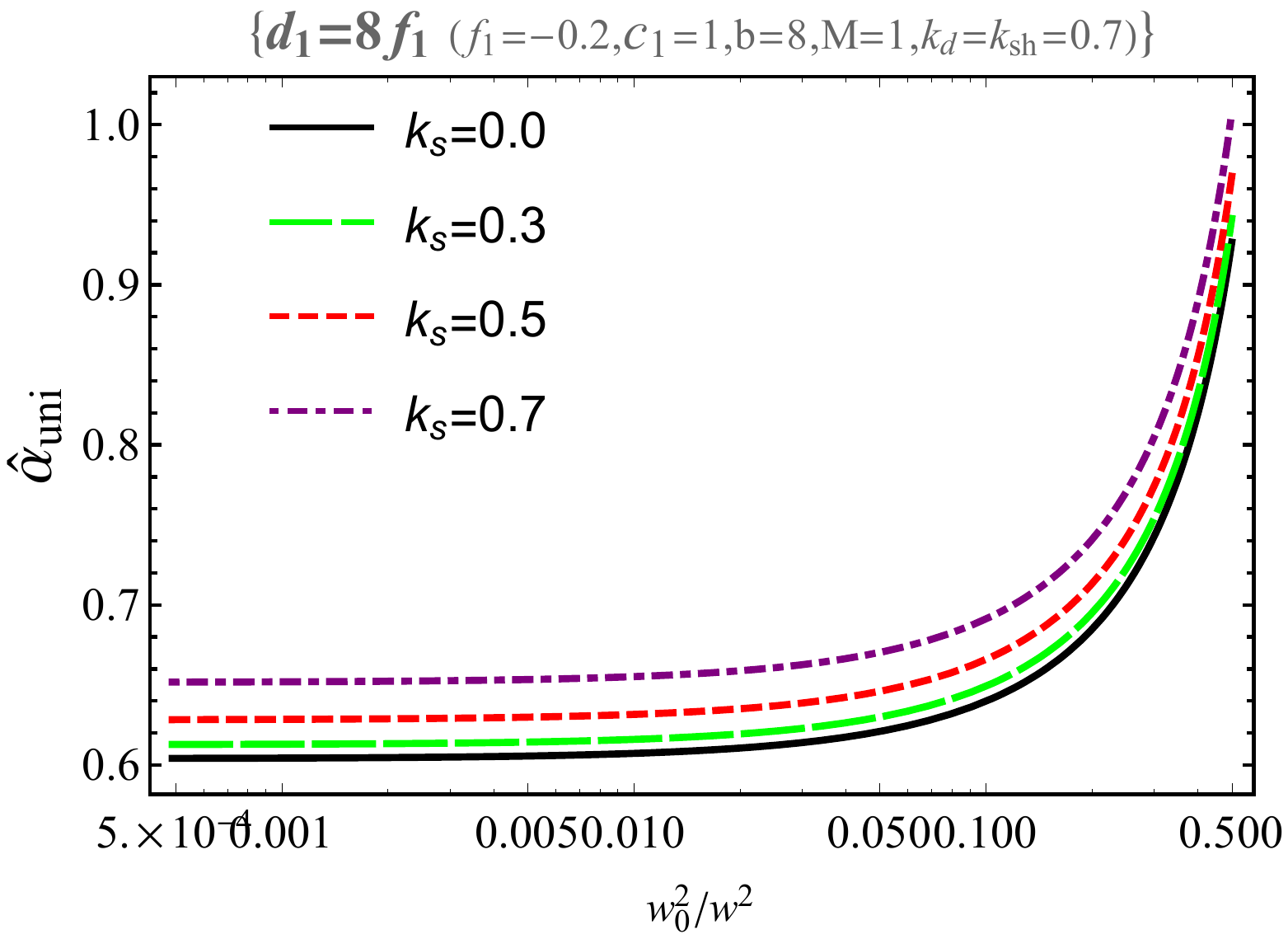}
    \includegraphics[scale=0.26]{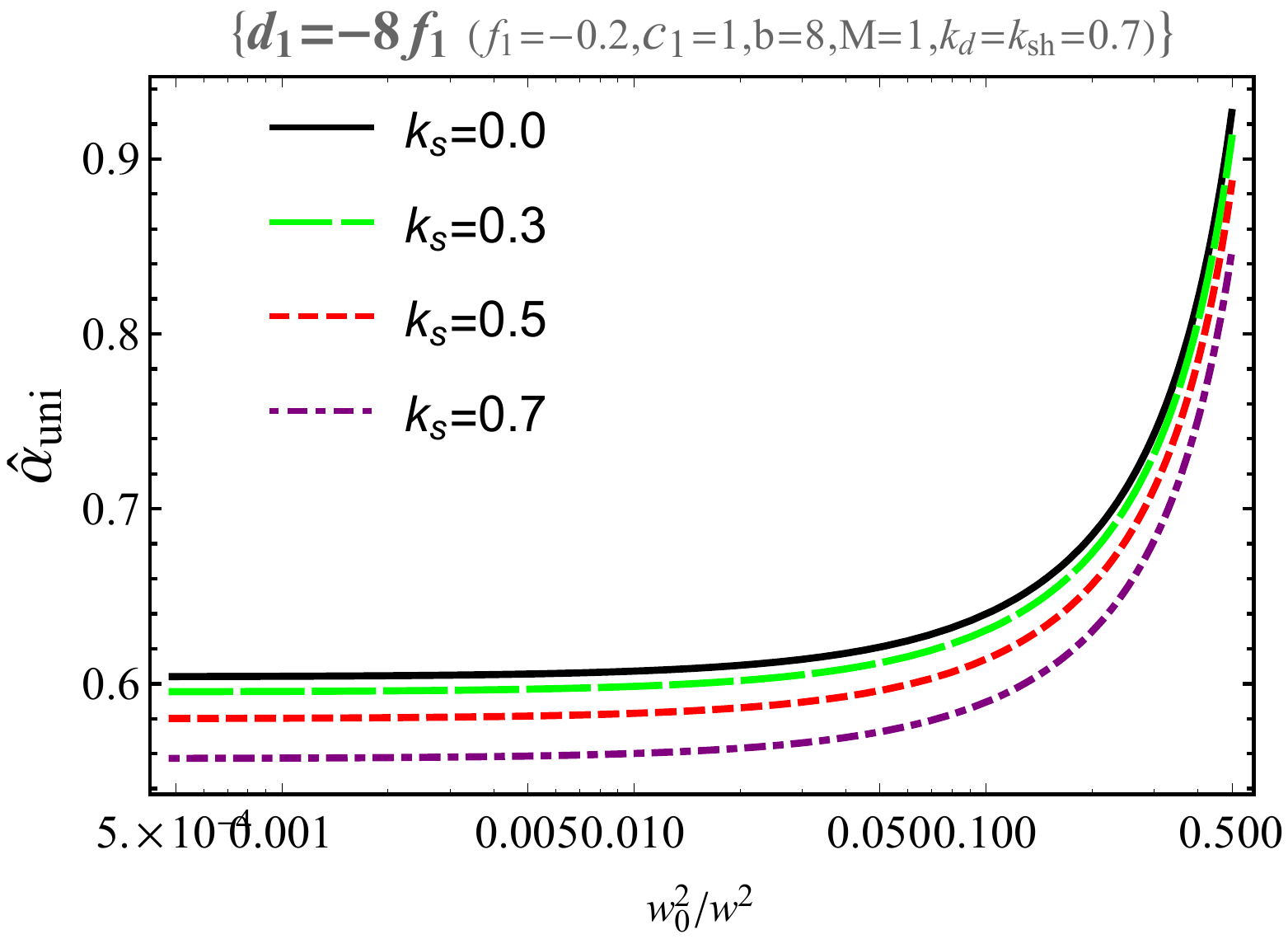}
    \includegraphics[scale=0.26]{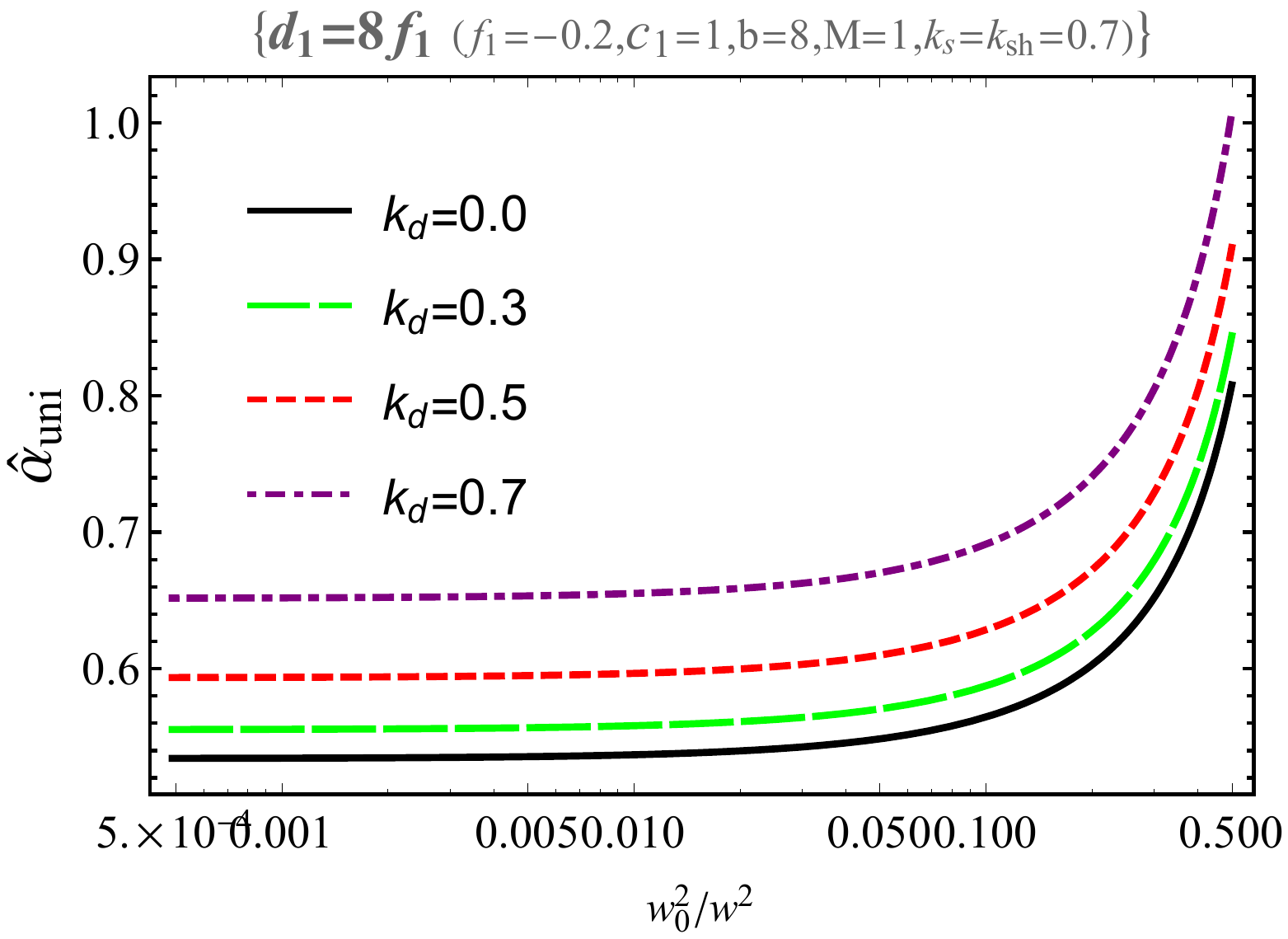}
    \includegraphics[scale=0.26]{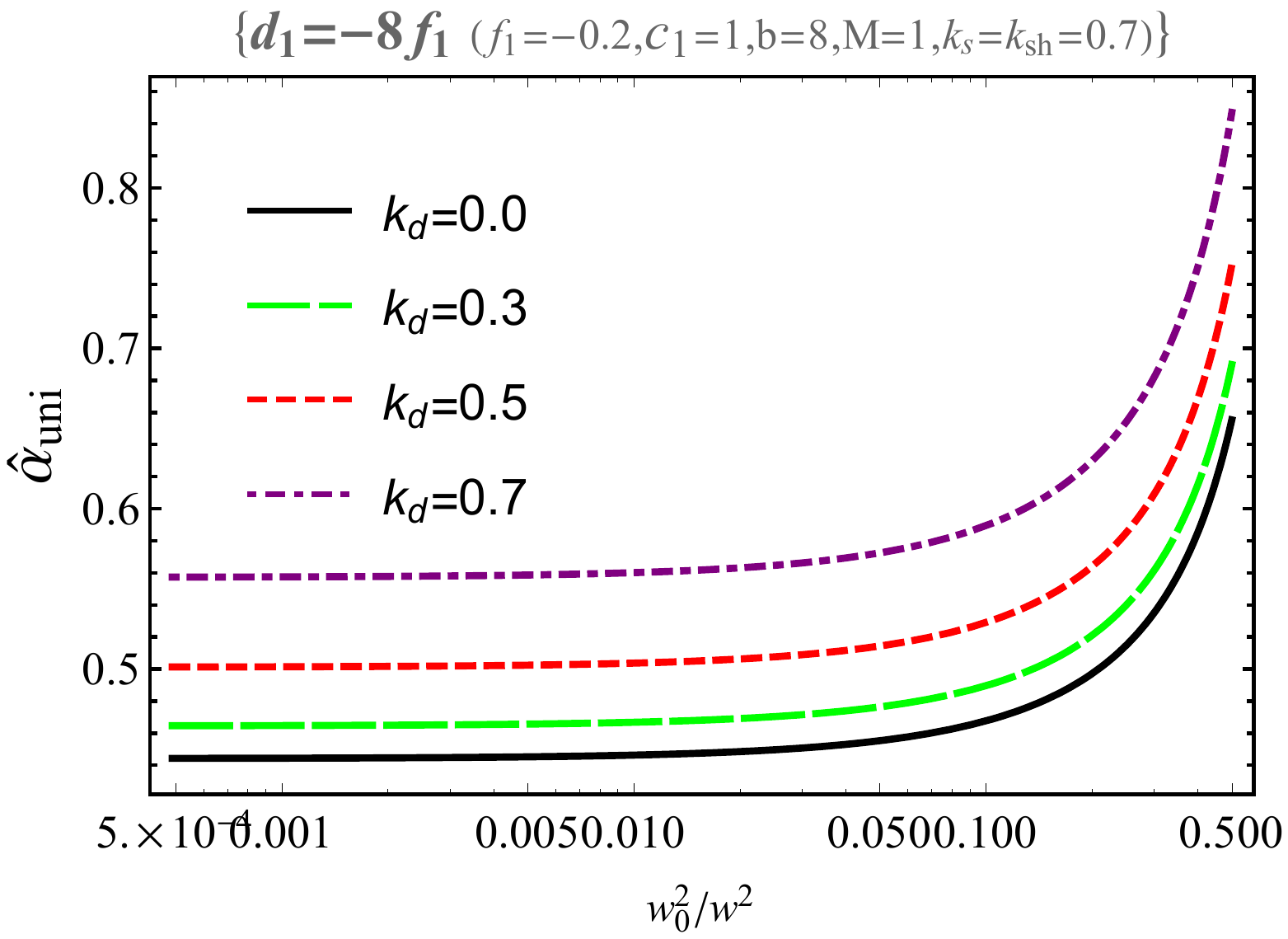}
    \includegraphics[scale=0.26]{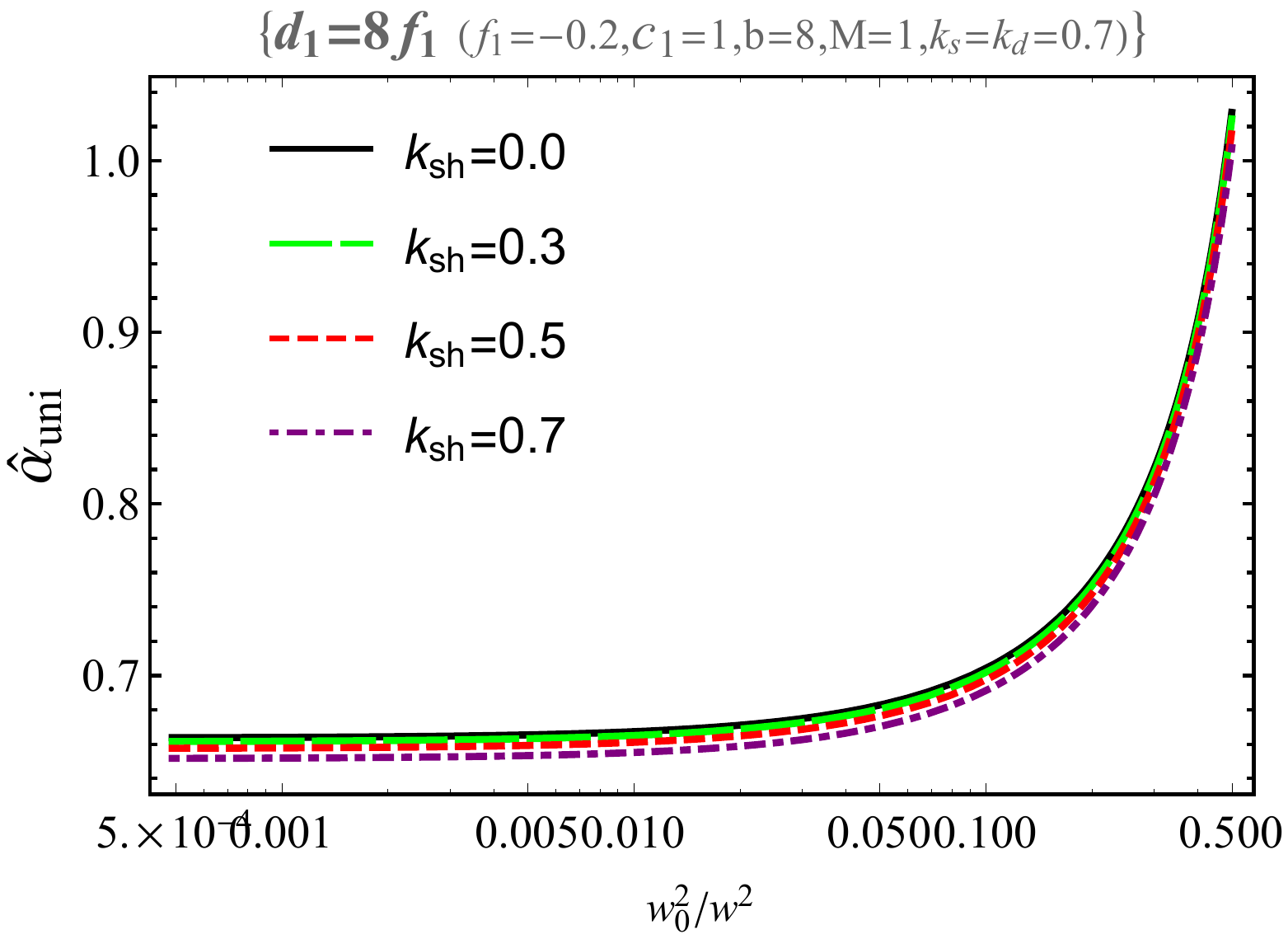}
    \includegraphics[scale=0.26]{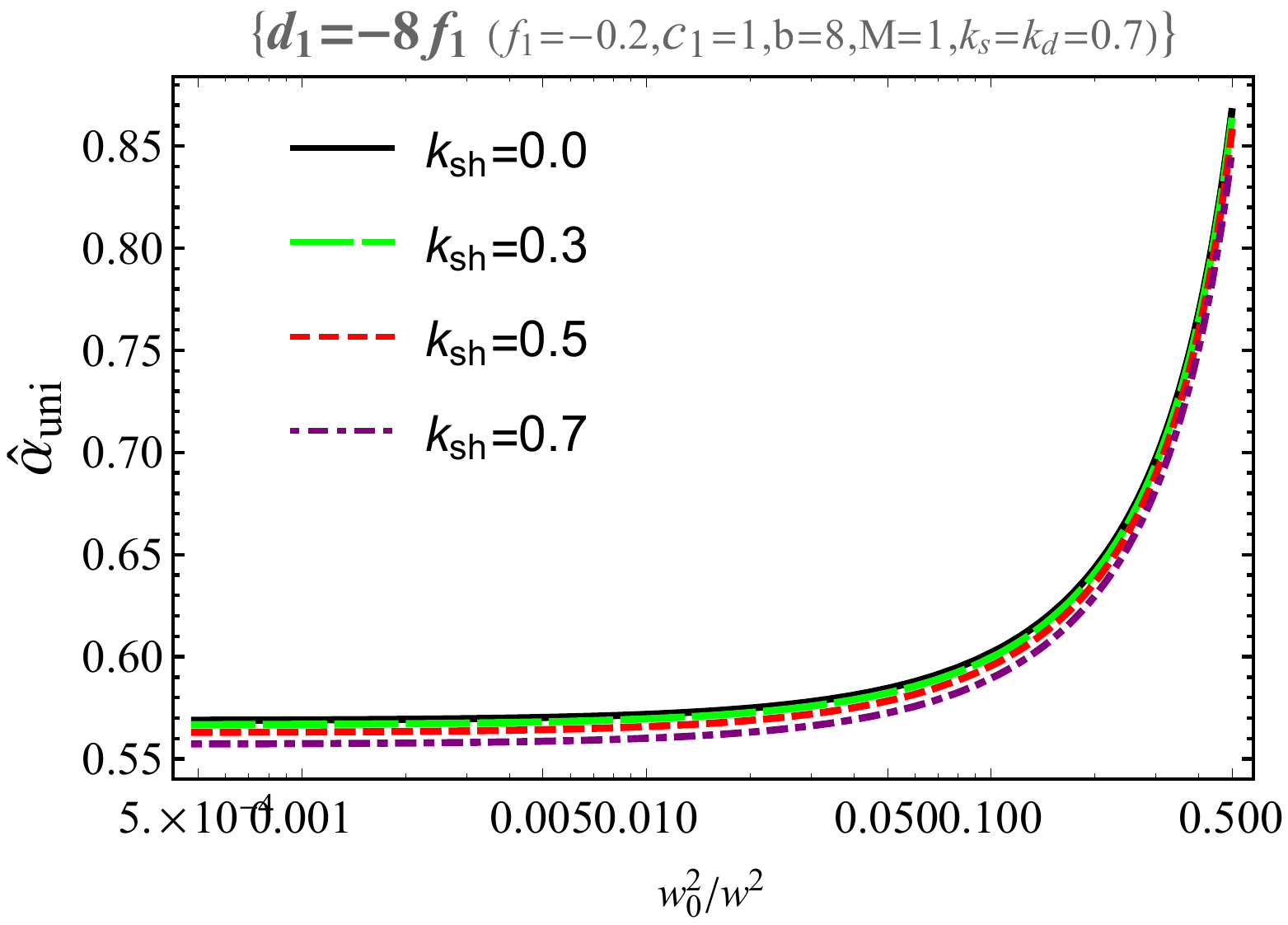}
    \includegraphics[scale=0.26]{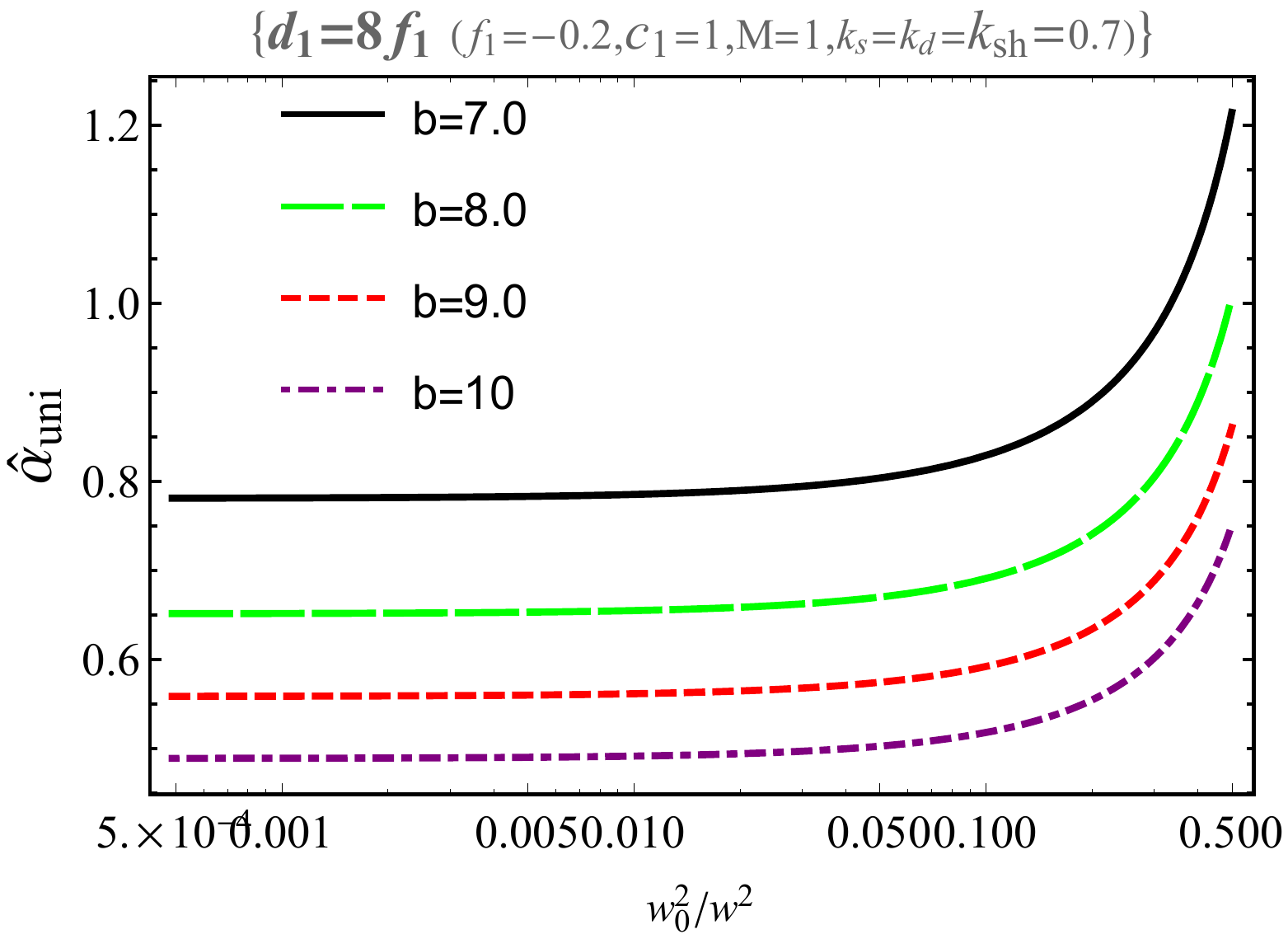}
    \includegraphics[scale=0.26]{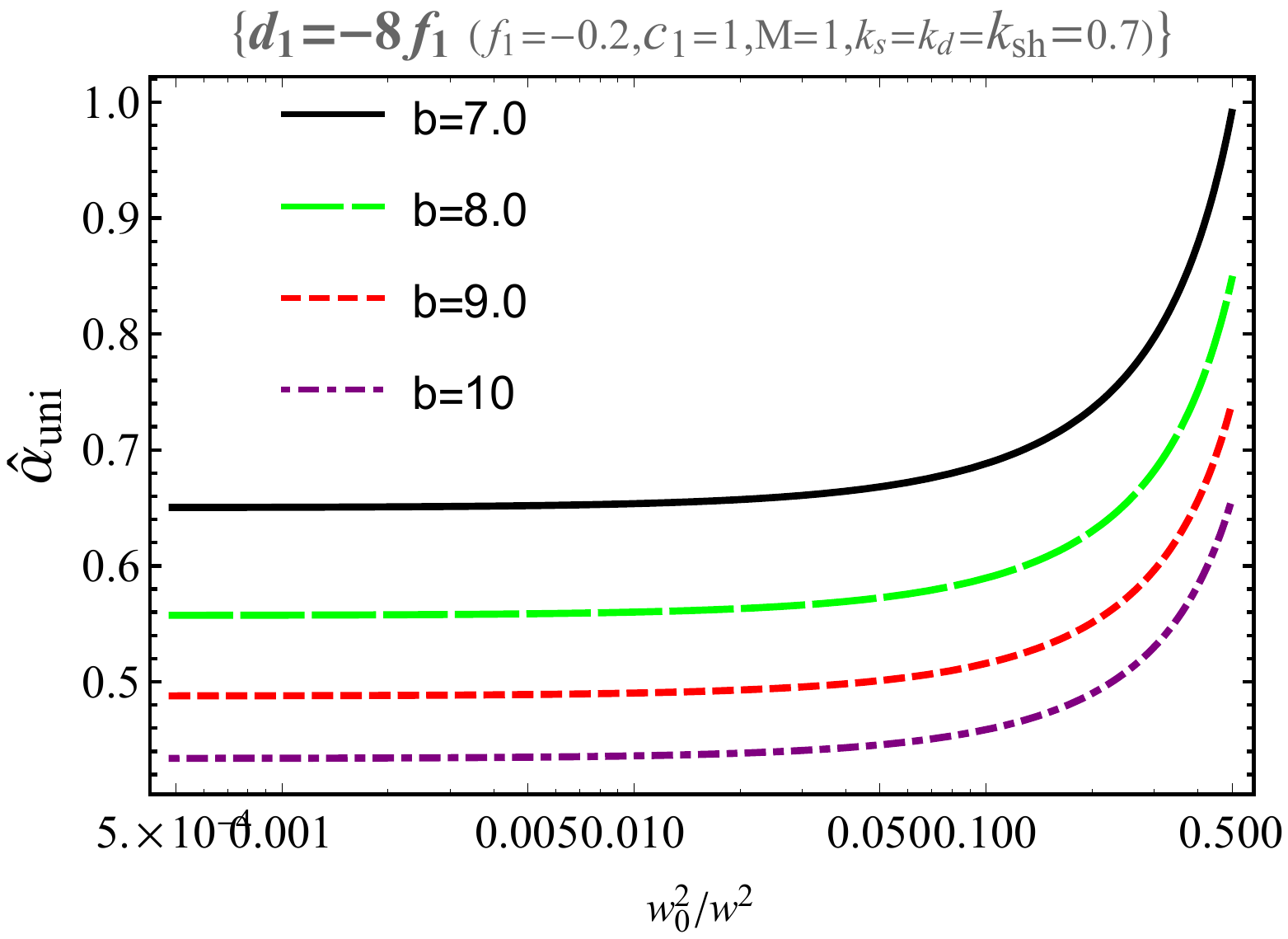}
    \includegraphics[scale=0.26]{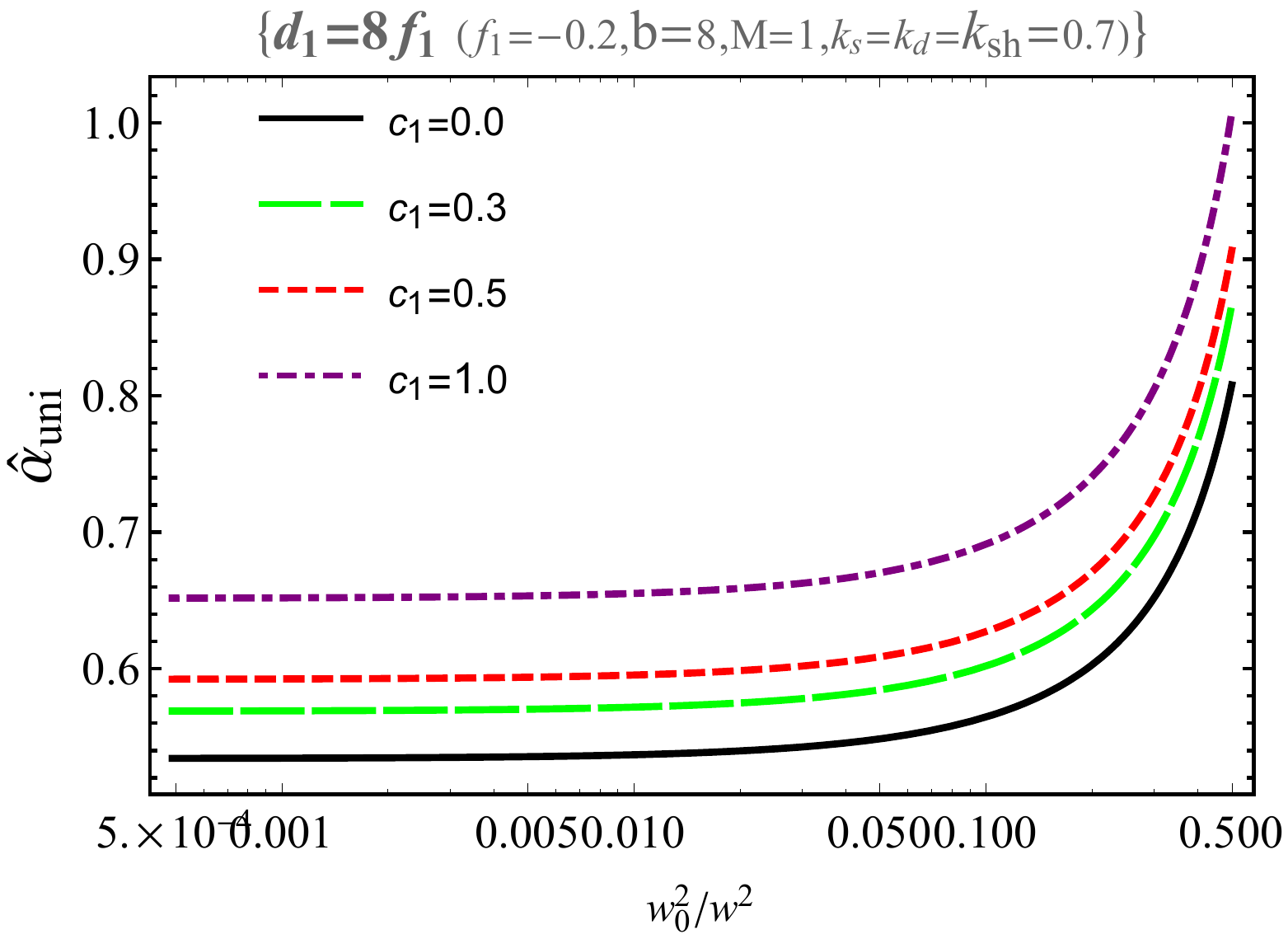}
    \includegraphics[scale=0.26]{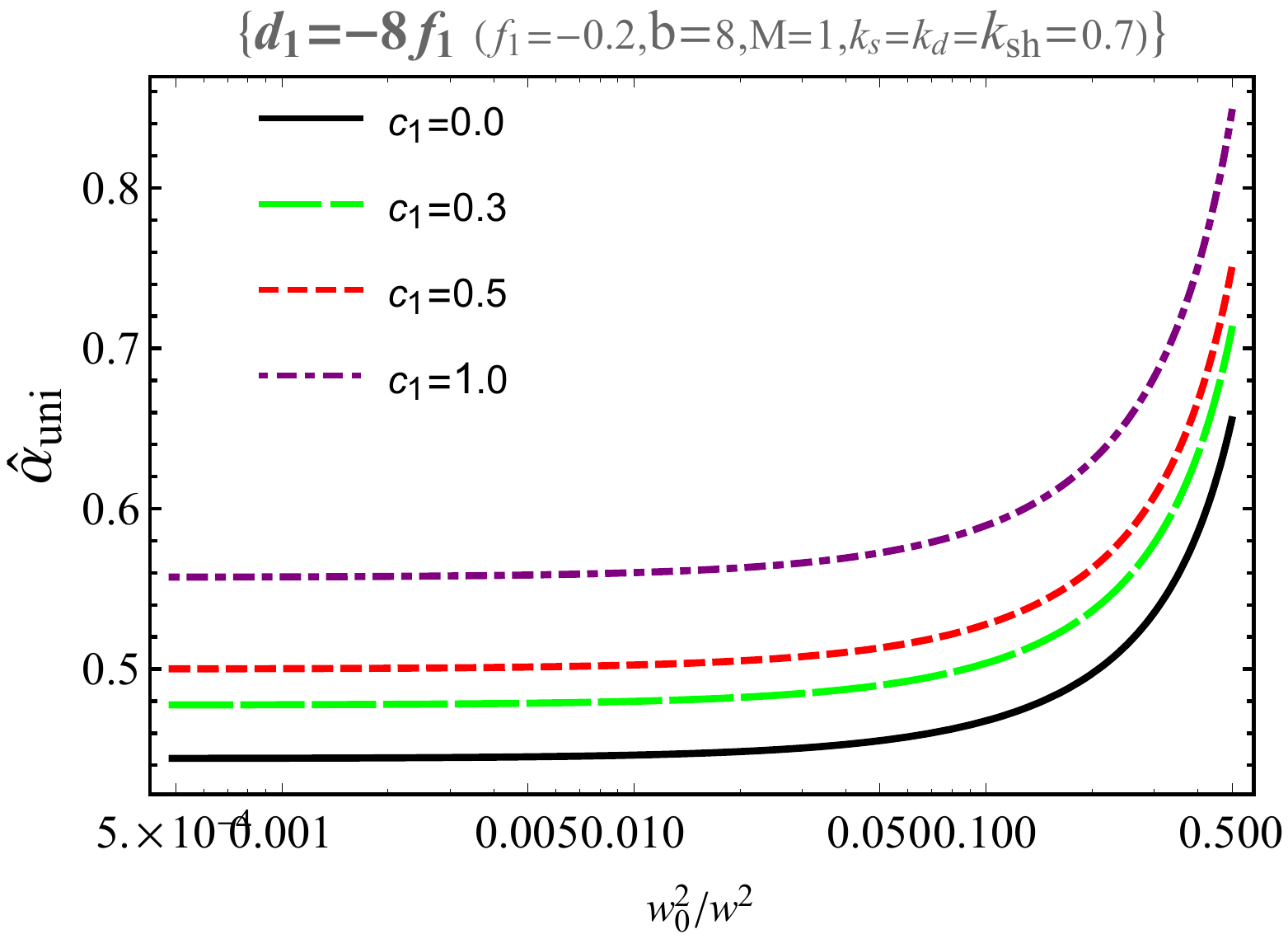}
    \caption{The deflection angle $\hat{\alpha}_{uni}$ in uniform plasma field for $d_1=8f_1$ (Left panel) and $d_1=-8f_1$ (Right panel) along $c_1$ taking different values of $f_1,\; k_s,\; k_d, \;\&\; k_{sh}.$}
    \label{plot:14}
    \end{figure}

The deflection angle for uniform plasma around the MAGBH turns out to be \cite{150}
\begin{equation}\label{h8}
  \hat{\alpha}_{uni}=\hat{\alpha}_{uni1}+\hat{\alpha}_{uni2}+\hat{\alpha}_{uni3}.
\end{equation}
Solving Eqs.(\ref{h3}), (\ref{h6}) and (\ref{h7}), we find the deflection angle in the uniform plasma given by
\begin{eqnarray}\label{40}
    \hat{\alpha}_{uni}&=&\frac{16 c_{1} k_{d}^2 R_s}{3 b^3}-\frac{4 d_{1} k_{s}^2 R_s}{3 b^3}+\frac{8 f_{1} k_{sh}^2 R_s}{3 b^3}+\frac{\pi  \sqrt{\frac{1}{b^2}} c_{1} k_{d}^2}{b}-\frac{\pi  \sqrt{\frac{1}{b^2}} d_{1} k_{s}^2}{4 b}+\frac{\pi  \sqrt{\frac{1}{b^2}} f_{1} k_{sh}^2}{2 b}-\frac{1}{1-\frac{w_0^2}{w^2}}\Big[-\frac{16 c_{1} k_{d}^2 R_s}{b^3}\nonumber\\&+&\frac{4 d_{1} k_{s}^2 R_s}{b^3}-\frac{8 f_{1} k_{sh}^2 R_s}{b^3}-\frac{2 \pi  \sqrt{\frac{1}{b^2}} c_{1} k_{d}^2}{b}+\frac{\pi  \sqrt{\frac{1}{b^2}} d_{1} k_{s}^2}{2 b}-\frac{\pi  \sqrt{\frac{1}{b^2}} f_{1} k_{sh}^2}{b}-\frac{12 \pi  \sqrt{\frac{1}{b^2}} c_{1}^2 k_{d}^4}{b^3}+\frac{6 \pi  \sqrt{\frac{1}{b^2}} c_{1} d_{1} k_{d}^2 k_{s}^2}{b^3}\nonumber\\&-&\frac{12 \pi  \sqrt{\frac{1}{b^2}} c_{1} f_{1} k_{d}^2 k_{sh}^2}{b^3}-\frac{3 \pi  \sqrt{\frac{1}{b^2}} d_{1}^2 k_{s}^4}{4 b^3}+\frac{3 \pi  \sqrt{\frac{1}{b^2}} d_{1} f_{1} k_{sh}^2 k_{s}^2}{b^3}-\frac{3 \pi  \sqrt{\frac{1}{b^2}} f_{1}^2 k_{sh}^4}{b^3}-\frac{R_s}{b}\Big]+\frac{3 \pi  \sqrt{\frac{1}{b^2}} c_{1}^2 k_{d}^4}{b^3}\nonumber\\&-&\frac{3 \pi  \sqrt{\frac{1}{b^2}} c_{1} d_{1} k_{d}^2 k_{s}^2}{2 b^3}+\frac{3 \pi  \sqrt{\frac{1}{b^2}} c_{1} f_{1} k_{d}^2 k_{sh}^2}{b^3}+\frac{3 \pi  \sqrt{\frac{1}{b^2}} d_{1}^2 k_{s}^4}{16 b^3}-\frac{3 \pi  \sqrt{\frac{1}{b^2}} d_{1} f_{1} k_{sh}^2 k_{s}^2}{4 b^3}+\frac{3 \pi  \sqrt{\frac{1}{b^2}} f_{1}^2 k_{sh}^4}{4 b^3}+\frac{R_s}{b}.
\end{eqnarray}
We also provide a graphical illustration of the deflection angle in Figs.~\ref{plot:13} and \ref{plot:14}. It is worthwhile to mention the following consequences:

\begin{itemize}
    \item  ($i$) In case of $d_1=8f_1$, an increase in $\hat{\alpha}_{uni}$ along $b$ by increasing $f_1,\;c_1,\;k_d,\;k_s,\;\&\;w_0^2/w^2$ and a decrease with increase in $k_{sh}$, ($ii$) also an increase in $\hat{\alpha}_{uni}$ along $b$ with a rise in $c_1,\;k_d,\;\&\;w_0^2/w^2$ and a fall with rise in $f_1,\;k_s,\;k_{sh}$ in case of $d_1=-8f_1$
    \item ($i$) an increase in $ \hat{\alpha}_{uni}$ along $c_1$ with an increase in $f_1,\;k_d,\;k_s,\;\&\;w_0^2/w^2$ and a decrease with an increase in $b,\;\&\;k_{sh}$ in case of $d_1=8f_1$, and ($ii$) an increase in $ \hat{\alpha}_{uni}$ along $c_1$ with an increase in $k_d,\;\&\;w_0^2/w^2$ and a decrease with rising values of $f_1,\;b,\;k_{s},\;k_{sh}$ in case of $d_1=-8f_1$.
    \item   ($i$) $ \hat{\alpha}_{uni}$ along $w_0^2/w^2$ increases with the increasing values of $f_1,\;c_1,\;k_d,\;k_{s}$ and decreases with increasing values of $b\;\&\;k_{sh}$ in case of $d_1=8f_1$, ($ii$) and $ \hat{\alpha}_{uni}$ along $w_0^2/w^2$ increases with the rising values of $c_1,\;\&\;k_d$ and falls with rise in $f_1,\;k_s,\;k_{sh}\;\&\;b$ in case of $d_1=-8f_1$
\end{itemize}

\subsection{Non uniform plasma}\label{A8.2}
\begin{figure}
    \centering
    \includegraphics[scale=0.26]{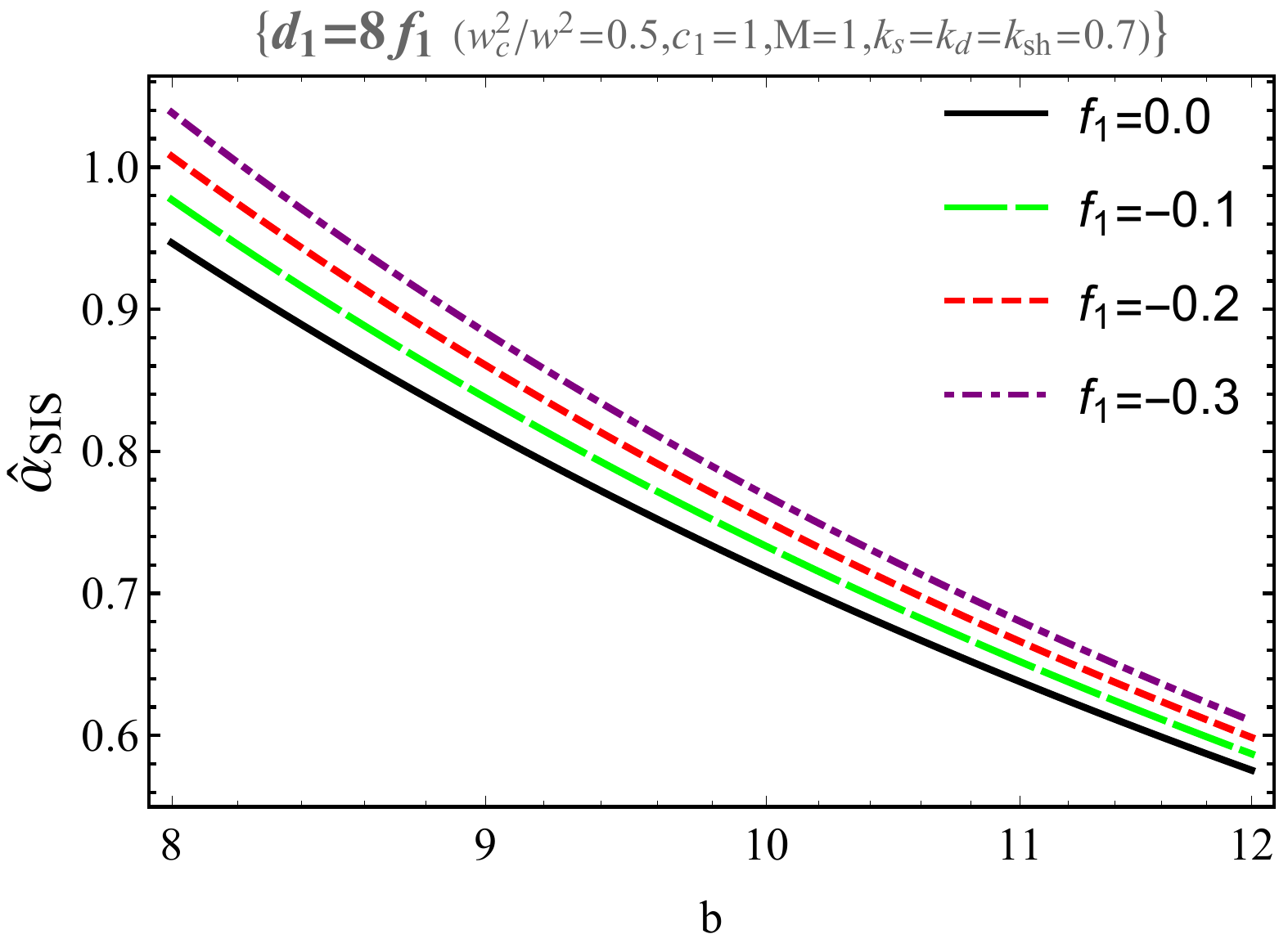}
    \includegraphics[scale=0.26]{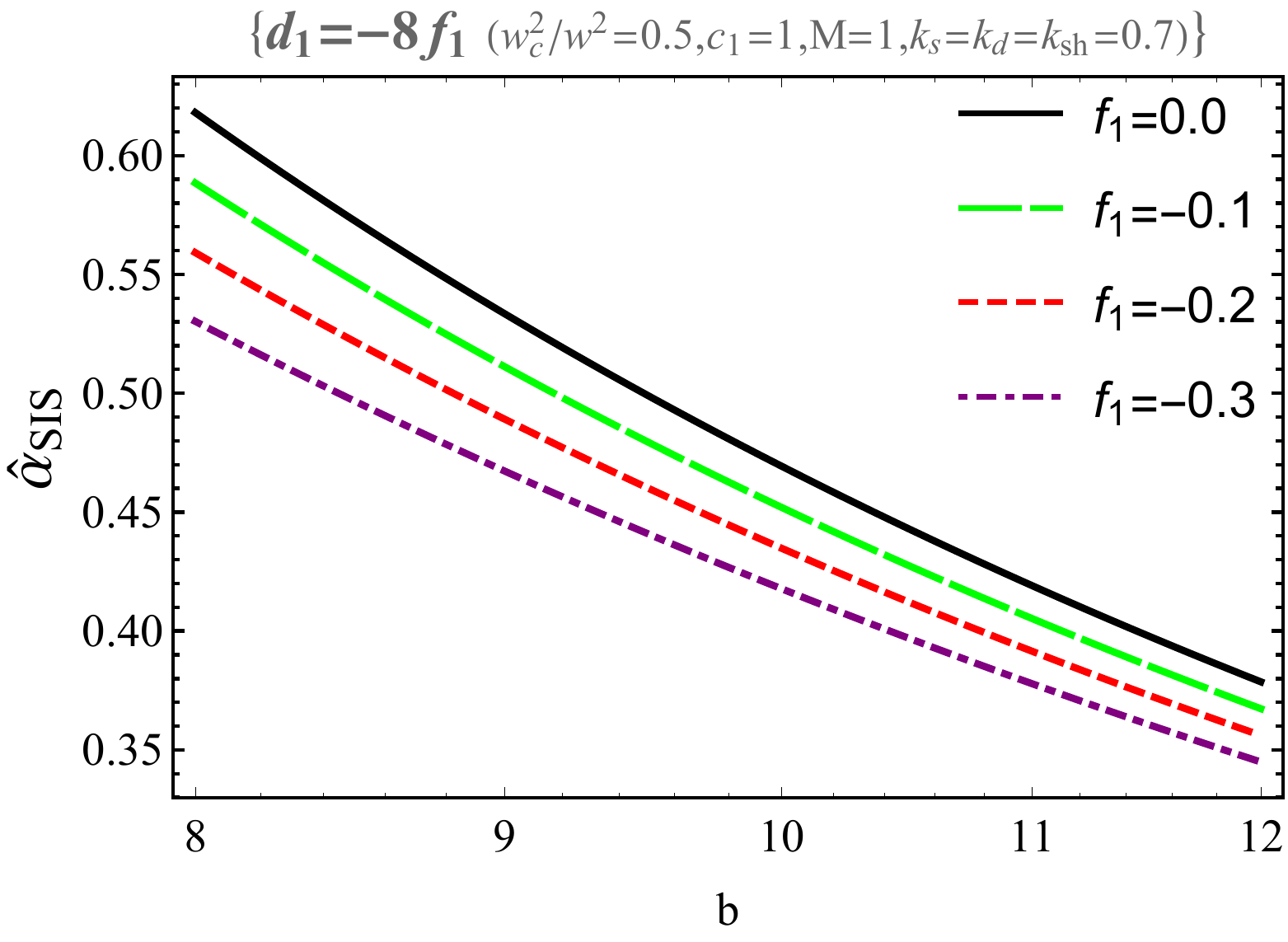}
    \includegraphics[scale=0.26]{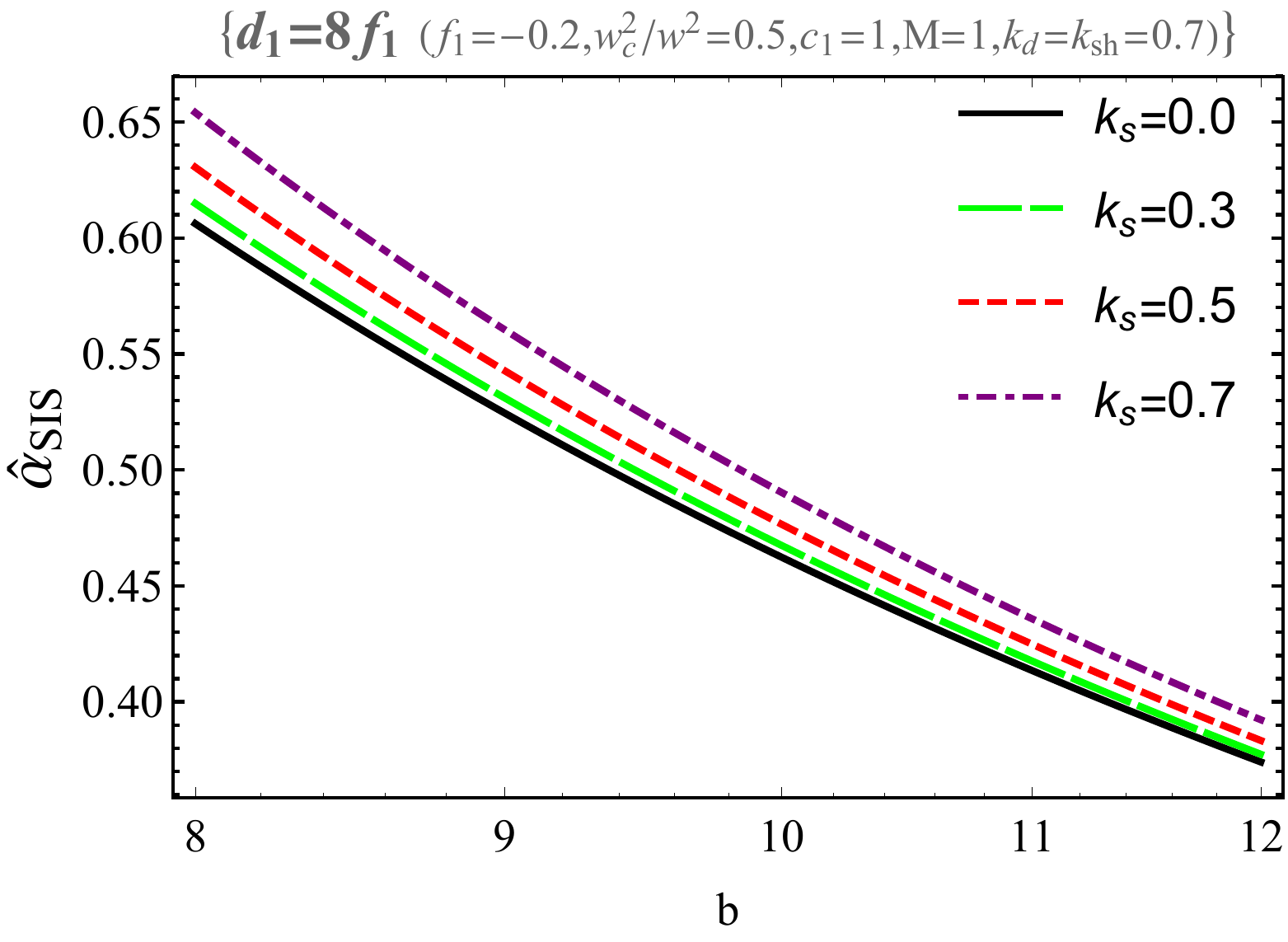}
    \includegraphics[scale=0.26]{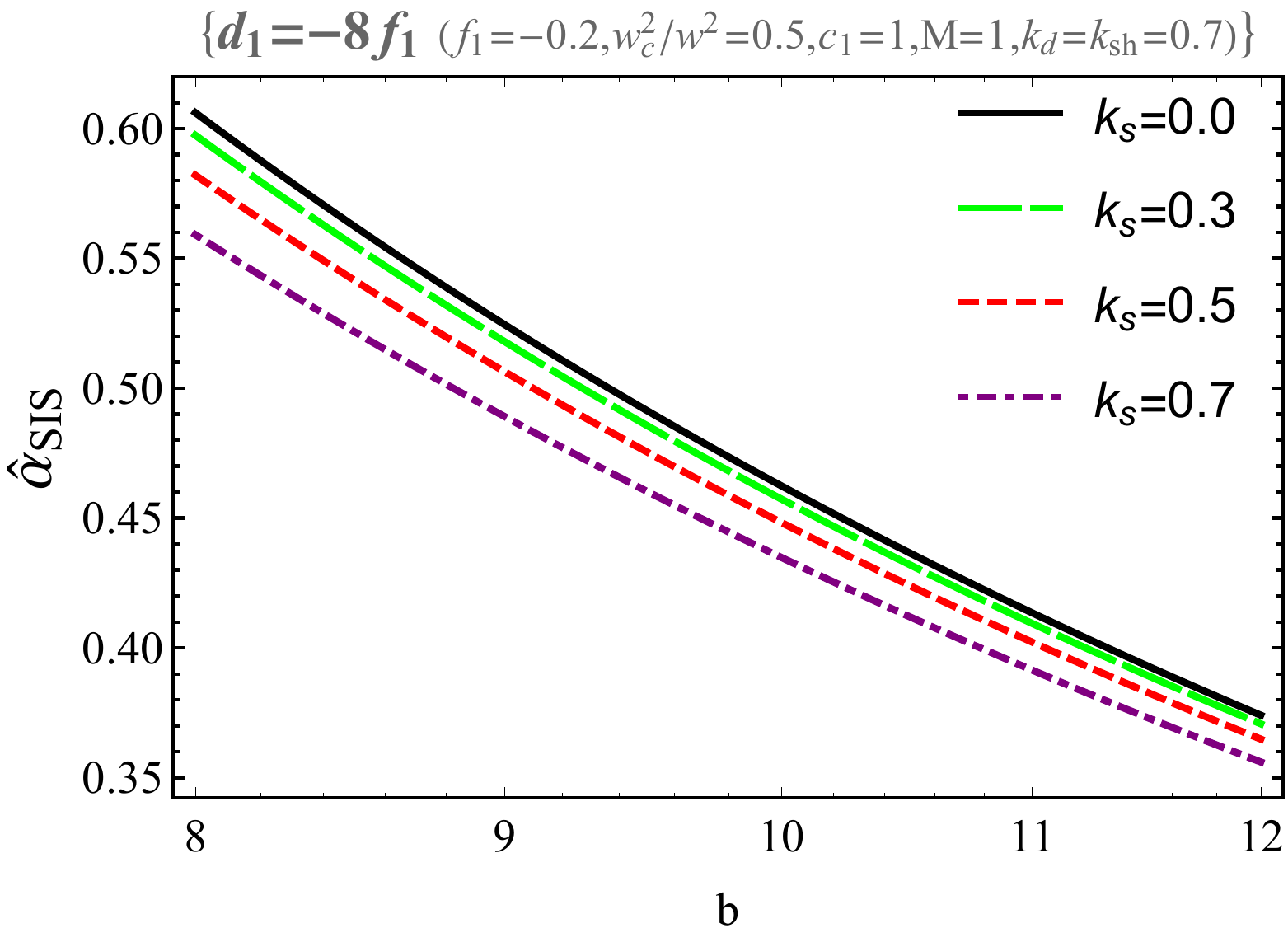}
    \includegraphics[scale=0.26]{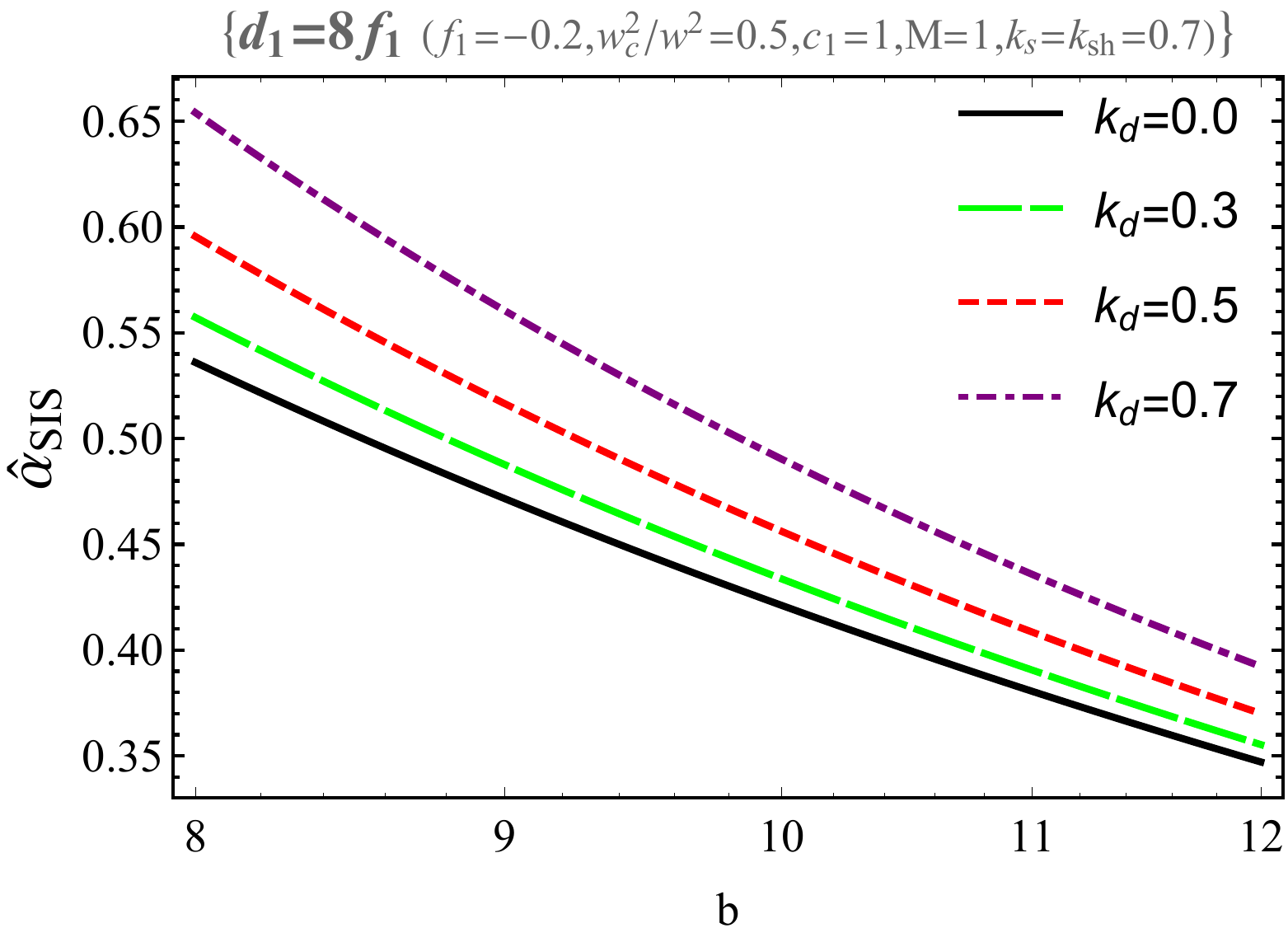}
    \includegraphics[scale=0.26]{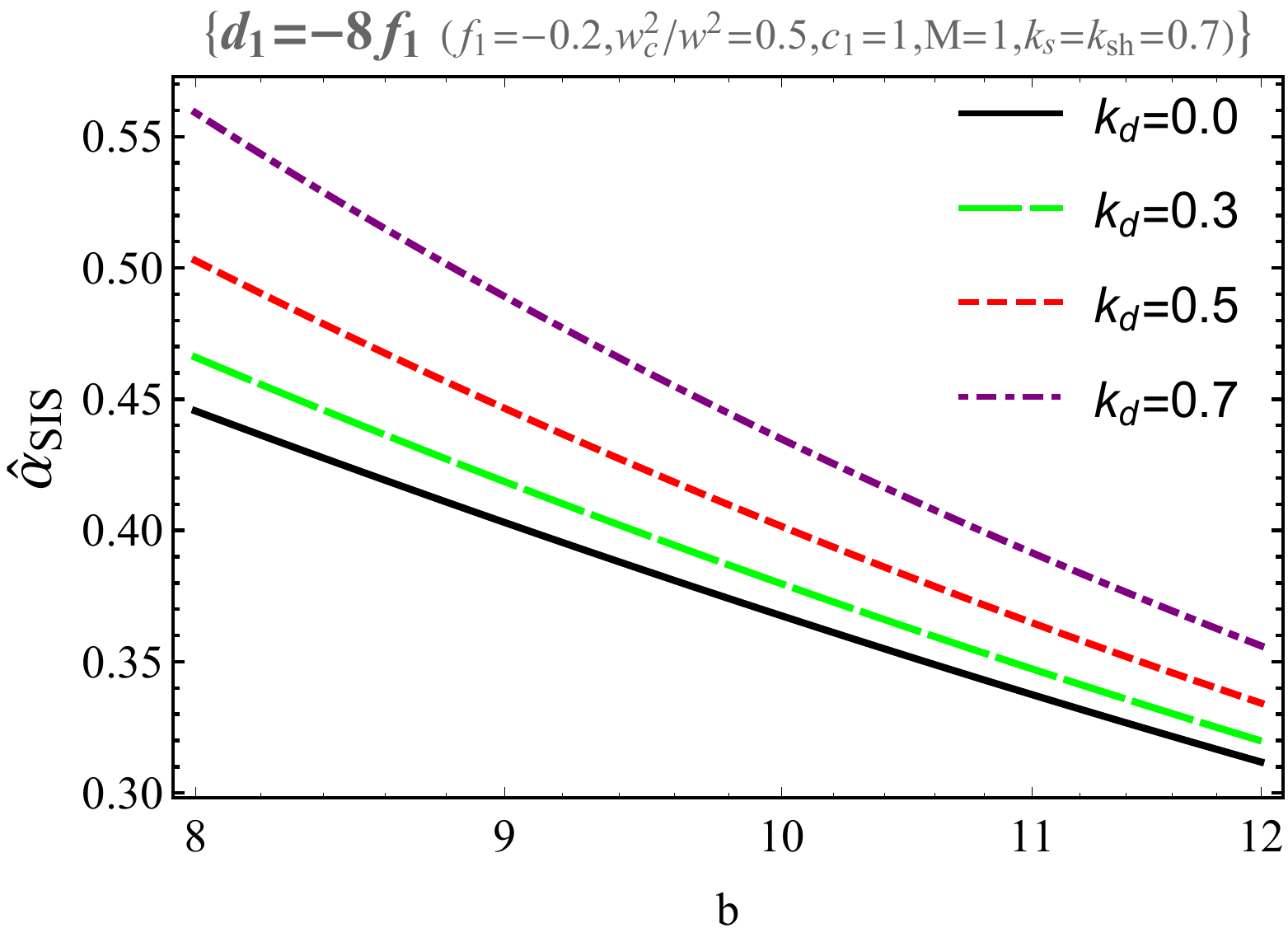}
    \includegraphics[scale=0.26]{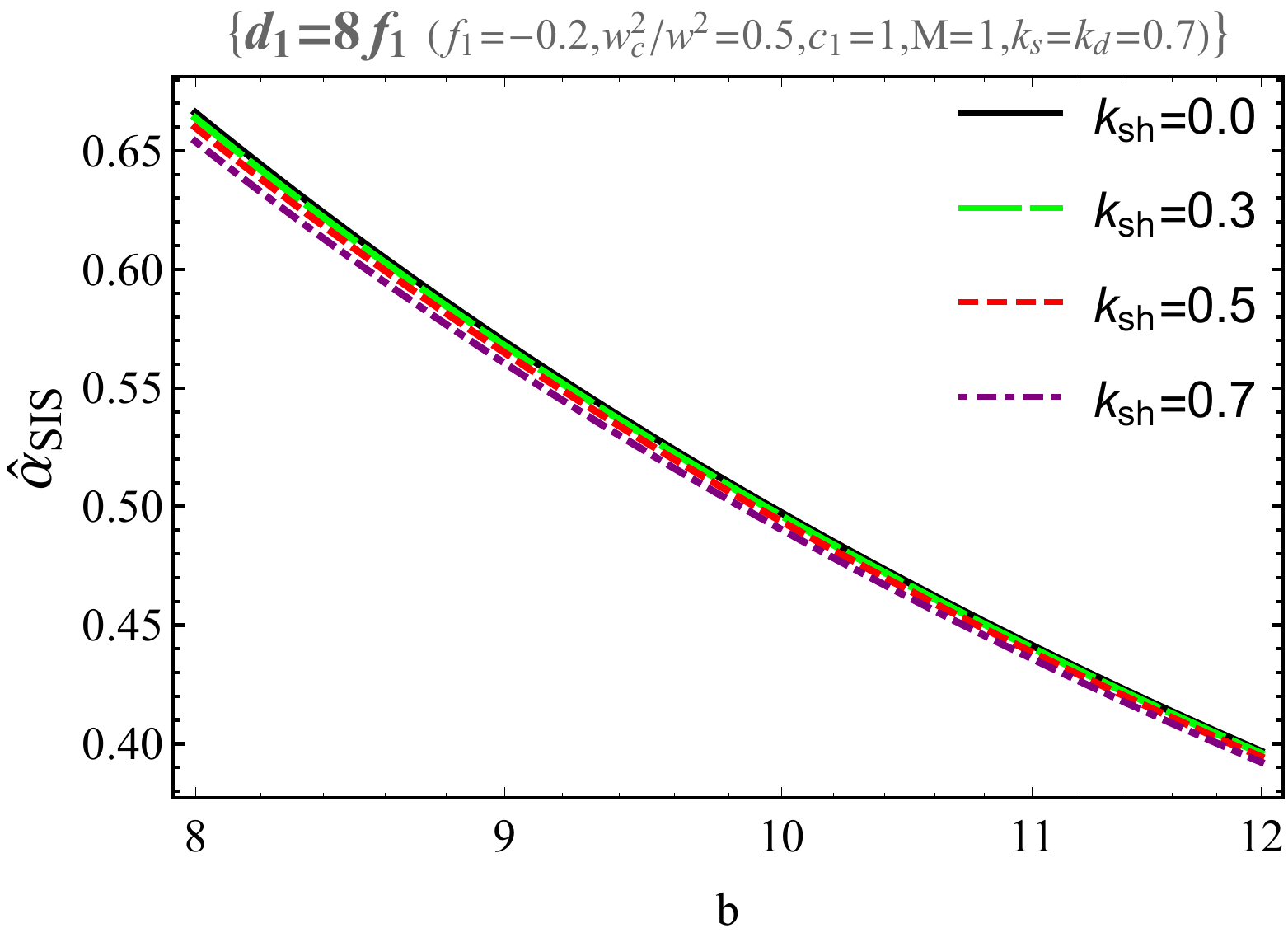}
    \includegraphics[scale=0.26]{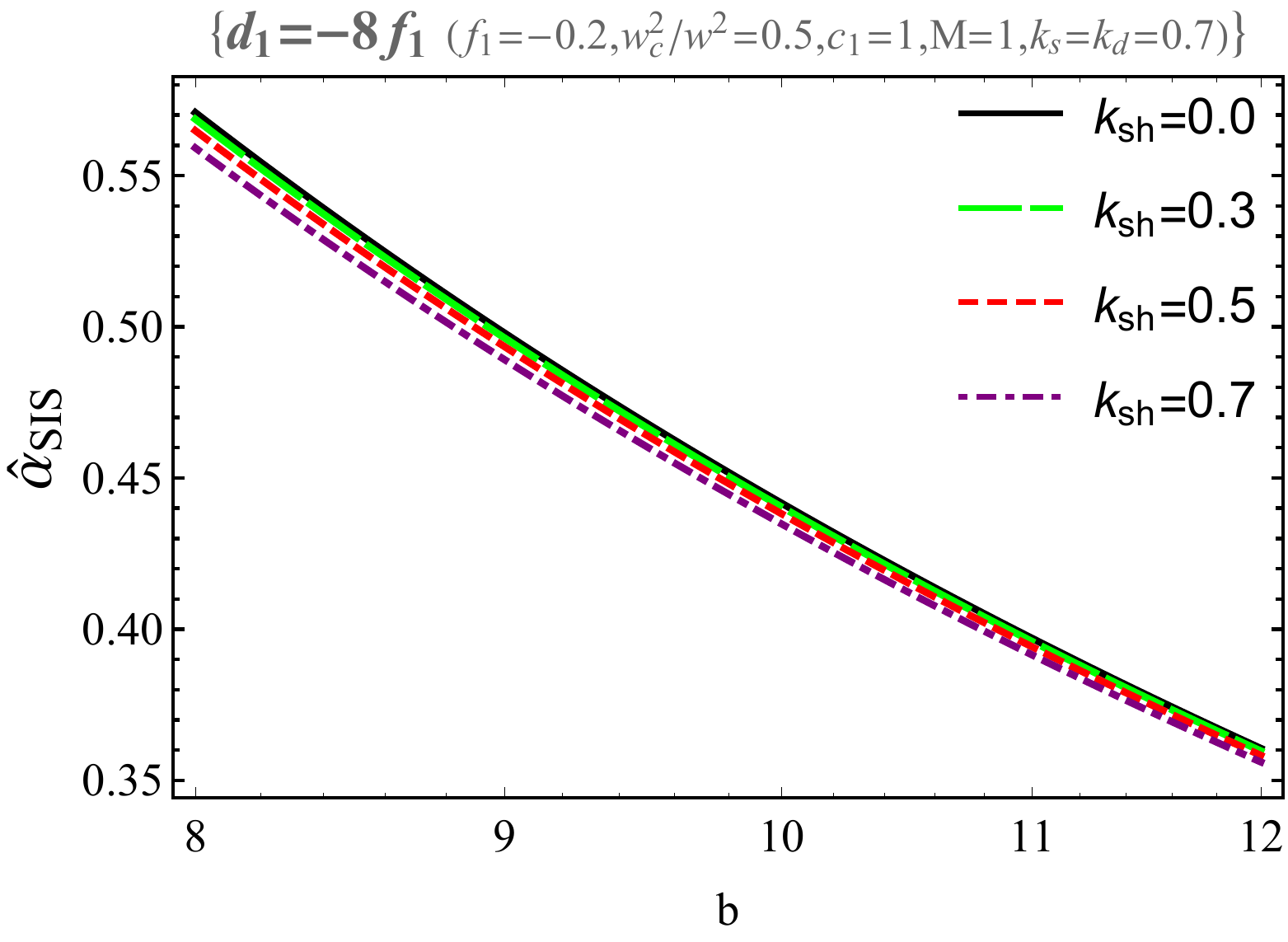}
    \includegraphics[scale=0.26]{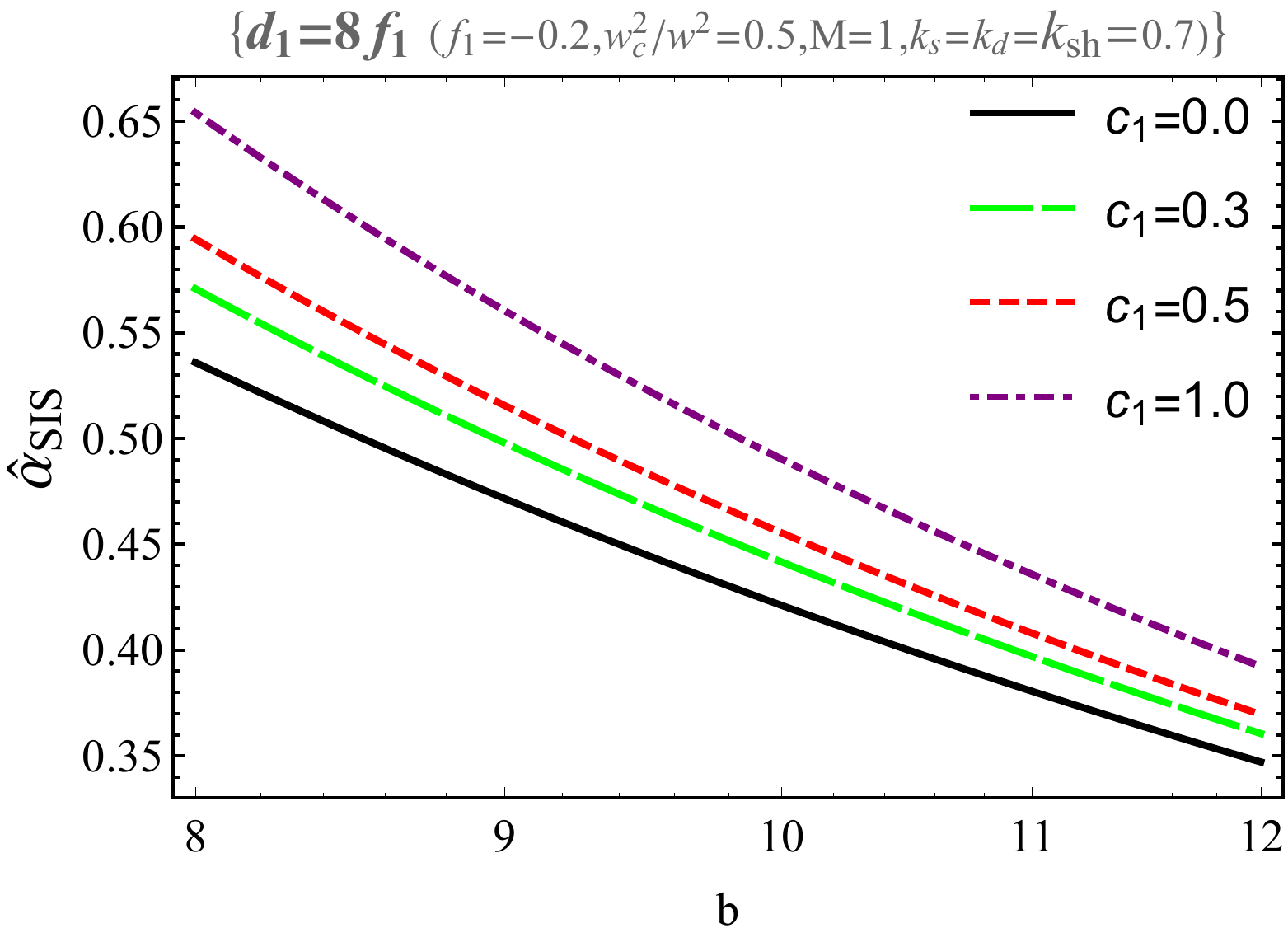}
    \includegraphics[scale=0.26]{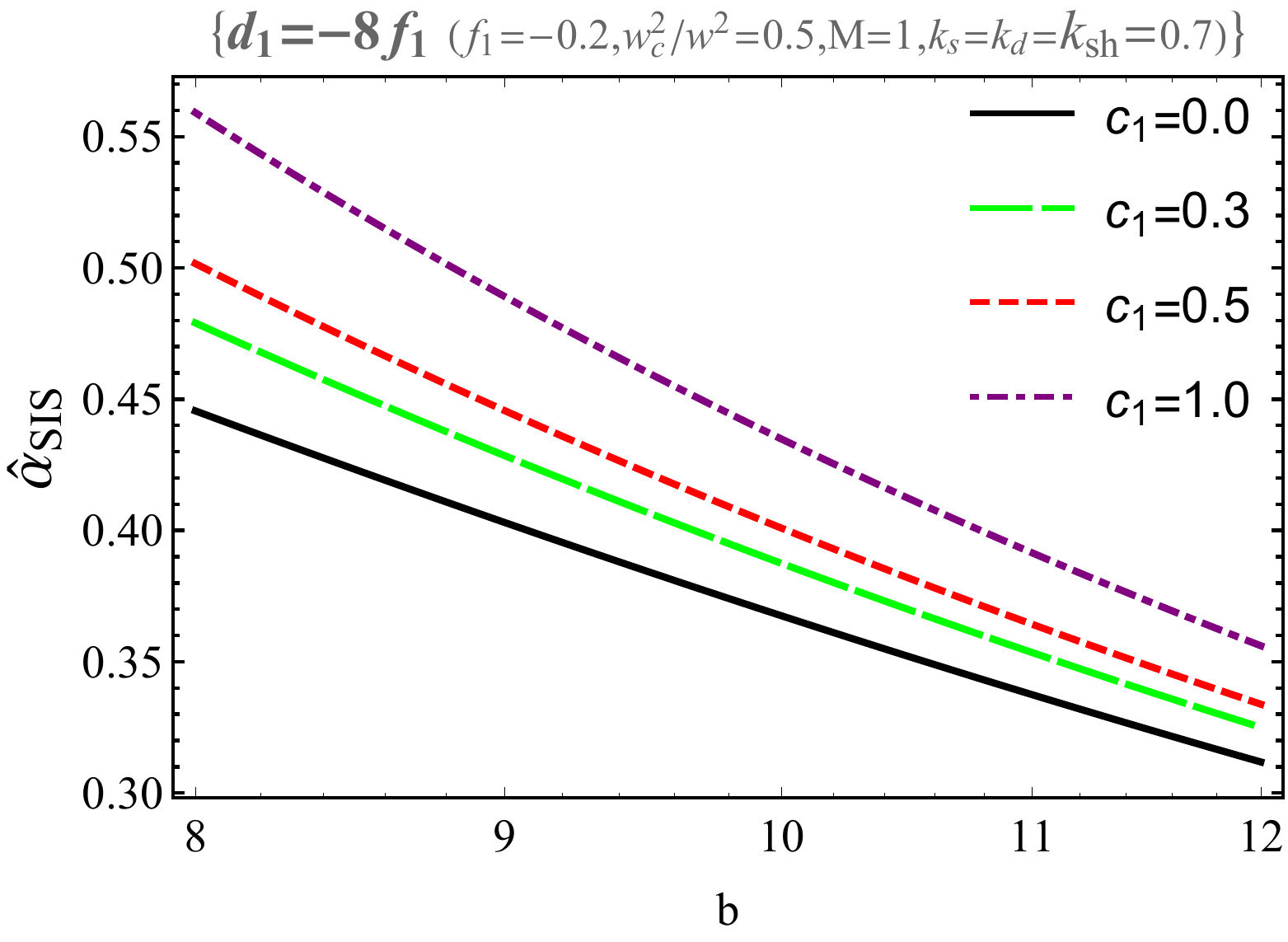}
    \includegraphics[scale=0.26]{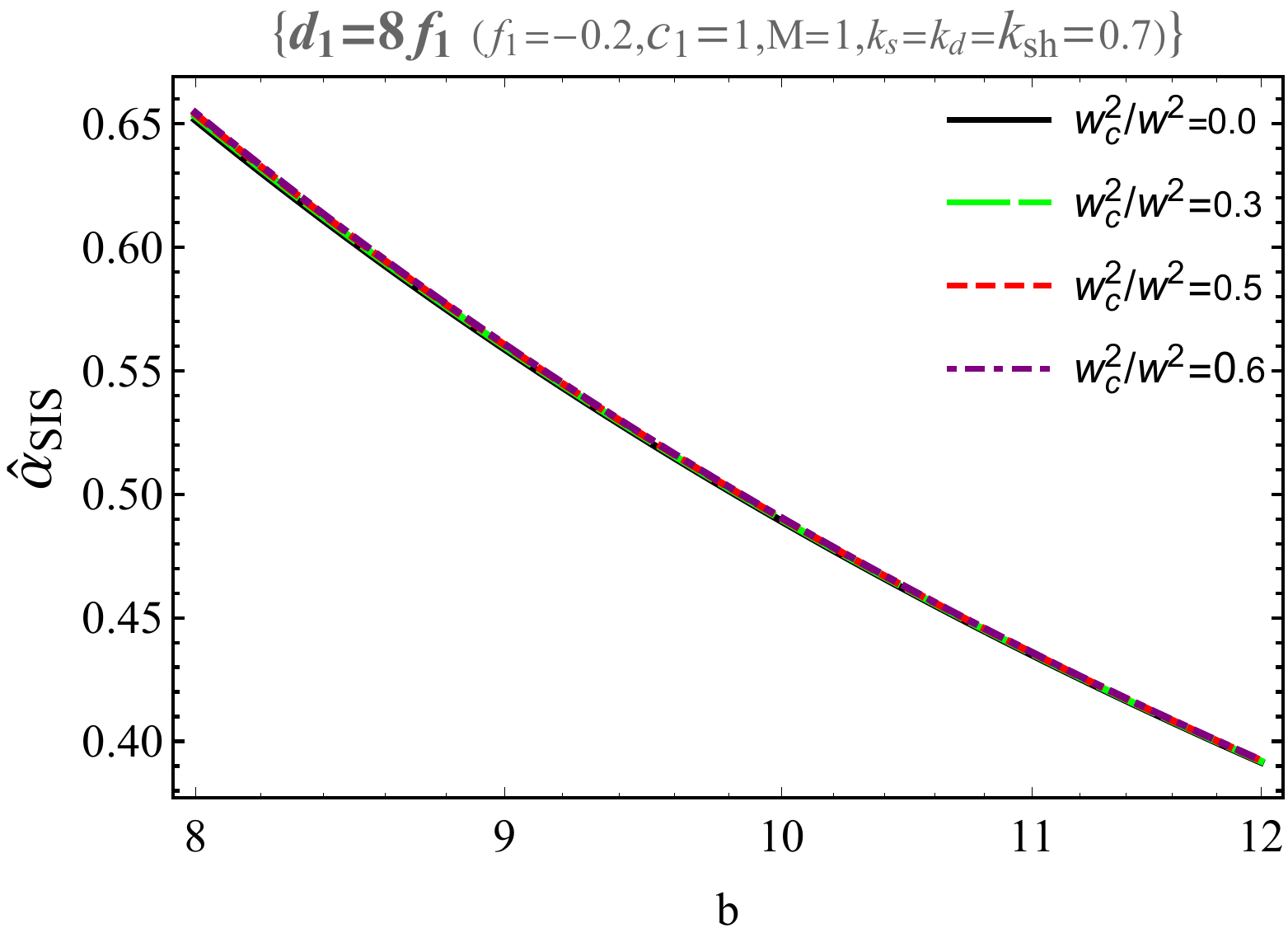}
    \includegraphics[scale=0.26]{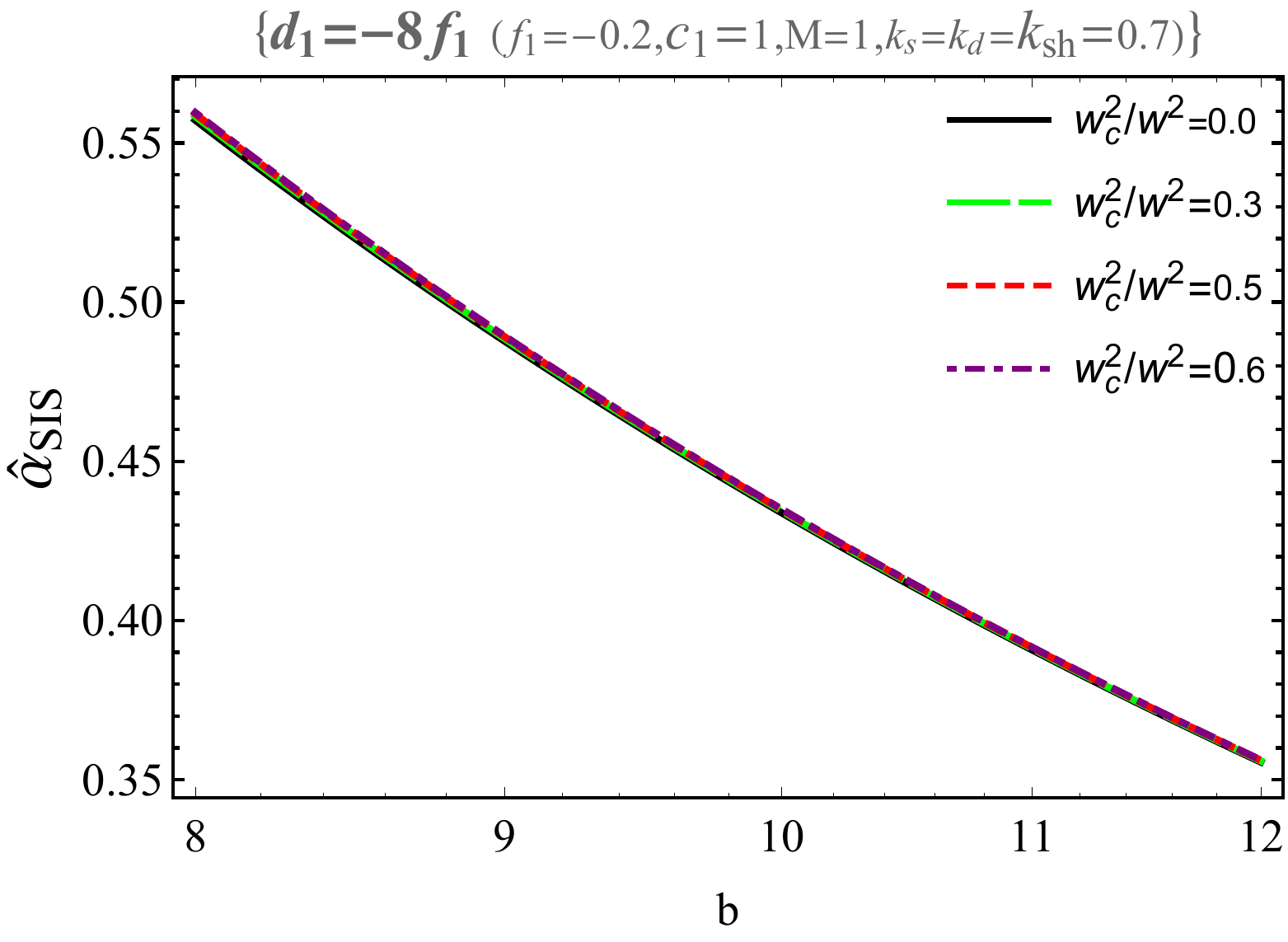}
    \includegraphics[scale=0.26]{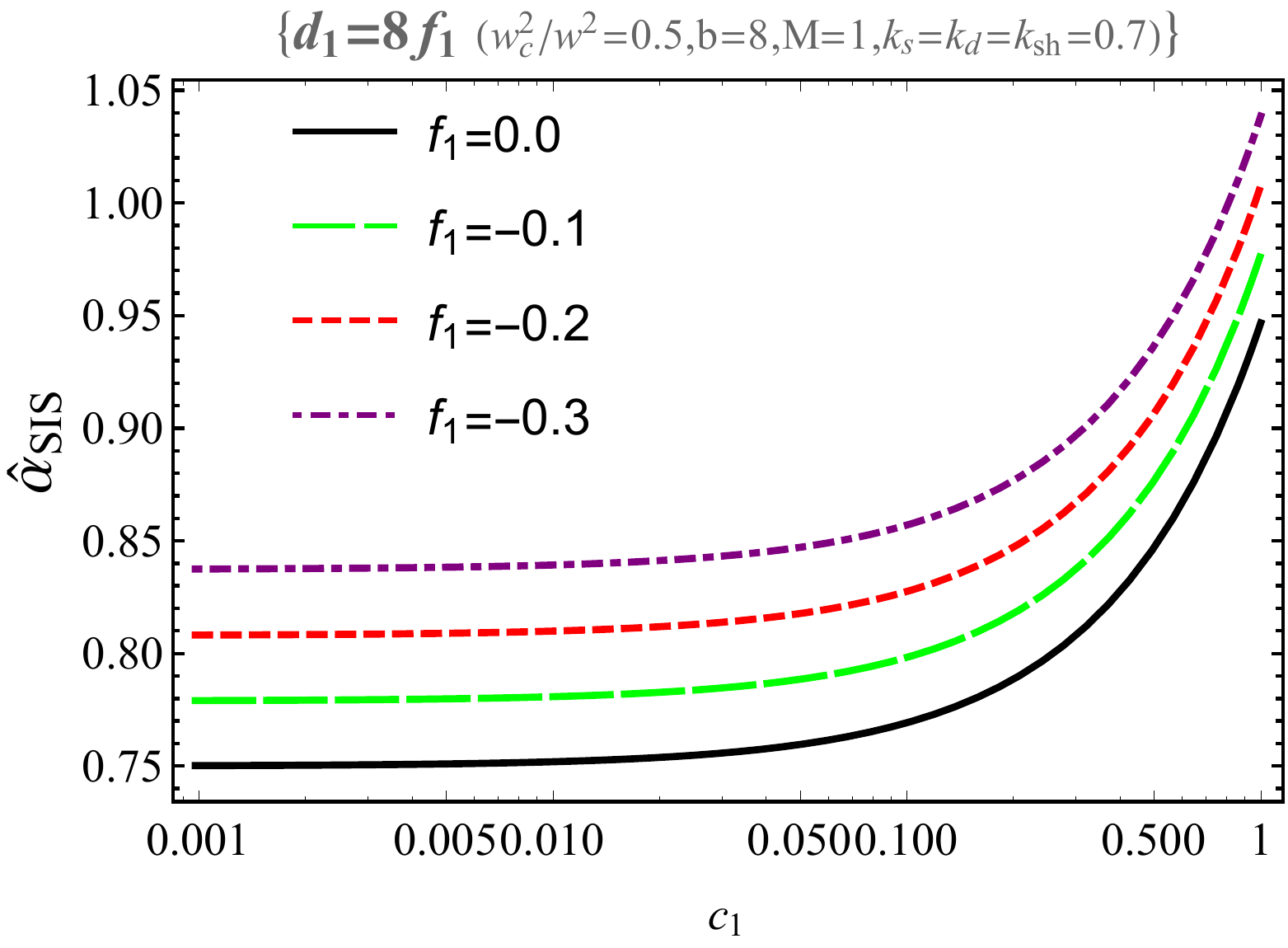}
    \includegraphics[scale=0.26]{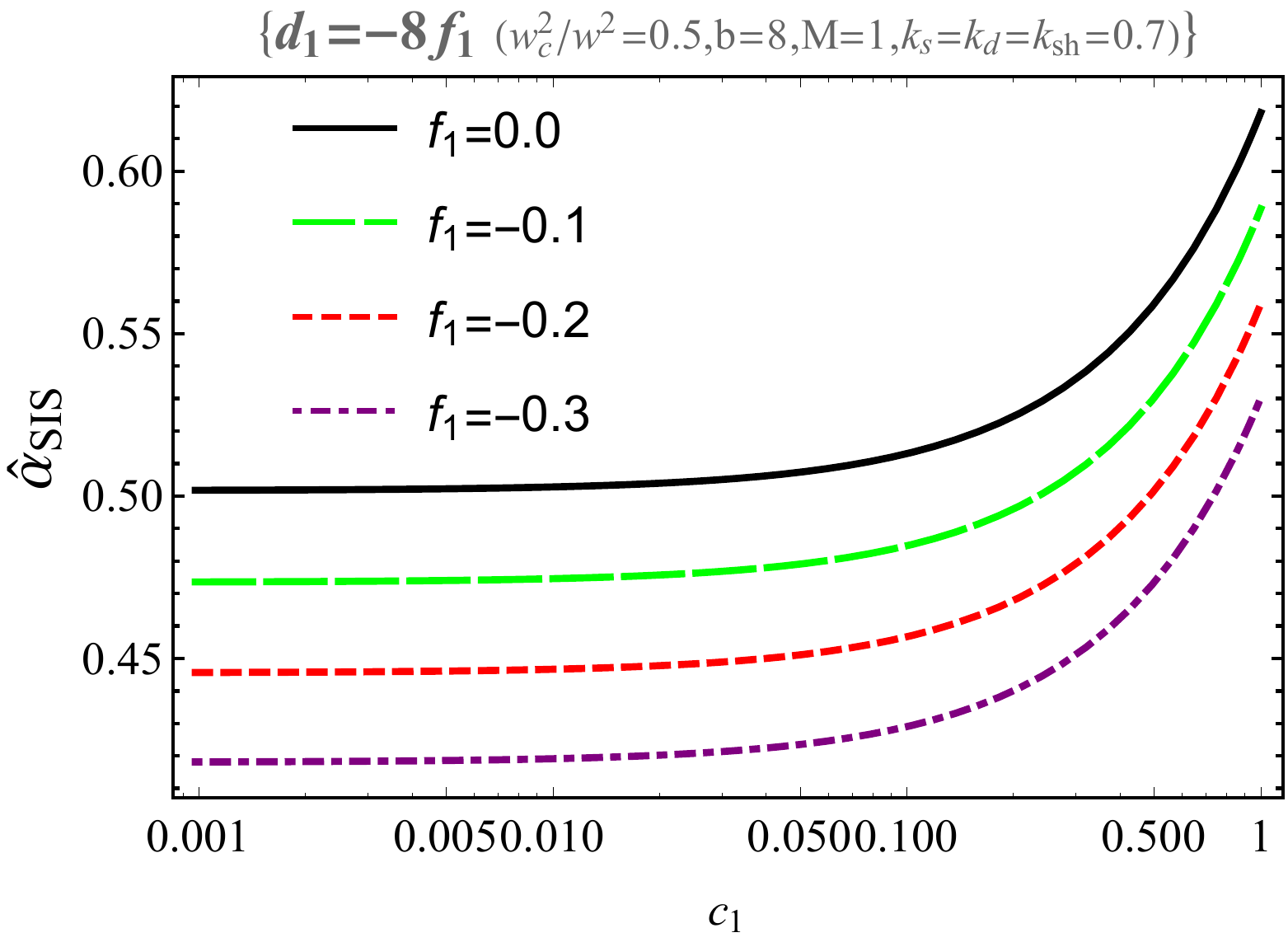}
    \includegraphics[scale=0.26]{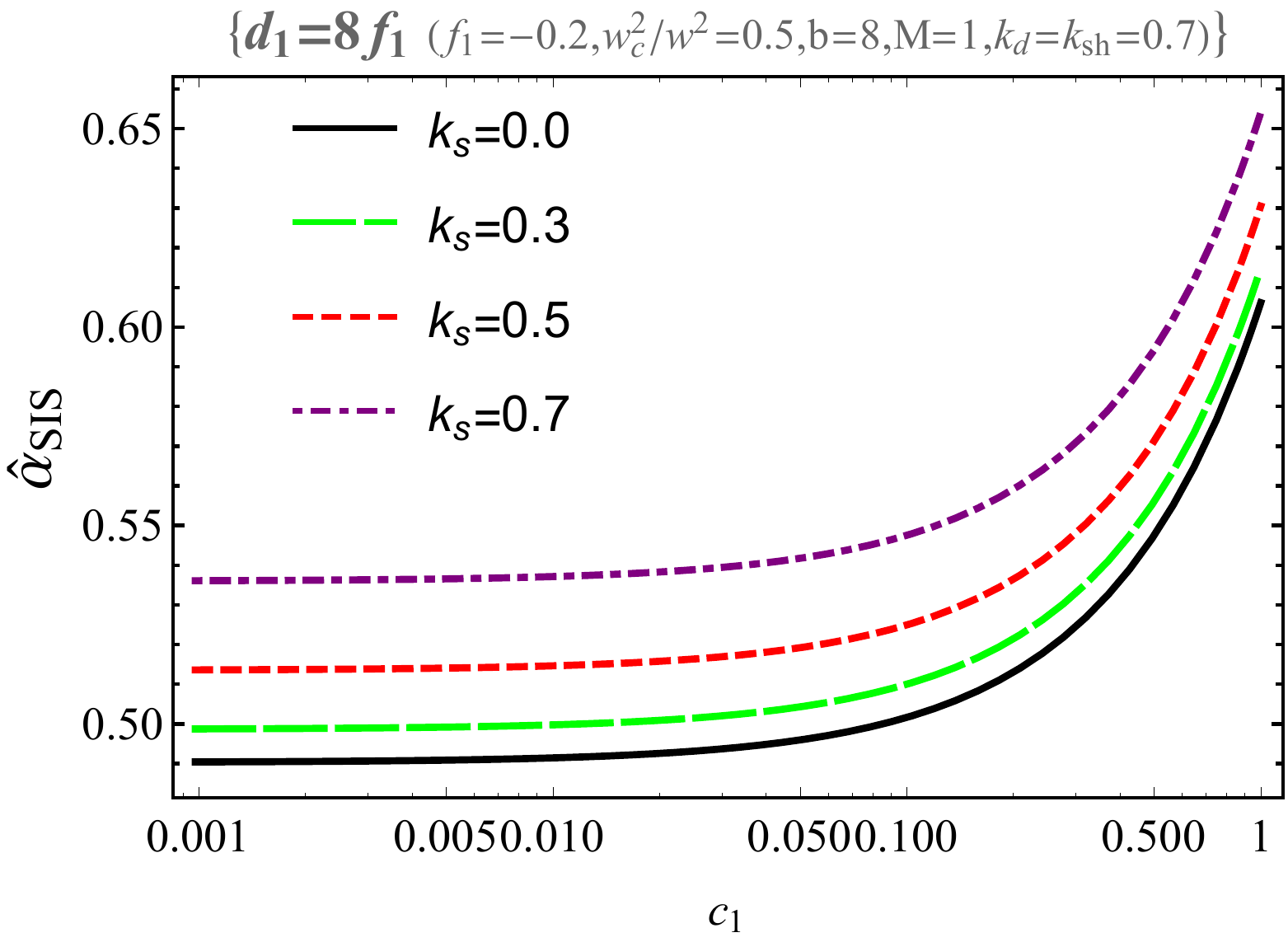}
    \includegraphics[scale=0.26]{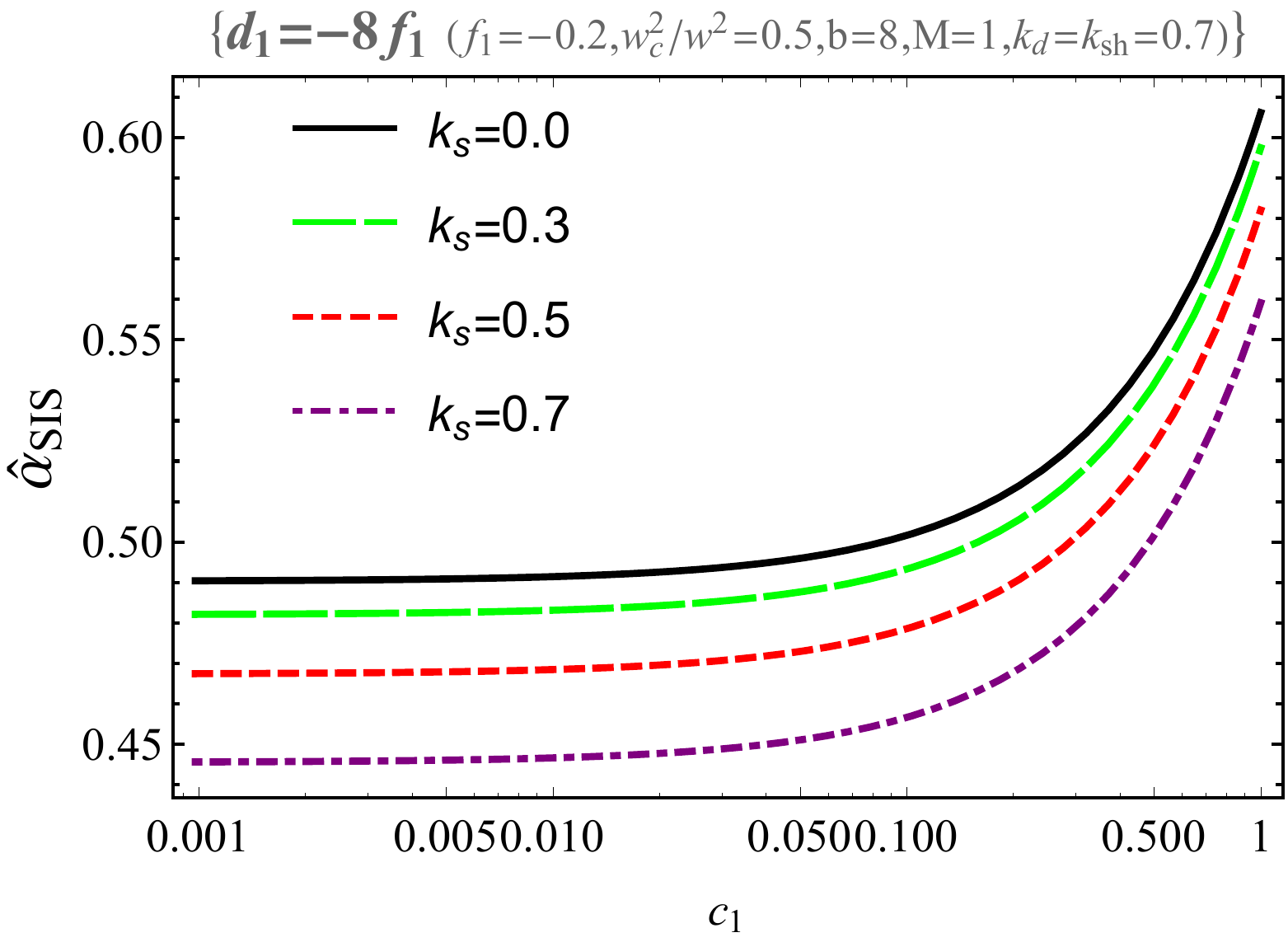}
    \includegraphics[scale=0.26]{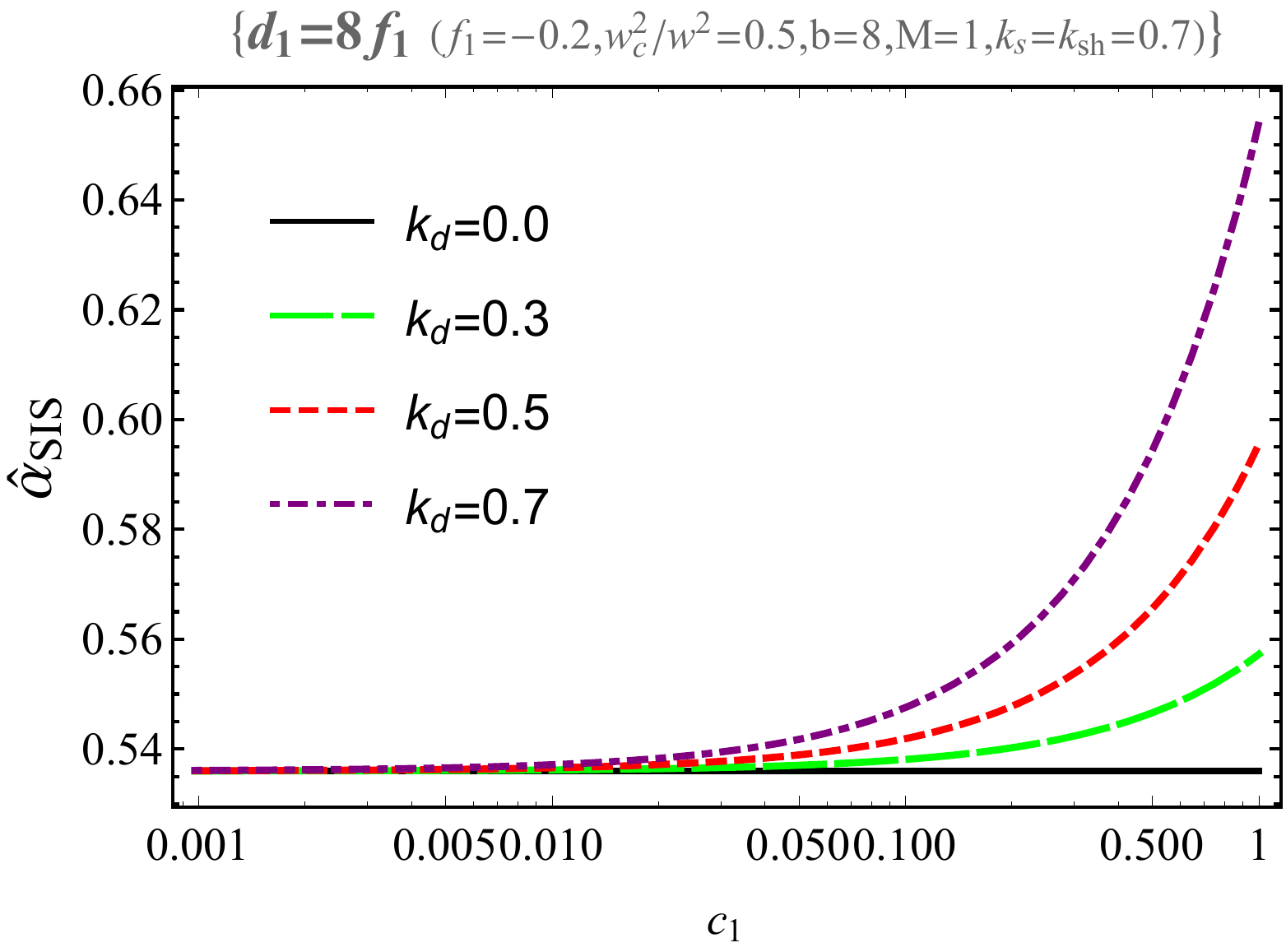}
    \includegraphics[scale=0.26]{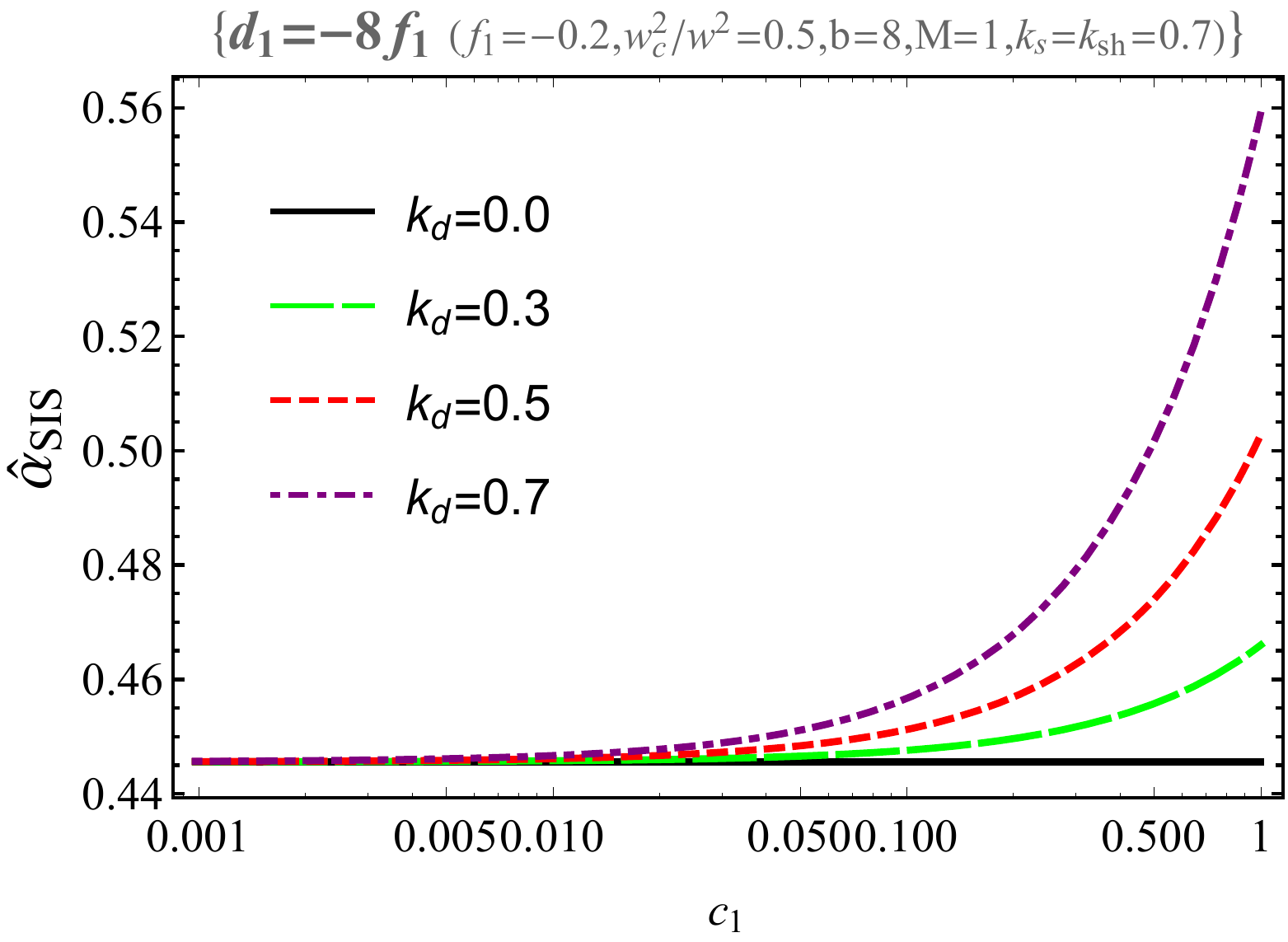}
    \includegraphics[scale=0.26]{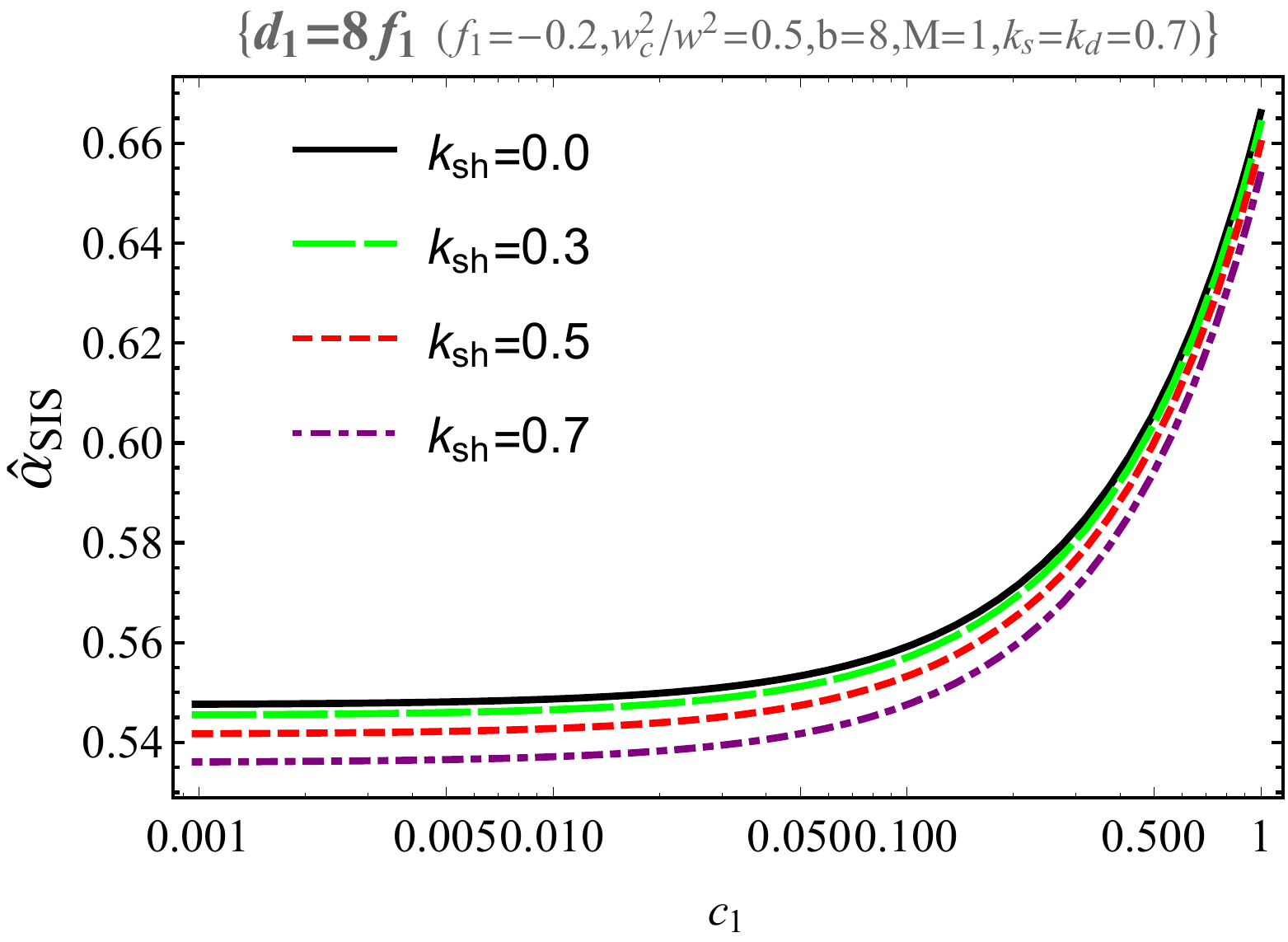}
    \includegraphics[scale=0.26]{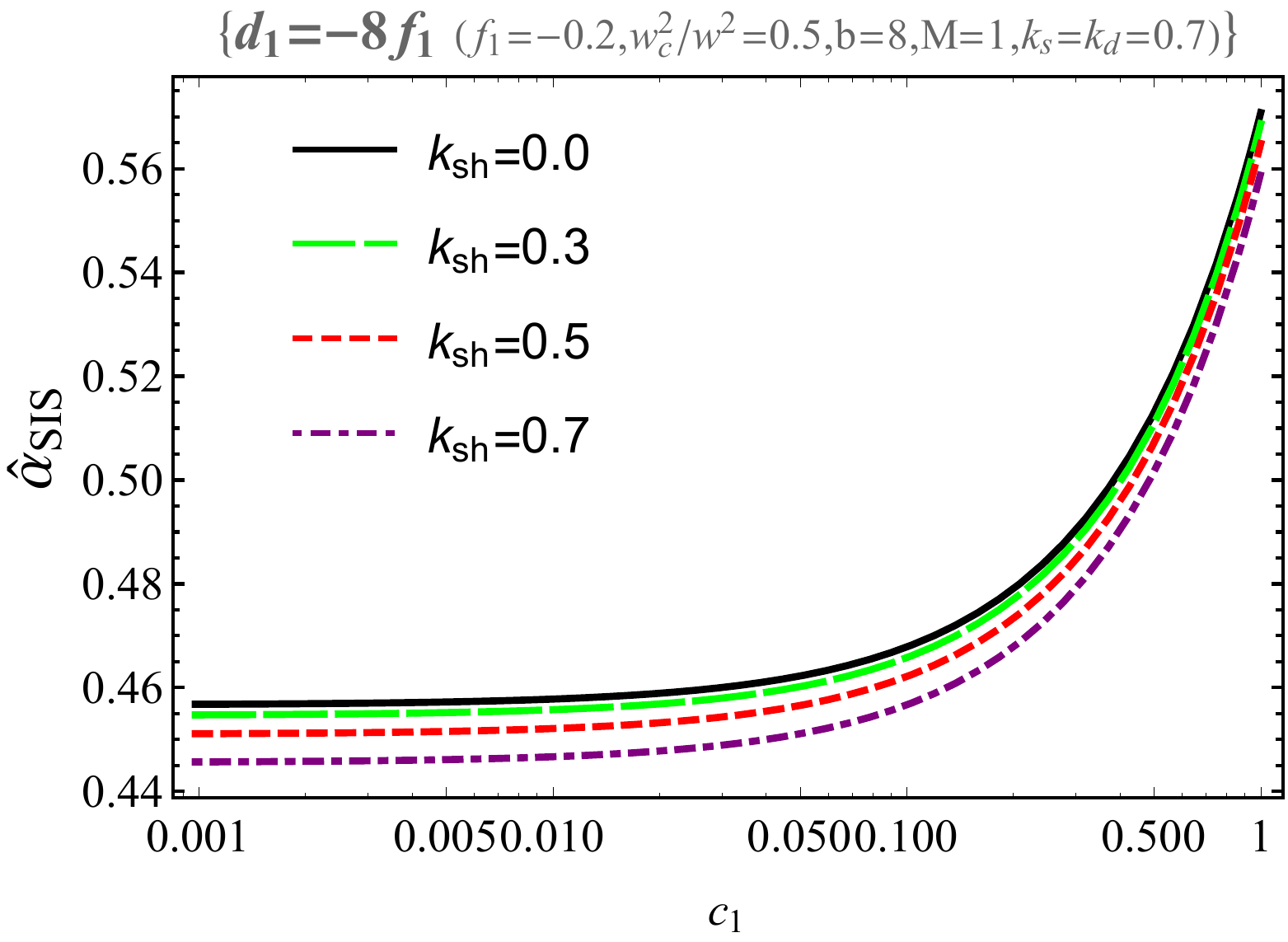}
    \includegraphics[scale=0.26]{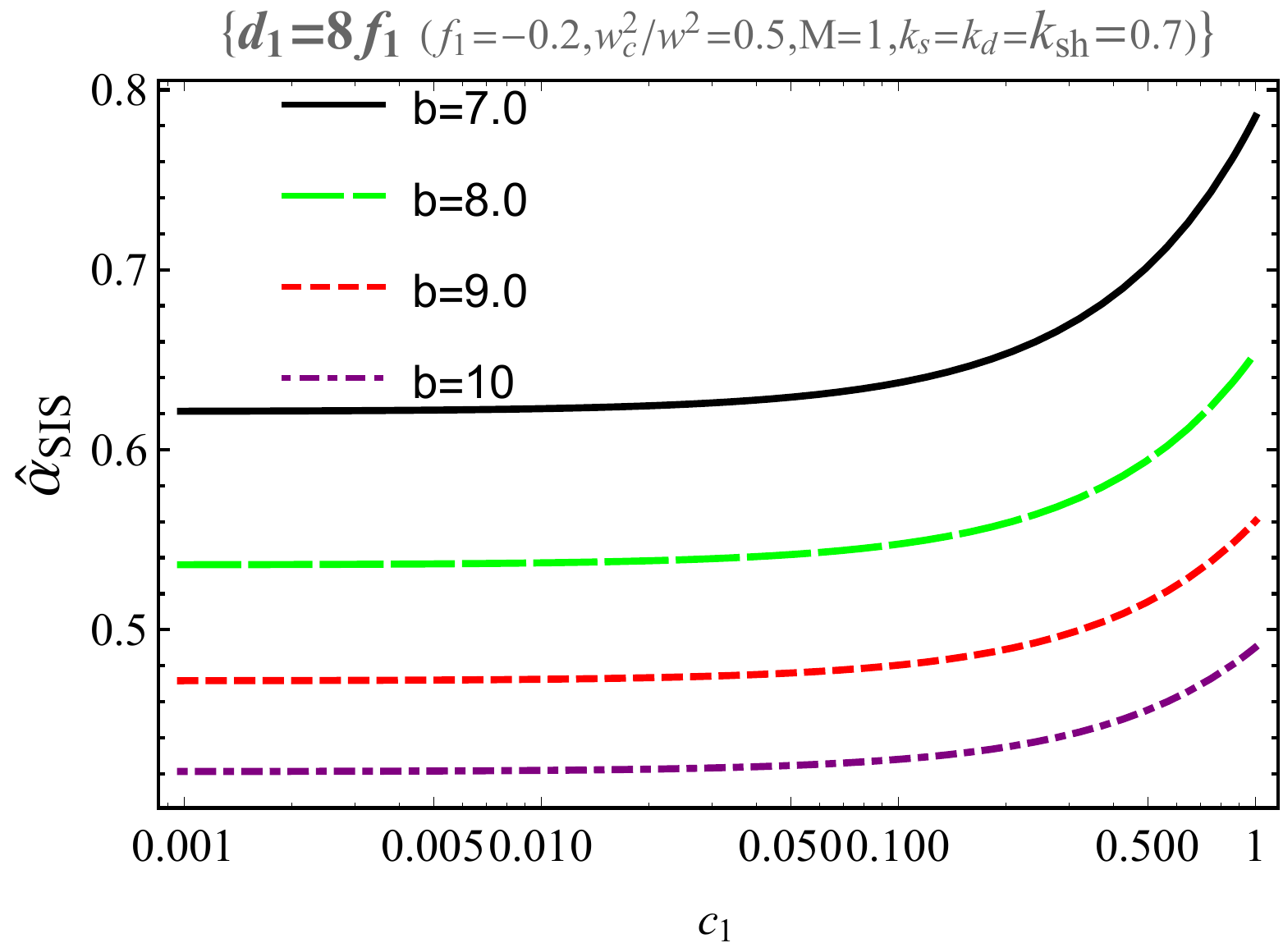}
    \includegraphics[scale=0.26]{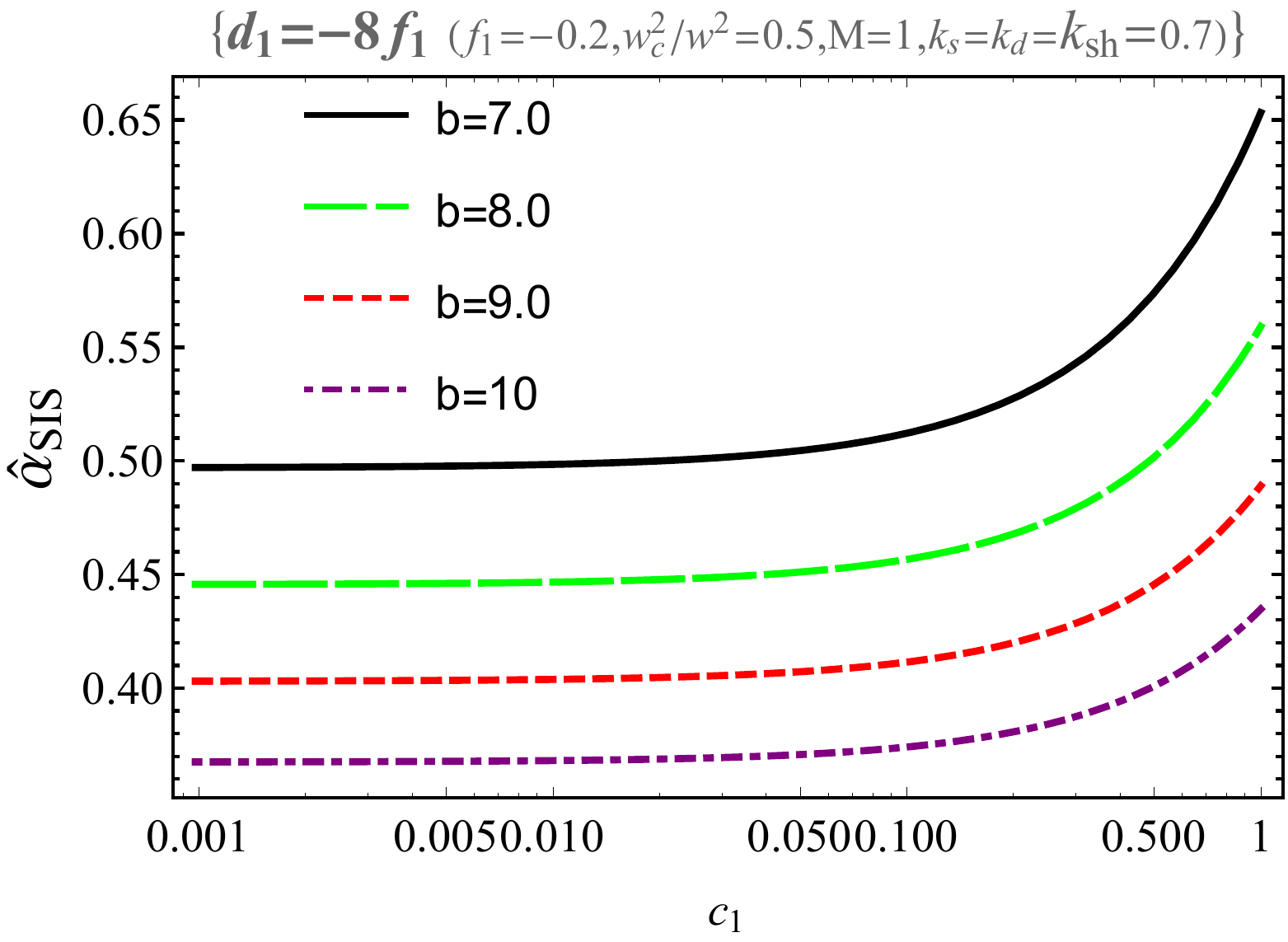}
    \includegraphics[scale=0.26]{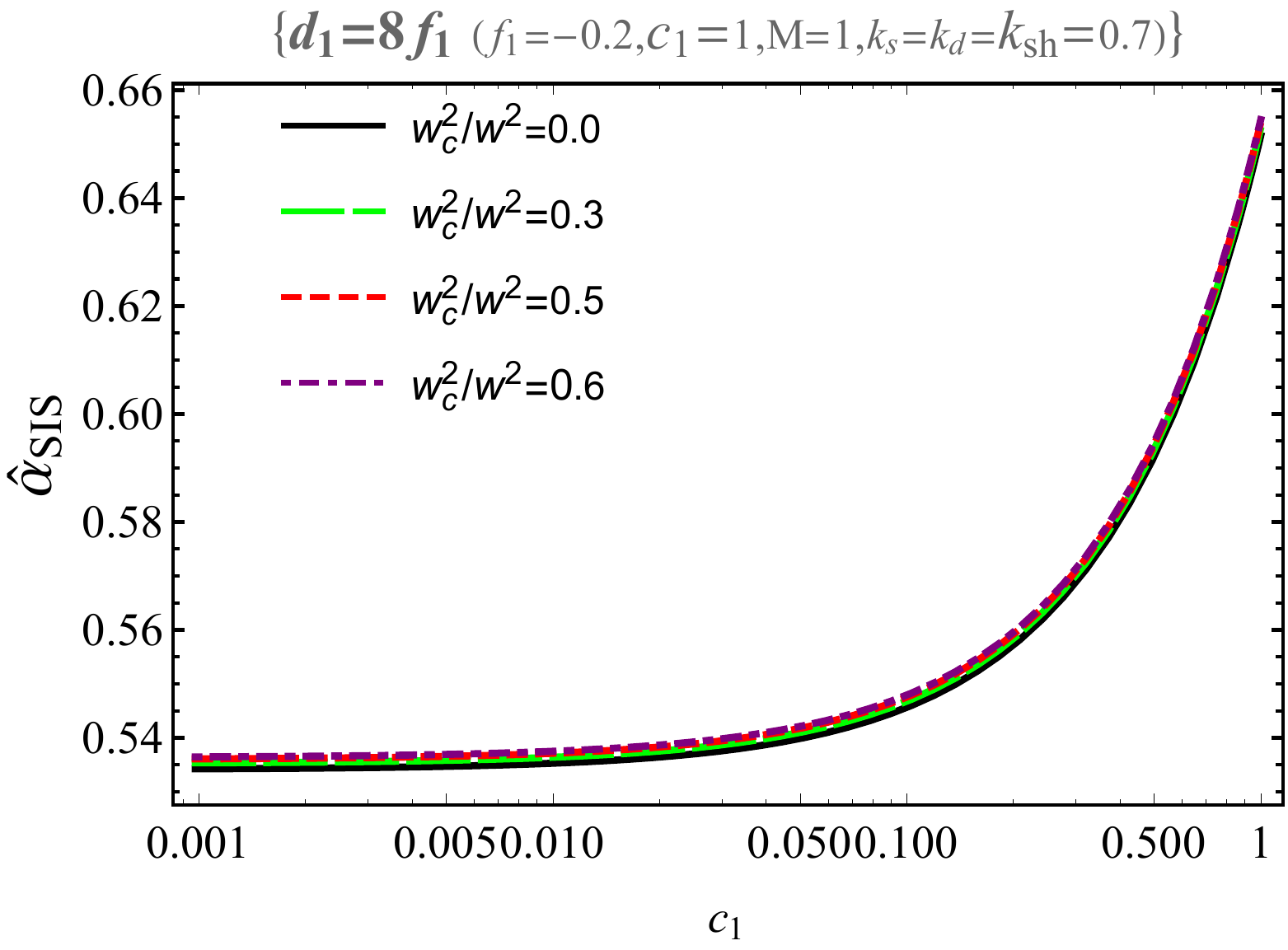}
    \includegraphics[scale=0.26]{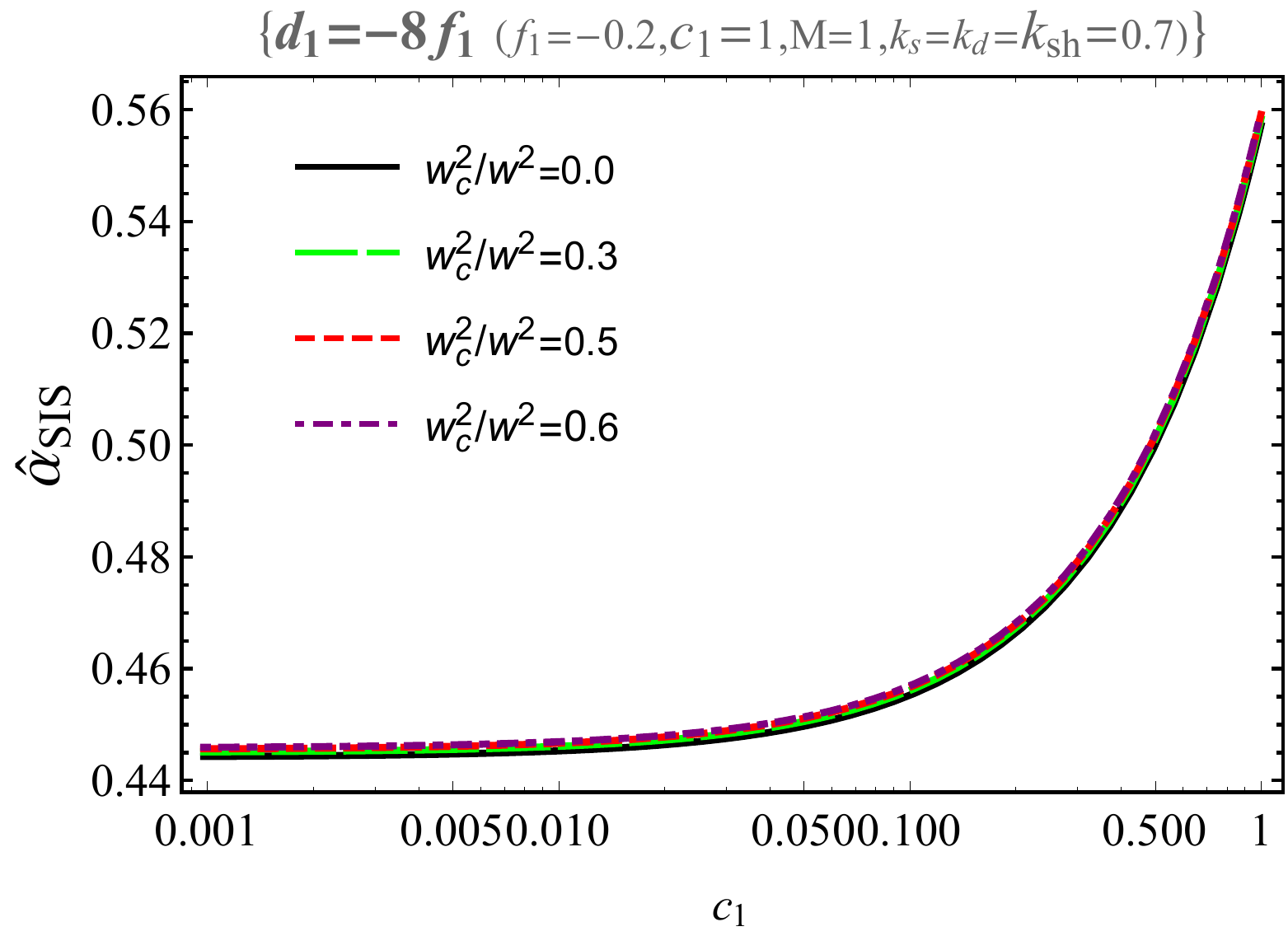}
    \caption{The deflection angle $\hat{\alpha}_{SIS}$ in $SIS$ plasma for $d_1=8f_1$ (Left panel) and $d_1=-8f_1$ (Right panel) along $c_1$ taking different values of $f_1,\; k_s,\; k_d, \;\&\; k_{sh}.$}
    \label{plot:15}
    \end{figure}

\begin{figure}
    \centering
    \includegraphics[scale=0.26]{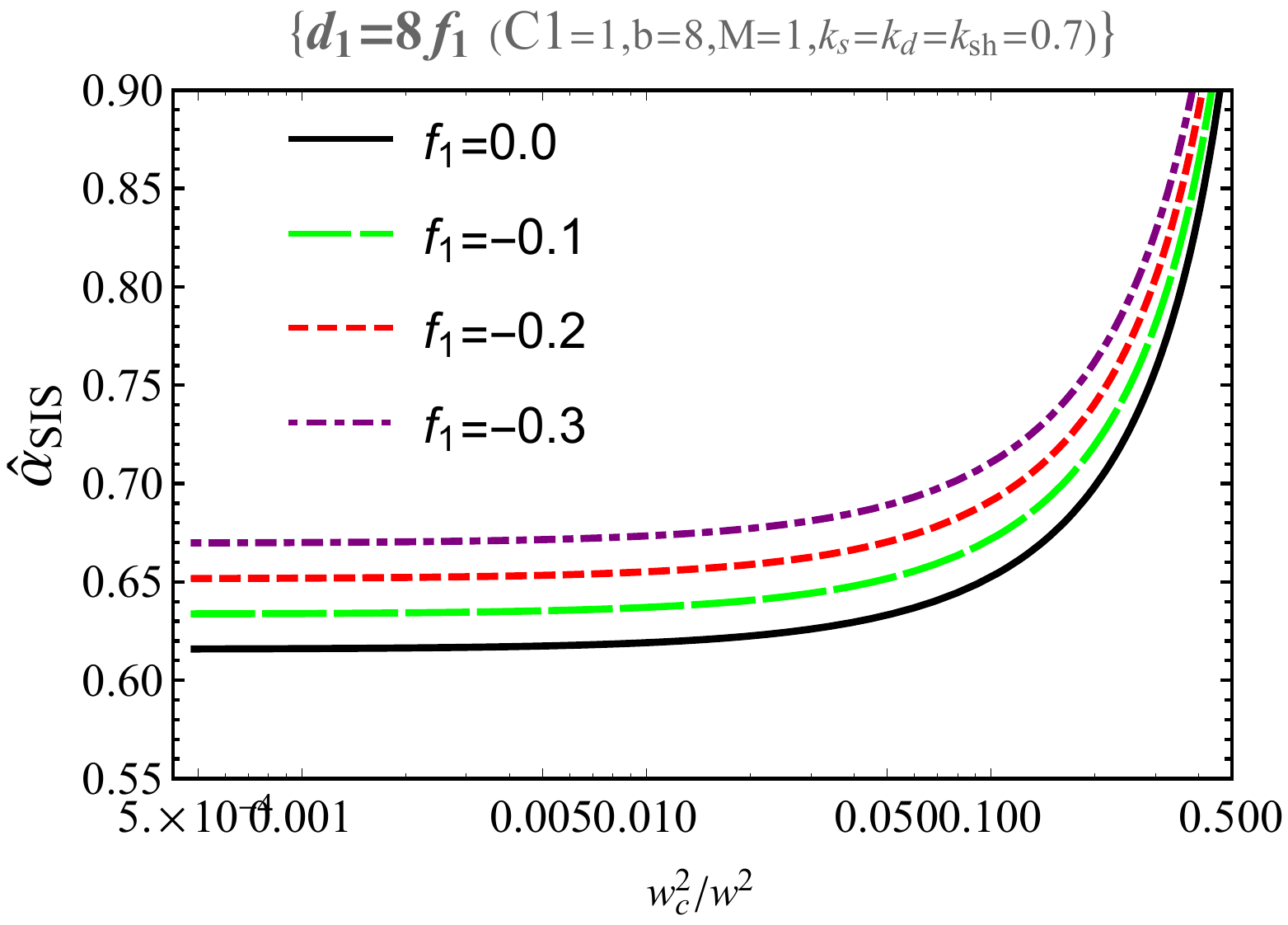}
    \includegraphics[scale=0.26]{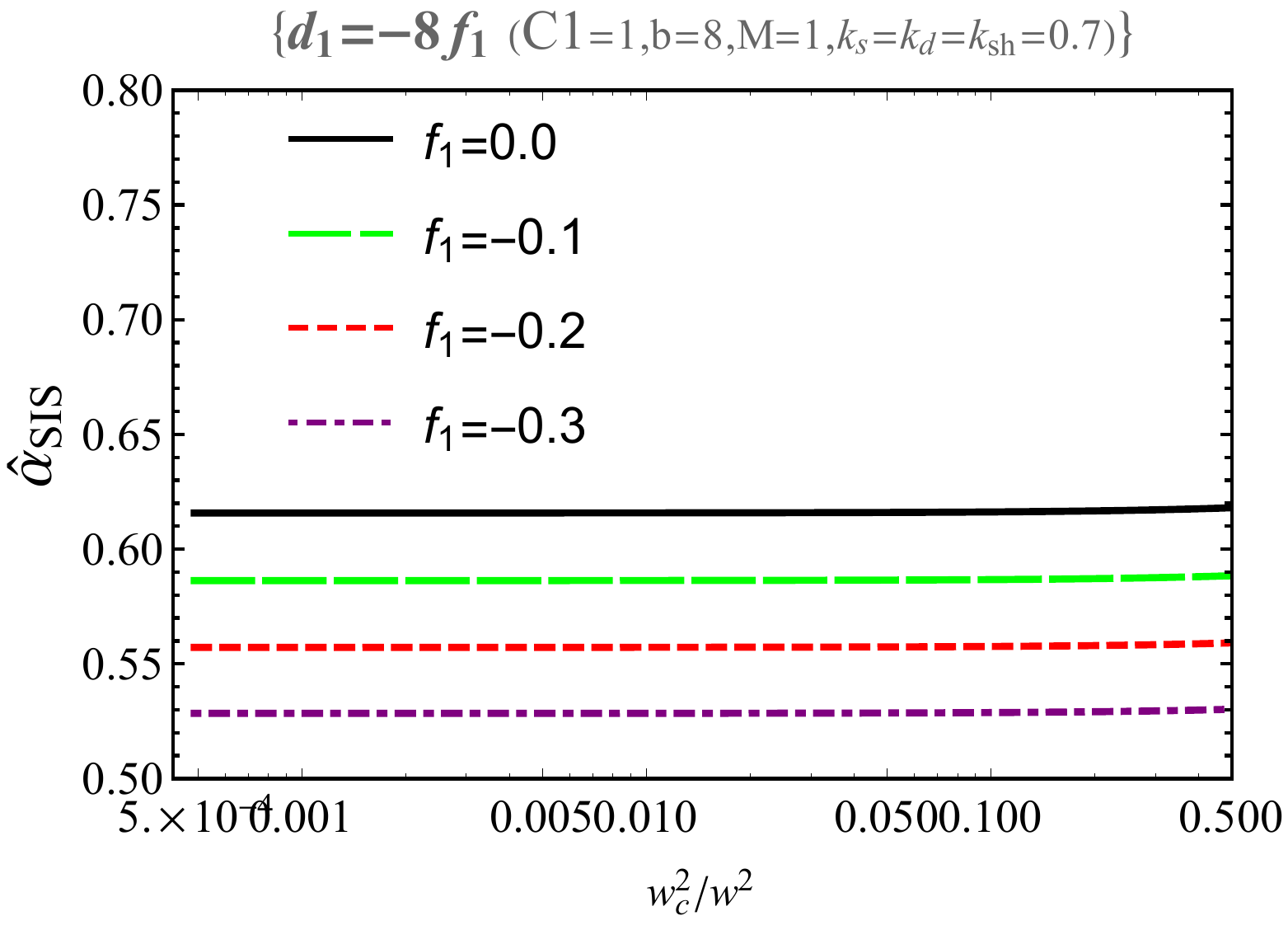}
    \includegraphics[scale=0.26]{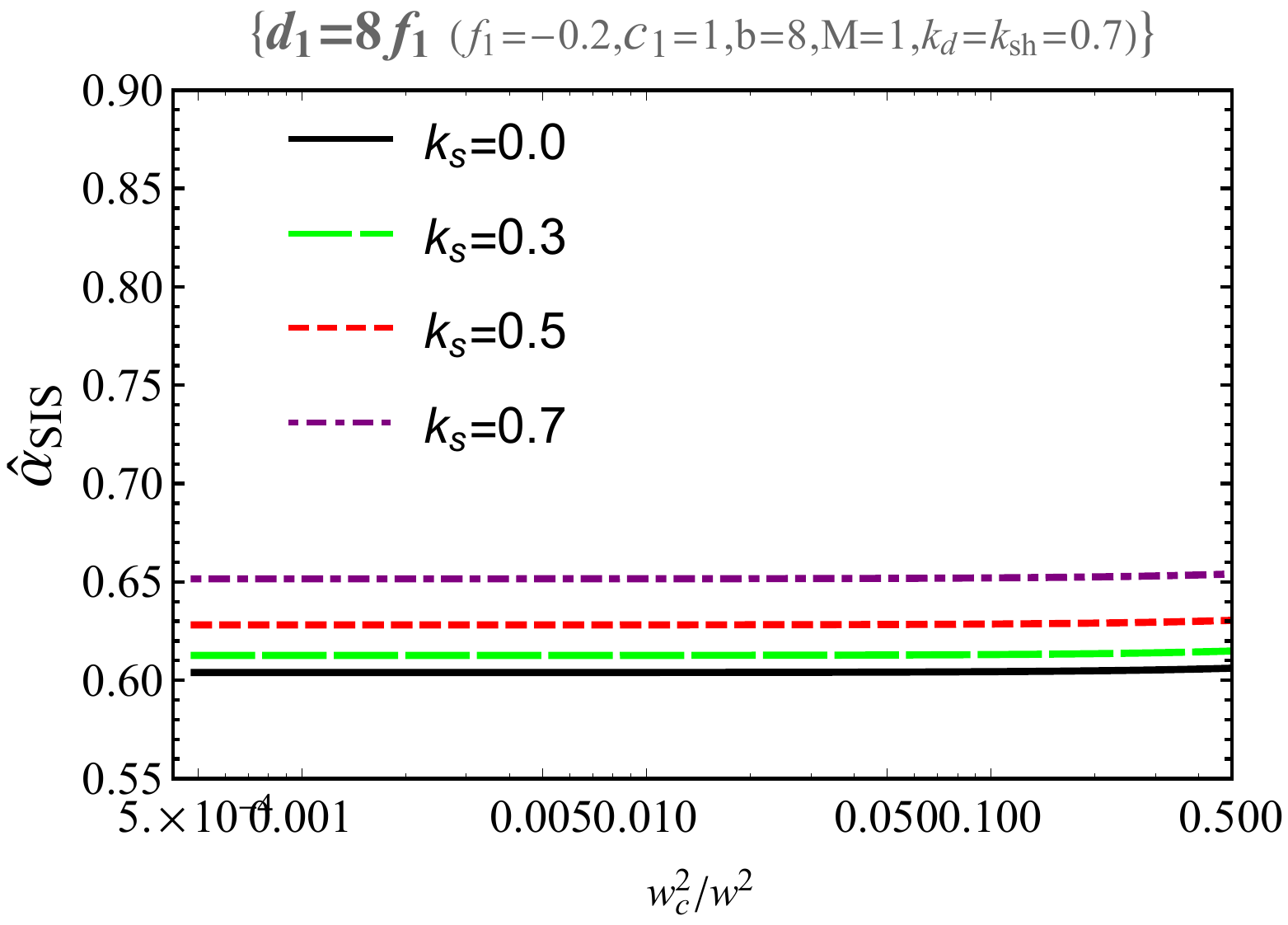}
    \includegraphics[scale=0.26]{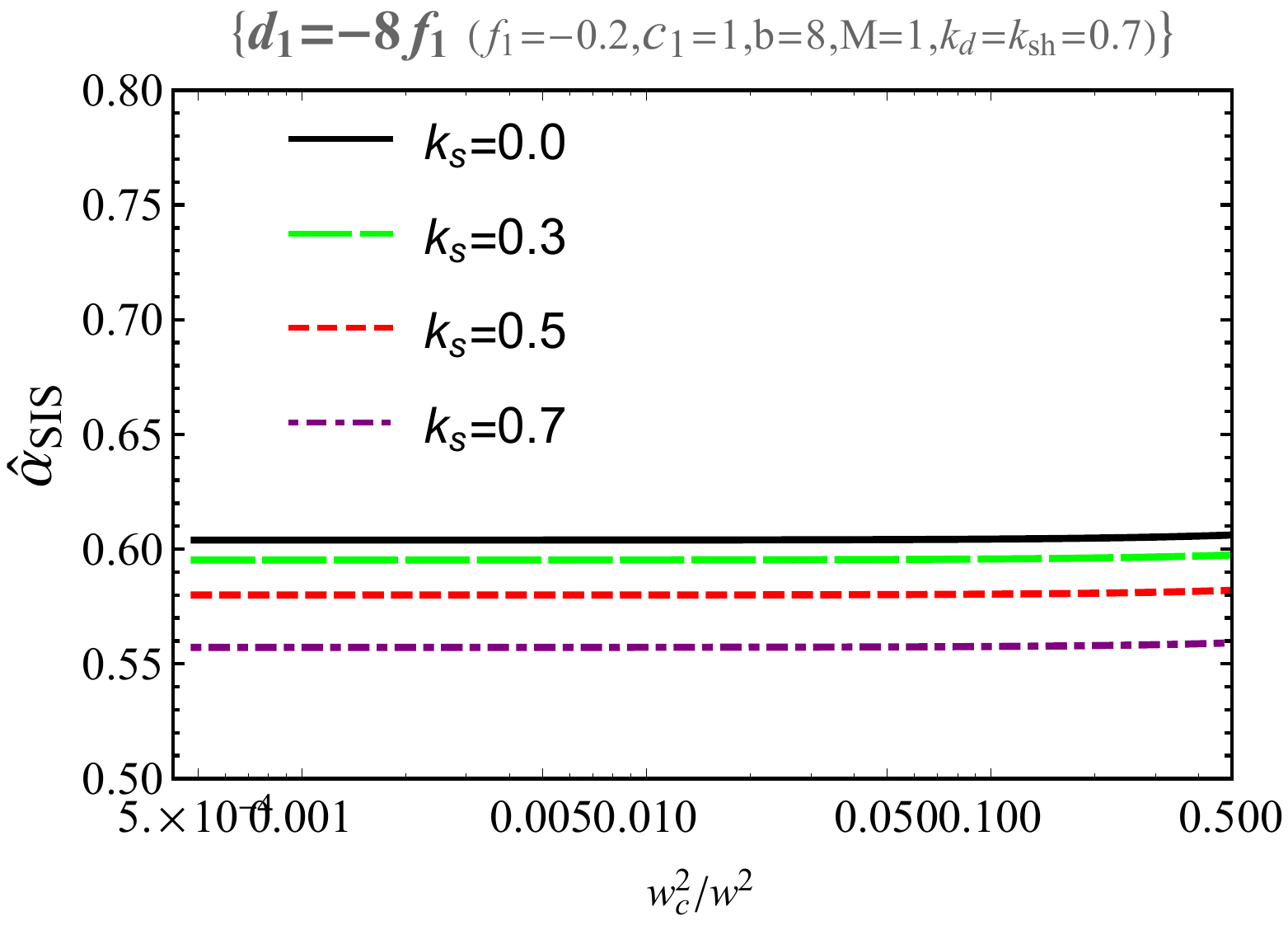}
    \includegraphics[scale=0.26]{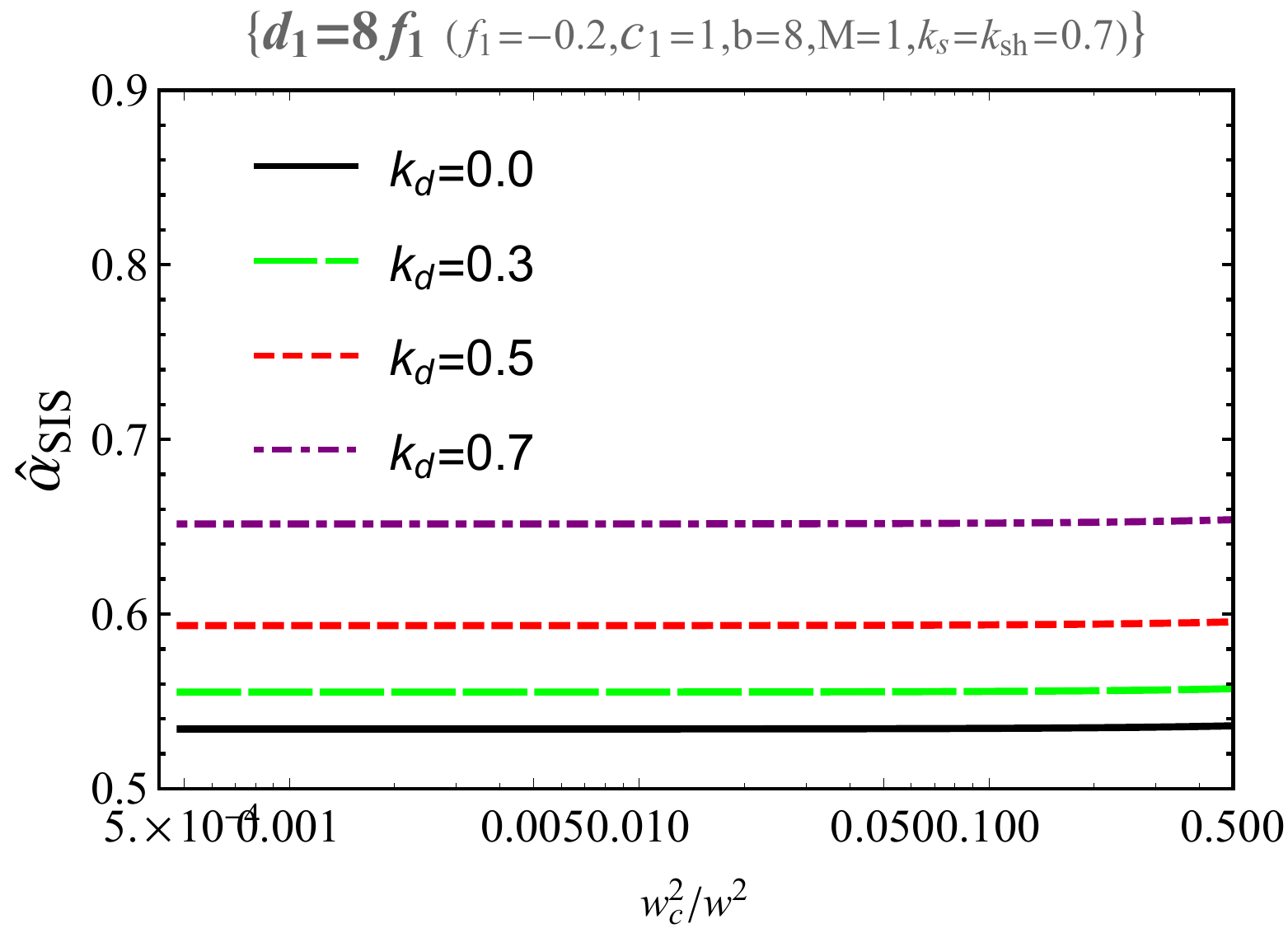}
    \includegraphics[scale=0.26]{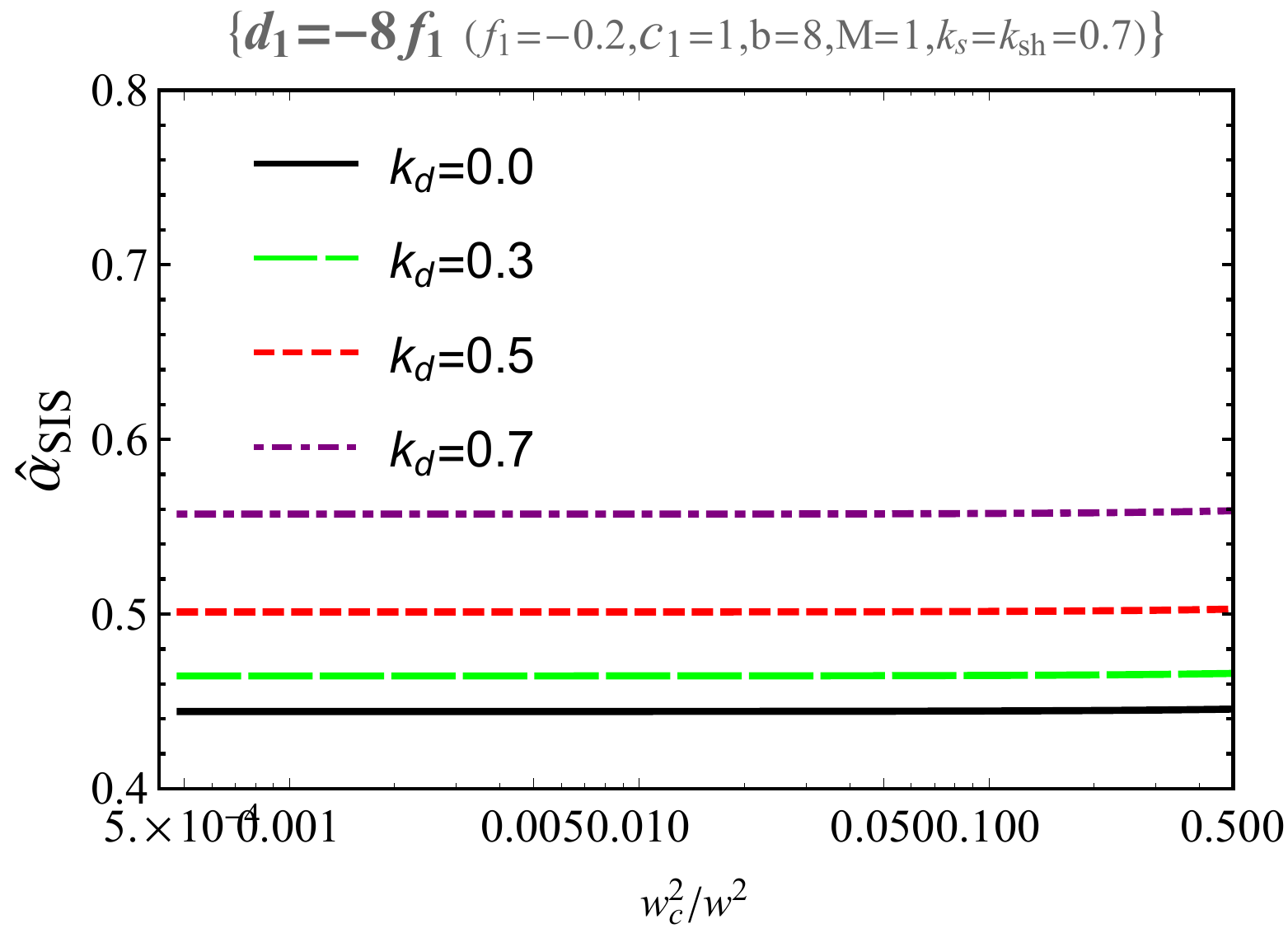}
    \includegraphics[scale=0.26]{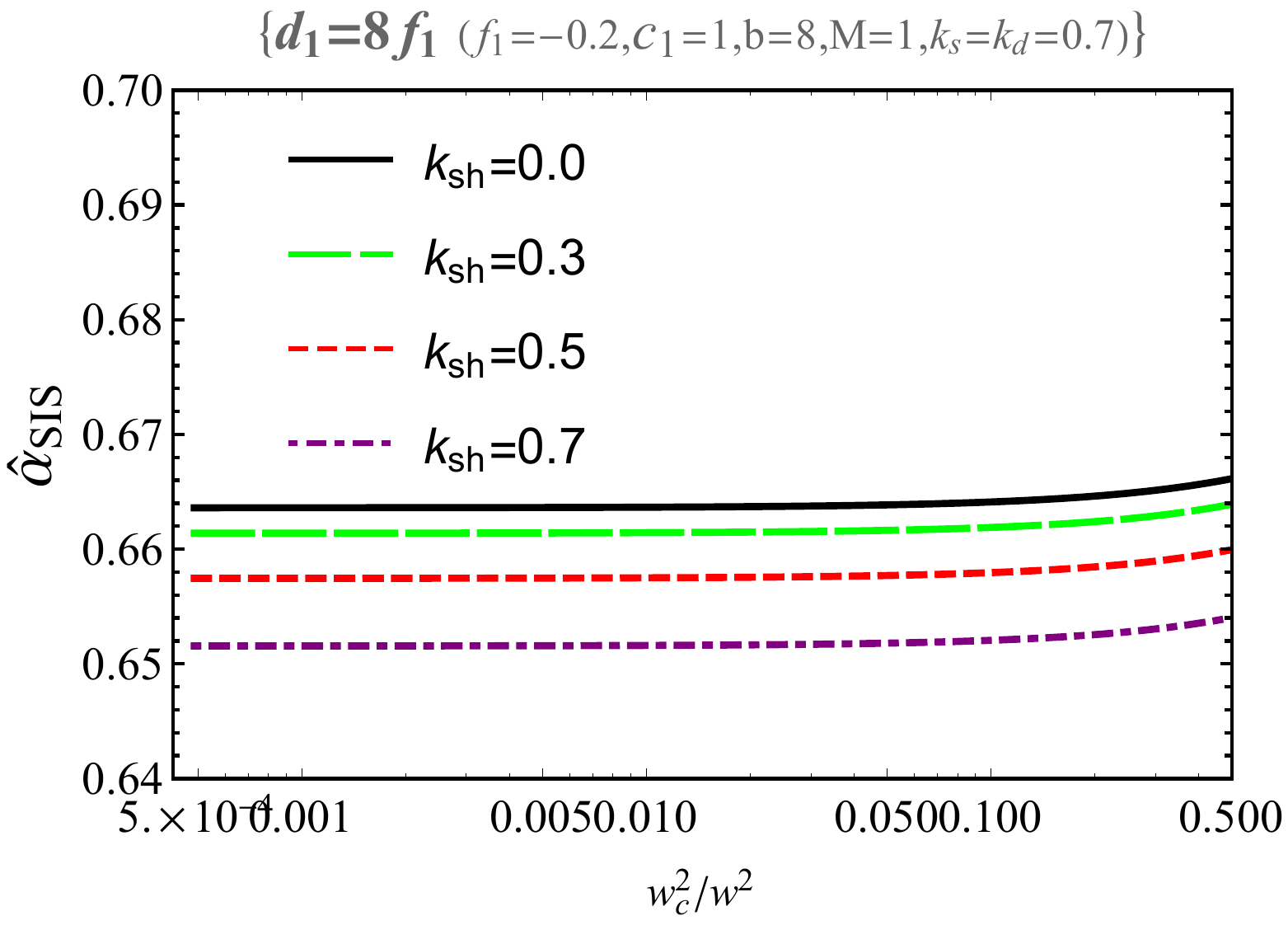}
    \includegraphics[scale=0.26]{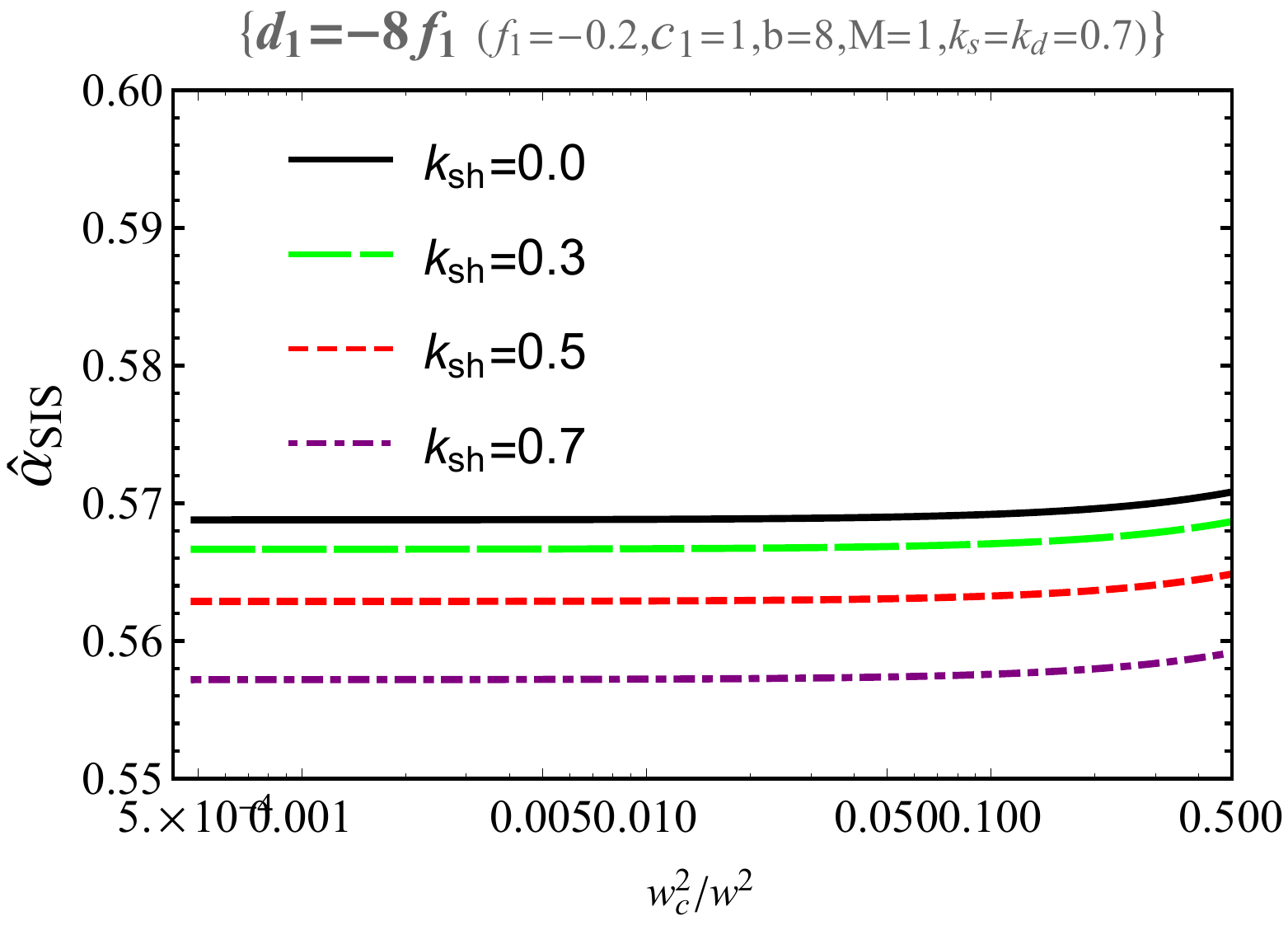}
    \includegraphics[scale=0.26]{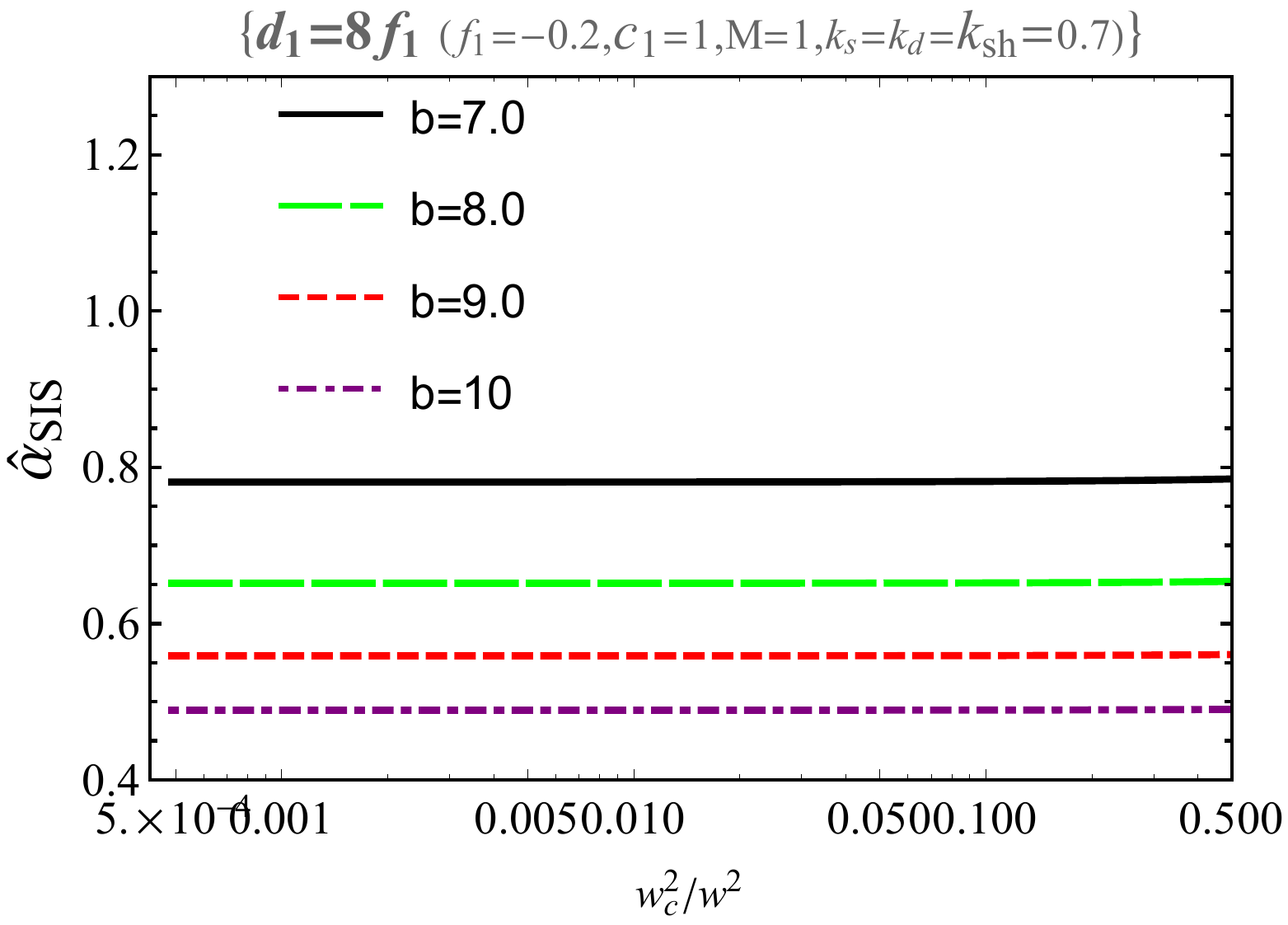}
    \includegraphics[scale=0.26]{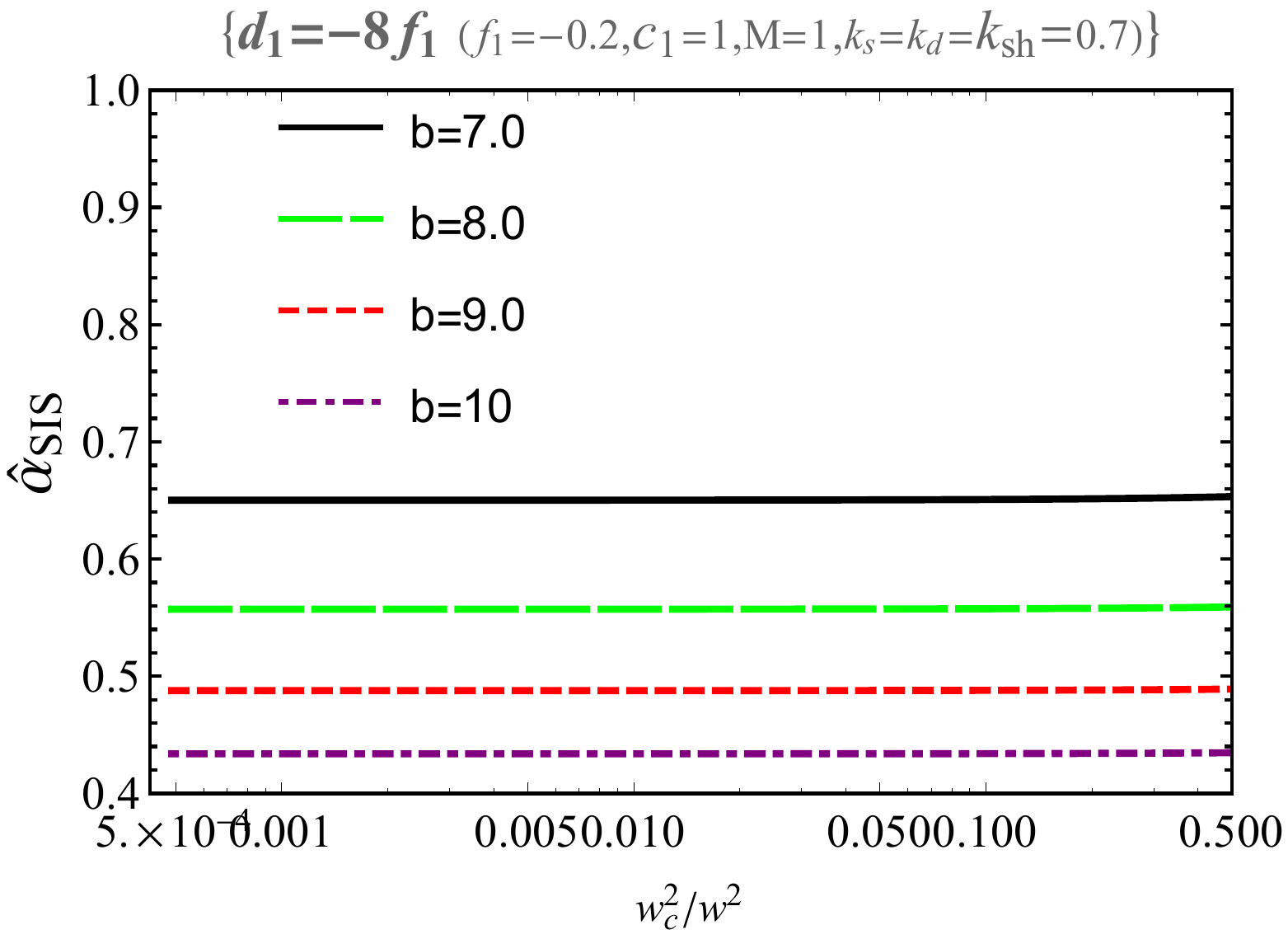}
    \includegraphics[scale=0.26]{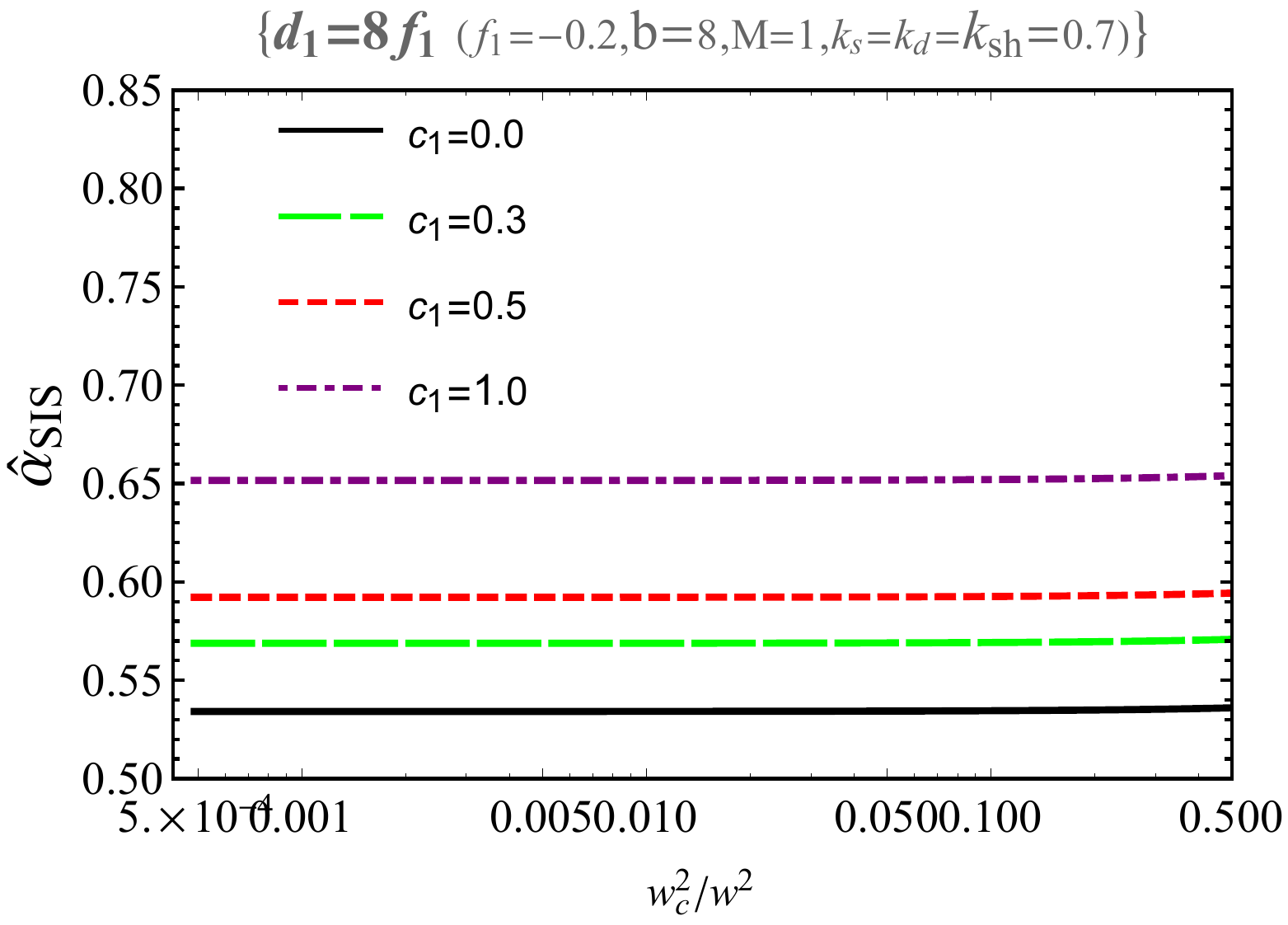}
    \includegraphics[scale=0.26]{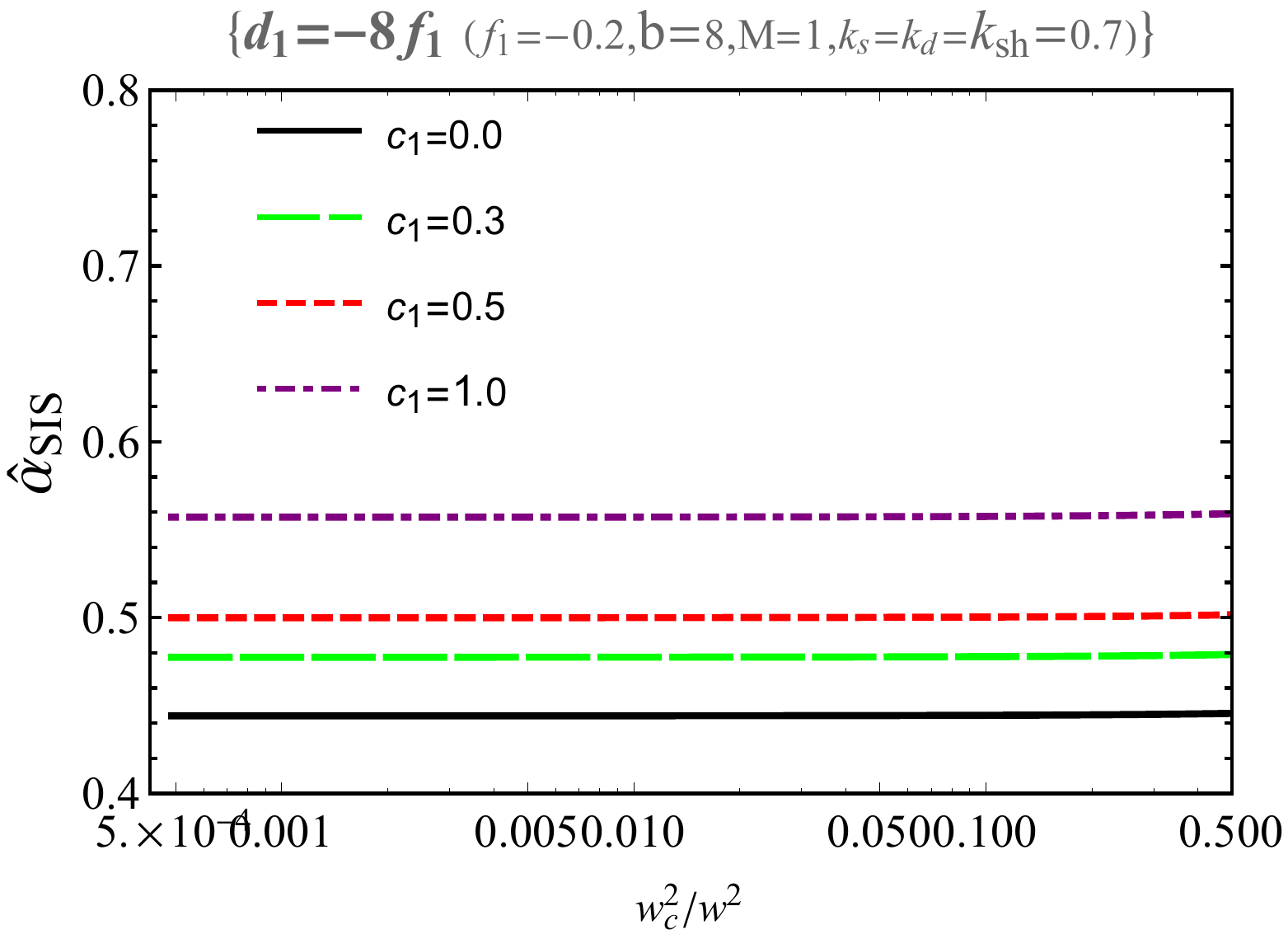}
    \caption{The deflection angle $\hat{\alpha}_{SIS}$ in $SIS$ plasma for $d_1=8f_1$ (Left panel) and $d_1=-8f_1$ (Right panel) along $c_1$ taking different values of $f_1,\; k_s,\; k_d, \;\&\; k_{sh}.$}
    \label{plot:16}
    \end{figure}

The non-uniform distribution for the plasma field is given by \cite{149}
\begin{equation}\label{pl}
    \rho(r)=\frac{\sigma^2_{\nu}}{2\pi r^2}\ ,
\end{equation}
where $\sigma^2_{\nu}$ defines the uni-dimensional velocity dispersion. We can write the non-uniform concentration of plasma field as \cite{149}
\begin{equation}\label{pll}
    N(r)=\frac{\rho(r)}{k m_p}\ ,
\end{equation}
where $m_p$ and $k$ denote the mass and dimensionless dark matter coefficient, respectively. The plasma frequency yields
\begin{equation}
    w^2_e=K_e N(r)=\frac{K_e \sigma^2_{\nu}}{2\pi k m_p r^2}\ .
\end{equation}
Here we are interested to explore the influence of non-uniform plasma (SIS) on the deflection angle around the MAGBH geometry. In this scenario, the deflection angle has a mathematical expression of the form \cite{150}
\begin{equation}
   \hat{\alpha}_{SIS}=\hat{\alpha}_{SIS1}+\hat{\alpha}_{SIS2}+\hat{\alpha}_{SIS3} \label{nonsis}\ .
\end{equation}
Using Eqs.(\ref{h3}), (\ref{h7}) and (\ref{nonsis}), one can write  the deflection angel for $SIS$ plasma as
\begin{eqnarray}
   \hat{\alpha}_{SIS}&=&\frac{64 c_{1} k_{d}^2 w_c^2 R_s^3}{5 \pi  b^5 w^2}-\frac{16 d_{1} k_{s}^2 w_c^2 R_s^3}{5 \pi  b^5 w^2}+\frac{32 f_{1} k_{sh}^2 w_c^2 R_s^3}{5 \pi  b^5 w^2}+\frac{2 w_c^2 R_s^3}{3 \pi  b^3 w^2}+\frac{64 c_{1} k_{d}^2 R_s}{3 b^3}-\frac{16 d_{1} k_{s}^2 R_s}{3 b^3}+\frac{32 f_{1} k_{sh}^2 R_s}{3 b^3}\nonumber\\&+&\frac{3 \pi  \sqrt{\frac{1}{b^2}} c_{1} k_{d}^2}{b}-\frac{3 \pi  \sqrt{\frac{1}{b^2}} d_{1} k_{s}^2}{4 b}+\frac{3 \pi  \sqrt{\frac{1}{b^2}} f_{1} k_{sh}^2}{2 b}+\frac{10 \sqrt{\frac{1}{b^2}} c_{1}^2 k_{d}^4 w_c^2 R_s^2}{b^5 w^2}-\frac{5 \sqrt{\frac{1}{b^2}} c_{1} d_{1} k_{d}^2 k_{s}^2 w_c^2 R_s^2}{b^5 w^2}\nonumber\\&+&\frac{10 \sqrt{\frac{1}{b^2}} c_{1} f_{1} k_{d}^2 k_{sh}^2 w_c^2 R_s^2}{b^5 w^2}+\frac{5 \sqrt{\frac{1}{b^2}} d_{1}^2 k_{s}^4 w_c^2 R_s^2}{8 b^5 w^2}-\frac{5 \sqrt{\frac{1}{b^2}} d_{1} f_{1} k_{sh}^2 k_{s}^2 w_c^2 R_s^2}{2 b^5 w^2}+\frac{5 \sqrt{\frac{1}{b^2}} f_{1}^2 k_{sh}^4 w_c^2 R_s^2}{2 b^5 w^2}\nonumber\\&+&\frac{3 \sqrt{\frac{1}{b^2}} c_{1} k_{d}^2 w_c^2 R_s^2}{2 b^3 w^2}-\frac{3 \sqrt{\frac{1}{b^2}} d_{1} k_{s}^2 w_c^2 R_s^2}{8 b^3 w^2}+\frac{3 \sqrt{\frac{1}{b^2}} f_{1} k_{sh}^2 w_c^2 R_s^2}{4 b^3 w^2}+\frac{15 \pi  \sqrt{\frac{1}{b^2}} c_{1}^2 k_{d}^4}{b^3}-\frac{15 \pi  \sqrt{\frac{1}{b^2}} c_{1} d_{1} k_{d}^2 k_{s}^2}{2 b^3}\nonumber\\&+&\frac{15 \pi  \sqrt{\frac{1}{b^2}} c_{1} f_{1} k_{d}^2 k_{sh}^2}{b^3}+\frac{15 \pi  \sqrt{\frac{1}{b^2}} d_{1}^2 k_{s}^4}{16 b^3}-\frac{15 \pi  \sqrt{\frac{1}{b^2}} d_{1} f_{1} k_{sh}^2 k_{s}^2}{4 b^3}+\frac{15 \pi  \sqrt{\frac{1}{b^2}} f_{1}^2 k_{sh}^4}{4 b^3}+\frac{2 R_s}{b}.
\end{eqnarray}
 The analytic form of plasma constant $w^2_c$ gives \cite{151}
 \begin{equation}
    w^2_c=\frac{K_e \sigma^2_{\nu}}{2\pi k m_p R^2_S}\ .
 \end{equation}
We plot $\hat{\alpha}_{SIS}$ in Figs.~\ref{plot:15} and \ref{plot:16}. It is found that the behavior of $SIS$ plasma deflection angel is quite similar to the uniform case. It should also be noted that $\hat{\alpha}_{SIS}<\hat{\alpha}_{uni}$.

     \begin{figure}
    \centering
    \includegraphics[scale=0.26]{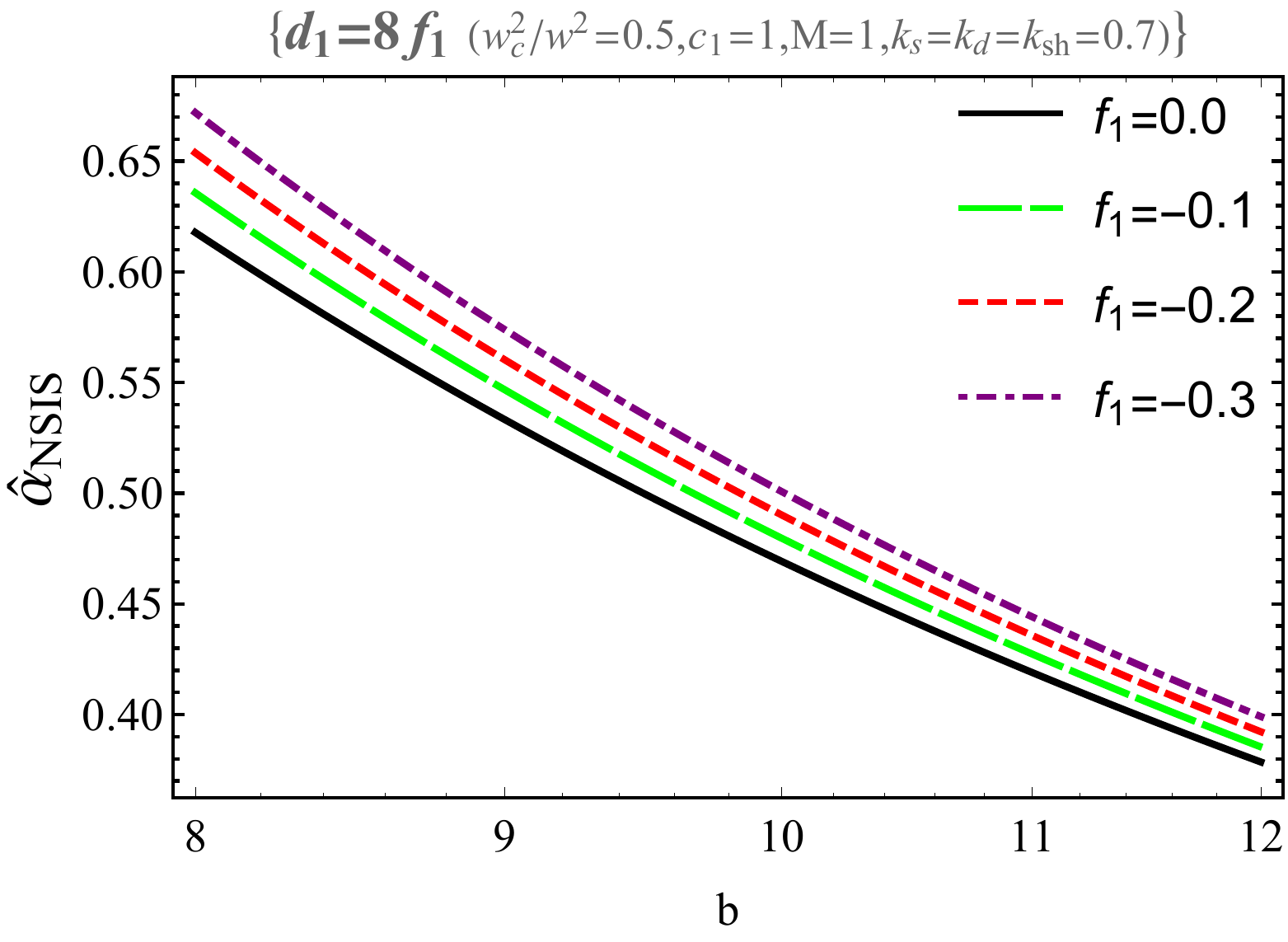}
    \includegraphics[scale=0.26]{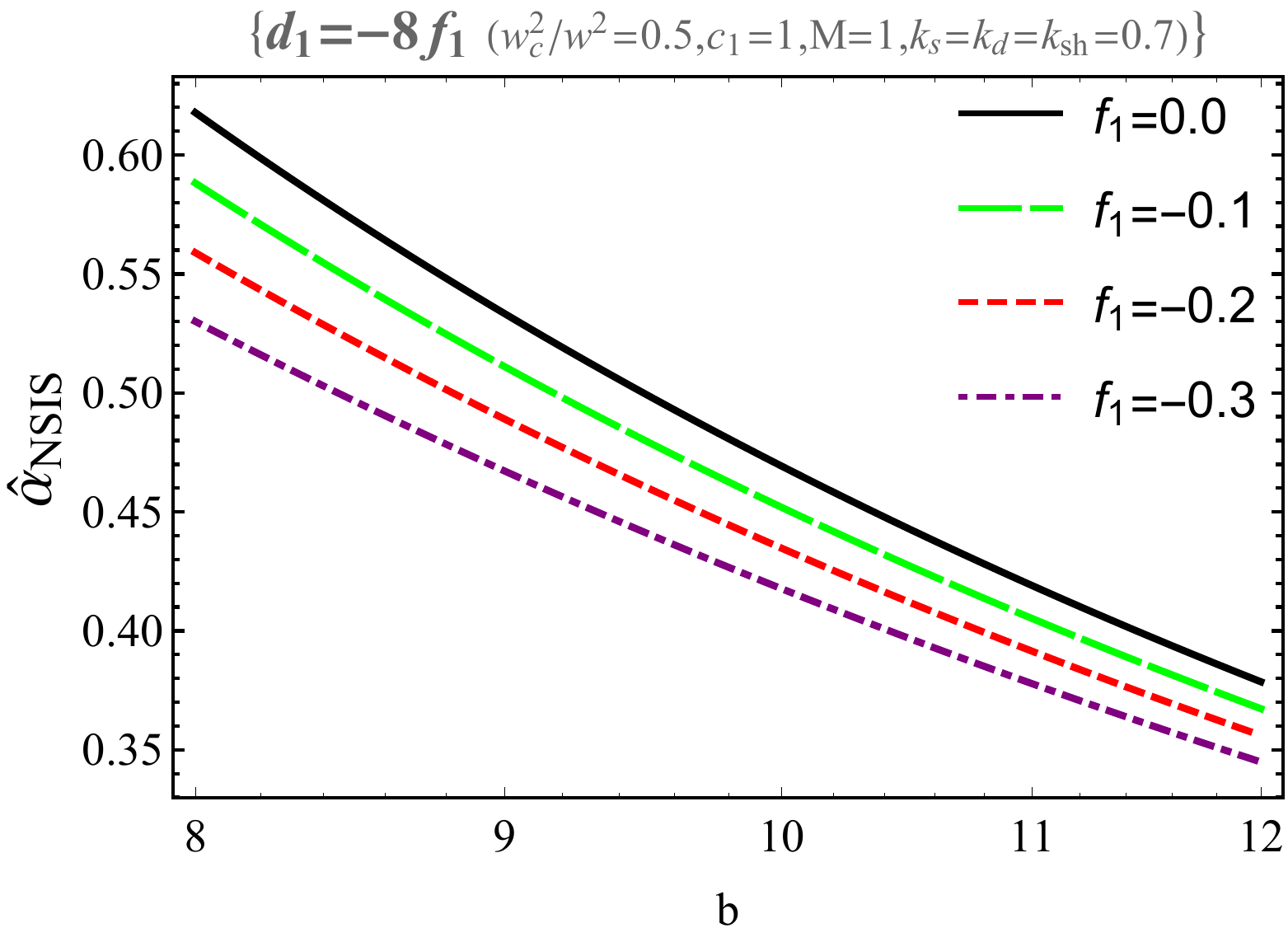}
    \includegraphics[scale=0.26]{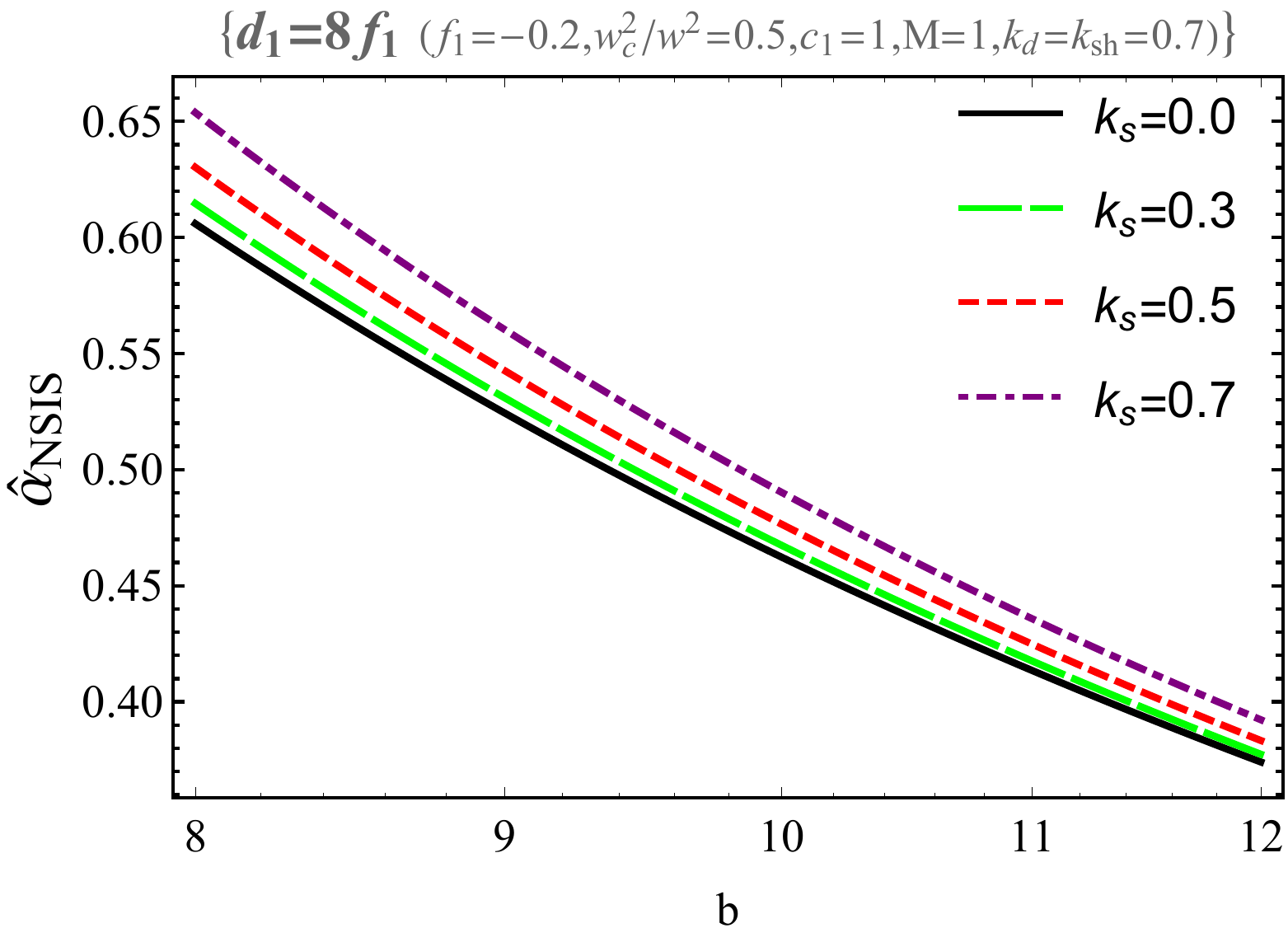}
    \includegraphics[scale=0.26]{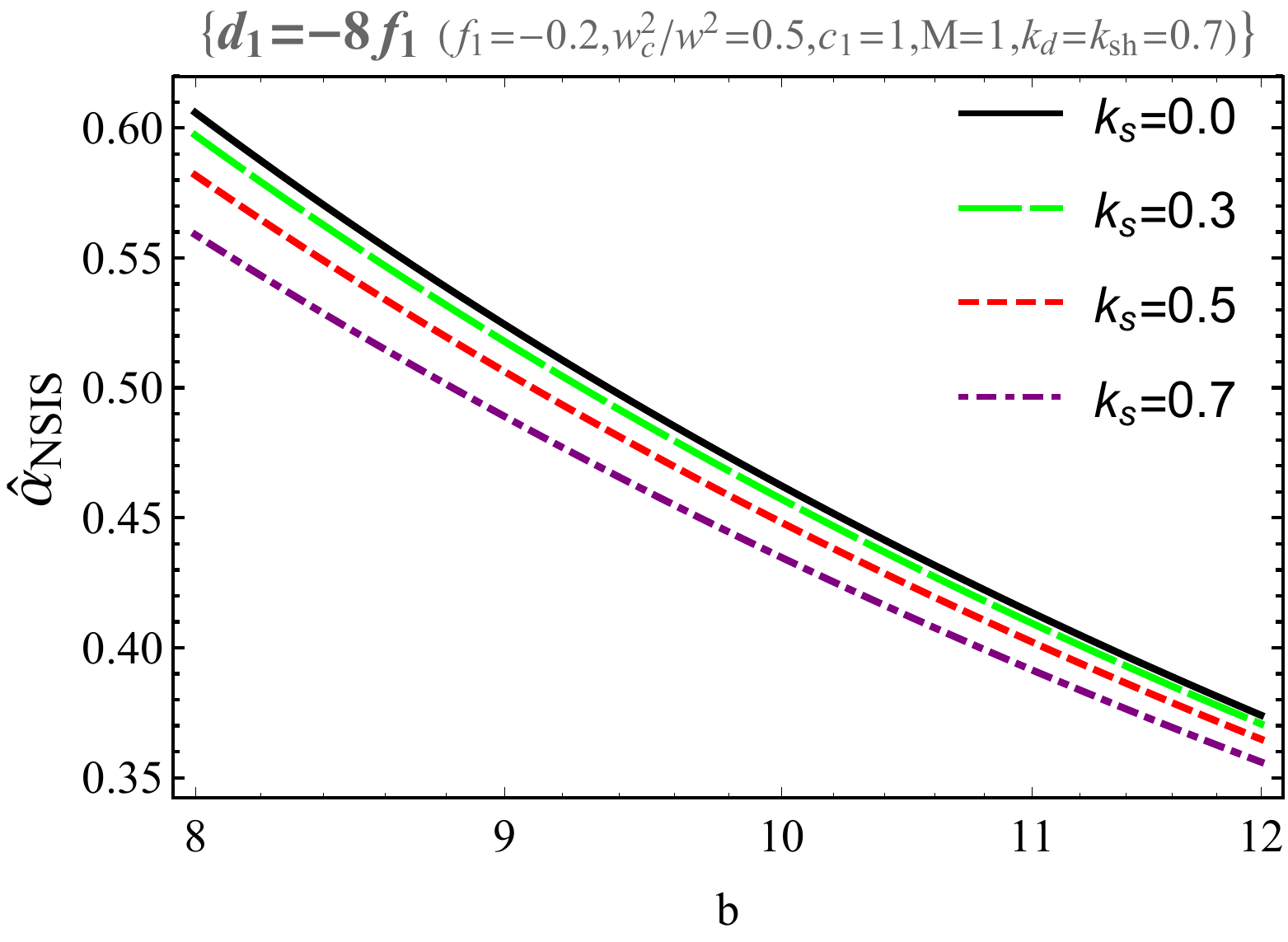}
    \includegraphics[scale=0.26]{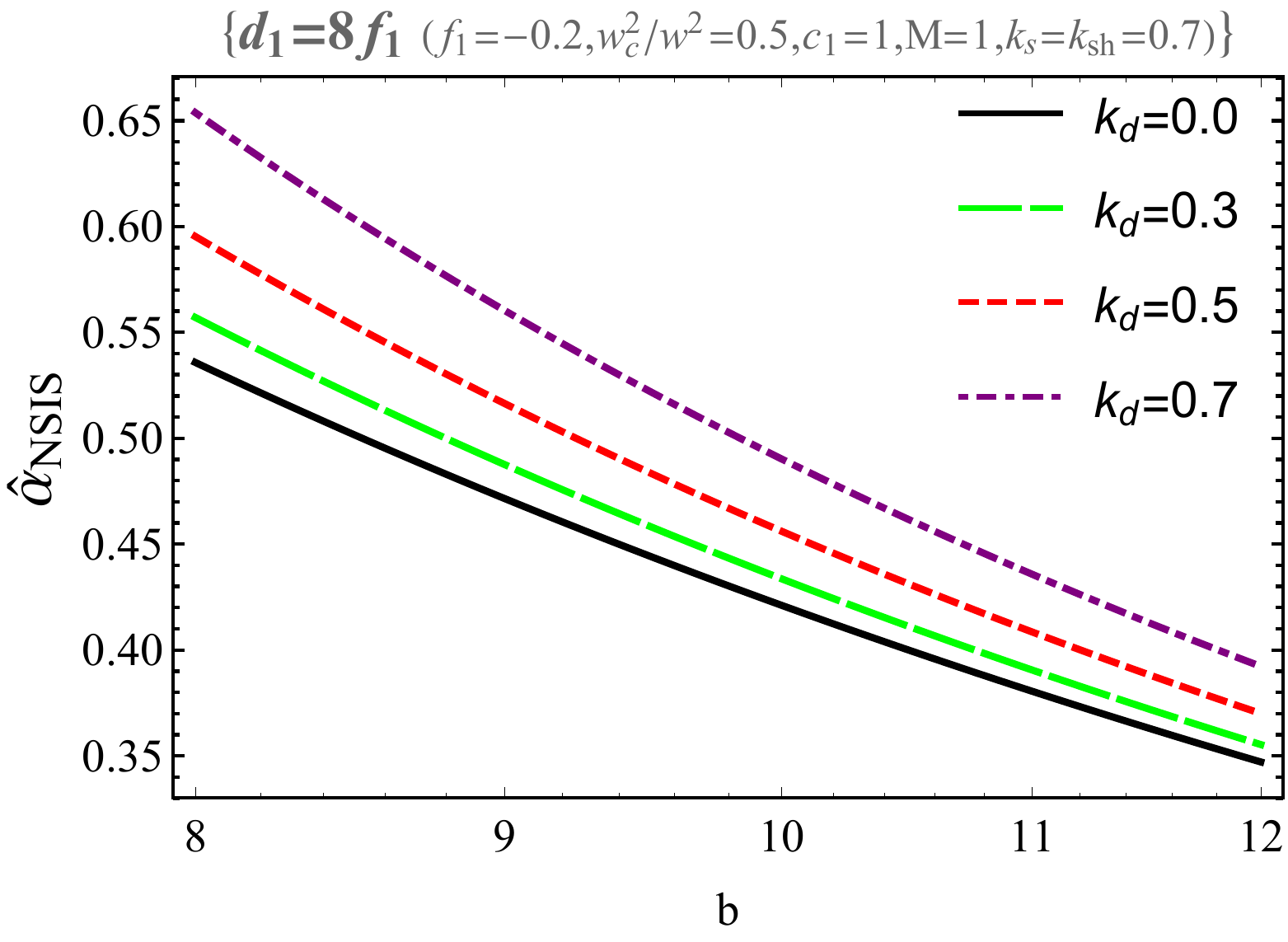}
    \includegraphics[scale=0.26]{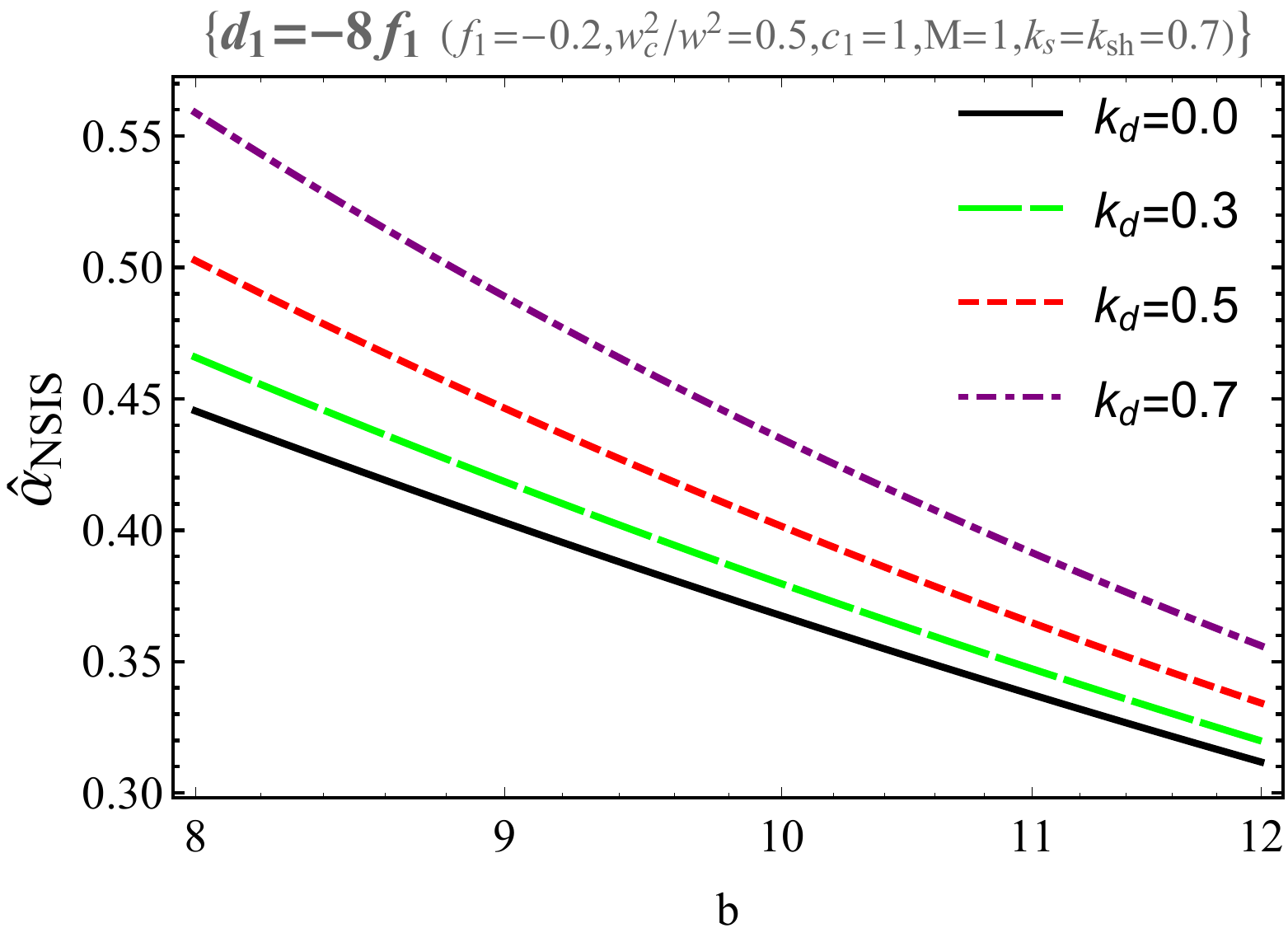}
    \includegraphics[scale=0.26]{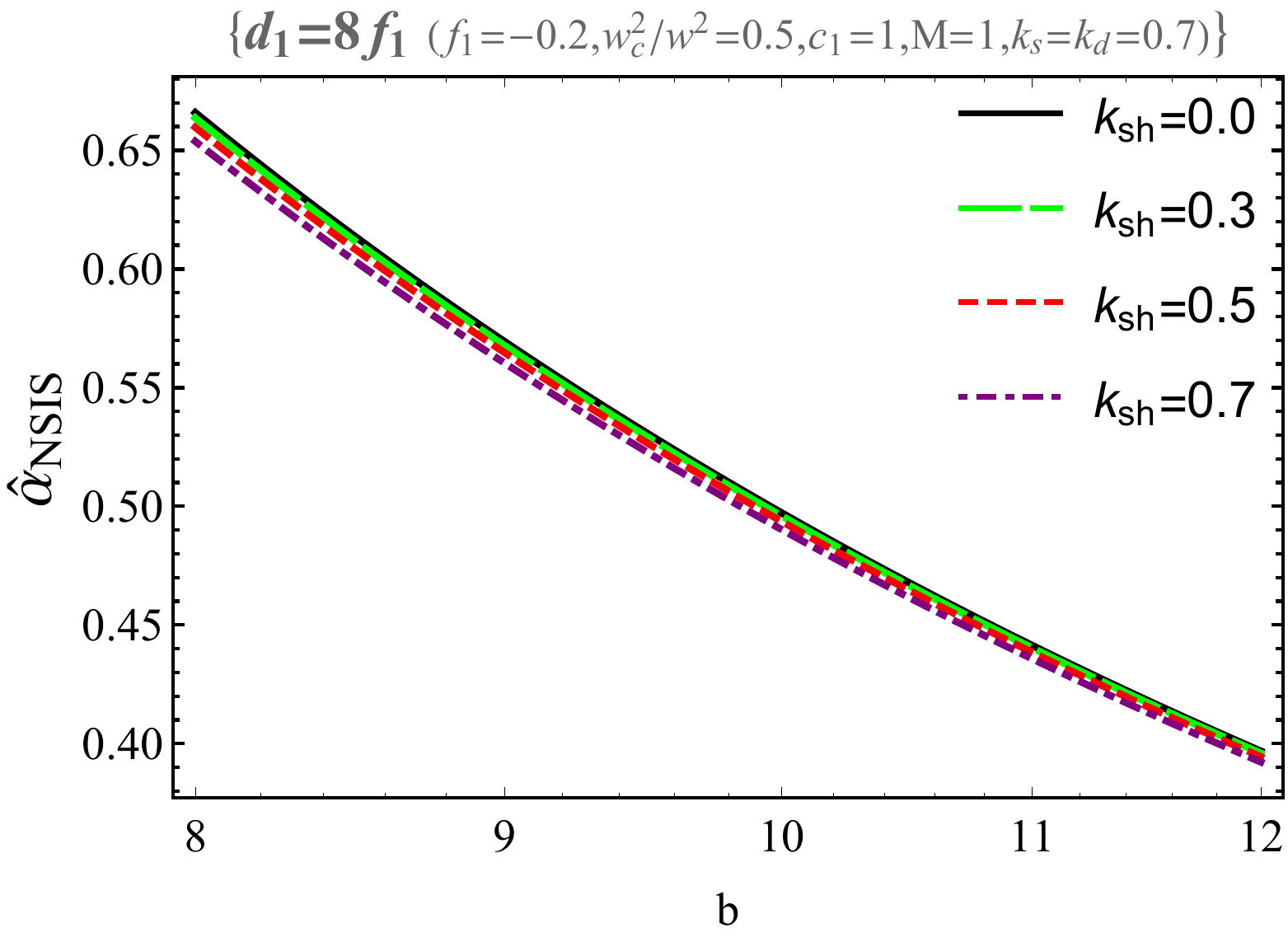}
    \includegraphics[scale=0.26]{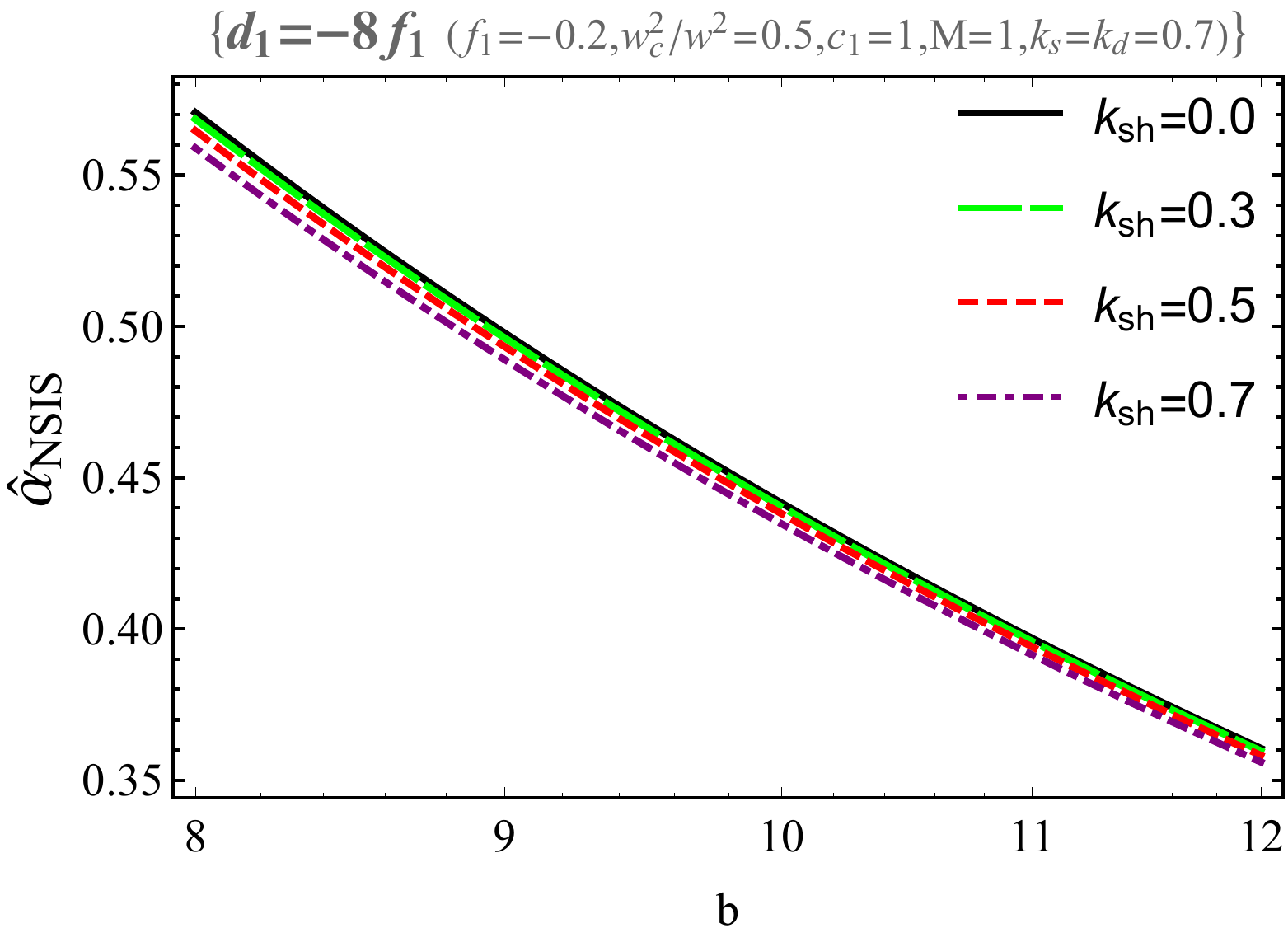}
    \includegraphics[scale=0.26]{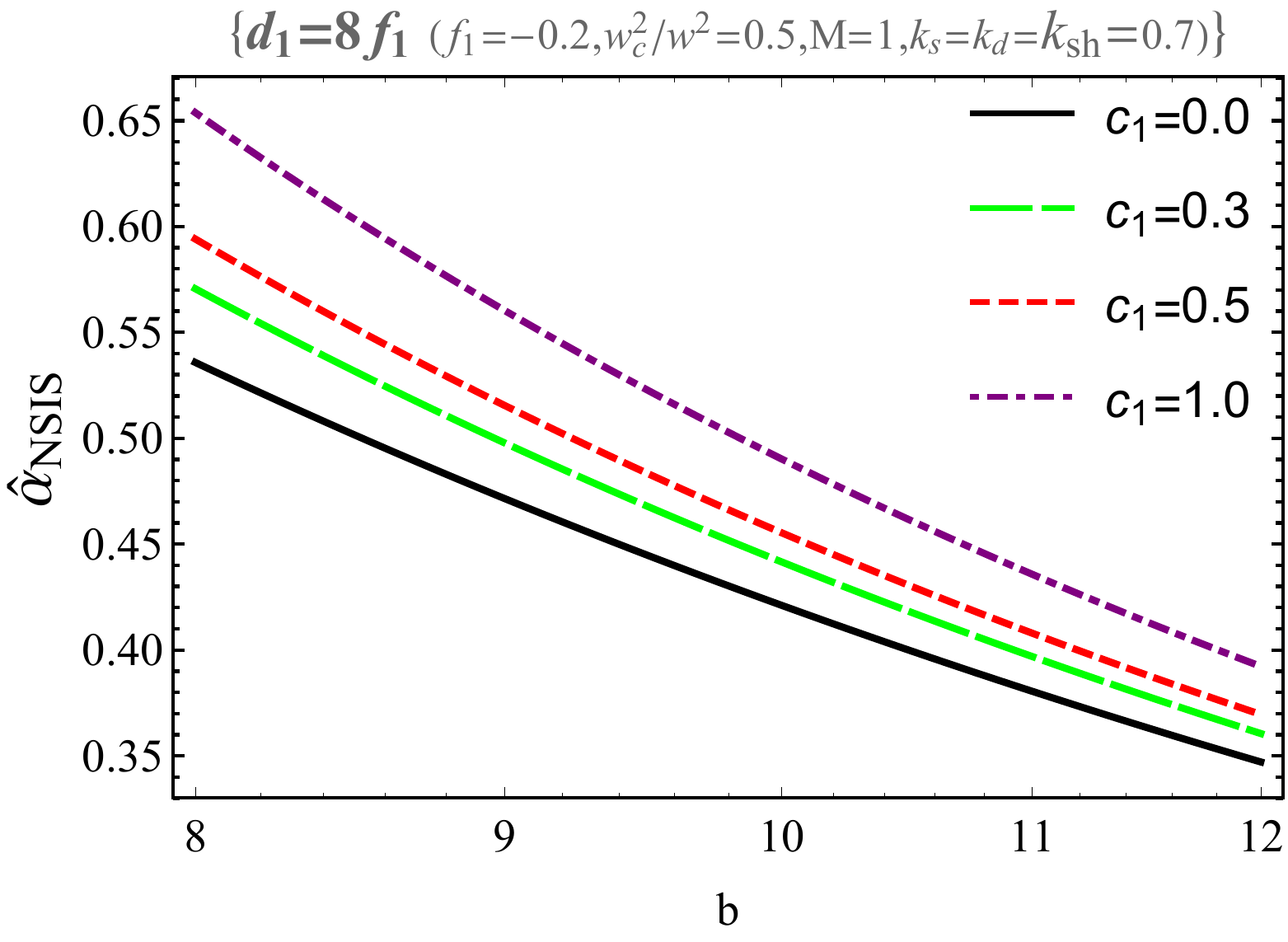}
    \includegraphics[scale=0.26]{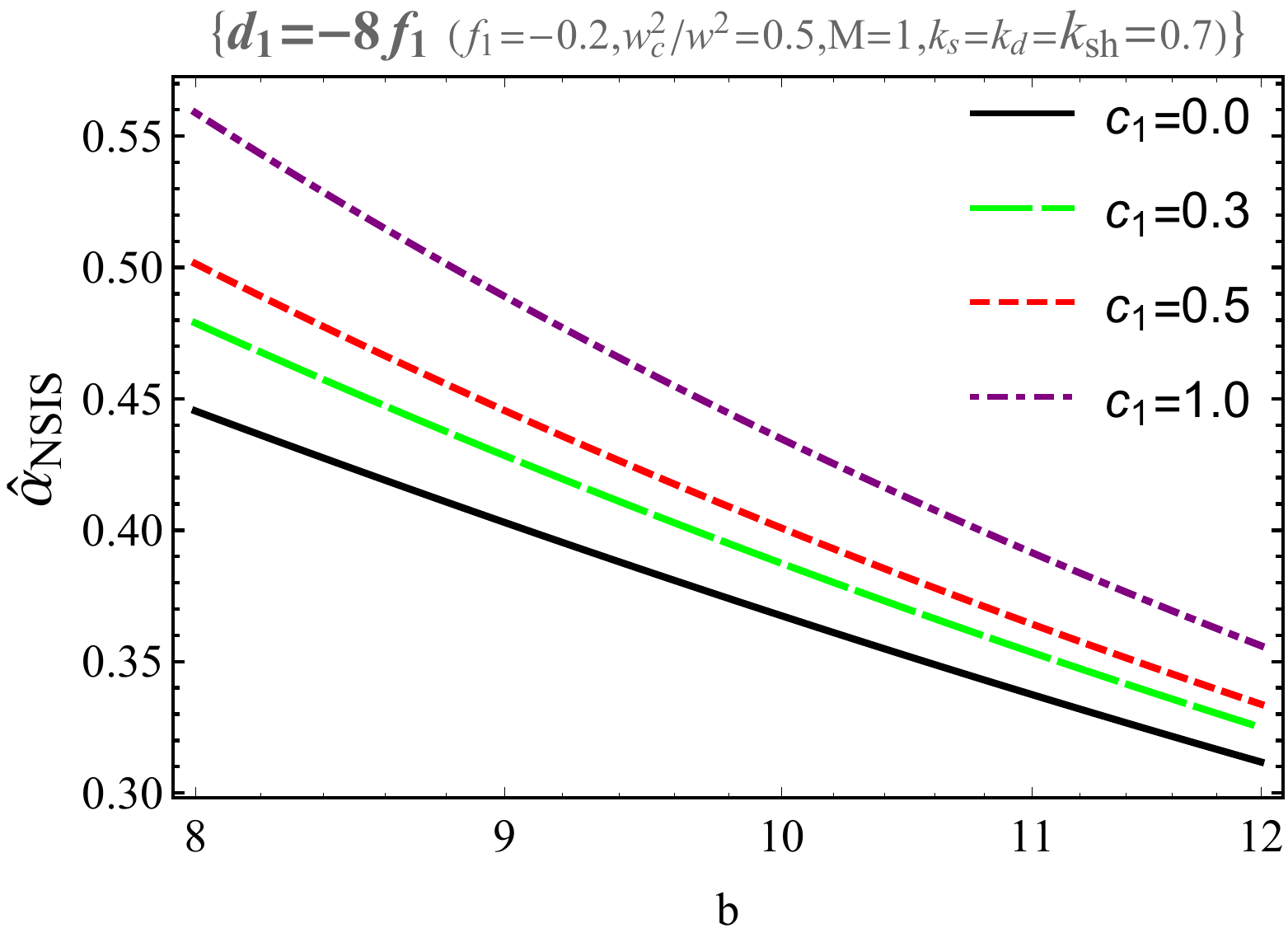}
    \includegraphics[scale=0.26]{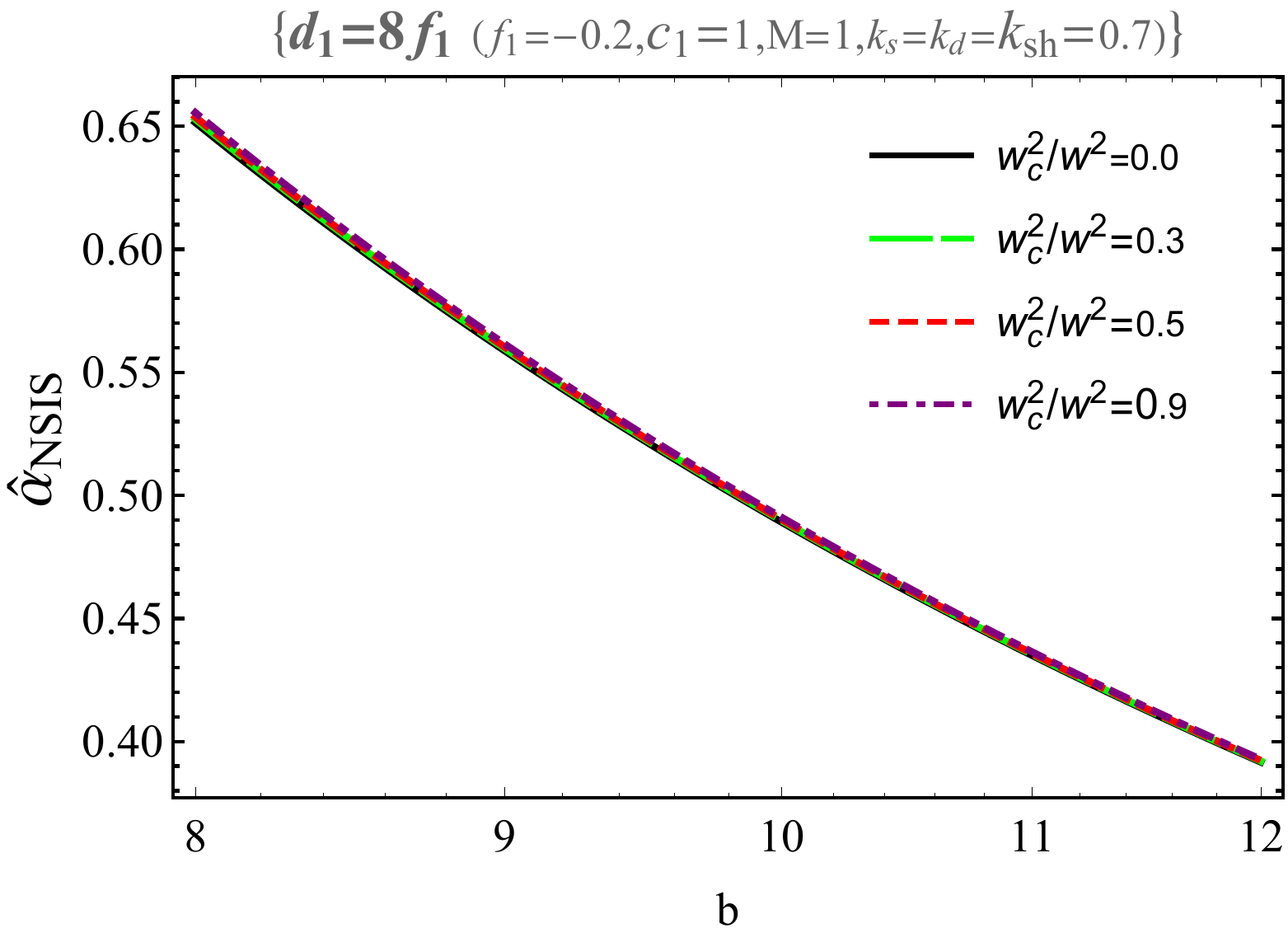}
    \includegraphics[scale=0.26]{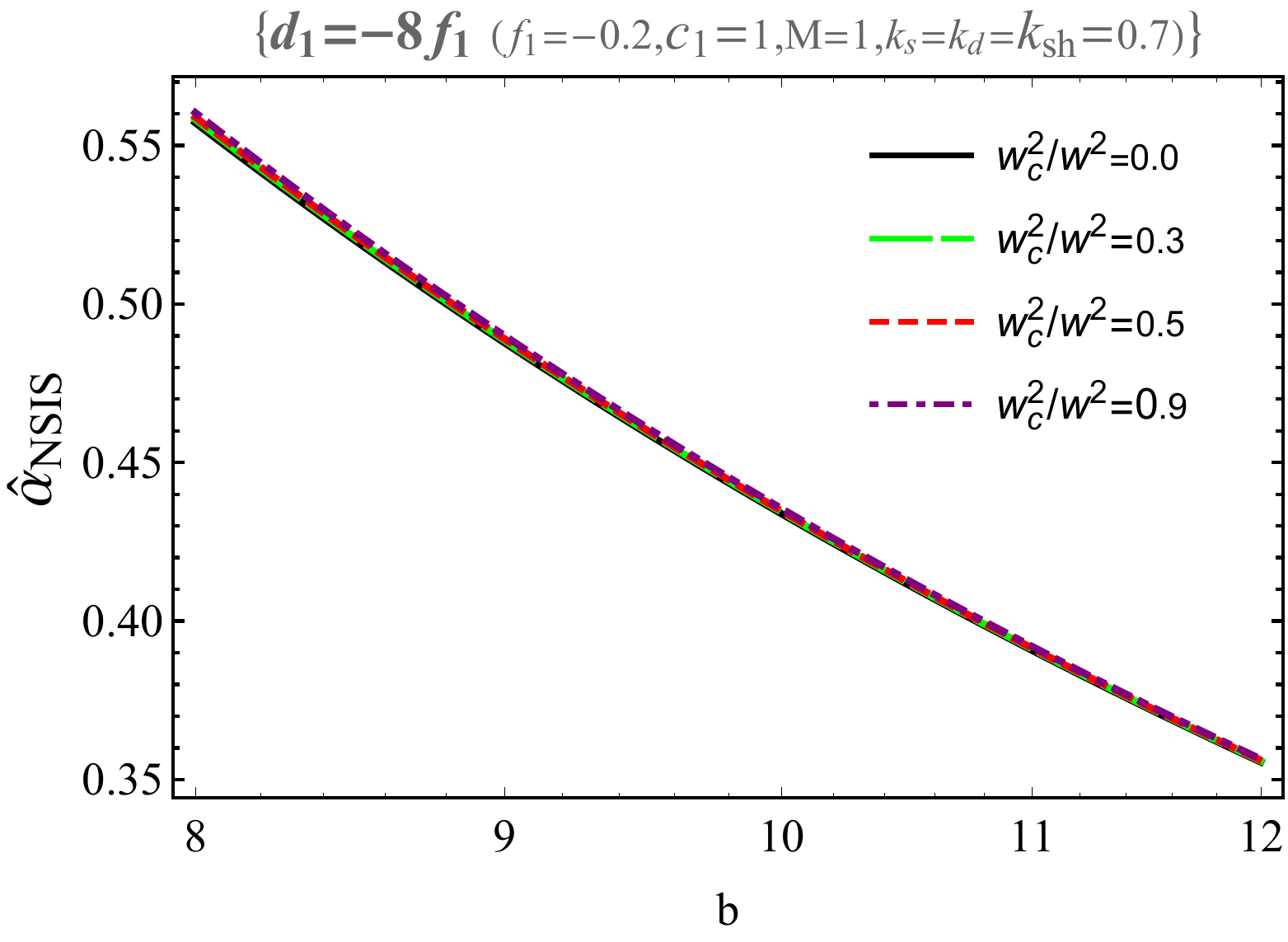}
    \includegraphics[scale=0.26]{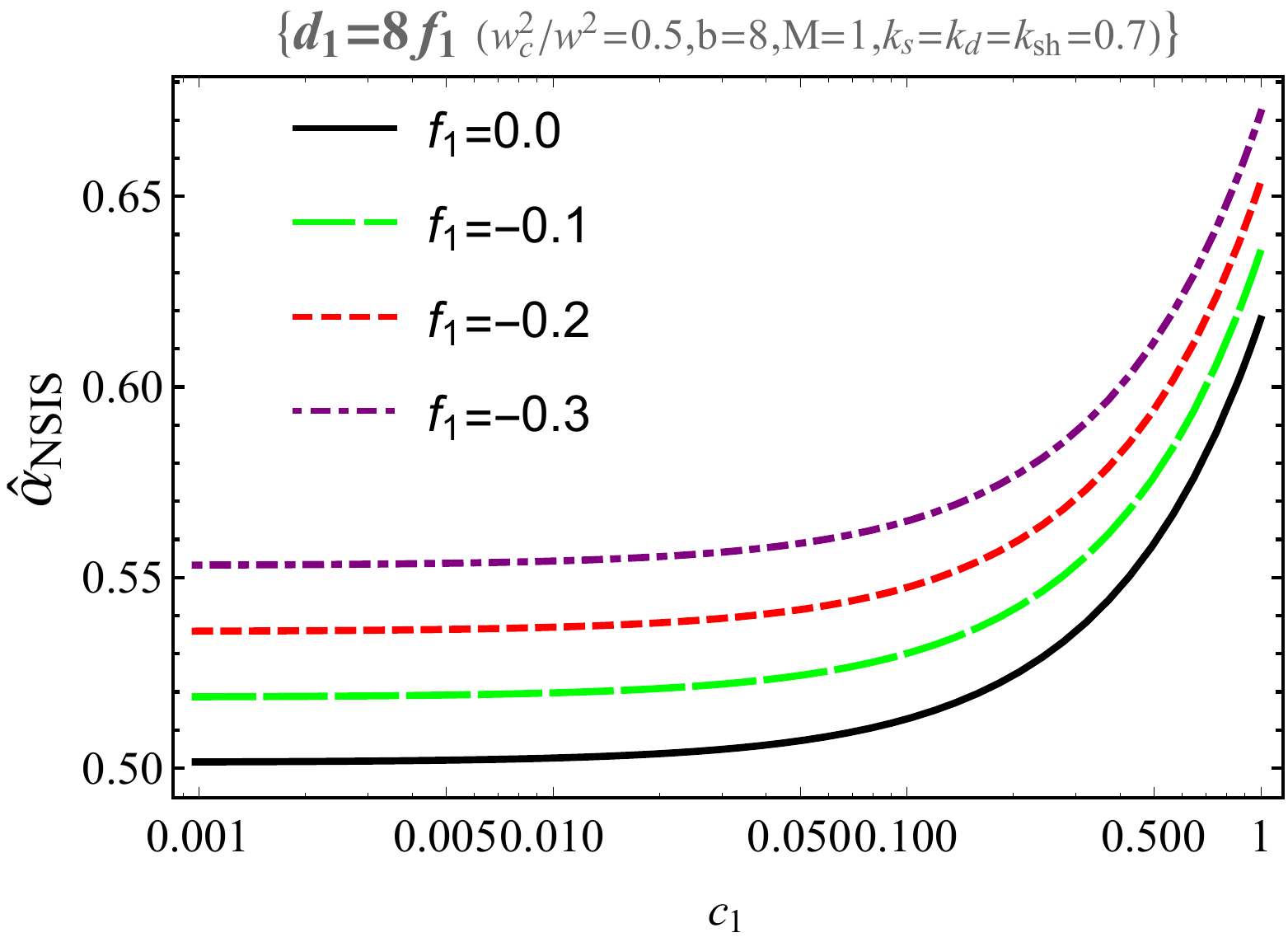}
    \includegraphics[scale=0.26]{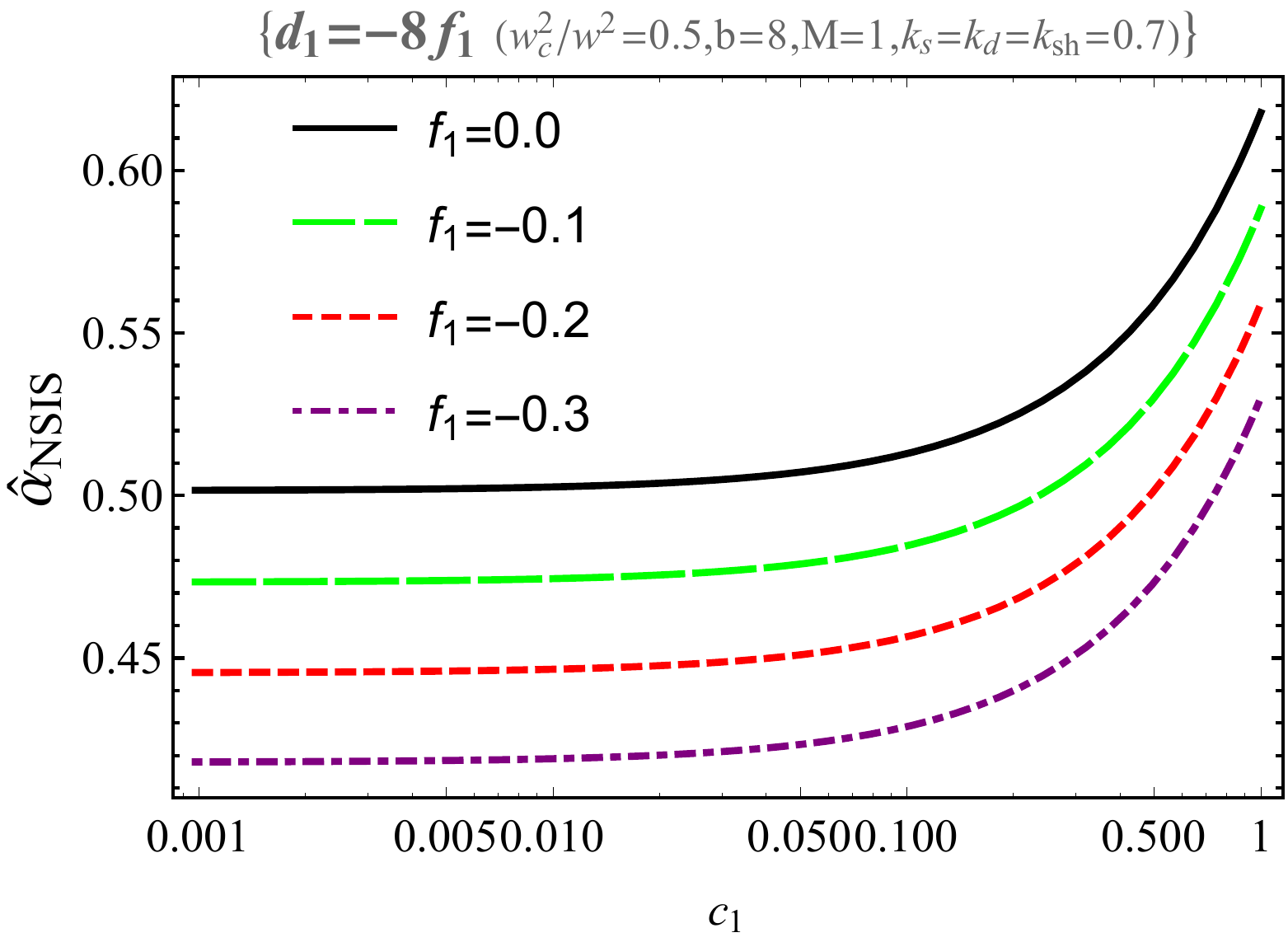}
    \includegraphics[scale=0.26]{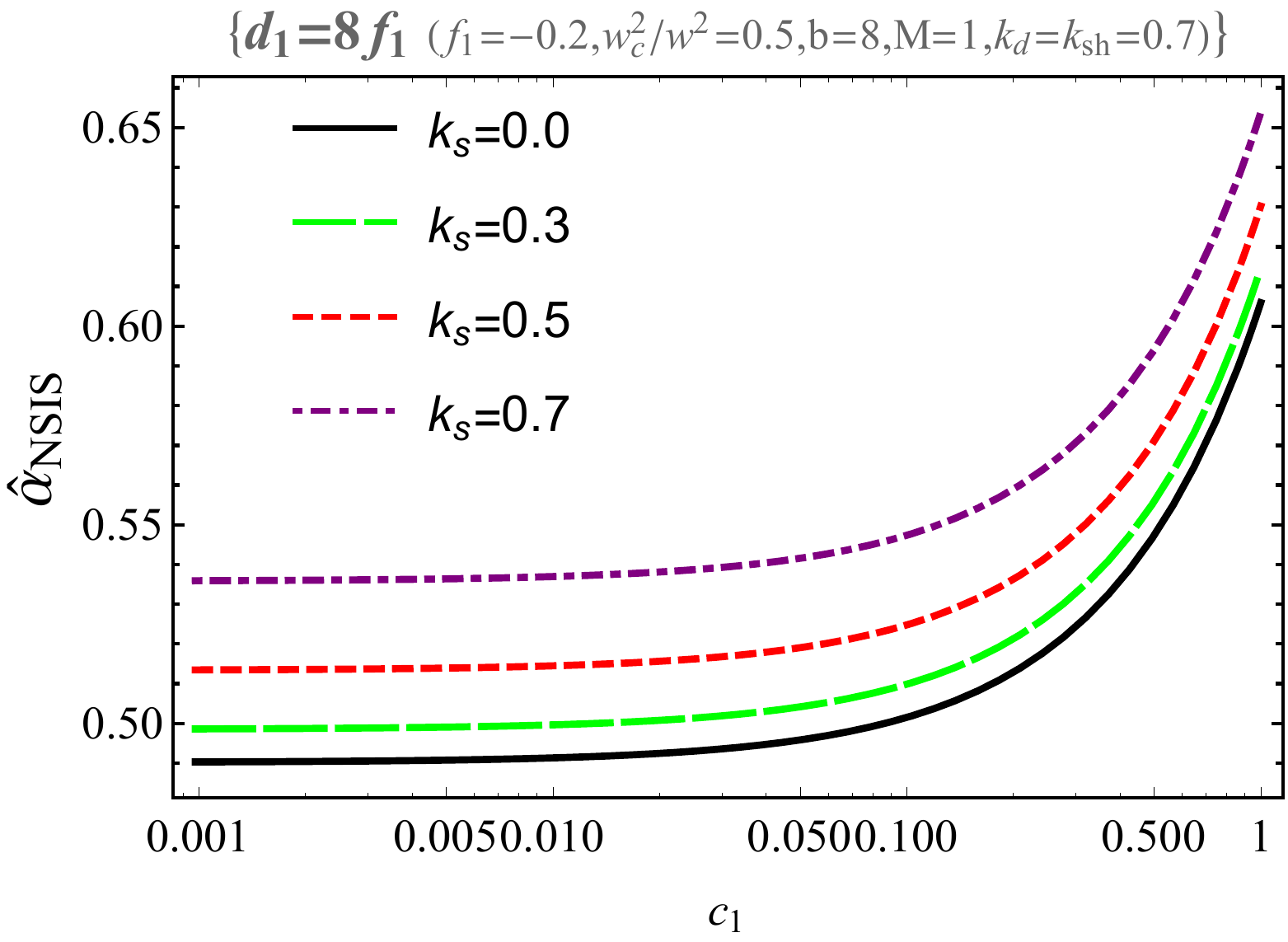}
    \includegraphics[scale=0.26]{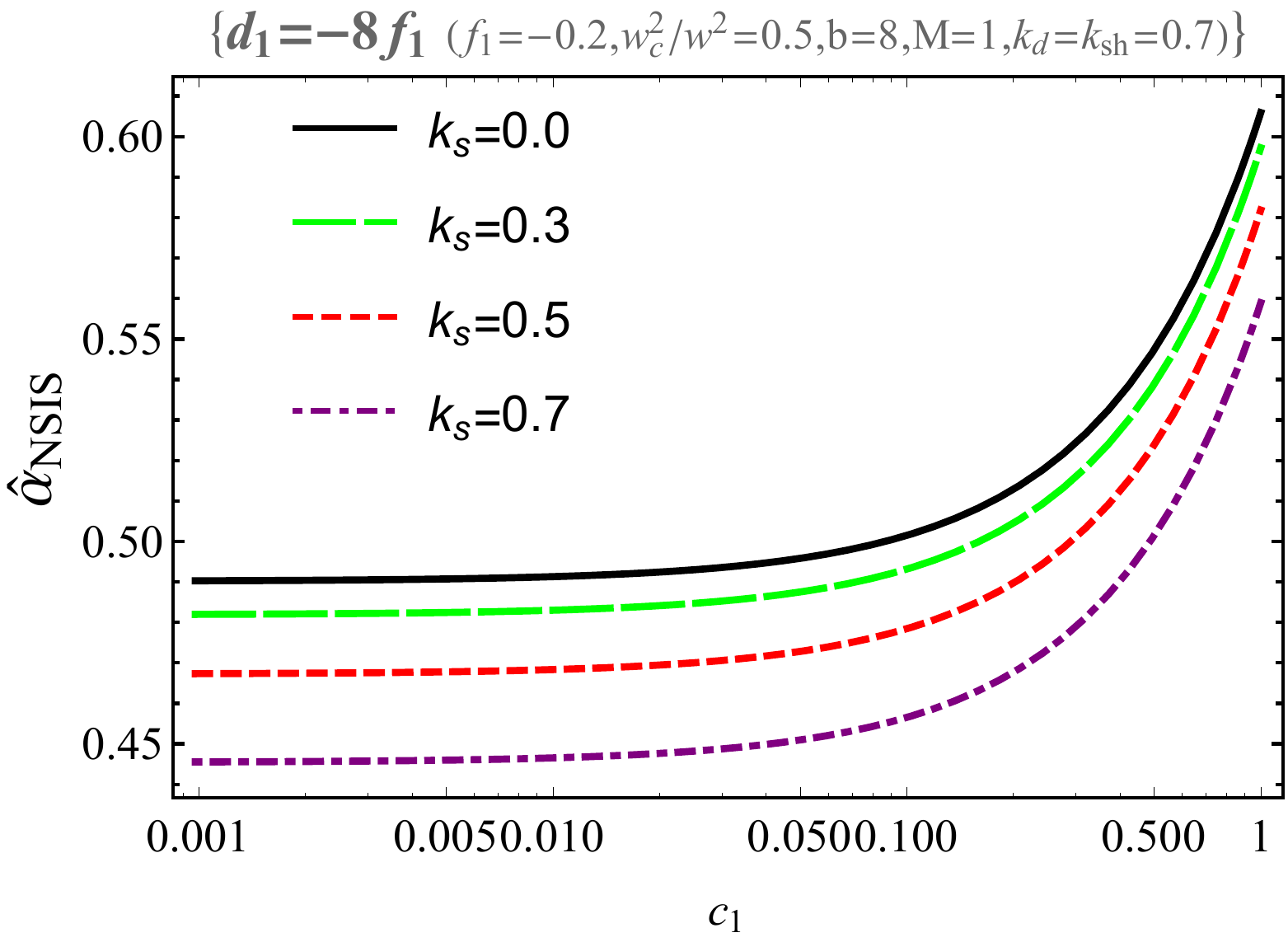}
    \includegraphics[scale=0.26]{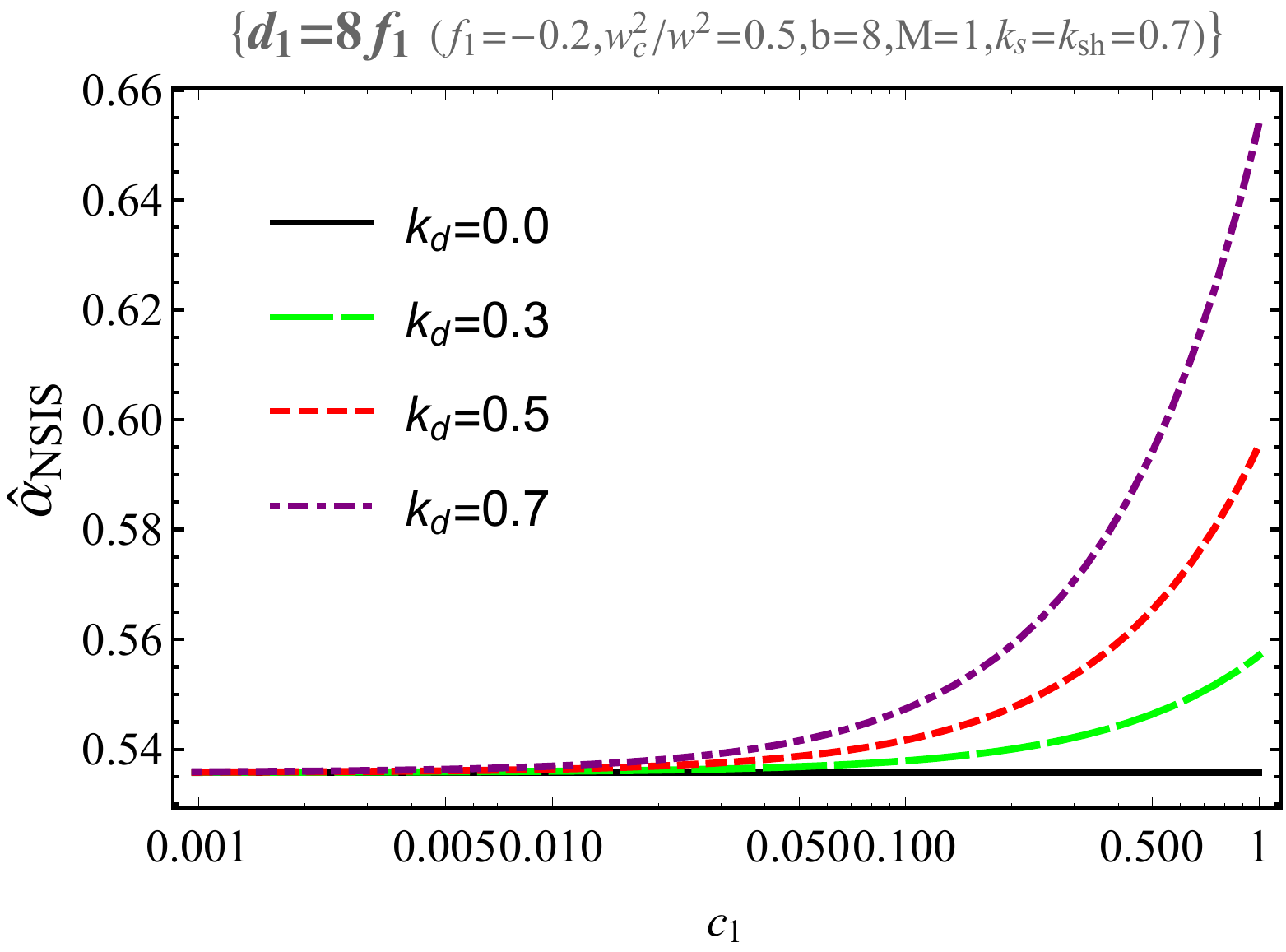}
    \includegraphics[scale=0.26]{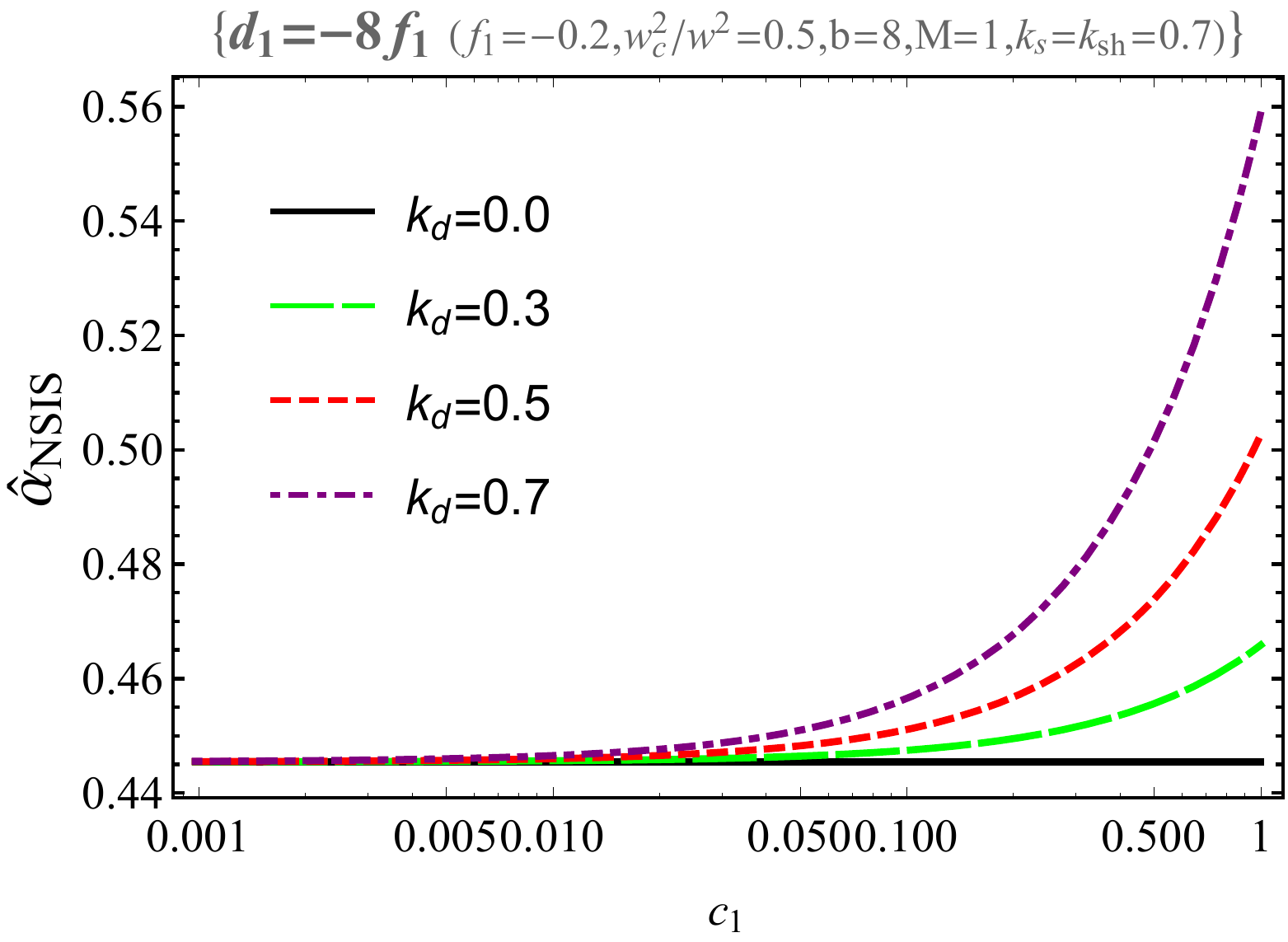}
    \includegraphics[scale=0.26]{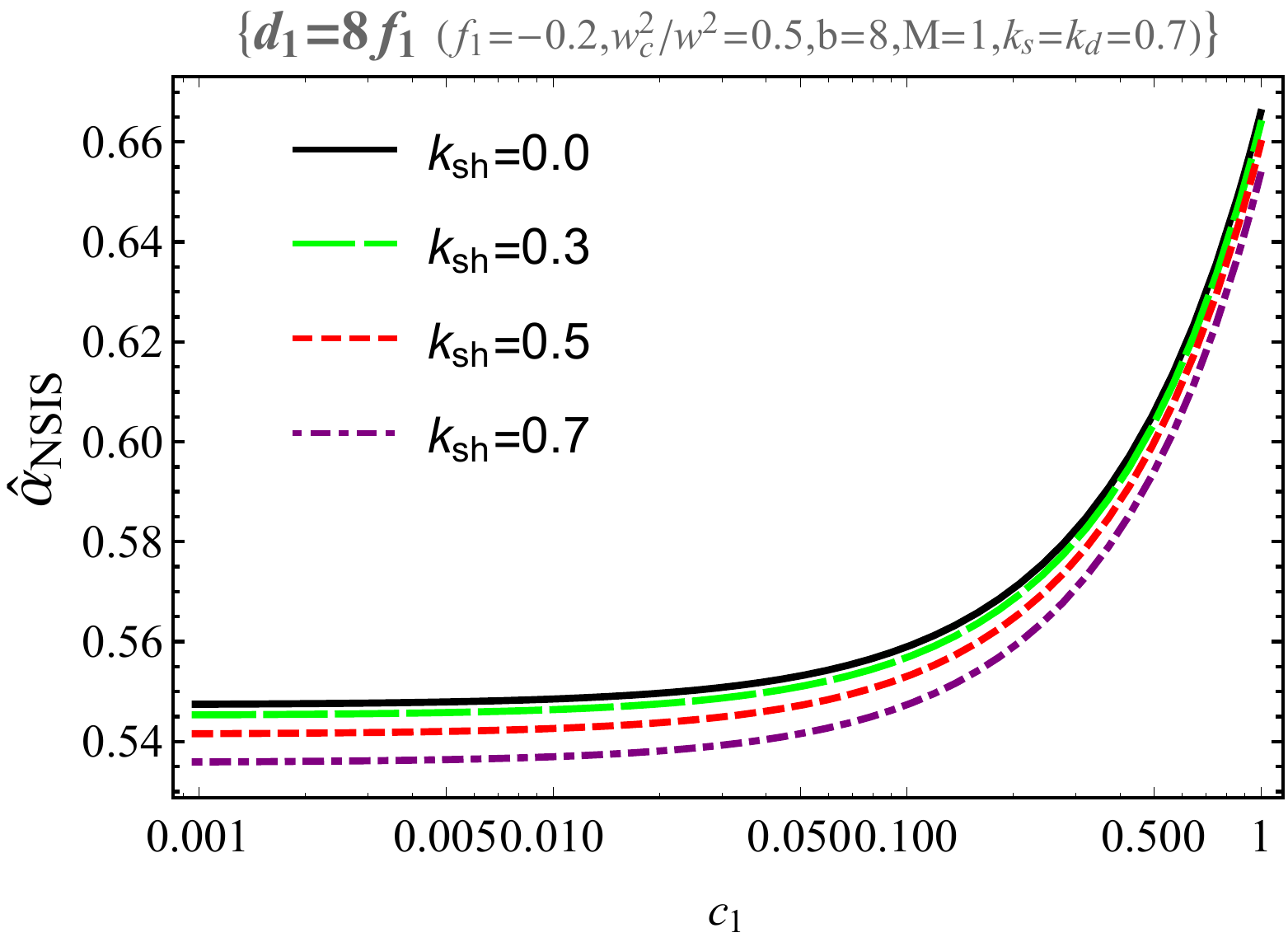}
    \includegraphics[scale=0.26]{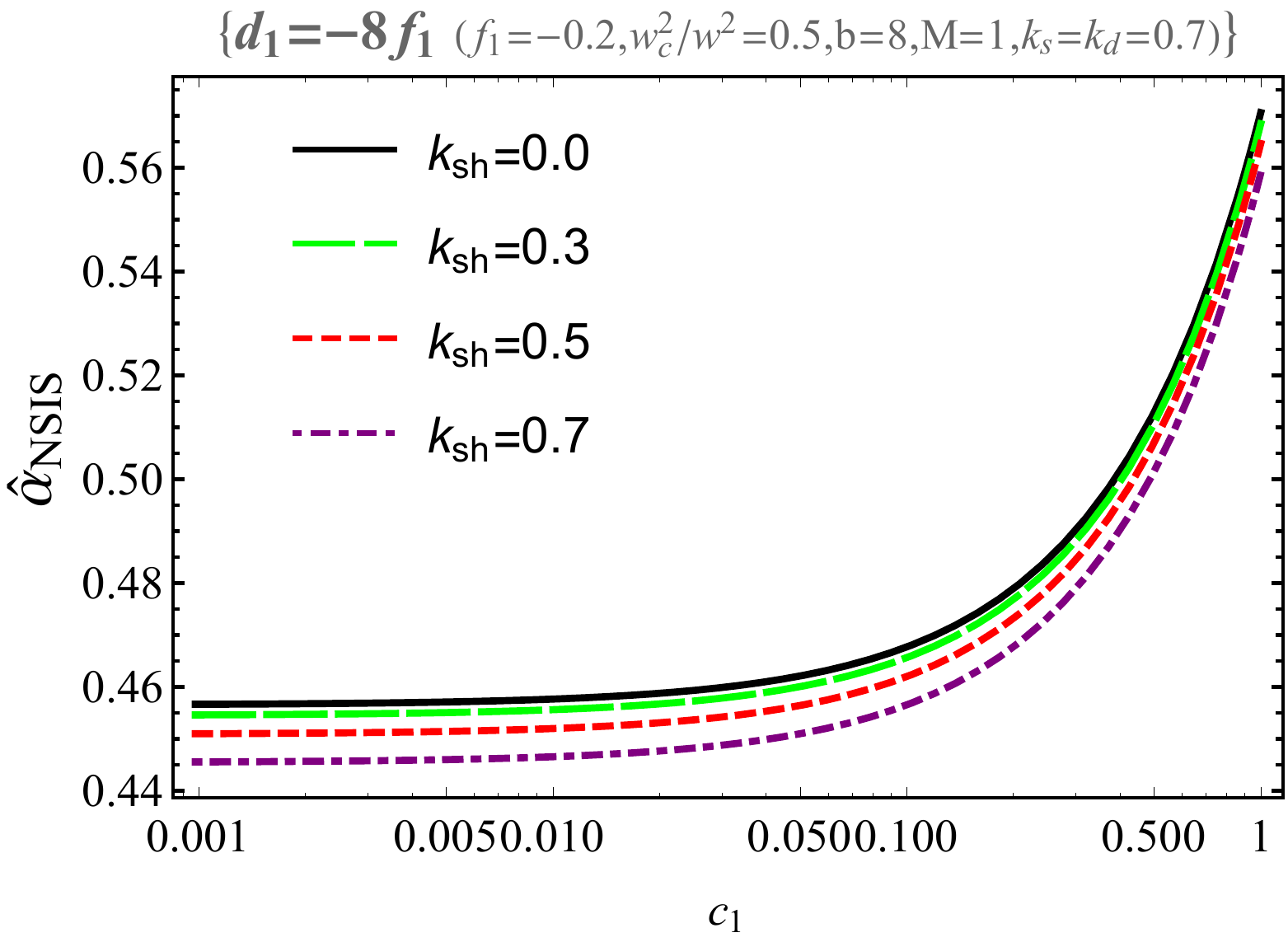}
    \includegraphics[scale=0.26]{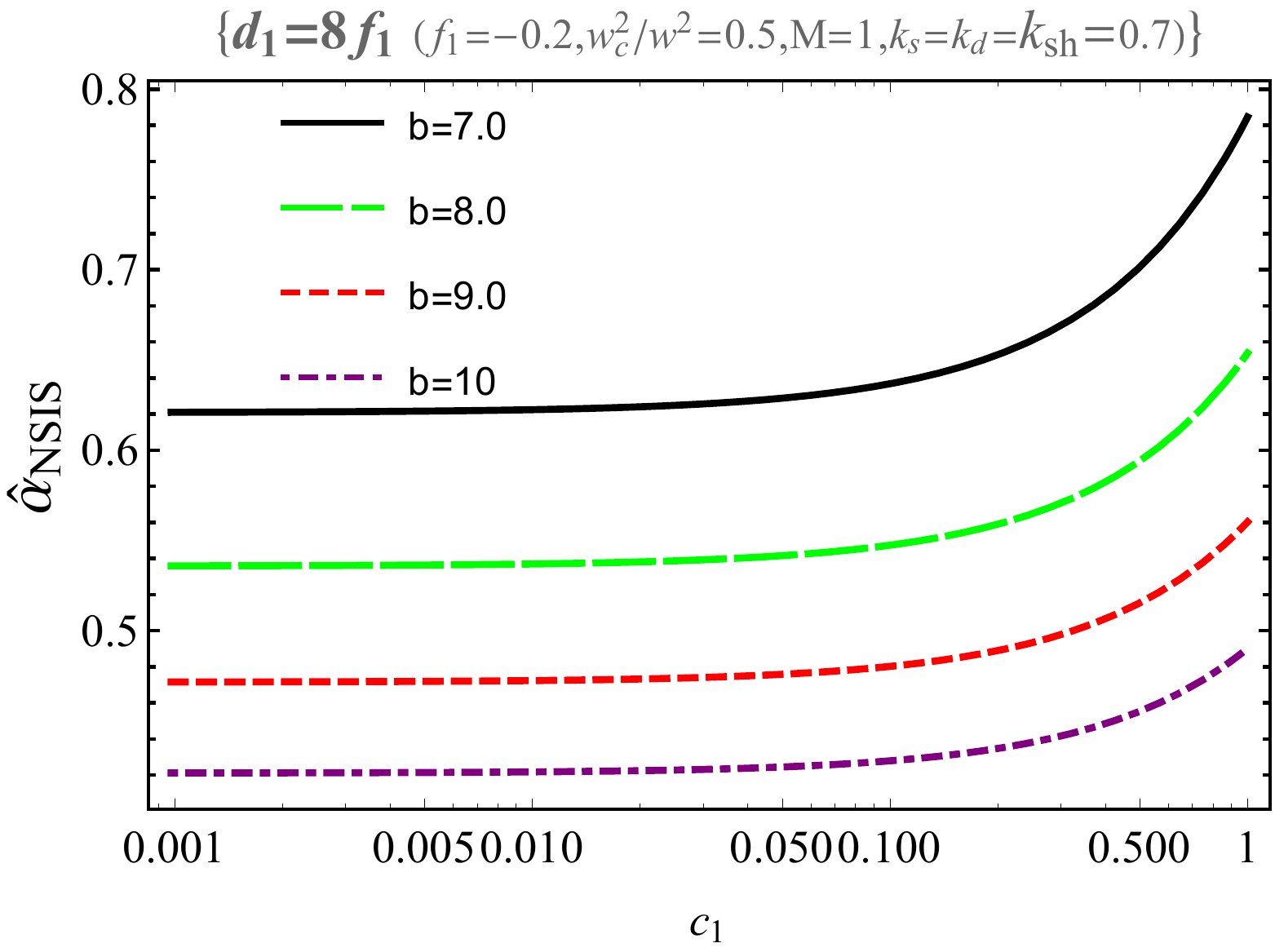}
    \includegraphics[scale=0.26]{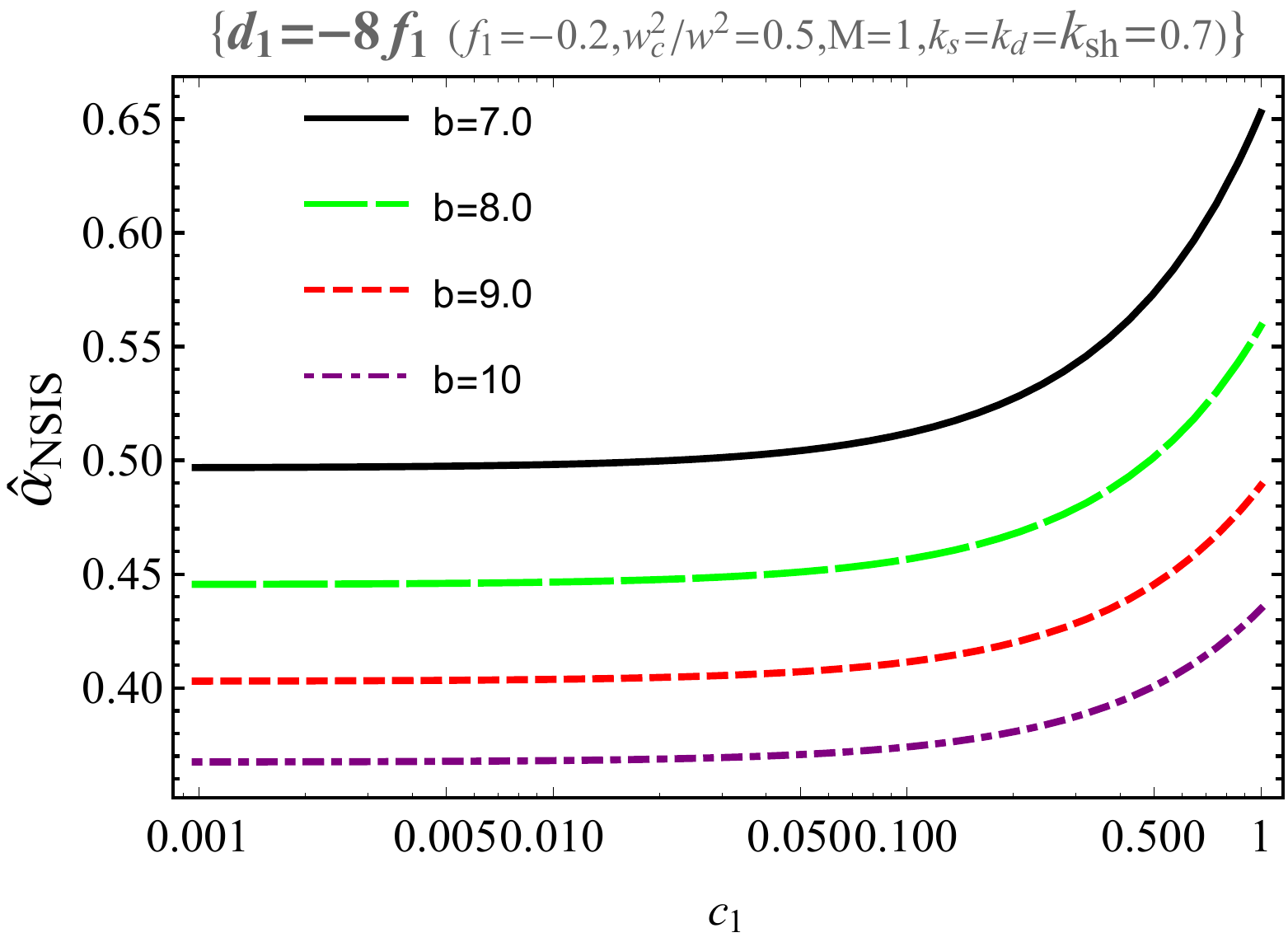}
    \includegraphics[scale=0.26]{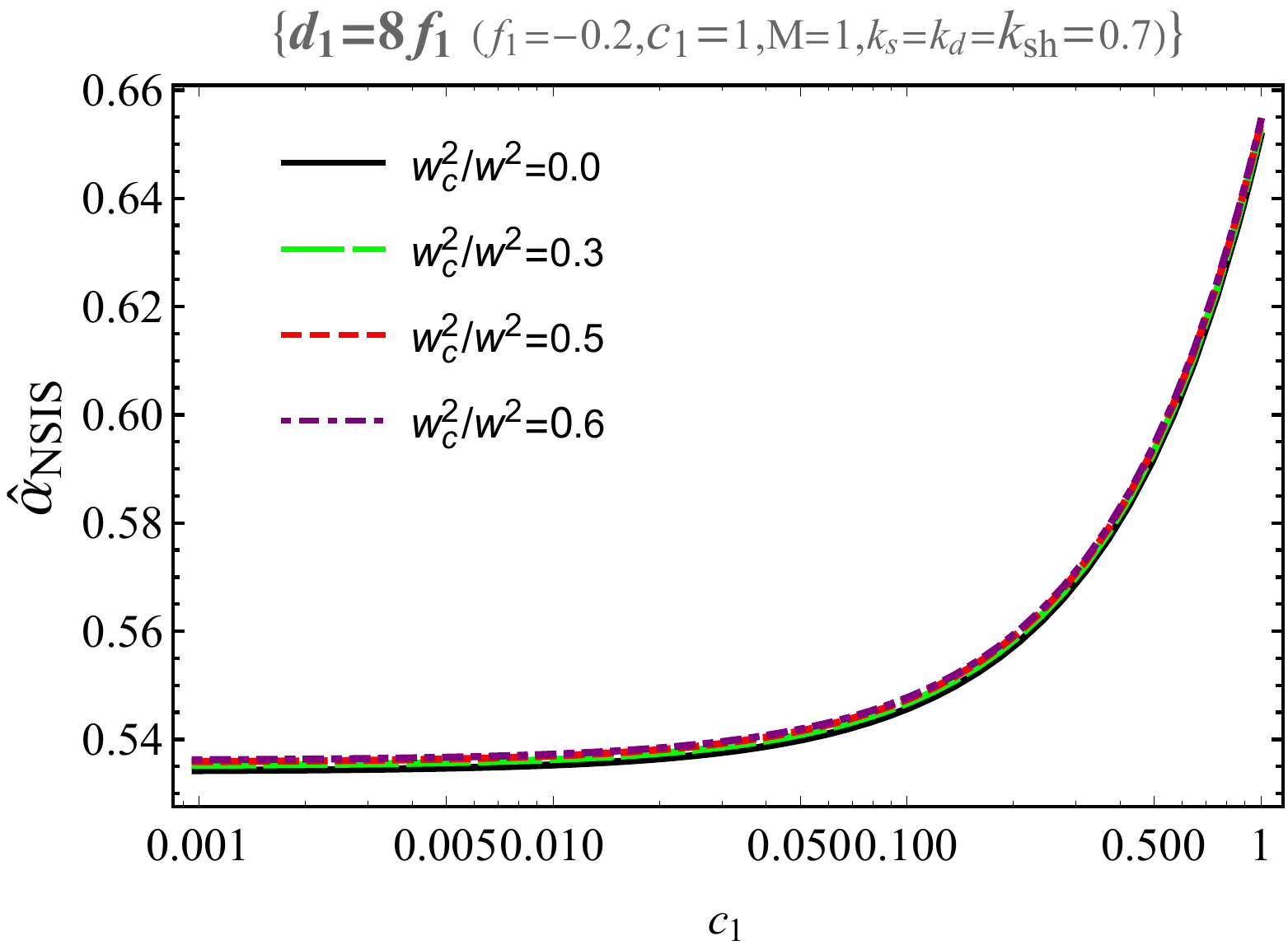}
    \includegraphics[scale=0.26]{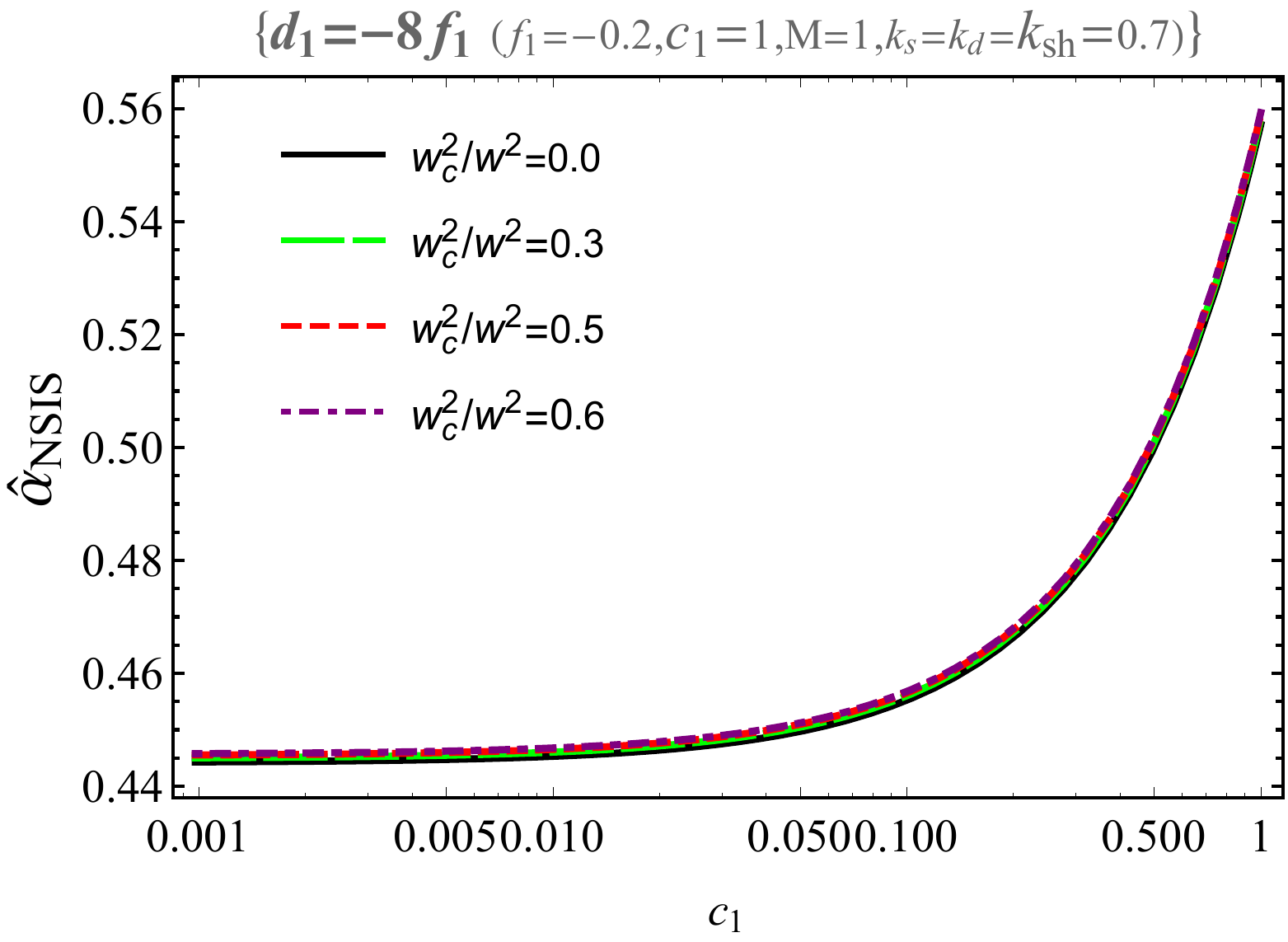}
    \caption{The deflection angle $\hat{\alpha}_{NSIS}$ in $NSIS$ plasma for $d_1=8f_1$ (Left panel) and $d_1=-8f_1$ (Right panel) along $c_1$ taking different values of $f_1,\; k_s,\; k_d, \;\&\; k_{sh}.$}
    \label{plot:17}
    \end{figure}
    \begin{figure}
    \centering
    \includegraphics[scale=0.26]{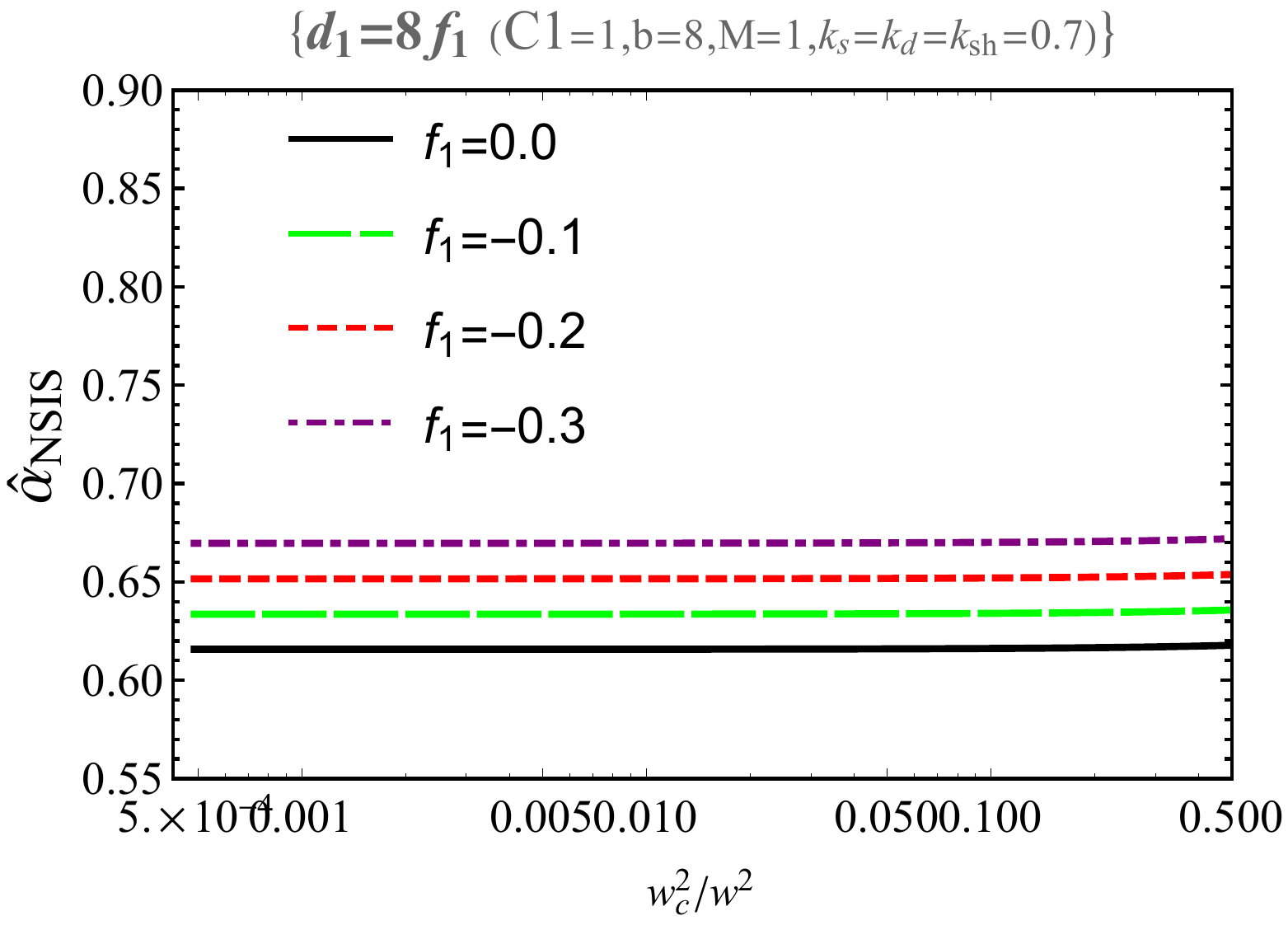}
    \includegraphics[scale=0.26]{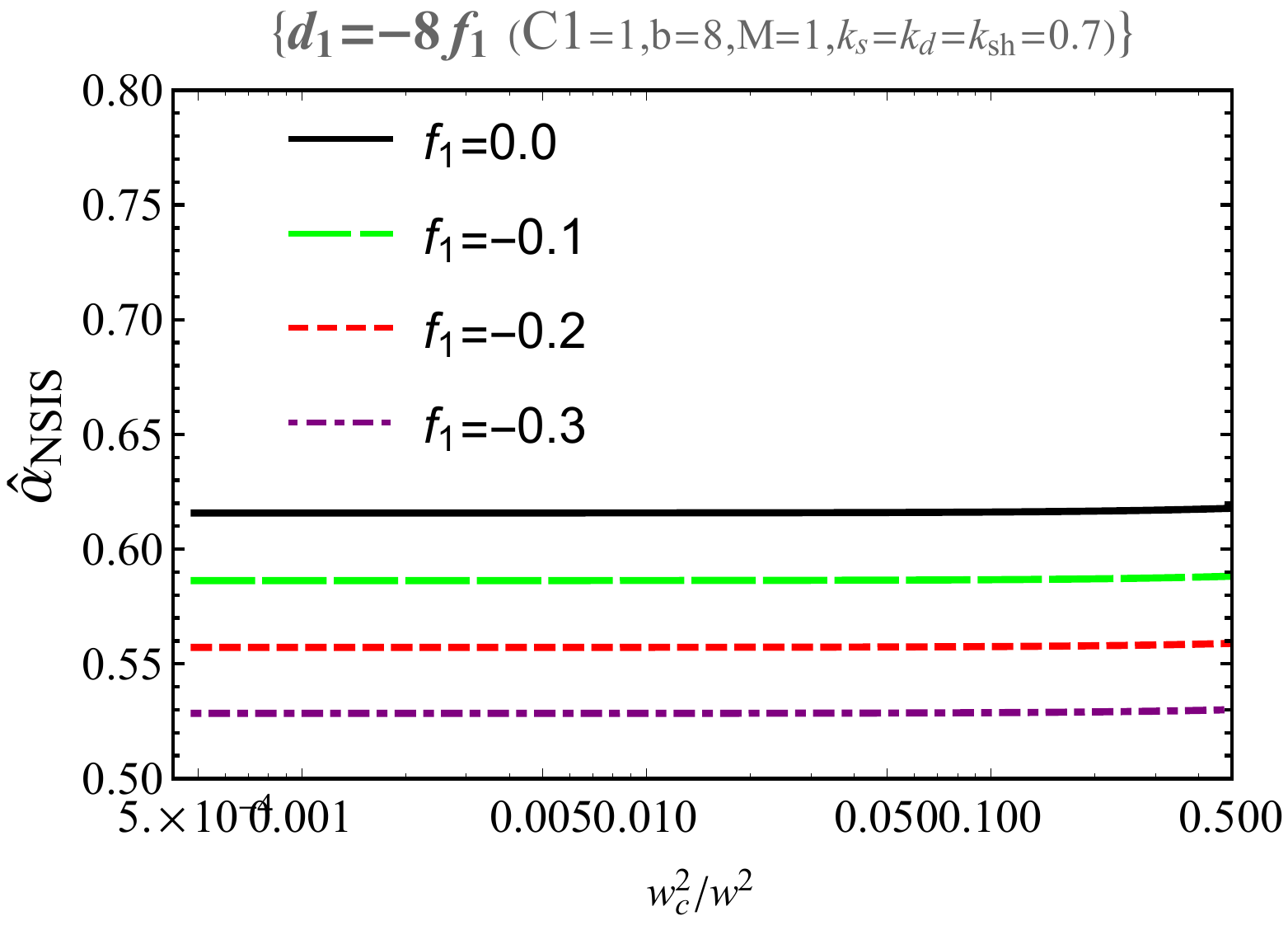}
    \includegraphics[scale=0.26]{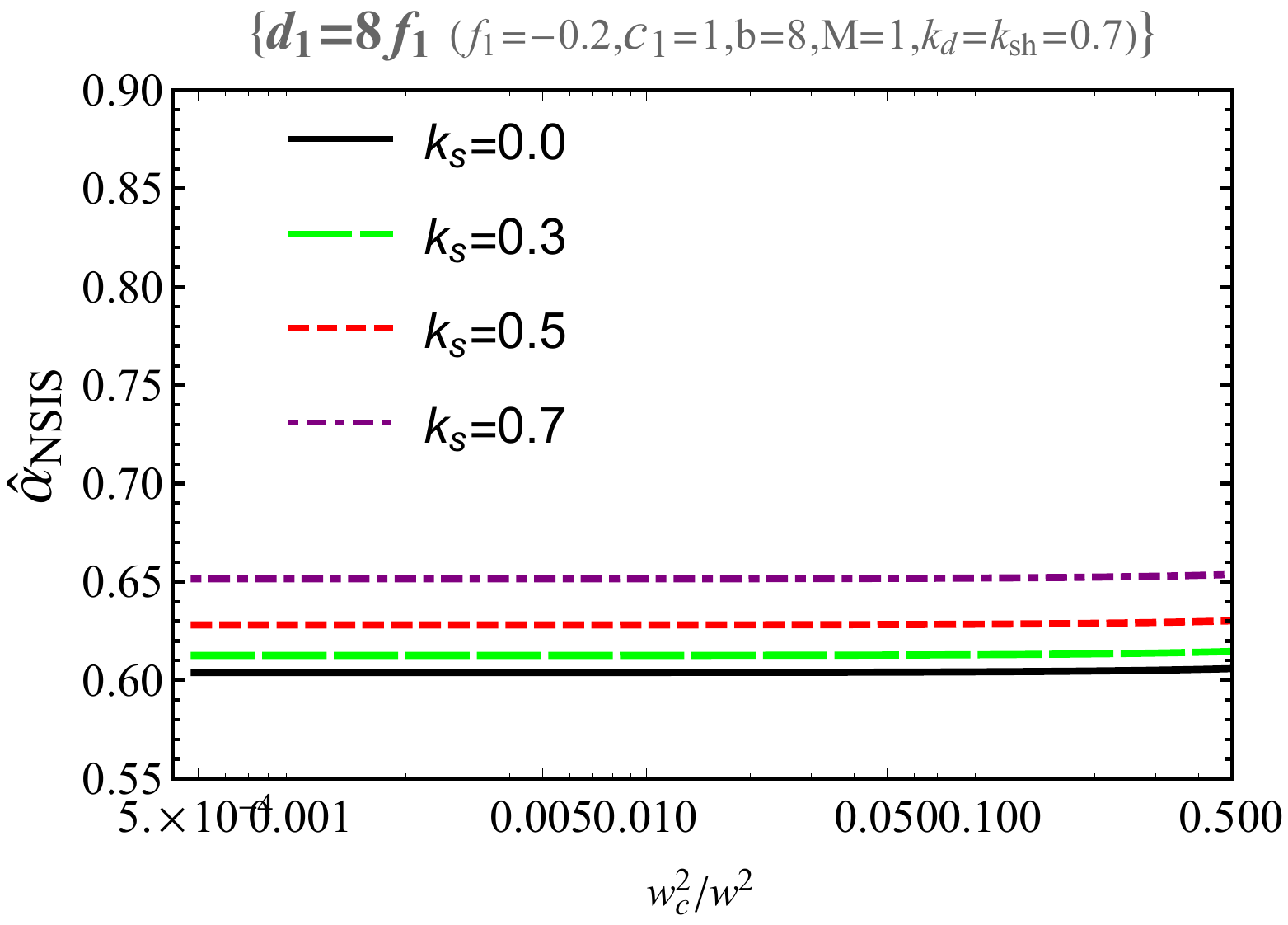}
    \includegraphics[scale=0.26]{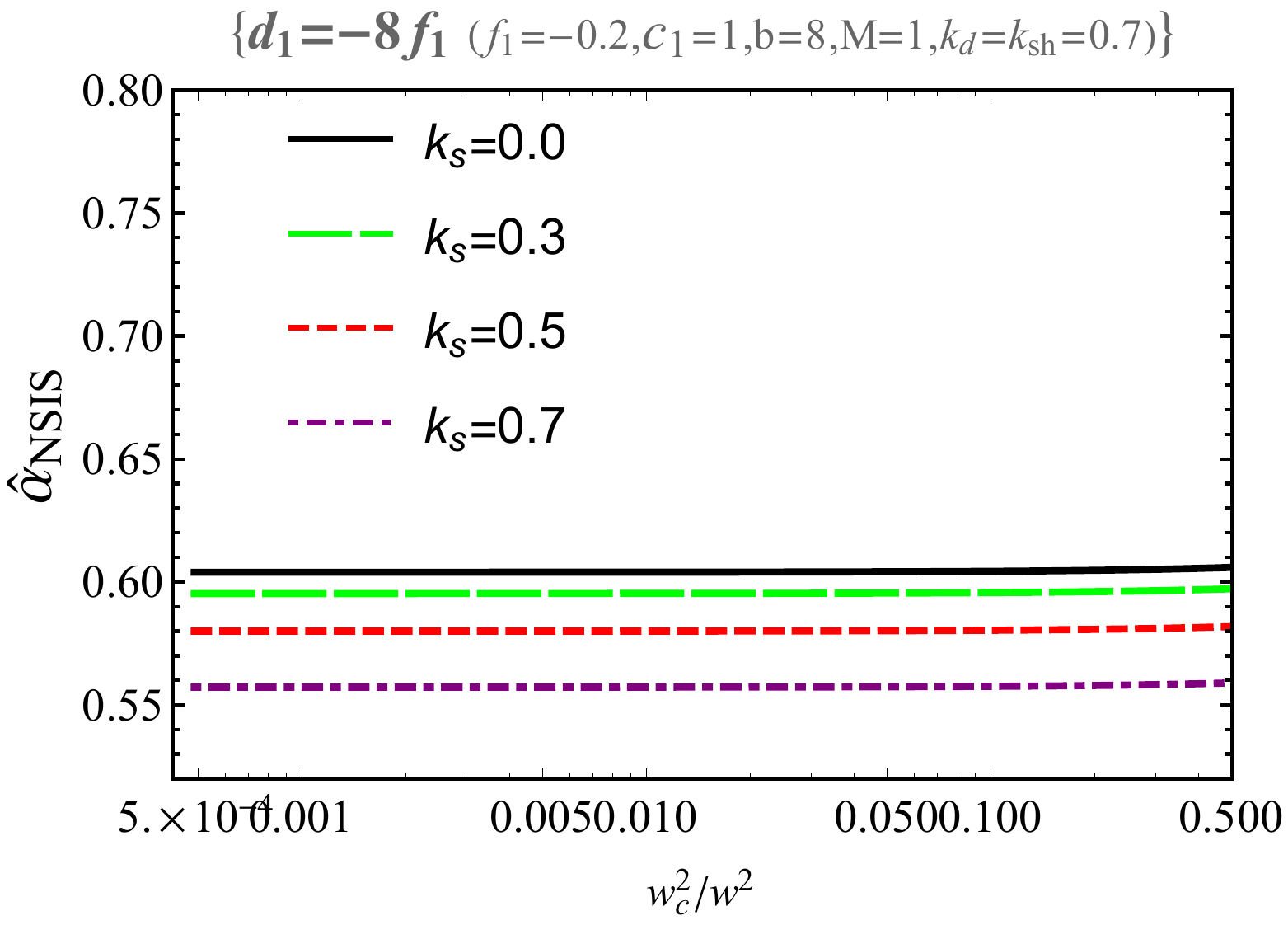}
    \includegraphics[scale=0.26]{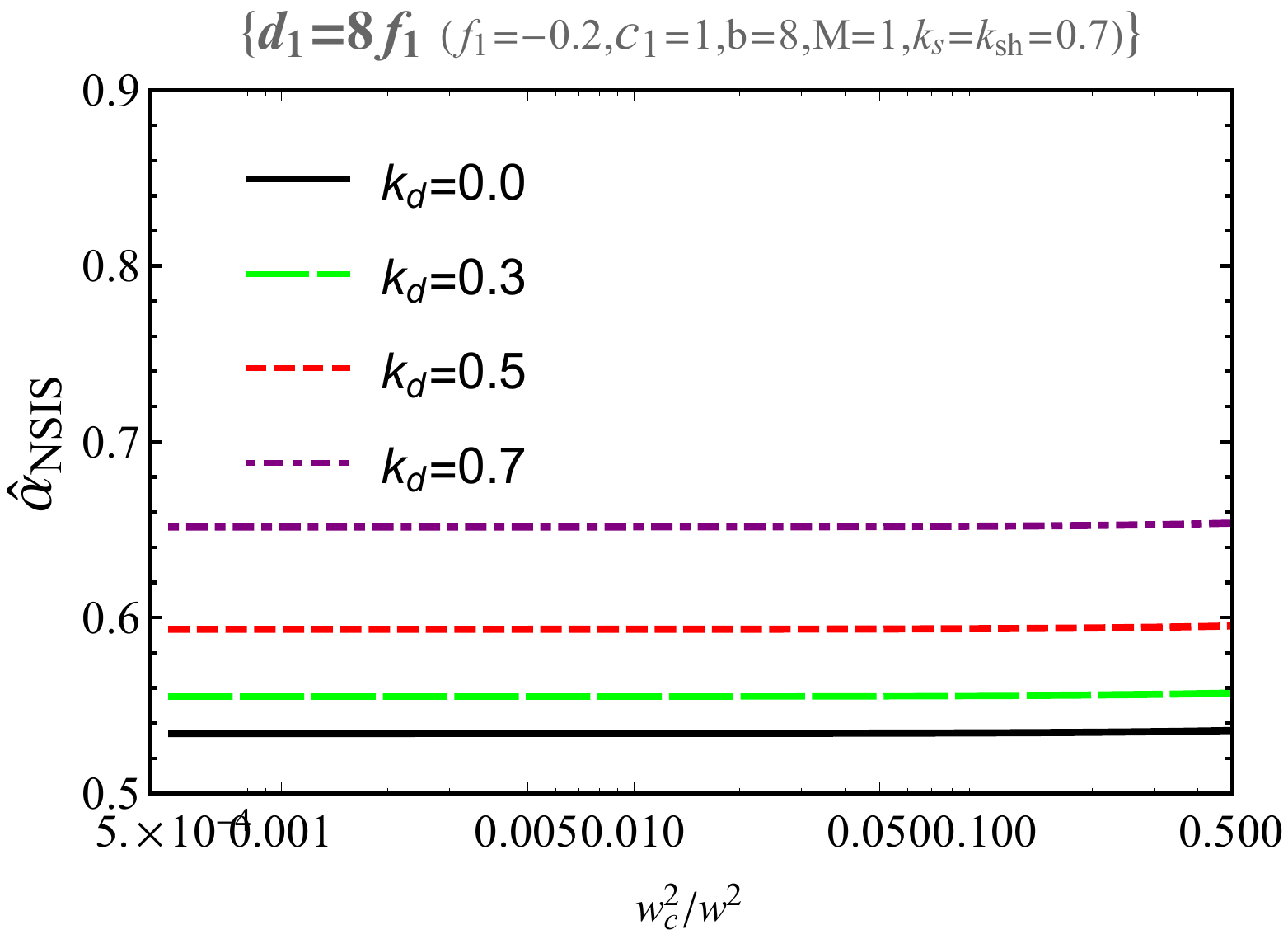}
    \includegraphics[scale=0.26]{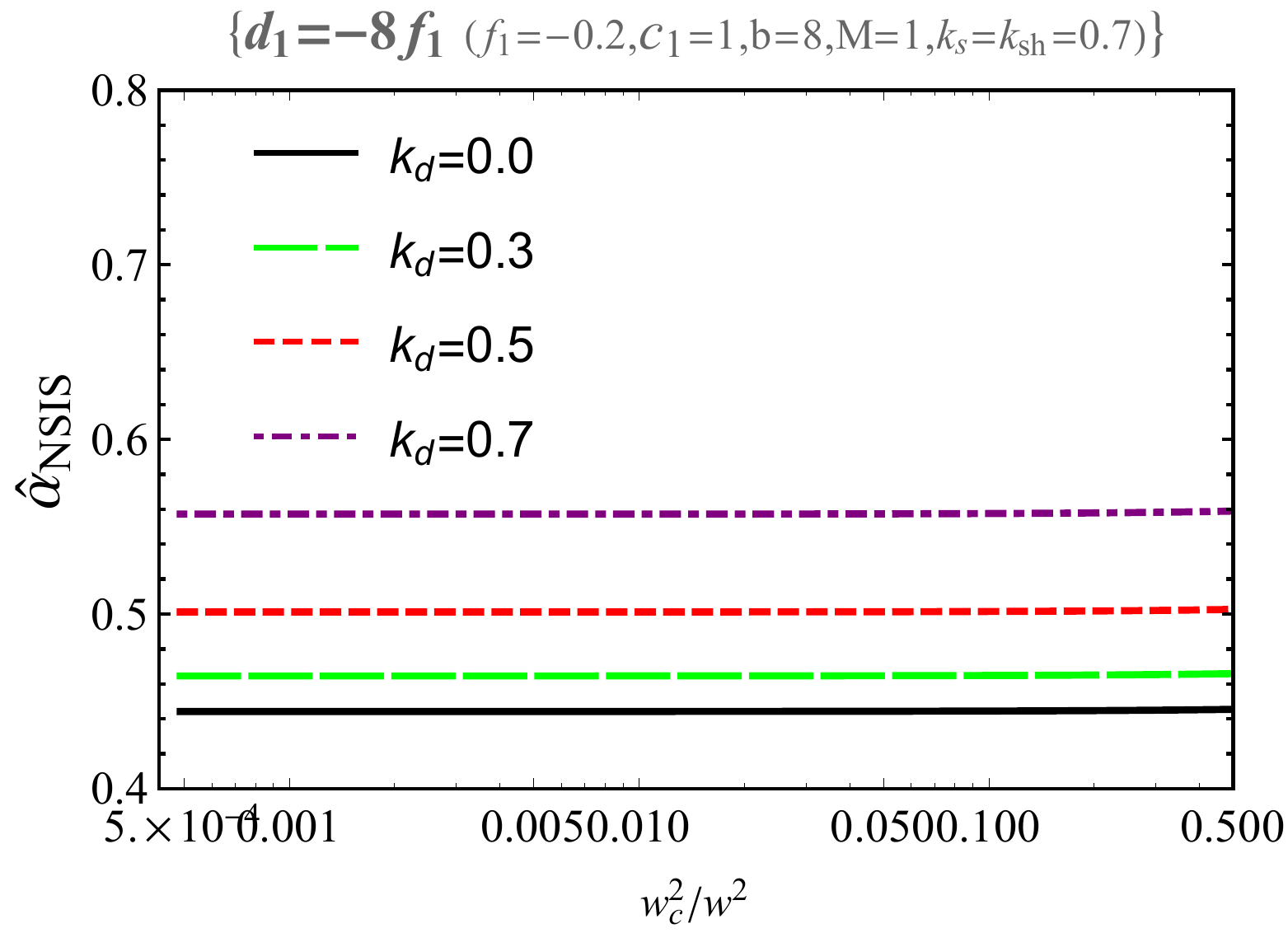}
    \includegraphics[scale=0.26]{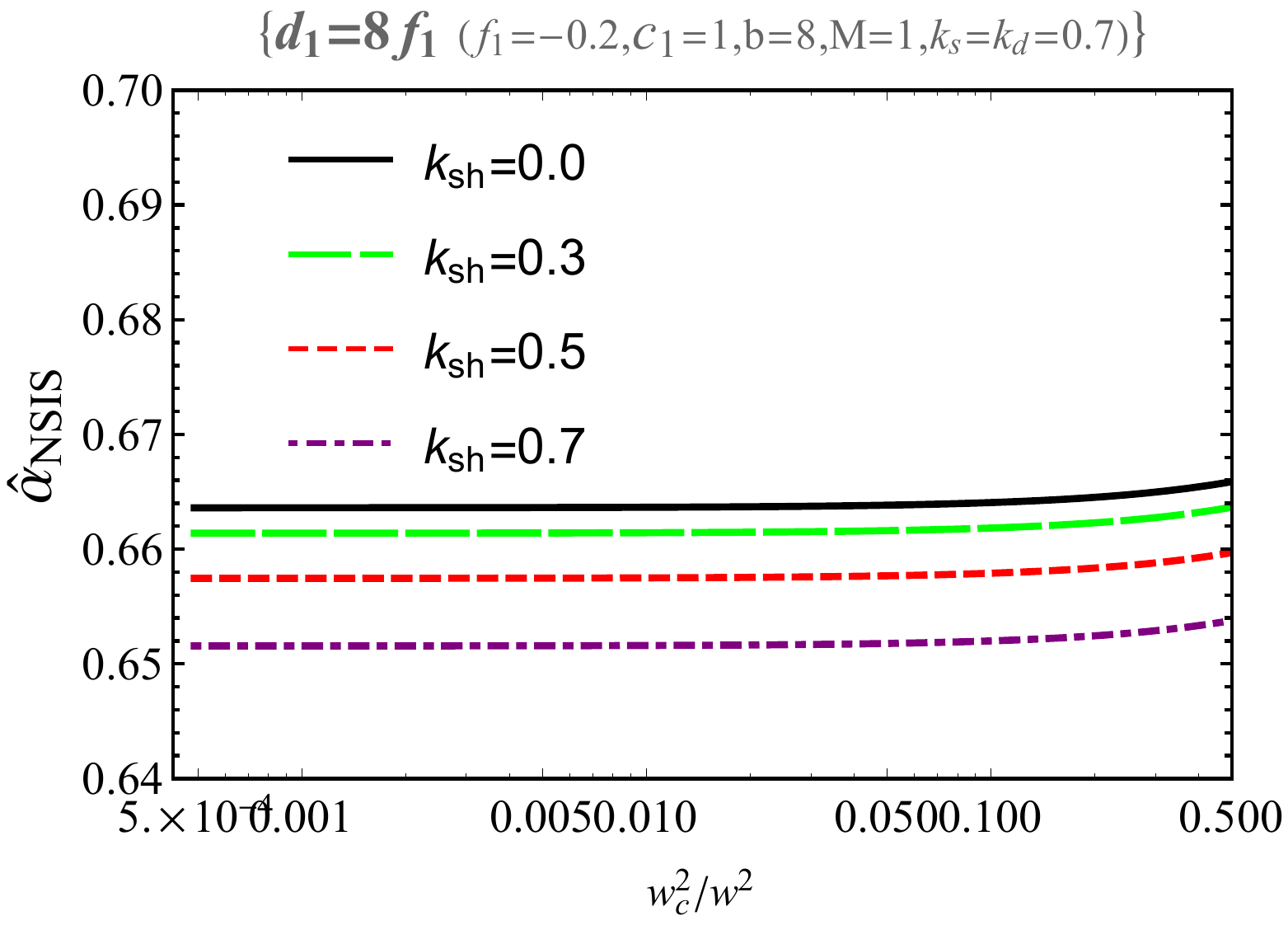}
    \includegraphics[scale=0.26]{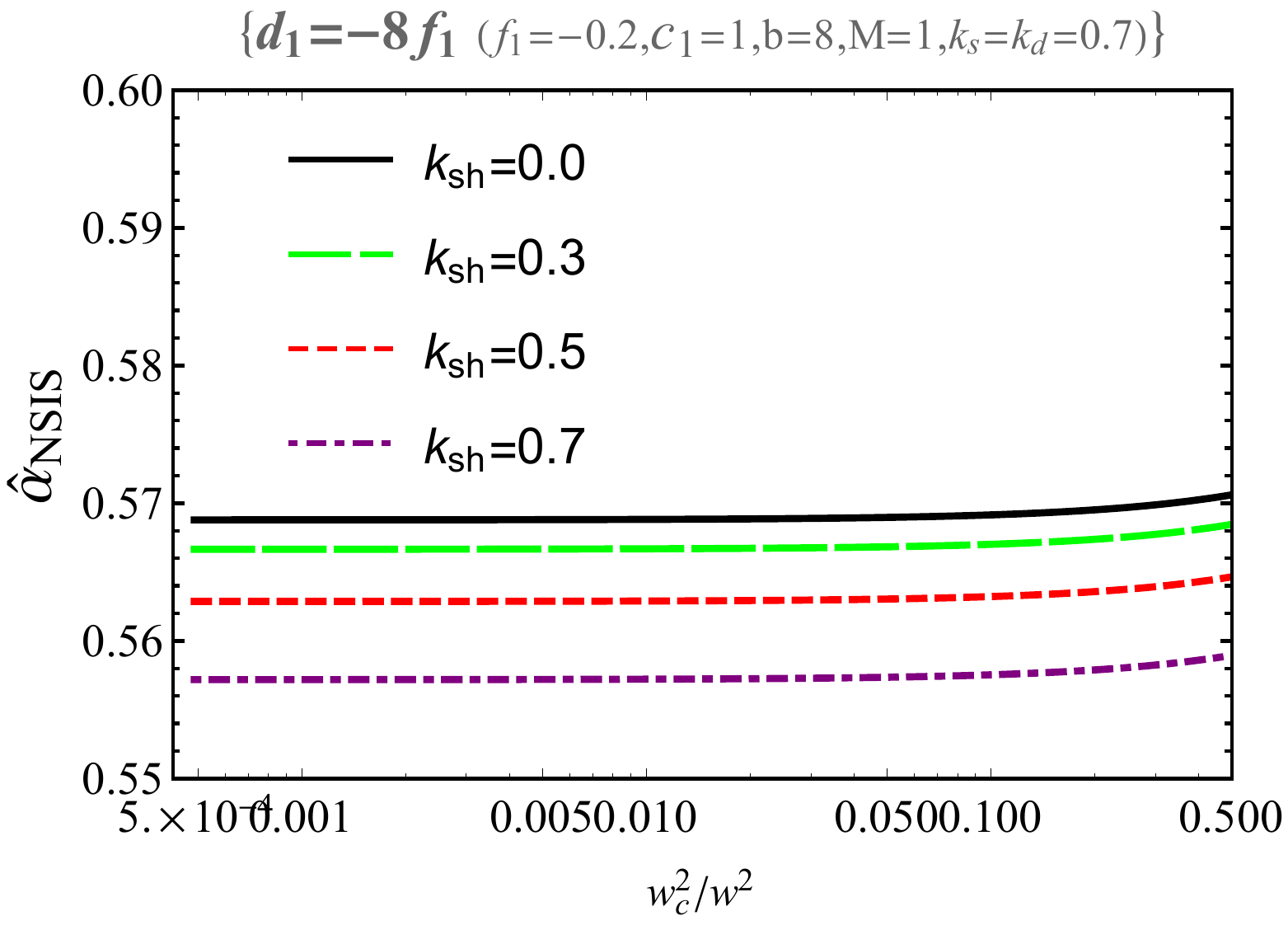}
    \includegraphics[scale=0.26]{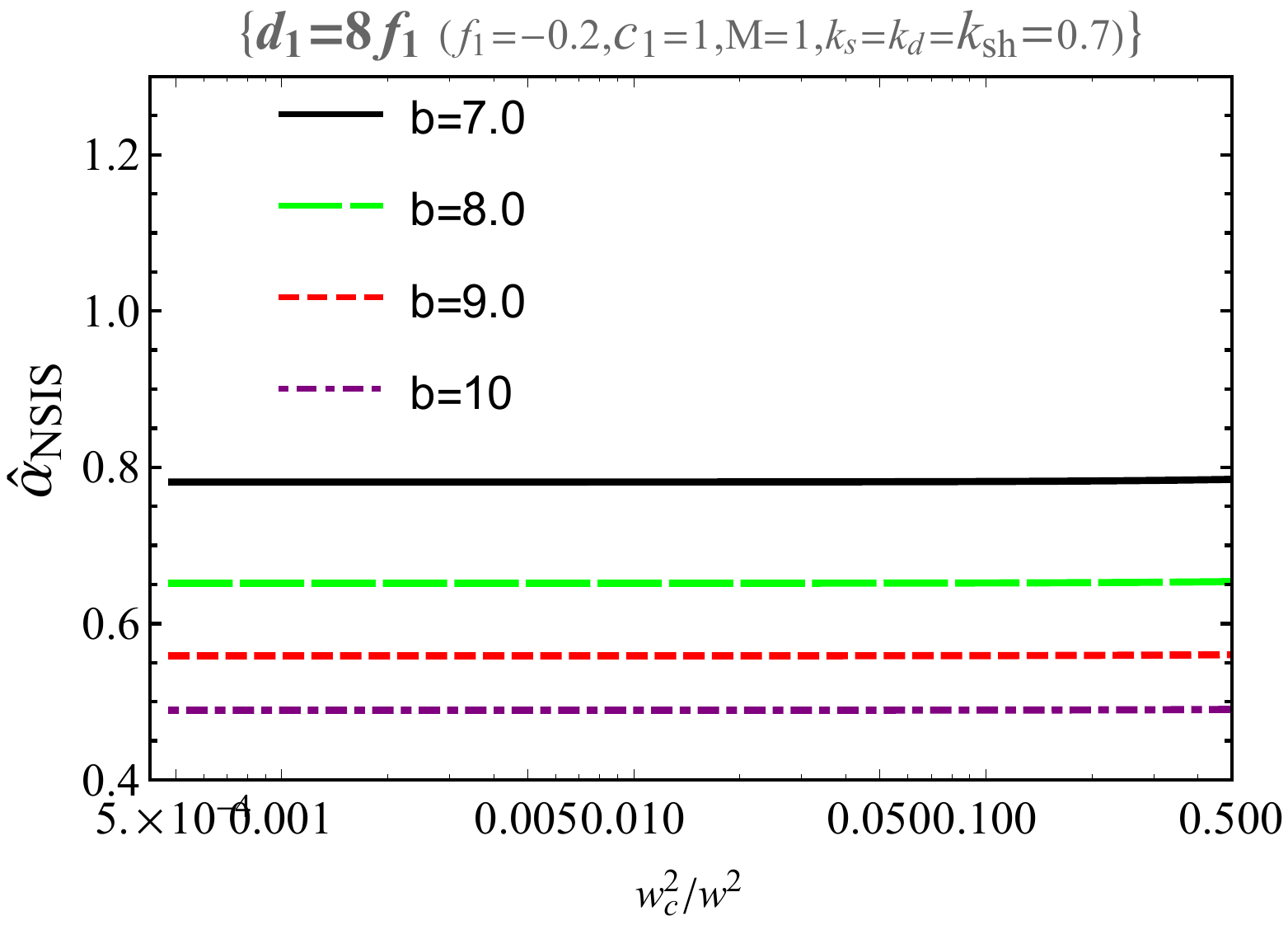}
    \includegraphics[scale=0.26]{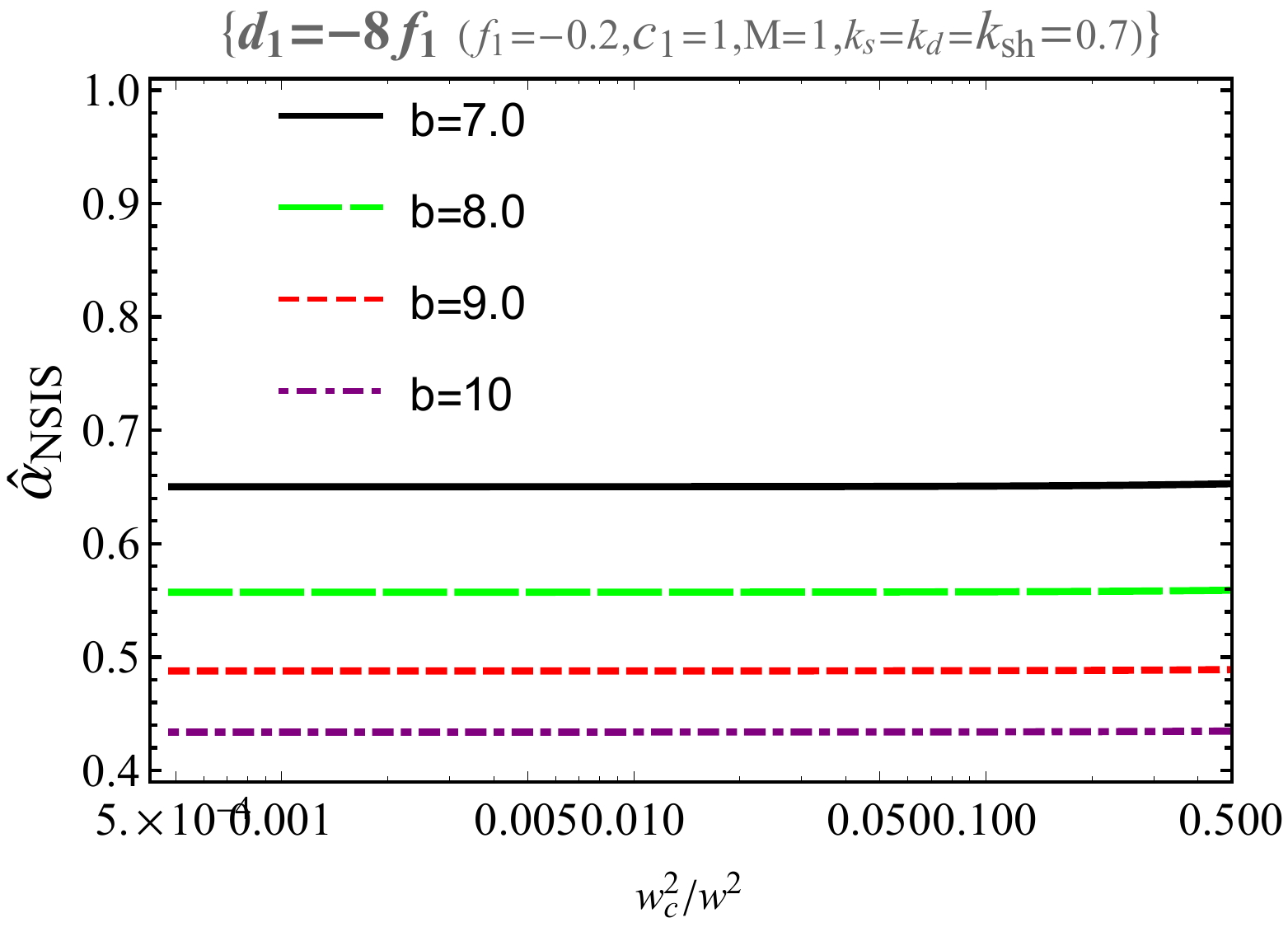}
    \includegraphics[scale=0.26]{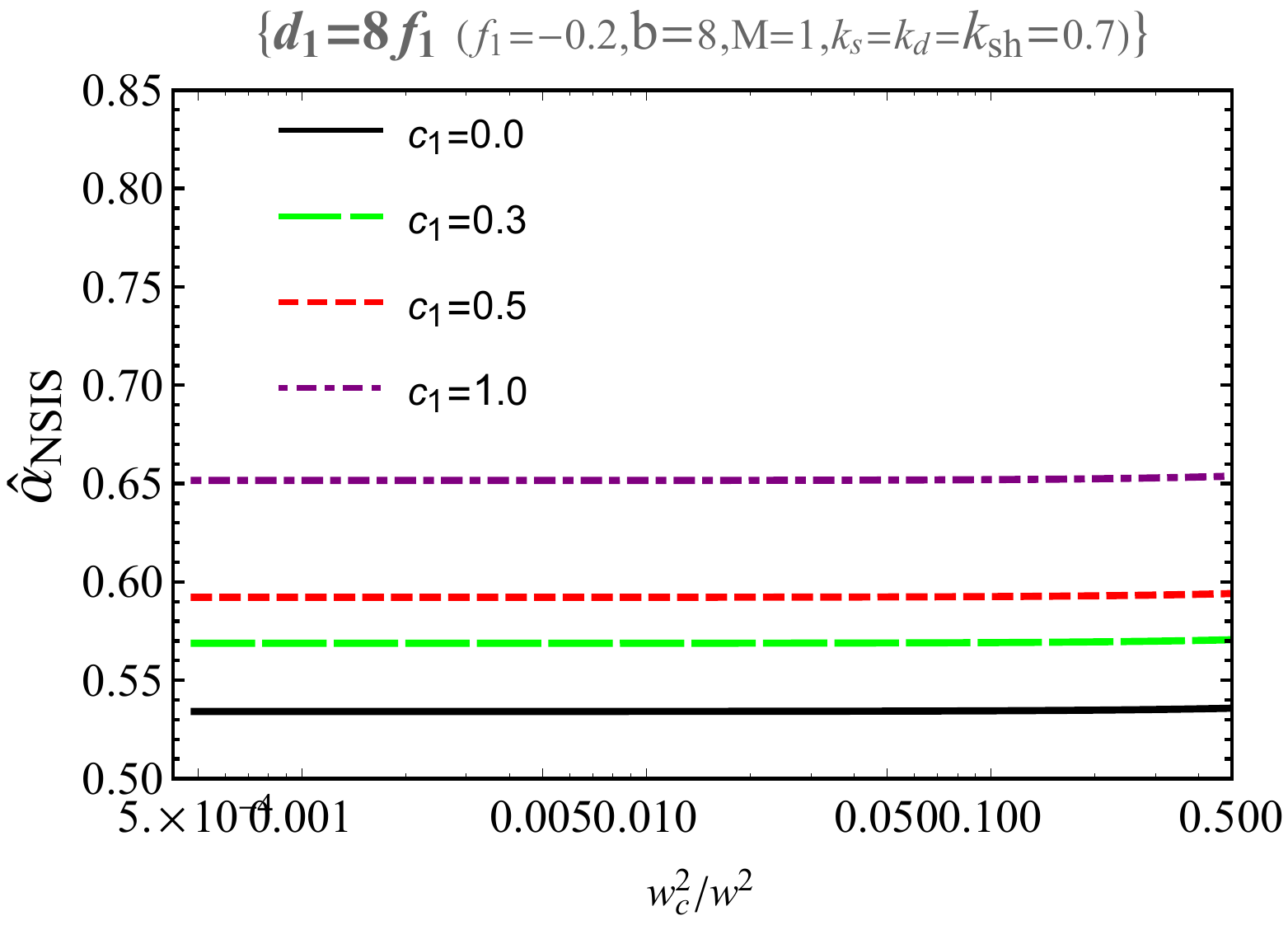}
    \includegraphics[scale=0.26]{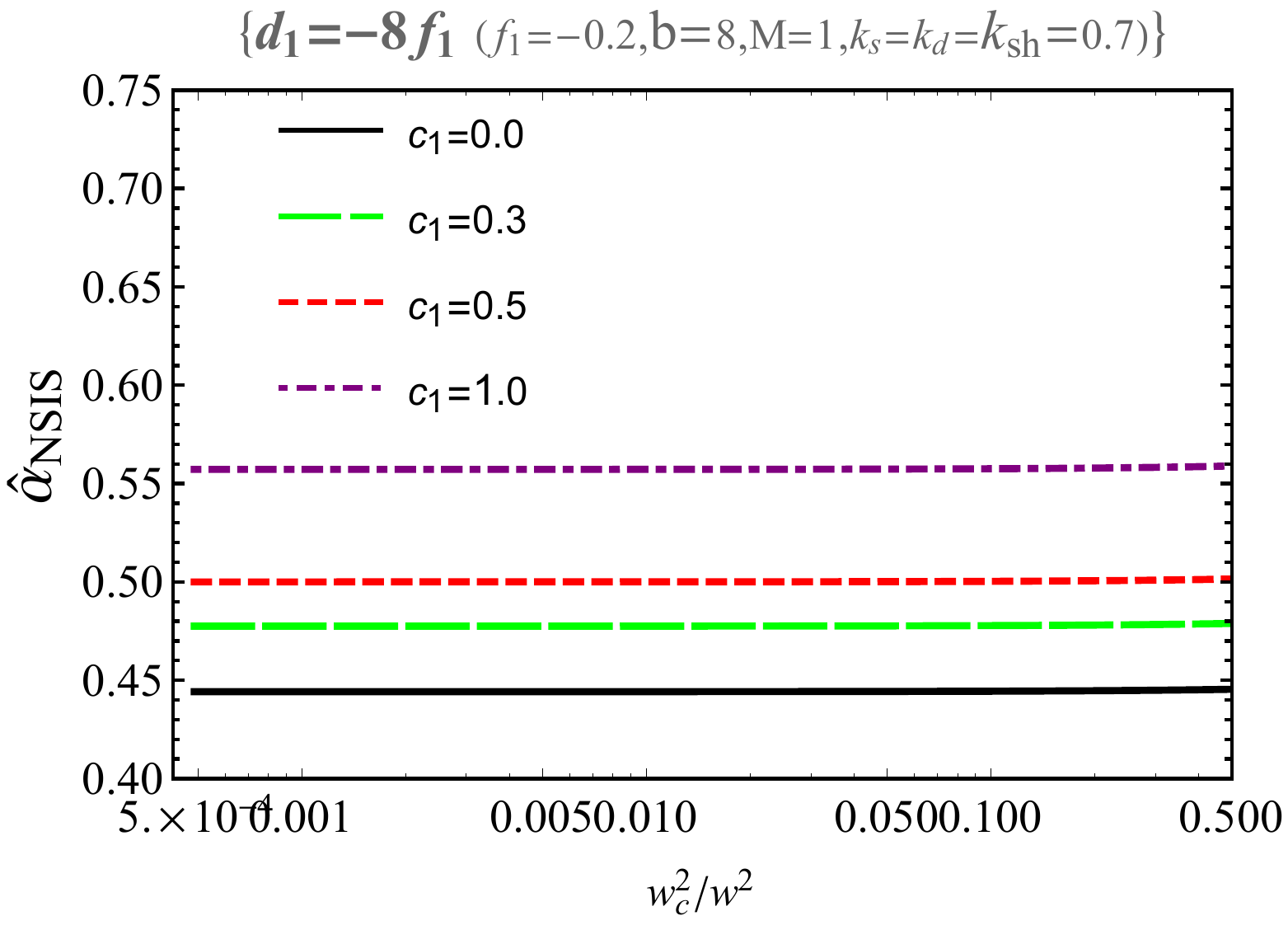}
    \caption{The deflection angle $\hat{\alpha}_{NSIS}$ for  $d_1=8f_1$ (Left panel) and $d_1=-8f_1$ (Right panel) along $c_1$ taking different values of $f_1,\; k_s,\; k_d, \;\&\; k_{sh}.$}
    \label{plot:18}
    \end{figure}
 \begin{figure}
    \centering
    \includegraphics[scale=0.53]{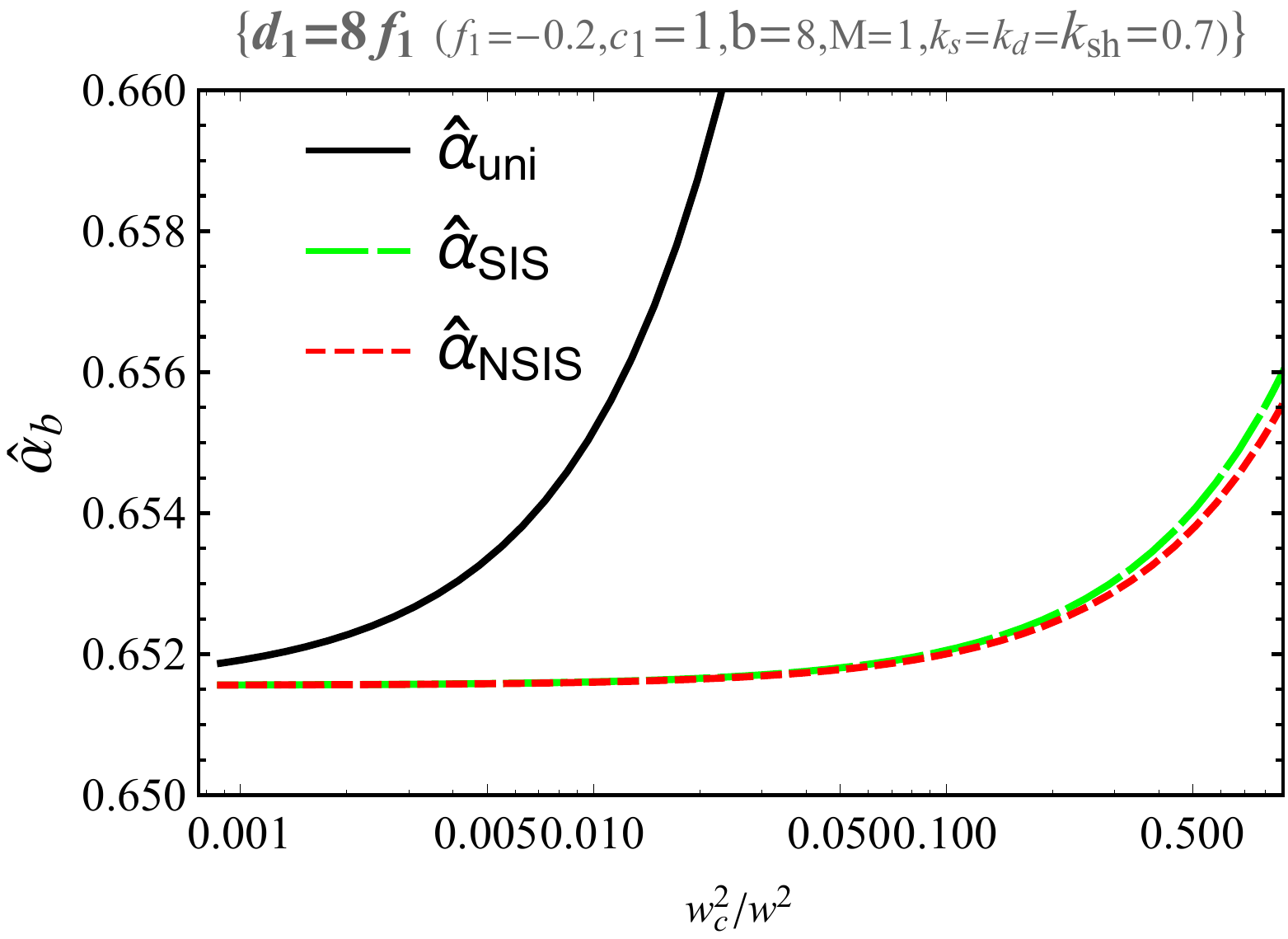}
    \includegraphics[scale=0.53]{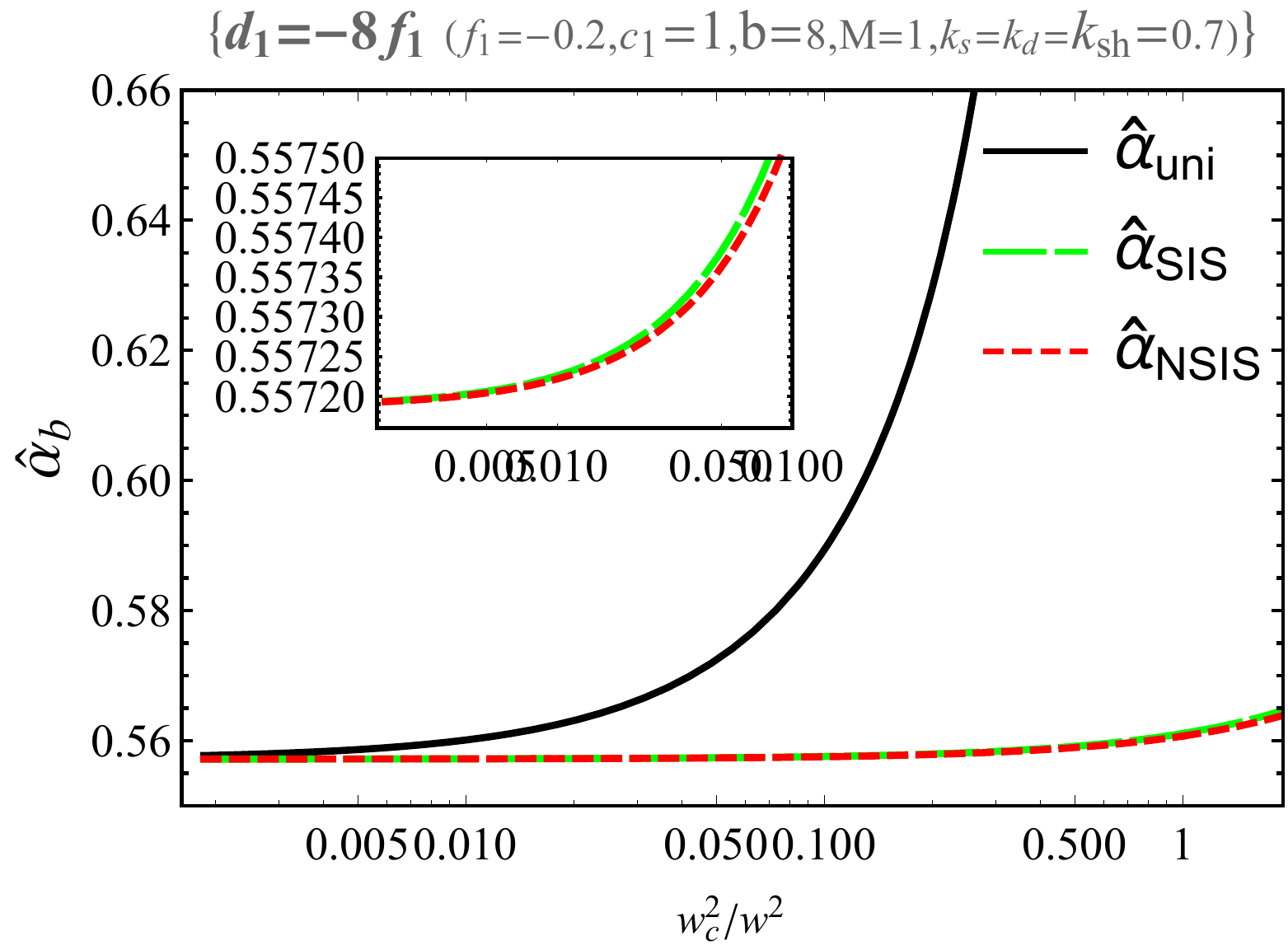}
    \includegraphics[scale=0.53]{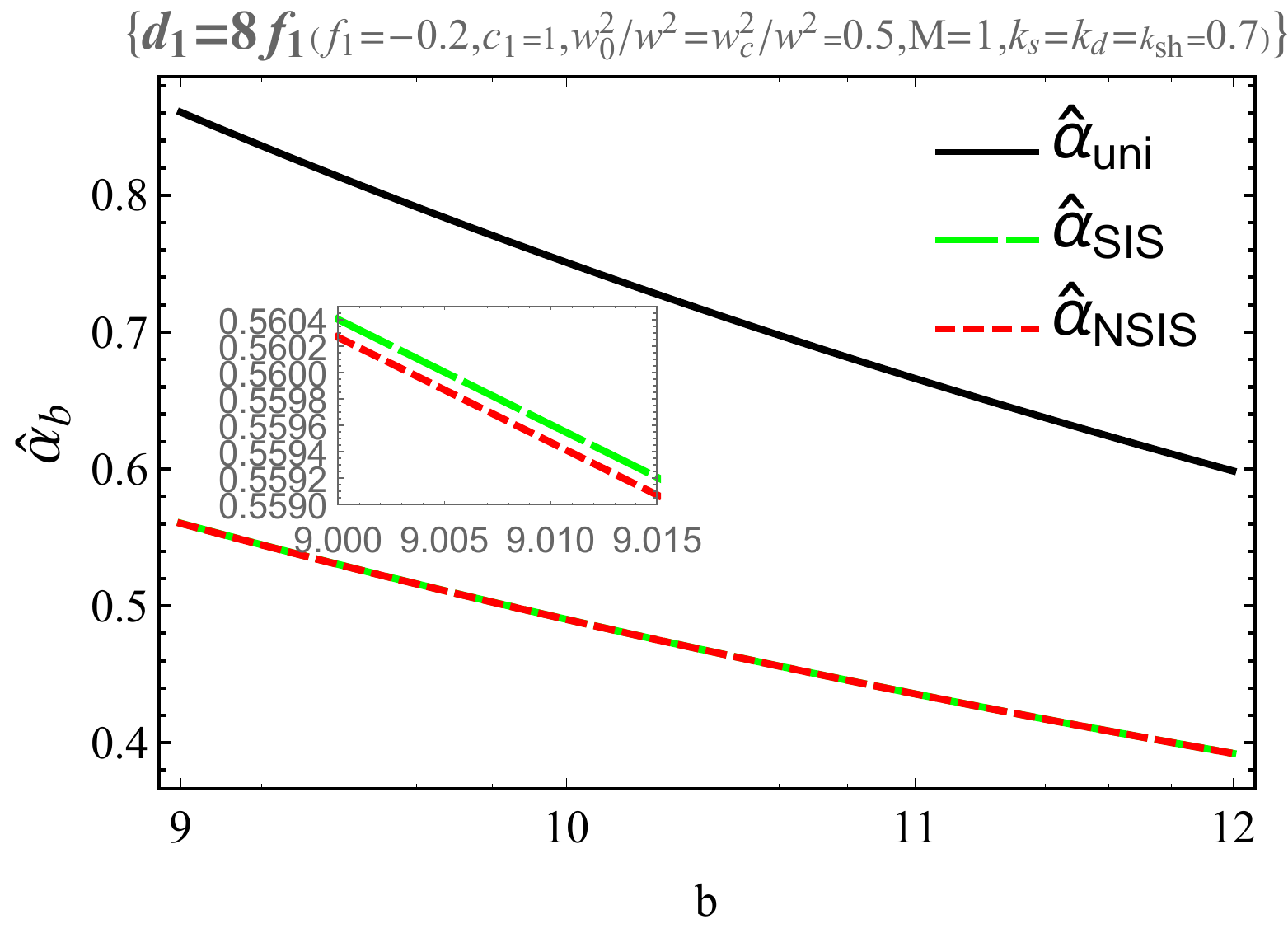}
    \includegraphics[scale=0.53]{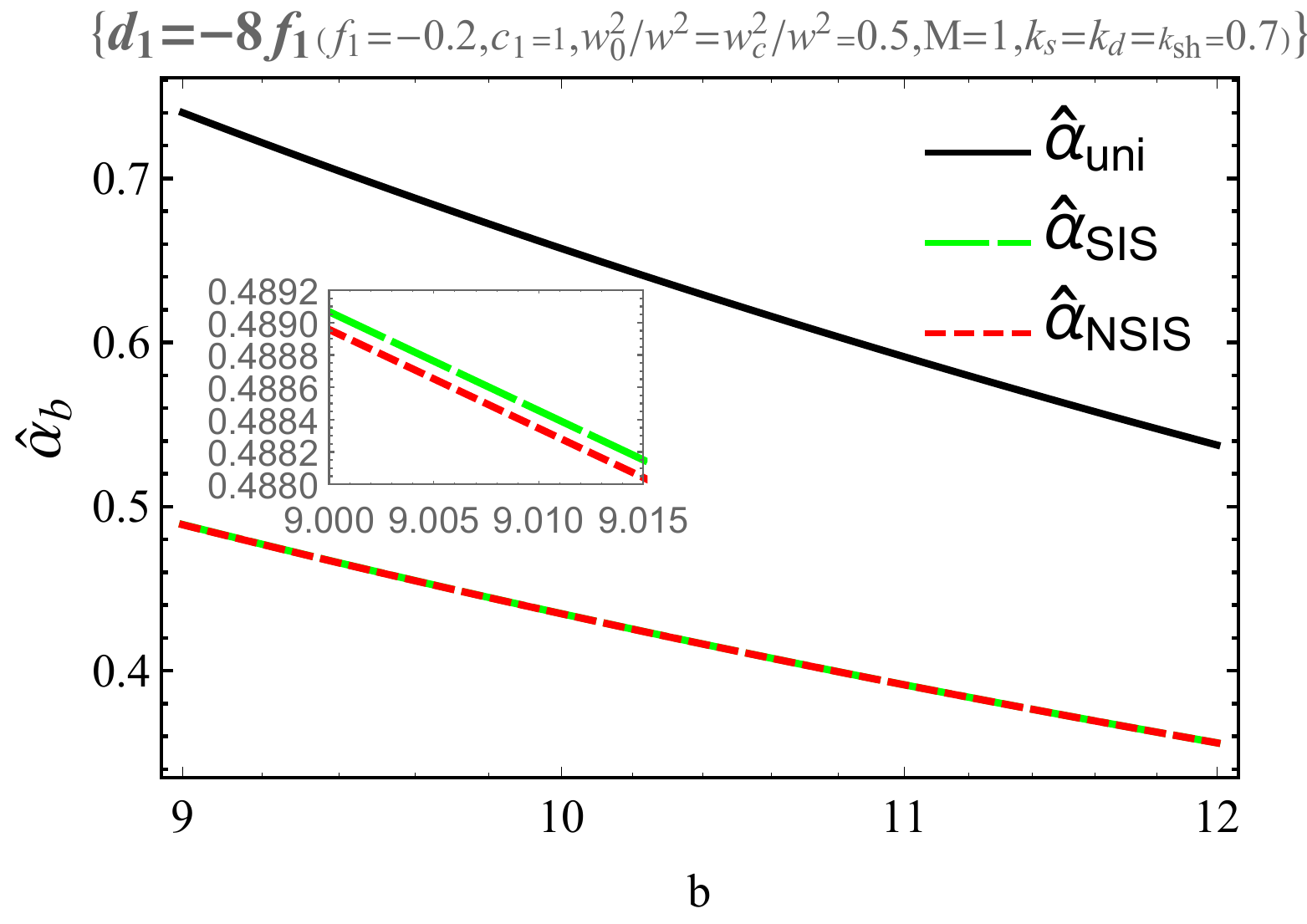}
    \includegraphics[scale=0.53]{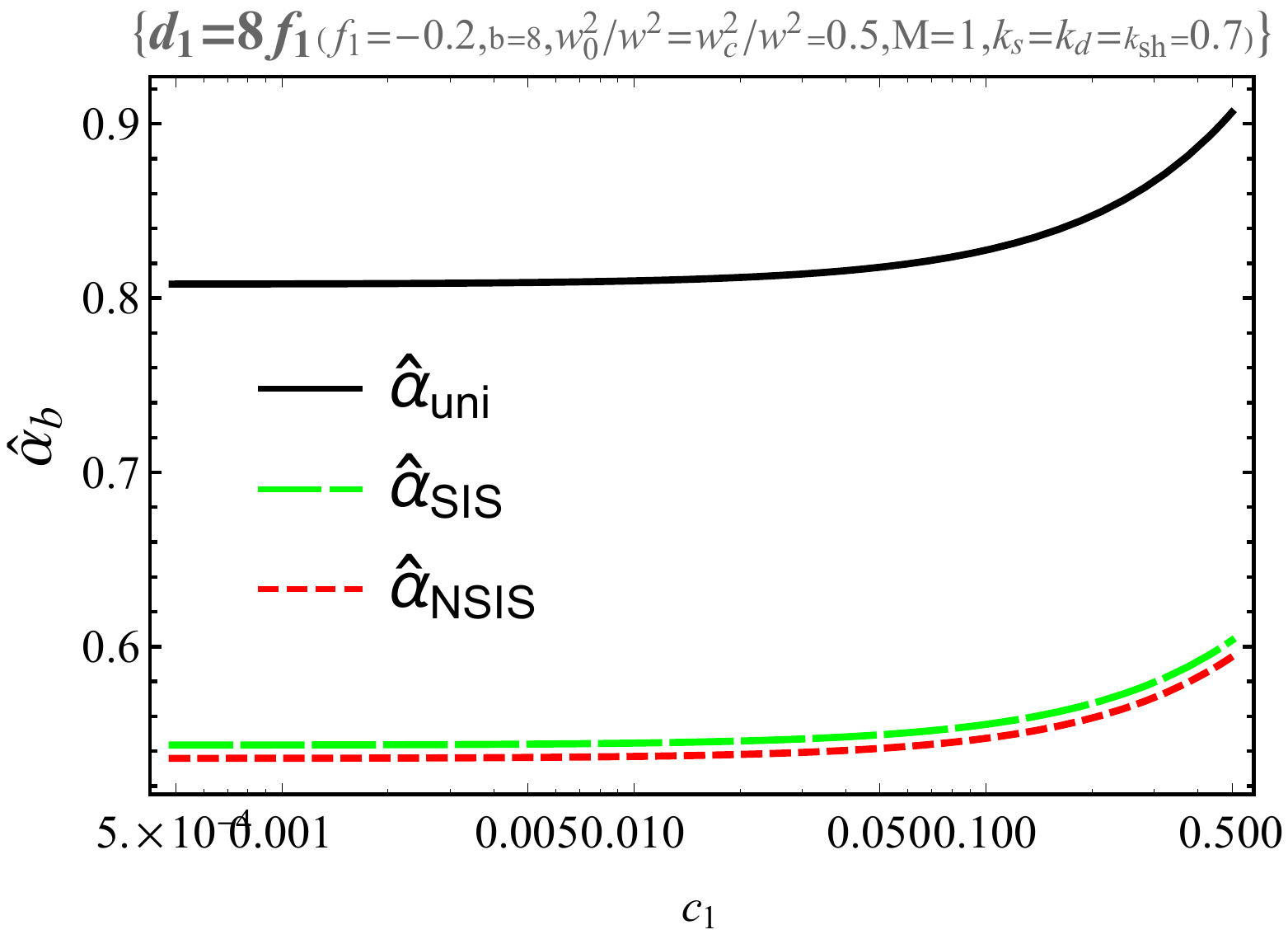}
    \includegraphics[scale=0.55]{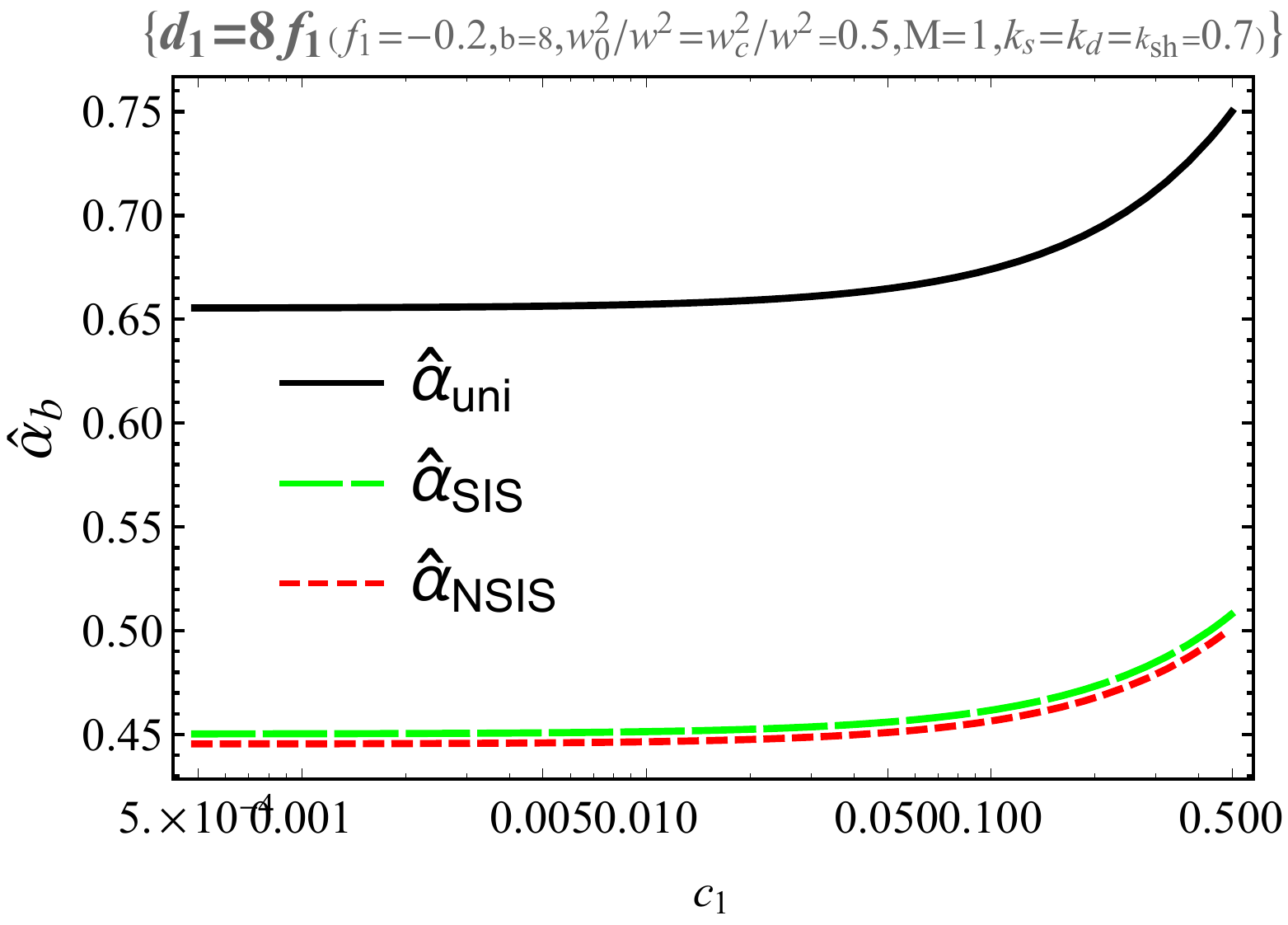}
    \caption{Comparison of deflection angles $\hat{\alpha}_{uni}>\hat{\alpha}_{SIS}>\hat{\alpha}_{NSIS}$ in uniform, $SIS$ and $NSIS$ plasma for $d_1=8f_1$ (Left panel) and $d_1=-8f_1$ (Right panel) along $c_1$ taking different values of $f_1,\; k_s,\; k_d, \;\&\; k_{sh}.$}
    \label{plot:19}
    \end{figure}

\subsection{Non-Singular Isothermal gas sphere}\label{A8.3}
We proceed with our discussion to study photon motion in a non-singular isothermal sphere (NSIS) plasma field, which is an appropriate approximation for the setup of physical analysis. Contrary to SIS plasma, the singularity restricted by the definite core comes up as the origin of gas cloud sphere whose density distribution can be elaborated as \cite{152,153}:
\begin{equation}\label{pa}
    \rho(r)=\frac{\sigma^2_v}{2\pi(r^2+r_c^2)}=\frac{\rho_0}{(1+\frac{r^2}{r_c^2})},\;\;\rho_0=\frac{\sigma^2_v}{2\pi r_c^2},
 \end{equation}
where $r_c$ notions the core radius. Using Eq.(\ref{pl}), the concentration of plasma for NSIS results in the following form:
\begin{equation}\label{pb}
    N(r)=\frac{\sigma^2_v}{2\pi k m_p (r^2+r_c^2)}.
 \end{equation}
Using Eqs.(\ref{pll}), (\ref{pa}) and (\ref{pb}), the plasma frequency $w_e$ yields
\begin{equation}\label{pc}
    w_e^2=\frac{K_e\sigma^2_v}{2\pi k m_p (r^2+r_c^2)}.
 \end{equation}
 The deviation of photons in NSIS plasma gravitational lensing field produces a deflection angle, which possesses all of its characteristics and can be computed for MAGBH as follows
\begin{eqnarray}
   \hat{\alpha}_{NSIS}&=&\frac{1}{48 b^3 \pi  r_{c}^6 \left(b^2+r_{c}^2\right)^{3/2} w^2}\Big[-9 \left(4 c_{1} k_{d}^2+2 f_{1} k_{sh}^2-d_{1} k_{s}^2\right) \left(4 b^2+20 c_{1} k_{d}^2+10 f_{1} k_{sh}^2-5 d_{1} k_{s}^2\right)\nonumber\\&& \pi ^2 r_{c}^6 w^2 \sqrt{\frac{1}{b^2}} \left(b^2+r_{c}^2\right)^{3/2}-32 \left(3 b^2+32 c_{1} k_{d}^2+16 f_{1} k_{sh}^2-8 d_{1} k_{s}^2\right) \pi  r_{c}^6 w^2 R_s \left(b^2+r_{c}^2\right)^{3/2}-24 r_{c}\nonumber\\&& \Big[\left(24 c_{1} k_{d}^2+12 f_{1} k_{sh}^2-6 d_{1} k_{s}^2-r_{c}^2\right) \log \left(\frac{r_{c}}{\sqrt{b^2+r_{c}^2}}+1\right) b^4+\Big[-24 c_{1} k_{d}^2-12 f_{1} k_{sh}^2+6 d_{1} k_{s}^2+r_{c}^2\Big]\nonumber\\&& \log \left(1-\frac{r_{c}}{\sqrt{b^2+r_{c}^2}}\right) b^4+2 r_{c} \Big[\left(-24 c_{1} k_{d}^2-12 f_{1} k_{sh}^2+6 d_{1} k_{s}^2+r_{c}^2\right) b^2+4 \left(4 c_{1} k_{d}^2+2 f_{1} k_{sh}^2-d_{1} k_{s}^2\right) \nonumber\\&& r_{c}^2\Big] \sqrt{b^2+r_{c}^2}\Big] R_s^3 w_c^2 \left(b^2+r_{c}^2\right)-12 \sqrt{\frac{1}{b^2}} \pi  \Big[4 \Big[-32 c_{1}^2 k_{d}^4+4 c_{1} \left(-8 f_{1} k_{sh}^2+4 d_{1} k_{s}^2+r_{c}^2\right) k_{d}^2-8 f_{1}^2 k_{sh}^4\nonumber\\&-&d_{1} k_{s}^2 \left(2 d_{1} k_{s}^2+r_{c}^2\right)+2 f_{1} k_{sh}^2 \left(4 d_{1} k_{s}^2+r_{c}^2\right)\Big] \sqrt{\frac{1}{b^2}} \sqrt{\frac{1}{b^2+r_{c}^2}} \sqrt{b^2+r_{c}^2} b^8+2 \Big[\sqrt{\frac{1}{b^2}} r_{c}^6+2 \Big[4 c_{1} k_{d}^2+2 f_{1} k_{sh}^2\nonumber\\&-&d_{1} k_{s}^2\Big] \sqrt{\frac{1}{b^2}} \sqrt{\frac{1}{b^2+r_{c}^2}} \sqrt{b^2+r_{c}^2} r_{c}^4-2 \left(4 c_{1} k_{d}^2+2 f_{1} k_{sh}^2-d_{1} k_{s}^2\right) \sqrt{b^2+r_{c}^2} \Big[8 c_{1} \sqrt{\frac{1}{b^2}} \sqrt{\frac{1}{b^2+r_{c}^2}} k_{d}^2\nonumber\\&-&2 \sqrt{\frac{1}{b^2}} d_{1} k_{s}^2 \sqrt{\frac{1}{b^2+r_{c}^2}}+4 f_{1} k_{sh}^2 \sqrt{\frac{1}{b^2}} \sqrt{\frac{1}{b^2+r_{c}^2}}+1\Big] r_{c}^2+4 \left(4 c_{1} k_{d}^2+2 f_{1} k_{sh}^2-d_{1} k_{s}^2\right)^2 \sqrt{b^2+r_{c}^2}\Big] b^6\nonumber\\&+&2 r_{c}^2 \left(32 c_{1}^2 k_{d}^4+4 c_{1} \left(8 f_{1} k_{sh}^2-4 d_{1} k_{s}^2-r_{c}^2\right) k_{d}^2+8 f_{1}^2 k_{sh}^4+d_{1} k_{s}^2 \left(2 d_{1} k_{s}^2+r_{c}^2\right)-2 f_{1} k_{sh}^2 \left(4 d_{1} k_{s}^2+r_{c}^2\right)\right) \nonumber\\&&\sqrt{b^2+r_{c}^2} b^4+r_{c}^4 \Big[-16 c_{1}^2 k_{d}^4+8 c_{1} \left(-2 f_{1} k_{sh}^2+d_{1} k_{s}^2+r_{c}^2\right) k_{d}^2-4 f_{1}^2 k_{sh}^4+4 f_{1} k_{sh}^2 \left(d_{1} k_{s}^2+r_{c}^2\right)\nonumber\\&-&d_{1} k_{s}^2 \left(d_{1} k_{s}^2+2 r_{c}^2\right)\Big] \sqrt{b^2+r_{c}^2} b^2+3 \left(4 c_{1} k_{d}^2+2 f_{1} k_{sh}^2-d_{1} k_{s}^2\right)^2 r_{c}^6 \sqrt{b^2+r_{c}^2}\Big] R_s^2 w_c^2\Big].
\end{eqnarray}
The deflection angle $\hat{\alpha}_{NSIS}$ for NSIS plasma is illustrated graphically in Figs.~\ref{17} and \ref{18}. Observing the varying behavior of the deflection angle in the $NSIS$ w.r.t $SIS$ and $UNI$ plasma fields seems difficult. It is also worthwhile to analyze that $\hat\alpha_{uni}>\hat\alpha_{SIS}>\hat\alpha_{NSIS}$ which has been plotted in Fig.~\ref{plot:19}.

\begin{figure}
    \centering
    \includegraphics[scale=0.53]{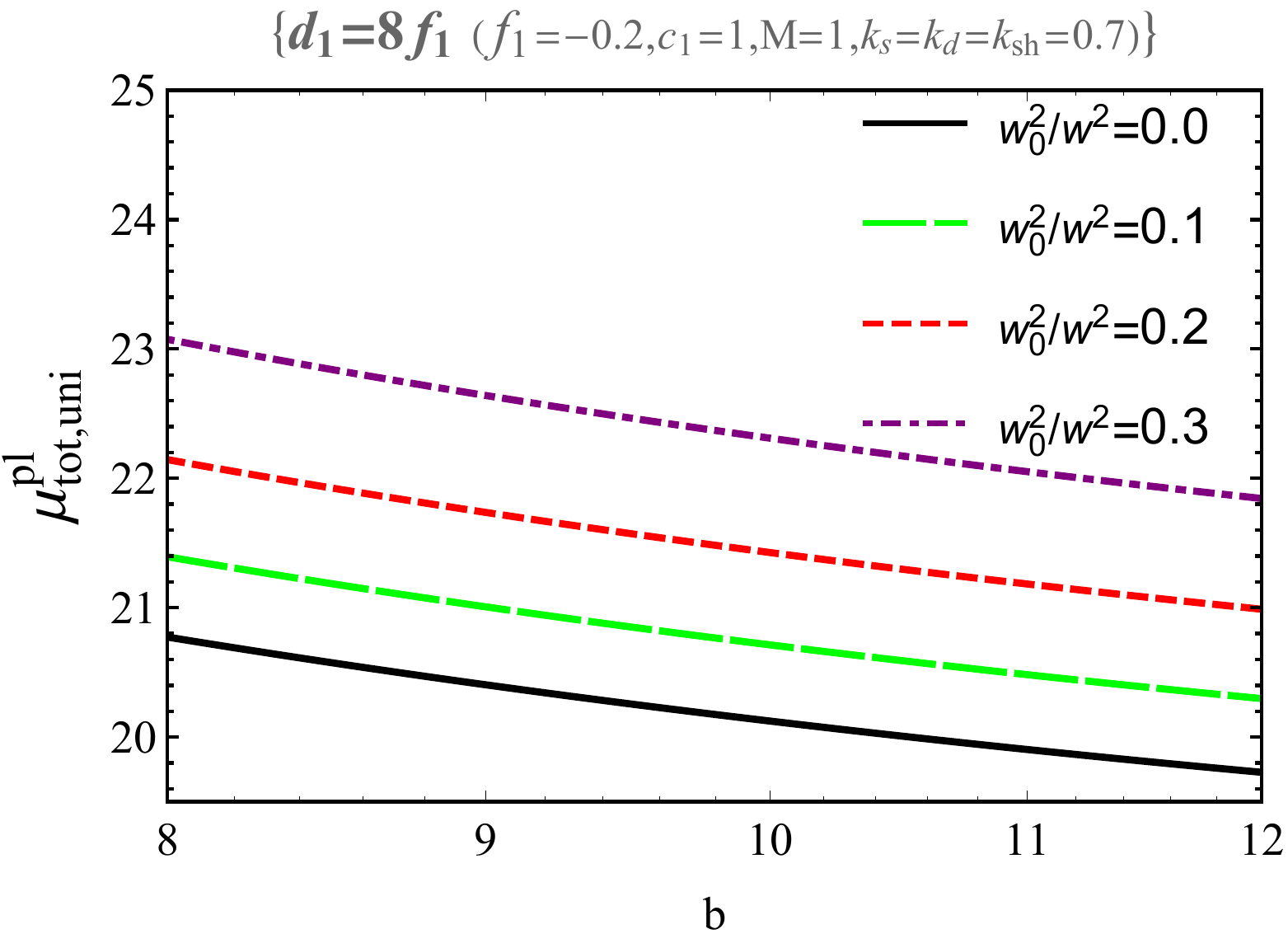}
    \includegraphics[scale=0.53]{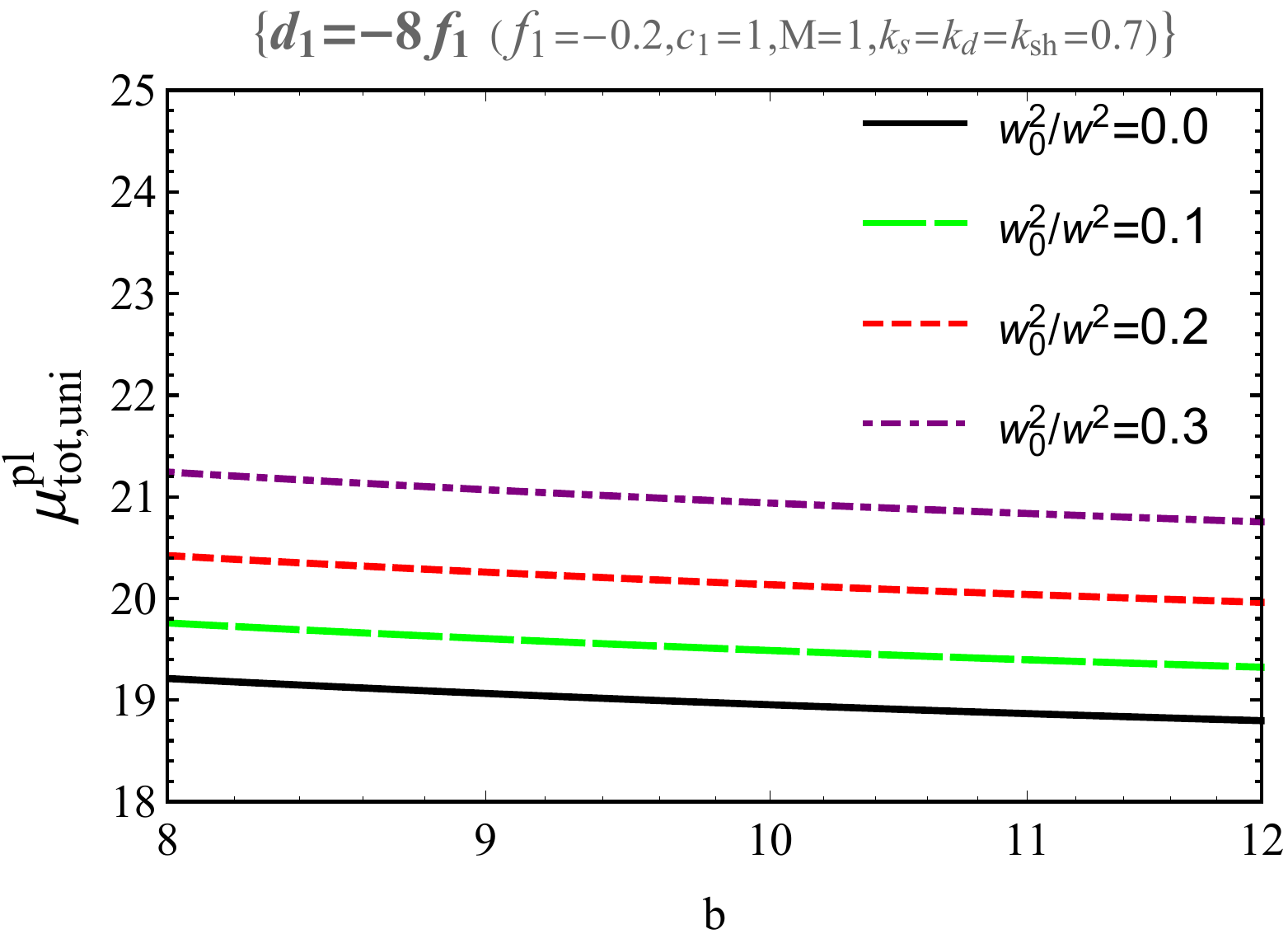}
    \includegraphics[scale=0.53]{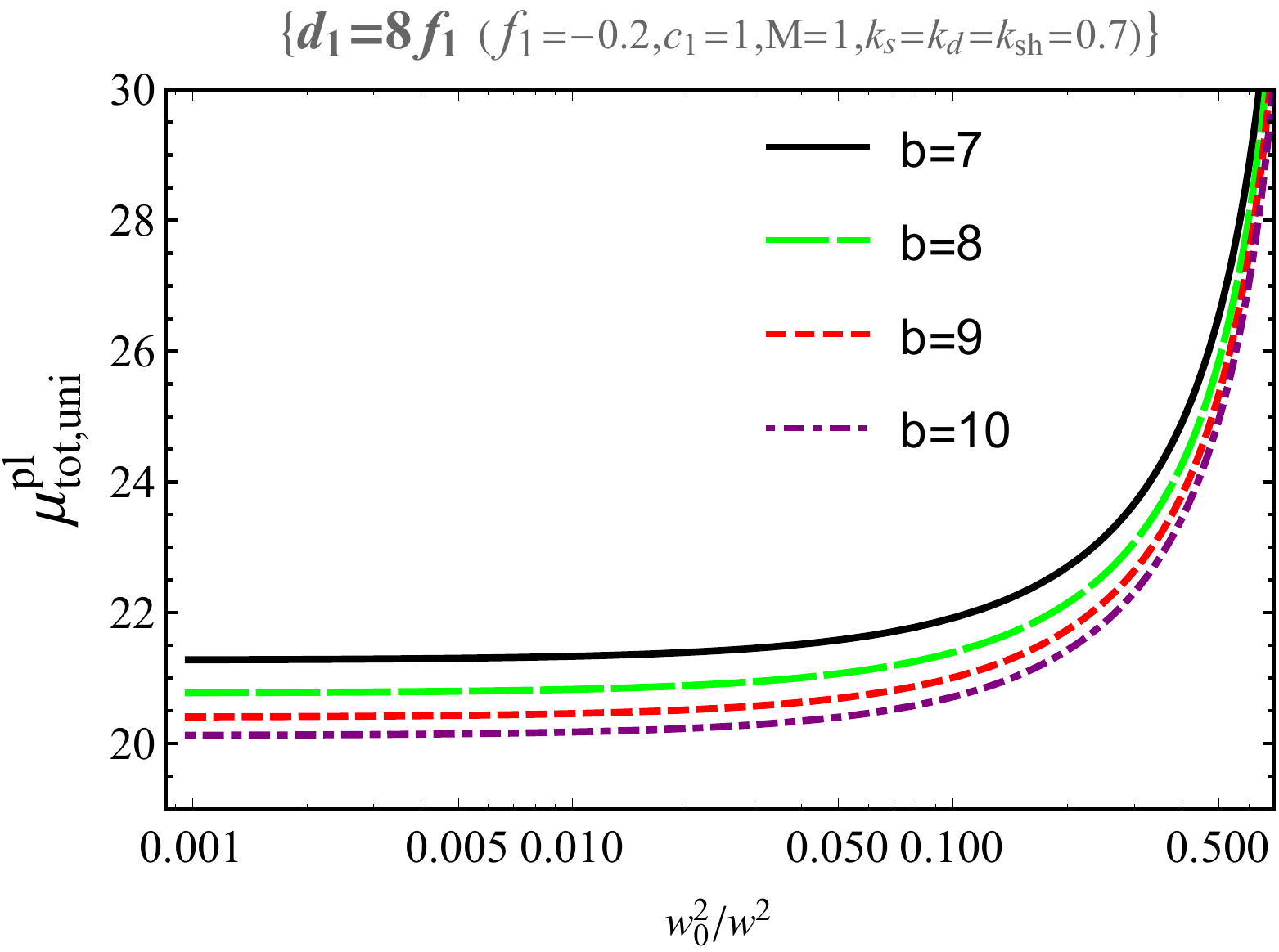}
    \includegraphics[scale=0.53]{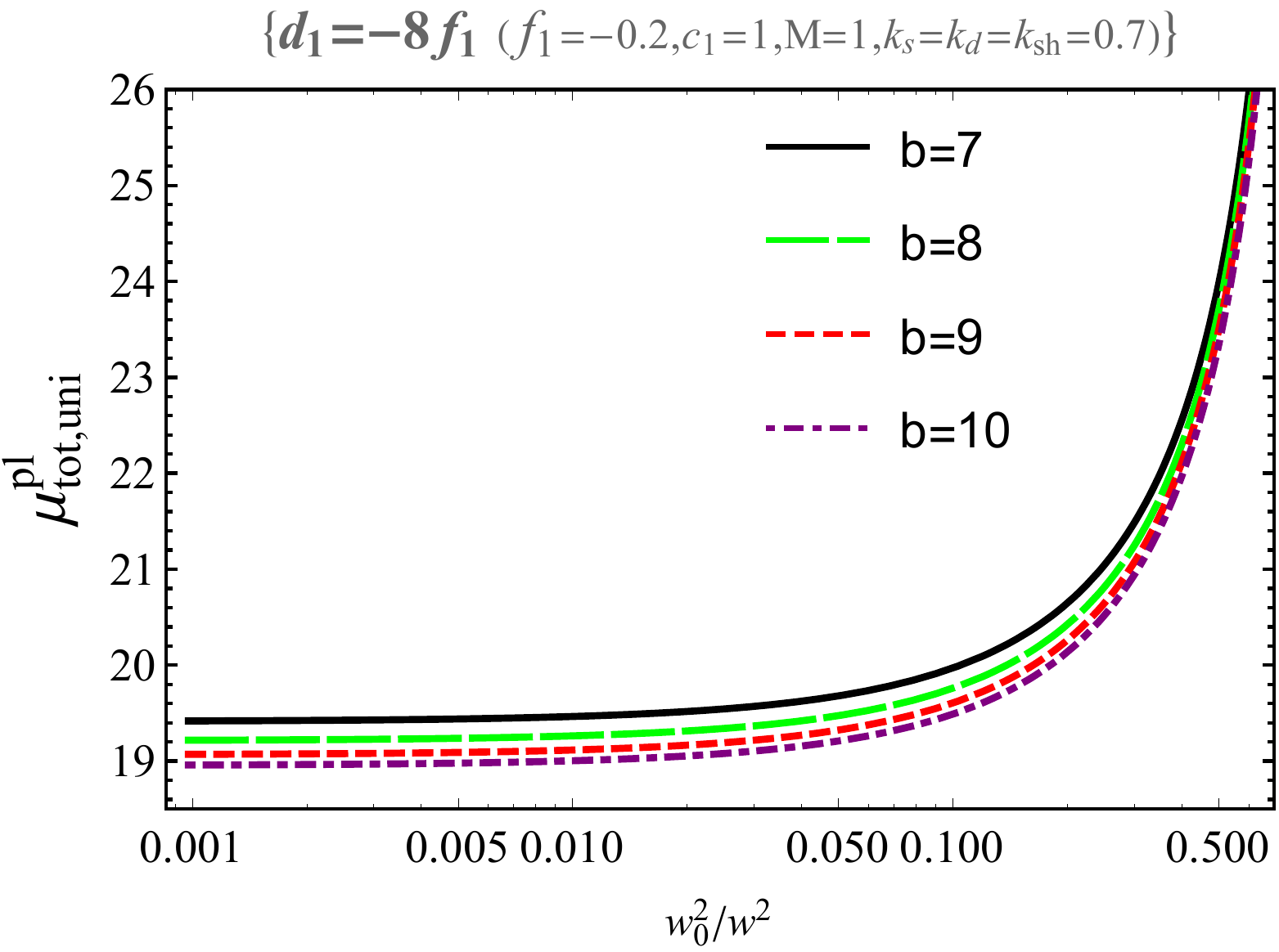}
\caption{Total Magnification of $\hat{\alpha}_{uni}$ for the cases $d_1=8f_1$ (Left panel) and $d_1=-8f_1$ (Right panel) along $c_1$ taking different values of $f_1,\; k_s,\; k_d, \;\&\; k_{sh}.$}
\label{plot:20}
\end{figure}
     \begin{figure}
    \centering
   \includegraphics[scale=0.53]{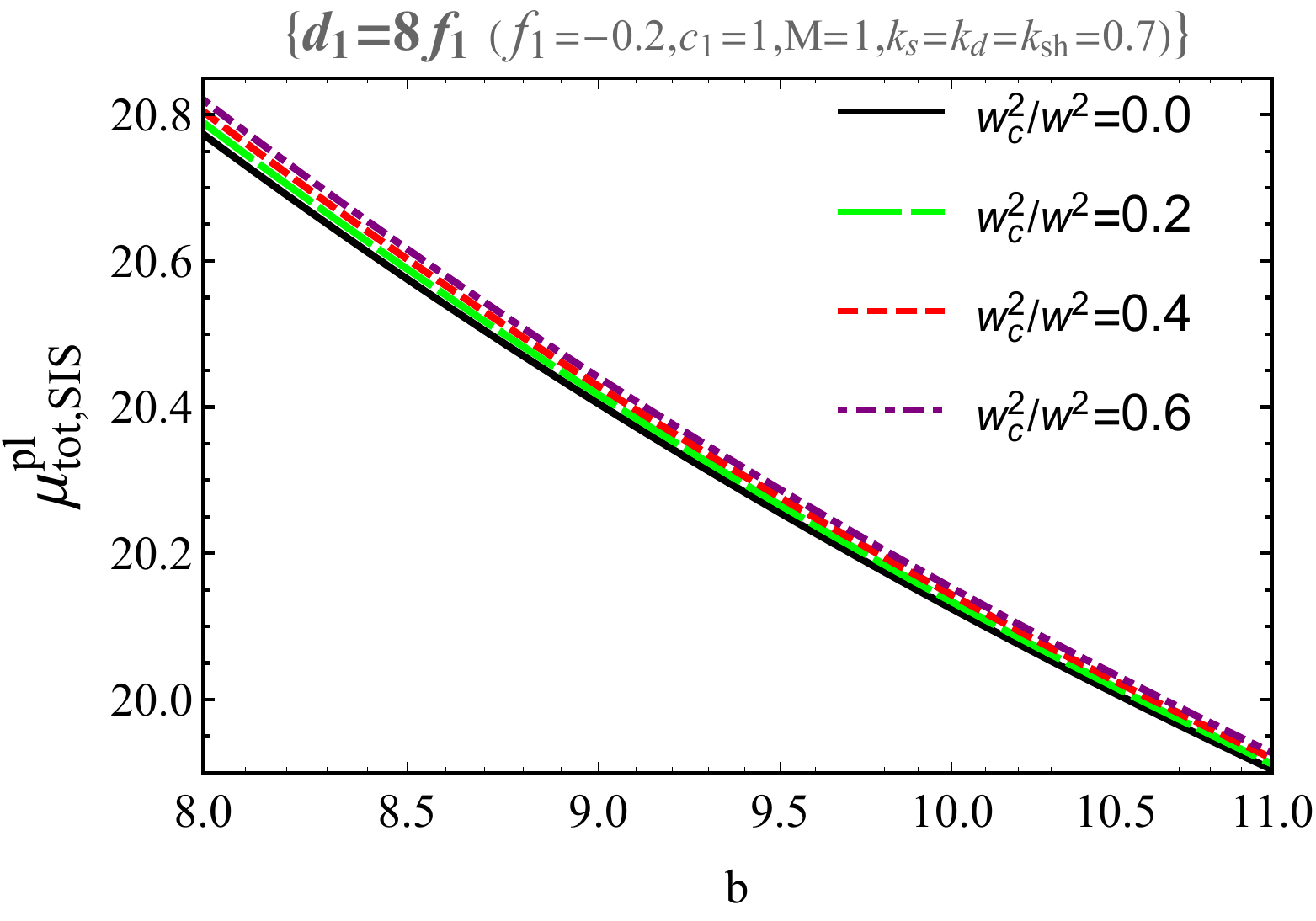}
    \includegraphics[scale=0.53]{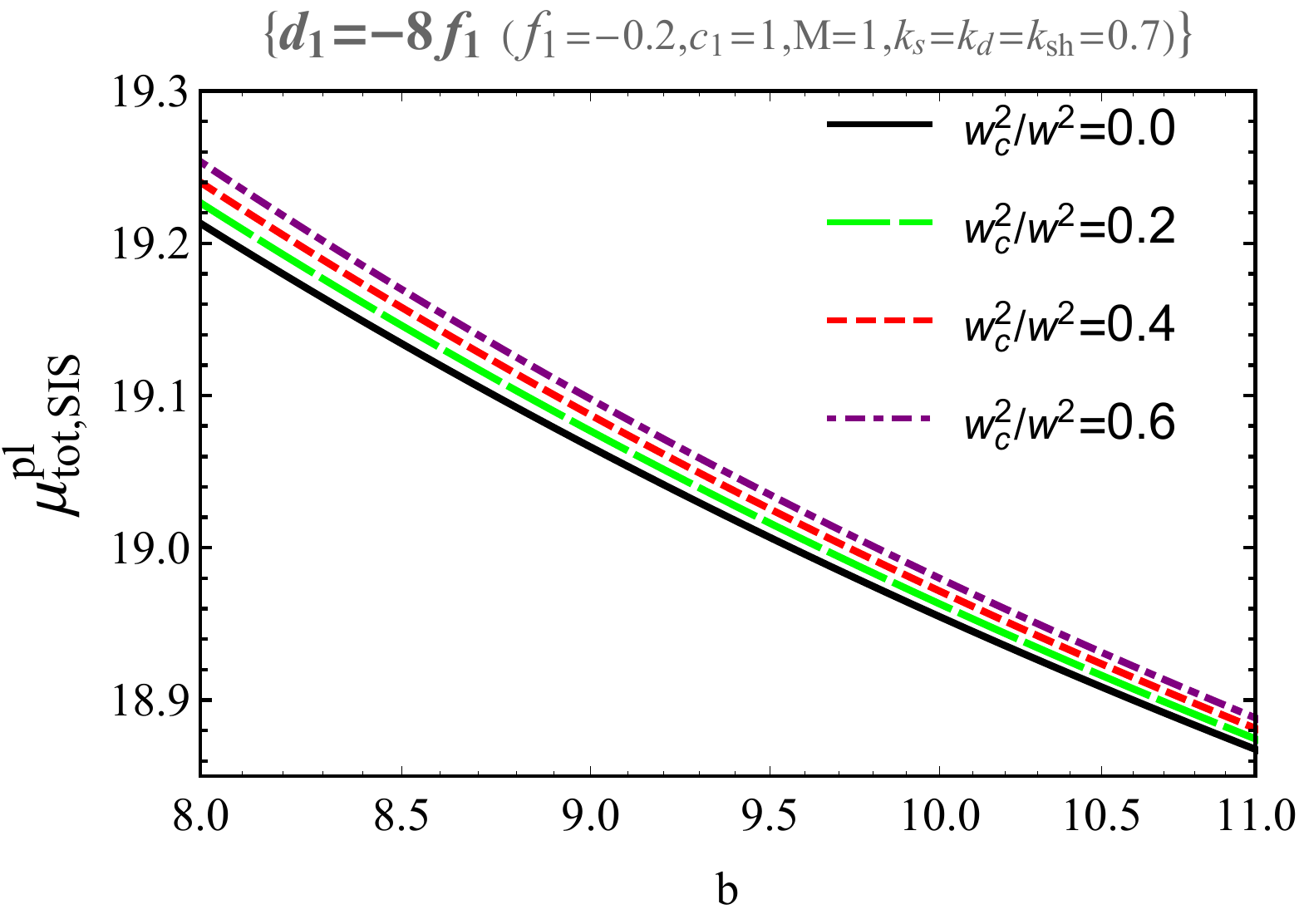}
    \includegraphics[scale=0.53]{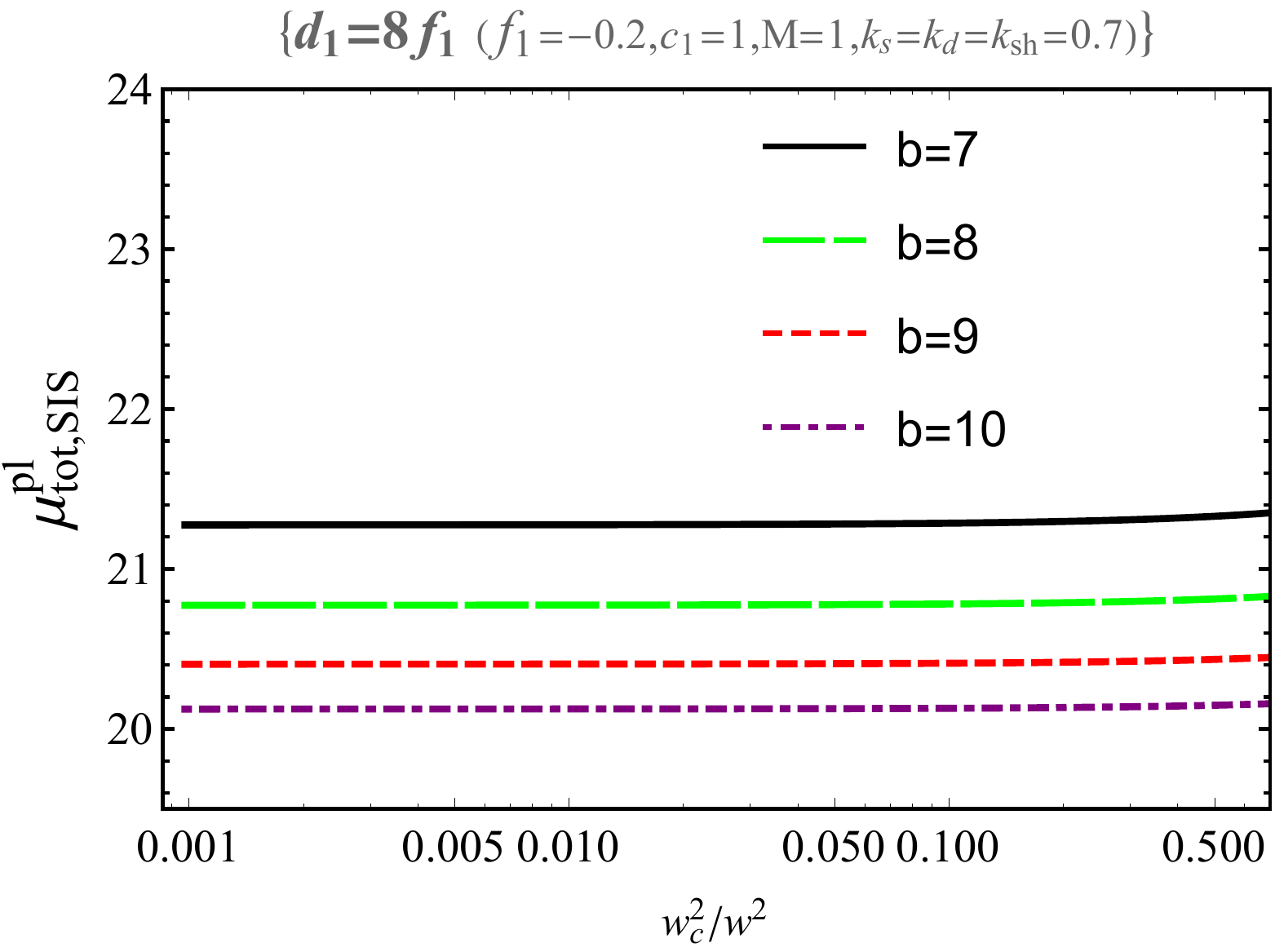}
    \includegraphics[scale=0.53]{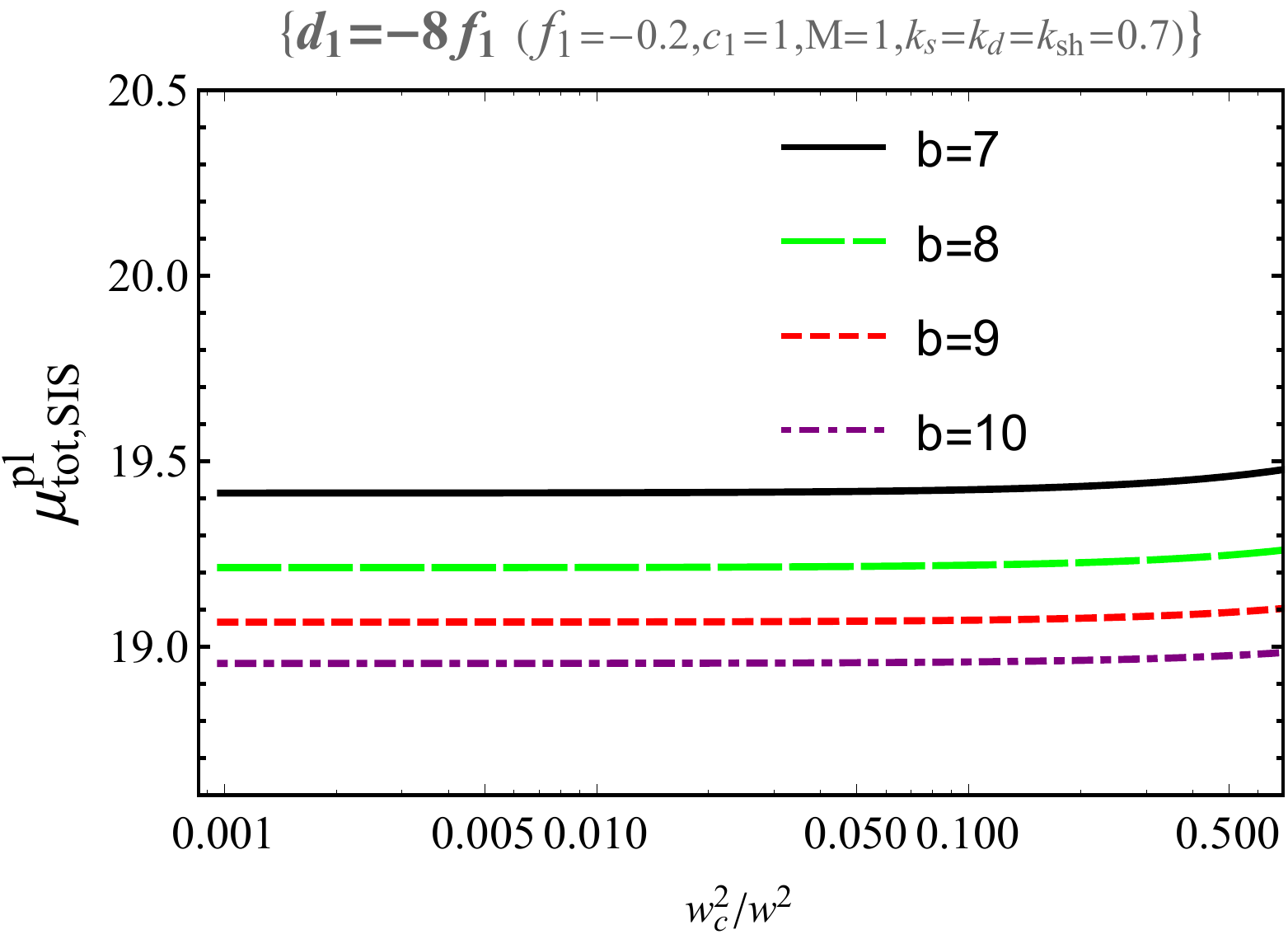}
    \caption{Total Magnification of $\hat{\alpha}_{SIS}$ for the cases $d_1=8f_1$ (Left panel) and $d_1=-8f_1$ (Right panel) along $c_1$ taking different values of $f_1,\; k_s,\; k_d, \;\&\; k_{sh}.$}
    \label{plot:21}
    \end{figure}
     \begin{figure}
    \centering
    \includegraphics[scale=0.53]{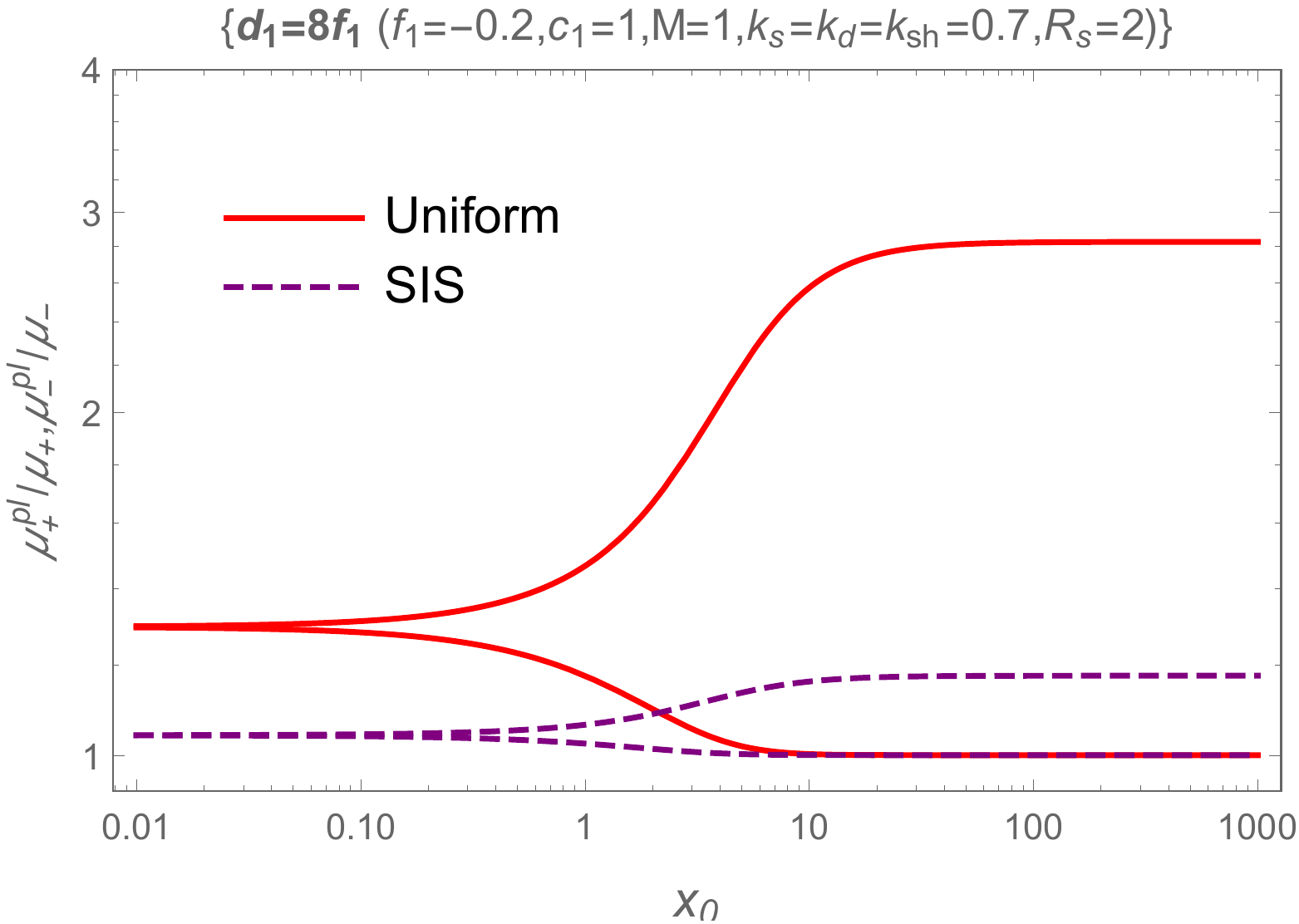}
    \includegraphics[scale=0.53]{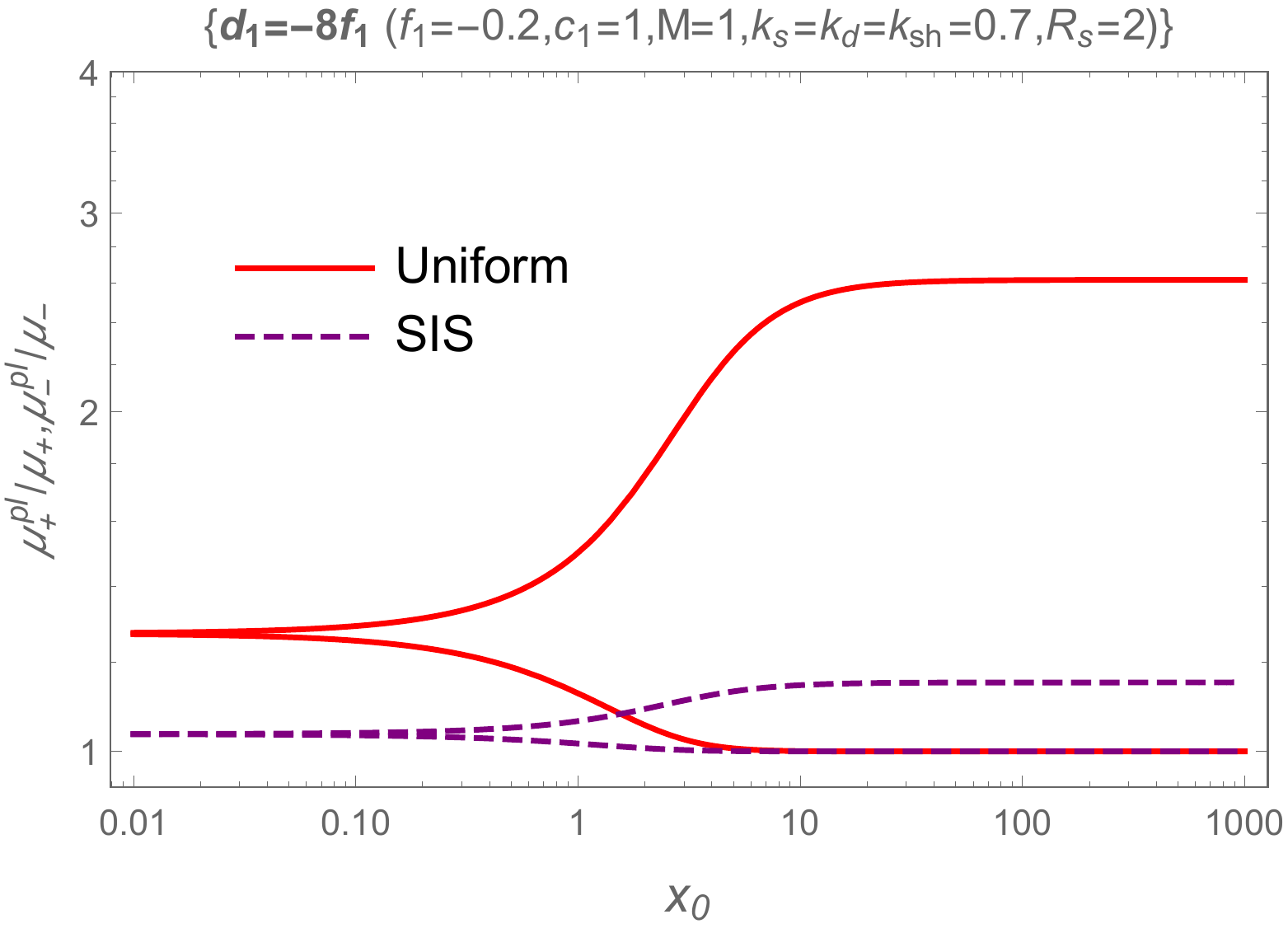}
    \includegraphics[scale=0.53]{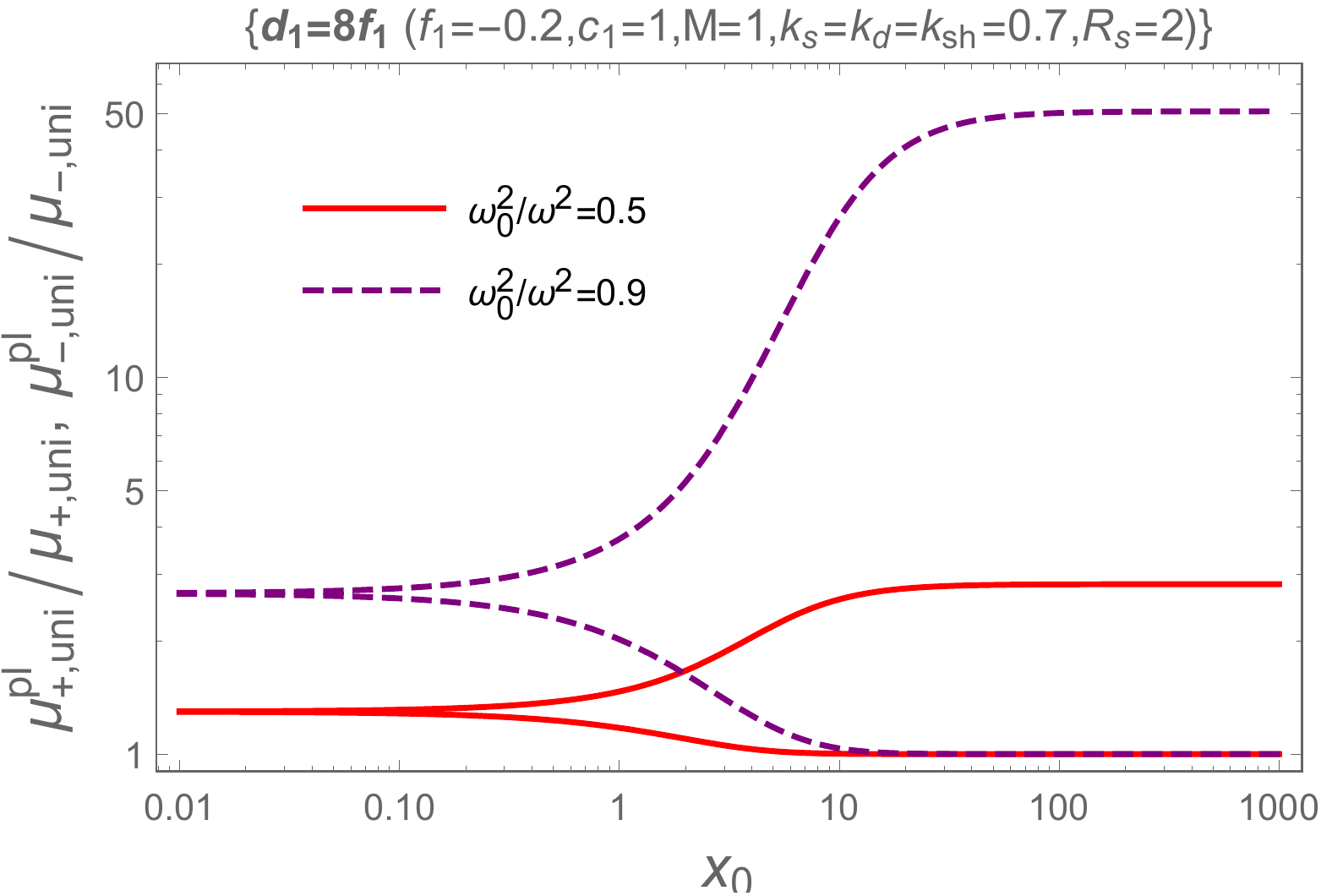}
    \includegraphics[scale=0.53]{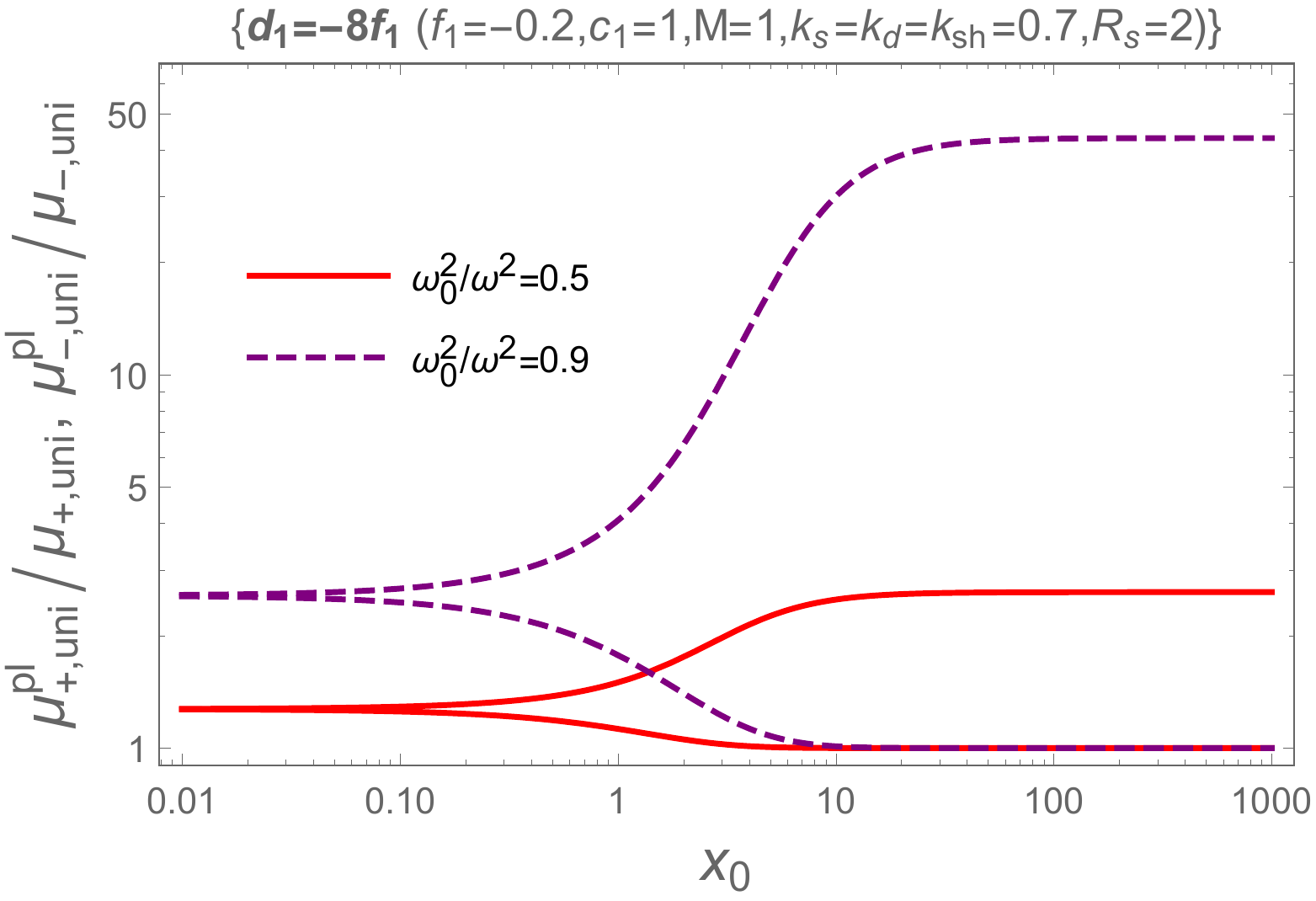}
    \includegraphics[scale=0.53]{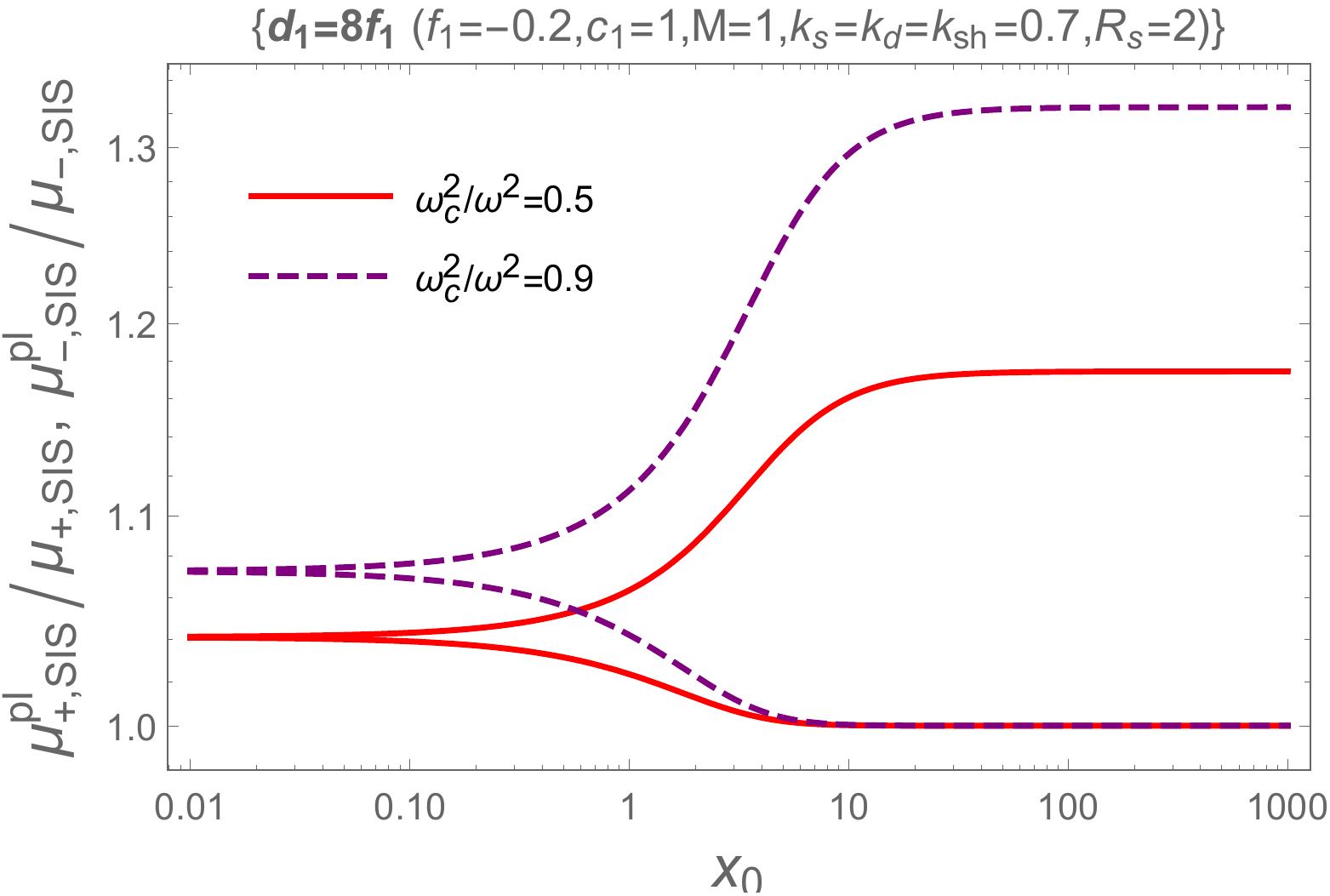}
    \includegraphics[scale=0.53]{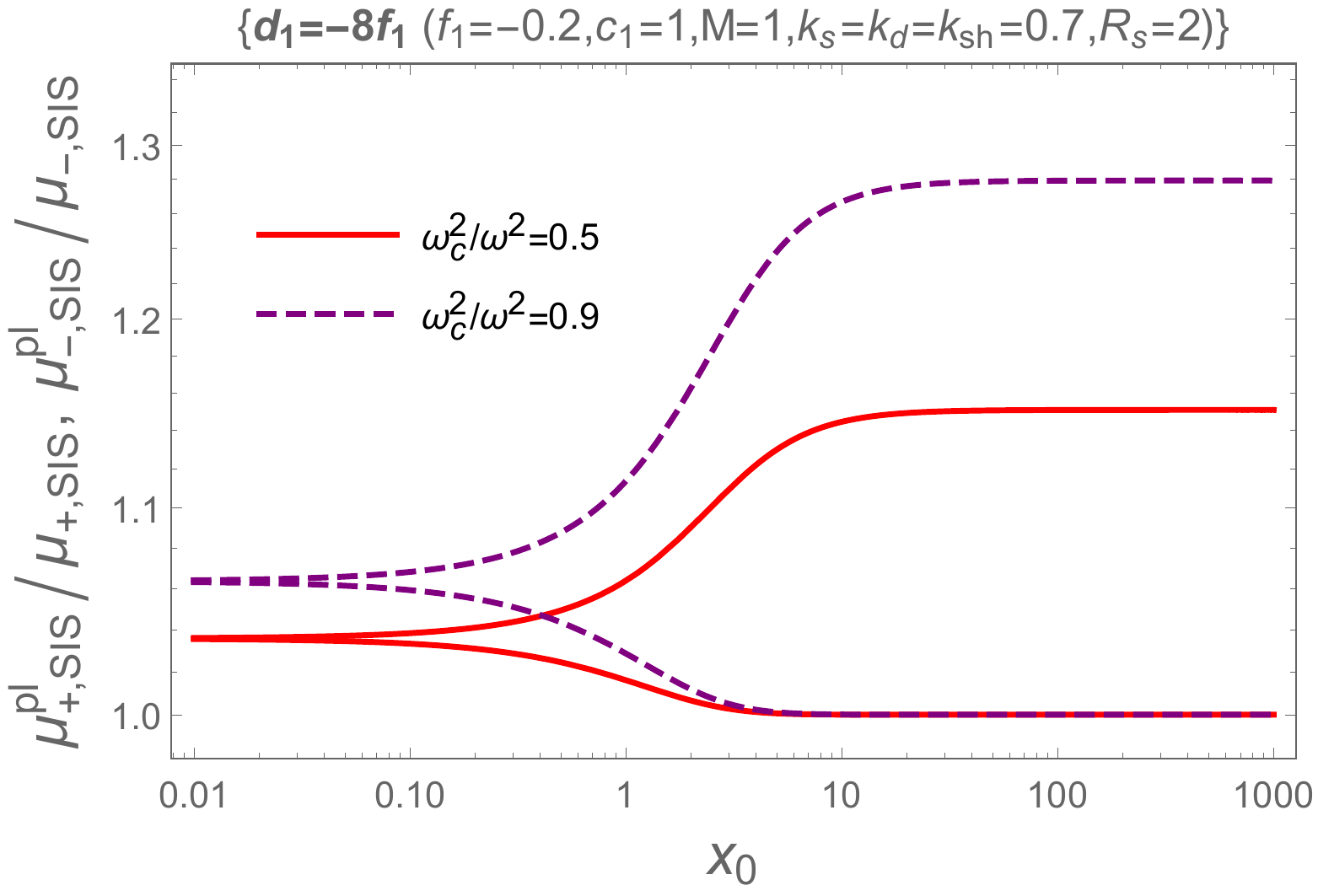}
       \caption{Brightness of source image for $d_1=8f_1$ (Left panel) and $d_1=-8f_1$ (Right panel) along $c_1$ taking different values of $f_1,\; k_s,\; k_d, \;\&\; k_{sh}.$}
    \label{plot:22}
    \end{figure}
    \section{Magnification of gravitationally lensed image}\label{A9}
    In this section, we are interested to study the image magnification and brightness of the image source
 in both uniform and non-uniform $SIS$ plasma fields. Here $D_s,\; D_d,\; \&\;D_{ds}$ denote the distances of the source to observer, lens to observer, and source to lens, respectively. Also,  $\beta\; \&\; \theta$ represent the angular posture of the source and image. Consequently, one can obtain the angular position (gravitational lensing) through the following formula \cite{151,154,155}
 \begin{equation}
    \theta D_{s}=\beta D_{s}+\hat{\alpha}D_{ds},
 \end{equation}
 which can also be written in terms of $\beta$ as
\begin{equation}
    \beta=\theta-\frac{D_{ds}}{D_{s}}\frac{\xi({\theta})}{D_{d}}\frac{1}{\theta},
 \end{equation}
where $\xi(\theta)=|\hat{\alpha}_{b}|b$ and $b = D_{d}\theta$ \cite{151}. The radius of Einstein's ring $R_s=D_d\theta_E$ refers to the radius of circular form of image. The Einstein's angle $\theta_E$ between the source and the images in a vacuum is given  by \cite{156}
\begin{equation}
    \theta_E= \sqrt{2R_s\frac{D_{ds}}{D_dD_s}}.
 \end{equation}
The magnification of brightness can be mathematically expressed as \cite{149,156}
\begin{equation}
    \sum \mu= \frac{I_{tot}}{I_{*}}=\sum_{K}\left|\left(\frac{\theta_k}{\beta}\right)\left(\frac{d\theta_k}{d\beta}\right)\right|,\;\;k=1,2,.....,j,
 \end{equation}
 where $I_{*},\;\&\;I_{tot}$ correspond to the notions of non-lensed brightness of the source and the total brightness of all the images, respectively. Magnification of the source is defined by \cite{146}
 \begin{eqnarray}
    \mu_{+}^{pl}&=& \frac{1}{4}\left(\frac{x}{\sqrt{x^2+4}}+\frac{\sqrt{x^2+4}}{x}+2\right),\label{b1}\\
\mu_{-}^{pl}&=& \frac{1}{4}\left(\frac{x}{\sqrt{x^2+4}}+\frac{\sqrt{x^2+4}}{x}-2\right),\label{b2}
 \end{eqnarray}
 where $x =\beta/\theta_0$ is a dimensionless entity \cite{151} and $\mu_{+}^{pl}$ and $\mu_{-}^{pl}$ correspond to the images in plasma field. We can derive an expression for the total magnification through Eqs.(\ref{b1}) and (\ref{b2}) as
 \begin{eqnarray}
    \mu_{tot}^{pl}=\mu_{+}^{pl}+\mu_{-}^{pl}= \frac{x^2+2}{x\sqrt{x^2+4}}.
 \end{eqnarray}
We intend to explore brightness of the source and the image magnification in the plasma field of MAGBH. To this end, we consider variation in density distribution of plasma field, i.e., $(i)$ uniform plasma and $(ii)$ non-uniform plasma.

\subsection{Uniform Plasma}\label{A9.1}
Here we apply the MAGBH geometry to unveil the influence of uniform plasma on the image magnification. The total magnification ($\mu_{tot}^{pl}$ and total deflection angel $\theta_{uni}^{ pl}$) can be calculated by the formula
\begin{equation}
    \mu_{tot}^{pl}=\mu_{+}^{pl}+\mu_{-}^{pl}=\frac{x_{uni}^2+2}{x_{uni}\sqrt{x_{uni}^2+4}},
 \end{equation}
 with
 \begin{eqnarray}
    \left(\mu_{+}^{pl}\right)_{uni}&=& \frac{1}{4}\Big[\frac{x_{uni}}{\sqrt{x_{uni}^2+4}}+\frac{\sqrt{x_{uni}^2+4}}{x}+2\Big],\label{b1}\\
\left(\mu_{-}^{pl}\right)_{uni}&=& \frac{1}{4}\Big[\frac{x_{uni}}{\sqrt{x_{uni}^2+4}}+\frac{\sqrt{x_{uni}^2+4}}{x_{uni}}-2\Big],\label{b2}
 \end{eqnarray}

\begin{equation}
    x_{uni}=\frac{\beta}{\left(\theta_{E}^{pl}\right)_{uni}},\;\; \theta^{pl}_{uni}=\theta _E \sqrt{\frac{b \alpha _b}{2 R_s}},
 \end{equation}
and
\begin{eqnarray}
    x_{uni}&=&x_0 \Big[\frac{b}{2 R_s} \Big[\frac{\pi  \sqrt{\frac{1}{b^2}} c_1 k_{d}^2}{b}+\frac{3 \pi  \sqrt{\frac{1}{b^2}} c_1^2 k_{d}^4}{b^3}+\frac{\pi  \sqrt{\frac{1}{b^2}} f_1 k_{sh}^2}{2 b}+\frac{3 \pi  \sqrt{\frac{1}{b^2}} c_1 f_1 k_{d}^2 k_{sh}^2}{b^3}+\frac{3 \pi  \sqrt{\frac{1}{b^2}} f_1^2 k_{sh}^4}{4 b^3}\nonumber\\&-&\frac{\pi  \sqrt{\frac{1}{b^2}} d_1 k_{s}^2}{4 b}-\frac{3 \pi  \sqrt{\frac{1}{b^2}} c_1 d_1 k_{d}^2 k_{s}^2}{2 b^3}-\frac{3 \pi  \sqrt{\frac{1}{b^2}} d_1 f_1 k_{sh}^2 k_{s}^2}{4 b^3}+\frac{3 \pi  \sqrt{\frac{1}{b^2}} d_1^2 k_{s}^4}{16 b^3}+\frac{R}{b}+\frac{16 c_1 k_{d}^2 R}{3 b^3}\nonumber\\&+&\frac{8 f_1 k_{sh}^2 R}{3 b^3}-\frac{4 d_1 k_{s}^2 R}{3 b^3}-\frac{1}{1-\frac{w_0^2}{w^2}}\Big[-\frac{16 c_1 k_{d}^2 R}{b^3}+\frac{4 d_1 k_{s}^2 R}{b^3}-\frac{8 f_1 k_{sh}^2 R}{b^3}-\frac{2 \pi  \sqrt{\frac{1}{b^2}} c_1 k_{d}^2}{b}\nonumber\\&+&\frac{\pi  \sqrt{\frac{1}{b^2}} d_1 k_{s}^2}{2 b}-\frac{\pi  \sqrt{\frac{1}{b^2}} f_1 k_{sh}^2}{b}-\frac{12 \pi  \sqrt{\frac{1}{b^2}} c_1^2 k_{d}^4}{b^3}+\frac{6 \pi  \sqrt{\frac{1}{b^2}} c_1 d_1 k_{d}^2 k_{s}^2}{b^3}-\frac{12 \pi  \sqrt{\frac{1}{b^2}} c_1 f_1 k_{d}^2 k_{sh}^2}{b^3}\nonumber\\&-&\frac{3 \pi  \sqrt{\frac{1}{b^2}} d_1^2 k_{s}^4}{4 b^3}+\frac{3 \pi  \sqrt{\frac{1}{b^2}} d_1 f_1 k_{sh}^2 k_{s}^2}{b^3}-\frac{3 \pi  \sqrt{\frac{1}{b^2}} f_1^2 k_{sh}^4}{b^3}-\frac{R}{b}\Big]\Big]\Big]^{-\frac{1}{2}},
 \end{eqnarray}
 where $x_0=\beta/\theta_E$. We provide a graphical analysis for the total magnification of uniform plasma as shown in Fig. (\ref{plot:20}. Magnification of the image source in $uni$-plasma field is also plotted in the middle panel of Fig. (\ref{plot:22}). It can be observed that magnification significantly rises with more concentration of plasma field.

 \subsection{Non-uniform plasma}\label{A9.2}
 Following the same procedure as mentioned above, one can explore the effects of $SIS$ plasma on the image magnification. We find the total magnification $\mu_{tot}^{pl}$ and total deflection angle $\theta_{SIS}^{ pl}$ for $SIS$ plasma field by the formula
 \begin{equation}
    \mu_{tot}^{pl}=\mu_{+}^{pl}+\mu_{-}^{pl}=\frac{x_{SIS}^2+2}{x_{SIS}\sqrt{x_{SIS}^2+4}},
 \end{equation}
 with
 \begin{eqnarray}
    \left(\mu_{+}^{pl}\right)_{SIS}&=& \frac{1}{4}\Big[\frac{x_{SIS}}{\sqrt{x_{SIS}^2+4}}+\frac{\sqrt{x_{SIS}^2+4}}{x}+2\Big],\label{b1}\\
\left(\mu_{-}^{pl}\right)_{SIS}&=& \frac{1}{4}\Big[\frac{x_{SIS}}{\sqrt{x_{SIS}^2+4}}+\frac{\sqrt{x_{SIS}^2+4}}{x_{SIS}}-2\Big],\label{b2}
 \end{eqnarray}

\begin{equation}
    x_{SIS}=\frac{\beta}{\left(\theta_{E}^{pl}\right)_{SIS}},\;\; \theta^{pl}_{uni}=\theta _E \sqrt{\frac{b \alpha _b}{2 R_s}},
 \end{equation}
and
\begin{eqnarray}
    x_{SIS}&=&x_0 \Big[\frac{b}{2 R_s} \Big[\frac{1}{240 \pi  b^5}\Big[\sqrt{\frac{1}{b^2}} \Big[45 \pi  b^2 \left(4 c_{1} k_{d}^2-d_{1} k_{s}^2+2 f_{1} k_{sh}^2\right) \left(\frac{2 R^2 w_c^2}{w^2}+20 \pi  c_{1} k_{d}^2-5 \pi  d_{1} k_{s}^2+10 \pi  f_{1} k_{sh}^2\right)\nonumber\\&+&\frac{6 R^2 w_c^2 \left(4 c_{1} k_{d}^2-d_{1} k_{s}^2+2 f_{1} k_{sh}^2\right) \left(100 \pi  \sqrt{\frac{1}{b^2}} c_{1} k_{d}^2-25 \pi  \sqrt{\frac{1}{b^2}} d_{1} k_{s}^2+50 \pi  \sqrt{\frac{1}{b^2}} f_{1} k_{sh}^2+128 R\right)}{\sqrt{\frac{1}{b^2}} w^2}\nonumber\\&+&480 \pi  \sqrt{\frac{1}{b^2}} b^6 R+20 b^4 \Big[\frac{8 \sqrt{\frac{1}{b^2}} R^3 w_c^2}{w^2}+4 \pi  c_{1} k_{d}^2 \left(64 \sqrt{\frac{1}{b^2}} R+9 \pi \right)-64 \pi  \sqrt{\frac{1}{b^2}} d_{1} k_{s}^2 R+2 \pi  f_{1} k_{sh}^2 \nonumber\\&&\left(64 \sqrt{\frac{1}{b^2}} R+9 \pi \right)-9 \pi ^2 d_{1} k_{s}^2\Big]\Big]\Big]\Big]\Big]^{-\frac{1}{2}},
 \end{eqnarray}
where $x_0 =\beta/\theta_E$. We plot the total image magnification in $SIS$ plasma in Fig.~\ref{plot:21} and in the lower panel of Fig.~\ref{plot:22}. We find that magnification increases in a higher concentration of $SIS$ plasma field. It is interesting to notice that the image magnification in uniform plasma is much higher as compared to the $SIS$ plasma field (see the upper panel of Fig.~\ref{plot:22}).

\section{Conclusion and summary}\label{A10}

The study in this article is equipped with the optical characteristics of BH in the paradigm of MAGBH geometry. This geometry is supplemented by the shear charge $k_{sh}$, spin charge $k_s$, dilation charge $k_d$, constant parameter $c_1$ and Lagrangian coefficients $d_1\;\&\;f_1$. We attempted to unveil the comprehensive impacts of these parameters on the optical features of MAGBH. We have computed horizon radius, inner stable circular orbit, photon sphere radius, BH shadows, quasi-periodic oscillations, the red-blue shift of photon particles, effective force, weak gravitational lensing, and image magnification. The results can be summarized as follows:

\begin{itemize}
    \item Figs.~\ref{plot:1}, \ref{plot:2} and \ref{plot:3} represent the horizon radius $r_h$, inner stable circular orbit $r_{ISCO}$ and photo orbit $r_{ph}$, respectively. One can examine the rising radial behavior along $c_1$ by increasing $f_1,\;k_s,\;k_d$ which declines with an increase in $k_{sh}$ for $d_1=8f_1$. On the other hand, a similar trend can be observed in the radial propagation by increasing $k_d$ and $f_1,\;k_s,\;k_{sh}$ for $d_1=-8f_1$. It is important to note that $r_{ISCO}>r_{ph}>r_h$.

    \item We have also plotted the shadows radius $R_{sh}$ as shown in Figs.~\ref{plot:4} and \ref{plot:5}. It can apparently be seen that an increase in shadow radius occurs along $c_1$ by increasing $f_1,\;k_d,\;\&\;k_s$. The respective radius tends to decrease by increasing $k_{sh}$ and $f_1,\;k_s,\;k_{sh}$ for the cases $d_1=8f_1$ and $d_1=-8f_1$, respectively.

    \item We plot the radial and tangential frequencies (Fig.~\ref{plot:6}). In Fig.~\ref{plot:7}, we plot the upper and lower frequencies of RP model while ER model frequencies are given in Figs.~\ref{plot:8}, \ref{plot:9}, and \ref{plot:10}. These graphical illustrations indicate variations in the frequencies by varying the parameters $f_1,\;c_1,\; d_1,\;k_d,\;k_s,\;k_{sh}$.

    \item The red-blue shift for both $d_1=8f_1\;\&\;d_1=-8f_1$ cases is plotted in Fig.~\ref{plot:11} for which an observer may find different horizons for the parameters  $f_1,\; d_1,\;k_d,\;k_s,\;k_{sh}$ along $c_1$.
    \item We have plotted the attractive behavior of the test particle through the effective force in Fig.~\ref{plot:12} along the BH radius in MAG.
    \item The deflection angle $\hat{\alpha}_b$ is also graphically illustrated for three plasma states, i.e., uniform plasma, $SIS$ plasma, and $NSIS$ (Figs.~\ref{plot:13}-\ref{plot:19}) whose outcomes yield
    \begin{itemize}
        \item   \textbf{The deflection angle $\hat{\alpha}_b$ along $b$:} ($i$) increases by increasing $f_1,\;c_1,\;k_d,\;k_s,\;\&\;w_0^2/w^2$ while it decreases with increase in $k_{sh}$ for $d_1=8f_1$, and ($ii$) rises with an increase in $c_1,\;k_d,\;\&\;w_0^2/w^2$ and falls with $f_1,\;k_s,\;k_{sh}$ for the case $d_1=-8f_1$.

   \item  \textbf{The deflection angle $\hat{\alpha}_b$ along $c_1$:} also has a direct relation with $f_1,\;k_d,\;k_s,\;\&\;w_0^2/w^2$ and $k_d,\;\&\;w_0^2/w^2$ for the cases $d_1=8f_1$ and $d_1=-8f_1$ but an inverse impact appears through $b,\;\&\;k_{sh}$ and $f_1,\;b,\;k_{s},\;k_{sh}$ for both cases.

  \item   \textbf{Furthermore, the deflection angle $\hat{\alpha}_b$ along $w_0^2/w^2$ (uniform case), $w_c^2/w^2$ ($SIS\;\&\;NSIS$):} increases with $f_1,\;c_1,\;k_d,\;k_{s}$ while it decreases with increasing values of $b\;\&\;k_{sh}$ in the case $d_1=8f_1$. Similar is the behavior for the case $d_1=-8f_1$.
    \end{itemize}
    It is necessary to emphasize that $\hat{\alpha}_{uni}>\hat{\alpha}_{SIS}>\hat{\alpha}_{NSIS}$.
    \item From lens equation the observed quantity of the Einstein angle and magnifications of images are also obtained in the Metfic-Affine gravity. We plot magnification of the image through deflection angle (for the uniform and non-uniform plasma cases) of the light rays in Figs.~\ref{plot:20}-\ref{plot:22}.
\end{itemize}

\section*{Acknowledgement}

The paper was funded by the National Natural Science Foundation of China 11975145. This research is partly supported by Research Grant F-FA-2021-510 of the Uzbekistan Ministry for Innovative Development.


\section*{Data Availability Statement} This manuscript has no associated data, or the data will not be deposited.
(There is no observational data related to this article. The
necessary calculations and graphic discussion can be made available
on request.)

\end{document}